# ABSTRACT NULL HYPERSURFACES AND CHARACTERISTIC INITIAL VALUE PROBLEMS IN GENERAL RELATIVITY

---

*Autor:*
Gabriel SÁNCHEZ PÉREZ

*Director:*
Prof. Marc MARS LLORET

Tesis doctoral presentada
para la obtención del título de
Doctor en Física Fundamental y Matemáticas

Departamento de Física Fundamental
Instituto Universitario de Física Fundamental y Matemáticas
Facultad de Ciencias

Universidad de Salamanca
Salamanca, España

Enero 2026

Autor:
Gabriel Sánchez Pérez

Director:
Prof. Marc Mars Lloret

Universidad de Salamanca, Departamento de Física Fundamental e IUFFyM

# CONTENTS

# ABSTRACT

This thesis is framed within the field of Mathematical Relativity and is organized into six chapters. After an introduction to the topic in Chapter 1, Chapter 2 reviews and further develops the formalism of hypersurface data, which provides the unifying framework for the entire thesis. In Chapter 3 we study the characteristic Cauchy problem from a fully detached perspective. Chapter 4 is devoted to the Killing initial data problem, also analyzed within this detached framework. In Chapter 5 we investigate the transverse (or asymptotic) expansion of the metric at a general null hypersurface. Finally, Chapter 6 addresses the geometry of conformal null infinity.

The hypersurface data formalism allows one to describe hypersurfaces of arbitrary causal character without the need of being embedded in any ambient manifold. This detached viewpoint is particularly well-suited for the formulation of initial value problems in General Relativity, as it allows the study of the initial data without a priori reference to the spacetime to be constructed. Within this framework, Chapter 3 provides a geometrization of the characteristic Cauchy problem via the notion of double null data, while Chapter 4 develops a general treatment of Killing and homothetic initial data on hypersurfaces of any causal character, with particular emphasis on null and characteristic settings.

When only a single null hypersurface is available, establishing existence and uniqueness of solutions to the Einstein equations becomes a subtle problem. This issue is analyzed in detail in Chapter 5, where general identities governing the transverse expansion of the metric at null hypersurfaces are derived and applied, in particular, to the study of Killing horizons. Finally, Chapter 6 considers the case in which the null hypersurface represents conformal null infinity. In this setting, we analyze the conformal Einstein equations in arbitrary dimension and characterize the free data at null infinity. We also find a conformal completion of the Fefferman–Graham ambient metric and analyze its asymptotic properties.

# DECLARACIÓN DEL SUPERVISOR

**Dr. D. Marc Mars Lloret**, Catedrático de Física Teórica en el Departamento de Física Fundamental de la Universidad de Salamanca,

CERTIFICA:

Que el trabajo de investigación que se recoge en la siguiente memoria titulada *Abstract null hypersurfaces and characteristic initial value problems in General Relativity*, presentada por D. Gabriel Sánchez Pérez para optar al Título de Doctor por la Universidad de Salamanca con la Mención de Doctorado Internacional, ha sido realizada en su totalidad bajo su dirección y autoriza su presentación.

En Salamanca, a 25 de febrero de 2026

Dr. D. Marc Mars Lloret,
Catedrático de Física Teórica,
Departamento de Física Fundamental,
Universidad de Salamanca.

# DECLARACIÓN SOBRE EL USO DE HERRAMIENTAS DE IA

**D. Gabriel Sánchez Pérez** declara explícitamente que en la elaboración de la presente tesis doctoral no se ha hecho uso de herramientas de Inteligencia Artificial generativa ni de sistemas automatizados para la redacción de textos, el desarrollo de argumentos científicos, la obtención de resultados, ni el análisis o interpretación de los mismos.

Asimismo, no se ha empleado Inteligencia Artificial para la toma de decisiones metodológicas ni para la producción de contenidos originales que formen parte del trabajo académico presentado. La totalidad de la investigación, incluyendo el planteamiento teórico, el desarrollo matemático, el análisis de resultados y la redacción del manuscrito, ha sido realizada directamente por el doctorando, respetando los principios éticos y de integridad académica establecidos por la normativa vigente.

En Salamanca, a 25 de febrero de 2026

D. Gabriel Sánchez Pérez
Departamento de Física Fundamental,
Universidad de Salamanca.

# PUBLICATIONS

This thesis is based on the following research papers [231–237].

1. M. Mars and G. Sánchez-Pérez, "Double null data and the characteristic problem in General Relativity", *Journal of Physics A: Mathematical and Theoretical*, vol. **56**, 2023.
2. M. Mars and G. Sánchez-Pérez, "Covariant definition of double null data and geometric uniqueness of the characteristic initial value problem", *Journal of Physics A: Mathematical and Theoretical*, vol. **56**, 2023.
3. M. Mars and G. Sánchez-Pérez, "Transverse expansion of the metric at null hypersurfaces I. Uniqueness and application to Killing horizons", *Journal of Geometry and Physics*, vol. **209**, 2025.
4. M. Mars and G. Sánchez-Pérez, "Transverse expansion of the metric at null hypersurfaces II. Existence results and application to Killing horizons", *Journal of Geometry and Physics*, vol. **217**, 2025.
5. M. Mars and G. Sánchez-Pérez, "Killing and homothetic initial data for general hypersurfaces", *Classical and Quantum Gravity*, vol. **42**, 2025.
6. M. Mars and G. Sánchez-Pérez, "Conformal characterization of the Fefferman-Graham ambient metric", accepted for publication in *Classical and Quantum Gravity*, arXiv:2510.21646.
7. M. Mars and G. Sánchez-Pérez, "Transverse expansion of the metric at null infinity", arXiv:2602.0506.

# AGRADECIMIENTOS

Quiero comenzar agradeciendo a mi director de tesis, Marc, su apoyo constante durante todos estos años. Gracias por tu paciencia, tu curiosidad infinita, tu dedicación y, sobre todo, por transmitirme esa pasión por el trabajo riguroso, minucioso y bien hecho. Sin ti, esta tesis doctoral no habría sido posible. También quiero agradecer a todas las personas del Departamento de Física Fundamental y del IUFFyM que me han acompañado durante este tiempo, y en particular a los miembros de mi grupo de investigación, GRACUS, con quienes he compartido tantos buenos momentos, no solo científicos. A Jose, Flor, Ivan, Fernando, Mario, Ana, y a tantos otros que inevitablemente me dejo en el tintero: gracias.

I am also deeply grateful to the Mathematical Physics group at the University of Edinburgh for their hospitality during my research stay, in particular to James, David, Virinchi, Gerben and Jelle. I would also like to thank all the researchers who have shown interest in my work and invited me to give seminars during my Ph.D. To all of them, thank you.

Esta tesis tampoco habría sido posible sin los amigos que he hecho durante el doctorado. En primer lugar, gracias a Miguel, Paz, Carlos, David Figueruelo, David Barba, Riccardo y Duvier por haber sido un apoyo fundamental durante mis primeros años de tesis. Gracias también a Pablo, Jorge, Samir, Liam y David por acogerme como a un matemático más, primero durante el máster y después durante la tesis. ¿Tenéis libre este finde para unos juegos de mesa? Y muchas gracias también a las nuevas generaciones de doctorandos, Alberto, Óscar, Jose, Diego, Manu y María. Dejo la facultad en muy buenas manos. ¡Ah! Por cierto, habéis perdido el juego. Quiero agradecer asimismo a los grandes amigos que la Física y las Matemáticas me han regalado durante todos estos años, como Dani, Albino, Marina y Sandra. Gracias por los viajes, las cenas, las fiestas, los festivales y todos los momentos compartidos. Gracias también a Ángel por nuestras rutas en bici, a Marcos por nuestros martes de Doctor Who, y a los amigos que conservo del San Agus, en especial a Vega.

Gracias, sobre todo, a mi familia, por el apoyo incondicional que me han brindado desde que, siendo muy pequeño, surgió en mí la curiosidad por la Física y las Matemáticas. Gracias a la mejor hermana que la vida me ha podido dar, Victoria; a mi madre, Rosa; a mi padre, Juan; y a todo el resto de mi familia. Quiero recordar de manera especial a mi tío Rafa, a mi abuelo y a mi abuela, quienes, por desgracia, no han podido verme convertirme en doctor. Y, en especial, gracias a ti, Sara, por tu bondad, tu cariño infinito, tu paciencia y por dejarme formar parte de tu vida. A todos vosotros, muchísimas gracias.

Finalmente, agradezco el apoyo económico del Ministerio de Ciencia, Innovación y Universidades a través del contrato FPU20/03751, así como de los proyectos de investigación PGC2018-096038-B-I00, PID2021-122938NB-I00 y PID2024-158938NB-I00.

*Sin ser*
*Me vuelvo duro como una roca*
*Si no puedo acercarme*
*Ni oir*
*Los versos que me dicta esa boca*
*Y ahora que ya no hay nada*
*Ni dar*
*La parte de dar que a mí me toca*
*Por eso no he dejado de andar*

La Ley Innata, Cuarto Movimiento,
Robe Iniesta

*A Sara*

# ABBREVIATIONS, NOTATION AND LIST OF SYMBOLS

## ABBREVIATIONS

Throughout this thesis the following abbreviations are employed.

- **CHD**. Characteristic hypersurface data, Definition 2.41.
- **CG**. Characteristic gauge, Definition 2.43.
- **DND**. Double null data, Definition 3.9.
- **KID**. Killing initial data, Chapter 4.
- **AKH data**. Abstract Killing horizon data, Definition 5.58.
- **EKH data**. Embedded Killing horizon data, Definition 5.61.
- **LHS** and **RHS** means left and right hand side, respectively.
- **w.r.t.** stands for "with respect to".

## NOTATION

Unless otherwise indicated, all manifolds in this thesis are assumed to be connected and smooth, and all embedded hypersurfaces are assumed to be two-sided. Depending on convenience, both index-free and abstract index notation are used to describe tensorial expressions. Spacetime indices are written using Greek letters $(\alpha, \beta, \dots)$, indices on a hypersurface are written in lowercase Latin letters from the beginning of the alphabet $(a, b, \dots, h)$, and indices at cross-sections of a hypersurface are written in uppercase Latin letters $(A, B, \dots)$. The indices $(i, j, k, \cdots)$ are employed in sums and as labels for normal pairs (Section 2.4). We use the symbol $d$ for the spacetime dimension, $\mathfrak{n} = d - 1$ for the dimension of a hypersurface, and $r = d - 2$ for the dimension of a codimension-two surface. As usual, square brackets enclosing indices indicate antisymmetrization, while parentheses indicate symmetrization. The symmetrized tensor product is written as $\otimes_s$. By $\mathcal{F}(\mathcal{M})$, $\mathfrak{X}(\mathcal{M})$ and $\mathfrak{X}^\star(\mathcal{M})$ we refer to, respectively, the sets of smooth functions, vector fields and one-forms on a manifold $\mathcal{M}$. The subset $\mathcal{F}^\star(\mathcal{M}) \subset \mathcal{F}(\mathcal{M})$ consists of nowhere vanishing functions on $\mathcal{M}$. For the tangent and cotangent spaces at a point we employ the usual notation $T_p\mathcal{M}$ and $T_p^\star\mathcal{M}$, while the corresponding bundles are written as $T\mathcal{M}$ and $T^\star\mathcal{M}$. Given an embedding $\iota : \mathcal{S} \hookrightarrow \mathcal{H}$, we say that $V$ is a vector field along $\iota(\mathcal{S})$ provided that $V|_{\iota(p)} \in T_{\iota(p)}\mathcal{H}$ for every $p \in \mathcal{S}$. The pullback of a function $f$ via a map $\phi$ is written as $\phi^\star f$, or simply as $f$ when the meaning is clear from the context. We also define the pullback of a vector field $X$ via a diffeomorphism $\phi$ by $\phi^\star X := (\phi^{-1})_\star X$. A $(p,q)$-tensor is a tensor with $p$ contravariant and $q$ covariant indices. Given any pair of $(2,0)$ and $(0,2)$ tensors $A^{ab}$ and $B_{cd}$, we use the notation $\text{tr}_A \boldsymbol{B} := A^{ab}B_{ab}$. We also employ $\text{grad}_h f$ to denote the gradient of a function $f$ w.r.t. the metric $h$, and $\Box_h$ for the Laplacian. Our convention for the curvature tensor of a connection $\mathcal{D}$ is

$$R(X,Y)Z = \mathcal{D}_X\mathcal{D}_Y Z - \mathcal{D}_Y\mathcal{D}_X Z - \mathcal{D}_{[X,Y]}Z,$$

and its Ricci tensor is the contraction between its contravariant index and its second covariant index. We employ the symbol $\nabla$ for the Levi-Civita connection of a metric $g$. A condition in parenthesis right next to an identity indicates that the identity has been obtained under the assumption that the condition holds. This is used mainly to indicate that a general expression has been particularized to the null case. Throughout this thesis we adopt the notation $\pounds_X^{(m)} T$ for the $m$-th Lie derivative of the tensor $T$ along $X$, and $X^{(m)}(f)$ for the $m$-th directional derivative of the function $f$ along $X$. When $m=1$ we also write $\pounds_X T$ and $X(f)$, while for $m=0$ both operators reduce to the identity. Finally, we say that an $(\mathfrak{n}+1)$-dimensional manifold $(\mathcal{M},g)$ with $\mathfrak{n}\geq 2$ satisfies the $\lambda$-vacuum equations provided that

$$\text{Ric}[g] = \lambda g, \qquad \text{or equivalently,} \qquad \text{Ein}[g] = \frac{1-\mathfrak{n}}{2}\lambda g. \tag{0.1}$$

When the emphasis is on the value of the cosmological constant, we shall say "$\Lambda$-vacuum" instead of "$\lambda$-vacuum", the relation between $\lambda$ and the cosmological constant $\Lambda$ being $\Lambda = \frac{\mathfrak{n}-1}{2}\lambda$. An equality with a symbol placed above (or below) the equality sign will have three different meanings depending on the context:

- The symbol $\stackrel{\mathcal{H}}{=}$ indicates that both sides of the equality are evaluated on a manifold $\mathcal{H}$.
- We use $\stackrel{(0.1)}{=}$ to indicate that the equation (0.1) has been used in order to obtain the right-hand side of the equality.
- Finally, the symbols $\stackrel{(m)}{=}$ and $\stackrel{[m]}{=}$ are specific to Chapters 5 and 6, and are appropriately introduced in Notations 5.19 and 5.23.

Finally, our definition of a Killing horizon in Section 5.5 allows the Killing vector to vanish on the horizon on a set with empty interior.

## LIST OF SYMBOLS

Here we provide a list of the symbols used throughout this thesis, together with their corresponding definitions, for the sake of clarity.

- **Chapter 2**. $\gamma_{ab}$, $\ell_a$, $\ell^{(2)}$, $\boldsymbol{\mathcal{A}}$ and $\text{Y}_{ab}$ are defined in Def. 2.1. The symbol $\underbracket{\boldsymbol{V}}$ is introduced right before (2.73). The tensors $P$, $n$ and $n^{(2)}$ are defined in (2.3)-(2.6), $\nu$ and $\boldsymbol{\theta}^a$ in (2.12), $\text{U}_{ab}$ in (2.15), $\text{K}_{ab}$ in (2.16), $\text{F}_{ab}$ in (2.17) and $\Pi_{ab}$ in (2.18). In (2.19) we introduce the tensors $\mathbf{s}$, $\mathbf{r}$ and $\kappa$. The connections $\overset{\circ}{\nabla}$ and $\overline{\nabla}$ are defined respectively in (2.38)-(2.39) and (2.60), and $\overset{\circ}{R}$, $\overline{R}$ denote their respective curvatures. The notation $\operatorname{div}_P \boldsymbol{t}$ is defined right after (2.99). The tensor $V^b{}_a$ is introduced in (2.54), $\mathcal{R}_{ab}$ in (2.95), $\boldsymbol{J}$ and $H$ are first written in (2.100), and $\mathbf{\Pi} \cdot \mathbf{\Pi}$ is defined in Prop. 2.13. The tensors $\mathcal{K}^{\mathrm{t}}$ and $\daleth$ are first defined in (2.121) and (2.122), respectively. The vector $\theta$ is defined in Remark 2.24, the tensors $\boldsymbol{\ell}_{\parallel}$, $\ell^{(2)}_{\parallel}$, $h^{\sharp}$, $\mathbf{Y}_{\parallel}$, $\mathbf{K}_{\parallel}$ in Notation 2.30, and the one-form $\boldsymbol{\omega}$ in Def. 2.37. Finally, the function $\mathfrak{z}$ is introduced in Def. 2.41, and $\overline{\Box}$ in Prop. 2.47.

- **Chapter 3**. The map $\Psi$ is defined in Def. 3.4 and the 4-tuple $\{\mathcal{D}, \underline{\mathcal{D}}, \phi, \sigma\}$ in Def. 3.9.

- **Chapter 4**. The tensors $\mathcal{K}[\eta]$, $\Sigma[\eta]$ and $\mathfrak{Q}[\eta]$ are introduced just at the beginning of Section 4.2. The notation ${}^{(i,j)}T$ in front of a tensor $T$ is introduced in Notation 4.5. $C$ and $\bar{\eta}$ are defined right before (4.24), and $\mathfrak{w}$, $\mathfrak{p}$ in Notation 4.8. The tensor $\mathfrak{S}$ is defined below Remark 4.10

- **Chapter 5**. $\mathbf{Y}^{(k)}$, $\mathbf{r}^{(k)}$, $\kappa^{(k)}$, $\mathcal{R}^{(k)}$, $\dot{\mathcal{R}}^{(k)}$ and $\ddot{\mathcal{R}}^{(k)}$ are introduced in (5.32)-(5.33). The tensors $\Sigma$ and $\underset{\smile}{\Sigma}$ are defined in (5.35) and (5.37), and $H^{\gamma\rho}_{\nu\beta}$, $F^{\rho\lambda\gamma}_{\nu\alpha\beta}$ are defined in (5.36)-(5.37). The tensors $\mathcal{O}^{(m)}_{ab}$, $\mathcal{O}^{(m)}_{a}$ and $\mathcal{O}^{(m)}$ appear first in Cor. 5.30. The tensors $A_\eta$, $X^a_\eta$, $\kappa$, $\mathfrak{w}$, $\mathfrak{p}$, $\mathfrak{q}$ and $\mathfrak{J}$ are introduced in (5.144), (5.145), (5.149) and (5.156), while $\mathfrak{w}^{(m)}$, $\mathfrak{p}^{(m)}$, $\mathfrak{q}^{(m)}$ and $\mathfrak{J}^{(m)}$ are defined in (5.162)-(5.163). The function $\alpha$ first appears right after Lemma 5.50, the function $\kappa_\star$ in (5.149), the tensor $\mathcal{P}^{(m)}$ is defined in (5.165), and the one-form $\boldsymbol{\tau}$ in (5.178). Finally, the symbol $\mathscr{K}$ denotes AKH data (Def. 5.58).

- **Chapter 6**. The scalar $q$ is introduced in (6.16), the tensors $\mathcal{Q}_{\alpha\beta}$ and $\mathcal{L}_\mu$ in (6.57) and (6.58), and the function $f$ right after them. The scalar functions $\{\sigma^{(k)}\}$ are defined right after (6.62). The tensors $\widetilde{\mathscr{R}}^{(m)}_{ab}$ and $\dot{\mathscr{R}}^{(m)}_{\mathcal{L}}$ are introduced in Prop. 6.18, and $\widetilde{\mathscr{R}}^{(m)}_{a}$ in Prop. E.4. Finally, the tensors $\mathcal{O}^{\mathscr{I}}$ and $\mathcal{O}^{\Sigma}$ are first mentioned in Remark 6.24 and defined in Def. 6.31.

1

# INTRODUCTION

## 1.1 CONTEXT, MOTIVATION AND AIM OF THIS THESIS

The year 1905 is often regarded as Albert Einstein's *annus mirabilis*, marking a turning point in the development of modern physics. In three foundational articles, Einstein addressed several fundamental problems, including the quantum nature of light [101], Brownian motion [100], and the development of the special theory of relativity [107]. This latter work introduced a radical reconceptualization of space and time, which were later unified by Hermann Minkowski [240, 241] into a new four-dimensional structure known as *spacetime*. By abandoning the notion of absolute simultaneity and postulating both the principle of relativity and the constancy of the speed of light, Einstein resolved longstanding inconsistencies between mechanics and electrodynamics, while at the same time raising profound questions about the nature of gravitation.

Over the following decade, Einstein sought to extend these ideas to the gravitational setting, guided by the equivalence principle and the insight that gravity should not be regarded as a force, but rather as a manifestation of the curvature of spacetime. This effort culminated in November 1915 with the presentation to the Prussian Academy of Sciences of what are now known as the Einstein field equations [102]

$$\mathrm{Ric} - \frac{1}{2}\,\mathrm{Scal}\, g = T.$$

These equations relate the geometry of spacetime, encoded in the Ricci tensor Ric, the scalar curvature Scal, and the metric tensor $g$, to the matter-energy content, represented by the energy-momentum tensor $T$.

The theory of General Relativity [104, 162] provides a geometric description of gravitation and successfully explains phenomena beyond the scope of Newtonian gravity, such as the anomalous precession of Mercury's orbit [103], the deflection of light by the Sun [96], and the gravitational redshift of light [5, 274]. Moreover, it predicts entirely new effects, including the emission of gravitational waves [105, 106, 286], whose first direct detection was achieved by the LIGO collaboration [1], the expansion of the Universe [267, 281], and the existence of black holes, whose shadows were recently observed by the Event Horizon Telescope [7–11, 85, 324]. From its earliest days, General Relativity has shown remarkable agreement with observational evidence, and it remains the most widely accepted theory of gravitation today.

One of the most striking consequences of General Relativity is the prediction of black holes. The first non-trivial exact solution of the Einstein field equations, found by Karl Schwarzschild [305], indeed describes a black hole, although for many years its physical interpretation was debated and often regarded as a mere mathematical artifact or a pathology of the theory. The Schwarzschild solution represents a non-rotating and uncharged black hole, and was later generalized to include electric charge [256, 278]. It was not until 1963 that a rotating solution, now known as the Kerr black hole, was discovered [191], followed shortly thereafter by its charged and rotating generalization [252]. A common feature of all these solutions is the presence of an event horizon, which acts as a causal barrier preventing even light from escaping, as well a singularity, which signal a breakdown of the predictive power of General Relativity (see [67, 149, 180] for some excellent reviews).

Understanding the nature of singularities has been one of the most subtle issues in the theory, in part due to the misleading role played by poorly chosen spacetime coordinates. For instance, the event horizon of the Schwarzschild black hole was long believed to be singular, until it was understood that, under an appropriate change of coordinates, it is in fact perfectly regular [97, 118, 198, 311]. In contrast, the so-called central singularity cannot be removed by any coordinate transformation, as it is characterized by the divergence of curvature invariants. In the late 1960s, Roger Penrose, and later Stephen Hawking, clarified this situation by showing that singularities (in the sense of existence of incomplete geodesics) are not limited to highly symmetric spacetimes, such as those with spherical or axial symmetry, but can also arise in generic, non-symmetric settings. In particular, Penrose's singularity theorem [249, 262, 307] establishes the *generic* formation of singularities under suitable geometric and topological assumptions. As a result, the presence of singularities is not an exceptional feature of General Relativity, but rather an intrinsic and widespread aspect of the theory.

Although the theory has not been refuted by any observational test to date, there are several reasons to believe that General Relativity cannot constitute a complete theory of gravitation. One such reason is its own prediction of singularities, as discussed above, which signal a breakdown of the classical description of spacetime. Another is the fundamental incompatibility between General Relativity and the Standard Model of particle physics, which successfully describes the remaining fundamental interactions in Physics [137]. These shortcomings have motivated numerous research directions, including the development of quantum theories of gravity, such as Loop Quantum Gravity [287] and String Theory [314], as well as the study of modified theories of gravity [308], which have been proposed in an effort to explain various cosmological phenomena.

Depending on the perspective adopted, General Relativity can be divided into several broad research areas, including numerical relativity [201], relativistic astrophysics [310], and cosmology [36]. The present thesis is framed within the field of mathematical relativity [182, 249], which addresses fundamental questions of the theory using rigorous mathematical methods and tools from differential geometry. Among the most important open problems in this area are the weak and strong Cosmic Censorship Conjectures [165, 266], the non-linear stability of the Kerr family of black hole solutions [194], the black hole uniqueness conjecture [176], as well as the Penrose inequality [225].

One of the central problems in mathematical relativity is the (standard) Cauchy, or initial value, problem. It consists in prescribing suitable initial data for the spacetime metric on a spacelike hypersurface, interpreted as an "instant of time", and in determining necessary and sufficient conditions under which such data give rise to a spacetime solving the Einstein field equations. The fundamental well-posedness result due to Choquet-Bruhat [58] establishes that prescribing the ambient metric and its first normal derivative on a spacelike hypersurface, subject to a system of constraint equations, uniquely determines (up to diffeomorphism) a local-in-time solution of the vacuum Einstein equations realizing this initial data. Owing to the geometric nature of the theory, the Einstein equations are invariant under diffeomorphisms: if $g$ is a solution and $\phi$ is a diffeomorphism, then the pullback metric $\phi^{\star}g$ is also a solution. This freedom prevents the Einstein equations from forming a hyperbolic system unless a coordinate gauge is fixed a priori. The standard approach to constructing a spacetime from spacelike initial data therefore consists in solving a reduced system of quasi-linear hyperbolic partial differential equations, known as the reduced Einstein equations, obtained by imposing harmonic (wave) coordinate conditions. One then verifies that the solution of the reduced system indeed satisfies the full Einstein equations by showing that the harmonic coordinate conditions are fulfilled. A detailed exposition of this construction can be found in [282] and will be reviewed in Section 3.1.

The original formulation of the initial value problem has evolved over time into a more abstract and geometrically natural framework. In this setting, rather than prescribing a spacetime metric and its normal derivative on a hypersurface, one considers an abstract $\mathfrak{n}$-dimensional manifold $\Sigma$ endowed with a Riemannian metric $h$ and a symmetric covariant 2-tensor field $K$, satisfying the so-called constraint equations [60]. Within this approach, the well-posedness of the initial value problem becomes a purely geometric statement, independent of any choice of coordinates, namely the existence of an $(\mathfrak{n}+1)$-dimensional Lorentzian manifold $(\mathcal{M}, g)$ and an embedding $\Phi\colon \Sigma \hookrightarrow \mathcal{M}$ such that $h$ arises as the induced metric and $K$ as the second fundamental form of $\Phi(\Sigma)$. In this sense, the initial data are completely detached from the spacetime one seeks to construct, a feature that is particularly satisfactory from a geometric viewpoint. Indeed, it places the initial data on the same conceptual and mathematical footing as the Einstein field equations themselves.

One of the aims of this thesis, addressed in Chapter 3, is to achieve an analogous geometrization for the characteristic initial value problem, in which data are prescribed not on a spacelike hypersurface, but on a pair of null hypersurfaces intersecting transversely. One motivation to study initial value problems on characteristic hypersurfaces is that, in a certain way, data on a null hypersurface represent the "present" of an observer better than data on a spacelike hypersurface. A fundamental result due to Rendall [279] establishes that, upon prescribing the spacetime metric on two such transversely intersecting null hypersurfaces, there exists a unique spacetime solution of the reduced Einstein equations in a future neighbourhood of their intersection, compatible with the given initial data. Moreover, by working in a suitable coordinate system, Rendall provides a procedure to reconstruct all components of the spacetime metric on the null hypersurfaces from a minimal set of freely specifiable data, in such a way that the resulting solution of the reduced system is indeed a solution of the vacuum Einstein equations. More recent works have extended the domain of existence to a full neighbourhood of the two null hypersurfaces [50, 78, 164, 208]. In addition, the characteristic initial value problem is known to be well posed when the data are prescribed on the future null cone of a point [57].

In order to achieve a geometrization of the characteristic Cauchy problem analogous to that of the spacelike case, it is necessary to formulate the initial data and the associated constraint equations independently of the spacetime one seeks to construct. While this separation is standard in the spacelike Cauchy problem, since the initial data are posed on a non-degenerate hypersurface, the situation for null hypersurfaces is considerably more subtle. Indeed, null hypersurfaces are intrinsically degenerate, and as a consequence no intrinsic Levi-Civita connection can be defined on them. Several approaches have been developed to overcome these difficulties. One possibility is to introduce an appropriate quotient structure and work on the resulting quotient space [138]. Another currently active approach is provided by Carrollian geometry [81], in which the fundamental geometric object is a triple $(\mathcal{N}, \boldsymbol{q}, n)$ consisting of a manifold $\mathcal{N}$, a degenerate metric $\boldsymbol{q}$, and a nowhere-vanishing vector field $n$ satisfying $\boldsymbol{q}(n, \cdot) = 0$.

There exists, however, a more general and systematic framework to treat abstract hypersurfaces of arbitrary causal character, and in particular null hypersurfaces, known as the *hypersurface data formalism* [221–223]. The fundamental object in this formalism is the notion of *metric hypersurface data*, denoted by $\{\mathcal{H}, \boldsymbol{\gamma}, \boldsymbol{\ell}, \ell^{(2)}\}$, where $\boldsymbol{\gamma}$ is a symmetric covariant 2-tensor, $\boldsymbol{\ell}$ is a one-form, and $\ell^{(2)}$ is a scalar function defined on $\mathcal{H}$. Associated with this metric hypersurface data there is a uniquely determined contravariant set $\{P, n, n^{(2)}\}$, consisting of a symmetric contravariant 2-tensor $P$, a vector field $n$, and a scalar function $n^{(2)}$. When the metric hypersurface data happens to be embedded in an ambient semi-Riemannian manifold $(\mathcal{M}, g)$, the objects $\{\boldsymbol{\gamma}, \boldsymbol{\ell}, \ell^{(2)}\}$ encode, respectively, the fully tangential, mixed tangential–transverse, and fully transverse components of the ambient metric $g$ restricted to $\mathcal{H}$. Moreover, metric hypersurface data determine a unique torsion-free connection $\overset{\circ}{\nabla}$ on $\mathcal{H}$ satisfying certain natural conditions. This structure can be further enlarged by an additional symmetric covariant 2-tensor field $\mathbf{Y}$ on $\mathcal{H}$, in which case the collection $\{\mathcal{H}, \boldsymbol{\gamma}, \boldsymbol{\ell}, \ell^{(2)}, \mathbf{Y}\}$ is referred to as *hypersurface data*. In the embedded setting, the tensor $\mathbf{Y}$ encodes information about the Lie derivative of the ambient metric $g$ along a transverse vector field $\xi$, also known as a *rigging* [223, 304], along $\mathcal{H}$. The large freedom in the choice of rigging is reflected, in the abstract setting, as a gauge freedom of the hypersurface data.

In Section 2.2 of Chapter 2, after reviewing the basic notions of hypersurface data, we show that for an embedded null hypersurface the purely tangential components of the ambient Ricci tensor can be expressed entirely in terms of hypersurface data. This result allows for a formulation of the constraint equations on a null hypersurface in a detached manner. Moreover, these constraint equations are shown to be gauge-covariant, in the sense that if they are satisfied in one gauge, then they are necessarily satisfied in any other. To define an abstract notion of two transverse null hypersurfaces suitable for a detached formulation of the characteristic initial value problem, it is necessary to identify the essential properties that two sets of null hypersurface data must satisfy in order to be simultaneously embeddable as null hypersurfaces with a common spacelike boundary. We find that this requires four compatibility conditions at the boundary, which admit a clear geometric interpretation as the agreement of the (i) induced metrics, (ii) first fundamental forms, (iii) torsion one-forms, and (iv) induced Ricci curvatures at the common spacelike boundary. These conditions are formulated in Chapter 3 in a fully diffeomorphism and gauge-covariant manner, leading to the central definition of *double null data* (Definition 3.9), which provides the desired detached counterpart of two transverse null hypersurfaces.

The compatibility conditions described above are, of course, necessary for a double null data set to be embeddable in an ambient spacetime. It is therefore natural to ask whether these conditions are also sufficient. We answer this question in the affirmative by showing that, given any double null data, there always exists a spacetime into which it can be embedded. Clearly, there are many spacetimes where a given double null data can be embedded and they are in general not solutions of any field equations. In this sense, the compatibility conditions are purely geometric in nature and must be satisfied independently of whether the ambient spacetime is required to solve the Einstein equations, or indeed any other system of field equations. To construct a solution of the Einstein equations from double null data satisfying the abstract constraint equations, we proceed, as in the spacelike case, by first solving the reduced Einstein equations and then showing that the resulting solution also satisfies the full system. Achieving this requires an abstract, coordinate-free formulation of the harmonic gauge condition. This is accomplished in Section 2.5 of Chapter 2. Given a collection of $\mathfrak{n}$ independent functions on an abstract null hypersurface, we construct a vector field depending only on these functions and on the hypersurface data, with the crucial property that, in the embedded case, it vanishes if and only if the chosen functions are harmonic with respect to the ambient metric. The gauge transformation properties of this vector field allow us to prove the existence of an essentially unique gauge in which it vanishes. This gauge, referred to as the *harmonic gauge*, is fully covariant and plays a central role in the resolution of the characteristic initial value problem within the present framework.

Given double null data expressed in the harmonic gauge and satisfying the constraint equations, we solve the reduced Einstein equations using the metric data as initial conditions. In order to show that the resulting solution of the reduced system is in fact a solution of the full Einstein field equations, we follow an approach that differs from that of Rendall. Whereas Rendall reconstructs the spacetime metric by integrating second-order ordinary differential equations for the unknown metric components in a specific coordinate system, in our setting the full metric is already prescribed as part of the initial data. This allows us to solve the reduced Einstein equations directly, without any preliminary reconstruction step. From the resulting spacetime solution, one can then construct embedded hypersurface data, which *a priori* need not coincide with the originally prescribed double null data. To complete the argument, it is therefore necessary to establish two facts: first, that the reconstructed hypersurface data agree with the original one, ensuring that the entire set of prescribed data (and not merely its metric component) has been embedded; and second, that the spacetime metric obtained in this way indeed satisfies the Einstein field equations. We achieve both objectives simultaneously by combining the abstract constraint equations with the harmonic gauge conditions. Our main existence result, Theorem 3.23, establishes that for any double null data (of arbitrary dimension) satisfying the abstract constraint equations, there exists a unique spacetime $(\mathcal{M}, g)$ solving the $\lambda$-vacuum Einstein equations in which the given double null data are realized as embedded hypersurface data.

Another central question addressed in Chapter 3 concerns the geometric uniqueness of the characteristic initial value problem at the abstract level. In the standard Cauchy problem, it is well known that two initial data sets $(\Sigma, h, K)$ and $(\Sigma', h', K')$ give rise to isometric spacetimes $(\mathcal{M}, g)$ and $(\mathcal{M}', g')$ whenever there exists an isometry between $(\Sigma, h)$ and $(\Sigma', h')$ that maps $K$ to $K'$ [323]. Establishing an analogous result in the characteristic setting requires first introducing an appropriate notion of isometry between abstract initial data sets. A crucial point in this construction is the presence of gauge freedom in the hypersurface data formalism: two sets of abstract hypersurface data $\mathcal{D}$ and $\mathcal{D}'$ related by a gauge transformation are geometrically indistinguishable. Consequently, the notion of isometric double null data must incorporate both a diffeomorphism between the underlying abstract hypersurfaces (together with their associated tensor fields) and an appropriate gauge transformation. With this definition in place, we prove in Theorem 3.30 that two double null data sets satisfying the constraint equations are isometric in the abstract sense if and only if the spacetimes they generate are isometric in the usual sense.

These results accomplish the two main objectives of Chapter 3. First, both the initial data and the associated constraint equations are formulated in a manner fully detached from the spacetime one seeks to construct, thereby providing a satisfactory geometrization of the characteristic initial value problem, analogous to that of the standard Cauchy problem. Second, they resolve, from an abstract and spacetime-independent perspective, the question of when two sets of characteristic initial data give rise to the same spacetime. Moreover, since in the resulting spacetime the initial data encode the full metric together with its first transverse derivative, our framework encompasses all possible choices of characteristic initial data derived from these quantities. There are many possible ways of trading prescribed data by constraints, each leading to a different formulation of the characteristic initial value problem. All such formulations, however, can be understood as particular instances of Theorem 3.23. In this sense, our results recover, and substantially extend, the analysis presented in [64].

Another central object in mathematical relativity is the study of spacetime symmetries, understood as transformations that leave certain geometric properties of the spacetime invariant. Of particular importance are Killing symmetries, which correspond to isometries of the spacetime and are associated for instance with equilibrium configurations of the gravitational field, and lead to conserved quantities along geodesic motion [323]. Homothetic symmetries, on the other hand, describe scale invariances and allow for the study of self-similar solutions of the Einstein equations [285], which arise, for instance, in critical gravitational collapse [151]. Homotheties also play a prominent role in the Fefferman–Graham approach to conformal geometry through the ambient space construction [112, 113]. Infinitesimally, Killing and homothetic symmetries are generated by vector fields $\eta$ satisfying, respectively, $\pounds_{\eta} g = 0$ and $\pounds_{\eta} g = \mu g$, with $\mu \in \mathbb{R}$.

A particularly interesting problem, which naturally connects with the discussion of the Cauchy problem above, is how to encode the existence of such symmetries at the level of initial data. Doing so allows, for instance, the reduction of a Lorentzian problem in dimension $\mathfrak{n}+1$ to a simpler problem in dimension $\mathfrak{n}$, a strategy that is often effective in addressing questions such as the existence and uniqueness of black hole equilibrium solutions (see e.g. [161, 181]). This connection between initial value problems in General Relativity and the existence of (homothetic) Killing vector fields in the resulting spacetime leads to the so-called (homothetic) Killing initial data (KID) problem, which has been extensively studied in the past.

For spacelike initial data, the classical approach [37, 84, 242] begins by prescribing the restriction of a candidate Killing vector field $\eta$ to the initial hypersurface $\mathcal{H}$ and imposing the so-called KID equations. These equations depend solely on the abstract initial data $(h, K)$ together with the tangential component of $\eta|_{\mathcal{H}}$, which is a vector field on $\mathcal{H}$, and its transversal component, which is a scalar. They ensure that, once the spacetime is constructed, the conditions $\Phi^{\star}(\pounds_{\eta} g) = 0$ and $\Phi^{\star}(\pounds_{\nu}\pounds_{\eta} g) = 0$ hold on $\mathcal{H}$, where $\nu$ denotes a normal vector field to $\mathcal{H}$. The next step consists in specifying $\pounds_{\nu}\eta|_{\mathcal{H}}$ so that the remaining components of $\pounds_{\eta} g$ vanish on $\mathcal{H}$. The vector field $\eta$ is then extended off $\mathcal{H}$ by solving a homogeneous wave-type equation with initial data $(\eta|_{\mathcal{H}}, \pounds_{\nu}\eta|_{\mathcal{H}})$. This equation is chosen to be compatible with the Killing condition and to satisfy two crucial properties: (i) that all remaining components of $\pounds_{\nu}\pounds_{\eta} g$ vanish identically on $\mathcal{H}$, and (ii) that $\pounds_{\eta} g$ itself satisfies a homogeneous wave-type equation. Combining these facts, one concludes that $\eta$ is a Killing vector field of the Cauchy development of $(\mathcal{H}, h, K)$. This strategy also extends to homothetic and conformal Killing vector fields [140].

In the null setting, the KID problem has been studied for initial data posed on light cones and on pairs of intersecting characteristic hypersurfaces, both for the Einstein equations [74] and for the conformal Einstein equations [258]. In the latter case, the analysis is restricted to four spacetime dimensions when one of the intersecting characteristic hypersurfaces corresponds to null infinity. The approach developed in [74] consists in considering two characteristic hypersurfaces embedded in a $\lambda$-vacuum spacetime and introducing spacetime coordinates adapted to these hypersurfaces. The initial values of the candidate Killing vector field $\eta$ are then given by its coordinate components in this adapted system. The extension of $\eta$ off the hypersurfaces is achieved by solving a wave-type equation, and sufficient conditions are derived to ensure that $\pounds_\eta g$ vanishes on the initial characteristic hypersurfaces. Since $\pounds_\eta g$ itself satisfies a homogeneous wave-type equation, this implies that $\eta$ is a Killing vector field of the spacetime metric $g$.

A key difference between the spacelike and null KID problems is that, in the former case, the KID equations can be formulated entirely in terms of abstract initial data, allowing one to determine whether a given data set gives rise to a spacetime admitting a Killing vector field *before* solving the Einstein equations. In the null case, however, the KID equations in principle depend explicitly on the ambient Levi-Civita connection and Riemann curvature tensor, which makes it necessary to first construct the spacetime solution and only then verify whether the KID conditions are satisfied. With the aim of removing this limitation, in Chapter 4 we address the KID problem for hypersurfaces of arbitrary causal character in a fully general and geometric framework.

To achieve this, in Section 4.2 we begin by considering a semi-Riemannian manifold $(\mathcal{M}, g)$ and an arbitrary vector field $\eta$. We develop a collection of fully general identities relating the deformation tensor of $\eta$, $\mathcal{K}[\eta]_{\mu\nu} = \pounds_\eta g_{\mu\nu}$, to the Lie derivative of the Levi-Civita connection, $\Sigma[\eta] = \pounds_\eta \nabla$, as well as to other geometric quantities associated with $g$. By pulling back these identities to a hypersurface $\mathcal{H}$, we obtain explicit relations connecting geometric quantities derived from $\mathcal{K}[\eta]_{\mu\nu}$ at $\mathcal{H}$ with the fully tangential components of $\Sigma[\eta]$ at $\mathcal{H}$, together with its one-transverse–two-tangent components, denoted by $\mathfrak{S}_{ab}$. These identities are fully gauge and diffeomorphism-covariant, do not assume any field equations for $g$, and hold for hypersurfaces of arbitrary causal character, as well as for any vector field $\eta$ (not necessarily Killing or homothety). Applying this formalism to spacelike hypersurfaces, in Section 4.4 we recover the homothetic KID problem for spacelike initial data in the language of hypersurface data. A key advantage of this approach is its gauge freedom, which allows one to adapt the rigging vector to specific requirements (in particular, the rigging need not be normal to $\mathcal{H}$).

Inspired by the ideas in [74], in Section 4.5 we particularize the identities to null hypersurfaces and show that (i) the tangential components of $\mathcal{K}[\eta]$ at $\mathcal{H}$ can be expressed purely in terms of hypersurface data, and (ii) the transverse–tangent ($\mathfrak{w}_a$) and transverse–transverse ($\mathfrak{p}$) components satisfy a hierarchical, inhomogeneous system of transport equations along the null generator $n$, with source terms determined by $\mathfrak{S}$ (Lemma 4.20). Moreover, we prove that certain contractions of $\mathfrak{S}$, namely $P^{ab}\mathfrak{S}_{ab}$ and $\mathfrak{S}_{ab}n^b$ (but not $\mathfrak{S}(n,n)$), also satisfy transport equations. Altogether, this yields a closed hierarchical system for the collection $\{\mathfrak{w},\mathfrak{p},\mathfrak{S}\}$. We show that this system becomes homogeneous provided that $\mathcal{K}_{ab}$ is pure trace (i.e. proportional to $\gamma_{ab}$) and that $\mathfrak{S}(n,n)=0$. These two conditions provide the natural analogue of the KID equations in the null setting. We exploit this fact in Theorem 4.27 to propagate $\{\mathfrak{w},\mathfrak{p},P^{ab}\mathfrak{S}_{ab},\mathfrak{S}_{ab}n^b\}$ from initial conditions prescribed on a cross-section, under the assumption that $\mathcal{H}$ admits a product topology.

A direct application of this formalism, developed in Section 4.6, concerns the characteristic KID problem. For two intersecting null hypersurfaces $\mathcal{H}$ and $\underline{\mathcal{H}}$ embedded in a $\lambda$-vacuum spacetime, the intersection surface $\mathcal{S}$ serves as the initial cross-section from which the fields $\{\mathfrak{w},\mathfrak{p},\mathfrak{S}\}$ are propagated using Theorem 4.27. We show that requiring $\mathfrak{S}(n,n)=0$ and $\mathcal{K}_{ab}$ to be pure trace on both null hypersurfaces, together with suitable conditions on the scalar $\mathfrak{w}(n)$ and the one-form $\mathfrak{S}_{ab}n^b$ at $\mathcal{S}$, ensures that the deformation tensor of $\eta$ satisfies $\pounds_\eta g=\mu g$ on $\mathcal{H}\cup\underline{\mathcal{H}}$, and hence throughout the spacetime. The crucial point, and the main difference with [74] (besides the fact that we deal with homotheties and not just Killings), is that all these conditions are formulated entirely in terms of double null data, and hence they can be ascertained, or imposed, *a priori* (i.e., before solving the Einstein equations). This result puts the characteristic KID problem on the same footing as the standard KID problem.

Other initial value problems with data prescribed on hypersurfaces combining features of the previous cases can be treated within the same framework. As an example, in Section 4.7 we study the homothetic KID problem for smooth spacelike–characteristic initial data. By imposing the standard KID equations on the spacelike region and the null KID equations on the characteristic region, both expressed in terms of detached hypersurface data, we show that the Cauchy development of the initial data admits a homothetic Killing vector field (Theorem 4.35). An analysis suggests that a similar conclusion holds when initial data are prescribed on a spacelike–characteristic hypersurface forming a corner [75, 91]. In this case, no additional compatibility conditions at the corner are required. We emphasize that results of this type, involving hypersurfaces of mixed causal character, rely crucially on a formalism capable of treating hypersurfaces of arbitrary causal character within a unified framework, and would not be possible in approaches restricted to a single causal type.

One of the fundamental differences between the standard (spacelike) and characteristic Cauchy problems is that, while in the former prescribing data on a single hypersurface suffices to obtain a well-posed problem, in the latter uniqueness typically fails unless data are given on two transverse null hypersurfaces. From a causal perspective, the absence of a second null hypersurface typically allows additional information from the past to influence the solution, thereby spoiling uniqueness (see Figure 1.1). However, in highly symmetric scenarios such as when the spacetime possesses Killing or homothetic vectors, data at a single null hypersurface suffices to determine, in a certain sense, a unique ambient spacetime solving the Einstein equations. This observation provides the conceptual link between the characteristic and KID problems discussed above and the second part of this thesis, namely analyzing the relation between the ambient manifold and the geometry at an embedded null hypersurface, with particular emphasis on how the higher-order transverse derivatives of the metric, referred to as the *transverse* or *asymptotic* expansion, are constrained by the ambient curvature and symmetries.

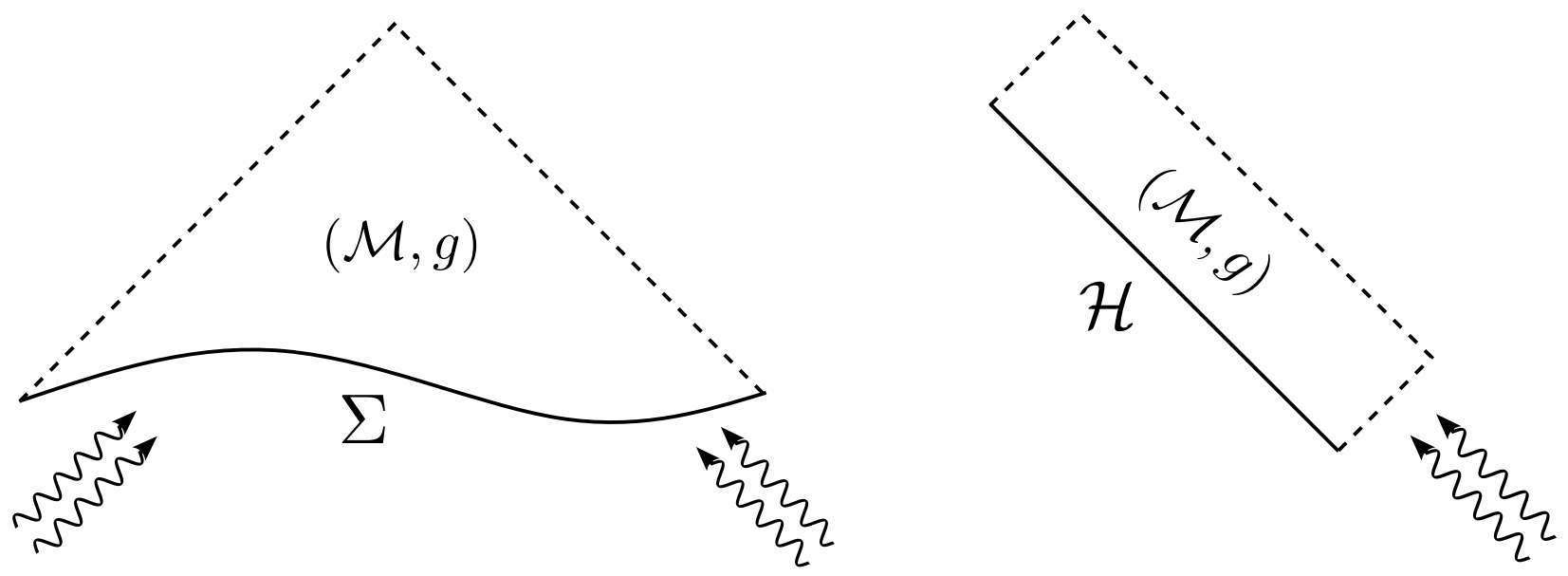

Figure 1.1: On the left figure, the spacetime $(\mathcal{M}, g)$ is uniquely determined by the initial data prescribed on the spacelike hypersurface $\Sigma$. On the right figure, data posed on a single null hypersurface $\mathcal{H}$ do not in general determine the spacetime uniquely, since additional information from the causal past and not encoded in the data on $\mathcal{H}$ may influence the solution.

For a better understanding of the situation, let us review three settings that have been considered in the literature: the homothetic horizon, the non-degenerate Killing horizon, and the degenerate Killing horizon cases. These examples illustrate, from different perspectives, how the ambient symmetries can effectively replace missing characteristic data. A more detailed discussion of these cases is given in Section 5.1.

In [113], C. Fefferman and R. Graham introduced the notion of the *ambient metric*, motivated by the observation that the light cone of a point in $(r+2)$-dimensional Minkowski spacetime encodes the conformal structure of the $r$-sphere. To generalize this idea to an arbitrary conformal class, their aim was to construct a homothetic horizon associated with the given conformal data and to embed it into an ambient manifold that is Ricci flat to infinite order. They showed that when the dimension of the conformal class is odd, the Einstein equations uniquely determine the full transverse expansion of the ambient metric along a null transverse coordinate at the horizon. In even dimensions, however, additional data are required to fully determine the expansion, and an *obstruction tensor* arises whose vanishing characterizes whether the ambient metric can be Ricci flat to all orders at the horizon. In both odd and even dimensions, the one-form $\boldsymbol{\eta}$ associated to the homothety is exact (as a one-form) up to the order determined by the equations, namely to infinite order in the odd case, and up to the order of the obstruction in the even case. Moreover, in even dimensions, if the free data satisfies a particular divergence condition and the obstruction tensor vanishes, $\boldsymbol{\eta}$ becomes exact to infinite order as well. Ambient metrics satisfying these conditions are known as *straight*. When the obstruction tensor does not vanish, the construction of an ambient metric as a formal series can still be performed by allowing logarithmic terms in the expansion. This leads to the so-called *generalized ambient metric*, which exhibits only finite differentiability at the horizon. Remarkably, in [285] it is shown that this generalized ambient metric is not merely Ricci flat to infinite order, but in fact *exactly* Ricci flat in the hyperbolic region. An analogous result in the elliptic region is also known to be true [152]. It is important to emphasize that in both the odd and even-dimensional cases the initial data are unconstrained, in sharp contrast with the spacelike and characteristic Cauchy problems described above. We review this construction in Subsection 5.1.1.

A second example in which curvature conditions in the ambient spacetime determine the transverse expansion is provided by non-degenerate Killing horizons. In [243], Moncrief showed that the transverse expansion of any three-dimensional non-degenerate Killing horizon in a Ricci-flat spacetime can be parametrized by six functions defined on the horizon. Since this analysis was carried out in a particular coordinate system, it is not straightforward to establish whether two such spacetimes are isometric to infinite order at the horizon. More recently, this issue was revisited from a geometric perspective in [195], where it was established that, for Ricci-flat spacetimes, the full expansion along a null transverse vector field is uniquely determined by *non-degenerate Killing horizon data*, namely a triple $(\mathcal{H}, \boldsymbol{\sigma}, \eta)$ consisting of a manifold $\mathcal{H}$, a Riemannian metric $\boldsymbol{\sigma}$, and a non-trivial Killing vector field $\eta$ on $(\mathcal{H}, \boldsymbol{\sigma})$ with non-vanishing constant norm w.r.t. $\boldsymbol{\sigma}$. While this characterization provides a fully geometric description of Killing horizons, it excludes the possibility that $\eta$ has zeros, and hence bifurcation surfaces on the horizon, and is restricted to the vacuum Einstein equations with $\Lambda = 0$. We revisit both approaches in Subsection 5.1.2.

The final example concerns degenerate Killing horizons. Given such a horizon, its *near-horizon limit* [199] is described by a function $F$, a one-form $\boldsymbol{\omega}$, and a Riemannian metric $h$ on the horizon cross-sections. In a recent work [188], the authors establish a uniqueness theorem for the extremal Schwarzschild–de Sitter spacetime. For a specific value of the mass parameter, this spacetime admits a degenerate Killing horizon with compact, maximally symmetric cross-sections, and therefore an associated near-horizon geometry. Using staticity and compactness arguments, one can show that the one-form $\boldsymbol{\omega}$ vanishes [33, 76, 328], reducing the near-horizon data to the pair $(h, F)$. By solving the $\Lambda$-vacuum Einstein equations order by order, leading to elliptic equations for the transverse expansion, the authors prove that all transverse derivatives of the metric at the horizon are uniquely determined by $(h, F)$ and coincide either with those of extremal Schwarzschild–de Sitter or with its near-horizon geometry (Nariai). As a consequence, in the real-analytic setting, the spacetime is shown to be locally isometric, in a neighbourhood of the horizon, to one of these two solutions. This analysis has been extended to the electrovacuum case in [187]. We summarize these results in Subsection 5.1.3.

Motivated by the preceding examples, Chapter 5 is devoted to the derivation of general identities relating the transverse derivatives of the ambient Ricci tensor to the transverse expansion of the metric at an arbitrary null hypersurface, and to the establishment of existence and uniqueness results for initial data prescribed on a single null hypersurface. In order to keep these identities as general as possible, we derive them without assumptions on the signature of the ambient metric, the spacetime dimension, or the topology of the hypersurface, and in a fully coordinate-free manner.

To encode higher-order information of the ambient metric within the hypersurface data formalism, we denote by $2\mathbf{Y}^{(k)}$ the pullback to $\mathcal{H}$ of the $k$-th Lie derivative of $g$ along a transverse vector field $\xi$. The collection $\{\mathbf{Y}^{(k)}\}_{k\geq 1}$ will be referred to as the *transverse* or *asymptotic expansion*. Our strategy to determine this expansion on a general null hypersurface proceeds as follows. First, we establish a general identity relating the $m$-th Lie derivative of the curvature and Ricci tensors along an arbitrary vector field $\xi$ to the Lie derivative of the corresponding connection along $\xi$. In the case of the Levi-Civita connection, this yields an identity expressing the $m$-th Lie derivative of the Ricci tensor in terms of Lie derivatives of the metric. Applying this identity to a null hypersurface $\mathcal{H}$ endowed with a transverse vector field $\xi$, we obtain explicit formulas for the leading-order terms of: (i) the purely transverse, (ii) mixed transverse–tangent, and (iii) purely tangent components of the $m$-th Lie derivative of the ambient Ricci tensor along $\xi$. These identities are explicitly written down in Corollary 5.30. A crucial feature is their geometric character: once $\xi$ is extended off $\mathcal{H}$ as a geodesic vector field, they depend only on the intrinsic metric hypersurface data and on the tensors $\{\mathbf{Y}^{(k)}\}_{k\geq 1}$.

Informally, the transverse expansion allows one to "reconstruct" the ambient metric order by order in a neighbourhood of the null hypersurface. To turn this formal expansion into an actual spacetime, we adapt a classical result due to Borel to the tensorial setting and prove that any null metric hypersurface data $\{\mathcal{H},\boldsymbol{\gamma},\boldsymbol{\ell},\ell^{(2)}\}$, together with a prescribed collection of tensors $\{\mathbb{Y}^{(k)}\}_{k\geq 1}$, gives rise to a smooth ambient manifold in which the data are embedded. More precisely, the resulting ambient metric satisfies that each tensor $2\mathbb{Y}^{(k)}$ coincides with $2\mathbf{Y}^{(k)}$, namely with the $k$-th transverse derivative of the ambient metric evaluated at the null hypersurface (Theorem 5.40). It is important to stress that, even if the formal Taylor series associated with $\mathbb{Y}^{(k)}$ fails to converge, the construction still yields a smooth ambient manifold.

A priori, the ambient manifold constructed this way does not satisfy any field equations. In order to identify necessary and sufficient conditions for the ambient metric to solve a prescribed system of equations, such as the Einstein equations, it is therefore essential to understand the structural content of identities (i)–(iii) above. First, the completely transverse component of the $m$-th Lie derivative of the Ricci tensor depends algebraically on the trace of the tensor $\mathbf{Y}^{(m+2)}$ with respect to $P$, which we denote by $\operatorname{tr}_P \mathbf{Y}^{(m+2)}$. Schematically, the identity reads

$$\Phi^\star \pounds_\xi^{(m)}(\operatorname{Ric}(\xi,\xi)) = -\operatorname{tr}_P \mathbf{Y}^{(m+2)} + \text{lower order terms.} \tag{1.1}$$

Second, the transverse–tangent components depend algebraically on the one-form $\mathbf{r}^{(m+2)} := \mathbf{Y}^{(m+2)}(n,\cdot)$, where $n$ denotes the null generator of the hypersurface. In schematic form,

$$\Phi^\star \pounds_\xi^{(m)}(\operatorname{Ric}(\xi,\cdot)) = \mathbf{r}^{(m+2)} + \text{lower order terms.} \tag{1.2}$$

Finally, the identity for the fully tangential components of the $m$-th Lie derivative of the Ricci tensor involves $\mathbf{Y}^{(m+1)}$ through a transport equation along the null generators of $\mathcal{H}$, and also depends on the scalar $\operatorname{tr}_P \mathbf{Y}^{(m+1)}$, the one-form $\mathbf{r}^{(m+1)}$, and their derivatives. Schematically, this identity takes the form

$$\Phi^\star \pounds_\xi^{(m)}(\operatorname{Ric}) = \pounds_n \mathbf{Y}^{(m+1)} + \text{lower order terms.} \tag{1.3}$$

Since $\operatorname{tr}_P \mathbf{Y}^{(m+1)}$ and $\mathbf{r}^{(m+1)}$ are themselves determined algebraically by derivatives of the Ricci tensor, identity (1.3) provides a mechanism to propagate the transverse expansion along the null direction, once the Ricci tensor is prescribed to all orders on $\mathcal{H}$. It is worth emphasizing that, for fixed $m$, the leading-order terms in identity (1.3) involve at most $m+1$ transverse derivatives of the metric, whereas the leading-order terms in identities (1.1)-(1.2) always depend on $m+2$ transverse derivatives of $g$ evaluated at $\mathcal{H}$.

By combining these identities with the aforementioned existence result, we obtain a set of necessary and sufficient conditions on the metric hypersurface data $\{\mathcal{H}, \boldsymbol{\gamma}, \boldsymbol{\ell}, \ell^{(2)}\}$ and on the abstract transverse expansion $\{\mathbb{Y}^{(k)}\}$ ensuring the existence of an ambient manifold that solves the $\Lambda$-vacuum Einstein equations to infinite order at the hypersurface (Theorem 5.42). These conditions may be interpreted as the "higher order" abstract constraint equations that must be satisfied on a null hypersurface. Our result holds without any restriction on the dimension or topology of $\mathcal{H}$. In particular, when the null hypersurface has a product topology, the constraint equations admit a natural reinterpretation as evolution equations along the null generators. This viewpoint allows us to reformulate the existence theorem entirely in terms of geometric data prescribed on a single cross-section of the hypersurface (Theorem 5.45). The proof is technically involved due to the fact that, at each order, there are more equations than quantities to be fixed. Thus, an important piece of the argument consists of showing that the redundant equations are automatically fulfilled.

When a symmetry generator $\eta$ is present, information encoded in its deformation tensor $\mathcal{K}[\eta] = \pounds_\eta g$ can be incorporated into identity (1.3). This leads to a new relation, which we call the *generalized master equation of order $m$* (cf. (5.170)), owing to its close connection with the generalized master equation introduced in [216] for the case $m=1$. A key feature of this equation is that its dependence on $\mathbf{Y}^{(m+1)}$ is no longer through a transport equation. Instead, it depends algebraically on $\mathbf{Y}^{(m+1)}$, as well as on the associated one-form $\mathbf{r}^{(m+1)}$ and the scalar $\text{tr}_P \mathbf{Y}^{(m+1)}$. When the surface gravity of $\eta$ does not vanish, this identity provides a systematic mechanism to identify the minimal amount of data on $\mathcal{H}$ required to determine the full transverse expansion in terms of the Ricci tensor and its derivatives at the hypersurface. Informally, the strategy is as follows: at each order, identities (1.1)-(1.2) at order $m-1$ determine $\text{tr}_P \mathbf{Y}^{(m+1)}$ and $\mathbf{r}^{(m+1)}$ algebraically. Substituting these quantities into the generalized master equation of order $m$ then allows one to recover the remaining components of $\mathbf{Y}^{(m+1)}$ in an entirely algebraic fashion.

We then apply the generalized master equation of order $m$ to the case of Killing horizons. For non-degenerate Killing horizons, we show that the full transverse expansion is uniquely determined by abstract data prescribed at the horizon together with the full tower of transverse derivatives of the ambient Ricci tensor evaluated at $\mathcal{H}$ (Theorem 5.72). This abstract data, referred to as *abstract Killing horizon data*, consists of metric hypersurface data together with a one-form $\boldsymbol{\tau}$ and a scalar function $\alpha$. Geometrically, $\boldsymbol{\tau}$ can be interpreted as the torsion one-form of the null hypersurface, while $\alpha$ encodes the proportionality between the Killing vector field (which may vanish) and the null generator $n$. This result extends the main theorem of [195] in several directions: first, by allowing for general field equations beyond the vacuum case with $\Lambda = 0$; second, by admitting zeroes of the Killing vector field, thereby including the possibility of bifurcation surfaces; and finally, by allowing for arbitrary ambient signature (compatible with null hypersurface data, of course).

The collection of transverse derivatives of the ambient Ricci tensor at $\mathcal{H}$ may be interpreted in two complementary ways. One possibility is to prescribe this collection as independent data on the hypersurface, for instance arising from an external matter model. Alternatively, one may view it as encoding functional relations between the metric hypersurface data and the transverse expansion itself. The simplest illustration of the latter viewpoint is provided by the $\Lambda$-vacuum equations, for which the Ricci tensor is proportional to the metric and hence its transverse derivatives are determined by the expansion. This structural property is formalized in Definition 5.73, and it allows us to prove that, whenever the ambient Ricci tensor satisfies this condition, the full transverse expansion at $\mathcal{H}$ is uniquely determined by abstract data at the horizon (Theorem 5.74). In the case of the $\Lambda$-vacuum Einstein equations, this result yields a characterization of all analytic $\Lambda$-vacuum spacetimes in a neighbourhood of a non-degenerate Killing horizon. As a direct application, we obtain a uniqueness theorem for the non-extremal Schwarzschild–de Sitter spacetime in the smooth setting (Theorem 5.80).

Finally, we apply our existence results to this scenario and show that every abstract Killing horizon data gives rise to an ambient manifold solving the $\Lambda$-vacuum Einstein equations to infinite order, for which the prescribed data is realized as a non-degenerate Killing horizon (Theorem 5.82). As before, the proof is technical because there are more equations to be fulfilled than variables to be constructed. We stress that no restriction is imposed on the topology or dimension of the horizon, and that Killing vector fields with zeroes are fully allowed, thus encompassing bifurcate horizons.

The identities obtained throughout this chapter, relating the geometry of a null hypersurface to that of the ambient manifold, are completely general and therefore applicable in a wide range of contexts. The final part of this thesis exploits these identities in the setting of null infinity and the conformal Einstein equations, with the goal of developing a coordinate-free and fully geometric framework for the analysis of gravitational radiation in General Relativity.

Studying gravitational radiation within General Relativity has always been a delicate task, since the metric that provides the notion of fall-off at infinity is itself the dynamical variable on which boundary conditions must be imposed. Since the early and initially controversial prediction of gravitational waves by Einstein and Rosen [105, 106, 286], there has been sustained interest in the analysis of gravitational radiation (see, for instance, the seminal works of Pirani, Trautman, and Lichnerowicz [204, 206, 271, 315, 316]). A major conceptual advance was later achieved by Sachs, who in a series of papers [99, 184, 290, 291, 293] proposed an invariant characterization of outgoing radiation based on the asymptotic fall-off behaviour of the Riemann tensor, known as the *peeling property*.

The study of gravitational radiation received a decisive impulse in the 1960s through the work of the Bondi–Pirani group at King's College [44, 218–220, 270], in particular through the analysis of axially symmetric radiating systems by Bondi, van der Burg, and Metzner [43], which was shortly afterwards generalized by Sachs [292]. Around the same time, Newman and Penrose developed the Newman–Penrose formalism and applied it to the study of gravitational radiation [251, 260, 265]. A connection between both approaches can be found in [318].

The Bondi–Sachs formalism is a coordinate-based approach in which a special coordinate system $\{r, u, x^A\}$ is constructed so as to be adapted to null infinity $\mathscr{I}$, which is formally located at $r \to \infty$ [14, 120, 209]. The hypersurfaces $u =$ const. are outgoing null hypersurfaces, while the intersections $u =$ const., $r =$ const. are topological 2-spheres. The metric coefficients in these coordinates are assumed to admit asymptotic expansions in powers of $1/r$, and can be determined order by order from the Einstein equations once suitable initial data are prescribed. Such data consist of the first fundamental form on a hypersurface $\{u = u_0 = \text{const.}\}$ intersecting $\mathscr{I}$ transversely, a certain $1/r$ coefficient of the metric at $\mathscr{I}$ encoding the radiation field, and a function and a one-form on $\{u = u_0\} \cap \mathscr{I}$ corresponding to the Bondi mass and angular momentum aspects. The evolution of the Bondi mass along $\mathscr{I}$ is governed by the Bondi mass-loss formula, which shows that the Bondi mass strictly decreases for radiating spacetimes.

A further crucial step was taken by Penrose [261, 263], who emphasized that the definition of asymptotic flatness and the characterization of gravitational radiation are most naturally and effectively addressed within the framework of conformal geometry. In this approach [142, 261, 263], the physical spacetime $(\widetilde{M}, \widetilde{g})$ is replaced by an unphysical spacetime $(\mathcal{M}, g, \Omega)$, endowed with a conformal gauge freedom, where infinity is attached as a null boundary $\mathscr{I} = \{\Omega = 0\}$ and is no longer regarded as a limit but as a genuine geometric object [27, 120, 130–132]. The physical and unphysical metrics are related by $\widetilde{g} = \Omega^{-2} g$. Informally, the conformal factor $\Omega$ behaves like $1/r$, so that the infinity of $\widetilde{g}$ is brought to a finite distance with respect to $g$.

There are several ways of defining asymptotic flatness within this conformal framework. One standard definition requires (i) the existence of an embedded null boundary $\mathscr{I}$ such that every null geodesic acquires both future and past endpoints on it, and (ii) the validity of the Einstein equations in a neighbourhood of $\mathscr{I}$. This definition is quite restrictive and already excludes spacetimes that one would like to regard as asymptotically flat, such as Schwarzschild. For this reason, the definition of asymptotic flatness is often relaxed e.g. by requiring the spacetime to be isometric in a neighbourhood of $\mathscr{I}$ to one satisfying conditions (i) and (ii) [264]. Other definitions aim at guaranteeing completeness properties of $\mathscr{I}$ in various senses, see e.g. [143]. The completeness of $\mathscr{I}$ is a particularly important property in the context of black holes.

In four dimensions, the requirement that every null geodesic acquires both future and past endpoints on $\mathscr{I}$ forces the topology of $\mathscr{I}$ to be $\mathbb{R} \times \mathbb{S}^2$ [254]. This has an additional consequence, namely the vanishing of the Weyl tensor at $\mathscr{I}$. Indeed, due to the Bianchi identity, the only potentially non-trivial components of the Weyl tensor at $\mathscr{I}$ correspond to a traceless divergence-free tensor on $\mathbb{S}^2$, which must vanish identically [116, 322]. As a result, most definitions of asymptotic flatness implicitly assume either the topology $\mathbb{R} \times \mathbb{S}^2$ or the vanishing of the Weyl tensor at $\mathscr{I}$. It is important to stress, however, that there exist smooth solutions of the Einstein equations with $\mathscr{I} \not\simeq \mathbb{R} \times \mathbb{S}^2$, for instance with $\mathscr{I} \simeq \mathbb{R} \times T^2$ [299]. In such cases, some null geodesics necessarily fail to reach $\mathscr{I}$. Moreover, the non-existence of non-trivial transverse-traceless tensors is a peculiarity of the two-dimensional sphere. Consequently, in higher dimensions (or already in four dimensions with non-spherical topology) it is far less reasonable to assume a priori that the Weyl tensor vanishes at null infinity.

There are several ways of characterizing radiation within the conformal framework. One possibility is through the news tensor, defined by subtracting from the ambient Schouten tensor a suitable background quantity, the so-called Geroch $\rho$-tensor [142], which in spherical topology is uniquely defined. Another approach is based on an equivalence class of connections on $\mathscr{I}$ and their curvature, yielding an analogous definition of the news [26]. In these approaches, the vanishing of the news tensor is equivalent to the vanishing of the radiative components of the rescaled Weyl tensor at $\mathscr{I}$, and is interpreted as the absence of radiation (see [142] and also [114] for a recent conformally covariant definition of news). Other approaches to the characterization of radiation include the use of the Bel–Robinson tensor (often interpreted as superenergy) [115, 116], tractor connections [90, 158, 160], and Carrollian geometry [159, 280].

Given a compactified spacetime solving the Einstein equations, it is relatively straightforward to analyze the geometry induced at its null infinity. The converse problem is far more subtle: given suitable boundary data at infinity, does there exist an asymptotically flat spacetime realizing them? Addressing this question needs the formulation of appropriate initial value problems and the proof of existence theorems for solutions of the Einstein equations. In the conformal setting, this requires rewriting the field equations in a form that remains regular on the unphysical spacetime, including at infinity. Restricted to four dimensions, such equations were obtained by Friedrich in the early 1980s [122, 123], and we review them in Section 6.2.

The first result concerning the asymptotic characteristic problem was established by Friedrich in the analytic setting [124], and later extended to the smooth case by Rendall [279] and by Kannar [186]. In these works, initial data are posed on an ingoing null hypersurface intersecting $\mathscr{I}^-$, together with data on the portion of $\mathscr{I}^-$ lying in the future of that hypersurface. The data on $\mathscr{I}^-$ encode the radiation, those on the null hypersurface encode the shear, and the data at their intersection encode the induced metric together with the mass and angular momentum aspects. This picture is fully consistent with Sachs' original analysis [289]. Later, following ideas of Luk [208], the domain of existence was enlarged to a full neighbourhood of $\mathcal{N} \cup \mathscr{I}$ [163]. To the best of our knowledge, the asymptotic characteristic problem has only been solved under the assumption that the Weyl tensor vanishes at $\mathscr{I}$. Another asymptotic initial value problem that has been extensively studied is the hyperboloidal one, where initial data are posed on a hyperboloidal hypersurface intersecting $\mathscr{I}$ transversely [22, 125, 127]. In general, hyperboloidal data lead to asymptotic expansions containing logarithmic terms [21], in agreement with other works such as [71, 190, 319–321, 325]. All the results described in this paragraph are restricted to dimension four. The hyperboloidal problem has also been investigated in higher even dimensions through the use of the bulk Fefferman–Graham obstruction tensor $\mathcal{O}^{\mathscr{F}\mathscr{G}}$ [19, 20, 185] as a suitable replacement of the Einstein field equations of the physical spacetime (any conformal class in even dimensions admitting a representative which is Einstein has necessarily vanishing obstruction tensor).

Motivated by the fact that comparatively fewer results are available in higher dimensions and non-spherical topologies, the aim of Chapter 6 is to analyze conformal null infinity in full generality, without assuming any restriction on the spacetime dimension, the topology of $\mathscr{I}$ beyond admitting a foliation by cross-sections, or the vanishing of the Weyl tensor at infinity. More precisely, given a conformal manifold admitting a null infinity, we study how the geometry of $\mathscr{I}$ is constrained by the requirement of asymptotic flatness, understood here as the validity of the conformal Einstein equations order by order at infinity.

As a first step, in Section 6.3 we consider the Fefferman-Graham ambient metric discussed in Subsection 5.1.1 and show that every straight ambient metric admits a conformal completion with a well-defined null infinity, and that in even dimensions the Fefferman-Graham obstruction tensor plays a prominent role at null infinity and can be identified from the geometric properties of $\mathscr{I}$. More specifically, the conformal Einstein equations fail to hold at infinity at a certain order of derivatives precisely when the Fefferman-Graham obstruction tensor of one (and hence any) cross-section of $\mathscr{I}$ does not vanish. We further study the conformal properties of the ambient metric and identify a necessary and sufficient set of conditions that completely characterize a Fefferman-Graham ambient metric from a conformal perspective. Finally, we show that the requirement of the homothety one-form $\boldsymbol{\eta}$ being exact can be relaxed, thereby extending the scope of the construction.

To deal with the general case one can employ the identities obtained in Chapter 5. However, the process is not straightforward because these relations are well adapted to the Einstein equations in the bulk, but not automatically to the study of null infinity. The reason is that the Einstein equations are naturally expressed in terms of the Ricci tensor, whereas the conformal Einstein equations involve additional terms involving the conformal factor. Consequently, the first step of the analysis is to rewrite these identities in a form suitable for the conformal field equations. The resulting identities depend not only on the tensors $\{\mathbf{Y}^{(k)}\}_{k\geq 1}$, but also on the transverse derivatives of the conformal factor $\Omega$ at $\mathscr{I}$, which we denote by $\{\sigma^{(k)}\}_{k\geq 1}$.

Once these identities have been established, we consider a conformal manifold $(\mathcal{M}, g, \Omega)$ and fix[1] the conformal gauge by imposing $|\nabla\Omega|^2 = 0$, which we refer to as a *conformal geodesic gauge*, and whose existence is proven in Section 6.3. Moreover, such gauge is uniquely determined once the value of $\Omega$ is prescribed on a hypersurface transverse to $\mathscr{I}$. Analyzing the conformal Einstein equations order by order in this gauge leads to the following conclusions:

1. At each order, the scalars $\operatorname{tr}_P \mathbf{Y}^{(k)}$, $\kappa^{(k+1)} := -\mathbf{Y}^{(k+1)}(n,n)$, and $\sigma^{(k+1)}$ satisfy a system of equations on $\mathscr{I}$ which, except for one value of $k$, can be solved to determine these quantities in terms of lower-order data and the value of $\sigma^{(k+1)}$ on a chosen cross-section $\Sigma \hookrightarrow \mathscr{I}$. The set of free functions $\{\sigma^{(k)}|_\Sigma\}$ encode the residual conformal freedom within the class of conformal geodesic gauge. For a very specific value of $k = m_1$, however, the system fails to be invertible, and instead we use another conformal equation, the so-called higher order Raychaudhuri equation, that allows us to determine $\operatorname{tr}_P \mathbf{Y}^{(m_1)}$, $\kappa^{(m_1+1)}$, and $\sigma^{(m_1+1)}$ provided an additional free function $\mathfrak{m}$ on $\Sigma$ is prescribed. As pointed out recently in [80], the null Raychaudhuri constraint gives rise to the Bondi mass-loss formula in four dimensions. Our result suggests that a similar conclusion might hold in higher dimensions. This is something that would be worth exploring in the future.

---

1 In order to compute the transverse expansion at $\mathscr{I}$, it is necessary to work with a specific representative of the conformal class, and hence to fix a conformal gauge.

2. Once $\operatorname{tr}_P \mathbf{Y}^{(k)}$, $\kappa^{(k+1)}$, and $\sigma^{(k+1)}$ have been fixed, the one-form $\mathbf{r}^{(k)} := \mathbf{Y}^{(k)}(n, \cdot)$ is completely determined by the conformal equations, except for a specific value of $k = m_2$, at which an additional one-form $\boldsymbol{\beta}$ on $\Sigma$ must be prescribed as free data in order to determine $\mathbf{r}^{(m_2)}$.

3. Finally, after determining $\operatorname{tr}_P \mathbf{Y}^{(k)}$, $\kappa^{(k+1)}$, $\sigma^{(k+1)}$, and $\mathbf{r}^{(k)}$, the remaining components of $\mathbf{Y}^{(k)}$, again for all values of $k$ except one, satisfy a transport equation along the null generator $n$, which can be integrated starting from an initial symmetric trace-free tensor $\mathcal{Y}^{(k)}$ on $\Sigma$. For even-dimensional spacetimes and for a specific value of $k = m_3$, however, $\mathbf{Y}^{(m_3)}$ satisfies no transport equation at all, and the components of $\mathbf{Y}^{(m_3)}$ not encoded in $\operatorname{tr}_P \mathbf{Y}^{(m_3)}$ or $\mathbf{r}^{(m_3)}$ must be prescribed as additional free data.

The number of degrees of freedom in the geometric and detached approach that we present agrees with the analysis in Bondi coordinates in dimension four [87, 289] and also in higher even dimensions [54, 280]. Furthermore, the fact that the tensor $\mathbf{Y}^{(m)}$ is only freely specifiable in even dimensions enforces the idea already pointed out in [171] that there are not smooth radiating odd dimensional spacetimes. This is why most of the definitions of asymptotic flatness in higher dimensions restrict to even dimensional spacetimes [166, 312].

The exceptional cases appearing in the second and third items give rise to two potential obstructions, analogous in spirit to the Fefferman–Graham obstruction tensor. These obstructions have already appeared in the literature in specific situations with other names (see e.g. [280] where they are denoted as the "Coulombian" and "radiative" anomalies). Let us see in more detail why these obstructions appear. The equation determining the one-form $\mathbf{r}^{(k)}$ at each order takes the schematic form

$$\pounds_n \mathbf{r}^{(k)} + \text{lower-order terms} = 0. \tag{1.4}$$

The initial condition for this transport equation is obtained from another conformal equation, which reads

$$(\mathfrak{n} - k)\, \mathbf{r}^{(k)}|_\Sigma + \text{lower-order terms} = 0, \tag{1.5}$$

where $\mathfrak{n}$ denotes the dimension of $\mathscr{I}$. For $k \neq \mathfrak{n}$, equation (1.5) uniquely determines $\mathbf{r}^{(k)}|_\Sigma$, which can then be used as initial data to integrate (1.4). For $k = \mathfrak{n}$, however, an initial condition for $\mathbf{r}^{(\mathfrak{n})}$ must be prescribed, which motivates the introduction of the free one-form $\boldsymbol{\beta}$. Moreover, if the lower-order terms in (1.5) do not vanish identically, the resulting spacetime cannot be smooth beyond this order. In this sense, equation (1.5) defines an obstruction tensor, which we also refer to as the *Coulombian obstruction tensor* and denote by $\mathcal{O}^\Sigma$. In Section 6.6 we derive a necessary and sufficient condition for the vanishing of this obstruction in four spacetime dimensions and relate it to the vanishing of the Weyl tensor at $\mathscr{I}$.

Concerning item 3, the recursive determination of $\mathbf{Y}^{(k)}$ is governed by another transport equation of the form

$$\left(\frac{\mathfrak{n}-1}{2}-k\right)\pounds_n\mathbf{Y}^{(k)}+\text{lower-order terms}=0. \tag{1.6}$$

For $k\neq\frac{\mathfrak{n}-1}{2}$, the tensor $\mathbf{Y}^{(k)}$ can be determined from an initial condition $\mathcal{Y}^{(k)}$ on $\Sigma$. When $k=\frac{\mathfrak{n}-1}{2}$ (which only occurs when $\mathfrak{n}$ is odd), $\mathbf{Y}^{(k)}$ satisfies no transport equation at all, and the components of $\mathbf{Y}^{(k)}$ not encoded in $\operatorname{tr}_P\mathbf{Y}^{(k)}$ or $\mathbf{r}^{(k)}$ must be prescribed as free data. Furthermore, the remainder term in (1.6) defines another obstruction tensor, which we call the *radiative obstruction tensor* and denote by $\mathcal{O}^{\mathscr{I}}$. If this tensor does not vanish identically, the conformal equations cannot be satisfied beyond this order. In Section 6.6 we show that this obstruction vanishes identically in four spacetime dimensions and that, in six dimensions, it is closely related to the Fefferman–Graham obstruction tensor. We also conjecture that this behaviour also emerges in higher dimensions (Conjecture 6.33). The fundamental reason that supports our conjecture is the analysis on the Fefferman-Graham ambient metric described above.

Once the free data have been identified, we prove a uniqueness theorem (Theorem 6.27), showing that any two asymptotically flat spacetimes sharing the same free data at $\mathscr{I}$ are necessarily isometric to infinite order. Our notion of asymptotic flatness is less restrictive than others commonly adopted in the literature, in particular all the ones described above, as it only requires the conformal field equations to be satisfied to infinite order at $\mathscr{I}$.

Having established that the free data fully characterize the geometry at null infinity, we prove the converse statement: given such free data on an abstract null hypersurface, together with the zeroth-order data that we call *$\mathscr{I}$-structure data*, there exists an asymptotically flat conformal spacetime realizing them (Theorem 6.30). The construction of the ambient spacetime is technically involved, since the conformal field equations are highly coupled order by order and, moreover, there are more equations than variables to be determined. A key part of the proof consists in showing that the redundant equations are automatically satisfied once the remaining ones are solved. At the core of this redundancy is the contracted Bianchi identity. While this redundancy is not an issue when the ambient spacetime is already given, it becomes a central aspect when the spacetime is to be constructed order by order, as it is the case in our existence theorem.

## 1.2 CONTENTS OF THE THESIS

This thesis is divided in five chapters.

In Chapter 2 we introduce the hypersurface data formalism, which provides the general geometric framework underlying this thesis. Section 2.1 reviews the main definitions and fundamental properties of hypersurface data. The constraint tensor, a central object throughout the thesis, is studied in detail in Section 2.2. In Section 2.3 we introduce the notion of extended hypersurface data, a tool that will play an important role in Chapters 4, 5, and 6. Section 2.4 is devoted to the analysis of non-degenerate hypersurfaces within this formalism, which are relevant both for the characteristic problem treated in Chapter 3 and for the study of Killing initial data in Chapter 4. Finally, Section 2.5 focuses on the harmonic gauge, a key ingredient in the analysis of the characteristic initial value problem developed in Chapter 3.

Chapter 3 addresses the characteristic Cauchy problem from a detached, spacetime-independent perspective. We begin in Section 3.1 with a brief overview of the main Cauchy problems in General Relativity. Section 3.2 introduces the notion of *double null data*, which constitutes the detached counterpart of two transversely intersecting null hypersurfaces. After establishing some of its basic properties, including the fact that any double null data can be embedded into an ambient spacetime, we prove in Section 3.3 the existence of $\Lambda$-vacuum developments for such data, and address the issue of uniqueness in Section 3.4. Finally, Section 3.5 introduces a related geometric structure, called *pre-double null data*, which allows us to connect our results with several standard formulations of the characteristic problem in the literature.

Chapter 4 is devoted to the Killing initial data (KID) problem. In Section 4.1 we review previous results on both the spacelike and characteristic KID problems. Section 4.2 recalls a collection of classical identities relating a vector field $\eta$ on a semi-Riemannian manifold $(\mathcal{M}, g)$ to its deformation tensor $\mathcal{L}_\eta g$ and to $\mathcal{L}_\eta \nabla$. These identities are then analyzed, in Section 4.3, in the context of embedded hypersurface data of arbitrary causal character. Focusing first on the non-null case, Section 4.4 revisits the spacelike homothetic KID problem within the hypersurface data formalism. Section 4.5 explores the null case, showing that the identities of Section 4.3 give rise to transport equations for certain components of $\mathcal{L}_\eta g$ and $\mathcal{L}_\eta \nabla$ along the null generators of the hypersurface. These equations play a central role in Section 4.6, where we develop a detached formulation of the homothetic KID problem on two intersecting null hypersurfaces. Finally, Section 4.7 presents an additional application of the formalism by addressing the smooth spacelike–characteristic problem.

In Chapter 5 we study the transverse expansion of the metric at a general null hypersurface. Section 5.1 reviews relevant results from the literature, including the cases of Killing and homothetic horizons. Motivated by these examples, Section 5.2 establishes general identities relating transverse derivatives of the ambient Ricci tensor to the transverse expansion of the metric. Using these identities, Section 5.3 proves general existence and uniqueness results for data prescribed on a single null hypersurface. In Section 5.4 we analyze the situation in which the ambient manifold admits a preferred vector field $\eta$ that is null, tangent to the hypersurface, and whose deformation tensor $\mathcal{L}_\eta g$ is known, encompassing (but not restricted to) Killing or homothetic vector fields. These results are then particularized to the case of Killing horizons in Section 5.5.

Chapter 6 is devoted to the study of null infinity. We begin in Section 6.1 with a brief historical overview, including the Bondi–Sachs formalism and Penrose's conformal completion. Section 6.2 reviews the tools of conformal geometry needed in the remainder of the chapter. In Section 6.3 we show that the Fefferman–Graham ambient metric, introduced in Subsection 5.1.1, admits a conformal completion with a null infinity, and we derive necessary and sufficient conditions characterizing the ambient metric from a conformal perspective. In Section 6.4 we write the conformal Einstein equations at null infinity order by order using the methods developed in Chapter 5, and in Section 6.5 we establish existence and uniqueness results for asymptotic initial data at $\mathscr{I}$. Finally, Section 6.6 discusses two potential obstruction tensors arising from the analysis of the conformal Einstein equations at null infinity.

After a final chapter that summarizes our conclusions and discusses future avenues of research, the thesis concludes with five appendices. Appendix A establishes several properties of two hypersurface data tensors that naturally arise in the study of the constraint tensor in Section 2.2. Appendix B collects a number of pullback identities used throughout the thesis. Appendix C reviews the construction of Gaussian null and Rácz–Wald coordinates and provides explicit expressions for certain tensors used in Chapter 6. Appendix D contains auxiliary computations omitted from the main text for clarity. Finally, Appendix E presents the quasi-Einstein equations to all orders at null infinity in full generality.

# 2

# HYPERSURFACE DATA FORMALISM

In this chapter we present the so-called hypersurface data formalism, which constitutes the general framework of this thesis. Originally developed in [221, 222, 226], with early ideas already appearing in [223], this formalism has proved useful in a variety of contexts, including the analysis of hypersurface perturbations [255], the matching of spacetimes [212–215], as well as in the developments presented in this thesis. This chapter is organized as follows. In Section 2.1 we review the main definitions and general properties of hypersurface data. The constraint tensor, a central object throughout the thesis, is the focus of Section 2.2. Section 2.3 explores the notion of extended hypersurface data, a tool that will prove useful in Chapters 4, 5, and 6. In Section 2.4, we analyze non-degenerate submanifolds within hypersurface data, which are relevant to both the characteristic problem (Chapter 3) and the analysis of Killing initial data (Chapter 4). Lastly, Section 2.5 focuses on the harmonic gauge, a tool that plays a key role in Chapter 3. The contents of Sections 2.2, 2.3, 2.4 and 2.5 have been published in [231, 232, 235].

## 2.1 PRELIMINARIES OF HYPERSURFACE DATA FORMALISM

In this section we review the so-called hypersurface data formalism, which as already mentioned is the general framework of this thesis. For further details see [221–223, 226].

### 2.1.1 *Abstract and embedded hypersurface data*

Let us start with the definition of (metric) hypersurface data.

**Definition 2.1.** *Let $\mathcal{H}$ be an $\mathfrak{n}$-dimensional manifold, $\gamma$ a symmetric (0,2)-tensor field, $\boldsymbol{\ell}$ a one-form and $\ell^{(2)}$ a scalar function on $\mathcal{H}$. We say that $\{\mathcal{H}, \gamma, \boldsymbol{\ell}, \ell^{(2)}\}$ defines* ***metric hypersurface data*** *provided that the (0,2) symmetric tensor $\boldsymbol{\mathcal{A}}|_p$ on $T_p\mathcal{H} \times \mathbb{R}$ defined by*

$$\mathcal{A}|_p\left((W,a),(V,b)\right) := \gamma|_p(W,V) + a\boldsymbol{\ell}|_p(V) + b\boldsymbol{\ell}|_p(W) + ab\ell^{(2)}|_p \tag{2.1}$$

*is non-degenerate at every $p \in \mathcal{H}$. A five-tuple $\{\mathcal{H}, \gamma, \boldsymbol{\ell}, \ell^{(2)}, \mathbf{Y}\}$, where $\mathbf{Y}$ is a (0,2) symmetric tensor field on $\mathcal{H}$, is called* ***hypersurface data****.*

The non-degeneracy of $\boldsymbol{\mathcal{A}}$ allows us to introduce its "inverse" $\boldsymbol{\mathcal{A}}^\sharp$ by means of $\boldsymbol{\mathcal{A}}^\sharp(\boldsymbol{\mathcal{A}}((V,a),\cdot),\cdot) = (V,a)$ for every $(V,a) \in \mathfrak{X}(\mathcal{H}) \otimes \mathcal{F}(\mathcal{H})$. From $\boldsymbol{\mathcal{A}}^\sharp$ one can define a $(2,0)$ symmetric tensor field $P$, a vector $n$ and a scalar $n^{(2)}$ on $\mathcal{H}$ by the decomposition

$$\mathcal{A}^\sharp\left((\boldsymbol{\alpha},a),(\boldsymbol{\beta},b)\right) = P(\boldsymbol{\alpha},\boldsymbol{\beta}) + an(\boldsymbol{\beta}) + bn(\boldsymbol{\alpha}) + abn^{(2)} \tag{2.2}$$

for every $(\boldsymbol{\alpha},a),(\boldsymbol{\beta},b) \in \mathfrak{X}^\star(\mathcal{H}) \times F(\mathcal{H})$. Equivalently, $P$, $n$ and $n^{(2)}$ can be defined in abstract index notation by

$$\gamma_{ab}n^b + n^{(2)}\ell_a = 0, \tag{2.3}$$

$$\ell_a n^a + n^{(2)}\ell^{(2)} = 1. \tag{2.4}$$

$$P^{ab}\ell_b + \ell^{(2)}n^a = 0, \tag{2.5}$$

$$P^{ac}\gamma_{cb} + \ell_b n^a = \delta^a_b. \tag{2.6}$$

An important property of the radical of $\boldsymbol{\gamma}$, defined as

$$\mathrm{Rad}(\boldsymbol{\gamma})|_p \coloneqq \{X \in T_p\mathcal{H} \ : \ \boldsymbol{\gamma}(X,\cdot) = 0\},$$

is that it is either empty or one-dimensional [222]. The points $p \in \mathcal{H}$ where $\mathrm{Rad}(\boldsymbol{\gamma})|_p \neq \emptyset$ are called **null points**, and the points $p \in \mathcal{H}$ where $\mathrm{Rad}(\boldsymbol{\gamma})|_p = \emptyset$ are called **non-null points**. Note that no assumptions on the signature of $\boldsymbol{\mathcal{A}}$ have been made (besides being non-degenerate). Moreover, it follows immediately from (2.3)-(2.6) that $p$ is a null point if and only if $n^{(2)}|_p = 0$, and then the radical of $\boldsymbol{\gamma}$ is given by $\mathrm{Rad}(\boldsymbol{\gamma})|_p = \langle n|_p \rangle$. At non-null points (i. e. where $n^{(2)} \neq 0$) it follows from the definition of $\boldsymbol{\mathcal{A}}$ that [222]

$$\det(\boldsymbol{\mathcal{A}}) = \det(\boldsymbol{\gamma})\Big(\ell^{(2)} - \gamma^\sharp(\boldsymbol{\ell},\boldsymbol{\ell})\Big), \tag{2.7}$$

where $\gamma^\sharp$ is the inverse of $\boldsymbol{\gamma}$. Hence the signature[1] of the tensor $\boldsymbol{\mathcal{A}}$ satisfies $\mathrm{sign}(\boldsymbol{\mathcal{A}}) = \mathrm{sign}(\boldsymbol{\gamma}) \sqcup \{\mathrm{sign}(\ell^{(2)} - \boldsymbol{\gamma}^\sharp(\boldsymbol{\ell},\boldsymbol{\ell}))\}$, where $\sqcup$ denotes the disjoint union. At null-points one has [211] $\mathrm{sign}(\boldsymbol{\mathcal{A}}) = \{-1,1\} \sqcup (\mathrm{sign}(\boldsymbol{\gamma}) \setminus \{0\})$.

Despite its name, the notion of (metric) hypersurface data does not view $\mathcal{H}$ as a hypersurface of another manifold. The connection between the abstract data and the standard notion of a hypersurface is as follows.

**Definition 2.2.** *Metric hypersurface data $\{\mathcal{H},\boldsymbol{\gamma},\boldsymbol{\ell},\ell^{(2)}\}$ is **$(\boldsymbol{\Phi},\boldsymbol{\xi})$-embedded** in a semi-Riemannian manifold $(\mathcal{M},g)$ if there exists an embedding $\Phi : \mathcal{H} \hookrightarrow \mathcal{M}$ and a vector field $\xi$ along $\Phi(\mathcal{H})$ everywhere transversal to $\Phi(\mathcal{H})$, called rigging, such that*

$$\Phi^\star(g) = \boldsymbol{\gamma}, \quad \Phi^\star\left(g(\xi,\cdot)\right) = \boldsymbol{\ell}, \quad \Phi^\star\left(g(\xi,\xi)\right) = \ell^{(2)}. \tag{2.8}$$

*Hypersurface data $\{\mathcal{H},\boldsymbol{\gamma},\boldsymbol{\ell},\ell^{(2)},\mathbf{Y}\}$ is $(\Phi,\xi)$-embedded provided that, in addition,*

$$\frac{1}{2}\Phi^\star\left(\pounds_\xi g\right) = \mathbf{Y}. \tag{2.9}$$

1 We take the signature of a quadratic form to be the *unordered* set of -1, 1 and 0 in its canonical expression.

In principle $\Phi^\star(\pounds_\xi g)$ requires an extension of $\xi$ off $\Phi(\mathcal{H})$, but the result is independent of this extension (it only depends on $\xi$ along $\Phi(\mathcal{H})$). Note that given metric hypersurface data embedded in $(\mathcal{M}, g)$, the signature of $\boldsymbol{\mathcal{A}}$ is the same as the signature of $g$. In the context of embedded data, $\Phi(\mathcal{H})$ being a null hypersurface is equivalent to $\boldsymbol{\gamma}$ being degenerate, and hence also equivalent to $n^{(2)} = 0$. Thus, hypersurface data satisfying $n^{(2)} = 0$ is called **null hypersurface data**. A submanifold $\mathcal{S} \hookrightarrow \mathcal{H}$ is said to be a cross-section provided $n$ is everywhere transverse to it. Note that, in the null case, (2.3) and (2.4) become

$$\gamma_{ab} n^b = 0, \qquad (n^{(2)} = 0) \quad (2.10) \qquad\qquad \ell_a n^a = 1. \qquad (n^{(2)} = 0) \quad (2.11)$$

Throughout this thesis, we will indicate the condition "$(n^{(2)} = 0)$" explicitly next to those equations that have been obtained under this assumption, but only in sections dealing with hypersurfaces of arbitrary causal character. When a section begins by specifying that $n^{(2)} = 0$, this will no longer be specified in any of the equations.

Let $\{\mathcal{H}, \boldsymbol{\gamma}, \boldsymbol{\ell}, \ell^{(2)}\}$ be metric hypersurface data $(\Phi, \xi)$-embedded in $(\mathcal{M}, g)$. By the transversality of $\xi$, given a basis $\{e_a\}$ of $T_p\mathcal{H}$, the set $\{\widehat{e}_a := \Phi_\star e_a, \xi\}$ is a basis of $T_{\Phi(p)}\mathcal{M}$. We write the dual basis as $\{\boldsymbol{\theta}^a, \boldsymbol{\nu}\}$, where $\boldsymbol{\nu}$ is the unique normal one-form satisfying $\boldsymbol{\nu}(\xi) = 1$. Raising indices we can introduce $\nu := g^\sharp(\boldsymbol{\nu}, \cdot)$ and $\theta^a := g^\sharp(\boldsymbol{\theta}^a, \cdot)$, which as a consequence of (2.3)-(2.6) are given in terms of $\{\xi, \widehat{e}_a\}$ by

$$\nu = n^{(2)} \xi + n^a \widehat{e}_a, \qquad \theta^a = P^{ab} \widehat{e}_b + n^a \xi. \tag{2.12}$$

For the sake of simplicity we will often abuse the notation and denote $\widehat{e}_a$ by $e_a$. From (2.2) the inverse metric $g^{\alpha\beta}$ at points on $\Phi(\mathcal{H})$ can be written in the basis $\{\xi, e_a\}$ as

$$g^{\alpha\beta} \stackrel{\Phi(\mathcal{H})}{=} P^{ab} e_a^\alpha e_b^\beta + n^a e_a^\alpha \xi^\beta + n^b e_b^\beta \xi^\alpha + n^{(2)} \xi^\alpha \xi^\beta. \tag{2.13}$$

Note that $g^\sharp(\boldsymbol{\nu}, \boldsymbol{\nu}) \stackrel{\Phi(\mathcal{H})}{=} n^{(2)}$. In the null case, $\nu = \Phi_\star n$ is the null normal (cf. (2.10)) and (2.13) becomes

$$g^{\alpha\beta} \stackrel{\Phi(\mathcal{H})}{=} P^{ab} e_a^\alpha e_b^\beta + n^a e_a^\alpha \xi^\beta + n^b e_b^\beta \xi^\alpha. \qquad (n^{(2)} = 0) \quad (2.14)$$

Given hypersurface data one can define the symmetric $(0, 2)$ tensor fields

$$\mathbf{U} := \frac{1}{2} \pounds_n \boldsymbol{\gamma} + \boldsymbol{\ell} \otimes_s dn^{(2)}, \quad (2.15) \qquad\qquad \mathbf{K} := \mathbf{U} + n^{(2)} \mathbf{Y}. \quad (2.16)$$

Note that in the null case one has $\mathbf{U} = \mathbf{K}$, so we will use them interchangeably without further notice in this situation. Using the first relation in (2.12) as well as the above definitions it is easy to see that when the data is embedded $\mathbf{K}$ corresponds to the second fundamental form of $\Phi(\mathcal{H})$ w.r.t. the normal one-form $\boldsymbol{\nu}$. It is also convenient to introduce the $(0, 2)$ tensors

$$\mathbf{F} := \frac{1}{2} d\boldsymbol{\ell}, \qquad (2.17) \qquad\qquad \mathbf{\Pi} := \mathbf{Y} + \mathbf{F} \qquad (2.18)$$

and the contractions

$$\mathbf{s} := \mathbf{F}(n, \cdot), \qquad \mathbf{r} := \mathbf{Y}(n, \cdot), \qquad \kappa := -\mathbf{r}(n). \tag{2.19}$$

Applying the Cartan identity $\pounds_n = \iota_n \circ d + d \circ \iota_n$ to the one-form $\boldsymbol{\ell}$, taking into account (2.17) and using (2.4) one has

$$\pounds_n \boldsymbol{\ell} = 2\mathbf{s} - d(n^{(2)} \ell^{(2)}). \tag{2.20}$$

From this and (2.3)-(2.4) the contraction of $\mathbf{U}$ with $n$ reads [222]

$$\mathbf{U}(n, \cdot) = n^{(2)} \left( -\mathbf{s} + \frac{1}{2} n^{(2)} d\ell^{(2)} \right) + \frac{1}{2} dn^{(2)}. \tag{2.21}$$

In the null case, these two relations become simply

$$\pounds_n \boldsymbol{\ell} = 2\mathbf{s}, \qquad (n^{(2)} = 0) \ (2.22) \qquad\qquad \mathbf{U}(n, \cdot) = 0. \qquad (n^{(2)} = 0) \ (2.23)$$

We conclude this subsection by recalling that, in the null case, given a 2-covariant, symmetric tensor field $T$, it can be decomposed uniquely as [226]

$$T_{ab} = \frac{\operatorname{tr}_P T}{\mathfrak{n} - 1} \gamma_{ab} + 2\ell_{(a} T_{b)c} n^c + T(n, n) \left( \frac{\ell^{(2)}}{\mathfrak{n} - 1} \gamma_{ab} - \ell_a \ell_b \right) + \widehat{T}_{ab}, \tag{2.24}$$

where $\widehat{\mathrm{T}}_{ab}$ is a tensor satisfying $P^{ab} \widehat{\mathrm{T}}_{ab} = 0$ and $\widehat{\mathrm{T}}_{ab} n^a = 0$. We will often call $\widehat{T}_{ab}$ the transverse part of $\mathrm{T}_{ab}$. The tensor $\widehat{\mathrm{T}}_{ab}$ lies on the kernel of the energy-momentum map $\mathcal{T}$, defined over any symmetric $(0, 2)$ tensor $V_{ab}$ by

$$\mathcal{T}(V)^{ab} := \Big( (n^a P^{bc} + n^b P^{ac}) n^d - P^{ab} n^c n^d - n^a n^b P^{cd} \Big) V_{cd}. \tag{2.25}$$

Similarly, for a one-form $\boldsymbol{\omega}$ we define

$$\widehat{\boldsymbol{\omega}} := \boldsymbol{\omega} - \boldsymbol{\omega}(n) \boldsymbol{\ell} \tag{2.26}$$

and call $\widehat{\boldsymbol{\omega}}$ the transverse part of $\boldsymbol{\omega}$, that satisfies $\widehat{\boldsymbol{\omega}}(n) = 0$.

### 2.1.2 *Gauge structure*

Given embedded hypersurface data the notion of rigging vector is not unique, since given any rigging $\xi$, any other vector of the form $\xi' = z(\xi + \Phi_\star V)$ with $(z, V) \in \mathcal{F}^\star(\mathcal{H}) \times \mathfrak{X}(\mathcal{H})$ is also transverse to $\Phi(\mathcal{H})$. This can be translated into the abstract setting by means of the following definition.

**Definition 2.3.** *Let* $\{\mathcal{H}, \boldsymbol{\gamma}, \boldsymbol{\ell}, \ell^{(2)}, \mathbf{Y}\}$ *be hypersurface data and* $(z, V) \in \mathcal{F}^\star(\mathcal{H}) \times \mathfrak{X}(\mathcal{H})$. *We define the* ***gauge*** *transformed hypersurface data with gauge parameters* $(z, V)$ *by*

$$\mathcal{G}_{(z,V)}(\boldsymbol{\gamma}) := \boldsymbol{\gamma}, \tag{2.27}$$

$$\mathcal{G}_{(z,V)}(\boldsymbol{\ell}) := z\left(\boldsymbol{\ell} + \boldsymbol{\gamma}(V, \cdot)\right), \tag{2.28}$$

$$\mathcal{G}_{(z,V)}\Big(\ell^{(2)}\Big) := z^2\Big(\ell^{(2)} + 2\boldsymbol{\ell}(V) + \boldsymbol{\gamma}(V, V)\Big), \tag{2.29}$$

$$\mathcal{G}_{(z,V)}(\mathbf{Y}) := z\mathbf{Y} + \boldsymbol{\ell} \otimes_s dz + \frac{1}{2}\pounds_{zV}\boldsymbol{\gamma}. \tag{2.30}$$

Transformations (2.27)-(2.30) induce the corresponding transformations on $P$ and $n$ [221]

$$\mathcal{G}_{(z,V)}(P) = P + n^{(2)} V \otimes V - 2V \otimes_s n, \tag{2.31}$$

$$\mathcal{G}_{(z,V)}(n) = z^{-1}(n - n^{(2)} V). \tag{2.32}$$

In the null case, they simplify to

$$\mathcal{G}_{(z,V)}(P) = P - 2V \otimes_s n, \qquad (n^{(2)} = 0) \quad (2.33) \qquad \mathcal{G}_{(z,V)}(n) = z^{-1} n. \qquad (n^{(2)} = 0) \quad (2.34)$$

The set of gauge transformations defines a group whose composition law and the inverse element are [222]

$$\mathcal{G}_{(z_1,V_1)} \circ \mathcal{G}_{(z_2,V_2)} = \mathcal{G}_{(z_1 z_2, V_2 + z_2^{-1} V_1)} \quad (2.35) \qquad \mathcal{G}^{-1}_{(z,V)} = \mathcal{G}_{(z^{-1}, -zV)}. \quad (2.36)$$

As expected from the geometric interpretation of $\mathbf{K}$ as the second fundamental form of $\Phi(\mathcal{H})$ w.r.t. $\nu$, its gauge transformation is given by [221]

$$\mathcal{G}_{(z,V)}(\mathbf{K}) = z^{-1}\mathbf{K}. \tag{2.37}$$

The abstract transformation laws (2.28)-(2.30) were obtained by studying the effect of change of rigging on embedded data, and then promoting these transformations into definitions. The following proposition, originally proven in [222], shows the converse, namely that under a gauge transformation the embedded data is still embedded with a change of rigging. We include a proof because it will be useful later when we prove Proposition 3.15 in Chapter 3.

**Proposition 2.4.** *Let* $\{\mathcal{H}, \boldsymbol{\gamma}, \boldsymbol{\ell}, \ell^{(2)}, \mathbf{Y}\}$ *be hypersurface data* $(\Phi, \xi)$*-embedded in* $(M, g)$. *For any gauge element* $(z, V)$, $\{\mathcal{H}, \mathcal{G}_{(z,V)}(\boldsymbol{\gamma}), \mathcal{G}_{(z,V)}(\boldsymbol{\ell}), \mathcal{G}_{(z,V)}(\ell^{(2)}), \mathcal{G}_{(z,V)}(\mathbf{Y})\}$ *is* $(\Phi, \xi')$*-embedded in* $(M, g)$, *with* $\xi'$ *given by* $\xi' := z(\xi + \Phi_\star V)$.

*Proof.* It is sufficient to check that $\Phi^\star(g(\xi', \cdot))$, $\Phi^\star(g(\xi', \xi'))$ and $\frac{1}{2}\Phi^\star(\pounds_{\xi'} g)$ agree with the right hand side of (2.28)-(2.30), respectively. For the first one,

$$\Phi^\star(g(\xi', \cdot)) = \Phi^\star\Big(g(z(\xi + \Phi_\star V), \cdot)\Big) = z\Big(\Phi^\star(g(\xi, \cdot)) + \Phi^\star(g(\Phi_\star V, \cdot))\Big) = z(\boldsymbol{\ell} + \boldsymbol{\gamma}(V, \cdot)),$$

where we used $\Phi^\star(g(\xi, \cdot)) = \boldsymbol{\ell}$. Similarly,

$$\Phi^\star(g(\xi', \xi')) = \Phi^\star\Big(g(z(\xi + \Phi_\star V), z(\xi + \Phi_\star V))\Big) = z^2(\ell^{(2)} + 2\boldsymbol{\ell}(V) + \boldsymbol{\gamma}(V, V))$$

and

$$
\begin{aligned}
\frac{1}{2}\Phi^{\star}(\pounds_{\xi'}g) &= \frac{1}{2}\Phi^{\star}\Big(\pounds_{z(\xi+\Phi_{\star}V)}g\Big)\\
&= \frac{1}{2}z\Phi^{\star}(\pounds_{\xi}g) + \Phi^{\star}\Big(g(\xi,\cdot)\otimes_s dz\Big) + \frac{1}{2}\Phi^{\star}\Big(\pounds_{z\Phi_{\star}V}g\Big)\\
&= z\mathbf{Y} + \boldsymbol{\ell}\otimes_s dz + \frac{1}{2}\pounds_{zV}\boldsymbol{\gamma}.
\end{aligned}
$$

□

### 2.1.3 *Metric hypersurface connection and induced connection*

Given metric hypersurface data $\{\mathcal{H},\boldsymbol{\gamma},\boldsymbol{\ell},\ell^{(2)}\}$ it is possible to define a torsion-free connection $\overset{\circ}{\nabla}$ on $\mathcal{H}$ by means of [221, 222]

$$\overset{\circ}{\nabla}_a\gamma_{bc} = -\ell_c \mathrm{U}_{ab} - \ell_b\mathrm{U}_{ac}, \qquad (2.38) \qquad\qquad \overset{\circ}{\nabla}_a\ell_b = \mathrm{F}_{ab} - \ell^{(2)}\mathrm{U}_{ab}. \qquad (2.39)$$

The following immediate consequence of (2.38) will be used several times throughout this thesis

$$2\overset{\circ}{\nabla}_{(a}\gamma_{b)c} - \overset{\circ}{\nabla}_c\gamma_{ab} = -2\ell_c\mathrm{U}_{ab}. \tag{2.40}$$

When the data is embedded in $(\mathcal{M},g)$, $\overset{\circ}{\nabla}$ is related with the Levi-Civita connection $\nabla$ of $g$ by [221, 223]

$$\nabla_{\Phi_{\star}X}\Phi_{\star}Y \overset{\Phi(\mathcal{H})}{=} \Phi_{\star}\overset{\circ}{\nabla}_X Y - \mathbf{Y}(X,Y)\Phi_{\star}n - (\mathbf{U}+n^{(2)}\mathbf{Y})(X,Y)\xi \qquad \forall X,Y\in\mathfrak{X}(\mathcal{H}). \tag{2.41}$$

Unless otherwise indicated, scalar functions related by $\Phi^{\star}$ are denoted with the same symbol. The action of $\overset{\circ}{\nabla}$ on the contravariant data $\{P,n\}$ is given by [221]

$$\overset{\circ}{\nabla}_a n^b = \mathrm{s}_a n^b + P^{bc}\mathrm{U}_{ac} - n^{(2)}\Big(n^b(d\ell^{(2)})_a + P^{bc}\mathrm{F}_{ac}\Big), \tag{2.42}$$

$$\overset{\circ}{\nabla}_a P^{bc} = -\Big(n^bP^{cd} + n^cP^{bd}\Big)\mathrm{F}_{ad} - n^bn^c(d\ell^{(2)})_a. \tag{2.43}$$

In the following lemma we compute all possible contractions of $\overset{\circ}{\nabla}_a\theta_b$ with $n$ using (2.42).

**Lemma 2.5.** *Let $\{\mathcal{H},\boldsymbol{\gamma},\boldsymbol{\ell},\ell^{(2)}\}$ be metric hypersurface data and $\boldsymbol{\theta}$ any one-form. Then,*

$$n^b\overset{\circ}{\nabla}_a\theta_b = \overset{\circ}{\nabla}_a(\boldsymbol{\theta}(n)) - \boldsymbol{\theta}(n)\mathrm{s}_a - P^{bc}\mathrm{U}_{ac}\theta_b + n^{(2)}\Big(\boldsymbol{\theta}(n)\overset{\circ}{\nabla}_a\ell^{(2)} + P^{bc}\mathrm{F}_{ac}\theta_b\Big), \tag{2.44}$$

$$n^b\overset{\circ}{\nabla}_b\theta_a = \pounds_n\theta_a - \boldsymbol{\theta}(n)\mathrm{s}_a - P^{bc}\mathrm{U}_{ac}\theta_b + n^{(2)}\Big(\boldsymbol{\theta}(n)\overset{\circ}{\nabla}_a\ell^{(2)} + P^{bc}\mathrm{F}_{ac}\theta_b\Big), \tag{2.45}$$

$$2n^b\overset{\circ}{\nabla}_{(a}\theta_{b)} = \pounds_n\theta_a + \overset{\circ}{\nabla}_a(\boldsymbol{\theta}(n)) - 2\Big(\boldsymbol{\theta}(n)\mathrm{s}_a + P^{bc}\mathrm{U}_{ac}\theta_b\Big) + 2n^{(2)}\Big(\boldsymbol{\theta}(n)\overset{\circ}{\nabla}_a\ell^{(2)} + P^{bc}\mathrm{F}_{ac}\theta_b\Big), \tag{2.46}$$

$$n^an^b\overset{\circ}{\nabla}_a\theta_b = \pounds_n(\boldsymbol{\theta}(n)) + n^{(2)}\boldsymbol{\theta}(n)n(\ell^{(2)}) + P^{bc}\theta_b\left(2n^{(2)}\mathrm{s}_c - \frac{1}{2}(n^{(2)})^2\overset{\circ}{\nabla}_c\ell^{(2)} - \frac{1}{2}\overset{\circ}{\nabla}_c n^{(2)}\right). \tag{2.47}$$

*Proof.* From (2.42) it follows

$$n^b\overset{\circ}{\nabla}_a\theta_b = \overset{\circ}{\nabla}_a(\boldsymbol{\theta}(n)) - \theta_b\overset{\circ}{\nabla}_a n^b = \overset{\circ}{\nabla}_a(\boldsymbol{\theta}(n)) - \boldsymbol{\theta}(n)\mathrm{s}_a - P^{bc}\mathrm{U}_{ac}\theta_b + n^{(2)}\Big(\boldsymbol{\theta}(n)\overset{\circ}{\nabla}_a\ell^{(2)} + P^{bc}\mathrm{F}_{ac}\theta_b\Big),$$

and

$$n^b\overset{\circ}{\nabla}_b\theta_a = \pounds_n\theta_a - \theta_b\overset{\circ}{\nabla}_a n^b = \pounds_n\theta_a - \boldsymbol{\theta}(n)\mathrm{s}_a - P^{bc}\mathrm{U}_{ac}\theta_b + n^{(2)}\Big(\boldsymbol{\theta}(n)\overset{\circ}{\nabla}_a\ell^{(2)} + P^{bc}\mathrm{F}_{ac}\theta_b\Big),$$

which are (2.44)-(2.45). Adding them, (2.46) follows. Another contraction with $n$ then gives

$$n^a n^b\overset{\circ}{\nabla}_a\theta_b = \pounds_n(\boldsymbol{\theta}(n)) - P^{bc}\mathrm{U}_{ac}\theta_b n^a + n^{(2)}\Big(\boldsymbol{\theta}(n)n(\ell^{(2)}) + P^{bc}\mathrm{s}_c\theta_b\Big),$$

which becomes (2.47) after using (2.21). □

From (2.43) one can easily compute the Lie derivative of the tensor $P$ along $n$.

**Lemma 2.6.** *Let $\{\mathcal{H},\boldsymbol{\gamma},\boldsymbol{\ell},\ell^{(2)}\}$ be metric hypersurface data. Then,*

$$\pounds_n P^{ab} = \Big(P^{ac}n^b + P^{bc}n^a\Big)\Big(n^{(2)}\overset{\circ}{\nabla}_c\ell^{(2)} - 2\mathrm{s}_c\Big) - 2P^{ac}P^{bd}\mathrm{U}_{cd} - n^a n^b n(\ell^{(2)}). \tag{2.48}$$

*As a consequence, for any (0,2) symmetric tensor field $T_{ab}$ it holds*

$$P^{ab}\pounds_n\mathrm{T}_{ab} = \pounds_n(\mathrm{tr}_P\,\mathbf{T}) + 4P(\boldsymbol{t},\mathbf{s}) + 2P^{ac}P^{bd}\mathrm{U}_{cd}T_{ab} + n(\ell^{(2)})\boldsymbol{t}(n) - 2n^{(2)}P(\boldsymbol{t},d\ell^{(2)}), \tag{2.49}$$

*where $\boldsymbol{t} := \mathbf{T}(n,\cdot)$. In the null case, these expressions become*

$$\pounds_n P^{ab} = -2\Big(P^{ac}n^b + P^{bc}n^a\Big)\mathrm{s}_c - 2P^{ac}P^{bd}\mathrm{U}_{cd} - n^a n^b n(\ell^{(2)}) \qquad (n^{(2)}=0) \tag{2.50}$$

*and*

$$P^{ab}\pounds_n\mathrm{T}_{ab} = \pounds_n(\mathrm{tr}_P\,\mathbf{T}) + 4P(\boldsymbol{t},\mathbf{s}) + 2P^{ac}P^{bd}\mathrm{U}_{cd}T_{ab} + n(\ell^{(2)})\boldsymbol{t}(n). \qquad (n^{(2)}=0) \tag{2.51}$$

*Proof.* Since $\overset{\circ}{\nabla}$ has no torsion, $\pounds_n P^{ab} = n^c\overset{\circ}{\nabla}_c P^{ab} - P^{cb}\overset{\circ}{\nabla}_c n^a - P^{ac}\overset{\circ}{\nabla}_c n^b$. Equation (2.48) is obtained after using (2.42) and (2.43). Inserting (2.48) into $P^{ab}\pounds_n T_{ab} = \pounds_n\,\mathrm{tr}_P\,\boldsymbol{T} - T_{ab}\pounds_n P^{ab}$ (2.49) also follows. □

The curvature tensor of $\overset{\circ}{\nabla}$ is denoted by $\overset{\circ}{R}{}^a{}_{bcd}$ and the Ricci tensor $\overset{\circ}{R}_{bc}$ is the contraction $\overset{\circ}{R}_{bc} = \overset{\circ}{R}{}^a{}_{bac}$. The $\nabla$-derivative of $\xi$ along tangent directions to $\mathcal{H}$ is given by [221]

$$e_a^\mu\nabla_\mu\xi^\beta \overset{\mathcal{H}}{=} (\mathrm{r}-\mathrm{s})_a\xi^\beta + P^{bc}(\mathrm{Y}_{ac}+\mathrm{F}_{ac})e_b^\beta + \frac{1}{2}\nu^\beta\overset{\circ}{\nabla}_a\ell^{(2)}, \tag{2.52}$$

or equivalently

$$e_a^\mu\nabla_\mu\xi^\beta \overset{\mathcal{H}}{=} \left(\mathrm{r}_a - \mathrm{s}_a + \frac{1}{2}n^{(2)}\overset{\circ}{\nabla}_a\ell^{(2)}\right)\xi^\beta + V^b{}_a e_b^\beta, \tag{2.53}$$

where

$$V^b{}_a := P^{bc}(\mathrm{Y}_{ac}+\mathrm{F}_{ac}) + \frac{1}{2}n^b\overset{\circ}{\nabla}_a\ell^{(2)}. \tag{2.54}$$

The following contractions of the tensor $V$ can be easily proven using (2.3)-(2.6),

$$\ell_b V^b{}_a = -\ell^{(2)} \left( \mathrm{r}_a - \mathrm{s}_a + \frac{1}{2} n^{(2)} \overset{\circ}{\nabla}_a \ell^{(2)} \right) + \frac{1}{2} \overset{\circ}{\nabla}_a \ell^{(2)}, \tag{2.55}$$

$$\gamma_{bc} V^b{}_a = \mathrm{Y}_{ac} + \mathrm{F}_{ac} - \left( \mathrm{r}_a - \mathrm{s}_a + \frac{1}{2} n^{(2)} \overset{\circ}{\nabla}_a \ell^{(2)} \right) \ell_c, \tag{2.56}$$

$$V^b{}_a n^a = P^{bc}(\mathrm{r}_c + \mathrm{s}_c) + \frac{1}{2} n(\ell^{(2)}) n^b. \tag{2.57}$$

In particular, from (2.53) and the decomposition of $\nu$ in (2.12),

$$\nu^\mu \nabla_\mu \xi^\beta \stackrel{\mathcal{H}}{=} \left( \frac{1}{2} n^{(2)} n(\ell^{(2)}) - \kappa \right) \xi^\beta + V^b{}_a n^a e_b^\beta + n^{(2)} \xi^\mu \nabla_\mu \xi,$$

and hence in the null case

$$\nu^\mu \nabla_\mu \xi^\beta \stackrel{\mathcal{H}}{=} -\kappa \xi^\beta + \left( P^{ab} (\mathrm{r} + \mathrm{s})_a + \frac{1}{2} n(\ell^{(2)}) n^b \right) e_b^\beta. \qquad (n^{(2)} = 0) \tag{2.58}$$

From the transformation laws for $\{\boldsymbol{\gamma}, \boldsymbol{\ell}, \ell^{(2)}\}$ and $\{P, n, n^{(2)}\}$ one can obtain the gauge behavior of the $\overset{\circ}{\nabla}$-connection [222]

$$\begin{aligned} \mathcal{G}_{(z,V)}(\overset{\circ}{\nabla}) - \overset{\circ}{\nabla} = & \frac{1}{2z} V \otimes \left( \pounds_{zn} \boldsymbol{\gamma} - n^{(2)} \pounds_{zV} \boldsymbol{\gamma} + 2z \boldsymbol{\ell} \otimes_s dn^{(2)} \right) \\ & + \frac{1}{2z} n \otimes \left( \pounds_{zV} \boldsymbol{\gamma} + 2 \boldsymbol{\ell} \otimes_s dz \right). \end{aligned} \tag{2.59}$$

There is another torsion-free connection that can be defined in terms of hypersurface data by means of

$$\overline{\nabla} := \overset{\circ}{\nabla} - n \otimes \mathbf{Y}. \tag{2.60}$$

While $\overset{\circ}{\nabla}$ only depends on metric hypersurface data, $\overline{\nabla}$ manifestly depends also on the tensor $\mathbf{Y}$. Relation (2.41) can be written in terms of $\overline{\nabla}$ using (2.16) as

$$\nabla_{\Phi_\star X} \Phi_\star Y \stackrel{\Phi(\mathcal{H})}{=} \Phi_\star \overline{\nabla}_X Y - \mathbf{K}(X,Y) \xi \qquad \forall X, Y \in \mathfrak{X}(\mathcal{H}). \tag{2.61}$$

This motivates the name *induced connection* for $\overline{\nabla}$ (this connection was first introduced by Schouten in [304] and was extensively studied in [223] where it was called "rigged connection"). Using (2.38)-(2.39) and (2.42)-(2.43) one has

$$\overline{\nabla}_a \gamma_{bc} = -2 \ell_{[b} \mathrm{K}_{c]a}, \tag{2.62}$$
$$\overline{\nabla}_a \ell_b = \mathrm{F}_{ab} + \mathrm{Y}_{ab} - \ell^{(2)} \mathrm{K}_{ab}, \tag{2.63}$$

and

$$\overline{\nabla}_a n^b = P^{bc} \left( \mathrm{K}_{ac} - n^{(2)} (\mathrm{Y}_{ac} + \mathrm{F}_{ac}) \right) - (\mathrm{r}_a - \mathrm{s}_a) n^b - n^{(2)} n^b (d\ell^{(2)})_a, \tag{2.64}$$

$$\overline{\nabla}_a P^{bc} = -P^{cd} (\mathrm{F}_{ad} + \mathrm{Y}_{ad}) n^b - P^{bd} (\mathrm{F}_{ad} + \mathrm{Y}_{ad}) n^c - n^c n^b (d\ell^{(2)})_a. \tag{2.65}$$

Throughout this thesis we will mainly work with null hypersurfaces ($n^{(2)} = 0$) and thus it is convenient to particularize (2.42) and (2.64) to this case,

$$\overset{\circ}{\nabla}_a n^b = \mathrm{s}_a n^b + P^{bc}\mathrm{U}_{ca}, \qquad (n^{(2)} = 0) \quad (2.66)$$

$$\overline{\nabla}_a n^b = P^{bc}\mathrm{U}_{ac} - (\mathrm{r}_a - \mathrm{s}_a) n^b. \qquad (n^{(2)} = 0) \quad (2.67)$$

Note that (2.67) together with (2.61) imply

$$\nabla_{e_a} \nu^\sigma = \Big( P^{cd}\mathrm{U}_{bd} - (\mathrm{r}-\mathrm{s})_b n^c \Big) e_c^\sigma, \qquad (n^{(2)} = 0) \quad (2.68)$$

In addition, from (2.66)-(2.67),

$$n^a \overset{\circ}{\nabla}_a n^b = 0, \qquad (n^{(2)} = 0) \quad (2.69) \qquad\qquad n^a \overline{\nabla}_a n^b = \kappa n^b, \qquad (n^{(2)} = 0) \quad (2.70)$$

and consequently (cf. (2.61))

$$\nu^\alpha \nabla_\alpha \nu^\beta \overset{\Phi(\mathcal{H})}{=} \kappa \nu^\beta, \qquad (n^{(2)} = 0) \quad (2.71)$$

so $\kappa$ measures the surface gravity of the null normal $\nu$. A direct consequence of (2.14) and (2.53) in the null case is

$$g^{\mu\rho}\nabla_\rho \xi^\beta \overset{\mathcal{H}}{=} \Big( P^{cd} e_d^\rho + n^c \xi^\rho \Big)\Big( (\mathrm{r}-\mathrm{s})_c \xi^\beta + V^b{}_c e_b^\beta \Big) + \nu^\rho \xi^\mu \nabla_\mu \xi^\beta. \qquad (n^{(2)} = 0) \quad (2.72)$$

It is also useful to compute the gauge transformation of $\mathbf{r}$ and $\kappa$ in the null case. For the former we simply contract (2.30) with $n'$ and use (2.32) and

$$n^b \pounds_{zV}\gamma_{ab} = z n^b \pounds_V \gamma_{ab} + n(z) \underset{\smile}{V}_a = z\gamma_{ab}\pounds_n V^b + n(z)\underset{\smile}{V}_a \overset{(2.15)}{=} z\pounds_n(\underset{\smile}{V}_a) - 2z\mathrm{U}_{ab}V^b + n(z)\underset{\smile}{V}_a,$$

where $\underset{\smile}{V}_a := \gamma_{ab}V^b$, to get

$$\mathcal{G}_{(z,V)}\mathbf{r}_a = \mathbf{r}_a + \frac{1}{2}\Big( \overset{\circ}{\nabla}_a \log|z| + n(\log|z|)(\ell_a + \underset{\smile}{V}_a) \Big) + \frac{1}{2}\pounds_n \underset{\smile}{V}_a - \mathrm{U}_{ab}V^b. \qquad (n^{(2)} = 0) \quad (2.73)$$

Another contraction with $n'$ yields

$$\mathcal{G}_{(z,V)}\kappa = z^{-1}\kappa + n(z^{-1}). \qquad (n^{(2)} = 0) \quad (2.74)$$

The gauge transformation of $\overline{\nabla}$ can be easily obtained from those of (2.59), (2.32) and (2.30) and is

$$\mathcal{G}_{(z,V)}(\overline{\nabla}) = \overline{\nabla} + V \otimes \mathbf{K}. \qquad (2.75)$$

The simple gauge behaviour of $\overline{\nabla}$ allows us to compute the gauge transformation of the one-form $\mathbf{\Pi}(\cdot, n) = \mathbf{r} - \mathbf{s}$ in the null case very easily as follows. From (2.63) and the property $\mathbf{K}(\cdot, n) = 0$ we have the equality $\Pi_{ab} n^b = n^b \overline{\nabla}_a \ell_b$. Using the transformation law of $\overline{\nabla}$ in (2.75),

$$\mathcal{G}_{(V,z)}(\Pi_{ab} n^b) = \mathcal{G}_{(V,z)}(n^b \overline{\nabla}_a \ell_b) = \mathcal{G}_{(V,z)}(n^b)(\overline{\nabla}_a \mathcal{G}_{(V,z)} \ell_b) = n^b \overline{\nabla}_a \ell_b + z^{-1} \overline{\nabla}_a z + n^b \overline{\nabla}_a (\gamma_{bc} V^c).$$

The last term is $n^b \overline{\nabla}_a (\gamma_{bc} V^c) = -\mathrm{K}_{ac} V^c$ because of (2.62) and recalling $\gamma_{ab} n^a = 0$ and $\ell_a n^a = 1$. So finally

$$\mathcal{G}_{(V,z)} \mathbf{\Pi}(\cdot, n) = \mathbf{\Pi}(\cdot, n) - \mathbf{K}(\cdot, V) + d \log |z|. \qquad (n^{(2)} = 0) \quad (2.76)$$

The transformation for the curvatures of $\overset{\circ}{\nabla}$ and $\overline{\nabla}$ are computed in the following lemma.

**Lemma 2.7.** *Let $\{\mathcal{H}, \boldsymbol{\gamma}, \boldsymbol{\ell}, \ell^{(2)}, \mathbf{Y}\}$ be hypersurface data, $(z, V)$ gauge parameters and $\overset{\circ}{R}$ and $\overline{R}$ the curvatures of $\overset{\circ}{\nabla}$ and $\overline{\nabla}$, respectively. Then,*

$$\mathcal{G}_{(z,V)}(\overset{\circ}{R}{}^a{}_{bcd}) = \overset{\circ}{R}{}^a{}_{bcd} + 2 \overset{\circ}{\nabla}_{[c} \overset{\circ}{S}{}^a{}_{d]b} + 2 \overset{\circ}{S}{}^a{}_{f[c} \overset{\circ}{S}{}^f{}_{d]b}, \qquad (2.77)$$

$$\mathcal{G}_{(z,V)}(\overline{R}{}^a{}_{bcd}) = \overline{R}{}^a{}_{bcd} + 2 \overline{\nabla}_{[c} (V^a \mathrm{K}_{d]b}) + 2 V^a V^f \mathrm{K}_{f[c} \mathrm{K}_{d]b}, \qquad (2.78)$$

*where the tensor $\overset{\circ}{S}$ is defined as the RHS of* (2.59).

*Proof.* Let ${}^{(1)}\nabla$ and ${}^{(2)}\nabla$ be two connections and let $S := {}^{(2)}\nabla - {}^{(1)}\nabla$. Then the curvatures of both connections are related by (see e.g. [323])

$${}^{(2)}R^a{}_{bcd} = {}^{(1)}R^a{}_{bcd} + 2 {}^{(1)}\nabla_{[c} S^a{}_{d]b} + 2 S^a{}_{f[c} S^f{}_{d]b}. \qquad (2.79)$$

Particularizing this first to $\overset{\circ}{\nabla}$ and second to $\overline{\nabla}$ and using (2.59) and (2.75) the result follows at once. □

The gauge transformations of the Ricci tensors of $\overset{\circ}{\nabla}$ and $\overline{\nabla}$ can be obtained from Lemma 2.7 after taking the trace. For later use, we will need the explicit behaviour of $\overset{\circ}{R}_{ab}$ under a transformation of the form $\mathcal{G}_{(z,0)}$.

**Proposition 2.8.** *Let $\{\mathcal{H}, \boldsymbol{\gamma}, \boldsymbol{\ell}, \ell^{(2)}\}$ be null metric hypersurface data and let $\overset{\circ}{R}_{ab}$ be the Ricci tensor of the metric connection $\overset{\circ}{\nabla}$. Let $z \in \mathcal{F}^{\star}(\mathcal{H})$ and $w = \log |z|$. Then,*

$$\begin{aligned}
\mathcal{G}_{(z,0)} \overset{\circ}{R}_{ab} = \overset{\circ}{R}_{ab} &+ \left( \operatorname{tr}_P \mathbf{U} + \frac{1}{2} n(w) \right) \ell_{(a} \overset{\circ}{\nabla}_{b)} w + \mathrm{s}_{(a} \left( \overset{\circ}{\nabla}_{b)} w - n(w) \ell_{b)} \right) \\
&+ \frac{1}{2} \ell_b \overset{\circ}{\nabla}_a n(w) - P^{cd} \mathrm{U}_{d(a} \ell_{b)} \overset{\circ}{\nabla}_c w - \frac{1}{2} \overset{\circ}{\nabla}_a \overset{\circ}{\nabla}_b w + \frac{1}{2} n(w) (\mathrm{F}_{ab} + \ell^{(2)} \mathrm{U}_{ab}) \qquad (2.80) \\
&- \frac{1}{4} \overset{\circ}{\nabla}_a w \overset{\circ}{\nabla}_b w - \frac{1}{4} (n(w))^2 \ell_a \ell_b.
\end{aligned}$$

*Proof.* Let $\nabla$ and $\nabla'$ be two connections and $S := \nabla' - \nabla$. By taking trace in (2.79) the difference of the Ricci tensors of the two connections is

$$R'_{ab} - R_{ab} = \nabla_c S^c{}_{ab} - \nabla_b S^c{}_{ca} + S^c{}_{dc} S^d{}_{ba} - S^c{}_{db} S^d{}_{ca}. \qquad (2.81)$$

Particularizing this identity to $\nabla = \overset{\circ}{\nabla}$ and $\nabla' = \mathcal{G}_{(z,0)}\overset{\circ}{\nabla} = \overset{\circ}{\nabla} + n \otimes \boldsymbol{\ell} \otimes_s dw$ (see (2.59)) gives (note that $S^c{}_{ab} = n^c \ell_{(a}\overset{\circ}{\nabla}_{b)} w$)

$$\overset{\circ}{R}{}'_{ab} - \overset{\circ}{R}_{ab} = \overset{\circ}{\nabla}_c \Big( n^c \ell_{(a}\overset{\circ}{\nabla}_{b)} w \Big) - \overset{\circ}{\nabla}_b \Big( n^c \ell_{(c}\overset{\circ}{\nabla}_{a)} w \Big) + n^c n^d \ell_{(d}\overset{\circ}{\nabla}_{c)} w \, \ell_{(b}\overset{\circ}{\nabla}_{a)} w - n^c n^d \ell_{(d}\overset{\circ}{\nabla}_{b)} w \, \ell_{(c}\overset{\circ}{\nabla}_{a)} w.$$

Using (2.66) and (2.39) the first term is

$$\begin{aligned}
\overset{\circ}{\nabla}_c \Big( n^c \ell_{(a}\overset{\circ}{\nabla}_{b)} w \Big) &= (\operatorname{tr}_P \mathbf{U}) \ell_{(a}\overset{\circ}{\nabla}_{b)} w + n^c \Big( F_{c(a} - \ell^{(2)} \mathrm{U}_{c(a} \Big) \overset{\circ}{\nabla}_{b)} w + n^c \ell_{(a|} \overset{\circ}{\nabla}_c \overset{\circ}{\nabla}_{|b)} w \\
&= (\operatorname{tr}_P \mathbf{U}) \ell_{(a}\overset{\circ}{\nabla}_{b)} w + \mathrm{s}_{(a}\overset{\circ}{\nabla}_{b)} w + n^c \ell_{(a}\overset{\circ}{\nabla}_{b)} \overset{\circ}{\nabla}_c w \\
&= (\operatorname{tr}_P \mathbf{U}) \ell_{(a}\overset{\circ}{\nabla}_{b)} w + \mathrm{s}_{(a}\overset{\circ}{\nabla}_{b)} w + \ell_{(a}\overset{\circ}{\nabla}_{b)} n(w) - n(w) \ell_{(a}\mathrm{s}_{b)} - P^{cd} \mathrm{U}_{d(b}\ell_{a)} \overset{\circ}{\nabla}_c w,
\end{aligned}$$

where in the second line we used that the Hessian is symmetric and in the last one equation (2.44) for $\theta_b = \overset{\circ}{\nabla}_b \omega$ and $n^{(2)} = 0$. Using again (2.39) the second term is given by

$$\begin{aligned}
\overset{\circ}{\nabla}_b \Big( n^c \ell_{(c}\overset{\circ}{\nabla}_{a)} w \Big) &= \frac{1}{2} \overset{\circ}{\nabla}_b \Big( \overset{\circ}{\nabla}_a w + n(w) \ell_a \Big) \\
&= \frac{1}{2} \overset{\circ}{\nabla}_b \overset{\circ}{\nabla}_a w + \frac{1}{2} \ell_a \overset{\circ}{\nabla}_b n(w) + \frac{1}{2} n(w) (\mathrm{F}_{ba} - \ell^{(2)} \mathrm{U}_{ba}).
\end{aligned}$$

The third term is simply given by

$$n^c n^d \ell_{(d}\overset{\circ}{\nabla}_{c)} w \, \ell_{(b}\overset{\circ}{\nabla}_{a)} w = n(w) \ell_{(b}\overset{\circ}{\nabla}_{a)} w.$$

Finally, the last term is

$$\begin{aligned}
n^c n^d \ell_{(d}\overset{\circ}{\nabla}_{b)} w \, \ell_{(c}\overset{\circ}{\nabla}_{a)} w &= \frac{1}{4} n^c n^d \Big( \ell_d \overset{\circ}{\nabla}_b w + \ell_b \overset{\circ}{\nabla}_d w \Big) \Big( \ell_c \overset{\circ}{\nabla}_a w + \ell_a \overset{\circ}{\nabla}_c w \Big) \\
&= \frac{1}{4} \overset{\circ}{\nabla}_b w \overset{\circ}{\nabla}_a w + \frac{1}{4} (n(w))^2 \ell_a \ell_b + \frac{1}{2} n(w) \ell_{(a}\overset{\circ}{\nabla}_{b)} w.
\end{aligned}$$

Combining everything (2.80) follows. □

Applying the previous result, we find that the gauge behaviour of the combination $\overset{\circ}{R}_{(ab)} + \mathrm{s}_a \mathrm{s}_b - \overset{\circ}{\nabla}_{(a}\mathrm{s}_{b)}$ is particularly simple. In particular, it is $\mathcal{G}_{(z,0)}$-invariant when $\mathbf{U} = 0$. This observation will be relevant in Section 5.5.

**Corollary 2.9.** *Let $\{\mathcal{H}, \boldsymbol{\gamma}, \boldsymbol{\ell}, \ell^{(2)}\}$ be null metric hypersurface, $z \in \mathcal{F}^\star(\mathcal{H})$ and $w = \log|z|$. Then,*

$$\begin{aligned}
\mathcal{G}_{(z,0)} \Big( \overset{\circ}{R}_{(ab)} + \mathrm{s}_a \mathrm{s}_b - \overset{\circ}{\nabla}_{(a}\mathrm{s}_{b)} \Big) = \overset{\circ}{R}_{(ab)} &+ \mathrm{s}_a \mathrm{s}_b - \overset{\circ}{\nabla}_{(a}\mathrm{s}_{b)} + (\operatorname{tr}_P \mathbf{U}) \, \ell_{(a}\overset{\circ}{\nabla}_{b)} w \\
&- P^{cd} \mathrm{U}_{d(a}\ell_{b)} \overset{\circ}{\nabla}_c w + n(w) \ell^{(2)} \mathrm{U}_{ab}.
\end{aligned}$$

*Proof.* We denote the gauge transformed data with a prime to simplify the notation. From $\mathbf{s}' = \mathbf{s} + \frac{1}{2} \left( n(w) \boldsymbol{\ell} - dw \right)$ (see [222]) it follows

$$\mathrm{s}'_a \mathrm{s}'_b = \mathrm{s}_a \mathrm{s}_b + n(w) \mathrm{s}_{(a}\ell_{b)} - \mathrm{s}_{(a}\overset{\circ}{\nabla}_{b)} w + \frac{1}{4} (n(w))^2 \ell_a \ell_b - \frac{1}{2} n(w) \ell_{(a}\overset{\circ}{\nabla}_{b)} w + \frac{1}{4} \overset{\circ}{\nabla}_a w \overset{\circ}{\nabla}_b w. \quad (2.82)$$

Using $\overset{\circ}{\nabla}{}' = \overset{\circ}{\nabla} + n \otimes \boldsymbol{\ell} \otimes_s dw$ gives

$$\begin{aligned}
\overset{\circ}{\nabla}{}'_{(a}\mathrm{s}'_{b)} &= \overset{\circ}{\nabla}_{(a}\mathrm{s}'_{b)} - \mathbf{s}'(n)\ell_{(a}\overset{\circ}{\nabla}_{b)}w \\
&= \overset{\circ}{\nabla}_{(a}\mathrm{s}_{b)} + \frac{1}{2}\overset{\circ}{\nabla}_{(a}\Big(n(w)\ell_{b)} - \overset{\circ}{\nabla}_{b)}w\Big) \\
&= \overset{\circ}{\nabla}_{(a}\mathrm{s}_{b)} + \frac{1}{2}\ell_{(b}\overset{\circ}{\nabla}_{a)}n(w) - \frac{1}{2}n(w)\ell^{(2)}\mathrm{U}_{ab} - \frac{1}{2}\overset{\circ}{\nabla}_a\overset{\circ}{\nabla}_b w,
\end{aligned} \tag{2.83}$$

where in the second line we used $\mathbf{s}'(n) = z\mathbf{s}'(n') = 0$ and in the third one we inserted (2.39) and used $\mathrm{F}_{(ab)} = 0$. Combining (2.82) and (2.83) with (2.80) the result follows. □

As proven in [223] the completely tangential components of the ambient Riemann tensor, as well as its 3-tangential, 1-transverse components, can be written in terms of hypersurface data as

$$R_{\alpha\beta\mu\nu}\xi^\alpha e^\beta_b e^\mu_c e^\nu_d \overset{\mathcal{H}}{=} A_{bcd}, \qquad R_{\alpha\beta\mu\nu}e^\alpha_a e^\beta_b e^\mu_c e^\nu_d \overset{\mathcal{H}}{=} B_{abcd}, \tag{2.84}$$

where $A$ and $B$ are the tensors on $\mathcal{H}$ defined by

$$A_{bcd} := \boldsymbol{\ell}_a \overline{R}^a{}_{bcd} + 2\ell^{(2)}\overline{\nabla}_{[d}\mathrm{K}_{c]b} + \mathrm{K}_{b[c}\overline{\nabla}_{d]}\ell^{(2)}, \tag{2.85}$$

$$B_{abcd} := \gamma_{af}\overline{R}^f{}_{bcd} + 2\overline{\nabla}_{[d}(\mathrm{K}_{c]b}\ell_a) + 2\ell^{(2)}\mathrm{K}_{b[c}\mathrm{K}_{d]a}. \tag{2.86}$$

Making the tensor $\mathbf{Y}$ fully explicit (i.e. using (2.16) and (2.60)), the expressions for $A$ and $B$ are

$$A_{bcd} = 2\overset{\circ}{\nabla}_{[d}\mathrm{F}_{c]b} + 2\overset{\circ}{\nabla}_{[d}\mathrm{Y}_{c]b} + \mathrm{U}_{b[d}\overset{\circ}{\nabla}_{c]}\ell^{(2)} + 2\mathrm{Y}_{b[d}(\mathrm{r}-\mathrm{s})_{c]} + n^{(2)}\mathrm{Y}_{b[d}\overset{\circ}{\nabla}_{c]}\ell^{(2)}, \tag{2.87}$$

$$B_{abcd} = \gamma_{af}\overset{\circ}{R}{}^f{}_{bcd} + 2\ell_a\overset{\circ}{\nabla}_{[d}\mathrm{U}_{c]b} + 2\mathrm{U}_{a[d}\mathrm{Y}_{c]b} + 2\mathrm{U}_{b[c}(\mathrm{Y}_{d]a} + \mathrm{F}_{d]a}) + 2n^{(2)}\mathrm{Y}_{b[c}\mathrm{Y}_{d]a}. \tag{2.88}$$

As a consequence, the contraction of the ambient Einstein tensor with the normal $\nu$ can be computed in terms of hypersurface data as follows [221, Prop. 4]

$$G^\alpha{}_\beta\nu_\alpha\xi^\beta \overset{\mathcal{H}}{=} P^{ac}A_{abc}n^b + \frac{1}{2}P^{ac}P^{bd}B_{abcd}, \tag{2.89}$$

$$G^\alpha{}_\beta\nu_\alpha e^\beta_c \overset{\mathcal{H}}{=} n^{(2)}P^{bd}A_{bcd} - A_{bcd}n^b n^d + P^{bd}B_{abcd}n^a. \tag{2.90}$$

## 2.2 THE CONSTRAINT TENSOR

In this section we introduce the notion of constraint tensor in the null case, that will play a fundamental role throughout this thesis. For the purposes of this work we restrict ourselves to the null case ($n^{(2)} = 0$). A definition of the constraint tensor for general hypersurfaces has been given in [211].

A basic observation from (2.84) and (2.14) is that all the tangential components of the ambient Ricci tensor can be written in terms of hypersurface data, because

$$g^{\mu\nu}R_{\alpha\mu\beta\nu}e^\alpha_a e^\beta_b \overset{\mathcal{H}}{=} B_{acbd}P^{cd} - (A_{bca} + A_{acb})n^c. \qquad (n^{(2)} = 0) \tag{2.91}$$

Since the RHS of this expression only depends on hypersurface data, one can define a tensor on $\mathcal{H}$, the so-called *constraint tensor*, that after inserting (2.85)-(2.86) becomes [232]

$$\begin{aligned}\mathcal{R}_{ab} := \Big(\gamma_{af}\overline{R}^{f}{}_{cbd} + 2\overline{\nabla}_{[d}(\mathrm{K}_{b]c}\ell_a) + 2\ell^{(2)}\mathrm{K}_{c[b}\mathrm{K}_{d]a}\Big)\, P^{cd} \\ - \Big(\ell_d\overline{R}^{d}{}_{bac} + 2\ell^{(2)}\overline{\nabla}_{[c}\mathrm{K}_{a]b} + \mathrm{K}_{b[a}\overline{\nabla}_{c]}\ell^{(2)} + \ell_d\overline{R}^{d}{}_{abc} + 2\ell^{(2)}\overline{\nabla}_{[c}\mathrm{K}_{b]a} + \mathrm{K}_{a[b}\overline{\nabla}_{c]}\ell^{(2)}\Big)\, n^c.\end{aligned} \tag{2.92}$$

By construction, when the data is embedded $\mathcal{R}_{ab}$ coincides with the pull-back of the ambient Ricci tensor into the null hypersurface. We show next that $\mathcal{R}_{ab}$ is symmetric and gauge invariant, as one can expect from the embedded data. The proof relies on the following gauge and symmetry properties of the tensors $A$ and $B$, all of them proved in Appendix A.

$$\mathcal{G}_{(z,V)}(A) = z(A + \iota_V B), \qquad \mathcal{G}_{(z,V)}(B) = B \tag{2.93}$$

$$\begin{gathered} A_{b(cd)} = 0, \qquad A_{[bcd]} = 0, \\ B_{a[bcd]} = 0, \qquad B_{(ab)cd} = B_{ab(cd)} = 0, \qquad B_{abcd} = B_{cdab}. \end{gathered} \tag{2.94}$$

Using (2.93) as well as (2.33)-(2.34) one shows

$$\begin{aligned}\mathcal{G}_{(z,V)}\mathcal{R}_{ab} &= \mathcal{G}_{(z,V)}\Big(B_{cadb}P^{cd} - n^c(A_{bac} + A_{abc})\Big) \\ &= \mathcal{R}_{ab} - V^c n^d B_{cadb} - V^d n^c B_{cadb} - V^d n^c B_{dbac} - n^c V^d B_{dabc} \\ &= \mathcal{R}_{ab},\end{aligned}$$

so $\boldsymbol{\mathcal{R}}$ is gauge-invariant. To prove the symmetry we use that $P$ is symmetric and also the property $B_{acbd} = B_{bdac}$ (see (2.94))

$$\mathcal{R}_{ab} = B_{acbd}P^{cd} - (A_{bca} + A_{acb})\, n^c = B_{bdac}P^{cd} - (A_{acb} + A_{bca})\, n^c = \mathcal{R}_{ba}.$$

Observe that the dependence of (2.92) on $\mathbf{Y}$ is not explicit, as it appears inside the connection $\overline{\nabla}$, and hence also on its curvature. The expression of the constraint tensor with all the dependence on $\mathbf{Y}$ explicit was obtained in [217]

$$\begin{aligned}\mathcal{R}_{ab} = \overset{\circ}{R}_{(ab)} - 2\pounds_n \mathrm{Y}_{ab} - (2\kappa + \mathrm{tr}_P\,\mathbf{U})\mathrm{Y}_{ab} + \overset{\circ}{\nabla}_{(a}(\mathrm{s}_{b)} + 2\mathrm{r}_{b)}) \\ - 2\mathrm{r}_a\mathrm{r}_b + 4\mathrm{r}_{(a}\mathrm{s}_{b)} - \mathrm{s}_a\mathrm{s}_b - (\mathrm{tr}_P\,\mathbf{Y})\mathrm{U}_{ab} + 2P^{cd}\mathrm{U}_{d(a}(2\mathrm{Y}_{b)c} + \mathrm{F}_{b)c}).\end{aligned} \tag{2.95}$$

This tensor possesses an interesting hierarchical structure that is crucial in this thesis and can be seen from the contraction of $\mathcal{R}_{ab}$ with $n^a$ and $P^{ab}$. The former was obtained in [217] and relies on the following identity

$$\overset{\circ}{R}_{(ab)}n^a = \frac{1}{2}\pounds_n \mathrm{s}_b - 2P^{ac}\mathrm{U}_{ab}\mathrm{s}_c + P^{ac}\overset{\circ}{\nabla}_c\mathrm{U}_{ab} - \overset{\circ}{\nabla}_b(\mathrm{tr}_P\,\mathbf{U}) + (\mathrm{tr}_P\,\mathbf{U})\mathrm{s}_b. \tag{2.96}$$

The result is

$$\mathcal{R}_{ab}n^a = -\pounds_n(\mathrm{r}_b - \mathrm{s}_b) - \overset{\circ}{\nabla}_b\kappa - (\mathrm{tr}_P\,\mathbf{U})(\mathrm{r}_b - \mathrm{s}_b) - \overset{\circ}{\nabla}_b(\mathrm{tr}_P\,\mathbf{U}) + P^{cd}\overset{\circ}{\nabla}_c\mathrm{U}_{bd} - 2P^{cd}\mathrm{U}_{bd}s_c. \tag{2.97}$$

Another contraction with $n^b$ gives

$$\mathcal{R}_{ab}n^a n^b = -\pounds_n(\operatorname{tr}_P \mathbf{U}) + (\operatorname{tr}_P \mathbf{U})\kappa - P^{ab}P^{cd}\mathrm{U}_{ac}\mathrm{U}_{bd}, \tag{2.98}$$

which is the hypersurface data version of the Raychaudhuri equation. For the trace of (2.95) with respect to $P$ we simply need to use (2.51) and get

$$\begin{aligned}\operatorname{tr}_P \boldsymbol{\mathcal{R}} = \operatorname{tr}_P \overset{\circ}{R} - 2\pounds_n(\operatorname{tr}_P \mathbf{Y}) - 2(\kappa + \operatorname{tr}_P \mathbf{U}) \operatorname{tr}_P \mathbf{Y} + \operatorname{div}_P(\mathbf{s} + 2\mathbf{r})\\ - 2P(\mathbf{r},\mathbf{r}) - 4P(\mathbf{r},\mathbf{s}) - P(\mathbf{s},\mathbf{s}) + 2\kappa n(\ell^{(2)}),\end{aligned} \tag{2.99}$$

where $\operatorname{div}_P \boldsymbol{t} := P^{ab}\overset{\circ}{\nabla}_a t_b$. In Chapter 3 we will find useful to use the following notation

$$H := -\frac{1}{2}\operatorname{tr}_P \boldsymbol{\mathcal{R}}, \qquad \boldsymbol{J} := \boldsymbol{\mathcal{R}}(n,\cdot). \tag{2.100}$$

From the gauge invariance of $\boldsymbol{\mathcal{R}}$ and the transformations (2.33) and (2.34) it follows

$$\begin{aligned}\mathcal{G}_{(z,V)}J_a &= \mathcal{G}_{(z,V)}(\mathcal{R}_{ab}n^b) = z^{-1}\mathcal{R}_{ab}n^b = z^{-1}J_a,\\ \mathcal{G}_{(z,V)}H &= -\frac{1}{2}\mathcal{G}_{(z,V)}(P^{ab}\mathcal{R}_{ab}) = -\frac{1}{2}P^{ab}\mathcal{R}_{ab} + \mathcal{R}_{ab}V^a n^b = H + J_a V^a.\end{aligned}$$

We collect these transformations in the following proposition [232].

**Proposition 2.10.** *Let $\mathcal{D} = \{\mathcal{H},\boldsymbol{\gamma},\boldsymbol{\ell},\ell^{(2)},\mathbf{Y}\}$ be null hypersurface data and $(z,V) \in \mathcal{F}^\star(\mathcal{H}) \times \mathfrak{X}(\mathcal{H})$. Then the tensors $\boldsymbol{\mathcal{R}}$, $H$ and $\boldsymbol{J}$ transform as:*

*1. $\mathcal{G}_{(z,V)}\mathcal{R}_{ab} = \mathcal{R}_{ab}$, 2. $\mathcal{G}_{(z,V)}H = H + V^a J_a$, 3. $\mathcal{G}_{(z,V)}J_a = z^{-1}J_a$.*

The constraint tensor presents an interesting hierarchical structure in the following sense. Suppose that $\mathcal{R}_{ab}$ is known e.g. by means of some field equations, or that it is given in terms of metric hypersurface data, e.g. for $\lambda$-vacuum $\mathcal{R}_{ab} = \lambda\gamma_{ab}$. It is then natural to analyze the existence and uniqueness of solutions to (2.95) regarded as an equation for the tensor $\mathbf{Y}$. From now on we restrict our discussion to $\lambda$-vacuum, i.e. $\mathcal{R}_{ab} = \lambda\gamma_{ab}$, but a similar analysis can be performed in other contexts.

Suppose $\{\mathcal{H},\boldsymbol{\gamma},\boldsymbol{\ell},\ell^{(2)}\}$ is null metric hypersurface data, that $\mathcal{H}$ admits a cross-section $\mathcal{S}$ and that we want to construct a tensor $\mathbf{Y}$ satisfying (i) $\mathcal{R}_{ab} = \lambda\gamma_{ab}$ and such that (ii) $\{\mathcal{H},\boldsymbol{\gamma},\boldsymbol{\ell},\ell^{(2)},\mathbf{Y}\}$ is hypersurface data. From (2.98), the equation

$$-\pounds_n(\operatorname{tr}_P \mathbf{U}) + (\operatorname{tr}_P \mathbf{U})\kappa - P^{ab}P^{cd}\mathrm{U}_{ac}\mathrm{U}_{bd} = 0$$

can be used to either construct $\kappa$ (at the points where $\operatorname{tr}_P \mathbf{U} \neq 0$). At first glance, it seems that inserting this value for $\kappa$ into (2.97) gives a transport equation for the one-form $\mathbf{r}$ along $n$ that can be integrated from an initial condition at $\mathcal{S}$, and that inserting these $\kappa$ and $\mathbf{r}$ into (2.95) yields a transport equation for $\mathrm{Y}_{ab}$ that can be solved from its value on $\mathcal{S}$. But how can one guarantee that $\mathbf{r}(n)$ agrees with $-\kappa$ and that $\mathbf{Y}(n,\cdot)$ agrees with $\mathbf{r}$? And how do we know that the gauge transformation of $\mathbf{Y}$ is the expected one (see (2.30))? We deal with these two questions in Theorem 2.11.

The hierarchical nature of the constraint tensor on a null hypersurface has been previously identified in a more general setting in [226], without imposing assumptions on the topology of $\mathcal{H}$. From a spacetime perspective, it is also well known that the components of the ambient Ricci tensor restricted to a null hypersurface satisfy a hierarchical system of equations, as seen for instance in the null structure equations written in double null coordinates [61, 62, 193], or in the decomposition of the Ricci tensor in Gaussian null coordinates [134, 245]. However, in these approaches the hierarchy is formulated at the level of individual components, in coordinates and within specific gauge choices, and it is not a priori clear that its solutions assemble into a genuine tensorial object intrinsic to the null hypersurface. The purpose of the following theorem is to show that, given null metric hypersurface data, the hierarchical system admits a unique solution that gives rise to a well-defined tensor $\mathbf{Y}$ on $\mathcal{H}$ satisfying the abstract constraint equation $\mathcal{R}_{ab} = \lambda\gamma_{ab}$. As the result is formulated within the hypersurface data formalism, it is in particular fully gauge and diffeomorphism covariant, and does not rely on the existence of an ambient spacetime.

**Theorem 2.11.** *Let $\{\mathcal{H}, \boldsymbol{\gamma}, \boldsymbol{\ell}, \ell^{(2)}\}$ be null metric hypersurface data admitting a cross-section $\iota : \mathcal{S} \hookrightarrow \mathcal{H}$ with metric $h := \iota^\star\boldsymbol{\gamma}$ and assume there exists a smooth function $f$ on $\mathcal{H}$ satisfying*

$$-\pounds_n(\operatorname{tr}_P \mathbf{U}) + (\operatorname{tr}_P \mathbf{U})f - P^{ab}P^{cd}\mathrm{U}_{ac}\mathrm{U}_{bd} = 0, \tag{2.101}$$

*with gauge behavior $\mathcal{G}_{(z,V)}f := z^{-1}(f + n(w))$, $w := \log|z|$. Choose any gauge-invariant real constant $\lambda$ and let $\chi$, $\boldsymbol{\beta}$ and $\mathcal{Y}_{AB}$ be, respectively, a function, a one-form and a $(0,2)$ symmetric tensor on $\mathcal{S}$ traceless w.r.t. $h$, with gauge transformations*

$$\mathcal{G}_{(z,V)}\beta_A := \beta_A + \frac{1}{2}\Big(n(w)\ell_A + \nabla_A w\Big) + \frac{1}{2}\pounds_n \underset{\smile}{V}_A - h^{BC}\mathrm{U}_{AB}V_C + \frac{1}{2}n(w)\underset{\smile}{V}_A, \tag{2.102}$$

$$\mathcal{G}_{(z,V)}\chi := z\chi - \ell^{(2)}n(z) + \frac{1}{2}P^{ab}\pounds_{zV}\gamma_{ab} - 2\boldsymbol{\beta}'(V_\parallel) + 2V_n f', \tag{2.103}$$

$$\begin{aligned}\mathcal{G}_{(z,V)}\mathcal{Y}_{AB} := z\mathcal{Y}_{AB} + \Big(&\ell_{(A}\nabla_{B)}z + \frac{1}{2}\pounds_{zV}\gamma_{AB} - 2z(\ell_{(A} + \underset{\smile}{V}_{(A})\beta'_{B)} \\ &- z^2 f'(\ell_A + \underset{\smile}{V}_A)(\ell_B + \underset{\smile}{V}_B)\Big)^{tf},\end{aligned} \tag{2.104}$$

*where $\underset{\smile}{V} := \boldsymbol{\gamma}(V,\cdot)$, $V_\parallel$ and $V_n$ are defined by $V|_\mathcal{S} = V_\parallel + V_n n$, $V_\parallel \in T\mathcal{S}$, $V_n \in \mathcal{F}(\mathcal{S})$, and "tf" stands for "trace free" w.r.t. $h$. Then, there exists a unique tensor $\mathbf{Y}$ on $\mathcal{H}$ satisfying $\mathbf{Y}(n,n) = -f$, $\operatorname{tr}_P \mathbf{Y}|_\mathcal{S} = \chi$, $\iota^\star(\mathbf{Y}(n,\cdot)) = \boldsymbol{\beta}$ and $(\iota^\star\widehat{\mathrm{Y}})^{tf}_{AB} = \mathcal{Y}_{AB}$ such that $\{\mathcal{H}, \boldsymbol{\gamma}, \boldsymbol{\ell}, \ell^{(2)}, \mathbf{Y}\}$ is null hypersurface data satisfying $\boldsymbol{\mathcal{R}} = \lambda\boldsymbol{\gamma}$.*

*Proof.* We first deal with the existence and uniqueness of $\mathbf{Y}$. The proof will be constructive and requires solving the equation $\mathcal{R}_{ab} = \lambda\gamma_{ab}$, namely

$$\begin{aligned}\lambda\gamma_{ab} = \overset{\circ}{R}_{(ab)} &- 2\pounds_n\mathrm{Y}_{ab} - (2\kappa + \operatorname{tr}_P \mathbf{U})\mathrm{Y}_{ab} + \overset{\circ}{\nabla}_{(a}(\mathrm{s}_{b)} + 2\mathrm{r}_{b)}) \\ &- 2\mathrm{r}_a\mathrm{r}_b + 4\mathrm{r}_{(a}\mathrm{s}_{b)} - \mathrm{s}_a\mathrm{s}_b - (\operatorname{tr}_P \mathbf{Y})\mathrm{U}_{ab} + 2P^{cd}\mathrm{U}_{d(a}(2\mathrm{Y}_{b)c} + \mathrm{F}_{b)c}),\end{aligned} \tag{2.105}$$

regarded as an equation for $\mathrm{Y}_{ab}$. As already anticipated, the difficulty is that this equation is a PDE due to the presence of $\overset{\circ}{\nabla}$-derivatives of $\mathbf{r}$. Nevertheless, we can circumvent this issue exploiting the decoupling property of the constraint tensor. Before integrating (2.105), we shall integrate the equations obtained by contracting it with $n^a$ and $P^{ab}$. A priori the quantities we obtain in this way are not related to $\mathbf{Y}$, so we give them completely different names. A posteriori we shall prove that this intermediate quantities are in fact the correct contractions.

Let us start with the contraction of (2.105) with $n$. That is, we solve the following equation for a one-form $\boldsymbol{c}$ (cf. (2.97))

$$0 = -\pounds_n(c_b - \mathrm{s}_b) - \overset{\circ}{\nabla}_b f - (\mathrm{tr}_P \mathbf{U})(c_b - \mathrm{s}_b) - \overset{\circ}{\nabla}_b(\mathrm{tr}_P \mathbf{U}) + P^{cd}\overset{\circ}{\nabla}_c \mathrm{U}_{bd} - 2P^{cd}\mathrm{U}_{bd} s_c, \quad (2.106)$$

with initial conditions $\iota^\star \boldsymbol{c} = \boldsymbol{\beta}$ and $\mathfrak{c} := -\boldsymbol{c}(n) = f$ on $\mathcal{S}$. Let us first prove that in fact $\mathfrak{c} = f$ everywhere. To that aim, the contraction of (2.106) with $n^b$ gives

$$\pounds_n(\mathfrak{c}) - \pounds_n(f) + (\mathrm{tr}_P \mathbf{U})\mathfrak{c} - n(\mathrm{tr}_P \mathbf{U}) - P^{ab}P^{cd}\mathrm{U}_{ac}\mathrm{U}_{bd} = 0,$$

and subtracting (2.101) it follows that

$$\pounds_n(\mathfrak{c} - f) + (\mathrm{tr}_P \mathbf{U})(\mathfrak{c} - f) = 0.$$

Since $\mathfrak{c} = f$ initially, then $\mathfrak{c} = f$ everywhere on $\mathcal{H}$. The second quantity we shall need is the would-be trace of $\mathbf{Y}$ w.r.t. $P$. We use $t$ for its replacement and solve the equation that comes from taking the trace of (2.105) with $P^{ab}$, namely

$$\begin{aligned} \lambda(\mathfrak{n} - 1) = \mathrm{tr}_P \overset{\circ}{R} - 2\pounds_n(t) - 2(f + \mathrm{tr}_P \mathbf{U})t + \mathrm{div}_P(\mathbf{s} + 2\boldsymbol{c}) - 2P(\boldsymbol{c}, \boldsymbol{c}) \\ - 4P(\boldsymbol{c}, \mathbf{s}) - P(\mathbf{s}, \mathbf{s}) + 2fn(\ell^{(2)}), \end{aligned} \quad (2.107)$$

where $\mathfrak{n}$ is the dimension of $\mathcal{H}$, with initial condition $t|_{\mathcal{S}} = \chi$. Having replaced $\mathbf{r}$ for $\boldsymbol{c}$ and $\mathrm{tr}_P \mathbf{Y}$ for $t$, we may integrate the following equation for $\mathbf{Y}$, obtained directly from (2.105) with the replacements above,

$$\begin{aligned} \lambda\gamma_{ab} = \overset{\circ}{R}_{(ab)} - 2\pounds_n \mathrm{Y}_{ab} - (2f + \mathrm{tr}_P \mathbf{U})\mathrm{Y}_{ab} + \overset{\circ}{\nabla}_{(a}(\mathrm{s}_{b)} + 2c_{b)}) - 2c_a c_b \\ + 4c_{(a}\mathrm{s}_{b)} - \mathrm{s}_a \mathrm{s}_b - t\mathrm{U}_{ab} + 2P^{cd}\mathrm{U}_{d(a}(2\mathrm{Y}_{b)c} + \mathrm{F}_{b)c}). \end{aligned} \quad (2.108)$$

This equation is no longer a PDE, but an ODE along each generator, so it admits a unique solution $\mathbf{Y}$ with initial conditions $\iota^\star(\mathbf{Y}(n, \cdot)) = \boldsymbol{\beta}$, $\mathbf{Y}(n, n)|_{\mathcal{S}} = -f$, $\mathrm{tr}_P \mathbf{Y}|_{\mathcal{S}} = t$ and $(\widehat{\mathrm{Y}}_{AB})^{tf}|_{\mathcal{S}} = \mathcal{Y}_{AB}$ (recall decomposition (2.24)). Having obtained a unique $\mathrm{Y}_{ab}$ on $\mathcal{H}$ let us now prove that $t = \mathrm{tr}_P \mathbf{Y}$ and $\boldsymbol{c} = \mathbf{r}$ on $\mathcal{H}$. Contracting (2.108) with $n^a$ and $P^{ab}$ re-spectively, using (2.46), (2.96), (2.50) and $\mathfrak{c} = f$ it follows that $\mathbf{r}$ and $\operatorname{tr}_P \mathbf{Y}$ satisfy the equations

$$
\begin{aligned}
0 = \pounds_n(\mathrm{s}_b + c_b - 2\mathrm{r}_b) - \mathring{\nabla}_b f - (\operatorname{tr}_P \mathbf{U})(\mathrm{r}_b - \mathrm{s}_b) - \mathring{\nabla}_b(\operatorname{tr}_P \mathbf{U}) + P^{cd}\mathring{\nabla}_c \mathrm{U}_{bd} \\
- 2P^{cd}\mathrm{U}_{bc}(\mathrm{s}_d + c_d - \mathrm{r}_d) + 2f(c_b - \mathrm{r}_b), \qquad (2.109) \\
\lambda(\mathfrak{n}-1) = \operatorname{tr}_P \mathring{R} - 2\pounds_n(\operatorname{tr}_P \mathbf{Y}) - (2f + \operatorname{tr}_P \mathbf{U})\operatorname{tr}_P \mathbf{Y} + \operatorname{div}_P(\mathbf{s} + 2\boldsymbol{c}) - 2P(\boldsymbol{c}, \boldsymbol{c}) \\
+ 4P(\mathbf{s}, \boldsymbol{c}) - 8P(\mathbf{s}, \mathbf{r}) - P(\mathbf{s}, \mathbf{s}) - t \operatorname{tr}_P \mathbf{U} + 2\kappa n(\ell^{(2)}). \qquad (2.110)
\end{aligned}
$$

Subtracting (2.109) from (2.106) gives

$$
2\pounds_n(c_b - \mathrm{r}_b) + (\operatorname{tr}_P \mathbf{U})(c_b - \mathrm{r}_b) - 2P^{cd}\mathrm{U}_{bc}(c_d - \mathrm{r}_d) + 2f(c_b - \mathrm{r}_b) = 0,
$$

which is a homogeneous transport equation for the one-form $\boldsymbol{c} - \mathbf{r}$, and since the initial conditions for $\boldsymbol{c}$ and $\mathbf{r}$ are the same, we conclude that $\boldsymbol{c} = \mathbf{r}$ on all $\mathcal{H}$. Subtracting (2.110) from (2.107) and using $\mathbf{r} = \boldsymbol{c}$ and $\kappa = f$ yields

$$
2\pounds_n(t - \operatorname{tr}_P \mathbf{Y}) + 2f(t - \operatorname{tr}_P \mathbf{Y}) + (\operatorname{tr}_P \mathbf{U})(t - \operatorname{tr}_P \mathbf{Y}) = 0,
$$

so by the same argument as above we conclude that $\operatorname{tr}_P \mathbf{Y} = t$ on $\mathcal{H}$. Hence we have constructed a tensor field $\mathrm{Y}_{ab}$ that satisfies equation (2.105) uniquely determined in terms of $f$, $\chi$, $\boldsymbol{\beta}$ and $\mathcal{Y}_{AB}$.

To prove the remaining claim of the theorem, namely that $\mathbf{Y}$ has the appropriate gauge behaviour, we shall use the gauge invariance of the constraint tensor and the uniqueness result we have just proved. Let $(z, V)$ be gauge parameters and define the tensor $\mathcal{G}_{(z,V)}\mathbf{Y} := z\mathbf{Y} + \boldsymbol{\ell} \otimes_s dz + \frac{1}{2}\pounds_{zV}\boldsymbol{\gamma}$. Let $\mathbf{Y}'$ be the tensor one would construct in a gauge-transformed data, namely the unique tensor satisfying (2.105) with all the quantities gauge-transformed and with initial conditions given by (2.103)-(2.104). By the gauge invariance of the constraint tensor it is clear that $\mathcal{G}_{(z,V)}\mathbf{Y}$ satisfies the same equation than $\mathbf{Y}'$, and by noting that[2] $\iota^\star(\mathbf{Y}'(n', \cdot)) = \mathcal{G}_{(z,V)}\boldsymbol{\beta}$, $\operatorname{tr}_{P'} \mathbf{Y}'|_{\mathcal{S}} = \mathcal{G}_{(z,V)}\chi$ and $(\iota^\star \widehat{\mathrm{Y}}'_{AB})^{tf} = \mathcal{G}_{(z,V)}\mathcal{Y}_{AB}$, the only possibility is $\mathbf{Y}' = \mathcal{G}_{(z,V)}\mathbf{Y}$, and the $\mathbf{Y}$ we have constructed satisfies the gauge transformation (2.30). Therefore $\{\mathcal{H}, \boldsymbol{\gamma}, \boldsymbol{\ell}, \ell^{(2)}, \mathbf{Y}\}$ is hypersurface data satisfying $\boldsymbol{\mathcal{R}} = \lambda\boldsymbol{\gamma}$. □

## 2.3 EXTENDED HYPERSURFACE DATA

As we saw in section 2.1 only the components of the form $R_{\alpha\beta\mu\nu}\xi^\alpha \widehat{e}_b^\beta \widehat{e}_c^\mu \widehat{e}_d^\nu$ and $R_{\alpha\beta\mu\nu}\widehat{e}_a^\alpha \widehat{e}_b^\beta \widehat{e}_c^\mu \widehat{e}_d^\nu$ can be written solely in terms of hypersurface data. The remaining ones, namely $R_{\alpha\beta\mu\nu}\xi^\alpha \widehat{e}_b^\beta \xi^\mu \widehat{e}_d^\nu$, require in addition second order transverse derivatives of the metric. In this section we extend the definition of hypersurface data to incorporate the sufficient in-

2 The gauge behaviors of $f$, $\boldsymbol{\beta}$, $\chi$ and $\mathcal{Y}_{AB}$ have been chosen so that they agree with those of $\kappa$, $\mathbf{r}$, $\operatorname{tr}_P \mathbf{Y}$ and $(\iota^\star \widehat{\mathbf{Y}})^{tf}$, respectively. The transformations of $\kappa$ and $\mathbf{r}$ have been written in (2.73)-(2.74), that of $\operatorname{tr}_P \mathbf{Y}$ follows from (2.33) and (2.30), and that of $(\iota^\star \widehat{\mathbf{Y}})^{tf}$ is a consequence of (2.24).

formation to be able to write down the full Riemann tensor at the hypersurface. We start by recalling a lemma from [224].

**Lemma 2.12.** *Let* $(\mathcal{M}, g)$ *be a semi-Riemannian manifold and* $\xi, X, Y \in \mathfrak{X}(\mathcal{M})$*. Then,*

$$\operatorname{Riem}(X,\xi,Y,\xi) = -\frac{1}{2}\left(\pounds^{(2)}_{\xi} g - \pounds_{\nabla_\xi \xi} g\right)(X,Y) + g(\nabla_X \xi, \nabla_Y \xi). \tag{2.111}$$

An immediate consequence of this expression is that on any embedded hypersurface $\Phi : \mathcal{H} \hookrightarrow \mathcal{M}$ the tensor constructed from second transverse derivatives by means of

$$\Phi^\star\left(\pounds^{(2)}_{\xi} g - \pounds_{\nabla_\xi \xi} g\right)$$

is independent of the extension of $\xi$ off $\Phi(\mathcal{H})$. We can now write down the pullback of $\operatorname{Riem}(\cdot,\xi,\cdot,\xi)$ in terms of this tensor and hypersurface data.

**Proposition 2.13.** *Let* $\{\mathcal{H}, \boldsymbol{\gamma}, \boldsymbol{\ell}, \ell^{(2)}, \mathbf{Y}\}$ *be hypersurface data* $(\Phi,\xi)$*-embedded in* $(\mathcal{M}, g)$*. Then, for any extension of* $\xi$ *off* $\Phi(\mathcal{H})$*,*

$$\Phi^\star\left(\operatorname{Riem}(\cdot,\xi,\cdot,\xi)\right) = -\frac{1}{2}\Phi^\star\left(\pounds^{(2)}_{\xi} g - \pounds_{\nabla_\xi \xi} g\right) + \boldsymbol{\Pi}\cdot\boldsymbol{\Pi} + \left(\mathbf{r} - \mathbf{s} + \frac{1}{4} n^{(2)} d\ell^{(2)}\right) \otimes_s d\ell^{(2)}, \tag{2.112}$$

*where* $(\boldsymbol{\Pi}\cdot\boldsymbol{\Pi})_{ab} := P^{cd}\Pi_{ac}\Pi_{bd}$*.*

*Proof.* From (2.53) together with (2.8) it follows

$$\begin{aligned} g_{\alpha\beta}\nabla_{e_a}\xi^\alpha \nabla_{e_b}\xi^\beta &= \ell^{(2)}\left(\mathrm{r}_a - \mathrm{s}_a + \frac{1}{2} n^{(2)} \overset{\circ}{\nabla}_a \ell^{(2)}\right)\left(\mathrm{r}_b - \mathrm{s}_b + \frac{1}{2} n^{(2)} \overset{\circ}{\nabla}_b \ell^{(2)}\right) \\ &\quad + 2\left(\mathrm{r}_{(a} - \mathrm{s}_{(a} + \frac{1}{2} n^{(2)} \overset{\circ}{\nabla}_{(a} \ell^{(2)}\right) \ell_c V^c{}_{b)} + \gamma_{cd} V^c{}_a V^d{}_b \\ &= P^{cd}\Pi_{ac}\Pi_{bd} + \left(\mathrm{r}_{(a} - \mathrm{s}_{(a} + \frac{1}{4} n^{(2)} \overset{\circ}{\nabla}_{(a} \ell^{(2)}\right) \overset{\circ}{\nabla}_{b)} \ell^{(2)}, \end{aligned}$$

where in the second equality we used (2.55)-(2.56). Equation (2.112) follows from this and (2.111). □

This result suggests the following detached definition [235].

**Definition 2.14.** *A set* $\{\mathcal{H}, \boldsymbol{\gamma}, \boldsymbol{\ell}, \ell^{(2)}, \mathbf{Y}, \mathbf{Z}^{(2)}\}$ *defines* ***extended hypersurface data*** *provided the set* $\{\mathcal{H}, \boldsymbol{\gamma}, \boldsymbol{\ell}, \ell^{(2)}, \mathbf{Y}\}$ *is hypersurface data and* $\mathbf{Z}^{(2)}$ *is a (0,2) symmetric tensor field on* $\mathcal{H}$*. Moreover, an extended hypersurface data set is said to be* $(\Phi,\xi)$*-embedded in a semi-Riemannian manifold* $(\mathcal{M}, g)$ *provided* $\{\mathcal{H}, \boldsymbol{\gamma}, \boldsymbol{\ell}, \ell^{(2)}, \mathbf{Y}\}$ *is* $(\Phi,\xi)$*-embedded in the sense of Definition 2.2 and, in addition,*

$$\mathbf{Z}^{(2)} = \frac{1}{2}\Phi^\star\left(\pounds^{(2)}_{\xi} g - \pounds_{\nabla_\xi \xi} g\right), \tag{2.113}$$

*where* $\xi$ *is any extension of the rigging off* $\Phi(\mathcal{H})$*.*

The fully tangential components of the ambient Ricci tensor are then computable in terms of extended hypersurface data,

$$
\begin{aligned}
R_{\alpha\beta}e^{\alpha}_{a}e^{\beta}_{b} = R_{ab} &= \left(P^{cd}e^{\mu}_{c}e^{\nu}_{d} + n^{c}e^{\mu}_{c}\xi^{\nu} + n^{d}e^{\nu}_{d}\xi^{\mu} + n^{(2)}\xi^{\mu}\xi^{\nu}\right)e^{\alpha}_{a}e^{\beta}_{b}R_{\alpha\mu\beta\nu} \\
&= P^{cd}B_{cadb} + (A_{bca} + A_{acb})n^{c} + n^{(2)}D_{ab},
\end{aligned} \tag{2.114}
$$

where (cf. (2.112))

$$
D_{ab} := -\mathrm{Z}^{(2)}_{ab} + P^{cd}\Pi_{ac}\Pi_{bd} + \left(\mathrm{r}_{(a} - \mathrm{s}_{(a} + \frac{1}{4}n^{(2)}\overset{\circ}{\nabla}_{(a}\ell^{(2)}\right)\overset{\circ}{\nabla}_{b)}\ell^{(2)}. \tag{2.115}
$$

Note that when $n^{(2)} = 0$ we recover (2.91). Solving for $n^{(2)}\mathbf{Z}^{(2)}$ in (2.114) one has

$$
\begin{aligned}
n^{(2)}\mathrm{Z}^{(2)}_{ab} &= P^{cd}B_{cadb} + 2A_{(a|c|b)}n^{c} - R_{ab} \\
&\quad + n^{(2)}\left(P^{cd}\Pi_{ac}\Pi_{bd} + \left(\mathrm{r}_{(a} - \mathrm{s}_{(a} + \frac{1}{4}n^{(2)}\overset{\circ}{\nabla}_{(a}\ell^{(2)}\right)\overset{\circ}{\nabla}_{b)}\ell^{(2)}\right).
\end{aligned}
$$

We now recall the following general identity[3] [211]

$$
\begin{aligned}
P^{cd}B_{cadb} + 2A_{(a|c|b)}n^{c} &= \overset{\circ}{R}_{(ab)} - 2\pounds_{n}\mathrm{Y}_{ab} - \left(2\kappa + \mathrm{tr}_{P}\,\mathbf{U} - n^{(2)}(n(\ell^{(2)}) - \mathrm{tr}_{P}\,\mathbf{Y})\right)\mathrm{Y}_{ab} \\
&\quad + \overset{\circ}{\nabla}_{(a}(\mathrm{s}+2\mathrm{r})_{b)} - 2\mathrm{r}_{a}\mathrm{r}_{b} + 4\mathrm{r}_{(a}\mathrm{s}_{b)} - \mathrm{s}_{a}\mathrm{s}_{b} - (\mathrm{tr}_{P}\,\mathbf{Y})\mathrm{U}_{ab} \\
&\quad + 2P^{cd}\mathrm{U}_{c(a}(2\mathrm{Y}+\mathrm{F})_{b)d} + n^{(2)}\left((\mathrm{s}-3\mathrm{r})_{(a}\overset{\circ}{\nabla}_{b)}\ell^{(2)} + P^{cd}\Pi_{ca}\Pi_{db}\right),
\end{aligned}
$$

and arrive at

$$
\begin{aligned}
n^{(2)}\mathrm{Z}^{(2)}_{ab} &= \overset{\circ}{R}_{(ab)} - R_{ab} - 2\pounds_{n}\mathrm{Y}_{ab} - \left(2\kappa + \mathrm{tr}_{P}\,\mathbf{U} - n^{(2)}(n(\ell^{(2)}) - \mathrm{tr}_{P}\,\mathbf{Y})\right)\mathrm{Y}_{ab} \\
&\quad + \overset{\circ}{\nabla}_{(a}(\mathrm{s}+2\mathrm{r})_{b)} - 2\mathrm{r}_{a}\mathrm{r}_{b} + 4\mathrm{r}_{(a}\mathrm{s}_{b)} - \mathrm{s}_{a}\mathrm{s}_{b} - (\mathrm{tr}_{P}\,\mathbf{Y})\mathrm{U}_{ab} + 2P^{cd}\mathrm{U}_{c(a}(2\mathrm{Y}+\mathrm{F})_{b)d} \\
&\quad + n^{(2)}\left(-2\mathrm{r}_{(a}\overset{\circ}{\nabla}_{b)}\ell^{(2)} + P^{cd}\left(\Pi_{ca}\Pi_{db} + \Pi_{ac}\Pi_{bd}\right) + \frac{1}{4}n^{(2)}\overset{\circ}{\nabla}_{(a}\ell^{(2)}\overset{\circ}{\nabla}_{b)}\ell^{(2)}\right).
\end{aligned} \tag{2.116}
$$

This expression shows, in particular, that given hypersurface data with $n^{(2)} \neq 0$ the second transversal derivative $\mathbf{Z}^{(2)}$ is uniquely determined in terms of the tangential components of the ambient Ricci tensor. This is of course a well-known result that dates back at least to the classic ADM equations in General Relativity [25]. Expression (2.116) is interesting because it provides second order derivatives of the metric along an arbitrary transversal direction $\xi$ in terms of hypersurface data (and the ambient Ricci tensor) in a fully gauge covariant manner.

In terms of $\mathbf{Z}^{(2)}$, equation (2.112) gets rewritten as

$$
\Phi^{\star}\left(\mathrm{Riem}(\xi,\cdot,\cdot,\xi)\right) = \mathbf{Z}^{(2)} - \mathbf{\Pi}\cdot\mathbf{\Pi} - \left(\mathbf{r} - \mathbf{s} + \frac{1}{4}n^{(2)}d\ell^{(2)}\right)\otimes_{s} d\ell^{(2)}. \tag{2.117}
$$

Once the full Riemann tensor at $\mathcal{H}$ is computable from hypersurface data and $\mathbf{Z}^{(2)}$, one can write down explicitly the full Ricci tensor on the abstract null hypersurface [235].

3 This is obtained from (2.87)-(2.88) when as many derivatives of $\mathbf{Y}$ as possible are expressed in terms of Lie derivatives along the direction $n$.

**Proposition 2.15.** *Let* $\mathcal{D} = \{\mathcal{H}, \boldsymbol{\gamma}, \boldsymbol{\ell}, \ell^{(2)}, \mathbf{Y}, \mathbf{Z}^{(2)}\}$ *be extended null hypersurface data* $(\Phi, \xi)$*-embedded in* $(\mathcal{M}, g)$*. Let* Ric *be the Ricci tensor of* $g$ *and* $\dot{\mathcal{R}} := \Phi^\star(\text{Ric}(\xi, \cdot))$, $\ddot{\mathcal{R}} := \Phi^\star(\text{Ric}(\xi, \xi))$*. Then,*

$$\ddot{\mathcal{R}} = -\operatorname{tr}_P \mathbf{Z}^{(2)} + \operatorname{tr}_P(\mathbf{\Pi} \cdot \mathbf{\Pi}) + P(\mathbf{r} - \mathbf{s}, d\ell^{(2)}), \tag{2.118}$$

$$\dot{\mathcal{R}}_c = -P^{ab} A_{abc} + \mathrm{z}^{(2)}_c - P^{ab}\Pi_{da}\Pi_{cb} n^d + \frac{1}{2}\kappa \overset{\circ}{\nabla}_c \ell^{(2)} - \frac{1}{2} n(\ell^{(2)})(\mathrm{r} - \mathrm{s})_c, \tag{2.119}$$

*where* $A_{abc}$ *is defined in* (2.87) *and* $\mathbf{z}^{(2)} := \mathbf{Z}^{(2)}(n, \cdot)$*.*

*Proof.* Let $\{e_a\}$ be a (local) basis of $\mathfrak{X}(\mathcal{H})$ and $\widehat{e}_a := \Phi_\star e_a$. From the definition of the Ricci tensor, the symmetries of the Riemann tensor and equations (2.14) and (2.117) it follows

$$\begin{aligned} \text{Ric}(\xi, \xi) &\overset{\mathcal{H}}{=} \left(P^{ab} \widehat{e}^\alpha_a \widehat{e}^\beta_b + n^a \widehat{e}^\alpha_a \xi^\beta + n^b \widehat{e}^\beta_b \xi^\alpha\right) R_{\alpha\mu\beta\nu} \xi^\mu \xi^\nu \\ &\overset{\mathcal{H}}{=} -P^{ab} \mathrm{Z}^{(2)}_{ab} + P^{ab} P^{cd} \Pi_{ca} \Pi_{db} + P^{ab} (\mathrm{r} - \mathrm{s})_a \overset{\circ}{\nabla}_b \ell^{(2)}. \end{aligned}$$

In the same way, using the first equation in (2.84), given $X \in \mathfrak{X}(\mathcal{H})$

$$\begin{aligned} \text{Ric}(\xi, \Phi_\star X) &\overset{\mathcal{H}}{=} \left(P^{ab} \widehat{e}^\alpha_a \widehat{e}^\beta_b + n^a \widehat{e}^\alpha_a \xi^\beta + n^a \widehat{e}^\beta_a \xi^\alpha\right) R_{\alpha\mu\beta\nu} \xi^\mu (\Phi_\star X)^\nu \\ &\overset{\mathcal{H}}{=} -P^{ab} A_{abc} X^c + \mathrm{Z}^{(2)}_{ab} n^a X^b - P^{cd} \Pi_{ac} \Pi_{bd} n^a X^b - n^a X^b (\mathrm{r} - \mathrm{s})_{(a} \overset{\circ}{\nabla}_{b)} \ell^{(2)}. \end{aligned}$$

□

## 2.4 NON-DEGENERATE SUBMANIFOLDS

In this section we study non-degenerate submanifolds of hypersurface data, so let us assume $\mathcal{D} = \{\mathcal{H}, \boldsymbol{\gamma}, \boldsymbol{\ell}, \ell^{(2)}, \mathbf{Y}\}$ is hypersurface data and $\mathcal{S} \overset{\iota}{\hookrightarrow} \mathcal{H}$ is a non-degenerate submanifold, i.e. $h := \iota^\star \boldsymbol{\gamma}$ is non-degenerate. For our purposes we restrict to codimension one submanifolds, but a similar strategy can be done for higher codimension. The idea is to identify, in terms of hypersurface data, a suitable notion of torsion and second fundamental forms of $\mathcal{S}$. Let us assume $\mathcal{D}$ is $(\Phi, \xi)$-embedded in $(\mathcal{M}, g)$. Given a vector field $t$ along $\Phi(\iota(\mathcal{S}))$, we define the vector field $T \in \mathfrak{X}(\mathcal{H})$ along $\iota(\mathcal{S})$ and the function $\alpha \in \mathcal{F}(\mathcal{S})$ by

$$t \overset{\Phi(\iota(\mathcal{S}))}{=} \alpha \xi + \Phi_\star T. \tag{2.120}$$

From Definition 2.1 it is clear that the condition of $t$ being normal to $\Phi(\iota(\mathcal{S}))$ is equivalent to

$$\alpha \boldsymbol{\ell}(X) + \boldsymbol{\gamma}(T, X) = 0 \qquad \forall X \in \mathfrak{X}(\mathcal{S}).$$

Given $X, Y \in \mathfrak{X}(\mathcal{S})$, the (ambient) second fundamental form $\mathrm{II}^t$ of $\Phi(\iota(\mathcal{S}))$ along $t$ is defined by

$$\mathrm{II}^t(X, Y) := \frac{1}{2} (\pounds_t g)(X, Y).$$

Inserting (2.120) one can check that $\mathrm{II}^t$ can be rewritten as

$$\mathrm{II}^t(X,Y) = \frac{1}{2}(\pounds_T\boldsymbol{\gamma})(X,Y) + \alpha\mathbf{Y}(X,Y) + (d\alpha \otimes_s \boldsymbol{\ell})(X,Y).$$

In addition to the second fundamental form, the extrinsic geometry of a codimension two surface is characterized by the torsion one-form. Given two normal vectors $t_i = \alpha_i\xi + T_i$ and $t_j = \alpha_j\xi + T_j$, the (ambient) torsion $T_{ij}$ is given by

$$T_{ij}(X) = g(t_i, \nabla_X t_j), \qquad \forall X \in \mathfrak{X}(\mathcal{S}).$$

As with the second fundamental form, this tensor can be also rewritten in terms of hypersurface data as follows. From (2.41) and (2.53), the vector $\nabla_X t_j$ can be written as (we omit the $\Phi_\star$ for simplicity)

$$\begin{aligned}\nabla_X t_j &= \Big(X(\alpha_j) - \mathbf{K}(X,T_j)\Big)\xi + \alpha_j\nabla_X\xi + \overset{\circ}{\nabla}_X T_j - \mathbf{Y}(X,T_j)n \\ &= \left(X(\alpha_j) - \mathbf{K}(X,T_j) + \alpha_j\left((\mathbf{r}-\mathbf{s})(X) + \frac{1}{2}n^{(2)}X(\ell^{(2)})\right)\right)\xi \\ &\quad + \alpha_j V^a{}_b X^b e_a + \overset{\circ}{\nabla}_X T_j - \mathbf{Y}(X,T_j)n.\end{aligned}$$

Using the definition of embedded data (Def. 2.2) and (2.55)-(2.56)

$$\begin{aligned}g(\xi,\nabla_X t_j) &= (X(\alpha_j) - \mathbf{U}(X,T_j))\,\ell^{(2)} + \frac{1}{2}\alpha_j X(\ell^{(2)}) + \boldsymbol{\ell}(\overset{\circ}{\nabla}_X T_j) - \mathbf{Y}(X,T_j), \\ g(T_i,\nabla_X t_j) &= (X(\alpha_j) - \mathbf{U}(X,T_j))\,\boldsymbol{\ell}(T_i) + \alpha_j\mathbf{\Pi}(X,T_i) + \boldsymbol{\gamma}(T_i,\overset{\circ}{\nabla}_X T_j),\end{aligned}$$

and therefore

$$\begin{aligned}g(t_i,\nabla_X t_j) = {}& \Big(X(\alpha_j) - \mathbf{U}(X,T_j)\Big)\Big(\ell^{(2)}\alpha_i + \boldsymbol{\ell}(T_i)\Big) + \frac{1}{2}\alpha_i\alpha_j X\Big(\ell^{(2)}\Big) + \alpha_i\boldsymbol{\ell}(\overset{\circ}{\nabla}_X T_j) \\ & - \alpha_i\mathbf{Y}(X,T_j) + \alpha_j\mathbf{\Pi}(X,T_i) + \boldsymbol{\gamma}(T_i,\overset{\circ}{\nabla}_X T_j).\end{aligned}$$

This motivates the following definition at the level of hypersurface data [231].

**Definition 2.16.** *Let $\{\mathcal{H},\boldsymbol{\gamma},\boldsymbol{\ell},\ell^{(2)},\mathbf{Y}\}$ be hypersurface data, $\iota : \mathcal{S} \hookrightarrow \mathcal{H}$ a non-degenerate submanifold, $\alpha \in \mathcal{F}(\mathcal{S})$ and $T \in \mathfrak{X}(\mathcal{H})$ a vector field along $\iota(\mathcal{S})$. We say that $\mathfrak{t} = (\alpha, T)$ is a normal pair provided $\alpha\boldsymbol{\ell}(X) + \boldsymbol{\gamma}(T,X) = 0$ for every $X \in \mathfrak{X}(\mathcal{S})$. We define the second fundamental form of $\mathfrak{t}$ by*

$$\mathcal{K}^{\mathfrak{t}} := \frac{1}{2}\pounds_T\boldsymbol{\gamma} + \alpha\mathbf{Y} + d\alpha \otimes_s \boldsymbol{\ell}. \tag{2.121}$$

*Given two normal pairs $\mathfrak{t}_i = (\alpha_i, T_i)$, $\mathfrak{t}_j = (\alpha_j, T_j)$, we define the torsion one-form of $\mathcal{S}$ by*[4]

$$\begin{aligned}\daleth_{ij}(X) := {}& \Big(X(\alpha_j) - \mathbf{U}(X,T_j)\Big)\Big(\ell^{(2)}\alpha_i + \boldsymbol{\ell}(T_i)\Big) + \frac{1}{2}\alpha_i\alpha_j X(\ell^{(2)}) + \alpha_i\boldsymbol{\ell}(\overset{\circ}{\nabla}_X T_j) \\ & - \alpha_i\mathbf{Y}(X,T_j) + \alpha_j\mathbf{\Pi}(X,T_i) + \boldsymbol{\gamma}(T_i,\overset{\circ}{\nabla}_X T_j).\end{aligned} \tag{2.122}$$

4 $\daleth$ (pronounced "dalet") is the fourth letter of the Hebrew alphabet.

The previous definition is abstract, in the sense that we do not require the data to be embedded in order to define the second fundamental form and the torsion. Of course, when the data happens to be embedded, the tensor (2.121) agrees with the (ambient) second fundamental form along $t[\mathfrak{t}] := \alpha\xi + T$, and (2.122) corresponds to the (ambient) torsion one-form w.r.t. $t_i[\mathfrak{t}_i] = \alpha_i\xi + T_i$ and $t_j[\mathfrak{t}_j] = \alpha_j\xi + T_j$. It is also clear that the space of normal pairs is two dimensional.

As expected from the spacetime interpretation, $\mathcal{K}^{\mathfrak{t}}$ has the following properties.

**Lemma 2.17.** *Let* $\mathfrak{t} = (\alpha, T)$ *be a normal pair,* $\mathcal{K}^{\mathfrak{t}}$ *the second fundamental form as in* (2.121) *and* $f$ *a scalar function. Then,*

$$\mathcal{K}^{f\mathfrak{t}} = f\mathcal{K}^{\mathfrak{t}}.$$

*Proof.* Directly from the definition (2.121),

$$\begin{aligned}
\mathcal{K}^{f\mathfrak{t}}(X,Y) &= \frac{1}{2}\left(\pounds_{fT}\gamma\right)(X,Y) + f\alpha\mathbf{Y}(X,Y) + \left(d\left(f\alpha\right)\otimes_s \boldsymbol{\ell}\right)(X,Y)\\
&= f\mathcal{K}^{\mathfrak{t}}(X,Y) + \left(df\otimes_s \gamma(T,\cdot)\right)(X,Y) + \left(\alpha df\otimes_s \boldsymbol{\ell}\right)(X,Y)\\
&= f\mathcal{K}^{\mathfrak{t}}(X,Y) + \left(df\otimes_s\left(\gamma(T,\cdot)+\alpha\boldsymbol{\ell}\right)\right)(X,Y).
\end{aligned}$$

Since $\mathfrak{t}$ is a normal pair, the term in parenthesis in the last line vanishes when contracted with tangent vectors, which proves the claim. □

**Lemma 2.18.** *Let* $\{\mathfrak{t}_i\}$ *be a basis of normal pairs and* $\hat{\mathfrak{t}}_i = \Omega^j{}_i\mathfrak{t}_j$ *a change of basis, with* $\Omega^j{}_i \in \mathcal{F}(\mathcal{S})$*. Then,*

$$\mathcal{K}^{\hat{\mathfrak{t}}_i} = \Omega^j{}_i\mathcal{K}^{\mathfrak{t}_i}.$$

*Proof.* By Lemma 2.17 it suffices to show that $\mathcal{K}^{\mathfrak{t}_1+\mathfrak{t}_2} = \mathcal{K}^{\mathfrak{t}_1} + \mathcal{K}^{\mathfrak{t}_2}$ for every two normal pairs $\mathfrak{t}_1 = (\alpha_1, T_1)$ and $\mathfrak{t}_2 = (\alpha_2, T_2)$. Directly from the definition of $\mathcal{K}$ in (2.121),

$$\begin{aligned}
\mathcal{K}^{\mathfrak{t}_1+\mathfrak{t}_2}(X,Y) &= \frac{1}{2}\left(\pounds_{T_1+T_2}\gamma\right)(X,Y) + (\alpha_1+\alpha_2)\mathbf{Y}(X,Y)\\
&\quad + \frac{1}{2}\left(X(\alpha_1+\alpha_2)\boldsymbol{\ell}(Y) + Y(\alpha_1+\alpha_2)\boldsymbol{\ell}(X)\right)\\
&= \mathcal{K}^{\mathfrak{t}_1}(X,Y) + \mathcal{K}^{\mathfrak{t}_2}(X,Y).
\end{aligned}$$

Finally, $\mathcal{K}^{\mathfrak{t}_1+\mathfrak{t}_2} = \mathcal{K}^{\mathfrak{t}_1} + \mathcal{K}^{\mathfrak{t}_2}$ together with $\mathcal{K}^{f\mathfrak{t}} = f\mathcal{K}^{\mathfrak{t}}$ proves $\mathcal{K}^{\hat{\mathfrak{t}}_i} = \Omega^j{}_i\mathcal{K}^{\mathfrak{t}_i}$. □

The spacetime picture also motivates the following definition.

**Definition 2.19.** *Let* $\mathcal{D} = \{\mathcal{H}, \gamma, \boldsymbol{\ell}, \ell^{(2)}, \mathbf{Y}\}$ *be hypersurface data and* $\iota : \mathcal{S} \hookrightarrow \mathcal{H}$ *a non-degenerate submanifold. Let* $(z, V)$ *be a gauge group element. The gauge transformation of a normal pair* $\mathfrak{t} = (\alpha, T)$ *of* $\mathcal{S}$ *is defined as*

$$\mathcal{G}_{(z,V)}(\mathfrak{t}) := \Big(\mathcal{G}_{(z,V)}(\alpha) = z^{-1}\alpha, \mathcal{G}_{(z,V)}(T) = T - \alpha V\Big).$$

Note that $\mathcal{G}_{(z,V)}(\mathfrak{t})$ is still a normal pair, since

$$\mathcal{G}_{(z,V)}\alpha\mathcal{G}_{(z,V)}\boldsymbol{\ell}(X) + \gamma(\mathcal{G}_{(z,V)}T, X) = \alpha\boldsymbol{\ell}(X) + \alpha\gamma(V,X) + \gamma(T,X) - \alpha\gamma(V,X) = 0.$$

Moreover, Def. 2.19 is a realization of the gauge group. Indeed, let $(z_1, V_1), (z_2, V_2)$ be gauge parameters. On the one hand,

$$\mathcal{G}_{(z_1,V_1)}\mathcal{G}_{(z_2,V_2)}(\alpha, T) = \mathcal{G}_{(z_1,V_1)}\Big(z_2^{-1}\alpha, T - \alpha V_2\Big) = \Big(z_1^{-1}z_2^{-1}\alpha, T - (z_2^{-1}V_1 + V_2)\alpha\Big),$$

and on the other, using (2.35),

$$\mathcal{G}_{(z_1,V_1)}\mathcal{G}_{(z_2,V_2)}(\alpha, T) = \mathcal{G}_{(z_1 z_2, V_2 + z_2^{-1}V_1)}(\alpha, T) = \Big(z_1^{-1}z_2^{-1}\alpha, T - (z_2^{-1}V_1 + V_2)\alpha\Big).$$

As expected from the spacetime picture, the second fundamental form along a normal pair is gauge invariant.

**Lemma 2.20.** *Let $\mathcal{D}$ be hypersurface data, $\mathcal{S}$ a non-degenerate submanifold and $\mathfrak{t}$ a normal pair. Then $\mathcal{K}^{\mathfrak{t}}$ is gauge invariant.*

*Proof.* As usual we denote the gauge-transformed quantities with primes. Let $\mathfrak{t} = (\alpha, T)$ and $\mathfrak{t}' = \Big(\alpha' = z^{-1}\alpha, T' = T - \alpha V\Big)$. Using the transformation laws (2.28) and (2.30),

$$\begin{aligned}
\mathcal{K}^{\mathfrak{t}'} &= \frac{1}{2}\pounds_{T'}\boldsymbol{\gamma} + \alpha'\mathbf{Y}' + d\alpha' \otimes_s \boldsymbol{\ell}' \\
&= \frac{1}{2}\pounds_{T}\boldsymbol{\gamma} - \frac{1}{2}\alpha\pounds_{V}\boldsymbol{\gamma} - d\alpha \otimes_s \boldsymbol{\gamma}(V,\cdot) + \alpha\mathbf{Y} + z^{-1}\alpha\boldsymbol{\ell} \otimes_s dz + \frac{1}{2}\alpha\pounds_{V}\boldsymbol{\gamma} \\
&\quad + z^{-1}\alpha dz \otimes_s \boldsymbol{\gamma}(V,\cdot) + d\alpha \otimes_s (\boldsymbol{\ell} + \boldsymbol{\gamma}(V,\cdot)) + z\alpha dz^{-1} \otimes_s (\boldsymbol{\ell} + \boldsymbol{\gamma}(V,\cdot)) \\
&= \frac{1}{2}\pounds_{T}\boldsymbol{\gamma} + \alpha\mathbf{Y} + d\alpha \otimes_s \boldsymbol{\ell},
\end{aligned}$$

and hence $\mathcal{K}^{\mathfrak{t}'} = \mathcal{K}^{\mathfrak{t}}$. □

As an example, consider the normal vector $\nu = n^{(2)}\xi + \Phi_\star n$. The normal pair associated with $\nu$ is then $\mathfrak{t}_n := (n^{(2)}, n)$, and obviously it satisfies $n^{(2)}\boldsymbol{\ell}(X) + \boldsymbol{\gamma}(n, X) = 0$ (see (2.3)). The (abstract) second fundamental form of $\mathfrak{t}$ is clearly $\mathcal{K}^{\mathfrak{t}} = \mathbf{K}_{\parallel}$. Another linearly independent normal pair is found in Proposition 2.22, but before we need an intermediate result.

**Lemma 2.21.** *Let $\{\mathcal{H}, \boldsymbol{\gamma}, \boldsymbol{\ell}, \ell^{(2)}\}$ be metric hypersurface data and $\iota : \mathcal{S} \hookrightarrow \mathcal{H}$ a non-degenerate submanifold with metric $h := \iota^\star\boldsymbol{\gamma}$. Let $N \in \mathfrak{X}(\mathcal{H})$ be a nowhere vanishing vector orthogonal to $\mathcal{S}$. Then,*

$$\det(\boldsymbol{\mathcal{A}}) = \Big(\boldsymbol{\gamma}(N,N)(\ell^{(2)} - \ell^{(2)}_{\sharp}) - (\boldsymbol{\ell}(N))^2\Big)\det(h).$$

*Furthermore, when $n^{(2)} \neq 0$ one has*

$$\ell^{(2)} - \gamma^{\sharp}(\boldsymbol{\ell}, \boldsymbol{\ell}) = \ell^{(2)} - \ell^{(2)}_{\sharp} - \frac{\boldsymbol{\ell}(N)^2}{\boldsymbol{\gamma}(N,N)}.$$

*Proof.* Let $\{e_A\}$ be a basis on $\mathcal{S}$ and complete it to a basis $\{e_A, N\}$ on $\mathcal{H}$. In this basis the matrix form of the tensor $\boldsymbol{\mathcal{A}}$ in Definition 2.1 is

$$\boldsymbol{\mathcal{A}} = \begin{pmatrix} h_{AB} & 0 & \ell_A \\ 0 & \gamma(N,N) & \boldsymbol{\ell}(N) \\ \ell_B^T & \boldsymbol{\ell}(N) & \ell^{(2)} \end{pmatrix},$$

where $\ell_A$ is thought as a column and $\ell_A^T$ as a row. Applying standard linear algebra properties the first result follows. The second statement follows after using (2.7) and $\det(\boldsymbol{\gamma}) = \boldsymbol{\gamma}(N,N)\det(h)$. □

**Proposition 2.22.** *Let $\{\mathcal{H}, \boldsymbol{\gamma}, \boldsymbol{\ell}, \ell^{(2)}\}$ be metric hypersurface data and $\iota : \mathcal{S} \hookrightarrow \mathcal{H}$ a non-degenerate submanifold with $h := \iota^\star\boldsymbol{\gamma}$ positive definite. Let $N \in \mathfrak{X}(\mathcal{H})$ be a nowhere vanishing vector orthogonal to $\mathcal{S}$ and define*

$$\beta := \beta_\perp N - \iota_\star \ell^\sharp,$$

*where $\beta_\perp$ solves the following second order algebraic equation*

$$(\beta_\perp)^2 \boldsymbol{\gamma}(N,N) + 2\beta_\perp \boldsymbol{\ell}(N) + \ell^{(2)} - \ell_\sharp^{(2)} = 0. \tag{2.123}$$

*If $n^{(2)} \neq 0$ assume also $n^{(2)}(\ell^{(2)} - \boldsymbol{\gamma}^\sharp(\boldsymbol{\ell}, \boldsymbol{\ell})) > 0$. Then $\mathfrak{t}_\beta := (1, \beta)$ is a normal pair linearly independent to $\mathfrak{t}_n = (n^{(2)}, n)$ and satisfies conditions $\boldsymbol{\mathcal{A}}((n^{(2)}, n), (1, \beta)) = 1$ and $\boldsymbol{\mathcal{A}}((1,\beta),(1,\beta)) = 0$.*

*Proof.* Let $(\alpha, \beta) \in \mathcal{F}(\mathcal{H}) \times \mathfrak{X}(\mathcal{H})$. From (2.3)-(2.4), condition $\boldsymbol{\mathcal{A}}((n^{(2)}, n), (\alpha, \beta)) = 1$ is equivalent to $\alpha = 1$. Since $N$ is orthogonal to $\mathcal{S}$ and $\iota^\star\boldsymbol{\gamma}$ is non-degenerate it follows that $N$ is transverse to $\mathcal{S}$, so we can decompose $\beta$ by means of $\beta = \beta_\perp N + \iota_\star \beta_{||}$, where $\beta_{||} \in \mathfrak{X}(\mathcal{S})$. By imposing that $(1, \beta)$ is a normal pair one obtains $\beta_{||} = -\ell^\sharp$, and finally condition $\boldsymbol{\mathcal{A}}\big((1,\beta),(1,\beta)\big) = 0$ yields

$$\begin{aligned} 0 &= \mathcal{A}\Big((1, \beta_\perp N - \ell^\sharp), (1, \beta_\perp N - \ell^\sharp)\Big) \\ &= \ell^{(2)} + 2\boldsymbol{\ell}(\beta_\perp N - \ell^\sharp) + \boldsymbol{\gamma}(\beta_\perp N - \ell^\sharp, \beta_\perp N - \ell^\sharp) \\ &= \ell^{(2)} + 2\beta_\perp \boldsymbol{\ell}(N) - 2\ell_\sharp^{(2)} + \big(\beta_\perp\big)^2 \boldsymbol{\gamma}(N,N) + \ell_\sharp^{(2)}, \end{aligned}$$

which is (2.123). □

**Remark 2.23.** *In the context of Proposition 2.22, when $n^{(2)} \neq 0$ the condition*

$$n^{(2)}\Big(\ell^{(2)} - \boldsymbol{\gamma}^\sharp(\boldsymbol{\ell}, \boldsymbol{\ell})\Big) > 0$$

*is necessary and sufficient to guarantee the signature of $\boldsymbol{\mathcal{A}}$ to be Lorentzian (see (2.7) and the comment below). In the null case there is no need to impose any condition on the sign of $\ell^{(2)} - \ell_\sharp^{(2)}$ because $\boldsymbol{\mathcal{A}}$ is automatically Lorentzian.*

**Remark 2.24.** *Let us check that equation* (2.123) *always admits real solutions. When* $\boldsymbol{\gamma}(N,N)=0$ *(i.e.* $n^{(2)}=0$*) equation* (2.123) *becomes*

$$2\beta_{\perp}\boldsymbol{\ell}(N)+\ell^{(2)}-\ell_{\sharp}^{(2)}=0,$$

*which always admits a unique solution* $\beta_{\perp}=-\frac{1}{2\boldsymbol{\ell}(N)}(\ell^{(2)}-\ell_{\sharp}^{(2)})$. *To distinguish the non-null and the null cases we shall denote the vector* $\beta$ *in the null case by* $\theta$, *which is therefore given by* $\theta=-\frac{1}{2}(\ell^{(2)}-\ell_{\sharp}^{(2)})n-\iota_{\star}\ell^{\sharp}$. *Since the space of normal pairs is two-dimensional and*

$$\boldsymbol{\mathcal{A}}(\mathfrak{t}_n,\mathfrak{t}_\theta)=\boldsymbol{\gamma}(n,\theta)+\boldsymbol{\ell}(n)=1,$$

*it follows that any normal pair which is also* $\boldsymbol{\mathcal{A}}$*-null lies either in span*$\{\mathfrak{t}_n\}$ *or span*$\{\mathfrak{t}_\theta\}$. *The fact that this vector is unique in the null case can be understood geometrically from the fact that* $(0,n)$ *is already null, so there is only one additional null and normal direction to* $\Phi(\iota(\mathcal{S}))$.

**Remark 2.25.** *When* $\boldsymbol{\gamma}(N,N)\neq 0$ *(i.e. when* $n^{(2)}\neq 0$*) equation* (2.123) *is a proper second order algebraic equation with discriminant given by (see Lemma 2.21)*

$$\Delta=\boldsymbol{\ell}(N)^2-\boldsymbol{\gamma}(N,N)(\ell^{(2)}-\ell_{\sharp}^{(2)})=-\boldsymbol{\gamma}(N,N)(\ell^{(2)}-\boldsymbol{\gamma}^{\sharp}(\boldsymbol{\ell},\boldsymbol{\ell})).$$

*In the same context as Proposition 2.22, sign*$(\boldsymbol{\gamma}(N,N))=-$*sign*$(n^{(2)})$, *and since we are assuming* $n^{(2)}(\ell^{(2)}-\boldsymbol{\gamma}^{\sharp}(\boldsymbol{\ell},\boldsymbol{\ell}))>0$ *it follows that* $\Delta>0$ *and hence equation* (2.123) *admits exactly two solutions (these corresponds geometrically with the two linearly independent null vectors emanating from any codimension-two spacelike surface).*

In the non-null case let us denote by $(1,\beta_+)$ and $(1,\beta_-)$ the two linearly independent normal pairs of Proposition 2.22. By construction, they are $\boldsymbol{\mathcal{A}}$-null (i.e. they satisfy $\boldsymbol{\mathcal{A}}\big((1,\beta_+),(1,\beta_+)\big)=\boldsymbol{\mathcal{A}}\big((1,\beta_-),(1,\beta_-)\big)=0$) and one also has

$$\boldsymbol{\mathcal{A}}\Big((1,\beta_-),(1,\beta_+)\Big)=2\Big(\ell^{(2)}-\boldsymbol{\gamma}^{\sharp}(\boldsymbol{\ell},\boldsymbol{\ell})\Big):=2\mathfrak{a}.$$

Clearly $\mathfrak{a}\neq 0$ (cf. (2.7)). In fact, one can check that $\mathcal{G}_{(z,V)}(\mathfrak{a})=z^2\mathfrak{a}$, so its sign is gauge-invariant, as expected. Therefore $\mathfrak{a}>0$ when $n^{(2)}>0$ and $\mathfrak{a}<0$ when $n^{(2)}<0$. This suggests that in both cases, one of the normal pairs of Prop. 2.22 always points towards $N$ and the other one towards $-N$. Indeed,

$$\boldsymbol{\mathcal{A}}\Big((0,N),(1,\beta_{\pm})\Big)=\boldsymbol{\ell}(N)+\boldsymbol{\gamma}(\beta_{\pm},N)=\pm\sqrt{-\mathfrak{a}\boldsymbol{\gamma}(N,N)}.$$

Since $\mathcal{S}$ is a codimension-one submanifold of $\mathcal{H}$ one has in principle four different torsion one-forms $\urcorner_{ij}$. However, not all them are independent.

**Proposition 2.26.** *Let* $\mathcal{D}=\{\mathcal{H},\boldsymbol{\gamma},\boldsymbol{\ell},\ell^{(2)},\mathbf{Y}\}$ *be hypersurface data,* $\{(\alpha_i,T_i)\}$ *a basis of normal pairs and*

$$M_{ij}:=\mathcal{A}\left((\alpha_i,T_i),(\alpha_j,T_j)\right)=\boldsymbol{\gamma}(T_i,T_j)+\alpha_i\boldsymbol{\ell}(T_j)+\alpha_j\boldsymbol{\ell}(T_i)+\alpha_i\alpha_j\ell^{(2)}. \tag{2.124}$$

*Then,*

$$\daleth_{ij} + \daleth_{ji} = dM_{ij}.$$

*Proof.* Directly from the definition of $\daleth$ in (2.122),

$$\begin{aligned}\daleth_{ij}(X) + \daleth_{ji}(X) = {} & X\left(\alpha_i\alpha_j\ell^{(2)}\right) + 2\left(X(\alpha_{(i}) - \mathbf{U}(X,T_{(i)})\right)\boldsymbol{\ell}(T_{j)}) - 2\ell^{(2)}\alpha_{(i}\mathbf{U}(X,T_{j)}) \\ & + 2\alpha_{(i}\boldsymbol{\ell}\left(\overset{\circ}{\nabla}_X T_{j)}\right) - 2\alpha_{(i}\mathbf{Y}(X,T_{j)}) + 2\boldsymbol{\gamma}\left(T_{(i},\overset{\circ}{\nabla}_X T_{j)}\right) + 2\alpha_{(i}\mathbf{\Pi}(X,T_{j)}).\end{aligned} \tag{2.125}$$

Using (2.38),

$$2\boldsymbol{\gamma}\left(T_{(i},\overset{\circ}{\nabla}_X T_{j)}\right) = X\left(\boldsymbol{\gamma}(T_i,T_j)\right) - (\overset{\circ}{\nabla}_X\boldsymbol{\gamma})(T_i,T_j) = X\left(\boldsymbol{\gamma}(T_i,T_j)\right) + 2\mathbf{U}(X,T_{(i)})\boldsymbol{\ell}(T_{j)}), \tag{2.126}$$

and from (2.39),

$$\begin{aligned}2\alpha_{(i}\boldsymbol{\ell}\left(\overset{\circ}{\nabla}_X T_{j)}\right) &= 2X\left(\alpha_{(i}\boldsymbol{\ell}(T_{j)})\right) - 2X(\alpha_{(i)})\boldsymbol{\ell}(T_{j)}) - 2\alpha_{(i}(\overset{\circ}{\nabla}_X\boldsymbol{\ell})(T_{j)}) \\ &= 2X\left(\alpha_{(i}\boldsymbol{\ell}(T_{j)})\right) - 2X(\alpha_{(i)})\boldsymbol{\ell}(T_{j)}) - 2\alpha_{(i}\mathbf{F}(X,T_{j)}) + 2\ell^{(2)}\alpha_{(i}\mathbf{U}(X,T_{j)}).\end{aligned} \tag{2.127}$$

Inserting (2.126) and (2.127) into (2.125) yields the result. □

In fact, since $\mathcal{S}$ is a codimension one submanifold of $\mathcal{H}$, there is only one independent torsion one-form on $\mathcal{S}$, that can be taken to be $\daleth_{12}$. In the embedded case, given two normal pairs $\mathfrak{t}_i$ and $\mathfrak{t}_j$, the function $M_{ij} = \mathcal{A}(\mathfrak{t}_i,\mathfrak{t}_j)$ as defined in (2.124) corresponds to the scalar product $g(t_i,t_j)$, where $t_i$ and $t_j$ are the ambient vector fields associated to $\mathfrak{t}_i$ and $\mathfrak{t}_j$, respectively. As expected from this geometric interpretation, the functions $M_{ij}$ are gauge invariant, as we show next at the abstract level.

**Lemma 2.27.** *Let $\{\mathfrak{t}_i\}$ be a basis of normal pairs of a non-degenerate submanifold $\mathcal{S}$ and $M_{ij} := \mathcal{A}(\mathfrak{t}_i,\mathfrak{t}_j)$. Then, the functions $M_{ij}$ are gauge invariant.*

*Proof.* Let $(z,V)$ be gauge parameters and write all gauge transformed quantities with a prime. Directly from (2.124) and Def. 2.19,

$$\begin{aligned}M'_{ij} &= \mathcal{A}'\left((z^{-1}\alpha_i, T_i - \alpha_i V),(z^{-1}\alpha_j, T_j - \alpha_j V)\right) \\ &= \boldsymbol{\gamma}\left(T_i - \alpha_i V, T_j - \alpha_j V\right) + z^{-1}\alpha_i\boldsymbol{\ell}'\left(T_j - \alpha_j V\right) + z^{-1}\alpha_j\boldsymbol{\ell}'\left(T_i - \alpha_i V\right) + z^{-2}\alpha_i\alpha_j\ell^{(2)\prime} \\ &= \boldsymbol{\gamma}\left(T_i,T_j\right) + \alpha_i\left(z^{-1}\boldsymbol{\ell}'(T_j) - \boldsymbol{\gamma}(V,T_j)\right) + \alpha_j\left(z^{-1}\boldsymbol{\ell}'(T_i) - \boldsymbol{\gamma}(V,T_i)\right) \\ &\quad + \alpha_i\alpha_j\left(z^{-2}\ell^{(2)\prime} + \boldsymbol{\gamma}(V,V) - 2z^{-1}\boldsymbol{\ell}'(V)\right) \\ &= \boldsymbol{\gamma}(T_i,T_j) + \alpha_i\boldsymbol{\ell}(T_j) + \alpha_j\boldsymbol{\ell}(T_i) + \alpha_i\alpha_j\ell^{(2)} \\ &= M_{ij},\end{aligned}$$

where in the fourth line we used (2.27)-(2.29). □

As expected from the geometric interpretation of the ambient torsion one-forms as the connection coefficients of the normal bundle connection, the transformation of the abstract torsion one-forms under a change of basis of normal pairs is as follows.

**Proposition 2.28.** *Let* $\{\mathfrak{t}_i\}$ *be a basis of normal pairs of* $\mathcal{S}$ *and* $\tilde{\mathfrak{t}}_j = \Omega^j{}_j \mathfrak{t}_i$ *with* $\Omega^j{}_j \in \mathcal{F}(\mathcal{S})$ *a change of basis. Then,*

$$\widetilde{\daleth}_{ij} = \Omega^k{}_i \Omega^l{}_j \daleth_{kl} + \Omega^k{}_i M_{kl} \, d\Omega^l{}_j. \tag{2.128}$$

*Proof.* Writing $\mathfrak{t}_i = (\alpha_i, T_i)$ and $\tilde{\mathfrak{t}}_i = (\widetilde{\alpha}_i, \widetilde{T}_i)$, the change of basis gives

$$\widetilde{\alpha}_i = \Omega^k{}_i \alpha_k, \qquad \widetilde{T}_i = \Omega^k{}_i T_k.$$

Inserting them in the expression of $\daleth_{ij}$ in (2.122) and noting that all the terms are multilinear except for $X(\widetilde{\alpha}_j)$ and $\overset{\circ}{\nabla}_X \widetilde{T}_j$, which become

$$X(\widetilde{\alpha}_j) = \Omega^l{}_j X(\alpha_l) + \alpha_l X(\Omega^l{}_j), \qquad \overset{\circ}{\nabla}_X \widetilde{T}_j = \Omega^l{}_j \overset{\circ}{\nabla}_X T_l + X(\Omega^l{}_j) T_l,$$

one immediately gets

$$\widetilde{\daleth}_{ij}(X) = \Omega^k{}_i \Omega^l{}_j \daleth_{kl} + \Omega^k{}_i \left( \alpha_k \alpha_l \ell^{(2)} + \alpha_k \boldsymbol{\ell}(T_l) + \alpha_l \boldsymbol{\ell}(T_k) + \boldsymbol{\gamma}(T_k, T_l) \right) X(\Omega^l{}_j),$$

which is (2.128) after recalling the definition of $M_{kl}$ in (2.124). □

**Corollary 2.29.** *Let* $\mathfrak{t}_1$ *and* $\mathfrak{t}_2$ *two linearly independent normal pairs and* $f$ *a scalar function. Under a change of basis* $\{\mathfrak{t}_1, \mathfrak{t}_2\} \longmapsto \{\tilde{t}_1 = f^{-1}\mathfrak{t}_1, \tilde{t}_2 = f\mathfrak{t}_2\}$*, the torsion one-form* $\daleth(\mathfrak{t}_1, \mathfrak{t}_2)$ *transforms as*

$$\widetilde{\daleth}_{12} = \daleth_{12} + M_{12} d \log |f|.$$

To conclude this section it remains to show that the torsion one-forms are gauge invariant. While this is expected from an ambient point of view, one must prove it at the level of the abstract data. One possible strategy to show this is simply to insert the transformation of Definition 2.19 into (2.122) and to check that $\daleth' = \daleth$. However, this is somewhat long and inefficient. We restrict the proof to the null case and postpone it to Proposition 2.40 below, where enough structure will have been introduced to make the proof much simpler.

### 2.4.1 *Non-degenerate submanifolds in the null case*

In this subsection we restrict our attention to the null case. Let $\mathcal{S} \hookrightarrow \mathcal{H}$ be a non-degenerate submanifold and $\{\mathcal{H}, \boldsymbol{\gamma}, \boldsymbol{\ell}, \ell^{(2)}, \mathbf{Y}\}$ be null hypersurface data. Throughout the thesis we will find useful to introduce the following notation.

**Notation 2.30.** *Let* $\{\mathcal{H}, \boldsymbol{\gamma}, \boldsymbol{\ell}, \ell^{(2)}, \mathbf{Y}\}$ *be null hypersurface data and* $\iota : \mathcal{S} \hookrightarrow \mathcal{H}$ *a non-degenerate submanifold. Let* $\boldsymbol{\omega}$ *any covariant tensor on* $\mathcal{H}$*. We use the notation* $\boldsymbol{\omega}_\parallel$ *for the pullback of* $\boldsymbol{\omega}$ *to* $\mathcal{S}$*, i.e.* $\boldsymbol{\omega}_\parallel := \iota^\star \boldsymbol{\omega}$*. In abstract index notation we shall simply use* $\omega_A$*, as the index already indicates that we refer to the pullback one-form. We will frequently use this notation for the tensors* $\boldsymbol{\ell}_\parallel$*,* $\mathbf{K}_\parallel$*,* $\mathbf{Y}_\parallel$*, etc. In addition, we define* $\ell^\sharp := h^\sharp(\boldsymbol{\ell}_\parallel, \cdot)$ *and* $\ell^{(2)}_\sharp := h(\ell^\sharp, \ell^\sharp)$*, where* $h := \iota^\star \boldsymbol{\gamma}$ *is the induced metric and* $h^\sharp$ *the inverse metric.*

It is clear that the vector $n$ must be transverse to $\mathcal{S}$ at every $p \in \mathcal{S}$ (otherwise $\mathcal{S}$ would be degenerate). Recall that when every integral curve of $n$ intersects $\mathcal{S}$ exactly once, we say that $\mathcal{S}$ is a cross-section of $\mathcal{H}$.

One immediate consequence of the transformations laws (2.28)-(2.29) is that one can always set the gauge so that $\ell^{(2)}$ and $\ell^\sharp$ take any value at $\mathcal{S}$.

**Lemma 2.31.** *Let $\{\mathcal{H},\boldsymbol{\gamma},\boldsymbol{\ell},\ell^{(2)},\mathbf{Y}\}$ be null hypersurface data and $\mathcal{S}\hookrightarrow\mathcal{H}$ a non-degenerate submanifold. Let $f\in\mathcal{F}(\mathcal{S})$ and $\boldsymbol{\alpha}\in\mathfrak{X}^\star(\mathcal{S})$. Then there always exists a gauge in which*

$$\boldsymbol{\ell}_\parallel=\boldsymbol{\alpha},\qquad \ell^{(2)}|_{\mathcal{S}}=f.$$

*Moreover, the freedom of this gauge is parametrized by the pair $(z,V)$ satisfying $V|_{\mathcal{S}}=0$.*

*Proof.* From (2.29) and applying $\iota^\star$ to (2.28),

$$\mathcal{G}_{(z,V)}\ell^{(2)}=z(\ell^{(2)}+2\boldsymbol{\ell}(V)+\boldsymbol{\gamma}(V,V)),\qquad \mathcal{G}_{(z,V)}\boldsymbol{\ell}_\parallel=z(\boldsymbol{\ell}_\parallel+\iota^\star(\boldsymbol{\gamma}(V,\cdot))).$$

Denoting $V_n$ and $V_\parallel$ by means of the decomposition $V|_{\mathcal{S}}=V_\parallel+V_n n$, $V_\parallel\in T\mathcal{S}$, $V_n\in\mathcal{F}(\mathcal{S})$, the transformations above become

$$\mathcal{G}_{(z,V)}\ell^{(2)}=z(\ell^{(2)}+2\boldsymbol{\ell}_\parallel(V_\parallel)+2V_n+h(V_\parallel,V_\parallel)),\qquad \mathcal{G}_{(z,V)}\boldsymbol{\ell}_\parallel=z(\boldsymbol{\ell}_\parallel+h(V_\parallel,\cdot)).$$

From the second one we can fix the value of $V_\parallel$ to set $\boldsymbol{\ell}_\parallel=\boldsymbol{\alpha}$ at $\mathcal{S}$, whereas by adjusting the value of $V_n$ in the first one we finally set $\mathcal{G}_{(z,V)}\ell^{(2)}=f$. The remaining gauge freedom are the transformations $(z,V)$ with $V|_{\mathcal{S}}=0$. □

At every point $p\in\mathcal{S}$ the tangent space $T_{\iota(p)}\mathcal{H}$ decomposes as

$$T_{\iota(p)}\mathcal{H}=T_p\mathcal{S}\oplus\langle n|_{\iota(p)}\rangle. \tag{2.129}$$

Given a (local) basis $\{e_A\}$ in $T_p\mathcal{S}$ with dual $\{\boldsymbol{\theta}^A\}$, the set $\{e_A,n\}$ is a (local) basis in $T_{\iota(p)}\mathcal{H}$, and we denote its dual by $\{\boldsymbol{\theta}^A,\boldsymbol{q}\}$, satisfying $\boldsymbol{\theta}^A(e_B)=\delta^A_B$, $\boldsymbol{q}(e_A)=0$ and $\boldsymbol{\theta}^A(n)=0$. Then, the cotangent space at $p$ can be written as

$$T^\star_{\iota(p)}\mathcal{H}=T^\star_p\mathcal{S}\oplus\langle\boldsymbol{q}\rangle.$$

By virtue of decomposition (2.129) the connection $\overline{\nabla}$ can be decomposed in terms of an induced torsion free-connection $\overline{\nabla}^{\mathcal{S}}$ and a symmetric tensor $\mathbf{Q}$ by means of

$$\overline{\nabla}_X Z=\overline{\nabla}^{\mathcal{S}}_X Z-\mathbf{Q}(X,Z)n,\qquad X,Z\in\mathfrak{X}(\mathcal{S}). \tag{2.130}$$

In order to determine $\mathbf{Q}$, we use that $\boldsymbol{q}(\overline{\nabla}^{\mathcal{S}}_X Z)=0$, so

$$\mathbf{Q}(X,Z)=-\boldsymbol{q}(\overline{\nabla}_X Z)=-\overline{\nabla}_X(\boldsymbol{q}(Z))+(\overline{\nabla}_X\boldsymbol{q})(Z)=(\overline{\nabla}_X\boldsymbol{q})(Z), \tag{2.131}$$

where in the last equality we used that $\boldsymbol{q}(Z)=0$. Recalling the notation $\boldsymbol{\ell}_\parallel:=\iota^\star\boldsymbol{\ell}\in T^\star\mathcal{S}$ we can decompose

$$\boldsymbol{q}=\boldsymbol{\ell}-\boldsymbol{\ell}_\parallel, \tag{2.132}$$

which follows at once by applying both sides to tangential vectors and to $n$. Here we are abusing the notation by identifying $\boldsymbol{\ell}_{\parallel}$ with the one-form on $\mathcal{H}$ that coincides with $\boldsymbol{\ell}_{\parallel}$ acting over tangent vectors and vanishes when it acts on $n$. Then,

$$\begin{aligned}(\overline{\nabla}_X \boldsymbol{q})(Z) &= (\overline{\nabla}_X \boldsymbol{\ell})(Z) - (\overline{\nabla}_X \boldsymbol{\ell}_{\parallel})(Z)\\ &= (\overline{\nabla}_X \boldsymbol{\ell})(Z) - X(\boldsymbol{\ell}_{\parallel}(Z)) + \boldsymbol{\ell}_{\parallel}\left(\overline{\nabla}^{\mathcal{S}}_X Z - \mathbf{Q}(X,Z)n\right)\\ &= (\overline{\nabla}_X \boldsymbol{\ell})(Z) - \left(\overline{\nabla}^{\mathcal{S}}_X \boldsymbol{\ell}_{\parallel}\right)(Z),\end{aligned}$$

since $\boldsymbol{\ell}_{\parallel}(n) = 0$. Inserting this together with (2.63) into (2.131), it follows

$$\mathbf{Q}(X,Z) = \mathbf{Y}(X,Z) + \mathbf{F}(X,Z) - \ell^{(2)}\mathbf{K}(X,Z) - \left(\overline{\nabla}^{\mathcal{S}}_X \boldsymbol{\ell}_{\parallel}\right)(Z). \tag{2.133}$$

Next we prove two intermediate results that will allow us to find the explicit relation between the Levi-Civita connection $\nabla^h$ of $h := \iota^\star \boldsymbol{\gamma}$ and $\overline{\nabla}^{\mathcal{S}}$.

**Proposition 2.32.** *Let $h = \iota^\star\boldsymbol{\gamma}$ be the induced metric on $\mathcal{S}$ and $\mathbf{K}_{\parallel} := \iota^\star \mathbf{K}$. Then,*

$$\overline{\nabla}^{\mathcal{S}}_X h = -2\boldsymbol{\ell}_{\parallel} \otimes_s \mathbf{K}_{\parallel}(X,\cdot) \qquad \forall X \in \mathfrak{X}(\mathcal{S}). \tag{2.134}$$

*Proof.* Let $V, X, Z \in \mathfrak{X}(S_{\underline{u}})$. Then,

$$\begin{aligned}\left(\overline{\nabla}^{\mathcal{S}}_X h\right)(V,Z) &= \overline{\nabla}^{\mathcal{S}}_X \left(h(V,Z)\right) - h(\overline{\nabla}^{\mathcal{S}}_X V, Z) - h(V, \overline{\nabla}^{\mathcal{S}}_X Z)\\ &= \overline{\nabla}_X \left(\boldsymbol{\gamma}(V,Z)\right) - \boldsymbol{\gamma}(\overline{\nabla}_X V, Z) - \boldsymbol{\gamma}(V, \overline{\nabla}_X Z)\\ &= \iota^\star(\overline{\nabla}_X \boldsymbol{\gamma})(V,Z),\end{aligned}$$

which, upon inserting (2.62), gives (2.134). □

In order to obtain a general relation between $\overline{\nabla}^{\mathcal{S}}$ and $\nabla^h$ we first determine how $\overline{\nabla}^{\mathcal{S}}$ behaves under a gauge transformation.

**Proposition 2.33.** *Let $\overline{\nabla}^{\mathcal{S}}$ be the induced connection on $\mathcal{S}$ and $(z,V) \in \mathcal{F}^\star(\mathcal{H}) \times \mathfrak{X}(\mathcal{H})$. Then,*

$$\mathcal{G}_{(z,V)}(\overline{\nabla}^{\mathcal{S}}) = \overline{\nabla}^{\mathcal{S}} + V_{\parallel} \otimes \mathbf{K}_{\parallel}, \qquad \mathcal{G}_{(z,V)}(\mathbf{Q}) = \mathbf{Q} - V_n \mathbf{K}.$$

*Proof.* Let $\overline{\nabla}' = \mathcal{G}_{(z,V)}\overline{\nabla}$. Since by (2.75) and (2.130)

$$\overline{\nabla}'_X Z = \overline{\nabla}_X Z + \mathbf{K}(X,Z)V = \overline{\nabla}^{\mathcal{S}}_X Z + \mathbf{K}(X,Z)V_{\parallel} + \left(\mathbf{K}(X,Z)V_n - \mathbf{Q}(X,Z)\right)n$$

for every $X, Z \in \mathfrak{X}(\mathcal{S})$, the result follows. □

Now we can find the relation between $\nabla^h$ and the induced connection $\overline{\nabla}^{\mathcal{S}}$ by making use of this gauge transformation.

**Proposition 2.34.** *Let $h = \iota^\star\boldsymbol{\gamma}$ be the induced metric on $\mathcal{S}$, $\nabla^h$ its Levi–Civita connection and $\overline{\nabla}^{\mathcal{S}}$ the induced one. Then,*

$$\overline{\nabla}^{\mathcal{S}} = \nabla^h + \ell^\sharp \otimes \mathbf{K}_{\parallel}.$$

*Proof.* Since $\overline{\nabla}^{\mathcal{S}}$ is torsion-free, by Proposition 2.32 the connection $\overline{\nabla}^{\mathcal{S}}$ coincides with $\nabla^h$ when $\boldsymbol{\ell}_\parallel = 0$. Given null hypersurface data $\mathcal{D} = \{\mathcal{H}, \boldsymbol{\gamma}, \boldsymbol{\ell}, \ell^{(2)}, \mathbf{Y}\}$, the transformed data $\mathcal{D}' = \mathcal{G}_{(1,-\ell^\sharp)}\mathcal{D}$ satisfies $\boldsymbol{\ell}'_\parallel = 0$, so by Proposition 2.33 $\nabla^h = \mathcal{G}_{(1,-\ell^\sharp)}(\overline{\nabla}^{\mathcal{S}}) = \overline{\nabla}^{\mathcal{S}} - \ell^\sharp \otimes \mathbf{K}_\parallel$, and hence the result follows. □

We now can relate the curvature tensor of $\overline{\nabla}^{\mathcal{S}}$ with that of $\overline{\nabla}$.

**Proposition 2.35** (Gauss identity)**.** *Let $X, Z, V, W \in \mathfrak{X}(\mathcal{S})$. Then,*

$$\boldsymbol{\gamma}(V, \overline{R}(X,W)Z) = h(V, R^{\mathcal{S}}(X,W)Z) - \mathbf{Q}(W,Z)\mathbf{K}_\parallel(X,V) + \mathbf{Q}(X,Z)\mathbf{K}_\parallel(W,V), \quad (2.135)$$

*where $R^{\mathcal{S}}$ is the curvature of $\overline{\nabla}^{\mathcal{S}}$ and $\mathbf{Q}$ is given by* (2.133)*.*

*Proof.* By definition of the curvature tensor and decomposition (2.130),

$$\begin{aligned}
\overline{R}(X,W)Z &= \overline{\nabla}_X \overline{\nabla}_W Z - \overline{\nabla}_W \overline{\nabla}_X Z - \overline{\nabla}_{[X,W]} Z \\
&= \overline{\nabla}_X \Big( \overline{\nabla}^{\mathcal{S}}_W Z - \mathbf{Q}(W,Z) n \Big) - \overline{\nabla}_W \Big( \overline{\nabla}^{\mathcal{S}}_X Z - \mathbf{Q}(X,Z) n \Big) - \overline{\nabla}^{\mathcal{S}}_{[X,W]} Z + \mathbf{Q}\left([X,W], Z\right) n \\
&= \overline{\nabla}^{\mathcal{S}}_X \overline{\nabla}^{\mathcal{S}}_W Z - \mathbf{Q}\Big(X, \overline{\nabla}^{\mathcal{S}}_W Z\Big) n - \overline{\nabla}^{\mathcal{S}}_X \left(\mathbf{Q}(W,Z)\right) n - \mathbf{Q}(W,Z) \overline{\nabla}_X n - (X \leftrightarrow W) \\
&\quad - \overline{\nabla}^{\mathcal{S}}_{[X,W]} Z + \mathbf{Q}\left([X,W], Z\right) n,
\end{aligned}$$

which, after identifying $R^{\mathcal{S}}(X,W)Z = \overline{\nabla}^{\mathcal{S}}_X \overline{\nabla}^{\mathcal{S}}_W Z - \overline{\nabla}^{\mathcal{S}}_W \overline{\nabla}^{\mathcal{S}}_X Z - \overline{\nabla}^{\mathcal{S}}_{[X,W]} Z$, and using that $\overline{\nabla}^{\mathcal{S}}$ is torsion-free, $[X,W] = \overline{\nabla}^{\mathcal{S}}_X W - \overline{\nabla}^{\mathcal{S}}_W X$, simplifies to

$$\begin{aligned}
\overline{R}(X,W)Z = R^{\mathcal{S}}(X,W)Z - \mathbf{Q}(W,Z)\overline{\nabla}_X n + \mathbf{Q}(X,Z)\overline{\nabla}_W n \\
+ \Big( (\overline{\nabla}^{\mathcal{S}}_W \mathbf{Q})(X,Z) - (\overline{\nabla}^{\mathcal{S}}_X \mathbf{Q})(W,Z) \Big) n.
\end{aligned}$$

The $\boldsymbol{\gamma}$-product of this expression with $V$ gives (2.135) after using (cf. (2.62) and recall $\boldsymbol{\gamma}(n,\cdot) = \mathbf{K}(n,\cdot) = 0$ and $\boldsymbol{\ell}(n) = 1$)

$$\boldsymbol{\gamma}(\overline{\nabla}_X n, V) = \overline{\nabla}_X(\boldsymbol{\gamma}(n,V)) - \boldsymbol{\gamma}(n, \overline{\nabla}_X V) - (\overline{\nabla}_X \boldsymbol{\gamma})(n,V) = \mathbf{K}(X,V).$$

□

#### 2.4.1.1 *Decomposition of the constraint tensor*

Our next aim is to exploit the hierarchical structure of the constraint tensor (2.95) in the null case and to decompose it into transport equations for the components of the tensor $\mathbf{Y}$. As before, let $\{e_A\}$ be a (local) basis of $T_p\mathcal{S}$ with dual basis $\{\boldsymbol{\theta}^A\}$, so that $\{n, e_A\}$ is a (local) basis of $T_p\mathcal{H} = T_p\mathcal{S} \oplus \langle n|_p \rangle$ with dual basis $\{\boldsymbol{q}, \boldsymbol{\theta}^A\}$. We define the following two contractions of $\boldsymbol{\mathcal{R}}$,

$$J_A := J_a e^a_A \qquad \text{and} \qquad \mathcal{R}_{AB} := \mathcal{R}_{ab} e^a_A e^b_B. \quad (2.136)$$

These objects (along with $H$ and $\boldsymbol{J}(n) := J_a n^a$, see (2.100)) are not all independent, as shown next.

**Proposition 2.36.** *Let $\{\mathcal{H}, \boldsymbol{\gamma}, \boldsymbol{\ell}, \ell^{(2)}, \mathbf{Y}\}$ be null hypersurface data and $\iota : \mathcal{S} \hookrightarrow \mathcal{H}$ a non-degenerate submanifold. Then, the following identity holds*

$$h^{AB}\mathcal{R}_{AB} - 2\boldsymbol{J}(\ell^\sharp) + (\ell^{(2)}_\sharp - \ell^{(2)})\boldsymbol{J}(n) + 2H = 0, \tag{2.137}$$

*where $\ell^\sharp$ and $\ell^{(2)}_\sharp$ are as in Notation 2.30.*

*Proof.* Since $\boldsymbol{\ell}(n) = \boldsymbol{q}(n)$ and $\boldsymbol{\ell}(e_A) = \ell_B \boldsymbol{\theta}^B(e_A)$, in the frame $\{n, e_A\}$ we can decompose

$$\boldsymbol{\ell} = \boldsymbol{q} + \ell_A \boldsymbol{\theta}^A, \tag{2.138}$$

$$P = P^{AB} e_A \otimes e_B + 2P^{n\,A} n \otimes_s e_A + P^{n\,n} n \otimes n. \tag{2.139}$$

Since in this basis $n^A = 0$ and $\boldsymbol{\gamma}_{n\,b} = 0$ it follows from (2.6) that $P^{AB}\boldsymbol{\gamma}_{BC} = \delta^A_C$, and hence we can identify the components $P^{AB}$ with the inverse metric $h^\sharp$, that is, $P^{AB} = h^{AB}$. To compute $P^{n\,A}$ and $P^{n\,n}$ we first note that (2.5), namely $P(\cdot, \boldsymbol{\ell}) = -\ell^{(2)} n$, gives $P(\boldsymbol{\theta}^A, \boldsymbol{\ell}) = 0$ and $P(\boldsymbol{\ell}, \boldsymbol{\ell}) = -\ell^{(2)}$ (by (2.11)). Hence,

$$P^{n\,A} = P(\boldsymbol{q}, \boldsymbol{\theta}^A) = P(\boldsymbol{\theta}^A, \boldsymbol{\ell} - \ell_B \boldsymbol{\theta}^B) = -P^{AB}\ell_B = -h^{AB}\ell_B = -\ell^A,$$

$$P^{n\,n} = P(\boldsymbol{q}, \boldsymbol{q}) = P(\boldsymbol{\ell}, \boldsymbol{\ell}) - 2P(\boldsymbol{\ell}, \ell_A \boldsymbol{\theta}^A) + P(\ell_A \boldsymbol{\theta}^A, \ell_B \boldsymbol{\theta}^B) = \ell^{(2)}_\sharp - \ell^{(2)}.$$

Inserting this decomposition into the definition of $H$ in (2.100),

$$-2H = P^{AB}\mathcal{R}_{AB} + 2P^{n\,A}\mathcal{R}_{A\,n} + P^{n\,n}\mathcal{R}_{n\,n} = h^{AB}\mathcal{R}_{AB} - 2\boldsymbol{J}(\ell^\sharp) + (\ell^{(2)}_\sharp - \ell^{(2)})\boldsymbol{J}(n),$$

so (2.137) is established. □

A by-product of the proof is that in a general gauge the tensor $P$ at $\mathcal{S}$ decomposes as

$$P = h^{AB} e_A \otimes e_B - 2\ell^A n \otimes_s e_A - (\ell^{(2)} - \ell^{(2)}_\sharp) n \otimes n. \tag{2.140}$$

In a gauge where $\ell^{(2)} = 0$ and $\ell^\sharp = 0$ (it always exists, by Lemma 2.31), this decomposition simplifies to

$$P = h^{AB} e_A \otimes e_B, \qquad (\ell^\sharp = \ell^{(2)} = 0) \tag{2.141}$$

so the previous identity reduces to

$$h^{AB}\mathcal{R}_{AB} + 2H = 0. \qquad (\ell^\sharp = \ell^{(2)} = 0) \tag{2.142}$$

Note that in particular (2.140) implies

$$\operatorname{tr}_P \mathbf{T} \stackrel{\mathcal{S}}{=} \operatorname{tr}_h \mathbf{T}_{\|} - 2\boldsymbol{t}(\ell^\sharp) - (\ell^{(2)} - \ell^{(2)}_\sharp)\boldsymbol{t}(n) \tag{2.143}$$

holds for every two covariant tensor $\mathbf{T}$, where $\boldsymbol{t} := \mathbf{T}(n, \cdot)$.

Next we compute the constraint tensors $H$, $\boldsymbol{J}(n)$, $J_A$ and $\mathcal{R}_{AB}$ in terms of the intrinsic geometry of $\mathcal{S}$. The first step is to define a set of tensors on $\mathcal{S}$ that capture the information of the hypersurface data. Some of such tensors have already been introduced before, but we recall their definition and introduce an additional one.

**Definition 2.37.** *Let $\{\mathcal{H},\boldsymbol{\ell},\ell^{(2)},\mathbf{Y}\}$ null hypersurface data and $\mathcal{S}$ a non-degenerate submanifold. We define the tensors $\mathbf{K}_{\|}$, $\mathbf{Y}_{\|}$, $\boldsymbol{\omega}$ and $\kappa$ on $\mathcal{S}$ by*

$$\mathbf{K}_{\|} := \iota^{\star}\mathbf{K}, \qquad \mathbf{Y}_{\|} := \iota^{\star}\mathbf{Y}, \qquad \boldsymbol{\omega} := \iota^{\star}\left(\mathbf{\Pi}(\cdot,n)\right), \qquad \kappa := -\mathbf{Y}(n,n).$$

As we show next, the motivation behind these definitions is that, when the data is embedded and written in a gauge in which $\boldsymbol{\ell}_{\|}=0$ and $\ell^{(2)}=0$ at $\mathcal{S}$, the tensors $\mathbf{K}_{\|}$ and $\mathbf{Y}_{\|}$ correspond to the two null second fundamental forms of $S_{\underline{u}}$ with respect to the vectors $\nu$ and $\xi$ respectively, while $\boldsymbol{\omega}$ is its torsion one-form.

**Remark 2.38.** *Let $\{\mathcal{H},\boldsymbol{\ell},\ell^{(2)},\mathbf{Y}\}$ null hypersurface data and $\mathcal{S}$ a non-degenerate submanifold. By Remark 2.24, $\mathfrak{t}_n=(0,n)$ and $\mathfrak{t}_\theta=(1,\theta)$ constitute a basis of $\boldsymbol{\mathcal{A}}$-null, normal pairs of $\mathcal{S}$. Using* (2.121) *we have $\mathcal{K}^{\mathfrak{t}_n}=\mathbf{K}_{\|}$ and $\mathcal{K}^{\mathfrak{t}_\theta}=\mathbf{Y}_{\|}+\frac{1}{2}\iota^{\star}\left(\pounds_{\theta}\gamma\right)$. From the definition of the torsion one-form (Def. 2.122) one also has*

$$\urcorner_{n\theta}(X) = -\mathbf{K}(X,\theta)+\mathbf{\Pi}(X,n) \qquad \Longrightarrow \qquad \urcorner_{n\theta}=\boldsymbol{\omega}-\mathbf{K}_{\|}(\ell^{\sharp},\cdot).$$

*Thus, whenever the data is written in a gauge in which $\boldsymbol{\ell}_{\|}=0$ and $\ell^{(2)}=0$ at $\mathcal{S}$, the tensors $\mathbf{K}_{\|}$ and $\mathbf{Y}_{\|}$ are the (null) second fundamental forms w.r.t. the normal pairs $\mathfrak{t}_n=(0,n)$ and $\mathfrak{t}_\theta=(1,\theta)$, respectively, whereas the tensor $\boldsymbol{\omega}$ is the torsion one-form w.r.t. the basis $\{\mathfrak{t}_n,\mathfrak{t}_\theta\}$.*

Using (2.140) one can rewrite (2.67) in terms of $\boldsymbol{\omega}$ as follows (recall the definition $\mathbf{\Pi}:=\mathbf{Y}+\mathbf{F}$ in (2.18))

$$\overline{\nabla}_{e_A}n = h^{BC}\mathrm{K}_{AB}e_C-\Big(\omega_A+\mathrm{K}_{AB}\ell^B\Big)n. \tag{2.144}$$

Once the motivation of the Definition 2.37 is clear, we compute the gauge transformation laws of the tensors $\mathbf{K}_{\|}$, $\mathbf{Y}_{\|}$, $\boldsymbol{\omega}$ and that of $\kappa$.

**Lemma 2.39.** *Let $\{\mathcal{H},\boldsymbol{\ell},\ell^{(2)},\mathbf{Y}\}$ null hypersurface data, $\mathcal{S}$ a non-degenerate submanifold and $(z,V)$ gauge parameters. Then,*

$$\mathcal{G}_{(z,V)}\mathbf{K}_{\|} = z^{-1}\mathbf{K}_{\|}, \tag{2.145}$$

$$\mathcal{G}_{(z,V)}\mathbf{Y}_{\|} = z\mathbf{Y}_{\|}+\boldsymbol{\ell}_{\|}\otimes_s dz+\frac{1}{2}\pounds_{zV_{\|}}h+V_n z\mathbf{K}_{\|}, \tag{2.146}$$

$$\mathcal{G}_{(z,V)}\boldsymbol{\omega} = \boldsymbol{\omega}-\mathbf{K}_{\|}(V_{\|},\cdot)+d(\log|z|), \tag{2.147}$$

$$\mathcal{G}_{(z,V)}\kappa = z^{-1}\kappa+n(z^{-1}), \tag{2.148}$$

*where we recall that we denote with the same symbol $z$ its pullback to $\mathcal{S}$.*

*Proof.* As usual we write gauge transformed objects with a prime. The first transformation law is a consequence of (2.37). For the second, taking the pullback of (2.30) to $\mathcal{S}$,

$$\mathbf{Y}'_{\|} = z\mathbf{Y}_{\|}+\boldsymbol{\ell}_{\|}\otimes_s \mathrm{d}z+\frac{1}{2}\iota^{\star}\left(\pounds_{zV}\gamma\right).$$

Writing $V = V_{\parallel} + V_n n$ and recalling that $\pounds_{fn}\gamma = f\pounds_n\gamma$ for any scalar $f$, the transformation law (2.146) follows because

$$\frac{1}{2}\iota^\star\left(\pounds_{zV}\gamma\right) = \frac{1}{2}\iota^\star(\pounds_{zV_{\parallel}}\gamma) + \frac{1}{2}zV_n\iota^\star\left(\pounds_n\gamma\right) = \frac{1}{2}\pounds_{zV_{\parallel}}h + zV_n\mathbf{K}_{\parallel},$$

where we have used the following standard property of the Lie derivative

$$\varphi^\star\left(\pounds_{\varphi_\star Z}T\right) = \pounds_Z\left(\varphi^\star T\right) \tag{2.149}$$

valid for any injective map $\varphi$, vector $Z$ and covariant tensor $T$. The transformation of $\boldsymbol{\omega}$ follows directly from (2.76). Finally, the transformation of $\kappa$ was already proven in (2.74). □

We are now ready to resume the task of showing the gauge invariance of the torsion one-forms in the null case (see the comment after Corollary 2.29).

**Proposition 2.40.** *Let $\mathcal{D}$ be null hypersurface data, $\mathcal{S}$ a non-degenerate submanifold and $\{\mathfrak{t}_i = (\alpha_i, T_i)\}$ a basis of normal pairs. Then, the torsion one-forms $\daleth_{ij}$ are gauge invariant.*

*Proof.* First we show that if $\daleth_{ij}$ are gauge invariant in a particular basis of normal pairs, they are also invariant in any basis. Let $\{\widetilde{t_i}\}$ (with $i = 1, 2$) be a basis of normal pairs and let $\tilde{\mathfrak{t}}_j = \Omega^i{}_j\mathfrak{t}_i$ with $\Omega^i{}_j \in \mathcal{F}(\mathcal{S})$ be a change of basis. From the transformation law in Def. 2.19,

$$\widetilde{\alpha}_j = \Omega^i{}_j\alpha_i, \quad \widetilde{T}_j = \Omega^i{}_jT_i \quad\Longrightarrow\quad z^{-1}\widetilde{\alpha}_j = \Omega^i{}_jz^{-1}\alpha_i, \quad \widetilde{T}_j - \widetilde{\alpha}_jV = \Omega^i{}_j\left(T_i - \alpha_iV\right),$$

which means that the gauge transformed basis $\{\mathfrak{t}'_i\}$ and $\{\tilde{\mathfrak{t}}'_i\}$ are also related by $\tilde{\mathfrak{t}}'_j = \Omega^i{}_j\mathfrak{t}'_i$. Thus, the functions $\Omega^i{}_j$ are gauge invariant. Moreover, from Lemma 2.27 the functions $M_{ij}$ are gauge invariant too. Then, from Proposition 2.28, if $\daleth_{ij}$ is gauge invariant, so it is $\widetilde{\daleth}_{ij}$. Hence it suffices to show the gauge invariance in the basis $\{\mathfrak{t}_n, \mathfrak{t}_\theta\}$, where the only nonzero torsion one-forms are (see Remark 2.38) $\daleth_{n\theta} = -\daleth_{\theta n} = \boldsymbol{\omega} - \mathbf{K}_{\parallel}(\ell^\sharp, \cdot)$. From Def. 2.19, given gauge parameters $(z, V)$ the transformed basis of normal pairs is $\mathcal{G}_{(z,V)}\left(\mathfrak{t}_n\right) = (0, n)$ and $\mathcal{G}_{(z,V)}\left(\mathfrak{t}_\theta\right) = \left(z^{-1}, \theta - V\right)$. Hence, applying (2.122) to the new gauge and recalling the definition of $\theta$ in Remark 2.24,

$$\begin{aligned}
\mathcal{G}_{(z,V)}\left(\daleth_{n\theta}\right)(X) &= z\left(X(z^{-1}) - \mathbf{K}'(X, \theta - V)\right) + z^{-1}\mathbf{\Pi}'(X, n) \\
&= -X\left(\log|z|\right) + \mathbf{K}(X, \ell^\sharp) + \mathbf{K}(X, V) + \boldsymbol{\omega}'(X) \\
&= \mathbf{K}(X, \ell^\sharp) + \boldsymbol{\omega}(X) = \daleth_{n\theta}(X),
\end{aligned}$$

where in the last equality we used the transformation law of $\boldsymbol{\omega}$ in (2.147). □

We conclude this subsection by writting down the constraint tensors $H$, $\boldsymbol{J}(n)$, $J_A$ and $\mathcal{R}_{AB}$ in terms of the quantities we introduced in Definition 2.37. The computation uses the form (2.92) of the constraint tensor written in terms of the connection $\overline{\nabla}$. This requires several somewhat lengthy contractions of the tensors $A$ and $B$ (2.85)-(2.86) which have been postponed to Appendix A in order not to interrupt the presentation. Using identities (A.13) and (A.14) of Lemma A.4 as well as $A_{b(ca)} = 0$ (cf. (2.94)), it follows that

$$\begin{aligned} J_c &= n^a \Big( P^{bd} B_{abcd} - (A_{cba} + A_{acb}) n^b \Big) \\ &= -\pounds_n (\mathrm{r}_c - \mathrm{s}_c) - \overline{\nabla}_c \kappa - 2P^{ab} \mathrm{K}_{cb} \mathrm{s}_a - (\operatorname{tr}_P \mathbf{K})(\mathrm{r}_c - \mathrm{s}_c) - \overline{\nabla}_c (\operatorname{tr}_P \mathbf{K}) + P^{ab} \overline{\nabla}_a \mathrm{K}_{bc}, \end{aligned} \quad (2.150)$$

which of course agrees with (2.97) (note that $\overline{\nabla}_a \mathrm{K}_{bc} = \overset{\circ}{\nabla}_a \mathrm{K}_{bc}$ because $\mathbf{K}(n, \cdot) = 0$, see (2.60)). Contracting it with $n^c$ yields to the abstract version of the Raychaudhuri equation (2.98) that we write in terms of $\mathbf{K}_\parallel$ as follows

$$J(n) = \pounds_n (\operatorname{tr}_h \mathbf{K}_\parallel) - \kappa \operatorname{tr}_h \mathbf{K}_\parallel + |\mathbf{K}_\parallel|^2. \quad (2.151)$$

where $|\mathbf{K}_\parallel|^2 := h^{AB} h^{CD} \mathrm{K}_{AC} \mathrm{K}_{BD}$. The contraction of (2.150) with $e^c_C$ is simply

$$J_C = -\pounds_n \omega_C - \nabla^h_C \kappa - 2h^{AB} \mathrm{K}_{CB} \mathrm{s}_A - (\operatorname{tr}_h \mathbf{K}_\parallel) \omega_C - \nabla^h_C (\operatorname{tr}_h \mathbf{K}_\parallel) + h^{AB} \nabla^h_A \mathrm{K}_{BC}, \quad (2.152)$$

which we have written it in terms of $\omega_C = \mathrm{r}_C - \mathrm{s}_C$.

Finally we compute the constraint tensor $\mathcal{R}_{AB}$, which from (2.91) is

$$\mathcal{R}_{AB} = \Big( B_{acbd} P^{cd} + (A_{bca} + A_{acb}) n^c \Big) e^a_A e^b_B. \quad (2.153)$$

For our purposes it suffices to compute this in a gauge where $\ell^{(2)} = 0$ and $\boldsymbol{\ell}_\parallel = 0$ at a single section $\mathcal{S} \hookrightarrow \mathcal{H}$ (note that by Lemma 2.31 this gauge always exists). The general expression of $\mathcal{R}_{AB}$ in an arbitrary gauge was obtained in [217]. The contractions of $A$ and $B$ in the right-hand side of (2.153) are computed in Lemma A.4, so inserting (A.15) and (A.16) and the symmetrized version of (A.16) into (2.153) we conclude

$$\begin{aligned} \mathcal{R}_{AB} \stackrel{\mathcal{S}}{=} R^h_{AB} &- 2\pounds_n \mathrm{Y}_{AB} - (2\kappa + \operatorname{tr}_h \mathbf{K}_\parallel) \mathrm{Y}_{AB} + (n(\ell^{(2)}) - \operatorname{tr}_h \mathbf{Y}_\parallel) \mathrm{K}_{AB} \\ &+ 4h^{CD} \mathrm{K}_{C(A} \mathrm{Y}_{B)D} + 2\nabla^h_{(A} \omega_{B)} + 4\nabla^h_{(A} \mathrm{s}_{B)} - 2\omega_A \omega_B. \end{aligned} \quad (2.154)$$

Recalling that $2H + h^{AB} R_{AB} = 0$ (2.142) and taking the trace of (2.154) w.r.t. $h$,

$$\begin{aligned} H \stackrel{\mathcal{S}}{=} -\frac{1}{2} R^h &+ \pounds_n (\operatorname{tr}_h \mathbf{Y}_\parallel) - \operatorname{div}_h(\boldsymbol{\omega}) - 2 \operatorname{div}_h(\mathrm{s}) + |\boldsymbol{\omega}|^2 \\ &+ (\operatorname{tr}_h \mathbf{K}_\parallel + \kappa) \operatorname{tr}_h \mathbf{Y}_\parallel - \frac{1}{2} n(\ell^{(2)}) \operatorname{tr}_h \mathbf{K}_\parallel, \end{aligned} \quad (2.155)$$

where we have defined $|\boldsymbol{\omega}|^2 := h^{AB} \omega_A \omega_B$ and $\operatorname{div}_h \boldsymbol{\omega} := h^{AB} \nabla^h_A \omega_B$. Note that the scalar $n(\ell^{(2)})$ does not necessarily vanish since $\ell^{(2)}$ is not assumed to be zero off $\mathcal{S}$.

#### 2.4.1.2 *Characteristic hypersurface data*

There are many relevant situations where a null hypersurface has a product topology and admits a foliation by non-degenerate submanifolds. In this subsection we study this situation in the context of hypersurface data [232].

**Definition 2.41.** *Let $\mathcal{D} = \{\mathcal{H}, \boldsymbol{\gamma}, \boldsymbol{\ell}, \ell^{(2)}, \mathbf{Y}\}$ be null hypersurface data. We say that the set $\mathcal{D}$ is "characteristic hypersurface data" (CHD) provided that*

1. *There exists $\underline{u} \in \mathcal{F}(\mathcal{H})$ satisfying $\mathfrak{z} := n(\underline{u}) \neq 0$. Such functions are called "foliation functions".*
2. *The leaves $\iota_{\underline{u}} : \mathcal{S}_{\underline{u}} := \{p \in \mathcal{H} : \ \underline{u}(p) = \underline{u}\}$ are all diffeomorphic.*

*Similarly one can define characteristic metric hypersurface data.*

Let $\mathcal{S}$ be the underlying topological space of each $\mathcal{S}_{\underline{u}}$. Then the topology of $\mathcal{H}$ is fixed to be a product of the form $\mathcal{H} \simeq \mathcal{I} \times \mathcal{S}$, where $\mathcal{I} \subset \mathbb{R}$ is an interval, and $\{\mathcal{S}_{\underline{u}}\}$ defines a foliation on $\mathcal{H}$. We will always assume that $\mathcal{S}$ is orientable, and hence also $\mathcal{H}$. Note that since $\operatorname{Rad}(\boldsymbol{\gamma})|_p = \langle n|_p \rangle$ the tensor $h := \iota_{\underline{u}}^{\star}\gamma$ is a metric on $\mathcal{S}_{\underline{u}}$, and hence each leaf $\mathcal{S}_{\underline{u}}$ is in particular a non-degenerate submanifold of $\mathcal{H}$. Therefore we can apply all the results developed so far to the context of CHD. For instance, relation (2.132) becomes

$$\mathrm{d}\underline{u} = \mathfrak{z}(\boldsymbol{\ell} - \boldsymbol{\ell}_{\parallel}) \tag{2.156}$$

after noting that $\boldsymbol{q} = \mathfrak{z}^{-1}\mathrm{d}\underline{u}$.

One consequence of Lemma 2.31 is that one can always set the gauge so that $\ell^{(2)} = 0$ and $\boldsymbol{\ell}_{\parallel} = 0$ at a given non-degenerate submanifold $\mathcal{S}$. While this obviously applies to any one leaf of a characteristic hypersurface data, in the following proposition we exploit the foliation structure of Definition 2.41 to show that in fact one can always set the gauge so that $\ell^{(2)} = 0$ and $\boldsymbol{\ell}_{\parallel} = 0$ everywhere. This class of gauges will be used frequently in Chapter 3.

**Proposition 2.42.** *Let $\mathcal{D}$ be CHD and $\underline{u}$ a foliation function. Then there exists $(z, V) \in \mathcal{F}^{\star}(\mathcal{H}) \times \mathfrak{X}(\mathcal{H})$ such that $\mathcal{D}' = \mathcal{G}_{(z,V)}\mathcal{D}$ satisfies the following properties*

1. *$\boldsymbol{\ell}'(X) = 0$ for all $X \in \mathfrak{X}(\mathcal{S})$,*
2. *$\ell'^{(2)} = 0$.*

*The gauge transformations respecting 1. and 2. are $\mathcal{G}_{(z,0)}$ for arbitrary $z \in \mathcal{F}^{\star}(\mathcal{H})$. Moreover, there exists a unique pair $(z, V) \in \mathcal{F}^{\star}(\mathcal{H}) \times \mathfrak{X}(\mathcal{H})$ such that, in addition,*

3. *$\mathfrak{z}' := n'(\underline{u}) = 1$.*

*Proof.* Directly from the transformations in Def. 2.3 and the decomposition $V = V_{\parallel} + V_n n$ the conditions $\boldsymbol{\ell}'(X) = 0$ and $\ell^{(2)\prime} = 0$ read

$$\boldsymbol{\ell}(X) + h(V_{\parallel}, X) = 0, \tag{2.157}$$

$$\ell^{(2)} + 2\boldsymbol{\ell}(V_{\parallel}) + 2V_n + h(V_{\parallel}, V_{\parallel}) = 0. \tag{2.158}$$

Equation (2.157) admits a unique solution for $V_{\parallel}$, which substituted in the equation (2.158) fixes completely $V_n$. Therefore, there always exist a gauge satisfying the conditions (1) and (2). Any transformation of the form $(z, 0)$ keeps these two conditions invariant, and by the uniqueness of $V$ it is clear that no other transformation does. To fulfill (3) simply apply an additional gauge transformation with $(z = \mathfrak{z}, V = 0)$, and use (2.34). Uniqueness of the gauge satisfying (1), (2), (3) is immediate from the argument. □

**Definition 2.43.** *A gauge in which (1) and (2) hold is called* ***characteristic gauge*** *(CG). The gauge satisfying (1), (2), (3) is called* ***adapted characteristic gauge*** *(ACG). We emphasize that the ACG is unique once the foliation function $\underline{u}$ is chosen. A change in $\underline{u}$ affects the corresponding ACG gauge.*

**Remark 2.44.** *When the data is $(\Phi, \xi)$-embedded in $(\mathcal{M}, g)$, the abstract conditions (1) and (2) are respectively equivalent to $\xi$ being orthogonal to the leaves and null.*

**Remark 2.45.** *By Proposition 2.32, in a CG the induced connection on the foliation, $\overline{\nabla}^{\mathcal{S}}$, coincides with the Levi–Civita connection, $\nabla^h$. Moreover $\boldsymbol{\ell}_{\parallel} = 0$ and hence $d\underline{u} = \mathfrak{z}\boldsymbol{\ell}$ by* (2.156)*. Then, $d\mathfrak{z} \wedge \boldsymbol{\ell} + \mathfrak{z} d\boldsymbol{\ell} = 0$, so $\mathbf{F} = -\frac{1}{2\mathfrak{z}} d\mathfrak{z} \wedge \boldsymbol{\ell}$ and hence $\mathrm{F}_{AB} = 0$, so $\Pi_{AB} = \mathrm{Y}_{AB}$. Moreover, from equation* (2.133)*, $Q_{AB} = \mathrm{Y}_{AB}$ in a CG and therefore the Gauss identity* (2.135) *takes the form*

$$\gamma(V, \overline{R}(X, W)Z) = h(V, R^h(X, W)Z) - \mathbf{Y}_{\parallel}(W, Z)\mathbf{K}_{\parallel}(X, V) + \mathbf{Y}_{\parallel}(X, Z)\mathbf{K}_{\parallel}(W, V). \tag{2.159}$$

We conclude this section by rewriting the transformation laws in Lemma 2.39 in the context of characteristic hypersurface data when the transformation takes place between characteristic gauges.

**Corollary 2.46.** *Let $\mathcal{D}$ be CHD written in a CG, $z \in \mathcal{F}^{\star}(\mathcal{H})$ and $\mathcal{D}' = \mathcal{G}_{(z,0)}\mathcal{D}$. Then,*

$$\mathbf{K}'_{\parallel} = z^{-1}\mathbf{K}_{\parallel}, \qquad \mathbf{Y}'_{\parallel} = z\mathbf{Y}_{\parallel}, \qquad \boldsymbol{\omega}' = \boldsymbol{\omega} + d(\log|z|), \qquad \kappa' = z^{-1}\kappa + n(z^{-1}).$$

## 2.5 THE HARMONIC GAUGE

In this section we introduce the so-called harmonic gauge, that will be needed in Chapter 3 to solve the characteristic Cauchy problem from a detached viewpoint. When solving any initial value problem in General Relativity one has to deal with the issue of the coordinates: since GR is a geometric theory the Einstein equations cannot have a unique solution. For that reason one has to "fix the gauge" by choosing an appropriate coordinate system in order to solve them. As we will see, the standard approach to deal with this issue, both in the classical Cauchy problem and in the characteristic one, is to write the Einstein equations in some well-chosen gauge (e.g. harmonic gauge). This yields a new system of geometric PDE, called reduced Einstein equations, which does have a well-posed initial value problem in a PDE sense, i.e., that there is a unique solution in some neighbourhood of the initial hypersurface(s) (see Section 3.1). This approach requires showing, at the very end, that the solution of the reduced system is in fact a solution of the full Einstein field equations. Thus, the first step is to translate the harmonic condition on the coordinates, namely $\Box_g x^{\mu} = 0$, into a condition on hypersurface data. We start with the embedded case and then promote the definition to the abstract level.

**Proposition 2.47.** *Let $\{\mathcal{H}, \boldsymbol{\gamma}, \boldsymbol{\ell}, \ell^{(2)}, \mathbf{Y}\}$ be null hypersurface data $(\Phi, \xi)$-embedded in $(\mathcal{M}, g)$ and let $f$ be a smooth function on $\mathcal{M}$. Then,*

$$\Box_g f \stackrel{\mathcal{H}}{=} \overline{\Box} f + 2n\left(\xi(f)\right) + \left(\operatorname{tr}_P \mathbf{K} + 2\kappa\right)\xi(f) - \left(2P^{bc}(\mathrm{r}_b + \mathrm{s}_b) + n(\ell^{(2)})n^c\right) e_c(f), \quad (2.160)$$

*where $\overline{\Box} f := P^{ab}\overline{\nabla}_a \overline{\nabla}_b f$.*

*Proof.* We use Proposition B.4 with $n^{(2)} = 0$ in Appendix B and replace $\overset{\circ}{\nabla}$ by $\overline{\nabla}$ by means of (2.60), which entails $\overset{\circ}{\nabla}_a \overset{\circ}{\nabla}_b f = \overline{\nabla}_a \overline{\nabla}_b f - n(f)\mathrm{Y}_{ab}$, and hence $\overset{\circ}{\Box} f = \overline{\Box} f - n(f)\operatorname{tr}_P \mathbf{Y}$. Inserting this into (B.11) the result follows. □

Let $\mathcal{D}$ be $\mathfrak{n}$-dimensional null hypersurface data and assume it is $(\Phi, \xi)$-embedded in $(\mathcal{M}, g)$. In order to write $\Box_g x^\mu = 0$ in terms of hypersurface data consider a set of $\mathfrak{n}$ independent functions $\{x^{\underline{a}}\}_{\underline{a}=1}^{\mathfrak{n}}$ on $\mathcal{H}$ and extend them to $\mathcal{M}$ in such a way that $\xi(x^{\underline{a}}) = 0$. Consider also a function $u$ on $\mathcal{M}$ satisfying $u|_{\mathcal{H}} = 0$ and $\xi(u) = 1$. Let $\Xi_c{}^{\underline{a}} := e_c(x^{\underline{a}})$ and $\Xi^c{}_{\underline{a}}$ its inverse. Since $\Xi_c{}^{\underline{a}}$ is a covector ($\underline{a}$ is *not* a tensorial index), $\Xi^c{}_{\underline{a}}$ is a vector and so it is $U^c := \Xi^c{}_{\underline{a}} \overline{\Box} x^{\underline{a}}$. In terms of this vector and by virtue of Proposition 2.47 the conditions $\Box_g x^{\underline{a}} = 0$ are equivalent to

$$2P^{bc}(\mathrm{r} + \mathrm{s})_b + n(\ell^{(2)})n^c = U^c, \quad (2.161)$$

which is a covariant equation. In the following theorem [232] we prove abstractly that given a set of independent functions $\{x^{\underline{a}}\}$ on $\mathcal{H}$ there exists essentially a unique gauge satisfying both $\operatorname{tr}_P \mathbf{K} + 2\kappa = 0$ and equation (2.161).

**Theorem 2.48.** *Let $\mathcal{D}$ be null hypersurface data admitting a cross-section $\mathcal{S} \hookrightarrow \mathcal{H}$, $\{x^{\underline{a}}\}$ a set of $\mathfrak{n}$ independent functions and $U^c := \Xi^c{}_{\underline{a}} \overline{\Box} x^{\underline{a}}$. Let $(z_0, V_0) \in \mathcal{F}^\star(\mathcal{S}) \times \mathfrak{X}(\mathcal{H})$. Then there exists a unique gauge satisfying the following conditions on $\mathcal{H}$,*

$$\operatorname{tr}_P \mathbf{K} + 2\kappa = 0, \quad (2.162)$$

$$2P\left(\mathbf{r} + \mathbf{s}, \cdot\right) + n(\ell^{(2)})n = U \quad (2.163)$$

*together with $(z, V)|_{\mathcal{S}} = (z_0, V_0)$. Any such gauge will be called "harmonic gauge with respect to $\{x^{\underline{a}}\}$" or simply "harmonic gauge" if the functions $\{x^{\underline{a}}\}$ are clear from the context.*

*Proof.* First consider equation (2.162) in the primed gauge. Taking into account transformation laws (2.33), (2.37) and (2.148) together with $\mathbf{K}(n, \cdot) = 0$,

$$\operatorname{tr}_{P'} \mathbf{K}' + 2\kappa' = z^{-1} \operatorname{tr}_P \mathbf{K} + 2z^{-1}\kappa + 2n(z^{-1}) = 0.$$

This is an ODE for $z$, so there exists a unique $z$ solving the equation with initial condition $z|_{\mathcal{S}} = z_0$. Moreover, since $z_0 \neq 0$, $z$ does not vanish in some neighbourhood of $\mathcal{S}$. The next task is to show that there is a gauge in which (2.163) holds. This requires determining the

gauge behavior of both sides in (2.163). Concerning the vector $U$ it suffices to study the transformation of $\overline{\square}F$ with $F \in \mathcal{F}(\mathcal{H})$. From equations (2.33) and (2.75),

$$\begin{aligned}\overline{\square}'F &= P'^{ab}\overline{\nabla}'_a\overline{\nabla}'_b F \\ &= \left(P^{ab} - V^a n^b - V^b n^a\right)\left(\overline{\nabla}_a\overline{\nabla}_b F - V(F)\mathrm{K}_{ab}\right) \\ &= P^{ab}\overline{\nabla}_a\overline{\nabla}_b F - V(F)P^{ab}\mathrm{K}_{ab} - V^a n^b\overline{\nabla}_a\overline{\nabla}_b F - V^b n^a\overline{\nabla}_a\overline{\nabla}_b F \\ &= \overline{\square}F - V(F)\operatorname{tr}_P\mathbf{K} - 2V\left(n(F)\right) + 2(\overline{\nabla}_V n)(F),\end{aligned}$$

where in the third line we used $\mathbf{K}(n,\cdot) = 0$ and in the last equality that $\overline{\nabla}$ is torsion-free. Hence, applying to $F = x^{\underline{a}}$,

$$U'^c = U^c - \Xi^c{}_a\left(V(x^{\underline{a}})\operatorname{tr}_P\mathbf{K} + 2V\left(n(x^{\underline{a}})\right) - 2(\overline{\nabla}_V n)(x^{\underline{a}})\right). \tag{2.164}$$

Concerning the LHS of (2.163) we write $\mathbf{r}+\mathbf{s}$ as $\overline{\nabla}_n\boldsymbol{\ell}$ (see (2.63) and recall $\mathbf{K}(n,\cdot) = 0$) and use transformations (2.34), (2.75) and $\mathbf{K}(n,\cdot) = 0$ as well as $\overline{\nabla}_n\boldsymbol{\gamma} = 0$ (cf. (2.60)). We analyze each term in (2.163) separately. For the first one we recall $\ell'_b = z^{-1}\left(\ell_b + \gamma_{ba}V^a\right)$ (see (2.28)) and compute

$$n'^d\overline{\nabla}'_d\ell'_b = z^{-1}n^d\overline{\nabla}_d\ell'_b = z^{-2}n(z)\ell'_b + n^d\overline{\nabla}_d\ell_b + \gamma_{ba}n^d\overline{\nabla}_d V^a.$$

Contracting with $P'^{bc}$ and using the primed versions of (2.5) and (2.6), namely

$$P'^{bc}\ell'_b = -\ell^{(2)\prime}n'^c, \qquad P'^{bc}\gamma_{ba} = \delta^c_a - \ell'_a n'^c,$$

as well as the transformation law (2.33),

$$\begin{aligned}2P'^{bc}n'^d\overline{\nabla}'_d\ell'_b &= 2z^{-2}n(z)P'^{bc}\ell'_b + 2P'^{bc}n^d\overline{\nabla}_d\ell_b + 2P'^{bc}\gamma_{ba}n^d\overline{\nabla}_d V^a \\ &= -2z^{-2}n(z)\ell^{(2)\prime}n'^c + 2P^{bc}n^d\overline{\nabla}_d\ell_b - 2V^b n^c n^d\overline{\nabla}_d\ell_b - 2V^c n^b n^d\overline{\nabla}_d\ell_b \\ &\quad + 2n^d\overline{\nabla}_d V^c - 2zn'^c\left(\ell_a n^d\overline{\nabla}_d V^a + \gamma_{ab}V^b n^d\overline{\nabla}_d V^a\right).\end{aligned} \tag{2.165}$$

To compute the term $n'(\ell^{(2)\prime})n'^c$ we insert $\ell^{(2)\prime} = z^2\left(\ell^{(2)} + 2\boldsymbol{\ell}(V) + \boldsymbol{\gamma}(V,V)\right)$ and use (2.34) to get

$$n'(\ell^{(2)\prime})n'^c = 2z^{-2}n(z)\ell^{(2)\prime}n'^c + z\left(n(\ell^{(2)}) + 2V^a n^d\overline{\nabla}_d\ell_a + 2\ell_a n^d\overline{\nabla}_d V^a + 2\boldsymbol{\gamma}_{ab}V^a n^d\overline{\nabla}_d V^b\right)n'^c. \tag{2.166}$$

Adding (2.165) and (2.166) and using $n^d n^c\overline{\nabla}_c\ell_d = -\kappa$,

$$2P'^{bc}n'^d\overline{\nabla}_d\ell'_b + n'(\ell^{(2)\prime})n'^c = 2P^{bc}n^d\overline{\nabla}_d\ell_b + n(\ell^{(2)})n^c + 2\kappa V^c + 2n^d\overline{\nabla}_d V^c. \tag{2.167}$$

Finally taking into account (2.164) and (2.167), equation (2.163) constitutes a first order ODE for $V$ with unique solution given $V_0$. □

**Remark 2.49.** *Observe that when the data is embedded and written in the harmonic gauge, Proposition 2.47 ensures that the functions* $\{u, x^{\underline{a}}\}$ *satisfy* $\Box_g x^{\underline{a}} = 0$ *and* $\Box_g u = 0$ *on* $\mathcal{H}$ *provided that* $\xi(x^{\underline{a}}) \stackrel{\mathcal{H}}{=} 0$ *and* $\xi(u)|_{\mathcal{H}}$ *is constant along each null generator of* $\mathcal{H}$*.*

In the following Lemma we show that one can exploit the remaining gauge freedom in Theorem 2.48 to set also $\ell^{(2)} = 0$, $\boldsymbol{\ell}_{\parallel} = 0$ on $\mathcal{S}$ (see Lemma 2.31).

**Lemma 2.50.** *Let* $\mathcal{D}$ *be null hypersurface data admitting a section* $\mathcal{S} \hookrightarrow \mathcal{H}$*,* $\{x^{\underline{a}}\}$ *a set of* $\mathfrak{n}$ *functionally independent functions and* $V^c := \Xi^c{}_{\underline{a}} \bar{\Box} x^{\underline{a}}$*. Then there exists a class of harmonic gauges satisfying additionally* $\ell^{(2)} = 0$ *and* $\boldsymbol{\ell}_{\parallel} = 0$ *on* $\mathcal{S}$*. The set of transformations keeping the previous conditions invariant is parametrized by* $z_0 \in \mathcal{F}^{\star}(\mathcal{S})$*.*

*Proof.* Let $z \in \mathcal{F}^{\star}(\mathcal{S})$ and $V = i_{\star} V_{\parallel} + V_n n \in \mathfrak{X}(\mathcal{H})$, $V_{\parallel} \in \mathfrak{X}(\mathcal{S})$, $V_n \in \mathcal{F}(\mathcal{S})$. Directly from the transformation laws (2.29) and (2.28),

$$z^2 \left( \ell^{(2)} + 2\boldsymbol{\ell}_{\parallel}(V_{\parallel}) + 2V_n + h\left(V_{\parallel}, V_{\parallel}\right) \right) = 0, \tag{2.168}$$

$$z\left( \boldsymbol{\ell}_{\parallel} + h(V_{\parallel}, \cdot) \right) = 0. \tag{2.169}$$

Equation (2.169) admits a unique solution for $V_{\parallel}$, which introduced into (2.168) fixes $V_n$. □

Let $\mathcal{D}$ be CHD $(\Phi, \xi)$-embedded in $(\mathcal{M}, g)$ and consider a set of $\mathfrak{n}$ independent functions adapted to the foliation, i.e., $\{x^{\underline{a}}\} = \{\underline{u}, x^A\}$, where $\{x^A\}$ is a set of $\mathfrak{n}-1$ functions on $\mathcal{H}$ satisfying $n(x^A) = 0$ and $n\left(\underline{u}\right) \neq 0$. We want to compute explicitly $\Box_g u$, $\Box_g x^A$ and $\Box_g \underline{u}$ in terms of the foliation tensors in the embedded case and then promote these expressions into abstract definitions on $\mathcal{H}$. The first one is immediate from Proposition 2.47 together with $\xi(u) = 1$ and the fact that $\operatorname{tr}_P \mathbf{K} = \operatorname{tr}_h \mathbf{K}_{\parallel}$,

$$\Box_g u = \operatorname{tr}_h \mathbf{K}_{\parallel} + 2\kappa. \tag{2.170}$$

For the other two we first prove an intermediate result.

**Proposition 2.51.** *Let* $\mathcal{D}$ *be CHD and* $\beta \in \mathcal{F}(\mathcal{H})$*. Then,*

$$\begin{aligned} \bar{\Box}\beta = {} & \Box_h \beta - \ell^{\sharp}(\beta) \operatorname{tr}_h \mathbf{K}_{\parallel} - 2\ell^{\sharp}\left(n(\beta)\right) + 2\mathrm{K}^{\sharp}_{\parallel}(\ell^{\sharp})(\beta) - (\ell^{(2)} - \ell^{(2)}_{\sharp}) n\left(n(\beta)\right) \\ & + \Big( \operatorname{tr}_h \mathbf{Y}_{\parallel} + (\kappa - \operatorname{tr}_h \mathbf{K}_{\parallel})(\ell^{(2)} - \ell^{(2)}_{\sharp}) - 2\boldsymbol{\omega}(\ell^{\sharp}) - 2\mathbf{K}_{\parallel}(\ell^{\sharp}, \ell^{\sharp}) - \operatorname{div}_h \boldsymbol{\ell}_{\parallel} \Big) n(\beta), \end{aligned} \tag{2.171}$$

$$P^{bc}(\mathrm{r}+\mathrm{s})_b e_c(\beta) = h^{\sharp}\left( d\beta, \boldsymbol{\omega} + \pounds_n \boldsymbol{\ell}_{\parallel} \right) + \kappa \ell^{\sharp}(\beta) + \Big( \kappa \Big( \ell^{(2)} - \ell^{(2)}_{\sharp} \Big) - (\boldsymbol{\omega} + \pounds_n \boldsymbol{\ell}_{\parallel})(\ell^{\sharp}) \Big) n(\beta). \tag{2.172}$$

*Proof.* From decomposition (2.140) and the fact that the Hessian of a function is symmetric,

$$\begin{aligned}
\overline{\Box}\beta &= P^{ab}\left(e_a\left(e_b(\beta)\right) - (\overline{\nabla}_{e_a}e_b)(\beta)\right)\\
&= h^{AB}\Big(e_A\left(e_B(\beta)\right) - (\overline{\nabla}^{\mathcal{S}}_{e_A}e_B)(\beta) + n(\beta)Q_{AB}\Big) - 2\ell^A\left(e_A\left(n(\beta)\right) - (\overline{\nabla}_{e_A}n)(\beta)\right)\\
&\quad - (\ell^{(2)} - \ell^{(2)}_{\sharp})\left(n\left(n(\beta)\right) - (\overline{\nabla}_n n)(\beta)\right)\\
&= h^{AB}\Big(e_A\left(e_B(\beta)\right) - (\overline{\nabla}^{h}_{e_A}e_B)(\beta) - \ell^{\sharp}(\beta)\mathrm{K}_{AB} + n(\beta)Q_{AB}\Big) - 2\ell^{\sharp}\left(n(\beta)\right)\\
&\quad + 2\mathrm{K}^C{}_A\ell^A e_C(\beta) - 2\Big(\boldsymbol{\omega}(\ell^{\sharp}) + \mathbf{K}_{\parallel}(\ell^{\sharp},\ell^{\sharp})\Big)n(\beta) - (\ell^{(2)} - \ell^{(2)}_{\sharp})\left(n\left(n(\beta)\right) - \kappa n(\beta)\right),
\end{aligned} \tag{2.173}$$

where in the second equality we used (2.130) and in the third one Proposition 2.34 and (2.144). The term $h^{AB}Q_{AB}$ can be computed from (2.133),

$$\begin{aligned}
h^{AB}Q_{AB} &= h^{AB}\Big(\mathrm{Y}_{AB} + \mathrm{F}_{AB} - \ell^{(2)}\mathrm{K}_{AB} - (\overline{\nabla}^{\mathcal{S}}_{e_A}\boldsymbol{\ell}_{\parallel})(e_B)\Big)\\
&= \operatorname{tr}_h \mathbf{Y}_{\parallel} - (\ell^{(2)} - \ell^{(2)}_{\sharp})\operatorname{tr}_h \mathbf{K}_{\parallel} - \operatorname{div}_h \boldsymbol{\ell}_{\parallel},
\end{aligned} \tag{2.174}$$

where in the second equality we used the fact that $\mathbf{F}$ is antisymmetric and Proposition 2.34. Introducing (2.174) into (2.173), (2.171) follows. In order to show (2.172) we employ again decomposition (2.140),

$$\begin{aligned}
e_c(\beta)P^{cb}(\mathrm{r}+\mathrm{s})_b = h^{BC}(\mathrm{r}+\mathrm{s})_B e_C(\beta) - n(\beta)(\mathbf{r}+\mathbf{s})(\ell^{\sharp})\\
+ \kappa\ell^{\sharp}(\beta) + \kappa(\ell^{(2)} - \ell^{(2)}_{\sharp})n(\beta).
\end{aligned}$$

Equation (2.172) is obtained after taking into account that (see (2.22))

$$\iota^{\star}(\mathbf{r}+\mathbf{s}) = \iota^{\star}(\mathbf{r}-\mathbf{s}+2\mathbf{s}) = \boldsymbol{\omega} + \pounds_n\boldsymbol{\ell}_{\parallel}.$$

□

From Propositions 2.47 and 2.51 one has the following Corollary.

**Corollary 2.52.** *Let $\mathcal{D}$ be CHD $(\Phi,\xi)$-embedded in $(\mathcal{M},g)$ and let $\underline{u}$ be a foliation function. Let $\{x^A\}$ be a set of functions satisfying $n(x^A)=0$ and extended off $\mathcal{H}$ by means of $\xi(\underline{u}) = \xi(x^A) = 0$. Then,*

$$\Box_g x^A = \Box_h x^A - (2\kappa + \operatorname{tr}_h \mathbf{K}_{\parallel})\ell^{\sharp}(x^A) + 2dx^A\left(\mathbf{K}^{\sharp}_{\parallel}(\ell^{\sharp}) - h^{\sharp}(\boldsymbol{\omega} + \pounds_n\boldsymbol{\ell}_{\parallel},\cdot)\right), \tag{2.175}$$

$$\Box_g \underline{u} = \mathfrak{z}\operatorname{tr}_h \mathbf{Y}_{\parallel} + \mathfrak{z}G\left(\boldsymbol{\ell}_{\parallel},\kappa,h,\ell^{(2)},\mathfrak{z}\right), \tag{2.176}$$

*where*

$$\begin{aligned}
G\left(\boldsymbol{\ell}_{\parallel},\kappa,h,\ell^{(2)},\mathfrak{z}\right) = -(\ell^{(2)} - \ell^{(2)}_{\sharp})\left(\kappa + \operatorname{tr}_h \mathbf{K}_{\parallel} + n(\log|\mathfrak{z}|)\right) - \operatorname{div}_h \boldsymbol{\ell}_{\parallel}\\
+ 2\left(\pounds_n\boldsymbol{\ell}\right)_{\parallel}(\ell^{\sharp}) - 2\mathbf{K}_{\parallel}(\ell^{\sharp},\ell^{\sharp}) - n(\ell^{(2)}) - 2\ell^{\sharp}(\log|\mathfrak{z}|).
\end{aligned} \tag{2.177}$$

Equations (2.170), (2.175) and (2.176) lead naturally to the definition of the following abstract functions on $\mathcal{H}$,

$$\Gamma^{u}_{\mathcal{H}} := \operatorname{tr}_h \mathbf{K}_{\parallel} + 2\kappa, \tag{2.178}$$

$$\Gamma^{A}_{\mathcal{H}} := \Box_h x^A - (2\kappa + \operatorname{tr}_h \mathbf{K}_{\parallel})\ell^{\sharp}(x^A) + 2\mathrm{d}x^A \left(\mathbf{K}^{\sharp}_{\parallel}(\ell^{\sharp})\right) - 2(\pounds_n \boldsymbol{\ell}_{\parallel} + \boldsymbol{\omega}) \left(\operatorname{grad}_h x^A\right), \tag{2.179}$$

$$\Gamma^{\underline{u}}_{\mathcal{H}} := \mathfrak{z} \operatorname{tr}_h \mathbf{Y}_{\parallel} + \mathfrak{z} G \left(\boldsymbol{\ell}_{\parallel}, \kappa, h, \ell^{(2)}, \mathfrak{z}\right). \tag{2.180}$$

These functions have the property that, whenever the data happens to be embedded, $\Box_g u = \Gamma^{u}_{\mathcal{H}}$, $\Box_g \underline{u} = \Gamma^{\underline{u}}_{\mathcal{H}}$ and $\Box_g x^A = \Gamma^{A}_{\mathcal{H}}$.

Let $\{x^{\underline{a}}\} = \{\underline{u}, x^A\}$ be functions on $\mathcal{H}$ satisfying (i) $n(\underline{u}) \neq 0$, (ii) $n(x^A) = 0$ and (iii) $\{x^A\}$ is a (local) coordinate system on $\mathcal{S}$. It is straightforward to check from (2.179)-(2.180) that when the data is written in the harmonic gauge of Lemma 2.50 w.r.t. $\{x^{\underline{a}}\}$, conditions (2.163) can be rewritten as equations involving the tensor $\mathbf{Y}$.

**Corollary 2.53.** *Let $\mathcal{D}$ be null hypersurface data admitting a cross-section $\mathcal{S}$ written in the harmonic gauge of Lemma 2.50 w.r.t. some functions $x^{\underline{a}} = \{\underline{u}, x^A\}$ satisfying conditions (i), (ii) and (iii) above. Then the following equations hold at $\mathcal{S}$*

$$\operatorname{tr}_h \mathbf{Y}_{\parallel} = n(\ell^{(2)}), \tag{2.181}$$

$$2\left(\pounds_n \boldsymbol{\ell} + \mathbf{r} - \mathbf{s}\right)\left(\operatorname{grad}_h x^A\right) = \Box_h x^A. \tag{2.182}$$

We conclude this section by showing that there exist linear combinations of the constraint tensors and the functions $\Gamma^{A}_{\mathcal{H}}$ and $\Gamma^{\underline{u}}_{\mathcal{H}}$ that are hierarchically independent of the tensor $\mathbf{Y}$. This will be crucial for Chapter 3. The explicit combinations are computed in the following lemma.

**Lemma 2.54.** *Let $\mathcal{D}$ be CHD with foliation function $\underline{u}$ and let $\{x^A\}$ be a set of functionally independent functions satisfying $n(x^A) = 0$. Express all the tensors in the coordinate basis $\{\underline{u}, x^A\}$. Then the following combination depends neither on $\boldsymbol{\omega}$ nor on $\mathbf{Y}_{\parallel}$,*

$$\mathcal{L}_A(J_B, \Gamma^{B}_{\mathcal{H}}) = J_A - \frac{1}{2} h_{AB} \ \pounds_n(\Gamma^{B}_{\mathcal{H}}) - \frac{1}{2}(\operatorname{tr}_h \mathbf{K}_{\parallel}) h_{AB} \Gamma^{B}_{\mathcal{H}}, \tag{2.183}$$

*and the combination*

$$\mathcal{L}\left(H, \Gamma^{\underline{u}}_{\mathcal{H}}\right) = H - \pounds_n(\mathfrak{z}^{-1}\Gamma^{\underline{u}}_{\mathcal{H}}) - \mathfrak{z}^{-1}(\operatorname{tr}_h \mathbf{K}_{\parallel} + \kappa)\Gamma^{\underline{u}}_{\mathcal{H}}, \tag{2.184}$$

*does not depend on $\mathbf{Y}_{\parallel}$.*

*Proof.* Let us start by finding the linear combination $\mathcal{L}_A\left(J_B, \Gamma^{B}_{\mathcal{H}},\right)$ which does not depend on the tensor $\boldsymbol{\omega}$ nor on $\mathbf{Y}_{\parallel}$. Note that the functions $\Gamma^{A}_{\mathcal{H}}$ defined above in the coordinates $\{x^A\}$ read (cf. (2.179))

$$\Gamma^{A}_{\mathcal{H}} = \Box_h x^A - (2\kappa + \operatorname{tr}_h \mathbf{K}_{\parallel})\ell^A + 4\left(\mathrm{K}^{\sharp}_{\parallel}(\ell^{\sharp})\right)^A - 2(\pounds_n \boldsymbol{\ell}_{\parallel})^A - 2\omega^A. \tag{2.185}$$

The expression of $J_A$ was found in (2.152) and takes the form

$$J_A = -\pounds_n \omega_A - (\operatorname{tr}_h \mathbf{K}_{\parallel})\omega_A + \cdots ,$$

where the dots represent terms involving solely metric data and $\kappa$. Thus, the combination $J_A - \frac{1}{2} h_{AB} \ \pounds_n(\Gamma^B_{\mathcal{H}})$ does not carry any derivative of $\boldsymbol{\omega}$ and its explicit form is

$$J_A - \frac{1}{2} h_{AB} \ \pounds_n(\Gamma^B_{\mathcal{H}}) = -(\operatorname{tr}_h \mathbf{K}_{\parallel})\omega_A + \cdots ,$$

and consequently the combination

$$\mathcal{L}_A(J_B, \Gamma^B_{\mathcal{H}}) = J_A - \frac{1}{2} h_{AB} \ \pounds_n(\Gamma^B_{\mathcal{H}}) - \frac{1}{2}(\operatorname{tr}_h \mathbf{K}_{\parallel}) h_{AB} \Gamma^B_{\mathcal{H}}$$

does not depend on $\boldsymbol{\omega}$ nor on $\mathbf{Y}_{\parallel}$.

Now we proceed analogously with the second identity, with the only difference that we prove it in the CG and then we show that it is actually true in any gauge. By (2.155) the constraint scalar $H$ reads

$$H = \pounds_n(\operatorname{tr}_h \mathbf{Y}_{\parallel}) + (\operatorname{tr}_h \mathbf{K}_{\parallel} + \kappa) \operatorname{tr}_h \mathbf{Y}_{\parallel} + \cdots ,$$

where the dots are now terms which do not depend on $\mathbf{Y}_{\parallel}$. Recalling the expression of $\Gamma^u_{\mathcal{H}}$ in (2.180), namely

$$\Gamma^u_{\mathcal{H}} = \mathfrak{z} \operatorname{tr}_h \mathbf{Y}_{\parallel} + \mathfrak{z} G\left(\boldsymbol{\ell}_{\parallel}, \kappa, h, \ell^{(2)}, \mathfrak{z}\right),$$

it is clear that the combination

$$\mathcal{L}\left(H, \Gamma^u_{\mathcal{H}}\right) = H - \pounds_n(\mathfrak{z}^{-1}\Gamma^u_{\mathcal{H}}) - \mathfrak{z}^{-1}(\operatorname{tr}_h \mathbf{K}_{\parallel} + \kappa)\Gamma^u_{\mathcal{H}}$$

does not depend on $\mathbf{Y}_{\parallel}$. From the transformation of $\mathbf{Y}_{\parallel}$ in (2.146), item 2. in Proposition 2.10 and the fact that the gauge transformation of $G$ does not add any extra $\mathbf{Y}_{\parallel}$, it will be still true that the combination $\mathcal{L}\left(H, \Gamma^u_{\mathcal{H}}\right)$ does not depend on $\mathbf{Y}_{\parallel}$ in any gauge since the constraint tensors $\boldsymbol{J}(n)$ and $J_A$ do not depend on $\mathbf{Y}_{\parallel}$ (see (2.151) and (2.152)). □

# 3

# DOUBLE NULL DATA AND THE CHARACTERISTIC CAUCHY PROBLEM

In this chapter, we present the characteristic Cauchy problem from a detached perspective. We begin in Section 3.1 with a brief review of the main Cauchy problems in General Relativity. Section 3.2 introduces the concept of *double null data*, which corresponds to the detached version of two transverse null hypersurfaces. After analyzing some of its fundamental properties, in Section 3.3 we address the existence of $\lambda$-vacuum developments for such data, and in Section 3.4 we turn to the question of uniqueness. The results of this chapter have been published in [231, 232].

## 3.1 CAUCHY PROBLEMS IN GENERAL RELATIVITY

It has been known since the foundational results of Choquet-Bruhat [58] and Choquet-Bruhat–Geroch [60] that the Einstein field equations can be formulated as an initial value (or Cauchy) problem. While in many other Cauchy problems in physics the field to be evolved is defined on a fixed background geometry (often Minkowski spacetime), General Relativity differs in that the dynamical variable is the spacetime geometry itself.

A further distinctive feature is that General Relativity is a generally covariant theory: if $g$ is a solution of the Einstein equations, then so it is $\phi^\star g$ for any diffeomorphism $\phi$. This gauge freedom is the geometric analogue of the $\mathrm{U}(1)$ gauge freedom for the electromagnetic potential $A_\mu$ in Maxwell's theory. As in that case, one can "fix the gauge" by imposing coordinate conditions that render the system into a well-posed PDE problem.

A particularly useful choice is the harmonic coordinate condition, in which the coordinates $\{x^\mu\}$ satisfy

$$\Gamma^\mu := -\Box_g x^\mu = -\nabla_\alpha \nabla^\alpha x^\mu = 0.$$

From the expression for the Christoffel symbols in terms of the metric $g$, namely

$$\Gamma^\mu_{\alpha\beta} = \frac{1}{2} g^{\mu\nu} \left( 2\partial_{(\alpha} g_{\beta)\nu} - \partial_\nu g_{\alpha\beta} \right),$$

it follows that $\Gamma^\mu = g^{\alpha\beta}\Gamma^\mu_{\alpha\beta}$. The components of the Ricci tensor can then be written as

$$R_{\alpha\beta} = R^H_{\alpha\beta} + g_{\mu(\alpha}\partial_{\beta)}\Gamma^\mu,$$

where

$$R^H_{\alpha\beta} = -\frac{1}{2}\Box_g g_{\alpha\beta} + F_{\alpha\beta}(g,\partial g)$$

is a quasi-linear wave operator acting on the metric components and $F_{\alpha\beta}(g,\partial g)$ is rational in $g$ and its first derivatives. In harmonic coordinates ($\Gamma^\mu = 0$), the Einstein vacuum equations reduce to

$$R^H_{\alpha\beta} = 0,$$

which is a quasi-linear hyperbolic system to which standard local existence and uniqueness results apply. In the following we recall how this framework applies to the spacelike and the characteristic Cauchy problems.

### 3.1.1 *Spacelike Cauchy problem*

Here we review the classic strategy to solve the spacelike Cauchy problem. For the details we refer the reader to the original references [58, 60] and also to [59, 148, 282, 310, 323]. Let $(\Sigma, h)$ be an $\mathfrak{n}$-dimensional Riemannian manifold and $K$ a $(0,2)$-symmetric tensor field. Assume they satisfy the constraint equations[1]

$$\begin{aligned} R^h - K_{ab}K^{ab} + (\operatorname{tr}_h K)^2 &= -(\mathfrak{n}-2)\lambda, \\ \nabla^h_b\Big(\mathrm{K}^b{}_a - (\operatorname{tr}_h K)\delta^b_a\Big) &= 0. \end{aligned} \tag{3.1}$$

Since all our considerations are local we work on an open set $M \subset \mathbb{R}^{\mathfrak{n}+1}$ with coordinates $\{x^\alpha\}$ and take $\Sigma = \{x^0 = 0\}$. The idea is to solve equation $R^H_{\alpha\beta} = 0$ for a metric $g$ from initial data on $\Sigma$ suitably chosen so that the coordinates $\{x^\alpha\}$ are harmonic w.r.t. $g$, because then $g$ is a solution to the Einstein equations. To achieve this, one chooses

$$g_{ab}|_\Sigma = h_{ab}, \qquad g_{0a}|_\Sigma = 0, \qquad g_{00}|_\Sigma = -1,$$

and

$$\pounds_n g_{ab}|_\Sigma = 2K_{ab},$$

while the components $\pounds_n g_{0a}|_\Sigma$ are determined from the harmonic gauge condition

$$0 = -\Gamma^\mu = \partial_\alpha g^{\alpha\mu} + \frac{1}{2}g^{\alpha\mu}g^{\rho\sigma}\partial_\alpha g_{\rho\sigma}.$$

This gives a unique metric $g$ on $M$ solving $R^H_{\alpha\beta} = 0$.

---

1 They are obtained from $G_{\mu\nu}n^\mu e^\nu_a = 0$ and $G_{\mu\nu}n^\mu n^\nu = \dfrac{\mathfrak{n}-1}{2}\lambda$ after using the Gauss and Codazzi identities, see e.g. [323].

To show that $(M, g)$ is a solution to the Einstein equations (and not just of the reduced system $R^H_{\mu\nu} = 0$) it suffices to prove that $\Gamma^\mu = 0$. The strategy is as follows. Firstly, one writes the Einstein tensor in terms of $R^H_{\mu\nu}$ and $\Gamma^\mu$ as

$$G_{\mu\nu} = R^H_{\mu\nu} - \frac{1}{2} R^H g_{\mu\nu} + g_{\alpha(\mu}\partial_{\nu)}\Gamma^\alpha - \frac{1}{2} g_{\mu\nu}\partial_\alpha \Gamma^\alpha.$$

Using that the constraint equations imply $G_{\mu\nu}n^\nu = 0$ on $\Sigma$ and the fact that $\Gamma^\mu = 0$ on $\Sigma$ by construction, one finds $\pounds_n \Gamma^\mu = 0$ on $\Sigma$. Secondly, one can show that assuming $R^H_{\mu\nu} = 0$, the contracted Bianchi identity $\nabla^\mu G_{\mu\nu} = 0$ yields a wave equation for $\Gamma^\mu$,

$$0 = \frac{1}{2} g_{\alpha\nu} \Box_g \Gamma^\alpha + \text{lower order terms in } \Gamma^\alpha.$$

By uniqueness of the wave equation (cf. [282]) one concludes that $\Gamma^\mu = 0$ everywhere, so $\{x^\alpha\}$ are harmonic coordinates w.r.t. $g$ and therefore $(M, g)$ satisfies the full Einstein equations. We can summarize the above discussion in the following standard result [58, 60]. Recall that by "$\lambda$-vacuum" we mean that equation (0.1) holds.

**Theorem 3.1.** *Let* $(\Sigma, h)$ *be a smooth Riemannian manifold and* $K$ *a smooth symmetric (0,2) tensor field satisfying the constraint equations* (3.1)*. Then, there exists a unique maximal and globally hyperbolic* $\lambda$*-vacuum development* $(M, g)$ *in which* $\Sigma$ *is embedded with* $h$ *and* $K$ *as the first and second fundamental forms, respectively. Furthermore, if* $(\Sigma, h, K)$ *and* $(\Sigma', h', K')$ *are isometric initial data sets with developments* $(M, g)$ *and* $(M', g')$*, then* $(M, g)$ *and* $(M', g')$ *are isometric.*

This result is detached, in the sense that the initial data set $(\Sigma, h, K)$ and its constraints do not involve any spacetime quantity *a priori.* It is only *a posteriori* (i.e., when the spacetime has been constructed) when $\Sigma$, $h$, $K$ and the constraints have something to do with the spacetime.

The well-posedness of the spacelike Cauchy problem has also been established for a wide range of Einstein-matter systems. Notable examples include the Einstein-Klein–Gordon and Einstein-Maxwell equations, both of which admit local existence and uniqueness results analogous to the vacuum case. More generally, any matter model whose field equations are themselves well posed, and whose energy-momentum tensor depends only on the fields, the metric, and their first derivatives, yields a coupled Einstein-matter system that is well-posed [323]. Other examples include certain perfect fluid models with specific equations of state (see e.g. [154] for a detailed discussion).

The spacelike Cauchy problem has served as the natural starting point for many developments in General Relativity, in particular with applications to cosmology and astrophysics. Among them are the formulation of the $3+1$ approach to Einstein's equations [12, 25, 144, 148], the mathematical analysis and construction of solutions to the constraint equations [41, 55, 88, 89, 147], and the definition of global quantities such as the ADM mass and angular momentum [25, 272, 277, 313, 323]. Closely related to these developments are the positive energy theorems, which provide fundamental structural results for the theory [296, 300–302, 326].

### 3.1.2 *Characteristic Cauchy problem*

We now turn to the characteristic Cauchy problem, which is the central topic of this chapter. In this setting, the initial data is prescribed on a pair of null hypersurfaces intersecting transversely along a codimension-two surface. The term characteristic arises because null hypersurfaces are precisely the characteristic surfaces of the Einstein equations. This problem is interesting for several reasons. First, in the characteristic case the constraint equations reduce to a hierarchical set of ordinary differential equations, making them considerably simpler than their spacelike counterparts. Second, the characteristic problem provides a natural framework for studying null infinity and gravitational radiation. Finally, the notion of an “instant of time” is more naturally adapted to null hypersurfaces: the information accessible to us from the “here and now” lies on our past light cone, rather than on a spacelike hypersurface.

The study of the characteristic problem has its origins in the 1960s, motivated largely by the desire to understand gravitational radiation [42, 263, 289]. A major step forward was taken by Friedrich in [122], where he analyzed the characteristic Cauchy problem using the contracted Bianchi identity and proved that the solutions to the problem are uniquely determined by some characteristic initial data set. The analysis was restricted to four dimensions but allowed him to treat both the standard and the asymptotic characteristic problem for the Einstein equations. In this review we will mainly follow Rendall’s approach [279], which was the first reference to prove that the problem is well-posed. We discuss alternative formulations at the end of the section.

Rendall’s method consists in reducing the characteristic problem to the standard Cauchy problem, a strategy we now review. Although his analysis in [279] is carried out in four dimensions, the arguments extend immediately to higher dimensions. Consider two hypersurfaces in $\mathbb{R}^n$, $N_1 := \{x^1 = -x^2\}$ and $N_2 := \{x^1 = x^2\}$, as well as a quasi-linear wave equations of the form

$$A^{ij}(x,\phi)\frac{\partial^2\phi}{\partial x^i \partial x^j} + F\left(x,\phi,\frac{\partial\phi}{\partial x^k}\right) = 0. \tag{3.2}$$

Let $\phi_0$ be a continuous function on $N_1 \cup N_2$ with smooth restriction to $N_1$ and $N_2$. Assume $N_1$ and $N_2$ are characteristic w.r.t. $A$ and that $\{x^1 = 0\}$ is spacelike (if not, interchange $x^1$ and $x^2$). Observe that all the transverse derivatives of $\phi_0$ can be computed on $N_1 \cup N_2$. To see this take for instance $N_1$. At the intersection $S := N_1 \cap N_2$, all the transverse derivatives of $\phi_0$ are just internal derivatives on $N_2$. By differentiating equation (3.2) order by order one can use the resulting equations to propagate such transverse derivatives along $N_1$ from their value at $S$, at least in a neighbourhood of $S$ in $N_1$. Rendall then uses this tower of derivatives on $N_1 \cup N_2$ to construct a function $\phi_1$ on $U$ (not necessarily a solution of (3.2)) using Whitney extension theorem. Abbreviating (3.2) as $A(\phi)\phi + F(\phi) = 0$, one sees that the function $\phi_1$ below is such that $A(\phi_1)\phi_1 + F(\phi_1)$ vanishes together with all its partial derivatives on $N_1 \cup N_2$. This allows us to construct a smooth function

$$\rho = \begin{cases} 0 & \text{for} \quad x^1 > x^2, x^1 > -x^2, \\ A(\phi_1)\phi_1 + F(\phi_1) & \text{elsewhere.} \end{cases}$$

Consider the equation

$$A(\phi_1 + \chi)(\phi_1 + \chi) + F(\phi_1 + \chi) = \rho$$

for an unknown function $\chi$. After defining $A_1(\chi) = A(\phi_1 + \chi)$ and $F_1(\chi) = A(\phi_1 + \chi)\phi_1 + F(\phi_1 + \chi) - \rho$, the previous equation becomes

$$A_1(\chi)\chi + F_1(\chi) = 0, \tag{3.3}$$

which is of the same form as (3.2). Consider the Cauchy problem for (3.3) with $\chi|_{\{x^1=0\}} = 0$. Since $A_1(\chi) = A(\phi_0)$ on $S$, it follows that the hypersurface $\{x^1 = 0\}$ is spacelike w.r.t. $A_1$ in some neighbourhood of $S$, and therefore the given Cauchy problem has a unique solution $\chi$ in a neighbourhood of $S$.

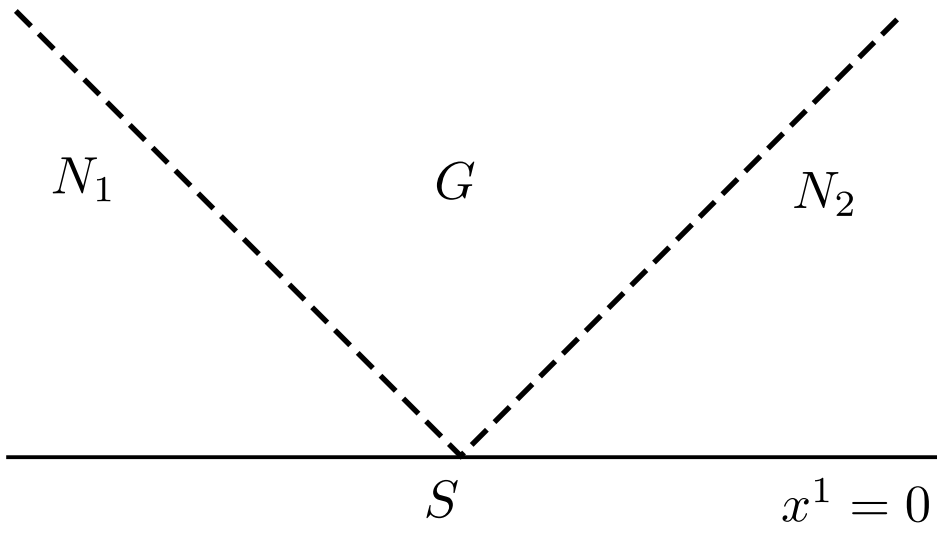

Figure 3.1

Consider now the region $G := \{x^1 \geq x^2,\ x^1 \geq -x^2\}$ as drawn in Figure 3.1 and let $U$ be a neighbourhood of $S$ in $\mathbb{R}^n$. Take a point $x \in U \setminus G$ and observe that there is a region $\Sigma$ of $\{x^1 = 0\}$ whose domain of dependence w.r.t. (3.2) contains $x$ and is disjoint from $G$. Note that in this domain $\chi = 0$ is a solution of (3.3), because $F_1(\chi = 0) = A(\phi_1)\phi_1 + F(\phi_1) - \rho$ vanishes at the complement of $G$. Since $\Sigma$ has the same domain of dependence w.r.t. (3.3) and $\chi = 0$, it follows that $\chi = 0$ is the unique solution of (3.3) in this domain of dependence with data $\chi|_{\{x^1=0\}} = 0$. Therefore, the solution $\chi$ of (3.3) with $\chi|_{\{x^1=0\}} = 0$ vanishes at the complement of $G$ in a neighbourhood of $S$. To conclude, define $\phi = \phi_1 + \chi$. It is clear that that $\phi$ solves (3.2) on $U \cap G$ and that $\phi|_{N_1 \cup N_2} = \phi_0$ because $\chi|_{N_1 \cup N_2} = 0$.

To prove uniqueness let $\widetilde{\phi}$ be another solution of (3.2) with the same initial data $\phi_0$. Then, the propagation equations along $N_1$ and $N_2$ imply that the function $\widetilde{\chi} := \widetilde{\phi} - \phi$ vanishes to infinite order on $N_1 \cup N_2$, and thus it can be extended smoothly to zero to the complement of $G$. Since both $\phi$ and $\widetilde{\phi}$ solve (3.2) with the same initial data, it follows that $\widetilde{\chi}$ satisfies the same equation as $\chi$ with the same initial data, so $\widetilde{\chi} = \chi$ on $G \cap U$ and therefore $\widetilde{\phi} = \phi$ on $G \cap U$. One can summarize this discussion as follows [279].

**Theorem 3.2.** *Let $\phi_0$ be a continuous function on $N_1 \cup N_2$ whose restrictions to $N_1$ and $N_2$ are smooth and such that $N_1 \cup N_2$ are characteristic hypersurfaces w.r.t. $A^{ij}(x, \phi_0)$. Then, there exists an open neighbourhood $U$ of $S$ and a unique smooth function $\phi$ on $U \cap G$ such that $\phi$ solves* (3.2) *and $\phi|_{N_1 \cup N_2} = \phi_0$.*

This result can also be extended to quasi-linear hyperbolic systems, such as the one obtained after writing the Einstein equations in harmonic coordinates [279]. The existence region can also be extended to a full neighbourhood of the two null hypersurfaces (see [50, 78]). Next we sketch how to construct initial data for the vacuum Einstein equations. For the details, we refer the reader to [208, 279].

First, one takes $\mathbb{R}^n$ with coordinates $\{u, v, x^A\}$, $A = 3, ..., n$, and considers the hyperplanes $N_1$ and $N_2$ given by $N_1 = \{u = 0\}$ and $N_2 = \{v = 0\}$. On $N_1 \cup N_2$ one specifies the conformal class $[g_{AB}]$ (we assume $g_{AB}$ is everywhere positive definite). The idea is to produce a spacetime $(M, g)$ for which $\{u, v, x^A\}$ are double null coordinates (see [208] for a review of this coordinate system). We only discuss the procedure on $N_1$, as the one on $N_2$ is analogous. In these coordinates, a vacuum spacetime satisfies

$$(g^{AB} g_{AB,u})_{,u} = \frac{1}{2} g^{AB}{}_{,u} g_{AB,u}. \tag{3.4}$$

One uses this equation as a second order transport equation for the conformal factor $\Omega$, which allows to determine $g_{AB}$ on $N_1$ from the values of $\Omega$, $\Omega_{,u}$ and $\Omega_{,v}$ at $S$ (see e.g. [208]). Assume $g_{uu} = g_{uA} = 0$. It then follows that $g^{uv} g_{uv} = 1$, $g^{vA} = g^{vv} = 0$ and that $g^{AB}$ is the inverse of $g_{AB}$. We want to prescribe $g_{vv}$, $g_{uv}$ and $g_{vA}$ on $N_1$ such that the harmonic coordinate gauge is satisfied to solve the reduced Einstein equations.

One starts by writing down condition $\Gamma^v = 0$ in coordinates, namely

$$g_{uv,u} = \frac{1}{4} g^{AB} g_{AB,u} g_{uv}. \tag{3.5}$$

Once the value of $g_{AB}$ have been determined, one solves the previous equation for $g_{uv}$ with initial condition $g_{uv} = -1$ on $S$. Then, it turns out that the reduced equation $R^H_{uu} = R_{uu} + g_{uv} \Gamma^v{}_{,u} = 0$ becomes a first order, homogeneous ODE for the quantity $g_{uu,v} - 2 g_{uv,u}$, so assuming it is zero on $S$, it then vanishes everywhere on $N_1$.

Next one wants to find a linear combination of $R_{uA}$ and $\Gamma^A$ that when set to zero gives a system of ODEs for $g_{vA}$. The combination $R_{uA} - \frac{1}{2} g_{AB} \Gamma^B{}_{,u}$ does, but it contains terms depending on transverse derivatives of the metric such as $g_{uA,v}$ that are still not under control. One can then modify this equation by including a term of the form $K_{AB} \Gamma^B$, where $K_{AB}$ is a complicated expression depending on $g_{AB}$, $g_{uv}$ and $g_{uu,v}$ (recall that the latter is already harmless). Then, by solving $R_{uA} - \frac{1}{2} g_{AB} \Gamma^B{}_{,u} + K_{AB} \Gamma^B = 0$ with initial data $g_{vA}|_S = 0$ (this comes from the corresponding argument on $N_2$) and $g_{vA,u}|_S$, one can determine $g_{vA}$ on $N_1$.

Moreover, the reduced Einstein equation $R^H_{uA} = R_{uA} + \frac{1}{2} g_{AB} \Gamma^B{}_{,u} = 0$ becomes a first order homogeneous equation for $\Gamma^B$. After noting that

$$\Gamma^B = g^{AB} g^{uv} (g_{uA,v} + g_{vA,u}) + \cdots , \tag{3.6}$$

where the dots are terms depending on harmless objects, it turns out that there is a unique way of choosing $g_{uA,v}$ on $S$ in terms of $g_{vA,u}|_S$ such that $\Gamma^B = 0$ on $S$, and hence everywhere on $N_1$. Note that now the terms $g_{uA,v}$ are known on $N_1$.

Finally, one wants to do a similar thing concerning $g_{vv}$, i.e. find a combination of $g^{AB} R_{AB}$ and $\Gamma^u$ that leads to an ODE for $g_{vv}$. The expression $g^{AB} R_{AB} - 2\Gamma^u{}_{,u}$ does, but it carries terms such as $g_{uv,v}$, $g_{uA,v}$ and $g_{AB,v}$. The terms $g_{uv,v}$ and $g_{uA,v}$ are already known on $N_1$, but the terms $g_{AB,v}$ do not. By following a similar strategy as before, and adding a term of the form $K\Gamma^u$, it turns out that the equation

$$g^{AB} R_{AB} - 2\Gamma^u{}_{,u} - K\Gamma^u = 0$$

provides a second order ODE for $g_{vv}$ that allows us to determine $g_{vv}$ on $N_1$ from $g_{vv}|_S = 0$ (again, this comes from the corresponding argument on $N_2$) and $g_{vv,u}|_S$. From the expression of $\Gamma^u$ in coordinates, namely

$$\Gamma^u = g^{uv} g^{uv} g_{vv,u} - \frac{1}{2} g^{uv} g^{AB} g_{AB,v} + \cdots ,$$

where again the dots are harmless terms, it follows that there is a unique way to prescribe $g_{vv,u}|_S$ such that $\Gamma^u|_S = 0$, and hence the reduced Einstein equation

$$g^{AB} R^H_{AB} = g^{AB} R_{AB} = 0$$

yields a homogeneous ODE for $\Gamma^u$, which proves $\Gamma^u = 0$ everywhere on $N_1$. The remaining part of the argument (i.e. the fact that $\Gamma^\mu = 0$ everywhere and then $(M, g)$ is a solution of the full Einstein equations) is analogous to the procedure described for the spacelike case. One can summarize this discussion in the following theorem [279].

**Theorem 3.3.** *Let $h_{AB}$, $A, B = 3, ..., n$ be continuous functions on $N_1 \cup N_2$ with smooth restrictions to $N_1$ and $N_2$ such that their corresponding matrix is symmetric, positive definite and of determinant 1. Let $\Omega$, $\Omega_{,u}$, $\Omega_{,v}$, $g_{vA,u}$ be given on $S$. Then, there exists an open neighbourhood $U$ of $S$ in the region $\{u \geq 0,\ v \geq 0\}$, a unique $\Omega$ on $(N_1 \cup N_2) \cap U$ and a unique Lorentzian metric $g$ on $U$ such that (i) $g_{AB} = \Omega h_{AB}$ on $N_1 \cup N_2$, (ii) $g$ satisfies the vacuum equations on $U$, (iii) $g$ induces the given data on $N_1 \cup N_2$ and (iv) the given coordinates on $\mathbb{R}^n$ are double null coordinates for $g$.*

This result can be generalized to $\lambda$-vacuum and to include matter fields admitting a well-posed characteristic problem and whose energy-momentum tensor has suitable dependence on the data, such as scalar, Maxwell or Yang-Mills fields (see [64]). We will analyze this further in Section 3.3.

It turns out that the torsion one-form can be written at $S$ as $\tau_A = \dfrac{1}{2} g^{uv}(g_{vA,u} - g_{uA,v})$. Taking into account that from (3.6) and $\Gamma^B = 0$ the term $g_{uA,v}$ on $S$ is fixed by $g_{vA,u}|_S$, Theorem 3.3 can be reformulated in terms of the torsion [64]. Another possibility is to prescribe all the metric coefficients at $N_1 \cup N_2$ as well as the surface gravity of the vectors $\partial_u$ and $\partial_v$ at $N_2$ and $N_1$, respectively, subject to the Raychaudhuri constraint equation. These possibilities, among others, are discussed in [64].

While Rendall's theorem only proves existence of vacuum spacetimes in a neighbourhood of the intersection $N_1 \cap N_2$, there has been much effort put in proving existence in a full neighbourhood of the characteristic hypersurfaces. In [208], Luk has shown that the maximal development of the data covers a full neighbourhood of $N_1 \cup N_2$ when $S$ is a two-sphere. The proof is based on energy-type estimates and relies in the null structure of Einstein field equations and the Bianchi identities in double null coordinates. This result was later improved in [50] to allow for general topologies but still in four dimensions (see also [164] for an analysis using the Newman–Penrose formalism). Finally, in a recent work [78] the authors extend the result to arbitrary topology and dimension. In this case, the proof is based on and observation by Senovilla [306] that the Bianchi identities lead to a symmetric hyperbolic system also in higher dimensions. The characteristic problem has also been analyzed in the presence of matter fields (see e.g. [247]) and when data is posed on the future light-cone of a point (see [57, 64] and the references therein).

We emphasize that none of the approaches mentioned in this section attempt to define and study the characteristic data as "detached" from the spacetime one wishes to construct. The main purpose of this chapter is to provide a full geometric formulation of the characteristic Cauchy problem completely detached form the spacetime to be constructed, as with the standard Cauchy problem. Such detached notion will be introduced in Section 3.2, and the existence and geometric uniqueness results are proven in Sections 3.3 and 3.4, respectively.

### 3.1.3 *Other Cauchy problems*

Apart from the spacelike and the characteristic problems, there are several other Cauchy problems studied in the literature. For instance, one can consider a compact spacelike hypersurface $\Sigma$ with boundary $\partial\Sigma$ along with the outgoing null hypersurface $\mathcal{H}$ emanating from $\partial\Sigma$. The spacelike-characteristic Cauchy problem with data on $\Sigma \cup \mathcal{H}$ has been analyzed in [75, 91]. It is very plausible that a similar result must be true when instead of considering a non-smooth "corner" $\mathcal{H} \cap \Sigma$, the transition from $\Sigma$ to $\mathcal{H}$ is smooth. As far as we known this result is not available in the literature. We make a conjecture along these lines in Section 4.7.

To conclude this introductory section we simply mention that there also exist boundary-initial value problems where data is posed on an initial spacelike hypersurface intersecting a timelike boundary (see e.g. [16–18, 129, 133]).

## 3.2 DOUBLE NULL DATA

In this section we introduce the main object of this chapter, namely the notion of *double null data*, which corresponds to the detached version of two transverse null hypersurfaces. In order to define such an object, the first step is being able to glue two null hypersurface data along their boundaries.

**Definition 3.4.** *Let $\mathcal{D}$ and $\underline{\mathcal{D}}$ be two null hypersurface data with boundaries $\mathcal{S} := \partial\mathcal{H}$ and $\underline{\mathcal{S}} := \partial\underline{\mathcal{H}}$ and $\phi : \mathcal{S} \longrightarrow \underline{\mathcal{S}}$ a diffeomorphism between them. An invertible linear map*

$$\Psi_p : T_p\mathcal{H} \oplus \mathbb{R} \longrightarrow T_{\phi(p)}\underline{\mathcal{H}} \oplus \mathbb{R}$$

*is called a $\partial$-isometry at $p \in \mathcal{S}$ provided that*[2]

1. $\Psi_p|_{T_p\mathcal{S}}\left((X,0)\right) = (\phi_\star|_p(X), 0)$ *for all* $X \in \mathfrak{X}(\mathcal{S})$,
2. $\Psi_p^\star \underline{\mathcal{A}}_{\phi(p)} = \mathcal{A}_p$.

*If conditions (1) and (2) hold at every $p \in \mathcal{S}$ we say that $\mathcal{D}$ and $\underline{\mathcal{D}}$ are $\partial$-isometric.*

We will always assume that $\mathcal{S}$ and $\underline{\mathcal{S}}$ are non-degenerate codimension one submanifolds of $\mathcal{H}$ and $\underline{\mathcal{H}}$, so we can talk about normal pairs on $\mathcal{S}$ and $\underline{\mathcal{S}}$. Given a normal pair $\mathfrak{t} = (T, \alpha)$ of $\mathcal{S}$, the image of $\mathfrak{t}$ under $\Psi$ is still a normal pair on $\underline{\mathcal{S}}$, since $\phi$ is a diffeomorphism and

$$0 = \mathcal{A}\left(\mathfrak{t}, (X,0)\right) = \left(\Psi^\star \underline{\mathcal{A}}\right)\left(\mathfrak{t}, (X,0)\right) = \underline{\mathcal{A}}\left(\Psi(\mathfrak{t}), (\phi_\star X, 0)\right) \qquad \forall X \in \mathfrak{X}(\mathcal{S}).$$

Moreover, if a normal pair $\mathfrak{t}$ of $\mathcal{S}$ is $\mathcal{A}$-null, the normal pair $\Psi(\mathfrak{t})$ of $\underline{\mathcal{S}}$ is also $\underline{\mathcal{A}}$-null, since

$$\underline{\mathcal{A}}\left(\Psi(\mathfrak{t}), \Psi(\mathfrak{t})\right) = \mathcal{A}\left(\mathfrak{t}, \mathfrak{t}\right) = 0.$$

In the following proposition we show that given an isometry $\phi : \mathcal{S} \longrightarrow \underline{\mathcal{S}}$ and a non-vanishing function $\sigma \in \mathcal{F}^\star(\underline{\mathcal{S}})$, a natural $\partial$-isometry can be constructed.

**Proposition 3.5.** *Let $\mathcal{D}$ and $\underline{\mathcal{D}}$ be two null hypersurface data and $\phi : (\mathcal{S}, h) \longrightarrow (\underline{\mathcal{S}}, \underline{h})$ an isometry. For each $\sigma \in \mathcal{F}^\star(\underline{\mathcal{S}})$, there exists a unique $\partial$-isometry $\Psi : T\mathcal{H} \oplus \mathcal{F}(\mathcal{S}) \longrightarrow T\underline{\mathcal{H}} \oplus \mathcal{F}(\underline{\mathcal{S}})$ determined by the condition $\underline{\mathcal{A}}\left((\underline{n}, 0), \Psi(n,0)\right) = \sigma$.*

*Proof.* The existence and uniqueness proof will be constructive, i.e., we will impose conditions (1) and (2) in Def. 3.4 and show that there is a unique candidate. We will then check that this candidate is indeed a $\partial$-isometry. Let $X \in \mathfrak{X}(\mathcal{S})$. We define $\Psi\left((X,0)\right) := (\phi_\star X, 0)$. In order to define the image of $(n, 0)$ and $(\theta, 1)$ under $\Psi$ first observe that since $(n,0)$ and $(\theta, 1)$ are

2 Given $V, W \in T_p\mathcal{H} \oplus \mathbb{R}$, we define $\left(\Psi^\star \underline{\mathcal{A}}_{\phi(p)}\right)(V,W) := \underline{\mathcal{A}}_{\phi(p)}\left(\Psi(V), \Psi(W)\right)$.

$\mathcal{A}$-null, $\Psi\left((n,0)\right)$ and $\Psi\left((\theta,1)\right)$ are $\underline{\mathcal{A}}$-null too, and since $\underline{\mathcal{A}}\left((\underline{n},0),(\underline{\theta},1)\right)=1$ (see Remark 2.24) we necessarily have $\Psi\left((n,0)\right)\in\operatorname{span}\left\{(\underline{\theta},1)\right\}$ and $\Psi\left((\theta,1)\right)\in\operatorname{span}\left\{(\underline{n},0)\right\}$. Imposing the condition $\underline{\mathcal{A}}\left((\underline{n},0),\Psi(n,0)\right)=\sigma$ the only option is

$$\Psi\left((n,0)\right)=\sigma(\underline{\theta},1). \tag{3.7}$$

To complete the determination of $\Psi$ we only need to find $\Psi\left((\theta,1)\right)$, which we already know is of the form $\Psi\left((\theta,1)\right)=\alpha(\underline{n},0)$. The proportionality function $\alpha$ is obtained from $\mathcal{A}\left((n,0),(\theta,1)\right)=1$ and its underlined version after imposing item (2) of Def. 3.4 as follows

$$1=\mathcal{A}\left((\theta,1),(n,0)\right)=\underline{\mathcal{A}}\left(\Psi\left((\theta,1)\right),\Psi((n,0))\right)=\alpha\sigma\underline{\mathcal{A}}\left((\underline{n},0),(\underline{\theta},1)\right)=\alpha\sigma.$$

Since $\sigma\neq 0$ by hypothesis we conclude

$$\Psi\left((\theta,1)\right)=\sigma^{-1}(\underline{n},0). \tag{3.8}$$

In matrix notation, the expression of $\Psi$ in the decomposition $T\mathcal{S}\oplus\operatorname{span}\left\{(n,0),(\theta,1)\right\}$ (and the corresponding one in the image) is[3]

$$\Psi=\begin{pmatrix}(\phi_\star) & \mathbf{0} & \mathbf{0}\\ \mathbf{0} & 0 & \sigma^{-1}\\ \mathbf{0} & \sigma & 0\end{pmatrix}. \tag{3.9}$$

It is now straightforward to check that this candidate to be a $\partial$-isometry is indeed a $\partial$-isometry. Indeed, conditions (1) and (2) of Def. 3.4 hold by construction and because $\phi$ is an isometry, and by expression (3.9) and the fact that $\phi$ is a diffeomorphism, we conclude that $\Psi$ is invertible. □

In (3.9) the expression of $\Psi$ in the basis $T\mathcal{S}\oplus\operatorname{span}\left\{(n,0),(\theta,1)\right\}$ (and the corresponding one in the image) was obtained. In some situations it may be interesting to have an expression for $\Psi$ in a more general basis.

**Proposition 3.6.** *Let $\mathcal{D}$ and $\underline{\mathcal{D}}$ be null hypersurface data, $\phi:\mathcal{S}\longrightarrow\underline{\mathcal{S}}$ an isometry and $\Psi$ the unique $\partial$-isometry satisfying $\underline{\mathcal{A}}\left((\underline{n},0),\Psi(n,0)\right)=\sigma\neq 0$. Let $v$, $\underline{v}$ be transverse vectors to $\mathcal{S}$ and $\underline{\mathcal{S}}$, respectively. Then the linear map $\Psi$ w.r.t. decompositions $T\mathcal{S}\oplus span\left\{(v,0),(0,1)\right\}$ and $T\mathcal{S}\oplus span\left\{(\underline{v},0),(0,1)\right\}$ is given by*

$$\Psi=\begin{pmatrix}(\phi_\star) & \phi_\star v_{\parallel}-v_n\sigma\left(\underline{\alpha}\underline{v}_n^{-1}\underline{v}_{\parallel}+\underline{\ell}^\sharp\right) & \phi_\star\ell^\sharp+\sigma\alpha\underline{\ell}^\sharp+\underline{v}_n^{-1}(\sigma\alpha\underline{\alpha}-\sigma^{-1})\underline{v}_{\parallel}\\ \mathbf{0} & \sigma v_n\underline{\alpha}\underline{v}_n^{-1} & \underline{v}_n^{-1}(\sigma^{-1}-\sigma\alpha\underline{\alpha})\\ \mathbf{0} & \sigma v_n & -\sigma\alpha\end{pmatrix}, \tag{3.10}$$

3 When a map is written in matrix notation we follow the convention that the entries of the $i$-th column are the coefficients of the image of the $i$-th vector in the basis.

*where* $v_n \in \mathcal{F}^\star(\mathcal{S})$, $\underline{v}_n \in \mathcal{F}^\star(\underline{\mathcal{S}})$, $v_\parallel \in \mathfrak{X}(\mathcal{S})$ *and* $\underline{v}_\parallel \in \mathfrak{X}(\underline{\mathcal{S}})$ *are univocally defined by the decompositions* $v = v_n n + v_\parallel$ *and* $\underline{v} = \underline{v}_n \underline{n} + \underline{v}_\parallel$, *and where we introduce the functions* $\alpha = -\frac{1}{2}(\ell^{(2)} - \ell_\sharp^{(2)})$ *and* $\underline{\alpha} = -\frac{1}{2}(\underline{\ell}^{(2)} - \underline{\ell}_\sharp^{(2)})$ *to simplify the notation.*

*Proof.* We only need to compute the second and third columns. Since $\Psi\left((v,0)\right) = v_n \Psi\left((n,0)\right) + \Psi((v_\parallel,0))$, employing (3.7) together with $\underline{\theta} = \underline{\alpha}\underline{n} - \underline{\ell}^\sharp$ (see Remark 2.24, and we omit the $\iota_\star$ in order not to overload the notation),

$$\Psi\left((v,0)\right) = v_n\sigma(\underline{\theta},1) + (\phi_\star v_\parallel, 0) = \left(v_n\sigma\underline{\alpha}\underline{\mu}^{-1}(\underline{v} - \underline{v}_\parallel) - v_n\sigma\underline{\ell}^\sharp + \phi_\star v_\parallel, v_n\sigma\right),$$

and hence the second column of (3.10) follows. The third column requires computing $\Psi\left((0,1)\right)$. Decomposing $(0,1) = (\theta,1) - (\alpha n, 0) + (\ell^\sharp, 0)$ and using (3.9) yields

$$\begin{aligned}\Psi(0,1) &= \left(\sigma^{-1}\underline{n} - \sigma\alpha\underline{\theta} + \phi_\star\ell^\sharp, -\sigma\alpha\right)\\ &= \left(\sigma^{-1}\underline{v}_n^{-1}(\underline{v} - \underline{v}_\parallel) - \sigma\alpha\underline{\alpha}\underline{v}_n^{-1}(\underline{v} - \underline{v}_\parallel) + \sigma\alpha\underline{\ell}^\sharp + \phi_\star\ell^\sharp, -\sigma\alpha\right),\end{aligned}$$

so the third column of (3.10) is established. □

Next we study how the map $\Psi$ changes under gauge transformations on $\mathcal{D}$ and $\underline{\mathcal{D}}$.

**Proposition 3.7.** *Let* $\mathcal{D}$ *and* $\underline{\mathcal{D}}$ *be null hypersurface data and* $\Psi$ *be a* $\partial$*-isometry. Under gauge transformations on* $\mathcal{D}$ *and* $\underline{\mathcal{D}}$ *with parameters* $(z,V)$ *and* $(\underline{z},\underline{V})$*, respectively,* $\Psi$ *transforms as*

$$\mathcal{G}_{(z,V)}\Psi = \underline{G}^{-1} \circ \Psi \circ G, \tag{3.11}$$

*where* $G$ *is the invertible linear map*

$$\begin{aligned} G : T\mathcal{H} \oplus \mathcal{F}(\mathcal{S}) &\longrightarrow T\mathcal{H} \oplus \mathcal{F}(\mathcal{S})\\ (X,a) &\longmapsto G\left((X,a)\right) := (X + azV, az) \end{aligned} \tag{3.12}$$

*and* $\underline{G} : T\underline{\mathcal{H}} \oplus \mathcal{F}(\underline{\mathcal{S}}) \longrightarrow T\underline{\mathcal{H}} \oplus \mathcal{F}(\underline{\mathcal{S}})$ *is defined identically but with all the quantities carrying an underline. As a consequence, the transformation law of the function* $\sigma := \underline{\mathcal{A}}\left((\underline{n},0), \Psi(n,0)\right)$ *is*

$$\mathcal{G}_{(z,V)}\sigma = z^{-1}\underline{z}^{-1}\sigma, \tag{3.13}$$

*where in order not to overload the notation we denote with the same symbol a function* $f \in \mathcal{F}(\partial\mathcal{H})$ *and* $f \circ \phi^{-1} \in \mathcal{F}(\partial\underline{\mathcal{H}})$*. This slight abuse of notation will be used repeatedly from now on when no confusion arises.*

*Proof.* Let $\mathcal{D}$ be hypersurface data. From Def. 2.3 one can write the transformation law for the tensor $\mathcal{A}$ in a compact way as

$$\mathcal{A}'\left((U,a),(W,b)\right) = \mathcal{A}\left(G(U,a), G(W,b)\right) \tag{3.14}$$

for all $(U,a),(W,b)\in T\mathcal{H}\oplus\mathcal{F}(\mathcal{S})$, where as usual we use a prime to denote gauge transformed quantities. To show this we use matrix notation in which vectors are represented as columns and covectors as rows. The map (3.12) gets rewritten in matrix form as

$$\begin{pmatrix} U+azV \\ az \end{pmatrix} = \begin{pmatrix} \mathbb{1} & zV \\ 0 & z \end{pmatrix}\begin{pmatrix} U \\ a \end{pmatrix}.$$

Define, therefore

$$(G) = \begin{pmatrix} \mathbb{1} & zV \\ 0 & z \end{pmatrix}, \qquad (\mathcal{A}) = \begin{pmatrix} \gamma & \boldsymbol{\ell}^T \\ \boldsymbol{\ell} & \ell^{(2)} \end{pmatrix}.$$

Then (3.14) can be written in matrix form as

$$(\mathcal{A}') = (G)^T(\mathcal{A})(G). \tag{3.15}$$

To prove this equality (and hence (3.14)) we compute

$$\begin{aligned}(G)^T(\mathcal{A})(G) &= \begin{pmatrix} \mathbb{1} & 0 \\ zV^T & z \end{pmatrix}\begin{pmatrix} \gamma & \boldsymbol{\ell}^T \\ \boldsymbol{\ell} & \ell^{(2)} \end{pmatrix}\begin{pmatrix} \mathbb{1} & zV \\ 0 & z \end{pmatrix} \\ &= \begin{pmatrix} \gamma & \boldsymbol{\ell}^T \\ z\left(\gamma(V,\cdot)+\boldsymbol{\ell}\right) & z\left(\boldsymbol{\ell}(V)+\ell^{(2)}\right) \end{pmatrix}\begin{pmatrix} \mathbb{1} & zV \\ 0 & z \end{pmatrix} \\ &= \begin{pmatrix} \gamma & z\left(\gamma(V,\cdot)+\boldsymbol{\ell}\right)^T \\ z\left(\gamma(V,\cdot)+\boldsymbol{\ell}\right) & z\left(\gamma(V,V)+2\boldsymbol{\ell}(V)+\ell^{(2)}\right) \end{pmatrix},\end{aligned}$$

which is precisely the matrix form of $\mathcal{A}'$ after taken into account the transformation laws (2.27)-(2.29). Item (2) of Def. 3.4 can be also written in matrix form as $(\Psi)^T(\underline{\mathcal{A}})(\Psi) = (\mathcal{A})$. To compute the gauge behaviour of $\Psi$ we impose $(\Psi')^T(\underline{\mathcal{A}}')(\Psi') = (\mathcal{A}')$ and use equation (3.15),

$$\begin{aligned}(\Psi')^T(\underline{\mathcal{A}}')(\Psi') = (\mathcal{A}') \quad &\Longleftrightarrow \quad (\Psi')^T(\underline{G})^T(\underline{\mathcal{A}})(\underline{G})(\Psi') = (G)^T(\mathcal{A})(G) \\ &\Longleftrightarrow \quad \left((G)^T\right)^{-1}(\Psi')^T(\underline{G})^T(\underline{\mathcal{A}})(\underline{G})(\Psi')(G)^{-1} = (\mathcal{A}),\end{aligned}$$

from where we conclude that $\Psi = \underline{G}\circ\Psi'\circ G^{-1}$ and hence $\Psi' = \underline{G}^{-1}\circ\Psi\circ G$, which is (3.11). The transformation of $\sigma$ follows from those of $\Psi$ and $\mathcal{A}$. Indeed, by (3.14),

$$\sigma' := \underline{\mathcal{A}}'\left(\Psi'(n',0),(\underline{n}',0)\right) = \underline{\mathcal{A}}\left((\underline{G}\circ\Psi')(n',0),\underline{G}(\underline{n}',0)\right),$$

and using (3.11) together with $G\left((n',0)\right) = z^{-1}(n,0)$ (and its underlined version, see (2.34)),

$$\begin{aligned}\sigma' &= \underline{\mathcal{A}}\left((\Psi \circ G)\,(n',0), \underline{G}(\underline{n}',0)\right)\\ &= \underline{\mathcal{A}}\left(\Psi\left(G(n',0)\right), \underline{G}(\underline{n}',0)\right)\\ &= z^{-1}\underline{z}^{-1}\underline{\mathcal{A}}\left(\Psi\left((n,0)\right), (\underline{n},0)\right)\\ &= z^{-1}\underline{z}^{-1}\sigma.\end{aligned}$$

□

**Remark 3.8.** *The transformation law of a normal pair in Def. 2.19 can be rewritten in terms of the map $G$ defined in* (3.12) *by* $\mathfrak{t}'_i = G^{-1}(\mathfrak{t}_i)$.

Once we have introduced the notion of $\partial$-isometry, the next step is to define two null and transverse hypersurfaces from an abstract point of view, i.e., without seeing them as embedded in any ambient spacetime. Let $\mathcal{D}$ and $\underline{\mathcal{D}}$ be embedded CHD with respective embeddings $\Phi$ and $\underline{\Phi}$ in a spacetime $(\mathcal{M}, g)$, and suppose $\Phi(\mathcal{S}) = \underline{\Phi}(\underline{\mathcal{S}})$. Let $\phi : \mathcal{S} \longrightarrow \underline{\mathcal{S}}$ be the induced diffeomorphism and let $\Psi$ be a $\partial$-isometry satisfying $\underline{\mathcal{A}}\left(\Psi(n,0), (\underline{n},0)\right) = \sigma \in \mathcal{F}(\underline{\mathcal{S}})$. By Defs. 2.1 and 2.2, the object $\underline{\mathcal{A}}\left(\Psi(n,0), (\underline{n},0)\right)$ can be thought as the scalar product $g(\nu, \underline{\nu})$ at $\Phi(\mathcal{S})$, where recall $\nu = \Phi_\star n$ and $\underline{\nu} = \underline{\Phi}_\star \underline{n}$. Thus, if one wants to define abstractly two null and transverse hypersurfaces with the same orientation for $\nu$ and $\underline{\nu}$, the function $\underline{\mathcal{A}}\left(\Psi(n,0), (\underline{n},0)\right)$ must be everywhere negative when $n$ and $\underline{n}$ point into the interior of $\mathcal{H}$ and $\underline{\mathcal{H}}$, respectively. Then by Prop. 3.5 the condition $\underline{\mathcal{A}}\left(\Psi(n,0), (\underline{n},0)\right) = \sigma \neq 0$ fixes uniquely the map $\Psi$. This allows for the identification (see Remark 2.24)

$$\nu \overset{\mathcal{S}}{=} \sigma(\underline{\xi} + \underline{\Phi}_\star \underline{\theta}), \qquad \underline{\nu} \overset{\mathcal{S}}{=} \sigma(\xi + \Phi_\star \theta), \tag{3.16}$$

where the vectors $\theta$ and $\underline{\theta}$ are given by

$$\theta = -\frac{1}{2}(\ell^{(2)} - \ell^{(2)}_\sharp) n - \ell^A e_A, \qquad \underline{\theta} = \frac{1}{2}(\underline{\ell}^{(2)} - \underline{\ell}^{(2)}_\sharp)\underline{n} - \underline{\ell}^A \underline{e}_A. \tag{3.17}$$

Moreover, if we want $\Phi(\mathcal{S})$ and $\underline{\Phi}(\underline{\mathcal{S}})$ to correspond to the same (codimension two) surface in the ambient spacetime, there are several necessary conditions that they need to fulfill. Firstly, their induced metrics have to agree. Secondly, their second fundamental forms and torsion one-forms also need to agree. And finally, the pullback of the ambient Ricci tensor into the two surfaces must agree too. The "zeroth order" condition, namely that the induced metrics coincide, can be simply written as $\phi^\star \underline{h} = h$. In order to write the "first order" conditions, i.e. the ones of the second fundamental forms and torsion one-forms, we can employ the tools developed in Section 2.4. Indeed, these conditions can be expressed abstractly thanks to the notion of normal pairs. Identifying the ambient vectors associated to a basis of normal pairs $\{\mathfrak{t}_i\}$ with the ambient vectors associated to the normal pairs $\{\Psi(\mathfrak{t}_i)\}$, the second fundamental forms $\mathcal{K}^{\mathfrak{t}_i}$ and the torsions $\urcorner(\mathfrak{t}_i, \mathfrak{t}_j)$ of $\mathcal{S}$ must agree with those of $\underline{\mathcal{S}}$, namely $\underline{\mathcal{K}}^{\Psi(\mathfrak{t}_i)}$ and $\urcorner(\Psi(\mathfrak{t}_i), \Psi(\mathfrak{t}_j))$. Finally, the "second order" condition, i.e. the one involving the ambient Ricci tensor, can be expressed abstractly thanks to the abstract tensor $\boldsymbol{\mathcal{R}}$ introduced in Section 2.2. Indeed, since $\Phi^\star \operatorname{Ric} = \boldsymbol{\mathcal{R}}$ and $\underline{\Phi}^\star \operatorname{Ric} = \underline{\boldsymbol{\mathcal{R}}}$, the pullback of the tensors $\boldsymbol{\mathcal{R}}$ and $\underline{\boldsymbol{\mathcal{R}}}$ on $\mathcal{S}$ and $\underline{\mathcal{S}}$ must also agree. This whole discussion motivates the following abstract and fully gauge-covariant definition of double null data.

**Definition 3.9.** *Let $\mathcal{H}$ and $\underline{\mathcal{H}}$ be two manifolds with boundaries $i : \mathcal{S} \hookrightarrow \mathcal{H}$ and $\underline{i} : \underline{\mathcal{S}} \hookrightarrow \underline{\mathcal{H}}$ and $\sigma \in \mathcal{F}(\underline{\mathcal{S}})$ everywhere negative. Let $\mathcal{D} = \{\mathcal{H}, \boldsymbol{\gamma}, \boldsymbol{\ell}, \ell^{(2)}, \mathbf{Y}\}$ and $\underline{\mathcal{D}} = \{\underline{\mathcal{H}}, \underline{\boldsymbol{\gamma}}, \underline{\boldsymbol{\ell}}, \underline{\ell}^{(2)}, \underline{\mathbf{Y}}\}$ be CHD with $\boldsymbol{\gamma}$ and $\underline{\boldsymbol{\gamma}}$ semi-positive definite, and restrict $n$ and $\underline{n}$ to point towards the interior of $\mathcal{H}$ and $\underline{\mathcal{H}}$, respectively. Let $\phi : \mathcal{S} \longrightarrow \underline{\mathcal{S}}$ be an isometry and define $\Psi$ as the unique $\partial$-isometry such that $\underline{\mathcal{A}}\left(\Psi(n,0), (\underline{n},0)\right) = \sigma$. Let $\{\mathsf{t}_i\}$ be a basis of normal pairs of $\mathcal{S}$. Then we say that the four-tuple $\{\mathcal{D}, \underline{\mathcal{D}}, \phi, \sigma\}$ is double null data (DND) provided that the following conditions hold at $\mathcal{S}$*

$$\phi^\star \underline{h} = h, \tag{3.18}$$

$$\Psi^\star \left(\underline{\mathcal{K}}^{\Psi(\mathsf{t}_i)}\right) = \mathcal{K}^{\mathsf{t}_i}, \tag{3.19}$$

$$\Psi^\star \left(\underline{⅂}(\Psi(\mathsf{t}_i), \Psi(\mathsf{t}_j))\right) = ⅂(\mathsf{t}_i, \mathsf{t}_j), \tag{3.20}$$

$$\phi^\star \left(\underline{i}^\star \underline{\boldsymbol{\mathcal{R}}}\right) = i^\star \boldsymbol{\mathcal{R}}. \tag{3.21}$$

Although (3.18) is redundant since $\phi : \mathcal{S} \longrightarrow \underline{\mathcal{S}}$ is already assumed to be an isometry, we write down it again for completeness. Note also that the above definition implies in particular that $\mathcal{H}$ and $\underline{\mathcal{H}}$ are of the same dimension (we refer to it as the dimension of the DND) and that $\boldsymbol{\gamma}$ and $\underline{\boldsymbol{\gamma}}$ have signature $(0, 1, ..., 1)$ at the boundaries $\partial\mathcal{H}$ and $\partial\underline{\mathcal{H}}$. The restriction on the signature is needed to show existence and uniqueness of $\lambda$-vacuum developments of the data (Theorems 3.23 and 3.30). In addition, since $\boldsymbol{\gamma}$ and $\underline{\boldsymbol{\gamma}}$ have exactly one degeneration direction at every point it follows that they have this signature everywhere. In the next remark we show that Def. 3.9 is well-defined, i.e., that the compatibility conditions are independent of the basis of normal pairs and also gauge invariant.

**Remark 3.10.** *Conditions $\phi^\star \underline{h} = h$ and $\phi^\star \left(\underline{i}^\star \underline{\boldsymbol{\mathcal{R}}}\right) = i^\star \boldsymbol{\mathcal{R}}$ are automatically gauge invariant by virtue of the transformations law* (2.27) *and item 1. in Prop. 2.10. Then it suffices to show the gauge invariance of* (3.19)*-*(3.20)*. We start with* (3.19)*. Let $(z, V)$ and $(\underline{z}, \underline{V})$ be gauge parameters and denote by $\mathsf{t}'_i$ the gauge transformed normal pair of $\mathsf{t}_i$. Firstly, the RHS of* (3.19) *is simply $\mathcal{K}^{\mathsf{t}'_i}$ as a direct consequence of the gauge invariance of the second fundamental form in Lemma 2.20. We need to show that the LHS can also be written with all objects gauge transformed. Let $G$, $\underline{G}$ be defined as in Prop. 3.7. By Lemma 2.20 the LHS of* (3.19) *can be written as*

$$\Psi^\star \left(\underline{\mathcal{K}}^{\Psi(\mathsf{t}_i)}\right) = \Psi^\star \left(\underline{\mathcal{K}}^{\underline{G}^{-1}(\Psi(\mathsf{t}_i))}\right),$$

*since the second fundamental form $\underline{\mathcal{K}}$ along a normal pair is gauge invariant and $\underline{G}^{-1}\left(\Psi(\mathsf{t}_i)\right)$ is the gauge transformation of the normal pair $\Psi(\mathsf{t}_i)$ with gauge parameters $(\underline{z}, \underline{V})$ (see Remark 3.8). Then, using $\underline{G}^{-1} \circ \Psi = \Psi' \circ G^{-1}$ (by Prop. 3.7),*

$$\Psi^\star \left(\underline{\mathcal{K}}^{\underline{G}^{-1}(\Psi(\mathsf{t}_i))}\right) = \Psi^\star \left(\underline{\mathcal{K}}^{\Psi'\left(G^{-1}(\mathsf{t}_i)\right)}\right) = \Psi^\star \left(\underline{\mathcal{K}}^{\left(\Psi'(\mathsf{t}'_i)\right)}\right),$$

*where in the last equality we used again $\mathsf{t}'_i = G^{-1}\mathsf{t}_i$. Noting that $\Psi\left((X,0)\right) = \Psi'\left((X,0)\right)$ for every $X \in \mathfrak{X}(\underline{\mathcal{S}})$, the LHS of equation* (3.19) *finally gets rewritten as*

$$\Psi'^\star \left(\underline{\mathcal{K}}^{\left(\Psi'(\mathsf{t}'_i)\right)}\right),$$

*which is exactly the LHS of* (3.19) *with all quantities gauge transformed. This establishes the gauge covariance of conditions* (3.19)*. Moreover, by Lemma 2.18,* (3.19) *are also independent of the basis of normal pairs.*

*The gauge invariance of conditions* (3.20) *can be shown in a similar way as in the case of the second fundamental form. Firstly, after a gauge transformation the RHS of* (3.20) *is simply* $\daleth(\mathfrak{t}'_i, \mathfrak{t}'_j)$*, by the gauge invariance of* $\daleth$ *in Prop. 2.40. To see that the LHS of* (3.20) *can be written with all quantities gauge transformed, we use again* $\underline{G}^{-1} \circ \Psi = \Psi' \circ G^{-1}$*,* $\mathfrak{t}'_i = G^{-1}\mathfrak{t}_i$ *and* $\Psi\left((X,0)\right) = \Psi'\left((X,0)\right)$*, so that*

$$\Psi^\star\left(\underline{\daleth}(\Psi(\mathfrak{t}_i), \Psi(\mathfrak{t}_j))\right) = \Psi^\star\left(\underline{\daleth}(\underline{G}^{-1}\left(\Psi(\mathfrak{t}_i)\right), \underline{G}^{-1}\left(\Psi(\mathfrak{t}_j)\right))\right) = \Psi'^\star\left(\underline{\daleth}(\Psi'(\mathfrak{t}'_i), \Psi'(\mathfrak{t}'_j))\right),$$

*where in the first equality we invoked again the gauge invariance of* $\daleth$ *(Prop. 2.40). Thus, the gauge covariance of* (3.20) *is also established. Finally, by Prop. 2.28 under a change of basis of normal pairs* $\hat{\mathfrak{t}}_i = \Omega_i^k \mathfrak{t}_k$*, the RHS of* (3.20) *transforms as*

$$\daleth(\hat{\mathfrak{t}}_i, \hat{\mathfrak{t}}_j) = \Omega_i^k \Omega_j^l \daleth(\mathfrak{t}_k, \mathfrak{t}_l) + \Omega_i^k M_{kl} d\Omega_j^l,$$

*whereas the transformation of the LHS is*

$$\begin{aligned}\Psi^\star\left(\underline{\daleth}(\Psi(\hat{\mathfrak{t}}_i), \Psi(\hat{\mathfrak{t}}_j))\right) = \Psi^\star\Big(&\left(\Omega_i^k \circ \phi^{-1}\right)\left(\Omega_j^l \circ \phi^{-1}\right)\underline{\daleth}(\Psi(\mathfrak{t}_k), \Psi(\mathfrak{t}_l)) \\ &+ \left(\Omega_i^k \circ \phi^{-1}\right)\underline{M}_{kl} d(\Omega_j^l \circ \phi^{-1})\Big) \\ = \Omega_i^k \Omega_j^l &\Psi^\star\left(\underline{\daleth}(\Psi(\mathfrak{t}_k), \Psi(\mathfrak{t}_l))\right) + \Omega_i^k M_{kl} d\Omega_j^l,\end{aligned}$$

*where we used* $\Psi^\star \underline{M}_{kl} = M_{kl}$ *(item (2) in Def. 3.4),* $\Psi(\hat{\mathfrak{t}}_i) = \Psi(\Omega_i^k \mathfrak{t}_k) = (\Omega_i^k \circ \phi^{-1})\Psi(\mathfrak{t}_k)$ *and that the pullback commutes with the exterior derivative. Since both sides in* (3.20) *transform in the same way, conditions* (3.20) *are invariant under change of basis of normal pairs.*

In the following Remark we write down the compatibility conditions in Definition 3.9 in the basis of normal pairs $\{\mathfrak{t}_n = (n, 0), \mathfrak{t}_\theta = (\theta, 1)\}$. We first state a straightforward consequence of Lemma 2.31 when two null hypersurfaces are present.

**Lemma 3.11.** *Let* $\mathcal{D}$ *and* $\underline{\mathcal{D}}$ *be null hypersurface data with respective boundaries* $\mathcal{S} \hookrightarrow \mathcal{H}$ *and* $\underline{\mathcal{S}} \hookrightarrow \underline{\mathcal{H}}$*. Then there exists a gauge in which the following conditions hold*

$$\begin{aligned}\boldsymbol{\ell}_\parallel|_{\mathcal{S}} &= \underline{\boldsymbol{\ell}}_\parallel|_{\underline{\mathcal{S}}} = 0, \\ \ell^{(2)}|_{\mathcal{S}} &= \underline{\ell}^{(2)}|_{\underline{\mathcal{S}}} = 0.\end{aligned} \tag{3.22}$$

*Moreover, the freedom of this gauge is parametrized by pairs* $(z, V)$ *and* $(\underline{z}, \underline{V})$ *satisfying* $V|_{\mathcal{S}} = \underline{V}|_{\underline{\mathcal{S}}} = 0$*.*

**Remark 3.12.** *Using Lemma 2.17 and Remark 2.38, the compatibility conditions* (3.19) *can be written in the basis of normal pairs* $\{\mathfrak{t}_n = (n,0), \mathfrak{t}_\theta = (\theta,1)\}$ *as*

$$\sigma\Psi^\star\left(\underline{\mathbf{Y}}_\parallel + \frac{1}{2}(\pounds_{\underline{\theta}}\underline{\gamma})_\parallel\right) = \mathbf{K}_\parallel, \tag{3.23}$$

$$\sigma^{-1}\Psi^\star\underline{\mathbf{K}}_\parallel = \mathbf{Y}_\parallel + \frac{1}{2}(\pounds_\theta\gamma)_\parallel. \tag{3.24}$$

*In addition, using again Remark 2.38 and Corollary 2.29,* (3.20) *can be written as*

$$\daleth(\mathfrak{t}_\theta,\mathfrak{t}_n) = (\Psi^\star\underline{\daleth})\left(\Psi(\mathfrak{t}_\theta),\Psi(\mathfrak{t}_n)\right) = (\Psi^\star\underline{\daleth})\left(\sigma^{-1}\underline{\mathfrak{t}}_n,\sigma\underline{\mathfrak{t}}_\theta\right) = \Psi^\star\left(\underline{\boldsymbol{\omega}} - \underline{\mathbf{K}}_\parallel(\underline{\ell}^\sharp,\cdot)\right) + d(\log|\sigma|),$$

*so employing Prop. 2.26 it yields*

$$\boldsymbol{\omega} + \Psi^\star\underline{\boldsymbol{\omega}} = \underline{\mathbf{K}}_\parallel(\underline{\ell}^\sharp,\cdot) - \mathbf{K}_\parallel(\ell^\sharp,\cdot) - d(\log|\sigma|). \tag{3.25}$$

*Restricting ourselves to the gauge of Lemma 3.11, these conditions become*

$$\sigma\Psi^\star\underline{\mathbf{Y}}_\parallel = \mathbf{K}_\parallel, \tag{3.26}$$

$$\sigma^{-1}\Psi^\star\underline{\mathbf{K}}_\parallel = \mathbf{Y}_\parallel, \tag{3.27}$$

$$\boldsymbol{\omega} + \Psi^\star\underline{\boldsymbol{\omega}} = -d(\log|\sigma|). \tag{3.28}$$

Double null data has gauge freedom on each component linked to each other by the behaviour of $\sigma$ as described in Prop. 3.7. Thus, we put forward the following definition.

**Definition 3.13.** *Let* $\{\mathcal{D},\underline{\mathcal{D}},\phi,\sigma\}$ *be DND and* $z \in \mathcal{F}^\star(\mathcal{H})$*,* $\underline{z} \in \mathcal{F}^\star(\underline{\mathcal{H}})$*,* $V \in \mathfrak{X}(\mathcal{H})$ *and* $\underline{V} \in \mathfrak{X}(\underline{\mathcal{H}})$*. The transformed double null data is given by* $\mathcal{G}_{(z,\underline{z},V,\underline{V})}\left(\{\mathcal{D},\underline{\mathcal{D}},\phi,\sigma\}\right) := \{\mathcal{D}',\underline{\mathcal{D}}',\phi,\sigma'\}$*, where* $\mathcal{D}'$ *and* $\underline{\mathcal{D}}'$ *are the transformed CHD in the sense of Definition 2.3 and*

$$\sigma' := z^{-1}\underline{z}^{-1}\sigma. \tag{3.29}$$

This definition guarantees that when $\{\mathcal{D},\underline{\mathcal{D}},\phi,\sigma\}$ is double null data, then $\mathcal{G}\left(\{\mathcal{D},\underline{\mathcal{D}},\phi,\sigma\}\right)$ is double null data too. This is a straightforward consequence of the gauge covariance of the compatibility conditions (see Remark 3.10). In order to connect the abstract definition of double null data with the geometric idea of two null and transverse hypersurfaces, it is necessary to extend the notion of embeddedness to the context of double null data.

**Definition 3.14.** *Let* $\{\mathcal{D},\underline{\mathcal{D}},\phi,\sigma\}$ *be DND and* $(\mathcal{M},g)$ *a spacetime. We say that* $\{\mathcal{D},\underline{\mathcal{D}},\phi,\sigma\}$ *is embedded in* $(\mathcal{M},g)$ *with riggings* $\xi,\underline{\xi}$ *and embeddings* $\Phi,\underline{\Phi}$*, respectively, provided that*

1. $\mathcal{D}$ *(resp.* $\underline{\mathcal{D}}$*) is* $(\Phi,\xi)$*-embedded (resp.* $(\underline{\Phi},\underline{\xi})$*-embedded) in* $(\mathcal{M},g)$ *in the sense of Def. 2.2 and* $\Phi(\mathcal{S}) = \underline{\Phi}(\underline{\mathcal{S}}) =: S$*.*

2. $\sigma(p) = g(\nu,\underline{\nu})|_{\Phi(p)}$ *for all* $p \in \underline{\mathcal{S}}$*, where* $\nu = \Phi_\star n$ *and* $\underline{\nu} = \underline{\Phi}_\star\underline{n}$*.*

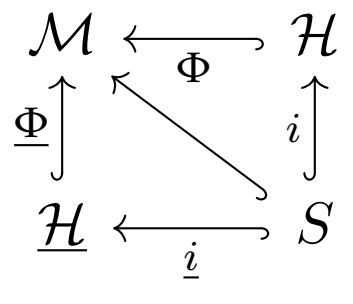

Definitions 2.2, 3.9 and 3.14 establishes our intended correspondence between embedded double null data and two transverse, null hypersurfaces. Moreover, since the quantity $\sigma = \underline{\mathcal{A}}\left(\Psi(n,0),(\underline{n},0)\right)$ is assumed to be negative when $n$ and $\underline{n}$ point into the interior of $\mathcal{H}$ and $\underline{\mathcal{H}}$, respectively, the two hypersurfaces have the same time-orientation.

The next proposition shows that a gauge transformation on a double null data does not affect its embeddedness properties on a spacetime. For general hypersurface data this is a known fact (see Proposition 2.4). Here we show that the extra structure involved in the double null data does not spoil this property.

**Proposition 3.15.** *Let $\{\mathcal{D},\underline{\mathcal{D}},\phi,\sigma\}$ be embedded double null data in a spacetime $(\mathcal{M},g)$ with embeddings $\Phi,\underline{\Phi}$ and riggings $\xi,\underline{\xi}$. For any pair of gauge parameters $(z,V)$ and $(\underline{z},\underline{V})$, the transformed data $\mathcal{G}_{(z,\underline{z},V,\underline{V})}\{\mathcal{D},\underline{\mathcal{D}},\phi,\sigma\}$ is embedded double null data in the same spacetime $(\mathcal{M},g)$, with the same embeddings $\Phi,\underline{\Phi}$ and with riggings*

$$\xi' = z(\xi + \Phi_\star V), \qquad \underline{\xi}' = \underline{z}(\underline{\xi} + \underline{\Phi}_\star \underline{V}).$$

*Proof.* From the definition of embedded double null data (Def. 3.14) we need to check three things. Firstly, that

$$\Phi^\star\left(g(\xi',\cdot)\right) = z\left(\boldsymbol{\ell} + \boldsymbol{\gamma}(V,\cdot)\right), \qquad \underline{\Phi}^\star\left(g(\underline{\xi}',\cdot)\right) = \underline{z}\left(\underline{\boldsymbol{\ell}} + \boldsymbol{\gamma}(\underline{V},\cdot)\right), \tag{3.30}$$

$$g(\xi',\xi') = z^2\left(\ell^{(2)} + 2\boldsymbol{\ell}(V) + \boldsymbol{\gamma}(V,V)\right), \qquad g(\underline{\xi}',\underline{\xi}') = \underline{z}^2\left(\underline{\ell}^{(2)} + 2\underline{\boldsymbol{\ell}}(\underline{V}) + \boldsymbol{\gamma}(\underline{V},\underline{V})\right), \tag{3.31}$$

$$\frac{1}{2}\Phi^\star(\pounds_{\xi'}g) = z\mathbf{Y} + \boldsymbol{\ell}\otimes_s \mathrm{d}z + \frac{1}{2}\pounds_{zV}\boldsymbol{\gamma}, \qquad \frac{1}{2}\underline{\Phi}^\star(\pounds_{\underline{\xi}'}g) = \underline{z}\underline{\mathbf{Y}} + \underline{\boldsymbol{\ell}}\otimes_s \mathrm{d}\underline{z} + \frac{1}{2}\pounds_{\underline{z}\underline{V}}\underline{\boldsymbol{\gamma}}. \tag{3.32}$$

Secondly, that

$$g(\nu',\underline{\nu}') = z^{-1}\underline{z}^{-1}g(\nu,\underline{\nu}), \tag{3.33}$$

and finally, that the compatibility conditions (3.18)-(3.21) still hold in the new gauge. That relations (3.30)-(3.32) hold follows from Proposition 2.4. Concerning (3.33), from $\xi' = z(\xi + \Phi_\star V)$ and $\underline{\xi}' = \underline{z}(\underline{\xi} + \underline{\Phi}_\star\underline{V})$ together with $g(\xi',\nu') = 1$ and $g(\underline{\xi}',\underline{\nu}') = 1$ it turns out that $\nu' = z^{-1}\nu$ and $\underline{\nu}' = \underline{z}^{-1}\underline{\nu}$, so (3.33) also holds. Finally, from equations (3.30)-(3.32) and Remark 3.10 one concludes that the compatibility conditions hold too. □

Let us summarize what we have done so far. In Definition 3.9 we have introduced a completely abstract object called double null data that captures the idea of two null and transverse hypersurfaces but without seeing them as embedded in any ambient spacetime. This object must satisfy certain compatibility conditions (3.18)-(3.21) at the "corner" that are clearly necessary for any DND to be able to be embedded. The rest of this section is devoted to show that these equations are not only necessary but also sufficient, i.e., that if (3.18)-(3.21) hold, then there exists a spacetime in which the double null data can be embedded (Theorem 3.19). Here we are not interested in solving any spacetime field equations yet, we simply want to find that there always exists a spacetime where the data can be embedded, with the aim of showing that we have not ignored any additional restriction on the data that might have been necessary.

The construction of such spacetime is based on the harmonic gauge (see Section 2.5). In Theorem 2.48 we have already proved the existence of a harmonic gauge associated to a set of $\mathfrak{n}$ functionally independent functions on $\mathcal{H}$. We now choose functions $\{x^{\underline{a}}\} = \{\underline{u}, x^A\}$ on $\mathcal{H}$ and $\{\underline{x}^{a}\} = \{u, \underline{x}^A\}$ on $\underline{\mathcal{H}}$ with the following properties:

$$
\left.\begin{array}{ll}
(i) & n(\underline{u}) \neq 0,\ \underline{u}|_{\mathcal{S}} = 0,\ n(x^A) = 0 \text{ and } \{x^A\} \text{ is a local coordinate system on } \mathcal{S} \\
(ii) & \underline{n}(u) \neq 0,\ u|_{\underline{\mathcal{S}}} = 0,\ \underline{n}(\underline{x}^A) = 0 \text{ and } \{\underline{x}^A\} \text{ is a local coordinate system on } \underline{\mathcal{S}} \\
(iii) & \underline{x}^A \circ \phi = x^A
\end{array}\right\} . \tag{3.34}
$$

Lemma 2.50 guarantees the existence of a harmonic gauge associated to the functions $\{\underline{u}, x^A\}$ (resp. $\{u, \underline{x}^A\}$) on $\mathcal{H}$ (resp. $\underline{\mathcal{H}}$) satisfying $\boldsymbol{\ell}_{\parallel} = 0$, $\ell^{(2)} = 0$ on $\mathcal{S}$ and $\underline{\boldsymbol{\ell}}_{\parallel} = 0$, $\underline{\ell}^{(2)} = 0$ on $\underline{\mathcal{S}}$. Moreover, the residual gauge freedom is parametrized by pairs $(z, \underline{z}) \in \mathcal{F}^{\star}(\mathcal{S}) \times \mathcal{F}^{\star}(\mathcal{S})$. One can exploit this freedom to fix the value of the functions $n(\underline{u})$ and $\underline{n}(u)$ at $\mathcal{S}$ and $\underline{\mathcal{S}}$, respectively.

**Lemma 3.16.** *Let $\{\mathcal{D}, \underline{\mathcal{D}}, \phi, \sigma\}$ be DND and consider two set of independent functions $\{\underline{u}, x^A\}$ and $\{u, \underline{x}^A\}$ satisfying conditions* (3.34)*. Then there exists a unique harmonic gauge w.r.t. $\{\underline{u}, x^A\}$ and $\{u, \underline{x}^A\}$ in $\mathcal{D}$ and $\underline{\mathcal{D}}$, respectively, in which $\boldsymbol{\ell}_{\parallel} = 0$, $\ell^{(2)} = 0$, $n(\underline{u}) = \sigma$ on $\mathcal{S}$ and $\underline{\boldsymbol{\ell}}_{\parallel} = 0$, $\underline{\ell}^{(2)} = 0$, $\underline{n}(u) = \sigma$ on $\underline{\mathcal{S}}$.*

*Proof.* The transformation law of the functions $n(\underline{u})$ and $\underline{n}(u)$ follow from that of $n$ (see (2.34)), $n'(\underline{u}) = z^{-1} n(\underline{u})$ and $\underline{n}'(u) = \underline{z}^{-1}\underline{n}(u)$, where again we use primes for the gauge transformed objects. Recalling the transformation of $\sigma$ in (3.13), namely $\sigma' = z^{-1}\underline{z}^{-1}\sigma$, one can choose $\underline{z} = \sigma n(\underline{u})^{-1}$ and $z = \sigma \underline{n}(u)^{-1}$ so that

$$
\sigma' = z^{-1}\underline{z}^{-1}\sigma = \underbrace{z^{-1} n(\underline{u})}_{n'(\underline{u})} = \underbrace{\underline{z}^{-1}\underline{n}(u)}_{\underline{n}'(u)} .
$$

□

Applying Corollary 2.53 it follows that when the DND is written in the unique gauge defined in the previous lemma, additional relations appear between the data at the boundary.

**Proposition 3.17.** *Let $\{\mathcal{D}, \underline{\mathcal{D}}, \phi, \sigma\}$ be DND written in the harmonic gauge of Lemma 3.16 w.r.t. $\{\underline{u}, x^A\}$ and $\{u, \underline{x}^A\}$. Then the following relations hold at $\mathcal{S}$ (we omit the pullback $\phi^{\star}$ for simplicity)*

$$
\sigma \underline{n}(\underline{\ell}^{(2)}) + 2\kappa = 0, \tag{3.35}
$$

$$
\sigma n(\ell^{(2)}) + 2\underline{\kappa}_n = 0, \tag{3.36}
$$

$$
2\boldsymbol{\omega} + d\log|\sigma| = \left(\pounds_{\underline{n}}\underline{\boldsymbol{\ell}}\right)_{\parallel} - \left(\pounds_{n}\boldsymbol{\ell}\right)_{\parallel} = 2\underline{\iota}^{\star}\underline{\mathrm{s}} - 2\iota^{\star}\mathrm{s}, \tag{3.37}
$$

$$
2\underline{\boldsymbol{\omega}} + d\log|\sigma| = \left(\pounds_{n}\boldsymbol{\ell}\right)_{\parallel} - \left(\pounds_{\underline{n}}\underline{\boldsymbol{\ell}}\right)_{\parallel} = 2\iota^{\star}\mathrm{s} - 2\underline{\iota}^{\star}\underline{\mathrm{s}}. \tag{3.38}
$$

*Proof.* Firstly, from equation (2.181) it follows $n(\ell^{(2)}) = \operatorname{tr}_h \mathbf{Y}_{\parallel}$ on $\mathcal{S}$, which together with (3.27) and (2.162) yields (3.36). Equation (3.35) is analogous. Secondly, from (2.182) and $\underline{x}^A \circ \phi = x^A$,

$$2\left(\pounds_n \boldsymbol{\ell} + \mathbf{r} - \mathbf{s}\right)(\operatorname{grad}_h x^A) = \Box_h x^A = \Box_{\underline{h}} \underline{x}^A = 2\left(\pounds_{\underline{n}} \underline{\boldsymbol{\ell}} + \underline{\mathbf{r}} - \underline{\mathbf{s}}\right)(\operatorname{grad}_{\underline{h}} \underline{x}^A).$$

Employing (3.28) and (2.22), equations (3.37) and (3.38) follow at once. □

In the following remark we compute the compatibility condition (3.21) in the gauge defined in Lemma 3.16. This requires writing the pullback of the tensor $\boldsymbol{\mathcal{R}}$ into a section where both $\ell^{(2)}$ and $\boldsymbol{\ell}_{\parallel}$ vanish (cf. (2.154)).

**Remark 3.18.** *Let us write down the condition $\underline{i}^{\star}\underline{\boldsymbol{\mathcal{R}}} = i^{\star}\boldsymbol{\mathcal{R}}$ on $\mathcal{S}$ in the gauge defined in Lemma 3.16 (we omit the pullback $\phi^{\star}$ for simplicity). In* (2.154) *we found that the pullback to $\mathcal{S}$ of the abstract constraint tensor $\boldsymbol{\mathcal{R}}$ in a gauge in which $\boldsymbol{\ell}_{\parallel} = 0$ and $\ell^{(2)} = 0$ on $\mathcal{S}$ is*

$$\begin{aligned}\mathcal{R}_{AB} = {} & R^h_{AB} - 2\pounds_n \mathrm{Y}_{AB} - (2\kappa + \operatorname{tr}_h \mathbf{K}_{\parallel})\mathrm{Y}_{AB} + (n(\ell^{(2)}) - \operatorname{tr}_h \mathbf{Y}_{\parallel})\mathrm{K}_{AB} \\ & + 4h^{CD}\mathrm{K}_{C(A}\mathrm{Y}_{B)D} + 2\nabla^h_{(A}\omega_{B)} + 4\nabla^h_{(A}\mathrm{s}_{B)} - 2\omega_A\omega_B,\end{aligned} \tag{3.39}$$

*and analogously the pullback of $\underline{\boldsymbol{\mathcal{R}}}$ into $\underline{\mathcal{S}}$ in the same gauge is*

$$\begin{aligned}\underline{\mathcal{R}}_{AB} = {} & R^{\underline{h}}_{AB} - 2\pounds_{\underline{n}} \underline{\mathrm{Y}}_{AB} - (2\underline{\kappa}_n - \operatorname{tr}_{\underline{h}} \underline{\mathbf{K}}_{\parallel})\underline{\mathrm{Y}}_{AB} + (\underline{n}(\underline{\ell}^{(2)}) - \operatorname{tr}_{\underline{h}} \underline{\mathbf{Y}}_{\parallel})\underline{\mathrm{K}}_{AB} \\ & + 4\underline{h}^{CD}\underline{\mathrm{K}}_{C(A}\underline{\mathrm{Y}}_{B)D} + 2\nabla^{\underline{h}}_{(A}\underline{\omega}_{B)} + 4\nabla^{\underline{h}}_{(A}\underline{\mathrm{s}}_{B)} - 2\underline{\omega}_A\underline{\omega}_B.\end{aligned} \tag{3.40}$$

*From condition* (3.18)*, namely that $h \stackrel{\mathcal{S}}{=} \underline{h}$, it follows $R^h_{AB} \stackrel{\mathcal{S}}{=} R^{\underline{h}}_{AB}$. From* (3.26)*-*(3.27)*, the terms $\operatorname{tr}_h \mathbf{K}_{\parallel}\ \mathrm{Y}_{AB}$, $\operatorname{tr}_h \mathbf{Y}_{\parallel}\ \mathrm{K}_{AB}$ and $4h^{CD}\mathrm{K}_{C(A}\mathrm{Y}_{B)D}$ from* (3.39) *are equal to the corresponding terms from* (3.40)*. Finally, from condition* (3.28) *we can substitute $\underline{\boldsymbol{\omega}}$ in* (3.40) *in terms of $\boldsymbol{\omega}$ and $d\log|\sigma|$. Then, the compatibility condition $\mathcal{R}_{AB} = \underline{\mathcal{R}}_{AB}$ in this gauge can be written as*

$$\begin{aligned}\pounds_n \mathrm{Y}_{AB} - \pounds_{\underline{n}} \underline{\mathrm{Y}}_{AB} = {} & -\left(\kappa - \frac{1}{2}\sigma\underline{n}(\underline{\ell}^{(2)})\right)\mathrm{Y}_{AB} + \left(\frac{1}{2}n(\ell^{(2)}) + \sigma^{-1}\underline{\kappa}_n\right)\mathrm{K}_{AB} \\ & + 2\nabla^h_{(A}\omega_{B)} + \nabla^h_A\nabla^h_B \log|\sigma| + 2\nabla^h_{(A}\mathrm{s}_{B)} - 2\nabla^h_{(A}\underline{\mathrm{s}}_{B)} \\ & + 2\omega_{(A}\nabla^h_{B)}\log|\sigma| + \nabla^h_A \log|\sigma| \nabla^h_B \log|\sigma|.\end{aligned} \tag{3.41}$$

*We now restrict further the gauge so that we are in the harmonic gauge of Lemma 3.16. Taking into account* (3.35)*-*(3.38) *the first and second lines of the RHS of* (3.41) *vanish. Replacing the term $2\omega_A$ of the third line by $2\underline{\mathrm{s}}_A - \nabla^h_A \log|\sigma| - 2\mathrm{s}_A$ (see* (3.37)*), equation* (3.41) *finally reads*

$$\pounds_n \mathrm{Y}_{AB} - \pounds_{\underline{n}} \underline{\mathrm{Y}}_{AB} = 2(\underline{\mathrm{s}}_{(A} - \mathrm{s}_{(A})\nabla^h_{B)} \log|\sigma|. \tag{3.42}$$

We now have the necessary ingredients to show that (3.18)-(3.21) is everything one needs to make sure that double null data can be embedded in some spacetime, i.e., that the definition is complete and we are not missing any extra conditions.

**Theorem 3.19.** *Let $\{\mathcal{D}, \underline{\mathcal{D}}, \phi, \sigma\}$ be double null data. Then there exists a spacetime $(\mathcal{M}, g)$, embeddings $\Phi : \mathcal{H} \hookrightarrow \mathcal{M}$, $\underline{\Phi} : \underline{\mathcal{H}} \hookrightarrow \mathcal{M}$ and vector fields $\xi, \underline{\xi}$ along $\Phi(\mathcal{H})$ and $\underline{\Phi}(\underline{\mathcal{H}})$, respectively, such that $\{\mathcal{D}, \underline{\mathcal{D}}, \phi, \sigma\}$ is embedded double null data in $(\mathcal{M}, g)$ with embeddings $\Phi$, $\underline{\Phi}$ and riggings $\xi, \underline{\xi}$, respectively.*

*Proof.* Let $\{\mathcal{D}, \underline{\mathcal{D}}, \phi, \sigma\}$ be double null data, $\mathcal{D} = \{\mathcal{H}, \boldsymbol{\gamma}, \boldsymbol{\ell}, \ell^{(2)}, \mathbf{Y}\}$ and $\underline{\mathcal{D}} = \{\underline{\mathcal{H}}, \underline{\boldsymbol{\gamma}}, \underline{\boldsymbol{\ell}}, \underline{\ell}^{(2)}, \underline{\mathbf{Y}}\}$. Consider a set of independent functions $\{\underline{u}, x^A\}$ and $\{u, \underline{x}^A\}$ satisfying the conditions of Lemma 3.16, namely (3.34). Henceforth the coordinates $u$ and $\underline{u}$ will be assumed to take positive values. We write the data in the gauge of Lemma 3.16 with respect to these functions. Consider the manifold $\mathbb{R} \times \mathbb{R} \times \mathcal{S}$ and use $u$ and $\underline{u}$ as the natural coordinates in the first and second factors, respectively. We shall work in the manifold with boundary $\mathcal{M} := (\{u \geq 0,\ \underline{u} \geq 0\} \subset \mathbb{R}^2) \times \mathcal{S}$. Any (local) coordinate system $\{x^A\}$ in $\mathcal{S}$ extends to a (local) coordinate system $\{u, \underline{u}, x^A\}$ in $\mathcal{M}$ such that $\{\underline{u}, x^A\}$ (resp. $\{u, x^A\}$) restricted to $\mathcal{H}$ (resp. $\underline{\mathcal{H}}$) are the given coordinates on $\mathcal{H}$ (resp. $\underline{\mathcal{H}}$), as well as $u|_{\mathcal{H}} = 0$ and $\underline{u}|_{\underline{\mathcal{H}}} = 0$. We define the embeddings

$$
\begin{aligned}
\Phi : \mathcal{H} &\longrightarrow \mathcal{M} & \underline{\Phi} : \underline{\mathcal{H}} &\longrightarrow \mathcal{M} \\
(\underline{u}, x^A) &\longmapsto (u = 0, \underline{u}, x^A) & (u, \underline{x}^A) &\longmapsto (u, \underline{u} = 0, \underline{x}^A).
\end{aligned} \tag{3.43}
$$

Thus, $\Phi(\mathcal{H}) = \{u = 0\}$ and $\underline{\Phi}(\underline{\mathcal{H}}) = \{\underline{u} = 0\}$. We denote by $S$ the intersection of $\Phi(\mathcal{H})$ and $\underline{\Phi}(\underline{\mathcal{H}})$, namely $S := \{u = \underline{u} = 0\} \subset \mathcal{M}$. Let $\xi$ and $\underline{\xi}$ be defined by $\xi := \partial_u$ and $\underline{\xi} := \partial_{\underline{u}}$ in the coordinate system we have introduced. In these coordinates we also have $n = \mathfrak{z}\partial_{\underline{u}}$ and $\underline{n} = \underline{\mathfrak{z}}\partial_u$, where $\mathfrak{z} := n(\underline{u})$ and $\underline{\mathfrak{z}} := \underline{n}(u)$. In order to prove the theorem we only need to construct a smooth metric $g$ on $\mathcal{M}$ inducing the given data on $\mathcal{H} \cup \underline{\mathcal{H}}$ (we do not write $\Phi$ or $\underline{\Phi}$ for simplicity) w.r.t. the riggings $\xi$ and $\underline{\xi}$, i.e.,

$$
\begin{aligned}
&g_{\underline{u}\,\underline{u}}|_{\mathcal{H}} = 0, \quad g_{u\,u}|_{\mathcal{H}} = \ell^{(2)}, \quad g_{\underline{u}\,\underline{u}}|_{\underline{\mathcal{H}}} = \underline{\ell}^{(2)}, \quad g_{u\,u}|_{\underline{\mathcal{H}}} = 0, \\
&g_{\underline{u}\,A}|_{\mathcal{H}} = 0, \quad g_{u\,A}|_{\mathcal{H}} = \ell_A, \quad g_{\underline{u}\,A}|_{\underline{\mathcal{H}}} = \underline{\ell}_A, \quad g_{u\,A}|_{\underline{\mathcal{H}}} = 0, \\
&\qquad g_{AB}|_{\mathcal{H}} = \gamma_{AB}, \quad g_{AB}|_{\underline{\mathcal{H}}} = \underline{\gamma}_{AB}, \\
&\qquad g_{u\,\underline{u}}|_{\mathcal{H}} = \mathfrak{z}^{-1}, \quad g_{u\,\underline{u}}|_{\underline{\mathcal{H}}} = \underline{\mathfrak{z}}^{-1},
\end{aligned} \tag{3.44}
$$

$$
g_{u\,\underline{u}}|_S = \sigma^{-1}, \tag{3.45}
$$

and

$$
\frac{1}{2}\left(\pounds_{\xi} g\right)_{ab} = \mathrm{Y}_{ab}, \qquad \frac{1}{2}(\pounds_{\underline{\xi}} g)_{ab} = \underline{\mathrm{Y}}_{ab}, \tag{3.46}
$$

where $\ell_A := \boldsymbol{\ell}(\partial_{x^A})$, $\gamma_{AB} := \boldsymbol{\gamma}(\partial_{x^A}, \partial_{x^B})$ and similarly on $\underline{\mathcal{H}}$. The conditions of the first line of (3.44) follow from the fact that $\boldsymbol{\gamma}(n, \cdot) = 0$ and $\Phi^\star\left(g(\xi, \xi)\right) = \ell^{(2)}$ (see (2.8)). The second line also follows from $\boldsymbol{\gamma}(n, \cdot) = 0$ as well as from $\Phi^\star\left(g(\xi, \cdot)\right) = \boldsymbol{\ell}$. The third line follows directly from $\Phi^\star g = \boldsymbol{\gamma}$. The fourth line follows from $1 = g(\xi, \nu) = g(\partial_u, \mathfrak{z}\partial_{\underline{u}}) = \mathfrak{z} g_{u\,\underline{u}}$ and its underlined version. These four lines constitute the metric part of the data. Condition (3.45) follows from $\sigma = g(\nu, \underline{\nu})|_S = \mathfrak{z}\underline{\mathfrak{z}} g_{u\,\underline{u}}$ and the fact that $\mathfrak{z} = \underline{\mathfrak{z}} = \sigma$ on $S$. Finally, conditions (3.46) guarantee that the metric $g$ also induces the $\mathbf{Y}$ tensor (see (2.9)).

In order to construct such $g$, our strategy is to extend the components of the hypersurface data tensors in these coordinates to all $\mathcal{M}$ and define the components of the metric $g$ in such a way that it induces the given data on $\mathcal{H} \cup \underline{\mathcal{H}}$. Thus, we introduce the notation $f^{\mathcal{H}}$ to denote the extension of the function $f \in \mathcal{F}(\mathcal{H})$ off $\mathcal{H}$ satisfying $\xi\,(f) = 0$. We define $f^{\underline{\mathcal{H}}}$ analogously, i.e., by extending the function $f \in \mathcal{F}(\underline{\mathcal{H}})$ by means of $\underline{\xi}(f) = 0$. Moreover, $f^S$ will denote the extension of $f \in \mathcal{F}(S)$ off $S$ satisfying $\xi\,(f) = \underline{\xi}\,(f) = 0$. Then we define the components of $g$ in this coordinate system as follows.

- Component $g_{\underline{u}\,\underline{u}}$: Let $g_{\underline{u}\,\underline{u}}$ be defined on $\mathcal{M}$ by $g_{\underline{u}\,\underline{u}} := (\underline{\ell}^{(2)})^{\underline{\mathcal{H}}} + 2(\mathrm{Y}^{\mathcal{H}}_{\underline{u}\,\underline{u}} - \mathrm{Y}^{S}_{\underline{u}\,\underline{u}})u$. Since $\mathrm{Y}^{\mathcal{H}}_{\underline{u}\,\underline{u}} = \mathrm{Y}^{S}_{\underline{u}\,\underline{u}}$ on $\underline{\mathcal{H}}$ we have $g_{\underline{u}\,\underline{u}}|_{\underline{\mathcal{H}}} = \underline{\ell}^{(2)}$. From $\underline{\ell}^{(2)} \stackrel{S}{=} 0$ its extension $(\underline{\ell}^{(2)})^{\underline{\mathcal{H}}}$ vanishes on $\mathcal{H}$ and hence $g_{\underline{u}\,\underline{u}}|_{\mathcal{H}} = 0$. Concerning the transverse derivative, we first note that for any function $f \in \mathcal{F}(\mathcal{H})$ it holds $\partial_{\underline{u}}(f^{\mathcal{H}}) = (\partial_{\underline{u}} f)^{\mathcal{H}}$ and similarly $\partial_u(f^{\underline{\mathcal{H}}}) = (\partial_u f)^{\underline{\mathcal{H}}}$ for any function $f \in \mathcal{F}(\underline{\mathcal{H}})$. Indeed, $\partial_u(\partial_{\underline{u}}(f^{\mathcal{H}})) = \partial_{\underline{u}}(\partial_u(f^{\mathcal{H}})) = 0$ and $\partial_u((\partial_{\underline{u}} f)^{\mathcal{H}}) = 0$ by construction, so the function $\partial_{\underline{u}}(f^{\mathcal{H}}) - (\partial_{\underline{u}} f)^{\mathcal{H}}$ is constant along each integral curve of $\partial_u$, and since on $\mathcal{H}$ $\partial_{\underline{u}}(f^{\mathcal{H}}) = (\partial_{\underline{u}} f)^{\mathcal{H}} = \partial_{\underline{u}} f$, we conclude that $\partial_{\underline{u}}(f^{\mathcal{H}}) = (\partial_{\underline{u}} f)^{\mathcal{H}}$ (and similarly $\partial_u(f^{\underline{\mathcal{H}}}) = (\partial_u f)^{\underline{\mathcal{H}}}$ on $\mathcal{M}$). Now, condition (3.35) together with $n = \mathfrak{z}\partial_{\underline{u}}$, $\underline{n} = \underline{\mathfrak{z}}\partial_u$ and $\mathfrak{z} = \underline{\mathfrak{z}}$ on $S$ gives $\partial_u \underline{\ell}^{(2)} \stackrel{S}{=} 2\mathrm{Y}_{\underline{u}\,\underline{u}}$. Therefore, all along $\mathcal{H}$ their extensions agree, $(\partial_u \underline{\ell}^{(2)})^{\underline{\mathcal{H}}} = \partial_u((\underline{\ell}^{(2)})^{\underline{\mathcal{H}}}) = 2\mathrm{Y}^{S}_{\underline{u}\,\underline{u}}$. Consequently,

$$\frac{1}{2}\left(\pounds_{\xi} g\right)_{\underline{u}\,\underline{u}} = \frac{1}{2}\partial_u g_{\underline{u}\,\underline{u}} = \frac{1}{2}\partial_u\Big((\underline{\ell}^{(2)})^{\underline{\mathcal{H}}}\Big) + \mathrm{Y}^{\mathcal{H}}_{\underline{u}\,\underline{u}} - \mathrm{Y}^{S}_{\underline{u}\,\underline{u}} \stackrel{\mathcal{H}}{=} \mathrm{Y}_{\underline{u}\,\underline{u}}.$$

- Component $g_{u\,u}$: Analogously, we define $g_{u\,u} := (\ell^{(2)})^{\mathcal{H}} + 2(\underline{\mathrm{Y}}^{\underline{\mathcal{H}}}_{u\,u} - \underline{\mathrm{Y}}^{S}_{u\,u})\underline{u}$. By symmetry of the construction this also induces the given data on $\mathcal{H}$ and $\underline{\mathcal{H}}$.

- Component $g_{u\,\underline{u}}$: Let $g_{u\,\underline{u}}$ be defined on $\mathcal{M}$ by $g_{u\,\underline{u}} := \left(\mathfrak{z}^{-1}\right)^{\mathcal{H}} + (\underline{\mathfrak{z}}^{-1})^{\underline{\mathcal{H}}} - \left(\sigma^{-1}\right)^{S}$. Since $\sigma = \mathfrak{z} = \underline{\mathfrak{z}}$ on $\mathcal{S}$, $g_{u\,\underline{u}}|_{\mathcal{H}} = \mathfrak{z}^{-1}$ and $g_{u\,\underline{u}}|_{\underline{\mathcal{H}}} = \underline{\mathfrak{z}}^{-1}$ (and they match on $S$ and fulfill condition (3.45)).

- Components $g_{\underline{u}\,A}$: Let $g_{\underline{u}\,A} := \underline{\ell}^{\underline{\mathcal{H}}}_A + 2(\mathrm{Y}^{\mathcal{H}}_{\underline{u}\,A} - \mathrm{Y}^{S}_{\underline{u}\,A})u$ on $\mathcal{M}$. Since $\mathrm{Y}^{\mathcal{H}}_{\underline{u}\,A} = \mathrm{Y}^{S}_{\underline{u}\,A}$ on $\underline{\mathcal{H}}$ we have $g_{\underline{u}\,A}|_{\underline{\mathcal{H}}} = \underline{\ell}_A$. From $\underline{\ell}_A = 0$ on $S$, its extension $\underline{\ell}^{\underline{\mathcal{H}}}_A$ also vanishes on $\mathcal{H}$, and then $g_{\underline{u}\,A}|_{\mathcal{H}} = 0$. In order to see that these $g_{\underline{u}\,A}$ induce the corresponding components of the tensor $\mathbf{Y}$ we start by writing equation (3.37) in the coordinate system $\{u, \underline{u}, x^A\}$, i.e., taking $n = \mathfrak{z}\partial_{\underline{u}}$ and $\underline{n} = \underline{\mathfrak{z}}\partial_u$. Using $\sigma \stackrel{S}{=} \underline{\mathfrak{z}}$,

$$\begin{aligned}
2\omega_A &\stackrel{S}{=} \underline{\mathfrak{z}}\partial_u \underline{\ell}_A + \underline{\boldsymbol{\ell}}(\partial_u)\partial_{x^A}(\underline{\mathfrak{z}}) - \partial_{x^A}\log|\underline{\mathfrak{z}}| - \mathfrak{z}\partial_{\underline{u}}\ell_A - \boldsymbol{\ell}(\partial_{\underline{u}})\partial_{x^A}(\mathfrak{z}) \\
&\stackrel{S}{=} \underline{\mathfrak{z}}\partial_u \underline{\ell}_A + \underline{\mathfrak{z}}^{-1}\partial_{x^A}(\underline{\mathfrak{z}}) - \partial_{x^A}\log|\underline{\mathfrak{z}}| - \mathfrak{z}\partial_{\underline{u}}\ell_A - \mathfrak{z}^{-1}\partial_{x^A}(\mathfrak{z}) \\
&\stackrel{S}{=} \underline{\mathfrak{z}}\partial_u \underline{\ell}_A - \partial_{x^A}\log|\underline{\mathfrak{z}}| - \mathfrak{z}\partial_{\underline{u}}\ell_A,
\end{aligned} \tag{3.47}$$

where in the first line we used the well-known formula $\pounds_{fX}\boldsymbol{\omega} = f\pounds_X\boldsymbol{\omega} + \boldsymbol{\omega}(X)df$ valid for any function $f$, vector $X$ and one-form $\boldsymbol{\omega}$, in the second line we used $\boldsymbol{\ell}(n) = \mathfrak{z}\boldsymbol{\ell}(\partial_{\underline{u}}) = 1$ and thus $\boldsymbol{\ell}(\partial_{\underline{u}}) = \mathfrak{z}^{-1}$ (and its underlined version) and in the third line $\underline{\mathfrak{z}}\partial_{x^A}(\underline{\mathfrak{z}}) \stackrel{S}{=} \mathfrak{z}\partial_{x^A}(\mathfrak{z})$ since $\mathfrak{z} = \underline{\mathfrak{z}}$ on $S$. The value of $\omega_A$ can be computed from (2.18) and (2.17), namely

$$2\omega_A = 2\mathfrak{z}\Pi_{A\underline{u}} = 2\mathfrak{z}\mathrm{Y}_{A\underline{u}} + 2\mathfrak{z}\mathrm{F}_{A\underline{u}} = 2\mathfrak{z}\mathrm{Y}_{A\underline{u}} + \mathfrak{z}\partial_{x^A}\mathfrak{z}^{-1} - \mathfrak{z}\partial_{\underline{u}}\ell_A, \tag{3.48}$$

where again we used $\ell_{\underline{u}} = \mathfrak{z}^{-1}$. Inserting (3.48) evaluated at $S$ into (3.47) and taking again into account that $\mathfrak{z} = \underline{\mathfrak{z}}$ on $\mathcal{S}$, it yields $2\mathrm{Y}_{\underline{u}A} = \partial_u\underline{\ell}_A$ on $S$ and therefore $2\mathrm{Y}^S_{\underline{u}A} = (\partial_u\underline{\ell}_A)^{\underline{\mathcal{H}}} = \partial_u\Big(\underline{\ell}^{\underline{\mathcal{H}}}_A\Big)$ on $\mathcal{H}$. Hence,

$$\frac{1}{2}\left(\pounds_\xi g\right)_{\underline{u}A} = \frac{1}{2}\partial_u g_{\underline{u}A} = \frac{1}{2}\partial_u\Big(\underline{\ell}^{\underline{\mathcal{H}}}_A\Big) + \mathrm{Y}^{\mathcal{H}}_{\underline{u}A} - \mathrm{Y}^S_{\underline{u}A} \stackrel{\mathcal{H}}{=} \mathrm{Y}_{\underline{u}A}.$$

- Components $g_{uA}$: Analogously, we define $g_{uA} := \ell^{\mathcal{H}}_A + 2(\underline{\mathrm{Y}}^{\underline{\mathcal{H}}}_{uA} - \mathrm{Y}^S_{uA})\underline{u}$ on $\mathcal{M}$, which by symmetry also induces the given data on $\mathcal{H}$ and $\underline{\mathcal{H}}$.

- Components $g_{AB}$: Let $h$ be the induced metric on $\mathcal{S}$. We define the functions $g_{AB}$ on $\mathcal{M}$ by means of

$$g_{AB} := \gamma^{\mathcal{H}}_{AB} + \underline{\gamma}^{\underline{\mathcal{H}}}_{AB} - h^S_{AB} + 2(\mathrm{Y}^{\mathcal{H}}_{AB} - \mathrm{Y}^S_{AB})u + 2(\underline{\mathrm{Y}}^{\underline{\mathcal{H}}}_{AB} - \underline{\mathrm{Y}}^S_{AB})\underline{u} - 2(\partial_u\underline{\mathrm{Y}}_{AB})^S u\underline{u}.$$

Since on $\mathcal{H}$ $\underline{\gamma}^{\underline{\mathcal{H}}}_{AB} = h^S_{AB}$ and $\underline{\mathrm{Y}}^{\underline{\mathcal{H}}}_{AB} = \underline{\mathrm{Y}}^S_{AB}$, and on $\underline{\mathcal{H}}$ $\gamma^{\mathcal{H}}_{AB} = h^S_{AB}$ and $\mathrm{Y}^{\mathcal{H}}_{AB} = \mathrm{Y}^S_{AB}$, we have $g_{AB}|_{\mathcal{H}} = \gamma_{AB}$ and $g_{AB}|_{\underline{\mathcal{H}}} = \underline{\gamma}_{AB}$. Moreover, from (2.16) and $\underline{n} = \underline{\mathfrak{z}}\partial_u$, we have $\partial_u\underline{\gamma}_{AB} = 2\underline{\mathfrak{z}}^{-1}\underline{\mathrm{K}}_{AB}$ on $\underline{\mathcal{H}}$, so in particular $\partial_u\underline{\gamma}_{AB} = 2\underline{\mathfrak{z}}^{-1}\underline{\mathrm{K}}_{AB}$ on $S$. Using $\underline{\mathfrak{z}} = \sigma$ and $\sigma^{-1}\underline{\mathrm{K}}_{AB} = \mathrm{Y}_{AB}$ on $\mathcal{S}$ (see (3.27)) it follows that $(\partial_u\underline{\gamma}_{AB})^{\underline{\mathcal{H}}} = \partial_u\Big(\underline{\gamma}^{\underline{\mathcal{H}}}_{AB}\Big) = 2\mathrm{Y}^S_{AB}$ on $\mathcal{H}$, and thus

$$\begin{aligned}
\frac{1}{2}\left(\pounds_\xi g\right)_{AB} &= \frac{1}{2}\partial_u g_{AB}\\
&= \frac{1}{2}\partial_u(\underline{\gamma}^{\underline{\mathcal{H}}}_{AB}) + \mathrm{Y}^{\mathcal{H}}_{AB} - \mathrm{Y}^S_{AB} + \partial_u(\underline{\mathrm{Y}}^{\underline{\mathcal{H}}}_{AB})\underline{u} - \left(\partial_u\underline{\mathrm{Y}}_{AB}\right)^S \underline{u}\\
&\stackrel{\mathcal{H}}{=} \frac{1}{2}\partial_u(\underline{\gamma}_{AB})^{\underline{\mathcal{H}}} + \mathrm{Y}^{\mathcal{H}}_{AB} - \mathrm{Y}^S_{AB}\\
&\stackrel{\mathcal{H}}{=} \mathrm{Y}_{AB},
\end{aligned}$$

where in the third equality we used $\partial_u\Big(\underline{\mathrm{Y}}^{\underline{\mathcal{H}}}_{AB}\Big) = \left(\partial_u\underline{\mathrm{Y}}_{AB}\right)^S$ on $\mathcal{H}$. Before computing $\pounds_{\underline{\xi}}g$ on $\underline{\mathcal{H}}$ we need to write equation (3.42) in the coordinate system $\{u, \underline{u}, x^A\}$. Since $[n, \partial_{x^A}] = [\mathfrak{z}\partial_{\underline{u}}, \partial_{x^A}] = -\partial_{x^A}\mathfrak{z}\ \partial_{\underline{u}}$ and $\mathfrak{z} = \sigma$ on $S$,

$$\begin{aligned}
\pounds_n\mathrm{Y}_{AB} &\stackrel{S}{=} \pounds_n\left(\mathrm{Y}_{AB}\right) + \partial_{x^A}(\log|\sigma|)\mathbf{Y}\left(n, \partial_{x^B}\right) + \partial_{x^B}(\log|\sigma|)\mathbf{Y}\left(\partial_{x^A}, n\right)\\
&\stackrel{S}{=} \mathfrak{z}\partial_{\underline{u}}\left(\mathrm{Y}_{AB}\right) + 2\mathrm{r}_{(A}\partial_{x^{B})}(\log|\sigma|).
\end{aligned}$$

Then, after using (3.37)-(3.38) equation (3.42) in this coordinate system becomes simply

$$\partial_{\underline{u}}\left(\mathrm{Y}_{AB}\right) \stackrel{S}{=} \partial_u\left(\underline{\mathrm{Y}}_{AB}\right), \tag{3.49}$$

and then the quantity $\pounds_{\underline{\xi}} g$ on $\underline{\mathcal{H}}$ is finally given by

$$\begin{aligned}\frac{1}{2}(\pounds_{\underline{\xi}} g)_{AB} &= \frac{1}{2}\partial_{\underline{u}} g_{AB} \\ &= \frac{1}{2}\partial_{\underline{u}}(\gamma^{\mathcal{H}}_{AB}) + \partial_{\underline{u}}(\mathrm{Y}^{\mathcal{H}}_{AB})u + \underline{\mathrm{Y}}^{\underline{\mathcal{H}}}_{AB} - \underline{\mathrm{Y}}^{S}_{AB} - (\partial_u \underline{\mathrm{Y}}_{AB})^S u \\ &\stackrel{\underline{\mathcal{H}}}{=} \frac{1}{2}\partial_u (\gamma_{AB})^{\mathcal{H}} + \underline{\mathrm{Y}}^{\underline{\mathcal{H}}}_{AB} - \underline{\mathrm{Y}}^{S}_{AB} \\ &\stackrel{\underline{\mathcal{H}}}{=} \underline{\mathrm{Y}}_{AB},\end{aligned}$$

where in the third line we used that, on $\underline{\mathcal{H}}$, $\partial_{\underline{u}}(\mathrm{Y}^{\mathcal{H}}_{AB}) = (\partial_{\underline{u}}\mathrm{Y}_{AB})^S = (\partial_u \underline{\mathrm{Y}}_{AB})^S$ (the second equality following from (3.49)), and in the fourth line that $(\partial_{\underline{u}}\gamma_{AB})^{\mathcal{H}} = \partial_{\underline{u}}(\gamma^{\mathcal{H}}_{AB}) = 2\underline{\mathrm{Y}}^S_{AB}$ on $\underline{\mathcal{H}}$.

Given that $g_{\mu\nu}$ fulfills all conditions (3.44)-(3.46), we conclude that $\{\mathcal{D}, \underline{\mathcal{D}}, \phi, \sigma\}$ is embedded double null data in $(\mathcal{M}, g)$ with embeddings $\Phi$, $\underline{\Phi}$ and riggings $\xi, \underline{\xi}$, respectively. □

## 3.3 EXISTENCE RESULT

In Section 3.2 we constructed the notion of double null data and we proved that every DND can be embedded in some spacetime $(\mathcal{M}, g)$. In this section we are interested in solving the characteristic Cauchy problem, i.e., we study necessary and sufficient conditions on $\{\mathcal{D}, \underline{\mathcal{D}}, \phi, \sigma\}$ so that there exists an ambient spacetime $(\mathcal{M}, g)$ solution to the $\lambda$-vacuum equations (0.1) where $\{\mathcal{D}, \underline{\mathcal{D}}, \phi, \sigma\}$ is embedded. Since we want the ambient spacetime to satisfy $\mathrm{Ric}[g] = \lambda g$, it is clear that conditions $\mathcal{R}_{ab} = \lambda\gamma_{ab}$ and $\underline{\mathcal{R}}_{ab} = \lambda\underline{\gamma}_{ab}$ on $\mathcal{D}$ and $\underline{\mathcal{D}}$ are necessary in order for $\{\mathcal{D}, \underline{\mathcal{D}}, \phi, \sigma\}$ to be embeddable in a $\lambda$-vacuum spacetime. The main result of this section is that these *constraint equations* are not only necessary, but also sufficient. We put forward the following definition for "development" of a double null data.

**Definition 3.20.** *Let $\{\mathcal{D}, \underline{\mathcal{D}}, \phi, \sigma\}$ be double null data. We say that a Lorentzian manifold $(\mathcal{M}, g)$ is a **development** of $\{\mathcal{D}, \underline{\mathcal{D}}, \phi, \sigma\}$ provided there exist embeddings $\Phi$, $\underline{\Phi}$ and riggings $\xi$, $\underline{\xi}$ such that $\{\mathcal{D}, \underline{\mathcal{D}}, \phi, \sigma\}$ is embedded DND in $(\mathcal{M}, g)$ in the sense of Def. 3.14 and $\Phi(\mathcal{H}) \cup \underline{\Phi}(\underline{\mathcal{H}}) = \partial\mathcal{M}$.*

As we already mentioned in the introduction, General Relativity is a geometric theory, so the Einstein equations as a PDE problem cannot have a unique solution. The standard approach to deal with this issue is to solve the reduced Einstein equations, namely (cf. [208, 279])

$$R^h_{\alpha\beta} := R_{\alpha\beta} + g_{\mu(\alpha}\Gamma^{\mu}{}_{,\beta)} = \lambda g_{\alpha\beta}. \tag{3.50}$$

This system admits a well-posed initial value problem essentially because its principal symbol is hyperbolic. In the following lemma we compute the tangent components of (3.50) on any embedded CHD in terms of the constraint tensors and the $\Gamma$-functions (see (2.178)-(2.180)).

**Lemma 3.21.** *Let $\mathcal{D}$ be $(\Phi, \xi)$-embedded CHD in $(\mathcal{M}, g)$ and $\{u, \underline{u}, x^A\}$ a coordinate system on $\mathcal{M}$ satisfying $\xi(x^A) \stackrel{\mathcal{H}}{=} \xi(\underline{u}) \stackrel{\mathcal{H}}{=} 0$, $\xi(u) \stackrel{\mathcal{H}}{=} 1$, $u|_{\mathcal{H}} = 0$, $n(x^A|_{\mathcal{H}}) = 0$ and that $\underline{u}|_{\mathcal{H}}$ is* *a foliation function of $\mathcal{D}$. Suppose also that $(\mathcal{M}, g)$ is a solution of the reduced Einstein equations* (3.50)*. Then,*

$$\boldsymbol{J}(n) + \pounds_n \left(\Gamma^u_{\mathcal{H}}\right) = 0, \tag{3.51}$$

$$J_A + \frac{1}{2}\partial_{x^A}\left(\Gamma^u_{\mathcal{H}}\right) + \frac{1}{2}\ell_A \; \pounds_n \left(\Gamma^u_{\mathcal{H}}\right) + \frac{1}{2} h_{AB} \; \pounds_n \left(\Gamma^B_{\mathcal{H}}\right) = 0, \tag{3.52}$$

$$\mathcal{R}_{AB} + h_{C(B} \; \partial_{x^{A)}} \Gamma^C_{\mathcal{H}} + \ell_{(A} \; \partial_{x^{B)}} \left(\Gamma^u_{\mathcal{H}}\right) = \lambda h_{AB}. \tag{3.53}$$

*Proof.* Contracting (3.50) with $\nu^\alpha \nu^\beta$ and taking into account that $\nu \stackrel{\mathcal{H}}{=} \mathfrak{z}\partial_{\underline{u}}$ is null and satisfies $g(\nu, \partial_{x^A}) \stackrel{\mathcal{H}}{=} 0$ it yields

$$R^h_{\alpha\beta}\nu^\alpha\nu^\beta \stackrel{\mathcal{H}}{=} R_{\alpha\beta}\nu^\alpha\nu^\beta + g(\partial_u, \nu) \; \pounds_\nu \left(\Gamma^u_{\mathcal{H}}\right) \stackrel{\mathcal{H}}{=} \boldsymbol{J}(n) + \pounds_n \left(\Gamma^u_{\mathcal{H}}\right),$$

where we used $R_{\alpha\beta}\nu^\alpha\nu^\beta \stackrel{\mathcal{H}}{=} \boldsymbol{J}(n)$ and that $\xi \stackrel{\mathcal{H}}{=} \partial_u$. Since $(\mathcal{M}, g)$ is a solution of (3.50), equation (3.51) follows. Similarly the "normal-tangent" components of (3.50) are

$$\begin{aligned}
R^h_{\alpha\beta}\nu^\alpha(\partial_{x^A})^\beta &\stackrel{\mathcal{H}}{=} R_{\alpha\beta}\nu^\alpha(\partial_{x^A})^\beta + \frac{1}{2} g(\partial_u, \nu)\partial_{x^A}\left(\Gamma^u_{\mathcal{H}}\right) \\
&\quad + \frac{1}{2} g(\partial_u, \partial_{x^A})\pounds_\nu \left(\Gamma^u_{\mathcal{H}}\right) + \frac{1}{2} g(\partial_{x^A}, \partial_{x^B}) \; \pounds_\nu(\Gamma^B_{\mathcal{H}}) \\
&\stackrel{\mathcal{H}}{=} J_A + \frac{1}{2}\partial_{x^A}\left(\Gamma^u_{\mathcal{H}}\right) + \frac{1}{2}\ell_A \; \pounds_n \left(\Gamma^u_{\mathcal{H}}\right) + \frac{1}{2}\pounds_n(\Gamma^B_{\mathcal{H}}) h_{BA},
\end{aligned}$$

where we used $g(\partial_u, \partial_{x^A}) \stackrel{\mathcal{H}}{=} g(\xi, \partial_{x^A}) \stackrel{\mathcal{H}}{=} \ell_A$ and $g(\partial_{x^A}, \partial_{x^B}) \stackrel{\mathcal{H}}{=} h_{AB}$. Finally the "tangent-tangent" component of (3.50) is

$$\begin{aligned}
R^h_{\alpha\beta}(\partial_{x^A})^\alpha(\partial_{x^B})^\beta = R_{\alpha\beta}(\partial_{x^A})^\alpha(\partial_{x^B})^\beta &+ \frac{1}{2} g(\partial_u, \partial_{x^A})\partial_{x^B}\Gamma^u_{\mathcal{H}} + \frac{1}{2} g(\partial_u, \partial_{x^B})\partial_{x^A}\Gamma^u_{\mathcal{H}} \\
&+ \frac{1}{2} g(\partial_{x^C}, \partial_{x^A})\partial_{x^B}\Gamma^C_{\mathcal{H}} + \frac{1}{2} g(\partial_{x^C}, \partial_{x^A})\partial_{x^B}\Gamma^C_{\mathcal{H}}
\end{aligned}$$

and after using $R^h_{\alpha\beta}(\partial_{x^A})^\alpha(\partial_{x^B})^\beta = \lambda h_{AB}$, equation (3.53) follows. □

Before moving on to the main result of this section we need an intermediate result, namely that conditions (3.35)-(3.38) are automatically fulfilled for embedded DND satisfying certain coordinate conditions on the intersection surface.

**Proposition 3.22.** *Let $\{\mathcal{D}, \underline{\mathcal{D}}, \phi, \sigma\}$ be embedded DND in $(\mathcal{M}, g)$ with riggings $\xi$ and $\underline{\xi}$. Consider a set of coordinates $\{u, \underline{u}, x^A\}$ on $\mathcal{M}$ satisfying*

1. $n(x^A|_{\mathcal{H}}) = 0$ *and* $\underline{n}(x^A|_{\underline{\mathcal{H}}}) = 0$,

2. $\mathfrak{z} := n\left(\underline{u}|_{\mathcal{H}}\right) \neq 0$ *and* $\underline{\mathfrak{z}} := \underline{n}\left(u|_{\underline{\mathcal{H}}}\right) \neq 0$,

3. $\underline{u}|_{\underline{\mathcal{H}}} = 0$ *and* $u|_{\mathcal{H}} = 0$,

4. $\xi = \partial_u$ *on* $\mathcal{H}$ *and* $\underline{\xi} = \partial_{\underline{u}}$ *on* $\underline{\mathcal{H}}$.

*Then, the following relations hold at $\mathcal{S}$,*

$$\underline{\mathfrak{z}}\,\underline{n}(\underline{\ell}^{(2)}) + 2\kappa = 0, \tag{3.54}$$

$$\mathfrak{z}\, n(\ell^{(2)}) + 2\underline{\kappa}_n = 0, \tag{3.55}$$

$$2\boldsymbol{\omega} + d\log|\underline{\mathfrak{z}}| = \left(\pounds_{\underline{n}}\underline{\boldsymbol{\ell}}\right)_{\parallel} - \left(\pounds_{n}\boldsymbol{\ell}\right)_{\parallel} = 2\underline{\iota}^{\star}\underline{\mathbf{s}} - 2\iota^{\star}\mathbf{s}, \tag{3.56}$$

$$2\underline{\boldsymbol{\omega}} + d\log|\mathfrak{z}| = \left(\pounds_{n}\boldsymbol{\ell}\right)_{\parallel} - \left(\pounds_{\underline{n}}\underline{\boldsymbol{\ell}}\right)_{\parallel} = 2\iota^{\star}\mathbf{s} - 2\underline{\iota}^{\star}\underline{\mathbf{s}}. \tag{3.57}$$

*Proof.* From items (1) and (2) we have $\nu \overset{\mathcal{H}}{=} \mathfrak{z}\partial_{\underline{u}}$ and $\underline{\nu} = \underline{\mathfrak{z}}\partial_u$ on $\underline{\mathcal{H}}$. This and (3) imply that we can replace $\nu \leftrightarrow \mathfrak{z}\partial_{\underline{u}}$, $\xi \leftrightarrow \partial_u$ (resp. $\underline{\nu} \leftrightarrow \underline{\mathfrak{z}}\partial_u$, $\underline{\xi} \leftrightarrow \partial_{\underline{u}}$) in any spacetime calculation at $\mathcal{H}$ (resp. $\underline{\mathcal{H}}$) that only involves tangential derivatives. We apply this without further warning. From $g(\xi,\nu) = 1$ and $g(\underline{\xi},\underline{\nu}) = 1$ it follows that $\mathfrak{z} \overset{\mathcal{S}}{=} \underline{\mathfrak{z}}$. Indeed, $1 = g(\xi,\nu) = \lambda g(\partial_u,\partial_{\underline{u}}) \overset{\mathcal{S}}{=} \lambda g\left(\underline{\mathfrak{z}}^{-1}\underline{\nu},\underline{\xi}\right) = \mathfrak{z}\underline{\mathfrak{z}}^{-1}$, so

$$2\mathbf{Y}(n,n) \overset{\mathcal{H}}{=} \left(\pounds_{\xi} g\right)(\nu,\nu) \overset{\mathcal{H}}{=} \left(\pounds_{\partial_u} g\right)\left(\mathfrak{z}\partial_{\underline{u}},\mathfrak{z}\partial_{\underline{u}}\right) \overset{\mathcal{S}}{=} \mathfrak{z}^2\partial_u\left(g(\partial_{\underline{u}},\partial_{\underline{u}})\right) \overset{\mathcal{S}}{=} \underline{\mathfrak{z}}\,\underline{n}(\underline{\ell}^{(2)}),$$

where in the third equality we used that $\mathfrak{z}\partial_{\underline{u}}$ is null on $\mathcal{S}$. Hence (3.54) (and by analogy (3.55)) follows. Next we prove (3.56) and (3.57). Let $X \in \mathfrak{X}(\mathcal{S})$ and consider any extension of it off $\mathcal{S}$. Equation (2.22) gives $2\mathbf{F}(X,n) = -2\mathbf{s}(X) = -\left(\pounds_n\boldsymbol{\ell}\right)(X)$, and from (2.9) one gets $2\mathbf{Y}(X,n) \overset{\mathcal{H}}{=} \left(\pounds_{\xi} g\right)(X,\nu)$ and thus

$$\begin{aligned} 2\mathbf{Y}(X,n) &\overset{\mathcal{S}}{=} \underline{\mathfrak{z}}\left(\partial_u\left(g(X,\partial_{\underline{u}})\right) - g\left(\pounds_{\partial_u}X,\partial_{\underline{u}}\right) - g\left(X,[\partial_u,\partial_{\underline{u}}]\right)\right)\\ &\overset{\mathcal{S}}{=} \pounds_{\underline{n}}\left(g(X,\partial_{\underline{u}})\right) - g\left(\underline{\mathfrak{z}}\pounds_{\partial_u}X,\partial_{\underline{u}}\right)\\ &\overset{\mathcal{S}}{=} \left(\pounds_{\underline{n}}\underline{\boldsymbol{\ell}}\right)(X) - X\left(\log|\underline{\mathfrak{z}}|\right), \end{aligned}$$

where we used $\mathfrak{z} \overset{\mathcal{S}}{=} \underline{\mathfrak{z}}$ and in the last equality we inserted $g(\partial_{\underline{u}},X) = g(\underline{\xi},X) = \underline{\boldsymbol{\ell}}(X)$ and used $\underline{\nu} = \underline{\mathfrak{z}}\partial_u$ so that $\underline{\mathfrak{z}}\pounds_{\partial_u}X = \pounds_{\underline{n}}X + X\left(\log|\underline{\mathfrak{z}}|\right)\underline{n}$ on $\underline{\mathcal{H}}$. Observe that the result is independent of the extension of $X$. Hence (3.56) (and similarly (3.57)) follows. □

Now we are ready to state the main theorem of this section.

**Theorem 3.23.** *Let $\{\mathcal{D},\underline{\mathcal{D}},\phi,\sigma\}$ be double null data as in Def. 3.9 satisfying the abstract constraint equations*

$$\boldsymbol{\mathcal{R}} = \lambda\boldsymbol{\gamma}, \qquad \underline{\boldsymbol{\mathcal{R}}} = \lambda\underline{\boldsymbol{\gamma}}, \tag{3.58}$$

*where $\boldsymbol{\mathcal{R}}$ is defined in* (2.92)*, $\underline{\boldsymbol{\mathcal{R}}}$ is its underlined version and $\lambda \in \mathbb{R}$. Then there exists a development $(\mathcal{M},g)$ of $\{\mathcal{D},\underline{\mathcal{D}},\phi,\sigma\}$ (possibly restricted if necessary) solution of the $\lambda$-vacuum Einstein equations. Moreover for any two such spacetimes $(\mathcal{M},g)$ and $(\widehat{\mathcal{M}},\widehat{g})$ there exist neighbourhoods of $\mathcal{H}\cup\underline{\mathcal{H}}$, $\mathcal{U}\subseteq\mathcal{M}$ and $\widehat{\mathcal{U}}\subseteq\widehat{\mathcal{M}}$, and a diffeomorphism $\varphi:\mathcal{U}\longrightarrow\widehat{\mathcal{U}}$, such that $\varphi^{\star}\widehat{g} = g$.*

*Proof.* We will use $\mathcal{R}_{ab} = \lambda\gamma_{ab}$ (and its underlined version) in the forms $\boldsymbol{J}(n) = 0$, $J_A = 0$ and $\mathcal{R}_{AB} = \lambda h_{AB}$ (see (2.100) and (2.136)). The first step of the proof is to solve the reduced Einstein equations. Let $\{\underline{u},x^A\}$ and $\{u,x^A\}$ be coordinates on $\mathcal{H}$ and $\underline{\mathcal{H}}$ satisfying $\underline{u}\geq 0$, $\mathfrak{z} := n(\underline{u}) \neq 0$, $n(x^A) = 0$ on $\mathcal{H}$ and $u \geq 0$, $\underline{\mathfrak{z}} := \underline{n}(u) \neq 0$, $\underline{n}(x^A) = 0$ on $\underline{\mathcal{H}}$, as well as $\mathcal{S} = \{u = \underline{u} = 0\}$. Consider a manifold $\mathcal{N}$ with coordinates $\{u, \underline{u}, x^A\}$ defined as $\mathcal{N} = (\{u \geq 0,\ \underline{u} \geq 0\} \subset \mathbb{R}^2) \times \mathcal{S}$ and two embeddings $\Phi : \mathcal{H} \hookrightarrow \mathcal{N}$ and $\underline{\Phi} : \underline{\mathcal{H}} \hookrightarrow \mathcal{N}$ such that $\Phi(\mathcal{H}) = \{u = 0\}$, $\underline{\Phi}(\underline{\mathcal{H}}) = \{\underline{u} = 0\}$, $\Phi(\mathcal{H} \cup \underline{\mathcal{H}}) = \mathcal{S}$ and $\{\underline{u}|_{\mathcal{H}}, x^A|_{\mathcal{H}}\}$ and $\{u|_{\mathcal{H}}, x^A|_{\mathcal{H}}\}$ are the given coordinates on $\mathcal{H}$ and $\underline{\mathcal{H}}$, respectively. Throughout the proof we identify $\mathcal{H}$ and $\underline{\mathcal{H}}$ with their images under $\Phi$ and $\underline{\Phi}$, respectively. We want to construct a metric $g$ solution of the reduced Einstein equations on some neighbourhood $\mathcal{M} \subseteq \mathcal{N}$ of $\mathcal{S}$. From Rendall's Theorem (Thm. 3.2) we need to provide initial data for the metric $g_{\mu\nu}$ on $\mathcal{H} \cup \underline{\mathcal{H}}$ continuous at $\mathcal{S}$ and with smooth restrictions on $\mathcal{H}$ and $\underline{\mathcal{H}}$. In order to do so we write $\{\mathcal{D}, \underline{\mathcal{D}}, \phi, \sigma\}$ in the gauge of Lemma 3.16 w.r.t. the functions $\{\underline{u}, x^A\}$ and $\{u, x^A\}$ on $\mathcal{H}$ and $\underline{\mathcal{H}}$, respectively, and we provide the following initial data on $\mathcal{H}$

$$g_{uu} = \ell^{(2)}, \qquad g_{u\underline{u}} = \mathfrak{z}^{-1}, \qquad g_{uA} = \ell_A, \qquad g_{\underline{u}\underline{u}} = 0, \qquad g_{\underline{u}A} = 0, \qquad g_{AB} = \gamma_{AB},$$

and on $\underline{\mathcal{H}}$,

$$g_{uu} = 0, \qquad g_{u\underline{u}} = \underline{\mathfrak{z}}^{-1}, \qquad g_{uA} = 0, \qquad g_{\underline{u}\underline{u}} = \underline{\ell}^{(2)}, \qquad g_{\underline{u}A} = \underline{\ell}_A, \qquad g_{AB} = \underline{\gamma}_{AB},$$

in the coordinates $\{u, \underline{u}, x^A\}$. Since $\{\mathcal{D}, \underline{\mathcal{D}}, \phi, \sigma\}$ is written in a gauge in which (3.22) and $\mathfrak{z} \stackrel{\mathcal{S}}{=} \underline{\mathfrak{z}}$ hold, the functions $g_{\mu\nu}$ are continuous on $\mathcal{H} \cup \underline{\mathcal{H}}$ and their restrictions to $\mathcal{H}$ and $\underline{\mathcal{H}}$ are smooth. Then from Theorem 3.2 there exists an open neighbourhood $U$ of $\mathcal{S}$ on $\mathcal{N}$ and a unique metric $g$ on $\mathcal{M} := U \cap \mathcal{N}$ solution of the reduced Einstein equations (3.50) such that the components of $g$ in the coordinates $\{u, \underline{u}, x^A\}$ on $U \cap (\mathcal{H} \cup \underline{\mathcal{H}})$ coincide with the given ones. By construction, $\Phi^\star(g(\partial_u, \cdot)) = \boldsymbol{\ell}$, $\Phi^\star(g(\partial_u, \partial_u)) = \ell^{(2)}$, $\underline{\Phi}^\star(g(\partial_{\underline{u}}, \cdot)) = \underline{\boldsymbol{\ell}}$ and $\underline{\Phi}^\star(g(\partial_{\underline{u}}, \partial_{\underline{u}})) = \underline{\ell}^{(2)}$, so the only riggings that have a chance to make the data embedded in the sense of Definition 2.2 are $\xi = \partial_u$ and $\underline{\xi} = \partial_{\underline{u}}$, respectively. Let $\widetilde{\mathbf{Y}}$ and $\widetilde{\underline{\mathbf{Y}}}$ be defined as $\widetilde{\mathbf{Y}} := \frac{1}{2}\Phi^\star(\mathcal{L}_\xi g)$ and $\widetilde{\underline{\mathbf{Y}}} := \frac{1}{2}\underline{\Phi}^\star(\mathcal{L}_{\underline{\xi}} g)$. Then $\widetilde{\mathcal{D}} = \{\mathcal{H}, \widetilde{\gamma} := \gamma, \widetilde{\boldsymbol{\ell}} := \boldsymbol{\ell}, \widetilde{\ell}^{(2)} := \ell^{(2)}, \widetilde{\mathbf{Y}}\}$ and $\widetilde{\underline{\mathcal{D}}} = \{\underline{\mathcal{H}}, \widetilde{\underline{\gamma}} := \underline{\gamma}, \widetilde{\underline{\boldsymbol{\ell}}} := \underline{\boldsymbol{\ell}}, \widetilde{\underline{\ell}}^{(2)} := \underline{\ell}^{(2)}, \widetilde{\underline{\mathbf{Y}}}\}$ are $(\Phi, \xi)$-embedded (resp. $(\underline{\Phi}, \underline{\xi})$-embedded) CHD in $(\mathcal{M}, g)$ as in Def. 2.2 (in what follows we denote with a tilde the expressions depending on $\widetilde{\mathcal{D}}$ and $\widetilde{\underline{\mathcal{D}}}$). By construction the metric part of $\{\widetilde{\mathcal{D}}, \widetilde{\underline{\mathcal{D}}}\}$ coincides with the one of $\{\mathcal{D}, \underline{\mathcal{D}}\}$ and $g(\widetilde{\nu}, \widetilde{\underline{\nu}}) = \sigma$, so $\{\widetilde{\mathcal{D}}, \widetilde{\underline{\mathcal{D}}}, \phi, \sigma\}$ is embedded DND (recall that the compatibility conditions must necessary hold because $\widetilde{\mathcal{D}}$ and $\widetilde{\underline{\mathcal{D}}}$ are already embedded with a common boundary). To prove that $\{\mathcal{D}, \underline{\mathcal{D}}, \phi, \sigma\}$ is also embedded DND we need to show that the original tensors $\mathbf{Y}$ and $\underline{\mathbf{Y}}$ coincide with the embedded ones $\widetilde{\mathbf{Y}}$ and $\widetilde{\underline{\mathbf{Y}}}$, respectively.

For the existence part of the theorem we need to prove two things: (1) that the solution of the reduced EFE is indeed a solution of the EFE and (2) that the tensors $\mathbf{Y}$ and $\underline{\mathbf{Y}}$ coincide with $\widetilde{\mathbf{Y}}$ and $\widetilde{\underline{\mathbf{Y}}}$, respectively. In order to prove (1) we will show that the coordinates are harmonic w.r.t. the metric $g$. To prove (2) we will write homogeneous ODE for the tensors $\mathbf{Y} - \widetilde{\mathbf{Y}}$ and $\underline{\mathbf{Y}} - \widetilde{\underline{\mathbf{Y}}}$ and show that they vanish on $\mathcal{S}$, and therefore everywhere. Both goals are achieved simultaneously.

From Propositions 3.22 and 3.17 applied to the embedded data $\{\widetilde{\mathcal{D}}, \widetilde{\underline{\mathcal{D}}}, \phi, \sigma\}$ and to the original one $\{\mathcal{D}, \underline{\mathcal{D}}, \phi, \sigma\}$, respectively, it follows that $\widetilde{\kappa} \stackrel{\mathcal{S}}{=} \kappa$, $\widetilde{\underline{\kappa}} \stackrel{\mathcal{S}}{=} \underline{\kappa}$, $\widetilde{\boldsymbol{\omega}} \stackrel{\mathcal{S}}{=} \boldsymbol{\omega}$ and $\widetilde{\underline{\boldsymbol{\omega}}} \stackrel{\mathcal{S}}{=} \underline{\boldsymbol{\omega}}$.

Moreover, since $\{\widetilde{\mathcal{D}}, \underline{\widetilde{\mathcal{D}}}, \phi, \sigma\}$ is DND it must satisfy the tilde version of conditions (3.18)-(3.21). Since $\{\mathcal{D}, \underline{\mathcal{D}}, \phi, \sigma\}$ also satisfy (3.18)-(3.21), and both sets of data share the same metric data, it follows that $\widetilde{\mathbf{Y}}_{\|} = \mathbf{Y}_{\|}$ and $\underline{\widetilde{\mathbf{Y}}}_{\|} = \underline{\mathbf{Y}}_{\|}$ at $\mathcal{S}$. In order to prove that these relations hold everywhere and not only on $\mathcal{S}$ the strategy is to prove that the quantities $\widetilde{\kappa} - \kappa$, $\widetilde{\boldsymbol{\omega}} - \boldsymbol{\omega}$ and $\widetilde{\mathbf{Y}}_{\|} - \mathbf{Y}_{\|}$ satisfy homogeneous transport equations along $n$. Let us start by considering the tilde version of equation (3.51) on $\mathcal{H}$, namely

$$\widetilde{\boldsymbol{J}}(n) + \pounds_n(\widetilde{\Gamma}^u_{\mathcal{H}}) = 0. \tag{3.59}$$

From the abstract constraint $\boldsymbol{J}(n) = 0$ (see (2.151)) it follows $\kappa \operatorname{tr}_h \mathbf{K}_{\|} = \pounds_n(\mathbf{K}_{\|}) + |\mathbf{K}_{\|}|^2$. Since the metric data from $\mathcal{D}$ and $\widetilde{\mathcal{D}}$ coincides,

$$\widetilde{\boldsymbol{J}}(n) = \pounds_n(\operatorname{tr}_h \mathbf{K}_{\|}) - \widetilde{\kappa} \operatorname{tr}_h \mathbf{K}_{\|} + |\mathbf{K}_{\|}|^2 = 2(\kappa - \widetilde{\kappa}) \operatorname{tr}_h \mathbf{K}_{\|}.$$

Recall that in the harmonic gauge $\operatorname{tr}_h \mathbf{K}_{\|} + 2\kappa = 0$ (see (2.162)) and therefore $\widetilde{\Gamma}^u_{\mathcal{H}} = \operatorname{tr}_h \mathbf{K}_{\|} + 2\widetilde{\kappa} = 2(\widetilde{\kappa} - \kappa)$. Hence equation (3.59) can be rewritten as

$$-(\kappa - \widetilde{\kappa}) \operatorname{tr}_h \mathbf{K}_{\|} + 2\pounds_n (\kappa - \widetilde{\kappa}) = 0,$$

which is a homogeneous ODE for $\kappa - \widetilde{\kappa}$. This together with $\kappa - \widetilde{\kappa} = 0$ at $\mathcal{S}$ implies $\kappa = \widetilde{\kappa}$ everywhere on $\mathcal{H}$, and then also $\widetilde{\Gamma}^u_{\mathcal{H}} = 0$. The corresponding argument applied to $\underline{\mathcal{H}}$ gives $\widetilde{\Gamma}^u_{\underline{\mathcal{H}}} = 0$ and $\underline{\kappa} = \underline{\widetilde{\kappa}}$ on $\underline{\mathcal{H}}$. Taking into account $\widetilde{\Gamma}^u_{\mathcal{H}} = 0$ the tilde version of equation (3.52) reads

$$\widetilde{J}_A + \frac{1}{2} \pounds_n(\widetilde{\Gamma}^B_{\mathcal{H}})\, h_{AB} = 0. \tag{3.60}$$

Since $J_A = 0$ and the functions $\Gamma^A_{\mathcal{H}}$ (2.179) vanish in the harmonic gauge in which $\mathcal{D}$ is written, the combination $\mathcal{L}_A(J_B, \Gamma^B_{\mathcal{H}})$ defined in (2.183) also vanishes. Recall that this particular combination depends neither on $\boldsymbol{\omega}$ nor in $\mathbf{Y}_{\|}$, so its tilde version also vanishes, $\mathcal{L}_A(\widetilde{J}_B, \widetilde{\Gamma}^B_{\mathcal{H}}) = 0$ (remember that we have already proved $\kappa = \widetilde{\kappa}$). This gives $\widetilde{J}_A$ in terms of $\widetilde{\Gamma}^A_{\mathcal{H}}$, which inserted into (3.60) yields a homogeneous ODE for the functions $\widetilde{\Gamma}^A_{\mathcal{H}}$. Since $\widetilde{\boldsymbol{\omega}} \stackrel{\mathcal{S}}{=} \boldsymbol{\omega}$ and $\Gamma^A_{\mathcal{H}} = 0$, expression (2.185) gives $\widetilde{\Gamma}^A_{\mathcal{H}} \stackrel{\mathcal{S}}{=} 0$ and hence $\widetilde{\Gamma}^A_{\mathcal{H}} = 0$ and $\boldsymbol{\omega} = \widetilde{\boldsymbol{\omega}}$ on $\mathcal{H}$. The same argument on $\underline{\mathcal{H}}$ proves $\widetilde{\Gamma}^A_{\underline{\mathcal{H}}} = 0$ and $\underline{\boldsymbol{\omega}} = \underline{\widetilde{\boldsymbol{\omega}}}$ on $\underline{\mathcal{H}}$. Finally consider the trace of the tilde version of (3.53) w.r.t. $h$, which taking into account $\widetilde{\Gamma}^u_{\mathcal{H}} = \widetilde{\Gamma}^A_{\mathcal{H}} = 0$ and (2.100) reads

$$\widetilde{H} = -\frac{\mathfrak{n} - 1}{2} \lambda, \tag{3.61}$$

where $\mathfrak{n}$ is the dimension of the DND. The abstract constraint equation $H = -\frac{\mathfrak{n}-1}{2}\lambda$ together with $\Gamma^u_{\mathcal{H}} = 0$ (because $\mathcal{D}$ is written in the harmonic gauge) imply that the combination $\mathcal{L}(\mathcal{H}, \Gamma^u_{\mathcal{H}})$ in (2.184) is equal to $-\frac{\mathfrak{n}-1}{2}\lambda$, and also is $\mathcal{L}(\widetilde{H}, \widetilde{\Gamma}^u_{\mathcal{H}})$ since this expression was constructed precisely so that it does not depend on $\mathbf{Y}_{\|}$ (recall that $\kappa = \widetilde{\kappa}$ and $\boldsymbol{\omega} = \widetilde{\boldsymbol{\omega}}$ already). Inserting (3.61) into the tilde version of (2.184) we get a homogeneous ODE for $\widetilde{\Gamma}^u_{\mathcal{H}}$. Since $\operatorname{tr}_h \widetilde{\mathbf{Y}}_{\|} \stackrel{\mathcal{S}}{=} \operatorname{tr}_h \mathbf{Y}_{\|}$ and $\Gamma^u_{\mathcal{H}} = 0$, expression (2.180) gives $\widetilde{\Gamma}^u_{\mathcal{H}} \stackrel{\mathcal{S}}{=} 0$. Hence $\widetilde{\Gamma}^u_{\mathcal{H}} = 0$ and $\operatorname{tr}_h \widetilde{\mathbf{Y}}_{\|} = \operatorname{tr}_h \mathbf{Y}_{\|}$ everywhere on $\mathcal{H}$. The corresponding argument on $\underline{\mathcal{H}}$ gives $\operatorname{tr}_h \underline{\widetilde{\mathbf{Y}}}_{\|} = \operatorname{tr}_h \underline{\mathbf{Y}}_{\|}$ and $\widetilde{\Gamma}^u_{\underline{\mathcal{H}}} = 0$ on $\underline{\mathcal{H}}$. It still remains to show that the trace-free part of $\widetilde{\mathbf{Y}}_{\|}$ and $\underline{\widetilde{\mathbf{Y}}}_{\|}$ agree with

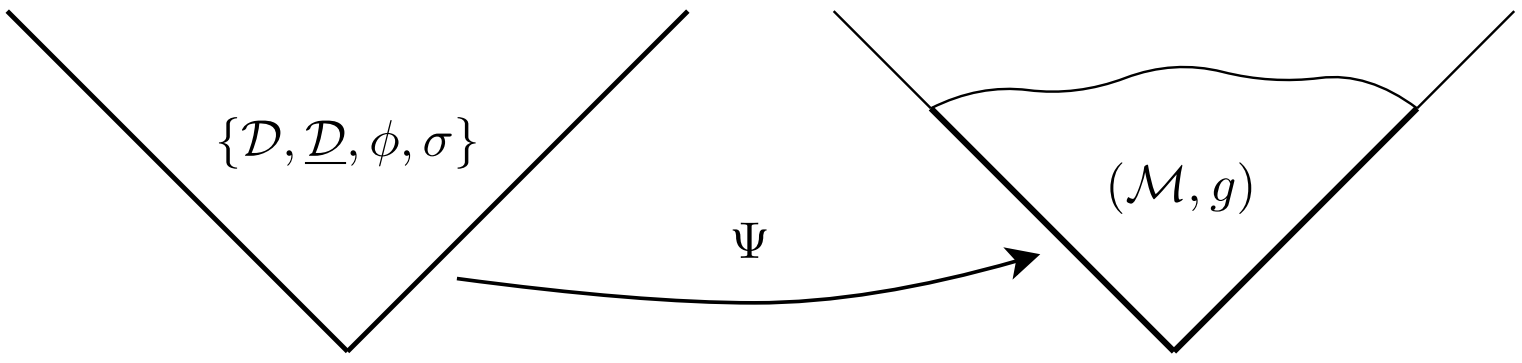

Figure 3.2: Embedded double null data $\{\mathcal{D}, \underline{\mathcal{D}}, \phi, \sigma\}$ with embedding $\Psi$ (in the sense of Def. 3.14) in a spacetime $(\mathcal{M}, g)$ solution of the $\Lambda$-vacuum Einstein field equations.

$\mathbf{Y}_{\parallel}$ and $\underline{\mathbf{Y}}_{\parallel}$, respectively.

The rest of the argument is standard (see Section 3.1). As a consequence of the Bianchi identity the functions $\Box_g u$, $\Box_g \underline{u}$ and $\Box_g x^A$ satisfy a homogeneous wave equation, which together with the fact that $\widetilde{\Gamma}^{\mu}_{\mathcal{H}} \stackrel{\mathcal{H}}{=} 0$ and $\widetilde{\Gamma}^{\mu}_{\underline{\mathcal{H}}} = 0$ on $\underline{\mathcal{H}}$ yields $\Box_g u = \Box_g \underline{u} = \Box_g x^A = 0$ everywhere on $\mathcal{M}$. Consequently $(\mathcal{M}, g)$ is indeed a solution of the $\lambda$-vacuum Einstein equations. As we already recalled, in order to show that $\{\mathcal{D}, \underline{\mathcal{D}}, \phi, \sigma\}$ is embedded DND in $(\mathcal{M}, g)$, as represented in Fig. 3.2, we still need to prove that the trace-free part of the tensors $\mathbf{Y}_{\parallel}$ and $\underline{\mathbf{Y}}_{\parallel}$ coincide with that of $\widetilde{\mathbf{Y}}_{\parallel}$ and $\widetilde{\underline{\mathbf{Y}}}_{\parallel}$, respectively. Since $\operatorname{Ric}(g) = \lambda g$ and the rest of the data coincides (including $\kappa$ and $\boldsymbol{\omega}$), the tensor $\widetilde{\mathbf{Y}}_{\parallel}$ satisfies the same equation as the original $\mathbf{Y}_{\parallel}$, namely $\mathcal{R}_{AB} = \lambda h_{AB}$ (see (2.154)), and thus the tensor $\mathbf{Y}_{\parallel} - \widetilde{\mathbf{Y}}_{\parallel}$ satisfies a homogeneous first order ODE, which together with $\mathbf{Y}_{\parallel} \stackrel{\mathcal{S}}{=} \widetilde{\mathbf{Y}}_{\parallel}$ yields $\mathbf{Y}_{\parallel} = \widetilde{\mathbf{Y}}_{\parallel}$ on $\mathcal{H}$. The same argument on $\underline{\mathcal{H}}$ proves $\underline{\mathbf{Y}}_{\parallel} = \widetilde{\underline{\mathbf{Y}}}_{\parallel}$ on $\underline{\mathcal{H}}$. Then the original tensors $\mathbf{Y}$, $\underline{\mathbf{Y}}$ coincide with the embedded ones $\widetilde{\mathbf{Y}}$, $\widetilde{\underline{\mathbf{Y}}}$, and consequently the original abstract constraint equations are the pullback of the Einstein $\lambda$-vacuum equations to $\mathcal{H} \cup \underline{\mathcal{H}}$.

In order to prove the uniqueness part of the theorem consider $\{\mathcal{D}, \underline{\mathcal{D}}, \phi, \sigma\}$ as embedded DND in another spacetime $(\widehat{\mathcal{M}}, \widehat{g})$ solution of the $\lambda$-vacuum equations with embeddings $\Phi, \underline{\Phi}$ and riggings $\xi, \underline{\xi}$. The aim is to show that there exist neighbourhoods of $\mathcal{H} \cup \underline{\mathcal{H}}$, $\mathcal{U} \subseteq \mathcal{M}$ and $\widehat{\mathcal{U}} \subseteq \widehat{\mathcal{M}}$, and a diffeomorphism $\varphi : \mathcal{U} \longrightarrow \widehat{\mathcal{U}}$, such that $\varphi^{\star}\widehat{g} = g$. By Theorem 3.2, for each set of independent functions $\{\underline{u}, x^A\}$ on $\mathcal{H}$ and $\{u, x^A\}$ on $\underline{\mathcal{H}}$ satisfying $\mathfrak{z} := n(\underline{u}) \neq 0$, $n(x^A) = 0$ on $\mathcal{H}$ and $\underline{\mathfrak{z}} := \underline{n}(u) \neq 0$, $\underline{n}(x^A) = 0$ on $\underline{\mathcal{H}}$, there exist an open neighbourhood $U$ of $\mathcal{S}$ and unique smooth functions $\{\widehat{u}, \widehat{\underline{u}}, \widehat{x}^A\}$ on $\widehat{\mathcal{U}} := U \cap \widehat{\mathcal{M}}$ such that $\Box_{\widehat{g}} \widehat{u} = \Box_{\widehat{g}} \widehat{\underline{u}} = \Box_{\widehat{g}} \widehat{x}^A = 0$ on $\widehat{\mathcal{U}}$ with the given functions on $\widehat{\mathcal{U}} \cap (\mathcal{H} \cup \underline{\mathcal{H}})$ as initial conditions and $\Phi(\mathcal{H}) = \{\widehat{u} = 0\}$, $\underline{\Phi}(\underline{\mathcal{H}}) = \{\widehat{\underline{u}} = 0\}$.

First we prove that when the data is written in the gauge of Lemma 3.16, the riggings are $\xi = \partial_{\widehat{u}}$ and $\underline{\xi} = \partial_{\widehat{\underline{u}}}$ on $\Phi(\mathcal{H})$ and $\underline{\Phi}(\underline{\mathcal{H}})$, respectively. Proposition 2.47 together with the fact that $\widehat{u}$ is harmonic w.r.t. $\widehat{g}$ imply $n\left(\xi(\widehat{u})\right) = 0$ on $\mathcal{H}$. Since the data is written in a gauge in which $\boldsymbol{\ell}_{\parallel} \stackrel{\mathcal{S}}{=} 0$ and $\ell^{(2)} \stackrel{\mathcal{S}}{=} 0$, relations $\underline{\nu} \stackrel{\mathcal{S}}{=} \sigma\xi$ and $\nu \stackrel{\mathcal{S}}{=} \sigma\underline{\xi}$ hold, so $\xi(\widehat{u}) \stackrel{\mathcal{S}}{=} \sigma^{-1}\underline{n}(\widehat{u}) = \sigma^{-1}\underline{\mathfrak{z}} = 1$, since $\sigma = \mathfrak{z} = \underline{\mathfrak{z}}$ on $\mathcal{S}$ in the gauge of Lemma 3.16. Consequently $\xi(\widehat{u}) = 1$ on $\mathcal{H}$, and by analogy, $\underline{\xi}(\widehat{\underline{u}}) = 1$ on $\underline{\mathcal{H}}$. Concerning the functions $\{\widehat{\underline{u}}, \widehat{x}^A\}$, Proposition 2.47, equations $\Box_{\widehat{g}} \widehat{\underline{u}} = \Box_{\widehat{g}} \widehat{x}^A = 0$ and the fact that the gauge is harmonic imply $n\left(\xi(\widehat{\underline{u}})\right) = 0$ and $n(\xi(\widehat{x}^A)) = 0$ on $\mathcal{H}$. By the same argument as before, $\xi(\widehat{\underline{u}}) \stackrel{\mathcal{S}}{=} \sigma^{-1}\underline{n}(\widehat{\underline{u}}) = 0$ and $\xi(\widehat{x}^A) \stackrel{\mathcal{S}}{=} \sigma^{-1}\underline{n}(\widehat{x}^A) = 0$, from where one concludes that $\xi(\widehat{\underline{u}}) = \xi(\widehat{x}^A) = 0$ everywhere on $\mathcal{H}$ (and similarly $\underline{\xi}(\widehat{u}) = \underline{\xi}(\widehat{x}^A) = 0$ on $\underline{\mathcal{H}}$). This proves $\xi = \partial_{\widehat{u}}$ and $\underline{\xi} = \partial_{\widehat{\underline{u}}}$ on $\Phi(\mathcal{H})$ and $\underline{\Phi}(\underline{\mathcal{H}})$, respectively. Then, from the definition of embedded data,

$$g_{\widehat{u}\widehat{u}} = \ell^{(2)}, \qquad g_{\widehat{u}\widehat{\underline{u}}} = \mathfrak{z}^{-1}, \qquad g_{\widehat{u}\widehat{A}} = \ell_A, \qquad g_{\widehat{\underline{u}}\widehat{\underline{u}}} = 0, \qquad g_{\widehat{\underline{u}}\widehat{A}} = 0, \qquad g_{\widehat{A}\widehat{B}} = \gamma_{AB}$$

on $\mathcal{H}$, and

$$g_{\widehat{u}\widehat{u}} = 0, \qquad g_{\widehat{u}\widehat{\underline{u}}} = \underline{\mathfrak{z}}^{-1}, \qquad g_{\widehat{u}\widehat{A}} = 0, \qquad g_{\widehat{\underline{u}}\widehat{\underline{u}}} = \underline{\ell}^{(2)}, \qquad g_{\widehat{\underline{u}}\widehat{A}} = \underline{\ell}_A, \qquad g_{\widehat{A}\widehat{B}} = \underline{\gamma}_{AB}$$

on $\underline{\mathcal{H}}$. After restricting $\widehat{\mathcal{U}}$ further, if necessary, there exists a neighbourhood $\mathcal{U} \subseteq \mathcal{M}$ and a diffeomorphism $\varphi : \mathcal{U} \longrightarrow \widehat{\mathcal{U}}$ defined by $x^\mu = \widehat{x}^\mu \circ \varphi$. By construction $\mathcal{U} \cap (\mathcal{H} \cup \underline{\mathcal{H}}) \neq \emptyset$. Since $0 = \varphi^\star(\Box_{\widehat{g}} \widehat{x}^\mu) = \Box_{\varphi^\star \widehat{g}} x^\mu$, the coordinates $x^\mu$ are harmonic w.r.t. $\varphi^\star \widehat{g}$. Moreover, from the fact that $\widehat{g}$ is a solution of the $\lambda$-vacuum equations $\operatorname{Ric}[\widehat{g}] = \lambda \widehat{g}$ it follows that

$$\varphi^\star \left(\operatorname{Ric}[\widehat{g}]\right) = \operatorname{Ric}[\varphi^\star \widehat{g}] = \lambda \varphi^\star \widehat{g},$$

and therefore $\varphi^\star \widehat{g}$ is a solution of the reduced equations in the coordinates $\{x^\mu\}$, just like $g$. In order to prove that $\varphi^\star \widehat{g}$ and $g$ are actually the same, by Theorem 3.2 we only need to show that their restrictions on $\mathcal{U} \cap (\mathcal{H} \cup \underline{\mathcal{H}})$ agree. This follows directly from the fact that the push-forward $\varphi_\star$ is the identity, because $\varphi_\star \partial_u = \partial_{\widehat{u}}$, $\varphi_\star \partial_{\underline{u}} = \partial_{\widehat{\underline{u}}}$ and $\varphi_\star \partial_{x^A} = \partial_{\widehat{x}^A}$, and therefore $(\varphi^\star \widehat{g})_{\mu\nu} = g_{\mu\nu}$ on $\mathcal{U} \cap (\mathcal{H} \cup \underline{\mathcal{H}})$. $\square$

**Remark 3.24.** *The argument in Theorem 3.23 can be immediately generalized to matter fields admitting a well-posed characteristic initial value problem in which their energy-momentum tensor on the hypersurfaces have the following dependence on the initial data: (i) $T_{nn}$ depends on the matter field, the metric data, and it is algebraic on $\kappa$; (ii) $T_{nA}$ depends on the matter field, the metric data, $\kappa$ and is algebraic on $\omega_A$; (iii) The combination $h^{AB} T_{AB} - \frac{\mathfrak{n}-1}{2}\left(P^{ab} T_{ab} + 2T(n,\xi)\right)$ depends on the matter field, the metric data, $\kappa$, $\omega_A$ and it is algebraic on $\operatorname{tr}_h \mathbf{Y}_\parallel$; (iv) $T_{AB}$ depends on the matter field, the metric data, $\kappa$, $\omega_A$, $\operatorname{tr}_h \mathbf{Y}_\parallel$ and it is algebraic on the trace-free part of $\mathrm{Y}_{AB}$. The third requirement follows from the fact that, from* (2.14),

$$g^{\mu\nu} T_{\mu\nu} \stackrel{\mathcal{H}}{=} \left(P^{ab} e_a^\mu e_b^\nu + 2n^a e_a^\mu \xi^\nu\right) T_{\mu\nu} \stackrel{\mathcal{H}}{=} P^{ab} T_{ab} + 2T(n,\xi).$$

**Remark 3.25.** *Theorems 3.19 and 3.23 establish a very clear hierarchy between the compatibility conditions and the constraint equations. The former are the necessary and sufficient conditions for a DND to be able to be embedded in some spacetime, whereas the later are necessary and sufficient for the DND to be embedded in a spacetime solution of the Einstein field equations.*

## 3.4 UNIQUENESS RESULT

Given two double null data, there arises the natural question of under which conditions their developments are the same (up to isometry). In this section we establish the necessary and sufficient conditions for two double null data to define two isometric spacetimes. We start with a definition to fix some notation.

**Definition 3.26.** *Let $\mathcal{D} = \{\mathcal{H}, \boldsymbol{\gamma}, \boldsymbol{\ell}, \ell^{(2)}, \mathbf{Y}\}$ be hypersurface data and $\psi : \widehat{\mathcal{H}} \longrightarrow \mathcal{H}$ a diffeomorphism. We define the pull-back hypersurface data $\psi^\star \mathcal{D}$ by*

$$\psi^\star \mathcal{D} := \left\{ \widehat{\mathcal{H}}, \widehat{\boldsymbol{\gamma}} := \psi^\star \boldsymbol{\gamma}, \widehat{\boldsymbol{\ell}} := \psi^\star \boldsymbol{\ell}, \widehat{\ell}^{(2)} := \psi^\star \ell^{(2)}, \widehat{\mathbf{Y}} := \psi^\star \mathbf{Y} \right\}.$$

From Def. 2.1 and the fact that $\psi$ is a diffeomorphism, it follows that $\psi^\star\mathcal{D}$ is still hypersurface data. Moreover, from (2.3)-(2.6) having a unique solution for $\{P, n, n^{(2)}\}$ given $\{\boldsymbol{\gamma}, \boldsymbol{\ell}, \ell^{(2)}\}$, it follows that $\widehat{P} = \psi^\star P$, $\widehat{n} = \psi^\star n$ and $\widehat{n}^{(2)} = \psi^\star n^{(2)}$. Thus, the causal character of $\mathcal{D}$ is the same as the one of $\psi^\star\mathcal{D}$, and in particular if $\mathcal{D}$ is null, so it is $\psi^\star\mathcal{D}$. The previous definition can be extended to the context of double null data as follows.

**Definition 3.27.** *Let $\{\mathcal{D}, \underline{\mathcal{D}}, \phi, \sigma\}$ be double null data and $\psi : \widehat{\mathcal{H}} \longrightarrow \mathcal{H}$, $\underline{\psi} : \widehat{\underline{\mathcal{H}}} \longrightarrow \underline{\mathcal{H}}$ diffeomorphisms. The pull-back double null data $\Omega^\star\{\mathcal{D}, \underline{\mathcal{D}}, \phi, \sigma\}$ is defined as*

$$\Omega^\star\{\mathcal{D}, \underline{\mathcal{D}}, \phi, \sigma\} := \left\{ \psi^\star\mathcal{D}, \underline{\psi}^\star\underline{\mathcal{D}}, \widehat{\phi}, \underline{\psi}|^\star_{\underline{\mathcal{S}}}(\sigma) \right\},$$

*where $\psi^\star\mathcal{D}$ and $\underline{\psi}^\star\underline{\mathcal{D}}$ are the pull-backs in the sense of Def. 3.26 and $\widehat{\phi}$ is the map that makes the following diagram commutative*

$$\begin{array}{ccccccc}
\widehat{\mathcal{S}} & \hookrightarrow & \widehat{\mathcal{H}} & \xrightarrow{\psi} & \mathcal{H} & \hookleftarrow & \mathcal{S} \\
\big\downarrow \widehat{\phi} & & & & & & \big\downarrow \phi \\
\widehat{\underline{\mathcal{S}}} & \hookrightarrow & \widehat{\underline{\mathcal{H}}} & \xrightarrow{\underline{\psi}} & \underline{\mathcal{H}} & \hookleftarrow & \underline{\mathcal{S}}
\end{array}$$

Since $\psi$ and $\underline{\psi}$ are diffeomorphisms, they preserve the boundaries $\mathcal{S}$ and $\underline{\mathcal{S}}$, and thus the map $\widehat{\phi} := \underline{\psi} \circ \phi \circ \psi^{-1} : \widehat{\mathcal{S}} \longrightarrow \widehat{\underline{\mathcal{S}}}$ is a diffeomorphism. Then, $\Omega^\star\{\mathcal{D}, \underline{\mathcal{D}}, \phi, \sigma\}$ is still double null data, since it satisfies Def. 3.9 with $\widehat{\phi}$ and $\widehat{\sigma} := \underline{\psi}|^\star_{\underline{\mathcal{S}}}(\sigma)$. In the following proposition we find the necessary conditions for two DND to define two isometric Lorentzian manifolds.

**Proposition 3.28.** *Let $\{\mathcal{D}, \underline{\mathcal{D}}, \phi, \sigma\}$ and $\{\widehat{\mathcal{D}}, \widehat{\underline{\mathcal{D}}}, \widehat{\phi}, \widehat{\sigma}\}$ be double null data satisfying the constraint equations* (3.58) *and let $(\mathcal{M}, g)$ and $(\widehat{\mathcal{M}}, \widehat{g})$ be respective developments. Suppose that there exists an isometry $\varphi : \mathcal{M} \longrightarrow \widehat{\mathcal{M}}$. Then there exist gauge parameters $(z, V)$ and $(\underline{z}, \underline{V})$ in $\mathcal{D}$ and $\underline{\mathcal{D}}$, respectively, and a map $\Omega^\star$ as in Def. 3.27 such that*

$$\Omega^\star\{\widehat{\mathcal{D}}, \widehat{\underline{\mathcal{D}}}, \widehat{\phi}, \widehat{\sigma}\} = \mathcal{G}_{(z,V)}\left(\{\mathcal{D}, \underline{\mathcal{D}}, \phi, \sigma\}\right).$$

*Proof.* Let $\widehat{x}^\mu = \{\widehat{u}, \widehat{\underline{u}}, \widehat{x}^A\}$ be coordinates on $\widehat{\mathcal{M}}$ whose restrictions on $\widehat{\mathcal{H}}$ and $\widehat{\underline{\mathcal{H}}}$ satisfy (3.34). Define the coordinates $x^\mu$ on $\mathcal{M}$ by $x^\mu := \widehat{x}^\mu \circ \varphi$. Let $\Phi, \underline{\Phi}$ and $\xi, \underline{\xi}$ the embeddings and the riggings of $\{\mathcal{D}, \underline{\mathcal{D}}, \phi, \sigma\}$ in $(\mathcal{M}, g)$, and $\widehat{\Phi}, \widehat{\underline{\Phi}}$ and $\widehat{\xi}, \widehat{\underline{\xi}}$ be the embeddings and the riggings of $\{\widehat{\mathcal{D}}, \widehat{\underline{\mathcal{D}}}, \widehat{\phi}, \widehat{\sigma}\}$ in $(\widehat{\mathcal{M}}, \widehat{g})$. Since $(\varphi^\star\widehat{\xi})(u) = \widehat{\xi}(u \circ \varphi^{-1}) = \widehat{\xi}(\widehat{u}) \neq 0$, there exist $z \in \mathcal{F}^\star(\mathcal{H})$ and $V \in \mathfrak{X}(\mathcal{H})$ such that $\varphi^\star\widehat{\xi} = z(\xi + \Phi_\star V)$ along $\Phi(\mathcal{H})$. Since $\Phi(\mathcal{H})$ is diffeomorphic to $\widehat{\Phi}(\widehat{\mathcal{H}})$ via $\varphi$ and both $\Phi$ and $\widehat{\Phi}$ are embeddings, there exists a diffeomorphism $\psi$ making the following diagram commutative

$$\begin{array}{ccc} \mathcal{H} & \xrightarrow{\psi} & \widehat{\mathcal{H}} \\ \downarrow \Phi & & \downarrow \widehat{\Phi} \\ \mathcal{M} & \xrightarrow{\varphi} & \widehat{\mathcal{M}} \end{array}$$

Then

$$\psi^\star \widehat{\boldsymbol{\gamma}} = \psi^\star \widehat{\boldsymbol{\Phi}}^\star \widehat{g} = \boldsymbol{\Phi}^\star \varphi^\star \widehat{g} = \boldsymbol{\Phi}^\star g = \boldsymbol{\gamma},$$

and

$$\psi^\star \widehat{\boldsymbol{\ell}} = \psi^\star \widehat{\boldsymbol{\Phi}}^\star (\widehat{g}(\widehat{\xi}, \cdot)) = \boldsymbol{\Phi}^\star \varphi^\star (\widehat{g}(\widehat{\xi}, \cdot)) = \boldsymbol{\Phi}^\star \left(g(z(\xi + \Phi_\star V), \cdot)\right) = z(\boldsymbol{\ell} + \boldsymbol{\gamma}(V, \cdot)).$$

Concerning $\widehat{\ell}^{(2)}$,

$$\begin{aligned} \psi^\star \widehat{\ell}^{(2)} &= \psi^\star \widehat{\boldsymbol{\Phi}}^\star (\widehat{g}(\widehat{\xi}, \widehat{\xi})) \\ &= \boldsymbol{\Phi}^\star \left( z^2 g(\xi, \xi) + 2z^2 g(\xi, \Phi_\star V) + z^2 g(\Phi_\star V, \Phi_\star V) \right) \\ &= z^2 (\ell^{(2)} + 2\boldsymbol{\ell}(V) + \boldsymbol{\gamma}(V, V)). \end{aligned}$$

Finally,

$$\psi^\star \widehat{\mathbf{Y}} = \frac{1}{2} \psi^\star \widehat{\boldsymbol{\Phi}}^\star \pounds_{\widehat{\xi}} \widehat{g} = \frac{1}{2} \boldsymbol{\Phi}^\star \varphi^\star \pounds_{\widehat{\xi}} \widehat{g} = \frac{1}{2} \boldsymbol{\Phi}^\star \left( \pounds_{z(\xi + \Phi_\star V)} g \right) = z \mathbf{Y} + dz \otimes_s \boldsymbol{\ell} + \frac{1}{2} \pounds_{zV} \boldsymbol{\gamma},$$

where the third equality holds because $\varphi^\star \widehat{\xi} = z(\xi + \Phi_\star V)$, $\varphi^\star \widehat{g} = g$ and $\varphi^\star(\pounds_{\widehat{\xi}} \widehat{g}) = \pounds_{\varphi^\star \widehat{\xi}} (\varphi^\star g)$ (this follows from the formula (2.149) particularized to $T = \widehat{g}$ and $\varphi_\star X = \widehat{\xi} \Longrightarrow X = \varphi^\star \widehat{\xi} = z(\xi + \Phi_\star V)$.). Thus, recalling Def. 2.2, $\psi^\star \widehat{\mathcal{D}} = \mathcal{G}_{(z,V)}(\mathcal{D})$. The same argument on $\underline{\mathcal{H}}$ proves that there exist gauge parameters $(\underline{z}, \underline{V})$ on $\underline{\mathcal{D}}$ such that $\underline{\psi}^\star \widehat{\underline{\mathcal{D}}} = \mathcal{G}_{(\underline{z}, \underline{V})}(\underline{\mathcal{D}})$. Finally, taking into account item 2. of Def 3.14,

$$\underline{\psi}|_{\mathcal{S}}^\star (\widehat{\sigma}) = \boldsymbol{\Phi}|_{\mathcal{S}}^\star \varphi^\star \left( \widehat{g}(\widehat{\nu}, \widehat{\underline{\nu}}) \right) = \boldsymbol{\Phi}|_{\mathcal{S}}^\star \left( g(z^{-1} \nu, \underline{z}^{-1} \underline{\nu}) \right) = z^{-1} \underline{z}^{-1} \sigma.$$

Comparing with (3.13), the result follows. □

The previous proposition motivates defining the notion of isometric double null data as follows.

**Definition 3.29.** *We say that two double null data* $\{\mathcal{D}, \underline{\mathcal{D}}, \phi, \sigma\}$ *and* $\{\widehat{\mathcal{D}}, \widehat{\underline{\mathcal{D}}}, \widehat{\phi}, \widehat{\sigma}\}$ *are isometric if there exist diffeomorphisms* $\psi : \mathcal{H} \longrightarrow \widehat{\mathcal{H}}$ *and* $\underline{\psi} : \underline{\mathcal{H}} \longrightarrow \widehat{\underline{\mathcal{H}}}$ *and gauge parameters* $(z, V)$ *and* $(\underline{z}, \underline{V})$ *in* $\mathcal{D}$ *and* $\underline{\mathcal{D}}$*, respectively, such that the pull-back double null data* $\Omega^\star \{\widehat{\mathcal{D}}, \widehat{\underline{\mathcal{D}}}, \widehat{\phi}, \widehat{\sigma}\}$ *satisfies*

$$\Omega^\star \{\widehat{\mathcal{D}}, \widehat{\underline{\mathcal{D}}}, \widehat{\phi}, \widehat{\sigma}\} = \mathcal{G}_{(z,V)} \left( \{\mathcal{D}, \underline{\mathcal{D}}, \phi, \sigma\} \right).$$

We conclude this chapter by proving that the necessary conditions of Prop. 3.28 are also sufficient. This result gives a geometric uniqueness statement of the characteristic problem of the Einstein field equations. Indeed, two isometric initial data are indistinguishable from a geometric point of view and thus they should have "the same" developments. The precise statement of the notion of uniqueness is given in the following theorem.

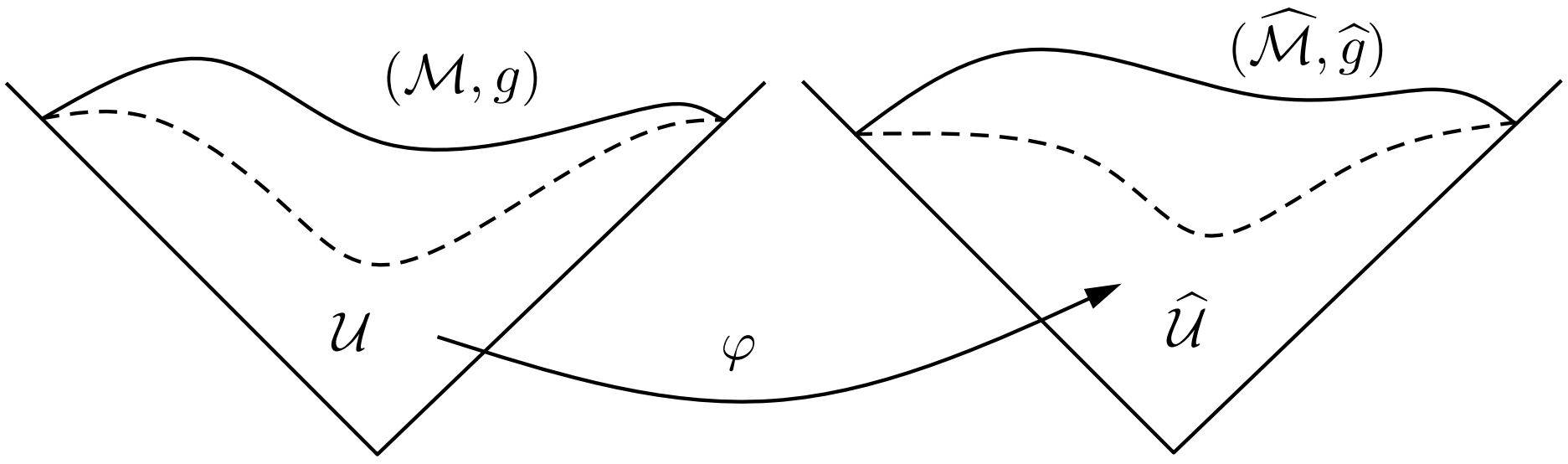

Figure 3.3: Given isometric DND, there exist isometric neighbourhoods $\mathcal{U}$ and $\widehat{\mathcal{U}}$ of the initial data.

**Theorem 3.30.** *Let* $\{\mathcal{D}, \underline{\mathcal{D}}, \phi, \sigma\}$*,* $\{\widehat{\mathcal{D}}, \widehat{\underline{\mathcal{D}}}, \widehat{\phi}, \widehat{\sigma}\}$ *be two isometric DND in the sense of Def. 3.29 with diffeomorphisms* $\psi : \mathcal{H} \longrightarrow \widehat{\mathcal{H}}$ *and* $\underline{\psi} : \underline{\mathcal{H}} \longrightarrow \widehat{\underline{\mathcal{H}}}$ *and satisfying the abstract constraint equations* (3.58)*, and let* $(\mathcal{M}, g)$*,* $(\widehat{\mathcal{M}}, \widehat{g})$ *be respective developments. Then there exist neighbourhoods* $\mathcal{U} \subseteq \mathcal{M}$ *and* $\widehat{\mathcal{U}} \subseteq \widehat{\mathcal{M}}$ *of* $\mathcal{H} \cup \underline{\mathcal{H}}$ *and* $\widehat{\mathcal{H}} \cup \widehat{\underline{\mathcal{H}}}$*, respectively, and a diffeomorphism* $\varphi : \mathcal{U} \longrightarrow \widehat{\mathcal{U}}$ *such that* $\varphi^\star \widehat{g} = g$*.*

*Proof.* We start by writing $\{\widehat{\mathcal{D}}, \widehat{\underline{\mathcal{D}}}, \widehat{\phi}, \widehat{\sigma}\}$ in the gauge of Lemma 3.16 w.r.t. some coordinates $\{\widehat{x}^{\underline{a}}\} = \{\widehat{u}, \widehat{x}^A\}$ and $\{\widehat{\underline{x}}^{\underline{a}}\} = \{\widehat{\underline{u}}, \widehat{\underline{x}}^A\}$ in $\widehat{\mathcal{H}}$ and $\widehat{\underline{\mathcal{H}}}$ satisfying (3.34), and $\{\mathcal{D}, \underline{\mathcal{D}}, \phi, \sigma\}$ in the gauge in which $\Omega^\star\{\widehat{\mathcal{D}}, \widehat{\underline{\mathcal{D}}}, \widehat{\phi}, \widehat{\sigma}\} = \{\mathcal{D}, \underline{\mathcal{D}}, \phi, \sigma\}$ holds. We want to show that $\Omega^\star\{\widehat{\mathcal{D}}, \widehat{\underline{\mathcal{D}}}, \widehat{\phi}, \widehat{\sigma}\}$ is written in the gauge of Lemma 3.16 w.r.t. the coordinates $\{x^{\underline{a}}\} := \{\widehat{x}^{\underline{a}} \circ \psi\}$ and $\{\underline{x}^{\underline{a}}\} := \{\widehat{\underline{x}}^{\underline{a}} \circ \underline{\psi}\}$. Let $\widehat{U}$ be the vector field defined in Thm. 2.48 w.r.t. $\{\widehat{x}^{\underline{a}}\}$. First we prove that $\psi^\star \widehat{U}$ is again the vector of Theorem 2.48 but w.r.t. $\{\psi^\star \widehat{x}^{\underline{a}}\}$ (and analogously on $\underline{\mathcal{H}}$). Let $\{\widehat{e}_c\}$ be a (local) basis of $\mathfrak{X}(\widehat{\mathcal{H}})$ and $\{e_c := \psi^\star \widehat{e}_c\}$ a (local) basis of $\mathfrak{X}(\mathcal{H})$. Since $\psi^\star\left(\widehat{e}_c(\widehat{x}^{\underline{a}})\right) = e_c(x^{\underline{a}})$, it follows $\psi^\star \widehat{\Xi}^c{}_{\underline{a}} = \Xi^c{}_{\underline{a}}$. Moreover, since $\psi^\star \widehat{\overline{\nabla}} = \overline{\nabla}$ (because $\Omega^\star\{\widehat{\mathcal{D}}, \widehat{\underline{\mathcal{D}}}, \widehat{\phi}, \widehat{\sigma}\} = \{\mathcal{D}, \underline{\mathcal{D}}, \phi, \sigma\}$), the pull-back of the Hessian of a function is the Hessian of the pullback of that function, and since $\psi^\star \widehat{P} = P$, it turns out that $\psi^\star\left(\widehat{P}^{ab} \widehat{\overline{\nabla}}_a \widehat{\overline{\nabla}}_b \widehat{x}^{\underline{a}}\right) = P^{ab} \overline{\nabla}_a \overline{\nabla}_b x^{\underline{a}}$, and thus $\psi^\star \widehat{U}^c = \psi^\star\left(\widehat{\Xi}^c{}_{\underline{a}} \widehat{\overline{\Box}} \widehat{x}^{\underline{a}}\right) = \Xi^c{}_{\underline{a}} \overline{\Box} x^{\underline{a}}$. Therefore $\Omega^\star\{\widehat{\mathcal{D}}, \widehat{\underline{\mathcal{D}}}, \widehat{\phi}, \widehat{\sigma}\}$ is written in a harmonic gauge w.r.t. $\{x^{\underline{a}}\}$ and $\{\underline{x}^{\underline{a}}\}$. Moreover, since $\psi^\star \widehat{\ell}^{(2)} = \ell^{(2)} = 0$, $\psi^\star \widehat{\boldsymbol{\ell}}_\parallel = \boldsymbol{\ell}_\parallel = 0$ on $\mathcal{S}$, $\underline{\psi}^\star \widehat{\underline{\ell}}^{(2)} = \underline{\ell}^{(2)} = 0$, $\underline{\psi}^\star \widehat{\underline{\boldsymbol{\ell}}}_\parallel = \underline{\boldsymbol{\ell}}_\parallel = 0$ on $\underline{\mathcal{S}}$, as well as $\underline{\psi}^\star_{\underline{\mathcal{S}}}(\widehat{\sigma}) = \underline{\psi}^\star_{\underline{\mathcal{S}}}\left(\widehat{\underline{n}}(\widehat{u})\right) = (\underline{\psi}^\star \widehat{\underline{n}})(u)$ on $\underline{\mathcal{S}}$ and similarly $\psi|^\star_{\underline{\mathcal{S}}}(\widehat{\sigma}) = (\psi^\star \widehat{n})\,(\underline{u})$ on $\mathcal{S}$ (we omit the $\phi^\star$ for simplicity), we conclude that the data $\Omega^\star\{\widehat{\mathcal{D}}, \widehat{\underline{\mathcal{D}}}, \widehat{\phi}, \widehat{\sigma}\}$ is written in the gauge of Lemma 3.16 w.r.t. the coordinates $\{x^{\underline{a}}\}$ and $\{\underline{x}^{\underline{a}}\}$.

Let $(\mathcal{M}, g)$ be a development of $\{\mathcal{D}, \underline{\mathcal{D}}, \phi, \sigma\}$ with embeddings $\Phi$, $\underline{\Phi}$ and riggings $\xi$, $\underline{\xi}$ and let $(\widehat{\mathcal{M}}, \widehat{g})$ be the same but everything with a "$\widehat{\ }$". Let $\{x^\mu\}$ be the harmonic coordinates on $(\mathcal{M}, g)$ restricting to the given ones at $\mathcal{H} \cup \underline{\mathcal{H}}$ and satisfying $\Phi(\mathcal{H}) = \{u = 0\}$, $\underline{\Phi}(\underline{\mathcal{H}}) = \{\underline{u} = 0\}$ (and the same with "$\widehat{\ }$"). A similar argument as the one in the proof of Theorem 3.23 shows that the rigging vectors are $\xi = \partial_u$, $\underline{\xi} = \partial_{\underline{u}}$ and $\widehat{\xi} = \partial_{\widehat{u}}$, $\widehat{\underline{\xi}} = \partial_{\widehat{\underline{u}}}$. Choosing suitable neighbourhoods $\mathcal{U} \subseteq \mathcal{M}$ of $\mathcal{H} \cup \underline{\mathcal{H}}$ and $\widehat{\mathcal{U}} \subseteq \widehat{\mathcal{M}}$ of $\widehat{\mathcal{H}} \cup \widehat{\underline{\mathcal{H}}}$ (see Figure 3.3), we can define the diffeomorphism $\varphi$ by $x^\mu = \widehat{x}^\mu \circ \varphi$, which by construction restricts to the given diffeomorphisms $\psi : \mathcal{H} \longrightarrow \widehat{\mathcal{H}}$ and $\underline{\psi} : \underline{\mathcal{H}} \longrightarrow \widehat{\underline{\mathcal{H}}}$. Since $(\widehat{\mathcal{M}}, \widehat{g})$ is a solution of the $\lambda$-vacuum equations, so it is $(\mathcal{M}, \varphi^\star \widehat{g})$. Moreover, $\Box_{\varphi^\star \widehat{g}} x^\mu = 0$, so both $g$ and $\varphi^\star \widehat{g}$ are a solution of the reduced equations in the coordinates $\{x^\mu\}$. By Theorem 3.2, in order to prove that $g = \varphi^\star \widehat{g}$ on $\mathcal{U}$, we only need to show that their restrictions to $\mathcal{H} \cup \underline{\mathcal{H}}$ agree, which follows at once after recalling $\Omega^{\star}\{\widehat{\mathcal{D}}, \widehat{\underline{\mathcal{D}}}, \widehat{\phi}, \widehat{\sigma}\} = \{\mathcal{D}, \underline{\mathcal{D}}, \phi, \sigma\}$ and using that $\varphi$ restricts to $\psi$ and $\underline{\psi}$ on $\mathcal{H}$ and $\underline{\mathcal{H}}$, respectively. □

**Remark 3.31.** *The geometric existence and uniqueness statements in Theorems 3.23 and 3.30 above refer to a neighbourhood of the intersection surface. This limitation arises because we rely on Rendall's approach (Theorem 3.2) to establish the results. As we already mentioned in the introductory section of this chapter, there are results of existence of the spacetime in a neighbourhood of the full initial hypersurfaces (see e.g. [50, 78, 208]). We expect that a suitable application of this result should allow us to extend our geometric existence and uniqueness results to the full double null data as well. This is however not immediate because the field variables and gauge fixing conditions are different (for instance, they usually involve the Weyl components and the Bianchi identities). I intend to analyze this in a future work.*

## 3.5 PRE-DOUBLE NULL DATA

In Section 3.1 we mentioned that there are many ways to prescribe initial data for the characteristic problem. The detached notion defined in Section 3.2 includes the full metric and its first transverse derivative as initial data (subject to the constraint equations if one wants to solve the characteristic problem). This encompasses all possible initial data constructed solely from the metric and its first derivative at the characteristic hypersurfaces. Then, one can choose which parts of the data are prescribed (and thus subject to the corresponding constraints) and what other parts are "reconstructed" from the remaining constraints, now seen as propagation equations. To exploit this, in this section we will show that only by prescribing the metric part of the data along with the torsion one-form at the bifurcation surface, one can reconstruct the full tensors $\mathbf{Y}$ and $\underline{\mathbf{Y}}$ by solving the constraint equations. This connects our abstract approach with the standard one analyzed in the literature (see e.g. [64]). We start by recalling a result from Section 2.2 (Theorem 2.11) whose proof relies on the hierarchical structure of the constraint tensor (2.95).

**Theorem 3.32.** *Let $\{\mathcal{H}, \boldsymbol{\gamma}, \boldsymbol{\ell}, \ell^{(2)}\}$ be null metric hypersurface data admitting a cross-section $\iota : \mathcal{S} \hookrightarrow \mathcal{H}$ with metric $h := \iota^{\star}\boldsymbol{\gamma}$ and assume there exists a smooth function $f$ on $\mathcal{H}$ satisfying*

$$-\pounds_n(\operatorname{tr}_P \mathbf{U}) + (\operatorname{tr}_P \mathbf{U})f - P^{ab}P^{cd}\mathrm{U}_{ac}\mathrm{U}_{bd} = 0, \tag{3.62}$$

*with gauge behavior $\mathcal{G}_{(z,V)}f := z^{-1}(f + n(\log|z|))$. Choose any gauge-invariant real constant $\lambda$ and let $\chi$, $\boldsymbol{\beta}$ and $\mathcal{Y}_{AB}$ be, respectively, a function, a one-form and a $(0,2)$ symmetric tensor on $\mathcal{S}$ traceless w.r.t. $h$, with gauge transformations*

$$\mathcal{G}_{(z,V)}\beta_A := \beta_A + \frac{1}{2}\Big(n(\log|z|)\ell_A + \nabla_A \log|z|\Big) + \frac{1}{2}\pounds_n \underset{\smile}{V}_A - h^{BC}\mathrm{U}_{AB}V_C + \frac{1}{2}n(\log|z|)\underset{\smile}{V}_A, \tag{3.63}$$

$$\mathcal{G}_{(z,V)}\chi := z\chi - \ell^{(2)}n(z) + \frac{1}{2}P^{ab}\pounds_{zV}\gamma_{ab} - 2\boldsymbol{\beta}'(V_{\parallel}) + 2V_n f', \tag{3.64}$$

$$\mathcal{G}_{(z,V)}\mathcal{Y}_{AB} := z\mathcal{Y}_{AB} + \Big(\ell_{(A}\nabla_{B)}z + \frac{1}{2}\pounds_{zV}\gamma_{AB} - 2z(\ell_{(A} + \underset{\smile}{V}_{(A)})\beta'_{B)} - z^2 f'(\ell_A + \underset{\smile}{V}_A)(\ell_B + \underset{\smile}{V}_B)\Big)^{tf}, \tag{3.65}$$

*where* $\underset{\smile}{\boldsymbol{V}} := \boldsymbol{\gamma}(V,\cdot)$*,* $V_{\|}$ *and* $V_n$ *are defined by* $V|_{\mathcal{S}} = V_{\|} + V_n n$*,* $V_{\|} \in T\mathcal{S}$*,* $V_n \in \mathcal{F}(\mathcal{S})$*, and "tf" stands for "trace free" (w.r.t. h). Then, there exists a unique tensor* $\mathbf{Y}$ *on* $\mathcal{H}$ *satisfying* $\mathbf{Y}(n,n) = -f$*,* $\operatorname{tr}_P \mathbf{Y}|_{\mathcal{S}} = \chi$*,* $\iota^\star(\mathbf{Y}(n,\cdot)) = \boldsymbol{\beta}$ *and* $(\iota^\star\widehat{\mathrm{Y}})^{tf}_{AB} = \mathcal{Y}_{AB}$ *such that* $\{\mathcal{H},\boldsymbol{\gamma},\boldsymbol{\ell},\ell^{(2)},\mathbf{Y}\}$ *is null hypersurface data satisfying* $\boldsymbol{\mathcal{R}} = \lambda\boldsymbol{\gamma}$*.*

Let $\{\mathcal{D},\underline{\mathcal{D}},\phi,\sigma\}$ be double null data. Taking into account the compatibility conditions (3.19)-(3.20), the previous result suggests that only by prescribing the metric part of $\mathcal{D}$ and $\underline{\mathcal{D}}$, two functions $f$ and $\underline{f}$ satisfying (3.62) on $\mathcal{H}$ and $\underline{\mathcal{H}}$, respectively, as well as the torsion one-form at $\mathcal{S}$, it is possible to reconstruct $\mathbf{Y}$ and $\underline{\mathbf{Y}}$ in a unique such that the compatibility conditions and the constraint equations hold. This motivates the following definition.

**Definition 3.33.** *Let* $\mathcal{H}$ *and* $\underline{\mathcal{H}}$ *be two manifolds with boundaries* $i : \mathcal{S} \hookrightarrow \mathcal{H}$ *and* $\underline{i} : \underline{\mathcal{S}} \hookrightarrow \underline{\mathcal{H}}$*,* $\boldsymbol{\omega}$ *a 1-form on* $\mathcal{S}$ *and* $f, \underline{f}$ *smooth functions on* $\mathcal{H}$ *and* $\underline{\mathcal{H}}$*, respectively. Let* $\mathcal{D}^{met} = \{\mathcal{H},\boldsymbol{\gamma},\boldsymbol{\ell},\ell^{(2)}\}$ *and* $\underline{\mathcal{D}}^{met} = \{\underline{\mathcal{H}},\underline{\boldsymbol{\gamma}},\underline{\boldsymbol{\ell}},\underline{\ell}^{(2)}\}$ *be characteristic metric hypersurface data and* $\phi : \mathcal{S} \longrightarrow \underline{\mathcal{S}}$ *an isometry. Then we say that* $\{\mathcal{D}^{met},\underline{\mathcal{D}}^{met}, f, \underline{f}, \boldsymbol{\omega}, \phi\}$ *is pre-double null data provided* $f$ *satisfies*

$$-\pounds_n(\operatorname{tr}_P \mathbf{U}) + (\operatorname{tr}_P \mathbf{U})f - P^{ab}P^{cd}\mathrm{U}_{ac}\mathrm{U}_{bd} = 0$$

*and* $\underline{f}$ *satisfies an analogous equation on* $\underline{\mathcal{H}}$*.*

Next we show that every pre-double null data gives rise to a canonical double null data satisfying the constraint equations.

**Proposition 3.34.** *Let* $\{\mathcal{D}^{met},\underline{\mathcal{D}}^{met}, f, \underline{f}, \boldsymbol{\omega}, \phi\}$ *be pre-double null data,* $\sigma \in \mathcal{F}(\underline{\mathcal{S}})$ *everywhere negative and* $\lambda$ *any real number. Then, there exists a unique double null data* $\{\mathcal{D},\underline{\mathcal{D}},\phi,\sigma\}$ *satisfying the constraint equations* $\boldsymbol{\mathcal{R}} = \lambda\boldsymbol{\gamma}$ *and* $\underline{\boldsymbol{\mathcal{R}}} = \lambda\underline{\boldsymbol{\gamma}}$ *on* $\mathcal{H}$ *and* $\underline{\mathcal{H}}$*, respectively, as well as* $f = \kappa$*,* $\underline{f} = \underline{\kappa}_n$ *and such that* $\boldsymbol{\omega}$ *is the torsion one-form on* $\mathcal{S}$*.*

*Proof.* The idea is to reconstruct the tensors $\mathbf{Y}$ and $\underline{\mathbf{Y}}$ on $\mathcal{H}$ and $\underline{\mathcal{H}}$ such that the compatibility conditions (3.19)-(3.20) and the constraint equations $\boldsymbol{\mathcal{R}} = \lambda\boldsymbol{\gamma}$ and $\underline{\boldsymbol{\mathcal{R}}} = \lambda\underline{\boldsymbol{\gamma}}$ hold. We give the argument on $\mathcal{H}$, and the one on $\underline{\mathcal{H}}$ is analogous. The strategy is (1) to prescribe the function $\chi$, one-form $\boldsymbol{\beta}$ and tensor $\mathcal{Y}_{AB}$ that come from the compatibility conditions (3.19)-(3.20) with the replacements $\operatorname{tr}_P \mathbf{Y}|_{\mathcal{S}} \to \chi$, $\iota^\star\mathbf{r} \to \boldsymbol{\beta}$ and $(\iota^\star\widehat{\mathrm{Y}}_{AB})^{tf} \to \mathcal{Y}_{AB}$; and (2) to apply Theorem 3.32 precisely with these values for $\chi$, $\boldsymbol{\beta}$ and $\mathcal{Y}_{AB}$. Recalling Remark 3.12 and the decomposition in (2.143) it is not hard to obtain the components $\operatorname{tr}_P \mathbf{Y}|_{\mathcal{S}}$, $\iota^\star\mathbf{r}$ and $(\iota^\star\widehat{\mathrm{Y}}_{AB})^{tf}$ from (3.19)-(3.20), which are

$$\begin{aligned}
\boldsymbol{\beta} &= \iota^\star\mathbf{s} - \boldsymbol{\omega} - \mathbf{K}_{\|}(\ell^\sharp,\cdot) - d(\log|\sigma|),\\
\chi &= \sigma^{-1}h^{AB}\underline{\mathrm{K}}_{AB} - \frac{1}{2}h^{AB}(\pounds_\theta\gamma)_{AB} - 2\boldsymbol{\beta}(\ell^\sharp) + (\ell^{(2)} - \ell^{(2)}_\sharp)f,\\
\mathcal{Y}_{AB} &= \Big(\sigma^{-1}\underline{\mathrm{K}}_{AB} - \frac{1}{2}(\widehat{\pounds_\theta\gamma})_{AB}\Big)^{tf}.
\end{aligned}$$

From these values of $\chi$, $\boldsymbol{\beta}$ and $\mathcal{Y}_{AB}$ (and the corresponding ones on the other side) we construct unique tensors $\mathbf{Y}$, $\underline{\mathbf{Y}}$ such that $\{\mathcal{D}, \underline{\mathcal{D}}, \phi, \sigma\}$ is double null data satisfying $\boldsymbol{\mathcal{R}} = \lambda\boldsymbol{\gamma}$ and $\underline{\boldsymbol{\mathcal{R}}} = \lambda\underline{\boldsymbol{\gamma}}$. By construction, $\boldsymbol{\omega}$ is the torsion one-form on $\mathcal{S}$. □

Note that by the this proposition, the existence and uniqueness Theorems 3.23 and 3.30 also extend to the context of pre-DND (and in general, to any geometric object constructed similarly by removing some part of the DND together with its corresponding constraints).

# 4

# HOMOTHETIC KID PROBLEM IN GENERAL RELATIVITY

In this chapter we introduce a novel formulation of the Killing initial data (KID) problem for hypersurfaces of arbitrary causal character. This problem consists of determining whether the Cauchy development of an initial data set will admit a Killing vector. Section 4.1 provides an overview of previous work on the spacelike and characteristic KID problems. In Section 4.2, we recall several classical identities relating a vector field $\eta$ on a semi-Riemannian manifold $(M, g)$ with its deformation tensor $\pounds_\eta g$ and with $\pounds_\eta \nabla$. These identities are then analyzed in the context of embedded hypersurface data of arbitrary causal character in Section 4.3. Focusing first on the non-null case, Section 4.4 revisits the spacelike homothetic KID problem in the language of hypersurface data. Section 4.5 turns to the null case, showing that the identities of Section 4.3 lead to transport equations for certain components of $\pounds_\eta g$ and $\pounds_\eta \nabla$ along the null generators of the hypersurface. These equations play a key role in Section 4.6, where we develop a detached formulation of the homothetic KID problem on two intersecting null hypersurfaces. Finally, Section 4.7 illustrates another application of the formalism by addressing the smooth spacelike-characteristic problem. The results in this chapter are published in [234].

## 4.1 PRELIMINARIES OF KILLING INITIAL DATA

Killing and homothetic vector fields are of an utmost importance in General Relativity, as they encode the symmetries of the spacetime, providing crucial insights into its physical and geometrical properties. Killing vectors are associated with isometries of the spacetime, leading to conserved quantities and corresponding, for instance, to equilibrium states of the gravitational field (see e.g. [154, 310, 323]). Homothetic vectors describe scale invariances, allowing the study of self-similar solutions of the Einstein equations [285], which arise, for instance, in critical gravitational collapse [151]. Thus, the identification of spacetimes possessing symmetries of this kind is of obvious physical interest.

One particularly interesting problem is how to encode the existence of a symmetry at the initial data level. This allows for instance to transform a problem from a Lorentzian setting in dimension $\mathfrak{n}+1$ into a simpler problem in dimension $\mathfrak{n}$. This translation is often the right setup to tackle several interesting problems, e.g. the existence and uniqueness of black holes in equilibrium (see [283, 298]). In this section we review previous works on the spacelike and characteristic KID problems.

### 4.1.1 *Spacelike Killing initial data*

Consider an initial data set $(\Sigma, h, K)$ (see Section 3.1), a function $C \in \mathcal{F}(\Sigma)$ and a vector field $\bar{\eta} \in \mathfrak{X}(\Sigma)$. The classical strategy [37, 84, 242] to solve the spacelike KID problem, i.e. to construct a Cauchy development of $(\Sigma, h, K)$ with a Killing vector $\eta$ such that $\eta|_\Sigma = Cn + \bar{\eta}$, is as follows (for simplicity we restrict the discussion to vacuum with $\lambda = 0$ and we postpone the derivation of all the equations to Section 4.2). First, one shows that if the Cauchy development $(M, g)$ admits a Killing $\eta$, it must satisfy

$$\Box_g \eta = 0. \tag{4.1}$$

Second, one can prove that if $(M, g)$ is vacuum and $\eta$ satisfies (4.1), its deformation tensor $\mathcal{K} := \pounds_\eta g$ fulfills the following wave equation

$$\Box_g \mathcal{K}_{\beta\nu} + 2R_{\sigma\beta\mu\nu}\mathcal{K}^{\sigma\mu} = 0. \tag{4.2}$$

Then, the strategy is to construct $\eta$ by solving equation (4.1) (which is well-posed by the standard theorems, see Section 3.1) from suitable initial data $\eta|_\Sigma$ and $\pounds_n \eta|_\Sigma$ such that both $\mathcal{K}$ and $\pounds_n \mathcal{K}$ vanish at $\Sigma$, because then by (4.2) $\mathcal{K}$ will vanish everywhere (and hence $\eta$ is a Killing vector).

A computation shows that the pullbacks of $\mathcal{K}$ and $\pounds_n \mathcal{K}$ to $\Sigma$ are

$$\begin{aligned} \mathcal{K}_{ab} &= -2CK_{ab} + 2\nabla^h_{(a}\bar{\eta}_{b)}, \\ (\pounds_n \mathcal{K})_{ab} &= C\left(-R^{(h)}_{ab} + 2h^{cd}K_{ac}K_{bd} - (\operatorname{tr}_h K)K_{ab}\right) - \pounds_{\bar{\eta}} K_{ab} + \nabla^h_a \nabla^h_b C, \end{aligned}$$

while the pullbacks of the remaining components of $\mathcal{K}$, namely $\mathcal{K}(n, \cdot)$ and $\mathcal{K}(n, n)$, are

$$\begin{aligned} (\Phi^\star \mathcal{K}(n, \cdot))_a &= -\nabla^h_a C + (\pounds_n \boldsymbol{\eta})_a - 2K_{ab}\bar{\eta}^b, \\ \Phi^\star \mathcal{K}(n, n) &= 2(\pounds_n \boldsymbol{\eta})(n) - 2g(\eta, \nabla_n n), \end{aligned}$$

where $\boldsymbol{\eta} := g(\eta, \cdot)$. This suggests that one must choose $\eta|_\Sigma = Cn + \bar{\eta}$ satisfying (these are the so-called *KID equations*)

$$0 = -2CK_{ab} + 2\nabla^h_{(a}\bar{\eta}_{b)}, \tag{4.3}$$

$$0 = C\left(-R^{(h)}_{ab} + 2h^{cd}K_{ac}K_{bd} - (\operatorname{tr}_h K)K_{ab}\right) - \pounds_{\bar{\eta}} K_{ab} + \nabla^h_a \nabla^h_b C, \tag{4.4}$$

as well as $\pounds_n\eta|_\Sigma$ satisfying

$$0 = -\nabla^h_a C + (\pounds_n \boldsymbol{\eta})_a - 2K_{ab}\bar{\eta}^b,$$
$$0 = 2(\pounds_n \boldsymbol{\eta})(n) - 2g(\eta, \nabla_n n),$$

since this implies $\mathcal{K}|_\Sigma = 0$ and $\Phi^\star(\pounds_n\mathcal{K}) = 0$. The remaining components of $\pounds_n\mathcal{K}$, i.e. those of the form $(\pounds_n\mathcal{K})(n,\cdot)$, are proven to be automatically zero provided $\eta$ satisfies (4.1). This discussion can be summarized in the following classical theorem [37, 84, 242].

**Theorem 4.1.** *Let* $(\Sigma, h, K)$ *be an initial data set satisfying the constraint equations* (3.1) *and* $C \in \mathcal{F}(\Sigma)$, $\bar{\eta} \in \mathfrak{X}(\Sigma)$ *satisfying the KID equations* (4.3)-(4.4). *Then, the Cauchy development* $(M, g)$ *of* $(\Sigma, h, K)$ *admits a Killing vector* $\eta$ *such that* $\eta|_\Sigma = Cn + \bar{\eta}$.

This existence result can be generalized to $\lambda$-vacuum and to include general matter, in the sense that the matter field itself is invariant along the Killing [84, 275, 276]. It has also been studied when more that one Killing is present [37]. Similar results are also known for general conformal Killing vectors [140] and conformal Killing spinors [98, 141].

### 4.1.2 *Characteristic Killing initial data*

The KID problem for the Einstein equations in two characteristic hypersurfaces, as well as the one for the lightcone, have been studied in [74]. In [258] one of the authors analyzes the case of the conformal Einstein equations (restricted to four dimensions if one of the characteristic hypersurfaces is null infinity). Here we discuss the strategy in [74] that the authors use to construct the Killing field.

Consider an $(\mathfrak{n}+1)$-dimensional vacuum spacetime $(M, g)$ with an adapted null coordinate system $\{u, v, x^A\}$, $A = 3, ..., \mathfrak{n}+1$, where $N_1 = \{u = 0\}$ and $N_2 = \{v = 0\}$ are two smooth null hypersurfaces that intersect along a smooth $(\mathfrak{n}-1)$-dimensional surface $S = \{u = v = 0\}$. Let $\eta|_{N_1 \cup N_2}$ be a continuous vector field on $N_1 \cup N_2$ with smooth restrictions on $N_1$ and $N_2$. The strategy is to construct a vector field $\eta$ on $M$ from suitable initial data $\eta|_{N_1\cup N_2}$ by solving (4.1) in such a way that $\mathcal{K} = \pounds_\eta g$ vanishes on $N_1 \cup N_2$, because since $\mathcal{K}$ satisfies (4.2), by Theorem 3.2 then $\mathcal{K}$ vanishes everywhere and hence $\eta$ is a Killing. We will mainly discuss the procedure on $N_1$ since the one on $N_2$ is analogous.

First of all, note that by imposing $(\nabla\eta)_{vv} = 0$, $(\nabla\eta)_{(vA)} = 0$ and $(\nabla\eta)_{(AB)} = 0$, with $\nabla$ being the ambient Levi-Civita connection, it follows that $\mathcal{K}_{vv} = \mathcal{K}_{Av} = \mathcal{K}_{AB} = 0$ on $N_1$. Let us define the tensor field $S_{\alpha\beta\mu} := \nabla_\alpha\nabla_\beta\eta_\mu - R^\sigma{}_{\alpha\beta\mu}\eta_\sigma$ (it is related with the tensor $\pounds_\eta\nabla$, as we will see in Section 4.2). It turns out that conditions $S_{vvu} = 0$, $S_{Avu} = 0$ and $g^{AB}S_{ABu} = 0$ imply that the transverse components of $\mathcal{K}$, namely $\mathcal{K}_{uv}$, $\mathcal{K}_{uA}$ and $\mathcal{K}_{uu}$, satisfy homogeneous transport equations along the null generator of $N_1$. Then, by imposing

1. $(\nabla\eta)_{vv} = 0$, $(\nabla\eta)_{(vA)} = 0$ and $(\nabla\eta)_{(AB)} = 0$ on $N_1$,
2. $(\nabla\eta)_{uu} = 0$, $(\nabla\eta)_{(uA)} = 0$ and $(\nabla\eta)_{(AB)} = 0$ on $N_2$,

together with

3. $S_{vvu} = 0$, $S_{Avu} = 0$ and $g^{AB}S_{ABu} = 0$ on $N_1$,

4. $S_{uuv} = 0$, $S_{Auv} = 0$ and $g^{AB}S_{ABv} = 0$ on $N_2$,

as well as $(\nabla\eta)_{(uv)} = 0$ on $S$, it follows that $\mathcal{K}|_{N_1\cup N_2} = 0$ and therefore $\eta$ is a Killing.

In fact, the conditions on $N_1$ and $N_2$ can be relaxed as follows. First, one observes that when $(\nabla\eta)_{vv} = 0$, $(\nabla\eta)_{(vA)} = 0$, $((\nabla\eta)_{(AB)})^{tf} = 0$ and $S_{vvu} = 0$ on $N_1$, the quantity $g^{AB}(\nabla\eta)_{AB}$ satisfies a second order, homogeneous transport equation along $\partial_v$. Then, it suffices to impose $g^{AB}(\nabla\eta)_{AB} = 0$ and $\nabla_{\partial_v}(g^{AB}(\nabla\eta)_{AB}) = 0$ on $S$, as well as $(\nabla\eta)_{vv} = 0$, $(\nabla\eta)_{(vA)} = 0$, $((\nabla\eta)_{(AB)})^{tf} = 0$ and $S_{vvu} = 0$ on $N_1$. Under the same conditions, $S_{Avu}$ and $g^{AB}S_{ABu}$ also satisfy first order homogeneous transport equations, so it suffices to impose that they vanish initially. And finally, $g^{AB}S_{ABu} = 0$ at $S$ follows automatically from the previous conditions as well as from $(\nabla\eta)_{(uv)} = 0$ and $(\nabla\eta)_{(uA)} = 0$ on $N_1$ and $(\nabla\eta)_{uu} = 0$ on $S$. Then, one finds an homogeneous system provided $(\nabla\eta)_{vv} = 0$, $(\nabla\eta)_{(vA)} = 0$, $((\nabla\eta)_{(AB)})^{tf} = 0$ and $S_{vvu} = 0$ hold on $N_1$ (and analogous conditions on $N_2$). This discussion can be summarized in the following theorem [74].

**Theorem 4.2.** *Consider a vacuum $(\mathfrak{n}+1)$-dimensional spacetime $(M, g)$ with an adapted null coordinate system $\{u, v, x^A\}$, $A = 3, ..., \mathfrak{n}+1$, where $N_1 = \{u = 0\}$ and $N_2 = \{v = 0\}$ are two smooth null hypersurfaces that intersect along a smooth $(\mathfrak{n}-1)$-dimensional surface $S = \{u = v = 0\}$. Let $\eta|_{N_1\cup N_2}$ be a continuous vector field on $N_1 \cup N_2$ with smooth restrictions on $N_1$ and $N_2$ satisfying*

$$\begin{aligned} &(\nabla\eta)_{vv} = 0, \quad (\nabla\eta)_{(vA)} = 0, \quad ((\nabla\eta)_{(AB)})^{tf} = 0, \quad S_{vvu} = 0 \quad \text{on } N_1, \\ &(\nabla\eta)_{uu} = 0, \quad (\nabla\eta)_{(uA)} = 0, \quad ((\nabla\eta)_{(AB)})^{tf} = 0, \quad S_{uuv} = 0 \quad \text{on } N_2, \end{aligned} \tag{4.5}$$

*as well as, on $S$,*

$$\begin{gathered} (\nabla\eta)_{(uv)} = 0, \qquad g^{AB}(\nabla\eta)_{AB} = 0, \\ \nabla_{\partial_u}(g^{AB}(\nabla\eta)_{AB}) = \nabla_{\partial_v}(g^{AB}(\nabla\eta)_{AB}) = 0, \qquad S_{Avu} = 0. \end{gathered} \tag{4.6}$$

*Then, there exists a Killing vector $\eta$ that coincides with $\eta|_{N_1\cup N_2}$ on the initial hypersurfaces.*

This problem has also been studied in scenarios where the characteristic problem is posed on a Killing horizon and there exist more that one Killing field along the null generators [227, 229]. To the best of our knowledge, the characteristic KID problem has not yet been fully explored in the presence of matter fields. In [275, 276], the author derives general conditions ensuring that the matter fields remain invariant under the action of a Killing vector, although these conditions are not explicitly formulated in terms of initial data.

A key difference between the spacelike and null KID problems is that, in the former, the KID equations are formulated entirely in terms of the abstract data (cf. Theorem 4.1), enabling to decide whether an initial data set gives rise to a spacetime admitting a Killing vector *before* solving the Einstein equations. In the null case, however, the KID equations depend on the ambient Levi-Civita connection and Riemann tensor (cf. equations (4.5)-(4.6)), making it necessary to first solve the Einstein equations and subsequently verify whether the KID equations are satisfied or not in a very particular coordinate system.

Motivated by this issue, in this chapter we find a collection of completely general identities relating the deformation tensor $\mathcal{K}[\eta] = \pounds_\eta g$ of an arbitrary vector field $\eta$ to the associated tensor $\Sigma[\eta] = \pounds_\eta \nabla$ on an abstract hypersurface of any causal character. These identities are fully gauge and diffeomorphism covariant and can be applied regardless the ambient field equations and the properties of $\eta$. As an application, in Section 4.6 we derive the (homothetic) KID equations for characteristic initial data and show that they are at the level of the initial data for the Einstein equations (i.e. double null data, see Chapter 3). This result puts the (homothetic) KID problem for two null hypersurfaces on equal footing with the standard KID problem for spacelike hypersurfaces, as well as generalizes Theorem 4.2 to the homothetic case.

## 4.2 GENERAL IDENTITIES FOR THE DEFORMATION TENSOR

In this section we review several well-known identities that relate the deformation tensor of a given vector field with the geometry of the manifold. For further details we refer the reader to [329]. Let $(\mathcal{M}, g)$ be a semi-Riemannian manifold, $\eta \in \mathfrak{X}(\mathcal{M})$ any vector field and $\nabla$ the Levi-Civita connection. We introduce the tensors $\mathcal{K}[\eta]$ and $\Sigma[\eta]$ by $\mathcal{K}[\eta] := \pounds_\eta g$ and $\Sigma[\eta] := \pounds_\eta \nabla$, or more specifically

$$\Sigma[\eta](V, W) := \pounds_\eta \nabla_V W - \nabla_{\pounds_\eta V} W - \nabla_V \pounds_\eta W, \qquad V, W \in \mathfrak{X}(\mathcal{M}).$$

The former is called "deformation tensor" of $\eta$. The latter can be expressed in terms of $\mathcal{K}[\eta]$ by means of [329]

$$\Sigma[\eta]^\mu{}_{\nu\beta} = \frac{1}{2} \left( \nabla_\nu \mathcal{K}[\eta]^\mu{}_\beta + \nabla_\beta \mathcal{K}[\eta]^\mu{}_\nu - \nabla^\mu \mathcal{K}[\eta]_{\nu\beta} \right), \tag{4.7}$$

or lowering the index,

$$\Sigma[\eta]_{\mu\nu\beta} = \frac{1}{2} \left( \nabla_\nu \mathcal{K}[\eta]_{\mu\beta} + \nabla_\beta \mathcal{K}[\eta]_{\mu\nu} - \nabla_\mu \mathcal{K}[\eta]_{\nu\beta} \right), \tag{4.8}$$

which immediately implies the symmetries

$$2\Sigma[\eta]_{(\mu\nu)\beta} = \nabla_\beta \mathcal{K}[\eta]_{\mu\nu}, \tag{4.9}$$

$$\Sigma[\eta]_{\mu[\nu\beta]} = 0. \tag{4.10}$$

The tensor $\Sigma[\eta]$ determines the commutation between covariant and Lie derivatives. Specifically, for any $(q,p)$ tensor field $A$ it holds that [329]

$$\begin{aligned}\pounds_\eta \nabla_\beta A^{\alpha_1\cdots\alpha_q}{}_{\mu_1\cdots\mu_p} = \nabla_\beta \pounds_\eta A^{\alpha_1\cdots\alpha_q}{}_{\mu_1\cdots\mu_p} + \sum_{j=1}^{q} A^{\alpha_1\cdots\alpha_{j-1}\sigma\alpha_{j+1}\cdots\alpha_q}{}_{\mu_1\cdots\mu_p}\Sigma[\eta]^{\alpha_j}{}_{\beta\sigma} \\ - \sum_{i=1}^{p} A^{\alpha_1\cdots\alpha_q}{}_{\mu_1\cdots\mu_{i-1}\sigma\mu_{i+1}\cdots\mu_p}\Sigma[\eta]^{\sigma}{}_{\beta\mu_i}.\end{aligned} \tag{4.11}$$

The tensors $\mathcal{K}[\eta]$ and $\Sigma[\eta]$ admit the following alternative expressions [329]

$$\mathcal{K}[\eta]_{\mu\nu} = 2\nabla_{(\mu}\eta_{\nu)}, \tag{4.12}$$
$$\Sigma[\eta]_{\mu\alpha\beta} = \nabla_\alpha\nabla_\beta\eta_\mu + R_{\alpha\nu\beta\mu}\eta^\nu. \tag{4.13}$$

A key object in this chapter is the vector field $\mathfrak{Q}[\eta]$ defined by

$$\mathfrak{Q}[\eta]^\mu := g^{\alpha\beta}\Sigma[\eta]^\mu{}_{\alpha\beta} = \Box_g\eta^\mu + R^\mu{}_\nu\eta^\nu, \tag{4.14}$$

the second equality being a consequence of (4.13). By (4.12) we note that $\mathfrak{Q}[\eta]$ can also be written as

$$\mathfrak{Q}[\eta]_\nu = \nabla_\mu\mathcal{K}[\eta]^\mu{}_\nu - \frac{1}{2}\nabla_\nu \operatorname{tr}_g \mathcal{K}[\eta]. \tag{4.15}$$

The tensors $\Sigma[\eta]$, $\mathcal{K}[\eta]$ and $\mathfrak{Q}[\eta]$ are related by the following identities.

**Lemma 4.3.** *Let $(\mathcal{M},g)$ be a semi-Riemannian manifold, $\eta\in\mathfrak{X}(\mathcal{M})$ and $\nabla$ the Levi-Civita connection. Then, the following identities hold*

$$\nabla_\mu\Sigma[\eta]^\mu{}_{\beta\nu} = \pounds_\eta R_{\beta\nu} + \frac{1}{2}\nabla_\beta\nabla_\nu \operatorname{tr}_g \mathcal{K}[\eta], \tag{4.16}$$
$$\Box_g\mathcal{K}[\eta]_{\beta\nu} - \mathcal{K}[\mathfrak{Q}[\eta]]_{\beta\nu} - 2R_{\sigma(\beta}\mathcal{K}[\eta]^\sigma{}_{\nu)} + 2R_{\sigma\beta\mu\nu}\mathcal{K}[\eta]^{\sigma\mu} + 2\pounds_\eta R_{\beta\nu} = 0. \tag{4.17}$$

*Proof.* We start by computing the trace and the divergence of $\Sigma[\eta]$. Using (4.7),

$$\Sigma[\eta]^\mu{}_{\mu\beta} = \frac{1}{2}\nabla_\beta\mathcal{K}[\eta]^\mu{}_\mu = \frac{1}{2}\nabla_\beta \operatorname{tr}_g\mathcal{K}[\eta], \tag{4.18}$$

$$\begin{aligned}\nabla_\mu\Sigma[\eta]^\mu{}_{\beta\nu} &= -\frac{1}{2}\nabla_\mu\nabla^\mu\mathcal{K}_{\beta\nu} + \nabla_\mu\nabla_{(\beta}\mathcal{K}[\eta]^\mu{}_{\nu)} \\ &= -\frac{1}{2}\nabla_\mu\nabla^\mu\mathcal{K}_{\beta\nu} + \nabla_{(\beta|}\nabla_\mu\mathcal{K}[\eta]^\mu{}_{|\nu)} + R^\mu{}_{\sigma\mu(\beta}\mathcal{K}[\eta]^\sigma{}_{\nu)} - R^\sigma{}_{(\nu|\mu|\beta)}\mathcal{K}[\eta]^\mu{}_\sigma,\end{aligned} \tag{4.19}$$

the last equality being a consequence of the Ricci identity applied to $\mathcal{K}[\eta]^\mu{}_\nu$. In order to prove (4.16) we use the well-known identity [329]

$$\pounds_\eta R_{\beta\nu} = \nabla_\mu\Sigma[\eta]^\mu{}_{\beta\nu} - \nabla_\nu\Sigma[\eta]^\mu{}_{\mu\beta}, \tag{4.20}$$

which combined with (4.18) gives (4.16). To prove (4.17) we insert (4.15) into (4.19),

$$\nabla_\mu\Sigma[\eta]^\mu{}_{\beta\nu} = -\frac{1}{2}\nabla_\mu\nabla^\mu\mathcal{K}[\eta]_{\beta\nu} + \nabla_{(\beta}\mathfrak{Q}[\eta]_{\nu)} + \frac{1}{2}\nabla_\beta\nabla_\nu \operatorname{tr}_g\mathcal{K} + R_{\sigma(\beta}\mathcal{K}[\eta]^\sigma{}_{\nu)} - R_{\sigma\beta\mu\nu}\mathcal{K}[\eta]^{\sigma\mu}.$$

Using (4.16) and taking into account $\mathcal{K}[\mathfrak{Q}[\eta]]_{\beta\nu} = 2\nabla_{(\beta}\mathfrak{Q}[\eta]_{\nu)}$, (4.17) follows at once. □

For any $\mu \in \mathbb{R}$, identity (4.17) can be rewritten as

$$\Box_g(\mathcal{K}[\eta]_{\beta\nu} - \mu g_{\beta\nu}) - \mathcal{K}[\mathfrak{Q}[\eta]]_{\beta\nu} - 2R_{\sigma(\beta}(\mathcal{K}[\eta]^{\sigma}{}_{\nu)} - \mu\delta^{\sigma}_{\nu)}) \\ +2R_{\sigma\beta\mu\nu}(\mathcal{K}[\eta]^{\sigma\mu} - \mu g^{\sigma\mu}) + 2\pounds_\eta R_{\beta\nu} = 0.$$

Whenever $R_{\mu\nu} = \lambda g_{\mu\nu}$ this identity becomes

$$\Box_g(\mathcal{K}[\eta]_{\beta\nu} - \mu g_{\beta\nu}) - \mathcal{K}[\mathfrak{Q}[\eta]]_{\beta\nu} - 2R_{\sigma(\beta}(\mathcal{K}[\eta]^{\sigma}{}_{\nu)} - \mu\delta^{\sigma}_{\nu)}) \\ +2R_{\sigma\beta\mu\nu}(\mathcal{K}[\eta]^{\sigma\mu} - \mu g^{\sigma\mu}) + 2\lambda(\mathcal{K}_{\beta\nu} - \mu g_{\beta\nu}) + 2\lambda\mu g_{\beta\nu} = 0. \tag{4.21}$$

When the vector $\eta$ satisfies $\mathfrak{Q}[\eta] = 0$ this becomes a homogeneous PDE for $\mathcal{K}[\eta]_{\beta\nu} - \mu g_{\beta\nu}$ provided $\mu\lambda = 0$. This gives a propagation equation that can be exploited to ensure that a vector field is a (proper) homothety in a Ricci flat manifold, or a Killing vector in a $\lambda$-vacuum manifold. Note that a homothety $\eta$ satisfying $\pounds_\eta g = \mu g$ with $\mu \neq 0$ cannot exist in an Einstein space with $\lambda \neq 0$, because homotheties satisfy $\Sigma[\eta] = 0$ and hence $0 = \pounds_\eta \operatorname{Ric} = \lambda\mu g$.

To make the computations in the following section as tractable as possible we shall make use of the following property of the tensor $\Sigma[\eta]$.

**Lemma 4.4.** *Let $(\mathcal{M}, g)$ be a semi-Riemannian manifold, $\eta, \zeta \in \mathfrak{X}(\mathcal{M})$ arbitrary vector fields, $\mathcal{K}[\eta] := \pounds_\eta g$, $\Sigma[\eta] := \pounds_\eta \nabla$ and define $\boldsymbol{\zeta} := g(\zeta, \cdot)$. Then, the following two identities hold*

$$\zeta^\alpha \Sigma[\eta]_{\alpha\beta\mu} = \nabla_{(\beta}(\pounds_\eta \boldsymbol{\zeta})_{\mu)} - \frac{1}{2}\pounds_\eta \pounds_\zeta g_{\beta\mu}, \tag{4.22}$$

$$\zeta^\alpha \Sigma[\eta]_{\alpha\beta\mu} = \frac{1}{2}\pounds_{k^{(\zeta)}} g_{\beta\mu} - \frac{1}{2}\pounds_\zeta \mathcal{K}[\eta]_{\beta\mu}, \tag{4.23}$$

*where $(k^{(\zeta)})^\mu := \mathcal{K}[\eta]^\mu{}_\beta \zeta^\beta$.*

*Proof.* From (4.11) applied to $A = \boldsymbol{\zeta}$ it follows

$$\pounds_\eta \nabla_\beta \boldsymbol{\zeta}_\mu = \nabla_\beta(\pounds_\eta \boldsymbol{\zeta})_\mu - \Sigma[\eta]^\alpha{}_{\beta\mu}\zeta_\alpha \qquad \Longrightarrow \qquad (\pounds_\eta \pounds_\zeta g)_{\beta\mu} = 2\nabla_{(\beta}(\pounds_\eta \boldsymbol{\zeta})_{\mu)} - 2\Sigma[\eta]^\alpha{}_{\beta\mu}\zeta_\alpha,$$

which proves (4.22). The second identity follows from the first after commuting $\pounds_\eta \pounds_\zeta = \pounds_\zeta \pounds_\eta + \pounds_{[\eta,\zeta]}$ and using

$$\nabla_{(\beta}\pounds_\eta (g_{\mu)\alpha}\zeta^\alpha) = \nabla_{(\beta}(\mathcal{K}[\eta]_{\mu)\alpha}\zeta^\alpha) + \nabla_{(\beta}([\eta,\zeta])_{\mu)} = \frac{1}{2}\pounds_{k^{(\zeta)}} g_{\beta\mu} + \frac{1}{2}\pounds_{[\eta,\zeta]} g_{\beta\mu}.$$

□

## 4.3 IDENTITIES ON ABSTRACT HYPERSURFACES

In this section we analyze the consequences of the identities in Section 4.2 on embedded hypersurface data $\{\mathcal{H}, \boldsymbol{\gamma}, \boldsymbol{\ell}, \ell^{(2)}, \mathbf{Y}\}$ in order to look for general conditions that any candidate of homothetic vector needs to satisfy at $\mathcal{H}$. The computations rely on general expressions for the pullback of ambient tensor fields into an arbitrary hypersurface. They are obtained in Appendix B, where the following notation is introduced.

**Notation 4.5.** *Given a* $(0,p)$ *tensor field* $T_{\alpha_1\cdots\alpha_p}$ *on* $\mathcal{M}$ *we use the standard notation* $T_{a_1\cdots a_p}$ *to denote its pullback to* $\mathcal{H}$. *Moreover, we introduce the notation* ${}^{(i)}T_{\alpha_1\cdots\alpha_{p-1}}$ *for* $\xi^\mu T_{\alpha_1\cdots\alpha_{i-1}\mu\alpha_i\cdots\alpha_{p-1}}$ *and* ${}^{(i)}T_{a_1\cdots a_{p-1}}$ *for the pullback of* ${}^{(i)}T_{\alpha_1\cdots\alpha_{p-1}}$ *to* $\mathcal{H}$. *In addition, we use* ${}^{(i,j)}T_{a_1\cdots a_{p-2}}$ *for the pullback to* $\mathcal{H}$ *of the tensor obtained by first the contraction of* $T$ *with* $\xi$ *in the* $j$*-th slot and then in the* $i$*-th slot of the resulting* $(0,p-1)$ *tensor, i.e.* ${}^{(i,j)}T = {}^{(i)}({}^{(j)}T)$. *This notation requires care with the order of indices (note that* ${}^{(1,1)}T = {}^{(1,2)}T$ *and* ${}^{(1,3)}T = {}^{(2,1)}T \neq {}^{(1,2)}T$*).*

From now on and for simplicity we drop the label "$[\eta]$" in the tensors $\mathcal{K}$, $\Sigma$ and $\mathfrak{Q}$. The structure of this section is as follows. First we compute the fully tangential components of $\mathcal{K}$ in terms of hypersurface data and $\eta|_{\mathcal{H}}$. Then, we relate the transverse-tangent and transverse-transverse components of $\mathcal{K}|_{\mathcal{H}}$ with the first transverse derivative of $\eta$ at $\mathcal{H}$. We then move on to the tensor $\Sigma$ and write down its fully tangential components as well as the pullbacks of $\Sigma$ contracted with one or several transverse vectors $\xi$. We conclude the section by expressing the tangential and transversal components of $\mathfrak{Q}$ in terms of objects defined on the hypersurface.

Given $\eta \in \mathfrak{X}(\mathcal{M})$ we define the vector $\bar{\eta} \in \mathfrak{X}(\mathcal{H})$ and the scalar $C \in \mathcal{F}(\mathcal{H})$ by means of $\eta|_{\Phi(\mathcal{H})} = C\xi + \Phi_\star\bar{\eta}$, where $\xi$ is the rigging of $\Phi(\mathcal{H})$ and $\Phi$ the embedding of $\mathcal{H}$ into $\mathcal{M}$. Using $\boldsymbol{\eta}_a$ to denote the pullback on $\mathcal{H}$ of $\boldsymbol{\eta} := g(\eta,\cdot)$ one easily has

$$\boldsymbol{\eta}_a = C\ell_a + \gamma_{ab}\bar{\eta}^b, \qquad \boldsymbol{\eta}(\xi) = C\ell^{(2)} + \boldsymbol{\ell}(\bar{\eta}). \tag{4.24}$$

In the next proposition we compute the fully tangent components of the tensor $\mathcal{K}$ and we show that they can be written solely in terms of hypersurface data and the pair $(C,\bar{\eta})$.

**Proposition 4.6.** *Let* $\{\mathcal{H},\boldsymbol{\gamma},\boldsymbol{\ell},\ell^{(2)},\mathbf{Y}\}$ *be hypersurface data* $(\Phi,\xi)$*-embedded in* $(\mathcal{M},g)$*,* $\eta \in \mathfrak{X}(\mathcal{M})$ *and* $\mathcal{K} := \pounds_\eta g$. *Define* $\eta \overset{\Phi(\mathcal{H})}{=} C\xi + \Phi_\star\bar{\eta}$. *Then,*

$$\mathcal{K}_{ab} = 2C\mathrm{Y}_{ab} + 2\ell_{(a}\overset{\circ}{\nabla}_{b)}C + \pounds_{\bar{\eta}}\gamma_{ab}. \tag{4.25}$$

*Proof.* Using Proposition B.5 with $T = g$ and $\zeta = \eta$ and taking into account $\mathcal{K} = \pounds_\eta g$, $\Phi^\star\pounds_\xi g = 2\mathbf{Y}$ and $\Phi^\star g = \boldsymbol{\gamma}$ (see Def. 2.2) the result follows at once. □

**Remark 4.7.** *When* $C = 1$ *and* $\bar{\eta} = 0$ *(i.e. when* $\eta|_{\mathcal{H}} = \xi$*) one gets* $\mathcal{K}_{ab} = 2\mathrm{Y}_{ab}$*, as required from the definition of embedded data regarding the tensor* $\mathbf{Y}$*. Similarly, when* $C = n^{(2)}$ *and* $\bar{\eta} = n$ *(and thus* $\eta|_{\mathcal{H}} = \nu$*) we obtain* $\mathcal{K}_{ab} = 2n^{(2)}\mathrm{Y}_{ab} + 2\mathrm{U}_{ab} = 2\mathrm{K}_{ab}$*, as required from the geometric interpretation of* $\mathbf{K}$ *as the second fundamental form of* $\Phi(\mathcal{H})$ *along the normal* $\nu$*.*

Unlike $\mathcal{K}$, the fully tangential components of $\Sigma$ cannot be written solely in terms of hypersurface data and $(C,\bar{\eta})$. Indeed, after contracting (4.8) (or (4.13)) with three tangent vectors and using Proposition B.1 one cannot avoid the appearance of the transverse components of the deformation tensor $\mathcal{K}$ that we introduce next.

**Notation 4.8.** *Given $\mathcal{K} := \pounds_\eta g$ and an embedded hypersurface $\Phi : \mathcal{H} \hookrightarrow \mathcal{M}$ with rigging $\xi$, we introduce*[1] *the one-form $\text{ש} := \Phi^\star(\mathcal{K}(\xi,\cdot))$ and the scalar $\text{ק} := \Phi^\star(\mathcal{K}(\xi,\xi))$.*

Up to terms involving hypersurface data and $(C,\bar{\eta})$, the tensors $\text{ש}_a$, ק are in one-to-one correspondence with the transverse derivative of the one-form $\boldsymbol{\eta}$. Indeed,

$$
\begin{aligned}
\text{ש}_a = 2e^\alpha_a\xi^\beta\nabla_{(\alpha}\boldsymbol{\eta}_{\beta)} &= e^\alpha_a\xi^\beta\nabla_\alpha\boldsymbol{\eta}_\beta + e^\alpha_a\xi^\beta\nabla_\beta\boldsymbol{\eta}_\alpha \\
&= \overset{\circ}{\nabla}_a(\boldsymbol{\eta}(\xi)) - 2e^\alpha_a\boldsymbol{\eta}_\beta\nabla_\alpha\xi^\beta + e^\alpha_a(\pounds_\xi\boldsymbol{\eta})_\alpha \\
&\overset{(2.53)}{=} \overset{\circ}{\nabla}_a(\boldsymbol{\eta}(\xi)) - 2\left(\mathrm{r}_a - \mathrm{s}_a + \frac{1}{2}n^{(2)}\overset{\circ}{\nabla}_a\ell^{(2)}\right)\boldsymbol{\eta}(\xi) - 2V^b{}_a\eta_b + (\pounds_\xi\boldsymbol{\eta})_a \\
&= \ell^{(2)}\overset{\circ}{\nabla}_a C + \overset{\circ}{\nabla}_a(\boldsymbol{\ell}(\bar{\eta})) + (\pounds_\xi\boldsymbol{\eta})_a - 2(\mathrm{Y}_{ac} + \mathrm{F}_{ac})\bar{\eta}^c,
\end{aligned}
\tag{4.26}
$$

where in the last line we inserted (4.24) and simplified after using (2.55)-(2.56). Similarly,

$$
\text{ק} = 2\xi^\alpha\xi^\beta\nabla_\alpha\boldsymbol{\eta}_\beta = 2\xi^\beta\left(\pounds_\xi\boldsymbol{\eta}_\beta - \eta_\alpha\nabla_\beta\xi^\alpha\right) = 2(\pounds_\xi\boldsymbol{\eta})(\xi) - 2g(\eta,\nabla_\xi\xi). \tag{4.27}
$$

Next we write down the pullback of the tensor $\Sigma$ in terms of hypersurface data and the values of $\mathcal{K}$ on the hypersurface.

**Proposition 4.9.** *Let $\{\mathcal{H},\boldsymbol{\gamma},\boldsymbol{\ell},\ell^{(2)},\mathbf{Y}\}$ be hypersurface data $(\Phi,\xi)$-embedded in $(\mathcal{M},g)$, $\eta \in \mathfrak{X}(\mathcal{M})$, $\mathcal{K} := \pounds_\eta g$ and $\Sigma := \pounds_\eta\nabla$. Then,*

$$
\Sigma_{abc} = \overset{\circ}{\nabla}_{(b}\mathcal{K}_{c)a} - \frac{1}{2}\overset{\circ}{\nabla}_a\mathcal{K}_{bc} + \mathrm{Y}_{bc}\mathcal{K}_{ad}n^d + \left(\mathrm{U}_{bc} + n^{(2)}\mathrm{Y}_{bc}\right)\text{ש}_a. \tag{4.28}
$$

*Proof.* Contracting (4.8) with three tangent vectors and applying (B.1) in Proposition B.1 to $T = \mathcal{K}$, (4.28) follows at once. □

**Remark 4.10.** *We observe that if $\mathcal{K} = \mu\gamma$ with $\mu \in \mathbb{R}$ then $\Sigma_{abc}$ vanishes identically at the points where $\text{ש} = \mu\boldsymbol{\ell}$. To show this we insert* (2.39) *and* (2.3) *into* (4.28) *and get*

$$
\Sigma_{abc} = -\mu\ell_a\mathrm{U}_{bc} - \mu n^{(2)}\ell_a\mathrm{Y}_{bc} + \mu\ell_a\left(\mathrm{U}_{bc} + n^{(2)}\mathrm{Y}_{bc}\right) = 0.
$$

Another key object in this chapter is the pullback tensor $\Phi^\star(\Sigma(\xi,\cdot,\cdot)) =: \mathfrak{S}$, i.e. the one-transverse–two-tangent components of $\Sigma$. This quantity is important because it arises in several other expressions involving the tensor $\nabla\mathcal{K}$ contracted with one transversal direction and two tangential ones. Let us first find these relationships. We start with $\Phi^\star({}^{(2)}\nabla\mathcal{K})$, for which we use (B.3) in Appendix B applied to $T = \mathcal{K}$ and $j = 1$ to get

$$
({}^{(2)}\nabla\mathcal{K})_{ab} = \overset{\circ}{\nabla}_a\text{ש}_b + \text{ש}(n)\mathrm{Y}_{ab} + \text{ק}\left(\mathrm{U}_{ab} + n^{(2)}\mathrm{Y}_{ab}\right) - \left(\mathrm{r}_a - \mathrm{s}_a + \frac{1}{2}n^{(2)}\overset{\circ}{\nabla}_a\ell^{(2)}\right)\text{ש}_b - V^c{}_a\mathcal{K}_{bc}. \tag{4.29}
$$

1 ש and ק are pronounced respectively "shin" and "kof", and are the 21st and 19th letters in the Hebrew alphabet.

Concerning $\Phi^\star({}^{(1)}\nabla\mathcal{K})$ we use (4.8), namely $\mathfrak{S}_{ab} = \xi^\mu e^\nu_a e^\beta_b \Sigma_{\mu\nu\beta} = ({}^{(2)}\nabla\mathcal{K})_{(ab)} - \frac{1}{2}({}^{(1)}\nabla\mathcal{K})_{ab}$, which combined with (4.29) gives

$$({}^{(1)}\nabla\mathcal{K})_{ab} = 2\overset{\circ}{\nabla}_{(a}\varpi_{b)} + 2\varpi(n)\mathrm{Y}_{ab} + 2\daleth\Big(\mathrm{U}_{ab} + n^{(2)}\mathrm{Y}_{ab}\Big) - 2\left(\mathrm{r}_{(a} - \mathrm{s}_{(a} + \frac{1}{2}n^{(2)}\overset{\circ}{\nabla}_{(a}\ell^{(2)}\right)\varpi_{b)}$$
$$- 2V^c{}_{(a}\mathcal{K}_{b)c} - 2\mathfrak{S}_{ab}. \tag{4.30}$$

For later use we shall need the contraction of $({}^{(2)}\nabla\mathcal{K})_{ab}$ and $({}^{(1)}\nabla\mathcal{K})_{ab}$ with $n^b$ in the null case. The former follows after using (2.44) in Lemma 2.5 with $\boldsymbol{\theta} = \varpi$ and $\mathrm{U}_{ab}n^b = 0$,

$$({}^{(2)}\nabla\mathcal{K})_{ab}n^b = \overset{\circ}{\nabla}_a(\varpi(n)) - P^{bc}\mathrm{U}_{ac}\varpi_b - V^c{}_a\mathcal{K}_{bc}n^b, \qquad (n^{(2)} = 0) \tag{4.31}$$

while the latter follows from (2.46) also in Lemma 2.5 with $\boldsymbol{\theta} = \varpi$,

$$({}^{(1)}\nabla\mathcal{K})_{ab}n^b = \pounds_n\varpi_a + \overset{\circ}{\nabla}_a(\varpi(n)) + \varpi(n)(\mathrm{r}_a - \mathrm{s}_a) - 2P^{bc}\mathrm{U}_{ac}\varpi_b + \kappa\varpi_a$$
$$- V^c{}_b n^b\mathcal{K}_{ac} - V^c{}_a\mathcal{K}_{bc}n^b - 2\mathfrak{S}_{ab}n^b. \qquad (n^{(2)} = 0) \tag{4.32}$$

It is also useful to write down the pullback of the Lie derivative of $\mathcal{K}$ along $\xi$. Lemma 4.4 [Eq. (4.23)] with $\zeta = \xi$ gives $\pounds_\xi\mathcal{K}_{\alpha\beta} = 2\nabla_{(\alpha}k^{(\xi)}_{\beta)} - 2\xi^\mu\Sigma_{\mu\alpha\beta}$. Pulling this back to $\mathcal{H}$ and using (B.1) in Proposition B.1 applied to $T_\mu = \mathcal{K}_{\mu\nu}\xi^\nu$ we arrive at

$$(\pounds_\xi\mathcal{K})_{ab} = 2\overset{\circ}{\nabla}_{(a}\varpi_{b)} + 2\varpi(n)\mathrm{Y}_{ab} + 2\daleth\Big(\mathrm{U}_{ab} + n^{(2)}\mathrm{Y}_{ab}\Big) - 2\mathfrak{S}_{ab}. \tag{4.33}$$

**Remark 4.11.** *We note that if $\varpi = \mu\boldsymbol{\ell}$, $\daleth = \mu\ell^{(2)}$ and $(\pounds_\xi\mathcal{K})_{ab} = 2\mu\mathrm{Y}_{ab}$, then $\mathfrak{S}_{ab}$ vanishes identically. The computation relies on* (4.33) *and uses* (2.4) *and* (2.39) *as follows*

$$2\mathfrak{S}_{ab} = 2\mu\overset{\circ}{\nabla}_{(a}\ell_{b)} + 2\mu\boldsymbol{\ell}(n)\mathrm{Y}_{ab} + 2\mu\ell^{(2)}\Big(\mathrm{U}_{ab} + n^{(2)}\mathrm{Y}_{ab}\Big) - 2\mu\mathrm{Y}_{ab}$$
$$= -2\mu\ell^{(2)}\mathrm{U}_{ab} + 2\mu(1 - n^{(2)}\ell^{(2)})\mathrm{Y}_{ab} + 2\mu\ell^{(2)}\Big(\mathrm{U}_{ab} + n^{(2)}\mathrm{Y}_{ab}\Big) - 2\mu\mathrm{Y}_{ab} = 0.$$

Note that (4.29) together with (4.9)-(4.10) provide all the contractions of $\Sigma_{\alpha\beta\mu}$ with one rigging and two tangential directions (in any order) in terms of hypersurface data and $\mathfrak{S}_{ab}$. Specifically, one has

$${}^{(2)}\Sigma_{ab} = {}^{(3)}\Sigma_{ab} = -\mathfrak{S}_{ab} + ({}^{(2)}\nabla\mathcal{K})_{ba}$$
$$= -\mathfrak{S}_{ab} + \overset{\circ}{\nabla}_b\varpi_a + \varpi(n)\mathrm{Y}_{ba} + \daleth\Big(\mathrm{U}_{ba} + n^{(2)}\mathrm{Y}_{ba}\Big)$$
$$- \left(\mathrm{r}_b - \mathrm{s}_b + \frac{1}{2}n^{(2)}\overset{\circ}{\nabla}_b\ell^{(2)}\right)\varpi_a - V^c{}_b\mathcal{K}_{ac}. \tag{4.34}$$

In particular, after using (2.21), (2.57) as well as (2.44)-(2.45) in Lemma 2.5 for $\boldsymbol{\theta} = \varpi$ we find that the contractions of ${}^{(2)}\Sigma_{ab}$ with $n^a$ and $n^b$ are (note that ${}^{(2)}\Sigma_{ab}$ is not symmetric)

$${}^{(2)}\Sigma_{ab}n^a = -\mathfrak{S}_{ab}n^a + \overset{\circ}{\nabla}_b(\varpi(n)) - P^{ac}\mathrm{U}_{bc}\varpi_a - V^c{}_b\mathcal{K}_{ac}n^a + \frac{1}{2}\daleth\overset{\circ}{\nabla}_b n^{(2)}$$
$$+ n^{(2)}\left(\frac{1}{2}(\varpi(n) + \daleth)\overset{\circ}{\nabla}_b\ell^{(2)} + P^{ac}\mathrm{F}_{bc}\varpi_a + \daleth(\mathrm{r}_b - \mathrm{s}_b)\right) \tag{4.35}$$

and

$$\begin{aligned}
{}^{(2)}\Sigma_{ab}n^b &= -\mathfrak{S}_{ab}n^b + \pounds_n \boldsymbol{\varpi}_a - P^{bc}\mathrm{U}_{ac}\boldsymbol{\varpi}_b + \boldsymbol{\varpi}(n)(\mathrm{r}_a - \mathrm{s}_a) + \left(\kappa - \frac{1}{2}n^{(2)}n(\ell^{(2)})\right)\boldsymbol{\varpi}_a \\
&\quad - \left(P^{cb}(\mathrm{r}_b + \mathrm{s}_b) + \frac{1}{2}n(\ell^{(2)})n^c\right)\mathcal{K}_{ac} + \frac{1}{2}\mathfrak{p}\overset{\circ}{\nabla}_a n^{(2)} \\
&\quad + n^{(2)}\left(\boldsymbol{\varpi}(n)\overset{\circ}{\nabla}_a\ell^{(2)} + P^{bc}\mathrm{F}_{ac}\boldsymbol{\varpi}_b + \mathfrak{p}\left(\mathrm{r}_a - \mathrm{s}_a + \frac{1}{2}n^{(2)}\overset{\circ}{\nabla}_a\ell^{(2)}\right)\right).
\end{aligned} \tag{4.36}$$

We now take on the task of obtaining the tensor $\mathfrak{S}_{ab}$ in terms of extended hypersurface data (as defined in Section 2.3) and the pair $(C, \bar{\eta})$.

**Proposition 4.12.** *Let* $\{\mathcal{H}, \boldsymbol{\gamma}, \boldsymbol{\ell}, \ell^{(2)}, \mathbf{Y}, \mathbf{Z}^{(2)}\}$ *be extended hypersurface data* $(\Phi, \xi)$*-embedded in* $(\mathcal{M}, g)$ *and* $\eta \in \mathfrak{X}(\mathcal{M})$*. Decomposing* $\eta \overset{\Phi(\mathcal{H})}{=} C\xi + \Phi_\star\bar{\eta}$*, then*

$$\begin{aligned}
\mathfrak{S}_{ab} &= -C\mathrm{Z}^{(2)}_{ab} - \pounds_{\bar{\eta}}\mathrm{Y}_{ab} + \left(\frac{1}{2}Cn(\ell^{(2)}) + \ell^{(2)}n(C) + (\pounds_{\bar{\eta}}\boldsymbol{\ell})(n)\right)\mathrm{Y}_{ab} + \overset{\circ}{\nabla}_{(a}\ell^{(2)}\overset{\circ}{\nabla}_{b)}C \\
&\quad + \frac{1}{2}\big(\bar{\eta}(\ell^{(2)}) + \mathfrak{p}\big)\big(\mathrm{U}_{ab} + n^{(2)}\mathrm{Y}_{ab}\big) + \frac{1}{2}C\overset{\circ}{\nabla}_a\overset{\circ}{\nabla}_b\ell^{(2)} + \ell^{(2)}\overset{\circ}{\nabla}_a\overset{\circ}{\nabla}_b C + \overset{\circ}{\nabla}_{(a}\pounds_{\bar{\eta}}\ell_{b)}.
\end{aligned} \tag{4.37}$$

*Proof.* From Lemma 4.4 [Eq. (4.22)] with $\zeta = \xi$ we can write

$$\mathfrak{S}_{ab} = e_a^\alpha e_b^\beta \xi^\mu \Sigma_{\mu\alpha\beta} = e_a^\alpha e_b^\beta \nabla_{(\alpha}(\pounds_\eta \boldsymbol{\xi})_{\beta)} - \frac{1}{2}e_a^\alpha e_b^\beta\Big(\pounds_\eta\pounds_\xi g\Big)_{\alpha\beta},$$

which after using Proposition B.1 [Eq. (B.1)] applied to $T = \boldsymbol{\xi}$ in the first term gives

$$\mathfrak{S}_{ab} = \overset{\circ}{\nabla}_{(a}(\pounds_\eta\boldsymbol{\xi})_{b)} + \mathrm{Y}_{ab}n^c(\pounds_\eta\boldsymbol{\xi})_c + \mathrm{K}_{ab}\xi^\mu(\pounds_\eta\boldsymbol{\xi})_\mu - \frac{1}{2}e_a^\alpha e_b^\beta(\pounds_\eta\pounds_\xi g)_{\alpha\beta}. \tag{4.38}$$

Hence to compute the first three terms it suffices to calculate $\Phi^\star\pounds_\eta\boldsymbol{\xi}$ and $(\pounds_\eta\boldsymbol{\xi})(\xi)$. Since the left-hand side of (4.38) does not depend on the extension of $\xi$ off $\Phi(\mathcal{H})$ we may assume without loss of generality $\nabla_\xi\xi = 0$ to simplify the computation. Applying (B.12) in Proposition B.5 with $\zeta = \eta$ and $T = \boldsymbol{\xi}$ together with

$$\Phi^\star\pounds_\xi\boldsymbol{\xi} = \Phi^\star\left(\nabla_\xi\boldsymbol{\xi} + \frac{1}{2}d(g(\xi,\xi))\right) = \frac{1}{2}d\ell^{(2)} \qquad (\nabla_\xi\xi = 0) \tag{4.39}$$

gives

$$(\pounds_\eta\boldsymbol{\xi})_b = \frac{1}{2}C\overset{\circ}{\nabla}_b\ell^{(2)} + \ell^{(2)}\overset{\circ}{\nabla}_b C + (\pounds_{\bar{\eta}}\boldsymbol{\ell})_b. \qquad (\nabla_\xi\xi = 0) \tag{4.40}$$

In order to compute $(\pounds_\eta\boldsymbol{\xi})(\xi)$ note that

$$(\pounds_\eta\boldsymbol{\xi})(\xi) = \Big(\eta^\mu\nabla_\mu\xi_\alpha + \xi_\mu\nabla_\alpha\eta^\mu\Big)\xi^\alpha = \frac{1}{2}\Big(\eta^\mu\nabla_\mu(\xi_\alpha\xi^\alpha) + \xi^\mu\xi^\alpha(\pounds_\eta g)_{\mu\alpha}\Big) \overset{\mathcal{H}}{=} \frac{1}{2}(\bar{\eta}(\ell^{(2)}) + \mathfrak{p}), \tag{4.41}$$

where in the last equality we used

$$\xi^\mu\nabla_\mu(\xi_\alpha\xi^\alpha) = \pounds_\xi(g(\xi,\xi)) = 2g(\nabla_\xi\xi,\xi) = 0. \qquad (\nabla_\xi\xi = 0). \tag{4.42}$$

For the last term in (4.38) we apply again Proposition B.5 [Eq. (B.12)] with $\zeta = \eta$ and $T = \pounds_\xi g$. Taking into account that now $\mathbf{Z}^{(2)} = \frac{1}{2}\Phi^\star(\pounds_\xi^{(2)} g)$ (cf. (2.113)) together with ${}^{(1)}T = \pounds_\xi \boldsymbol{\xi}$ and (4.39) it follows

$$e_a^\alpha e_b^\beta (\pounds_\eta \pounds_\xi g)_{\alpha\beta} = 2C\mathrm{Z}^{(2)}_{ab} + \overset{\circ}{\nabla}_{(a}\ell^{(2)}\overset{\circ}{\nabla}_{b)}C + 2\pounds_{\bar\eta}\mathrm{Y}_{ab}. \tag{4.43}$$

Inserting (4.40), (4.41) and (4.43) into (4.38) and using (2.16) the result is established. □

Having understood the contraction of $\nabla\mathcal{K}$ and $\Sigma$ with one transversal and two tangential directions, we proceed with the computation of the contraction of $\Sigma$ with more than one $\xi$. From (4.8),

$$\begin{aligned}
\Sigma_{\mu\alpha\beta}\xi^\alpha\xi^\beta &= \xi^\alpha\xi^\beta\nabla_\alpha\mathcal{K}_{\mu\beta} - \frac{1}{2}\xi^\alpha\xi^\beta\nabla_\mu\mathcal{K}_{\alpha\beta}\\
&= \xi^\beta\Big(\pounds_\xi\mathcal{K}_{\mu\beta} - \mathcal{K}_{\mu\rho}\nabla_\beta\xi^\rho - \mathcal{K}_{\rho\beta}\nabla_\mu\xi^\rho\Big) - \frac{1}{2}\nabla_\mu(\mathcal{K}(\xi,\xi)) + \mathcal{K}_{\alpha\beta}\xi^\alpha\nabla_\mu\xi^\beta\\
&= \xi^\beta\pounds_\xi\mathcal{K}_{\mu\beta} - \mathcal{K}_{\mu\rho}\xi^\beta\nabla_\beta\xi^\rho - \frac{1}{2}\nabla_\mu(\mathcal{K}(\xi,\xi)),
\end{aligned} \tag{4.44}$$

and as consequence,

$$\Sigma_{\mu\alpha\beta}\xi^\mu\xi^\alpha\xi^\beta = \frac{1}{2}\xi^\mu\xi^\beta\pounds_\xi\mathcal{K}_{\mu\beta} - \mathcal{K}_{\mu\rho}\xi^\mu\xi^\beta\nabla_\beta\xi^\rho. \tag{4.45}$$

Similarly from (4.10) and (4.8),

$$\Sigma_{\mu\alpha\beta}\xi^\mu\xi^\alpha = \Sigma_{\mu\beta\alpha}\xi^\mu\xi^\alpha = \frac{1}{2}\xi^\mu\xi^\alpha\nabla_\beta\mathcal{K}_{\mu\alpha} = \frac{1}{2}\nabla_\beta(\mathcal{K}(\xi,\xi)) - \mathcal{K}_{\rho\mu}\xi^\mu\nabla_\beta\xi^\rho. \tag{4.46}$$

Defining $\mathfrak{w}^{(2)} := \Phi^\star(\pounds_\xi\mathcal{K}(\xi,\cdot))$, $\mathfrak{p}^{(2)} := (\pounds_\xi\mathcal{K})(\xi,\xi)$ and $\beta, a_\parallel$ by $\nabla_\xi\xi \stackrel{\mathcal{H}}{=} \beta\xi + \Phi_\star a_\parallel$ it follows that the pullbacks of (4.44)-(4.46) are

$$({}^{(2,3)}\Sigma)_c = \mathfrak{w}^{(2)}_c - \beta\mathfrak{w}_c - a_\parallel^b\mathcal{K}_{bc} - \frac{1}{2}\overset{\circ}{\nabla}_c\mathfrak{p}, \tag{4.47}$$

$${}^{(1,2,3)}\Sigma = \frac{1}{2}\mathfrak{p}^{(2)} - \beta\mathfrak{p} - a_\parallel^b\mathfrak{w}_b, \tag{4.48}$$

$$({}^{(1,2)}\Sigma)_c = ({}^{(1,3)}\Sigma)_c = \frac{1}{2}\overset{\circ}{\nabla}_c\mathfrak{p} - \left(\mathrm{r}_c - \mathrm{s}_c + \frac{1}{2}n^{(2)}\overset{\circ}{\nabla}_c\ell^{(2)}\right)\mathfrak{p} - \mathfrak{w}_a V^a{}_c, \tag{4.49}$$

where in the last one we used (2.53). We shall also need the contraction of the first and third ones with $n^c$. The first is immediate, and for the second one we use (2.57). The result is

$$({}^{(2,3)}\Sigma)_c n^c = \mathfrak{w}^{(2)}(n) - \beta\mathfrak{w}(n) - a_\parallel^b\mathcal{K}_{bc}n^c - \frac{1}{2}n(\mathfrak{p}), \tag{4.50}$$

$$({}^{(1,2)}\Sigma)_c n^c = ({}^{(1,3)}\Sigma)_c n^c = \frac{1}{2}n(\mathfrak{p}) + \left(\kappa - \frac{1}{2}n^{(2)}n(\ell^{(2)})\right)\mathfrak{p} - P^{ab}(\mathrm{r}+\mathrm{s})_a\mathfrak{w}_b - \frac{1}{2}n(\ell^{(2)})\mathfrak{w}(n). \tag{4.51}$$

Observe that, from $\nu = \Phi_\star n + n^{(2)}\xi$, these expressions imply

$$\Sigma(\nu,\xi,\xi) \stackrel{\mathcal{H}}{=} \mathfrak{w}^{(2)}(n) - \beta\mathfrak{w}(n) - a_{\|}^{b}\mathcal{K}_{bc}n^{c} - \frac{1}{2}n(\mathfrak{p}) + n^{(2)}\left(\frac{1}{2}\mathfrak{p}^{(2)} - \beta\mathfrak{p} - a_{\|}^{b}\mathfrak{w}_b\right), \tag{4.52}$$

$$\begin{aligned}\Sigma(\xi,\nu,\xi) = \Sigma(\xi,\xi,\nu) \stackrel{\mathcal{H}}{=} \frac{1}{2}n(\mathfrak{p}) + \left(\kappa - \frac{1}{2}n^{(2)}n(\ell^{(2)})\right)\mathfrak{p} - P^{ab}(\mathrm{r}+\mathrm{s})_a\mathfrak{w}_b \\ - \frac{1}{2}n(\ell^{(2)})\mathfrak{w}(n) + n^{(2)}\left(\frac{1}{2}\mathfrak{p}^{(2)} - \beta\mathfrak{p} - a_{\|}^{b}\mathfrak{w}_b\right).\end{aligned} \tag{4.53}$$

We conclude this section by computing the tangential and transverse components of $\mathfrak{Q}_\mu$ on the hypersurface. The result will play a crucial role both in the non-null and in the null cases considered below.

**Proposition 4.13.** *Let $\{\mathcal{H},\boldsymbol{\gamma},\boldsymbol{\ell},\ell^{(2)},\mathbf{Y}\}$ be hypersurface data $(\Phi,\xi)$-embedded in $(\mathcal{M},g)$, $\eta\in\mathfrak{X}(\mathcal{M})$, $\mathcal{K} := \pounds_\eta g$, $\Sigma := \pounds_\eta\nabla$ and let $\mathfrak{Q}_c := e_c^\mu\mathfrak{Q}_\mu$ and $\mathfrak{q} := \xi^\mu\mathfrak{Q}_\mu$, where $\mathfrak{Q}$ is defined in* (4.14)*. Then,*

$$\begin{aligned}\mathfrak{Q}_c = {} & -2\mathfrak{S}_{ca}n^a + 2\pounds_n\mathfrak{w}_c - 2P^{ab}\mathrm{U}_{bc}\mathfrak{w}_a + (2\kappa + \operatorname{tr}_P\mathbf{U})\mathfrak{w}_c + 2\mathfrak{w}(n)\Big(\mathrm{r}_c - \mathrm{s}_c + n^{(2)}\overset{\circ}{\nabla}_c\ell^{(2)}\Big) \\ & + P^{ab}\overset{\circ}{\nabla}_a\mathcal{K}_{bc} - \frac{1}{2}P^{ab}\overset{\circ}{\nabla}_c\mathcal{K}_{ab} + (\operatorname{tr}_P\mathbf{Y} - n(\ell^{(2)}))\mathcal{K}_{cd}n^d - 2P^{db}(\mathrm{r}_b+\mathrm{s}_b)\mathcal{K}_{dc} + \mathfrak{p}\overset{\circ}{\nabla}_c n^{(2)} \\ & + n^{(2)}\Bigg(\mathfrak{w}_c^{(2)} - \beta\mathfrak{w}_c - a_{\|}^{b}\mathcal{K}_{bc} + (\operatorname{tr}_P\mathbf{Y} - n(\ell^{(2)}))\mathfrak{w}_c - \frac{1}{2}\overset{\circ}{\nabla}_c\mathfrak{p} + 2P^{ab}\mathrm{F}_{cb}\mathfrak{w}_a \\ & \qquad + 2\left(\mathrm{r}_c - \mathrm{s}_c + \frac{1}{2}n^{(2)}\overset{\circ}{\nabla}_c\ell^{(2)}\right)\mathfrak{p}\Bigg)\end{aligned} \tag{4.54}$$

*and*

$$\begin{aligned}\mathfrak{q} = {} & \operatorname{tr}_P\mathfrak{S} + n(\mathfrak{p}) + 2\kappa\mathfrak{p} - 2P^{ab}(\mathrm{r}+\mathrm{s})_a\mathfrak{w}_b - n(\ell^{(2)})\mathfrak{w}(n) \\ & + n^{(2)}\left(\frac{1}{2}\mathfrak{p}^{(2)} - (n(\ell^{(2)}) + \beta)\mathfrak{p} - a_{\|}^{b}\mathfrak{w}_b\right),\end{aligned} \tag{4.55}$$

*where $\mathcal{K}$, $\mathfrak{w}$, $\mathfrak{p}$, $\mathfrak{S}$, $\beta$, $a_{\|}$, $\mathfrak{w}^{(2)}$ and $\mathfrak{p}^{(2)}$ are as before.*

*Proof.* In order to prove the first identity we contract the definition $\mathfrak{Q}_\mu = g^{\alpha\beta}\Sigma_{\mu\alpha\beta}$ with $e_c^\mu$ and insert (2.13) to get (recall that $\Sigma$ is symmetric in the last two indices)

$$\mathfrak{Q}_c = \left(P^{ab}e_a^\alpha e_b^\beta + 2n^b e_b^\beta\xi^\alpha + n^{(2)}\xi^\alpha\xi^\beta\right)e_c^\mu\Sigma_{\mu\alpha\beta} = P^{ab}\Sigma_{cab} + 2({}^{(2)}\Sigma)_{cb}n^b + n^{(2)}({}^{(2,3)}\Sigma)_c.$$

Identity (4.54) follows after inserting (4.28), (4.36) and (4.47). To prove (4.55) we contract $\mathfrak{Q}_\mu = g^{\alpha\beta}\Sigma_{\mu\alpha\beta}$ with $\xi^\mu$,

$$\mathfrak{q} = \left(P^{ab}e_a^\alpha e_b^\beta + 2n^b e_b^\beta\xi^\alpha + n^{(2)}\xi^\alpha\xi^\beta\right)\xi^\mu\Sigma_{\mu\alpha\beta} = P^{ab}\mathfrak{S}_{ab} + 2({}^{(1,2)}\Sigma)_b n^b + n^{(2)}({}^{(1,2,3)}\Sigma), \tag{4.56}$$

and substitute (4.51) and (4.48). □

## 4.4 NON-NULL CASE

In this section we revisit the KID problem for spacelike Cauchy data in the language of hypersurface data. The existence results are known, both in the case of Killing vectors [37, 84, 242] and in the homothetic case [140] (see Section 4.1 for a review). However, the form of the Killing initial data equations that we find in this section are more flexible as they are written in an arbitrary gauge. As we shall discuss below, this form can be useful in practical applications. We reproduce the existence argument in the present setting for completeness. First we restate the classic well-posedness theorem of the Einstein field equations (Theorem 3.1) in the language of hypersurface data. Throughout this section we will assume $\gamma$ to be positive definite and $\mathcal{A}$ of Lorentzian signature. Recall that by "$\lambda$-vacuum" we mean that equation (0.1) holds.

**Theorem 4.14.** *Let $\{\mathcal{H},\gamma,\boldsymbol{\ell},\ell^{(2)},\mathbf{Y}\}$ be hypersurface data of dimension $\mathfrak{n}$, $\lambda\in\mathbb{R}$ and assume the following constraint equations (cf.* (2.89)-(2.90)*)*

$$P^{ac}A_{abc}n^b+\frac{1}{2}P^{ac}P^{bd}B_{abcd}=-\frac{\mathfrak{n}-1}{2}\lambda, \tag{4.57}$$

$$n^{(2)}P^{bd}A_{bcd}-A_{bcd}n^bn^d+P^{bd}B_{abcd}n^a=0, \tag{4.58}$$

*where $A$ and $B$ are as in* (2.87)-(2.88)*. Then there exists a unique maximal and globally hyperbolic $\lambda$-vacuum spacetime $(\mathcal{M},g)$, embedding $\Phi:\mathcal{H}\hookrightarrow\mathcal{M}$ and rigging $\xi$ such that $\{\mathcal{H},\gamma,\boldsymbol{\ell},\ell^{(2)},\mathbf{Y}\}$ is $(\Phi,\xi)$-embedded on $(\mathcal{M},g)$.*

**Remark 4.15.** *For hypersurface data of Lorentzian signature with $\gamma$ positive definite there always exists a gauge satisfying $\boldsymbol{\ell}=0$ and $\ell^{(2)}=-1$ (and hence $P^{ab}=\gamma^{ab}$, $n=0$, $n^{(2)}=-1$) [221]. This gauge corresponds geometrically to a choice of rigging $\xi$ that is unit and normal to $\Phi(\mathcal{H})$. Since in this case $\nu=-\xi$ (see* (2.12)*) the tensor $\mathbf{Y}$ coincides with minus the second fundamental form $\mathbf{K}$ of $\Phi(\mathcal{H})$ w.r.t. the normal $\boldsymbol{\nu}$. In addition, $\overset{\circ}{\nabla}$ agrees with the Levi-Civita connection of $\gamma$, $\nabla^{(\gamma)}$. Therefore (cf.* (2.87)-(2.88)*) $A_{bcd}=-2\nabla^{(\gamma)}_{[d}\mathrm{K}_{c]b}$ and $B_{abcd}=R^{(\gamma)}_{abcd}-2\mathrm{K}_{b[c}\mathrm{K}_{d]a}$, so the constraint equations* (4.57)-(4.58) *simplify to*

$$R^{(\gamma)}-\mathrm{K}_{ab}\mathrm{K}^{ab}+(\operatorname{tr}_\gamma\mathbf{K})^2=-(\mathfrak{n}-2)\lambda, \tag{4.59}$$

$$\nabla^{(\gamma)}_b\left(\mathrm{K}^b{}_a-(\operatorname{tr}_\gamma\mathrm{K})\delta^b{}_a\right)=0, \tag{4.60}$$

*which are the standard constraint equations for spacelike hypersurfaces (see* (3.1)*). Although* (4.57)-(4.58) *are certainly more complicated than* (4.59)-(4.60)*, they are useful because they have the key property of being fully gauge covariant, and hence one can adapt the gauge to simplify the problem at hand. For instance, if one wants to solve the KID problem with a Killing vector that is transverse to the initial hypersurface (but not necessarily normal), one can fix the gauge so that $C=1$ and $\bar{\eta}=0$, and hence $\mathbf{Y}=0$ (see Remark 4.7). Geometrically this gauge corresponds to choosing $\xi=\eta|_{\Phi(\mathcal{H})}$.*

To construct the candidate to Killing/homothetic vector the strategy is to integrate the equation one gets by setting $\mathfrak{Q}=0$ and $\operatorname{Ric}=\lambda g$ in (4.14). The initial data $\eta|_{\mathcal{H}}$ and $\mathcal{L}_\xi\eta|_{\mathcal{H}}$ are to be selected so that $\mathcal{K} \stackrel{\mathcal{H}}{=} \mu g$ and $\pounds_\xi \mathcal{K} \stackrel{\mathcal{H}}{=} \mu \pounds_\xi g$ with $\mu$ a constant restricted to satisfy $\lambda\mu = 0$. Since (4.21) is then a homogeneous wave equation with trivial initial data it follows that $\mathcal{K} = \mu g$ everywhere on $(\mathcal{M}, g)$. This approach requires first of all being able to write down $\mathcal{K}_{ab}$ and $(\pounds_\xi \mathcal{K})_{ab}$ in terms of hypersurface data in order to ensure $\mathcal{K}_{ab} = \mu\gamma_{ab}$ and $(\pounds_\xi \mathcal{K})_{ab} = 2\mu \mathrm{Y}_{ab}$ at the abstract level. The former is simply (see (4.25))

$$\mu\gamma_{ab} = 2C\mathrm{Y}_{ab} + 2\ell_{(a}\overset{\circ}{\nabla}_{b)}C + \pounds_{\bar{\eta}}\gamma_{ab}. \tag{4.61}$$

For the latter we first insert (4.37) into (4.33) and use (2.116) to get the (still general) identity

$$\begin{aligned}
\Phi^\star(\pounds_\xi \mathcal{K})_{ab} = {} & \frac{2C}{n^{(2)}}\Big(\overset{\circ}{R}_{(ab)} - R_{ab} - 2\pounds_n \mathrm{Y}_{ab} - \Big(2\kappa + \mathrm{tr}_P\,\mathbf{U} - n^{(2)}\big(n(\ell^{(2)}) - \mathrm{tr}_P\,\mathbf{Y}\big)\Big)\mathrm{Y}_{ab} \\
& + \overset{\circ}{\nabla}_{(a}(\mathrm{s} + 2\mathrm{r})_{b)} - 2\mathrm{r}_a\mathrm{r}_b + 4\mathrm{r}_{(a}\mathrm{s}_{b)} - \mathrm{s}_a\mathrm{s}_b - (\mathrm{tr}_P\,\mathbf{Y})\mathrm{U}_{ab} + 2P^{cd}\mathrm{U}_{c(a}(2\mathrm{Y} + \mathrm{F})_{b)d}\Big) \\
& + 2C\Big(-2\mathrm{r}_{(a}\overset{\circ}{\nabla}_{b)}\ell^{(2)} + P^{cd}(\Pi_{ca}\Pi_{bd} + \Pi_{ac}\Pi_{db}) + \frac{1}{4}n^{(2)}\overset{\circ}{\nabla}_a\ell^{(2)}\overset{\circ}{\nabla}_b\ell^{(2)} - \frac{1}{2}n(\ell^{(2)})\mathrm{Y}_{ab} \\
& - \frac{1}{2}\overset{\circ}{\nabla}_a\overset{\circ}{\nabla}_b\ell^{(2)}\Big) - 2\ell^{(2)}\Big(n(C)\mathrm{Y}_{ab} + \overset{\circ}{\nabla}_a\overset{\circ}{\nabla}_b C\Big) - 2\overset{\circ}{\nabla}_{(a}C\overset{\circ}{\nabla}_{b)}\ell^{(2)} + (\mathfrak{p} - \bar{\eta}(\ell^{(2)}))\mathrm{K}_{ab} \\
& + 2\pounds_{\bar{\eta}}\mathrm{Y}_{ab} - 2\left((\pounds_{\bar{\eta}}\boldsymbol{\ell})(n) - \boldsymbol{\varpi}(n)\right)\mathrm{Y}_{ab} - 2\overset{\circ}{\nabla}_{(a}\pounds_{\bar{\eta}}\ell_{b)} + 2\overset{\circ}{\nabla}_{(a}\boldsymbol{\varpi}_{b)}.
\end{aligned} \tag{4.62}$$

Thus, the equation one must impose at the initial data level is the same one but replacing $(\pounds_\xi\mathcal{K})_{ab}$ by $2\mu\mathrm{Y}_{ab}$, $\mathfrak{p}$ by $\mu\ell^{(2)}$, $\boldsymbol{\varpi}$ by $\mu\boldsymbol{\ell}$ and $R_{ab}$ by $\lambda\gamma_{ab}$, i.e.

$$\begin{aligned}
2\mu\mathrm{Y}_{ab} = {} & \frac{2C}{n^{(2)}}\Big(\overset{\circ}{R}_{(ab)} - \lambda\gamma_{ab} - 2\pounds_n \mathrm{Y}_{ab} - \Big(2\kappa + \mathrm{tr}_P\,\mathbf{U} - n^{(2)}\big(n(\ell^{(2)}) - \mathrm{tr}_P\,\mathbf{Y}\big)\Big)\mathrm{Y}_{ab} \\
& + \overset{\circ}{\nabla}_{(a}(\mathrm{s} + 2\mathrm{r})_{b)} - 2\mathrm{r}_a\mathrm{r}_b + 4\mathrm{r}_{(a}\mathrm{s}_{b)} - \mathrm{s}_a\mathrm{s}_b - (\mathrm{tr}_P\,\mathbf{Y})\mathrm{U}_{ab} + 2P^{cd}\mathrm{U}_{c(a}(2\mathrm{Y} + \mathrm{F})_{b)d}\Big) \\
& + 2C\Big(-2\mathrm{r}_{(a}\overset{\circ}{\nabla}_{b)}\ell^{(2)} + P^{cd}(\Pi_{ca}\Pi_{bd} + \Pi_{ac}\Pi_{db}) + \frac{1}{4}n^{(2)}\overset{\circ}{\nabla}_a\ell^{(2)}\overset{\circ}{\nabla}_b\ell^{(2)} - \frac{1}{2}n(\ell^{(2)})\mathrm{Y}_{ab} \\
& - \frac{1}{2}\overset{\circ}{\nabla}_a\overset{\circ}{\nabla}_b\ell^{(2)}\Big) - 2\ell^{(2)}\Big(n(C)\mathrm{Y}_{ab} + \overset{\circ}{\nabla}_a\overset{\circ}{\nabla}_b C\Big) - 2\overset{\circ}{\nabla}_{(a}C\overset{\circ}{\nabla}_{b)}\ell^{(2)} - (\mu\ell^{(2)} + \bar{\eta}(\ell^{(2)}))\mathrm{K}_{ab} \\
& + 2\pounds_{\bar{\eta}}\mathrm{Y}_{ab} - 2\left((\pounds_{\bar{\eta}}\boldsymbol{\ell})(n) - \mu\right)\mathrm{Y}_{ab} - 2\overset{\circ}{\nabla}_{(a}\pounds_{\bar{\eta}}\ell_{b)}.
\end{aligned} \tag{4.63}$$

The hypersurface data version of the existence theorem for homothetic/Killing initial data is the following.

**Theorem 4.16.** *Let $\{\mathcal{H}, \boldsymbol{\gamma}, \boldsymbol{\ell}, \ell^{(2)}, \mathbf{Y}\}$ be hypersurface data and $\lambda \in \mathbb{R}$ such that the constraint equations* (4.57)-(4.58) *hold. Assume that $(C, \bar{\eta}) \in \mathcal{F}(\mathcal{H}) \times \mathfrak{X}(\mathcal{H})$ satisfy* (4.61) *and* (4.63) *with $\mu \in \mathbb{R}$ such that $\lambda\mu = 0$. Then there exists a unique maximal, globally hyperbolic $\lambda$-vacuum spacetime $(\mathcal{M}, g)$, embedding $\Phi : \mathcal{H} \hookrightarrow \mathcal{M}$, rigging $\xi$ and vector field $\eta$ such that $\{\mathcal{H}, \boldsymbol{\gamma}, \boldsymbol{\ell}, \ell^{(2)}, \mathbf{Y}\}$ is $(\Phi, \xi)$-embedded on $(\mathcal{M}, g)$, $\eta|_{\Phi(\mathcal{H})} = C\xi + \Phi_\star\bar{\eta}$ and $\pounds_\eta g = \mu g$ on $\mathcal{M}$.*

*Proof.* By Theorem 4.14 there exists a unique maximal, globally hyperbolic $\lambda$-vacuum spacetime $(\mathcal{M}, g)$, embedding $\Phi : \mathcal{H} \hookrightarrow \mathcal{M}$ and rigging $\xi$ such that $\{\mathcal{H}, \boldsymbol{\gamma}, \boldsymbol{\ell}, \ell^{(2)}, \mathbf{Y}\}$ is $(\Phi, \xi)$-embedded on $(\mathcal{M}, g)$. Let us extend $\xi$ off $\Phi(\mathcal{H})$ by $\nabla_\xi \xi = 0$ and construct the (unique) vector field $\eta \in \mathfrak{X}(\mathcal{M})$ by integrating (cf. (4.14))

$$\Box_g \eta^\mu + \lambda\eta^\mu = 0$$

with initial conditions $\eta|_{\mathcal{H}} = C\xi + \Phi_\star \bar{\eta}$ and

$$(\pounds_\xi \boldsymbol{\eta})_a = \mu \ell_a - \ell^{(2)} \overset{\circ}{\nabla}_a C - \overset{\circ}{\nabla}_a(\boldsymbol{\ell}(\bar{\eta})) + 2(\mathrm{Y}_{ac} + \mathrm{F}_{ac})\bar{\eta}^c, \qquad (\pounds_\xi \boldsymbol{\eta})(\xi) = \frac{1}{2}\mu \ell^{(2)}. \tag{4.64}$$

By the identities (4.26) and (4.27) it follows that $\mathfrak{w} = \mu \boldsymbol{\ell}$ and $\mathfrak{p} = \mu \ell^{(2)}$. Equation (4.61) guarantees (by identity (4.25)) that $\mathcal{K}_{ab} = \mu \gamma_{ab}$, and hence $\mathcal{K} \overset{\mathcal{H}}{=} \mu g$. Therefore, condition (4.63) along with identity (4.62) implies $(\pounds_\xi \mathcal{K})_{ab} \overset{\mathcal{H}}{=} 2\mu \mathrm{Y}_{ab}$. Observe also that $\mathfrak{S} = 0$ (see Remark 4.11). To make sure that $\pounds_\xi \mathcal{K} = \mu \pounds_\xi g$ on the initial hypersurface $\Phi(\mathcal{H})$ it only remains to show that $\mathfrak{w}^{(2)}_c = \mu \Phi^\star((\pounds_\xi g)(\xi, \cdot))_c$ and $\mathfrak{p}^{(2)} = \Phi^\star((\pounds_\xi g)(\xi,\xi))$. To prove this we insert $\mathcal{K}_{ab} = \mu\gamma_{ab}$, $\mathfrak{w}_a = \mu \ell_a$, $\mathfrak{p} = \mu \ell^{(2)}$ and $\mathfrak{S} = 0$ into (4.54) with $\mathfrak{Q}_c = 0$ (and $\beta = a_\parallel = 0$) and using (2.3)-(2.6) and (2.38) gives

$$\begin{aligned}
0 &= 2\mu \pounds_n \ell_c + 2\mu \ell^{(2)} \mathrm{U}_{bc} n^b + (2\kappa + \operatorname{tr}_P \mathbf{U})\mu \ell_c + 2\mu(1 - n^{(2)}\ell^{(2)})\left(\mathrm{r}_c - \mathrm{s}_c + n^{(2)} \overset{\circ}{\nabla}_c \ell^{(2)}\right) \\
&\quad - 2\mu P^{ab} \mathrm{U}_{a(b}\ell_{c)} + \mu P^{ab}\mathrm{U}_{c(a}\ell_{b)} - 2\mu(\delta^b_c - n^b \ell_c)(\mathrm{r}_b + \mathrm{s}_b) + \mu \ell^{(2)} \overset{\circ}{\nabla}_c n^{(2)} \\
&\quad + n^{(2)}\left(\mathfrak{w}^{(2)}_c - \frac{1}{2}\mu \overset{\circ}{\nabla}_c \ell^{(2)} + 2\mu \ell^{(2)} \mathrm{s}_c + \left(2\mathrm{r}_c - 2\mathrm{s}_c + n^{(2)}\overset{\circ}{\nabla}_c \ell^{(2)}\right)\mu \ell^{(2)}\right) \\
&= 2\mu \pounds_n \ell_c + 2\mu \ell^{(2)} \mathrm{U}_{bc} n^b - 4\mu \mathrm{s}_c + \mu \ell^{(2)} \overset{\circ}{\nabla}_c n^{(2)} \\
&\quad + n^{(2)}\left(\mathfrak{w}^{(2)}_c + \frac{3}{2}\mu \overset{\circ}{\nabla}_c \ell^{(2)} + 2\mu \ell^{(2)} \mathrm{s}_c - \mu n^{(2)} \ell^{(2)} \overset{\circ}{\nabla}_c \ell^{(2)}\right).
\end{aligned}$$

Using now (2.20) and (2.21) this equation becomes

$$n^{(2)}\left(\mathfrak{w}^{(2)}_c - \frac{1}{2}\mu \overset{\circ}{\nabla}_c \ell^{(2)}\right) = 0 \qquad \Longrightarrow \qquad \mathfrak{w}^{(2)}_c = \frac{1}{2}\mu \overset{\circ}{\nabla}_c \ell^{(2)}. \tag{4.65}$$

By the equality $\Phi^\star((\pounds_\xi g)(\xi,\cdot)) = \Phi^\star(\pounds_\xi \boldsymbol{\xi}) = \frac{1}{2}\overset{\circ}{\nabla}\ell^{(2)}$ valid when $\nabla_\xi \xi = 0$ (see (4.39)) we conclude $\mathfrak{w}^{(2)} = \mu \Phi^\star((\pounds_\xi g)(\xi,\cdot))$, as needed. Similarly, inserting $\mathfrak{S} = 0$, $\mathfrak{p} = \mu \ell^{(2)}$ and $\mathfrak{w} = \mu \boldsymbol{\ell}$ into (4.55) with $\mathfrak{q} = 0$ (and $\beta = a_\parallel = 0$) gives

$$n^{(2)} \mathfrak{p}^{(2)} = 0 \qquad \Longrightarrow \qquad \mathfrak{p}^{(2)} = 0. \tag{4.66}$$

Since $\Phi^\star((\pounds_\xi g)(\xi,\xi)) = 0$ (see (4.42)) we conclude $\mathfrak{p}^{(2)} = \Phi^\star((\pounds_\xi g)(\xi,\xi))$, and therefore $\pounds_\xi \mathcal{K}|_{\Phi(\mathcal{H})} = \mu \pounds_\xi g|_{\Phi(\mathcal{H})}$. Equation (4.21) with $\mathfrak{Q} = 0$, namely

$$\Box_g(\mathcal{K}_{\beta\nu} - \mu g_{\beta\nu}) - 2R_{\sigma(\beta}(\mathcal{K}^\sigma{}_{\nu)} - \mu \delta^\sigma_{\nu)}) + 2R_{\sigma\beta\mu\nu}(\mathcal{K}^{\sigma\mu} - \mu g^{\sigma\mu}) + 2\lambda(\mathcal{K}_{\beta\nu} - \mu g_{\beta\nu}) = 0,$$

then proves $\mathcal{K} = \mu g$ everywhere on $\mathcal{M}$, and hence $\eta$ is a homothety or Killing vector (depending on the value of $\mu$). □

As already mentioned in the introduction, the existence result for the spacelike KID problem can be generalized to include general matter [84, 275, 276], in the sense that the matter field itself is invariant along the Killing vector.

**Remark 4.17.** *Equations* (4.61) *and* (4.63) *are the hypersurface data version of the standard KID equations. Indeed, if one chooses the gauge as in Remark 4.15 these equations become*

$$0 = -2C\mathrm{K}_{ab} + 2\nabla^{(\gamma)}_{(a}\bar{\eta}_{b)} - \mu\gamma_{ab}, \tag{4.67}$$

$$0 = C\left(-R^{(\gamma)}_{ab} + \lambda\gamma_{ab} + 2\gamma^{cd}\mathrm{K}_{ac}\mathrm{K}_{bd} - (\mathrm{tr}_\gamma\,\mathbf{K})\mathrm{K}_{ab}\right) - \pounds_{\bar{\eta}}\mathrm{K}_{ab} + \frac{1}{2}\mu\mathrm{K}_{ab} + \nabla^{(\gamma)}_a\nabla^{(\gamma)}_b C, \tag{4.68}$$

*which are the standard KID equations (see* (4.3)-(4.4) *for the Killing case and [140] for homotheties). The advantage of* (4.61) *and* (4.63) *w.r.t.* (4.67)-(4.68) *is that the gauge is fully free, so it can be adapted to one's convenience. For instance, if one wants to construct a Killing/homothetic vector transverse to the initial hypersurface (but not necessarily normal) it may be easier to choose the gauge such that* $C = 1$ *and* $\bar{\eta} = 0$*, because then* $\mathbf{Y} = 2\mu\gamma$ *and the constraint and KID equations take a much simpler form. As already mentioned before, geometrically this gauge corresponds to choosing* $\xi = \eta|_{\Phi(\mathcal{H})}$.

**Remark 4.18.** *The KID equations* (4.61) *and* (4.63) *are at the level of the initial data, in the sense that they do not involve more data that the one in Theorem 4.14 (apart from the pair* $(C, \bar{\eta})$*, of course).*

## 4.5 NULL HYPERSURFACES

In this section we find necessary conditions on a null hypersurface to be embeddable in a manifold $(\mathcal{M}, g)$ with a homothetic vector field $\eta$, i.e. $\mathcal{K} = \mu g$. We will then analyse their sufficiency in two specific cases, namely the characteristic problem and the smooth spacelike-characteristic problem. Other Cauchy problems (such as the spacelike-characteristic with corner) could be treated similarly.

In [74] the authors perform a similar analysis for Killing vectors and derive a set of equations on an embedded null hypersurface that relate the would-be Killing vector on the hypersurface, the spacetime Levi-Civita connection and Riemann tensor, as well as the induced metric on $\mathcal{H}$ (see Section 4.1). Then they study their sufficiency in two specific scenarios, namely the light-cone case and two null hypersurfaces intersecting in a spacelike surface. The equations they find, that we have recalled in Theorem 4.2 above, are formulated in a particular coordinate system and with a specific choice of transverse vector, while in this section we find analogous equations in a gauge covariant and coordinate-free manner. More importantly, in Section 4.6 we will show that the equations can be written solely in terms of the initial data for the characteristic $\lambda$-vacuum equations (double null data), thus suppressing any reference to the ambient space one wishes to construct. Consequently, we will be able to reformulate the existence theorem of homothetic/Killing vectors for the characteristic problem in a fully detached way. This puts the Killing initial data problem in the characteristic case in an equal footing to the KID problem in the standard Cauchy case, where the KID equations are written in terms of initial value quantities, with a priori no reference to the spacetime to be constructed.

In any of the two cases we will examine, the initial data for the wave equation in the null subset of the hypersurface consists only of the value of the field, and not any of its transverse derivatives (recall Theorem 3.2). This makes the analysis of the KID conditions different to the spacelike case, as now well-posedness of equations $\mathfrak{Q} = 0$ (cf. (4.14)) and (4.21) only requires initial data for $\eta|_{\mathcal{H}}$ and $\mathcal{K}|_{\mathcal{H}}$ (and not for their first transverse derivative). The first immediate consequence is that one cannot use (4.26)-(4.27) to achieve $\mathfrak{w} = \mu\boldsymbol{\ell}$ and $\mathfrak{p} = \mu\ell^{(2)}$ as in the non-null case. Instead, one must prove that $\mathfrak{w} = \mu\boldsymbol{\ell}$ and $\mathfrak{p} = \mu\ell^{(2)}$ hold by some other means. The strategy is to find transport equations for the tensors $\mathfrak{w} - \mu\boldsymbol{\ell}$ and $\mathfrak{p} - \mu\ell^{(2)}$. To do that we start by writing down (4.54)-(4.55) with $n^{(2)} = 0$,

$$\begin{aligned}\mathfrak{Q}_c &= -2\mathfrak{S}_{ac}n^a + 2\pounds_n\mathfrak{w}_c - 2P^{ab}\mathrm{U}_{cb}\mathfrak{w}_a + (2\kappa + \operatorname{tr}_P\mathbf{U})\mathfrak{w}_c + 2\mathfrak{w}(n)(\mathrm{r}-\mathrm{s})_c \\ &\quad + P^{ab}\overset{\circ}{\nabla}_a\mathcal{K}_{bc} - \frac{1}{2}P^{ab}\overset{\circ}{\nabla}_c\mathcal{K}_{ab} + (\operatorname{tr}_P\mathbf{Y} - n(\ell^{(2)}))\mathcal{K}_{cd}n^d - 2P^{bd}(\mathrm{r}+\mathrm{s})_b\mathcal{K}_{dc}, \qquad (4.69)\\ \mathfrak{q} &= \operatorname{tr}_P\mathfrak{S} + \pounds_n(\mathfrak{p}) + 2\kappa\mathfrak{p} - 2P^{ab}(\mathrm{r}+\mathrm{s})_a\mathfrak{w}_b - n(\ell^{(2)})\mathfrak{w}(n), \qquad (4.70)\end{aligned}$$

as well as the contraction of (4.69) with $n^c$ (we shall use that $\mathrm{U}_{cb}n^c = 0$ repeatedly),

$$\begin{aligned}\mathfrak{Q}_c n^c &= -2\mathfrak{S}_{bc}n^bn^c + 2\pounds_n(\mathfrak{w}(n)) + (\operatorname{tr}_P\mathbf{U})\mathfrak{w}(n) + P^{ab}n^c\overset{\circ}{\nabla}_a\mathcal{K}_{bc} - \frac{1}{2}P^{ab}n^c\overset{\circ}{\nabla}_c\mathcal{K}_{ab} \\ &\quad + (\operatorname{tr}_P\mathbf{Y} - n(\ell^{(2)}))\mathcal{K}_{cd}n^cn^d - 2P^{bd}(\mathrm{s}+\mathrm{r})_b\mathcal{K}_{dc}n^c.\end{aligned} \qquad (4.71)$$

As we show explicitly in the following lemma, the contraction of $\mathfrak{S}_{ab}$ with $n^b$ only depends on hypersurface data, the pair $(C, \bar{\eta})$ and $\operatorname{Ric}(\xi, \cdot)$. This will be crucial in Section 4.6 to show that the KID equations are at the level of the initial data.

**Lemma 4.19.** *Let $\{\mathcal{H}, \gamma, \boldsymbol{\ell}, \ell^{(2)}, \mathbf{Y}\}$ be null hypersurface data $(\Phi, \xi)$-embedded in $(\mathcal{M}, g)$. Let $\eta \in \mathfrak{X}(\mathcal{M})$ and define $\eta \overset{\mathcal{H}}{=} C\xi + \Phi_\star\bar{\eta}$. Then,*

$$\begin{aligned}\mathfrak{S}_{ab}n^b &= -C\Big(\operatorname{Ric}(\xi, e_a) + P^{bc}A_{bca} + P^{bc}(\mathrm{r}_b + \mathrm{s}_b)(\mathrm{Y}_{ac} + \mathrm{F}_{ac}) - \frac{1}{2}\kappa\overset{\circ}{\nabla}_a\ell^{(2)} - 2\overset{\circ}{\nabla}_a(n(\ell^{(2)})) \\ &\qquad + 2P^{bc}\mathrm{U}_{ab}\overset{\circ}{\nabla}_c\ell^{(2)}\Big) - n^b\pounds_{\bar{\eta}}\mathrm{Y}_{ab} + \Big(\ell^{(2)}n(C) + (\pounds_{\bar{\eta}}\boldsymbol{\ell})(n)\Big)(\mathrm{r}_a - \mathrm{s}_a) \\ &\quad + \frac{1}{2}n(\ell^{(2)})\overset{\circ}{\nabla}_aC + \frac{1}{2}n(C)\overset{\circ}{\nabla}_a\ell^{(2)} + \ell^{(2)}\Big(\overset{\circ}{\nabla}_a(n(C)) - P^{bc}\mathrm{U}_{ab}\overset{\circ}{\nabla}_cC\Big) + \frac{1}{2}\pounds_n\pounds_{\bar{\eta}}\ell_a \\ &\quad + \frac{1}{2}\overset{\circ}{\nabla}_a((\pounds_{\bar{\eta}}\boldsymbol{\ell})(n)) - P^{bc}\mathrm{U}_{ab}\pounds_{\bar{\eta}}\ell_c \qquad (4.72)\end{aligned}$$

*and*

$$\begin{aligned}\mathfrak{S}_{ab}n^an^b &= -C\left(\operatorname{Ric}(\xi, \nu) + P^{bc}A_{bca}n^a + P(\mathbf{r}+\mathbf{s}, \mathbf{r}+\mathbf{s}) - \frac{1}{2}\kappa n(\ell^{(2)}) - 2\pounds_n^{(2)}\ell^{(2)}\right) \\ &\quad - n^an^b\pounds_{\bar{\eta}}\mathrm{Y}_{ab} - \Big(\ell^{(2)}n(C) + (\pounds_{\bar{\eta}}\boldsymbol{\ell})(n)\Big)\kappa + n(\ell^{(2)})n(C) \\ &\quad + \ell^{(2)}\pounds_n^{(2)}C + \pounds_n((\pounds_{\bar{\eta}}\boldsymbol{\ell})(n)), \qquad (4.73)\end{aligned}$$

*where $A$ is the tensor introduced in* (2.87).

*Proof.* The starting point is (4.37), which after contraction with $n^b$ gives

$$\mathfrak{S}_{ab}n^b = -C\mathrm{z}^{(2)}_a - n^b\pounds_{\bar{\eta}}\mathrm{Y}_{ab} + \left(\frac{1}{2}Cn(\ell^{(2)}) + \ell^{(2)}n(C) + (\pounds_{\bar{\eta}}\boldsymbol{\ell})(n)\right)\mathrm{r}_a + \frac{1}{2}n(\ell^{(2)})\overset{\circ}{\nabla}_a C$$
$$+\frac{1}{2}n(C)\overset{\circ}{\nabla}_a\ell^{(2)} + \frac{1}{2}Cn^b\overset{\circ}{\nabla}_a\overset{\circ}{\nabla}_b\ell^{(2)} + \ell^{(2)}n^b\overset{\circ}{\nabla}_a\overset{\circ}{\nabla}_b C + n^b\overset{\circ}{\nabla}_{(a}\pounds_{\bar{\eta}}\ell_{b)},$$

where recall that $\mathbf{z}^{(2)}$ is defined as $\mathrm{z}^{(2)}_a := \mathrm{Z}^{(2)}_{ab}n^b$. We already know from the identity (2.119) that $\mathbf{z}^{(2)}$ is expressible in terms of hypersurface data and Ricci tensor components. Inserting it into the above equation and using

$$n^b\overset{\circ}{\nabla}_a\overset{\circ}{\nabla}_b f = \overset{\circ}{\nabla}_a n(f) - (\overset{\circ}{\nabla}_b f)\overset{\circ}{\nabla}_a n^b \overset{(2.42)}{=} \overset{\circ}{\nabla}_a(n(f)) - P^{bc}\mathrm{U}_{ab}\overset{\circ}{\nabla}_c f - n(f)\mathrm{s}_a$$

together with (2.46) in Lemma 2.5 applied to $\theta_b = \pounds_\eta \ell_b$, equation (4.72) follows. To obtain (4.73) it suffices to contract (4.72) with $n^a$ again. □

In the following lemma we show that (4.69)-(4.71) possess a hierarchical structure that eventually leads to transport equations for $\mathfrak{w}$ and $\mathfrak{p}$.

**Lemma 4.20.** *Assume $\mathcal{K}$ is of the form $\mathcal{K} = \mu\gamma$ for some $\mu \in \mathbb{R}$. Then,*

1. $\mathfrak{Q}_c n^c = -2\mathfrak{S}_{ac}n^a n^c + 2\pounds_n(\mathfrak{w}(n) - \mu) + (\mathfrak{w}(n) - \mu)\operatorname{tr}_P \mathbf{U}$.
2. *If, moreover, $\mathfrak{w}(n) = \mu$, then*
$$\mathfrak{Q}_c = -2\mathfrak{S}_{ac}n^a + 2\pounds_n(\mathfrak{w}_c - \mu\ell_c) - 2P^{ab}\mathrm{U}_{bc}(\mathfrak{w}_a - \mu\ell_a) + (2\kappa + \operatorname{tr}_P \mathbf{U})(\mathfrak{w}_c - \mu\ell_c).$$
3. *If $\mathfrak{w} = \mu\boldsymbol{\ell}$, then $\mathfrak{q} = \operatorname{tr}_P \mathfrak{S} + \pounds_n(\mathfrak{p} - \mu\ell^{(2)}) + 2\kappa(\mathfrak{p} - \mu\ell^{(2)})$.*

*Proof.* The three items are consequences of the identities in Lemma 4.13, or more specifically, of their simplification (4.69)-(4.70) in the case $n^{(2)} = 0$. First note that under the hypothesis $\mathcal{K} = \mu\gamma$ together with $\pounds_n\ell_c = 2\mathrm{s}_c$, $P^{ab}\mathrm{U}_{bc}\ell_a = 0$ (cf. (2.22) and (2.23)), (2.40) and (2.6), expression (4.69) can be written as

$$\mathfrak{Q}_c = -2\mathfrak{S}_{ca}n^c + 2\pounds_n(\mathfrak{w}_c - \mu\ell_c) - 2P^{ab}\mathrm{U}_{bc}(\mathfrak{w}_a - \mu\ell_a) + (2\kappa + \operatorname{tr}_P \mathbf{U})(\mathfrak{w}_c - \mu\ell_c)$$
$$+ 2(\mathfrak{w}(n) - \mu)(\mathrm{r} - \mathrm{s})_c.$$

The contraction of this with $n^c$ gives item 1., and the substitution $\mathfrak{w}(n) = \mu$ gives item 2. Finally, the last item is immediate from (4.70) after using (2.5). □

**Remark 4.21.** *Actually, item 3. in this lemma does not require the assumption $\mathcal{K} = \mu\gamma$, but presented in this way we emphasize the hierarchical structure that shall be used below.*

When the candidate to Killing field $\eta$ is constructed by solving $\mathfrak{Q} = 0$ in the bulk spacetime, then the identities in Lemma 4.20 become homogeneous transport equations on $\mathcal{H}$ for $\mathfrak{w}$ and $\mathfrak{p}$ provided one already knew that $\mathcal{K}_{ab} = \mu\gamma_{ab}$, $\mathfrak{S}_{bc}n^b = 0$ and $\operatorname{tr}_P \mathfrak{S} = 0$. So it is necessary to find conditions to guarantee that these three equations hold on the null hypersurface. They will be obtained as corollaries of Proposition 4.24 below, where we show that the tensor $\mathfrak{S}$ satisfies a transport equation along $n$. Before proving it we start with two preliminary results.

**Lemma 4.22.** *Let $\{\mathcal{H},\boldsymbol{\gamma},\boldsymbol{\ell},\ell^{(2)}\}$ be null metric hypersurface data $(\Phi,\xi)$-embedded in $(\mathcal{M},g)$ and $\mathfrak{Z}\in\mathfrak{X}(\mathcal{M})$. Then,*

$$\mathfrak{Z}^\alpha \overset{\Phi(\mathcal{H})}{=} \xi^\alpha n^a \mathfrak{Z}_a + P^{ab}e_a^\alpha \mathfrak{Z}_b + \nu^\alpha \boldsymbol{\mathfrak{Z}}(\xi),$$

*where $\boldsymbol{\mathfrak{Z}} = g(\mathfrak{Z},\cdot)$ and $\mathfrak{Z}_a := \Phi^\star(\boldsymbol{\mathfrak{Z}})_a$.*

*Proof.* Using (2.13) with $n^{(2)}=0$ it follows

$$\mathfrak{Z}^\alpha = g^{\alpha\beta}\mathfrak{Z}_\beta = \left(P^{ab}e_a^\alpha e_b^\beta + \xi^\alpha\nu^\beta + \xi^\beta\nu^\alpha\right)\mathfrak{Z}_\beta = P^{ab}e_a^\alpha\mathfrak{Z}_b + \xi^\alpha n^a\mathfrak{Z}_a + \nu^\alpha\boldsymbol{\mathfrak{Z}}(\xi).$$

□

**Lemma 4.23.** *Let $\{\mathcal{H},\boldsymbol{\gamma},\boldsymbol{\ell},\ell^{(2)},\mathbf{Y},\mathbf{Z}^{(2)}\}$ be extended null metric hypersurface data $(\Phi,\xi)$-embedded in $(\mathcal{M},g)$, $\eta\in\mathfrak{X}(\mathcal{M})$ and $\mathcal{K}:=\pounds_\eta g$. Assume $\xi$ and $\nu$ are extended off $\Phi(\mathcal{H})$ by $\nabla_\xi\xi=0$ and $\pounds_\xi\nu=0$. Then,*

$$\begin{aligned}
(\pounds_{k^{(\nu)}}\pounds_\xi g)_{ab} &= 2\mathcal{K}(n,n)\mathrm{Z}^{(2)}_{ab} + \overset{\circ}{\nabla}_{(a}\ell^{(2)}\overset{\circ}{\nabla}_{b)}(\mathcal{K}(n,n)) + 2\pounds_X \mathrm{Y}_{ab}\\
&\quad + 2\boldsymbol{\varpi}(n)\pounds_n\mathrm{Y}_{ab} + 4\mathrm{r}_{(a}\overset{\circ}{\nabla}_{b)}(\boldsymbol{\varpi}(n)), \qquad (4.74)\\
(\pounds_{[\xi,k^{(\nu)}]}g)_{ab} &= 2\Big(\chi + 2\pounds_n(\boldsymbol{\varpi}(n)) - 2\mathfrak{S}(n,n)\Big)\mathrm{Y}_{ab} + 2\ell_{(a}\overset{\circ}{\nabla}_{b)}\Big(\chi + 2\pounds_n(\boldsymbol{\varpi}(n)) - 2\mathfrak{S}(n,n)\Big)\\
&\quad + \pounds_W\gamma_{ab} - 2\pounds_\Xi\gamma_{ab} + 2\Sigma(\nu,\xi,\xi)\mathrm{U}_{ab}, \qquad (4.75)
\end{aligned}$$

*where $\Xi^a := P^{ab}\mathfrak{S}_{bc}n^c$, $X^a := P^{ab}\mathcal{K}_{bc}n^c$, $\chi := -2\mathbf{r}(X) - \frac{1}{2}n(\ell^{(2)})\mathcal{K}(n,n)$ and*

$$\begin{aligned}
W^a &:= P^{ab}\left(\pounds_n\boldsymbol{\varpi}_b + \overset{\circ}{\nabla}_b(\boldsymbol{\varpi}(n)) - 2\boldsymbol{\varpi}(n)\mathrm{s}_b - 2P^{dc}\mathrm{U}_{bd}\boldsymbol{\varpi}_c - V^c{}_b\mathcal{K}_{dc}n^d - (\mathrm{Y}_{cb}+\mathrm{F}_{cb})X^c\right)\\
&\quad + \frac{1}{2}\Big(n(\mathfrak{p}) - X(\ell^{(2)}) - n(\ell^{(2)})\boldsymbol{\varpi}(n)\Big)n^a.
\end{aligned}$$

***Remark:*** *The two terms involving Lie derivative of $\gamma_{ab}$ in (4.75) could be grouped together, but as it will become clear below, our interest is to keep the terms depending on $\mathfrak{S}_{ab}$ separated from the rest.*

*Proof.* To show (4.74) we use Lemma 4.22 with $\mathfrak{Z}=k^{(\nu)}$ together with $g(k^{(\nu)},e_a)=\mathcal{K}_{ab}n^b$ and $g(k^{(\nu)},\xi)=\boldsymbol{\varpi}(n)$ to write

$$k^{(\nu)} = \mathcal{K}(n,n)\xi + \left(P^{ab}\mathcal{K}_{bc}n^c + \boldsymbol{\varpi}(n)n^a\right)e_a = \mathcal{K}(n,n)\xi + X + \boldsymbol{\varpi}(n)n, \qquad (4.76)$$

and hence by applying Proposition B.5 to $T=\pounds_\xi g$ with $\zeta_t=\mathcal{K}(n,n)$ and $\zeta_\parallel = X+\boldsymbol{\varpi}(n)n$ and using (2.113), (4.39) and (2.9) in the respective terms,

$$\begin{aligned}
(\pounds_{k^{(\nu)}}\pounds_\xi g)_{ab} &= 2\mathcal{K}(n,n)\mathrm{Z}^{(2)}_{ab} + \overset{\circ}{\nabla}_{(a}\ell^{(2)}\overset{\circ}{\nabla}_{b)}(\mathcal{K}(n,n)) + 2\pounds_X\mathrm{Y}_{ab} + 2\pounds_{\boldsymbol{\varpi}(n)n}\mathrm{Y}_{ab}\\
&= 2\mathcal{K}(n,n)\mathrm{Z}^{(2)}_{ab} + \overset{\circ}{\nabla}_{(a}\ell^{(2)}\overset{\circ}{\nabla}_{b)}(\mathcal{K}(n,n)) + 2\pounds_X\mathrm{Y}_{ab} + 2\boldsymbol{\varpi}(n)\pounds_n\mathrm{Y}_{ab} + 4\mathrm{r}_{(a}\overset{\circ}{\nabla}_{b)}(\boldsymbol{\varpi}(n)).
\end{aligned}$$

In order to prove (4.75) we need the commutator $[\xi, k^{(\nu)}]$. It is easier to compute the one-form $[\xi, k^{(\nu)}]_\alpha$ and then use Lemma 4.22,

$$\begin{aligned}
[\xi, k^{(\nu)}]_\alpha &= \xi^\mu \nabla_\mu (\mathcal{K}_{\alpha\beta}\nu^\beta) - \mathcal{K}^\mu{}_\beta \nu^\beta \nabla_\mu \xi_\alpha \\
&= \xi^\mu \nu^\beta \nabla_\mu \mathcal{K}_{\alpha\beta} + \mathcal{K}_{\alpha\beta}\xi^\mu \nabla_\mu \nu^\beta - \mathcal{K}^\mu{}_\beta \nu^\beta \nabla_\mu \xi_\alpha \\
&= \xi^\mu \nu^\beta \nabla_\mu \mathcal{K}_{\alpha\beta} + \mathcal{K}_{\alpha\beta}\nu^\mu \nabla_\mu \xi^\beta - \Big(X^b + \mathfrak{w}(n) n^b\Big) e^\mu_b \nabla_\mu \xi_\alpha, \qquad (4.77)
\end{aligned}$$

where in the third line we used $\pounds_\xi \nu = 0$, inserted the expression for $\mathcal{K}^\mu{}_\beta \nu^\beta = (k^{(\nu)})^\mu$ in (4.76) and used $\nabla_\xi \xi = 0$. Contracting (4.77) with $\xi^\alpha$ and using the symmetries (4.9)-(4.10),

$$\begin{aligned}
[\xi, k^{(\nu)}]_\alpha \xi^\alpha &= \Sigma(\nu,\xi,\xi) + \Sigma(\xi,\xi,\nu) + \mathcal{K}_{\alpha\beta}\xi^\alpha \nu^\mu \nabla_\mu \xi^\beta - \Big(X^b + \mathfrak{w}(n) n^b\Big) e^\mu_b \xi^\alpha \nabla_\mu \xi_\alpha \\
&\overset{\Phi(\mathcal{H})}{=} \Sigma(\nu,\xi,\xi) + \frac{1}{2}\Big( n(\mathfrak{p}) - X(\ell^{(2)}) - \mathfrak{w}(n) n(\ell^{(2)})\Big),
\end{aligned}$$

where in the second line we used (4.46) contracted with $\nu$. Note that the scalar $\Sigma(\nu,\xi,\xi)$ on $\Phi(\mathcal{H})$ cannot be written in terms of extended hypersurface data and $\mathcal{K}|_{\Phi(\mathcal{H})}$ (see (4.52) and observe the appearance of $\mathfrak{w}^{(2)}$). Now, the contraction of (4.77) with $e^\alpha_a$ gives, after using (2.58),

$$\begin{aligned}
[\xi, k^{(\nu)}]_\alpha e^\alpha_a &= ({}^{(1)}\nabla \mathcal{K})_{ab} n^b + \mathcal{K}_{\alpha\beta} e^\alpha_a \left( -\kappa \xi^\beta + V^c{}_b n^b e^\beta_c \right) - \Big(X^b + \mathfrak{w}(n) n^b\Big) e^\mu_b e^\alpha_a \nabla_\mu \xi_\alpha \\
&= \pounds_n \mathfrak{w}_a + \overset{\circ}{\nabla}_a(\mathfrak{w}(n)) - 2\mathfrak{w}(n)\mathrm{s}_a - 2P^{bc}\mathrm{U}_{ac}\mathfrak{w}_b - V^c{}_a \mathcal{K}_{bc} n^b - (\mathrm{Y}_{ba} + \mathrm{F}_{ba}) X^b - 2\mathfrak{S}_{ab} n^b,
\end{aligned}$$

where in the second line we inserted (4.32) and used $e^\mu_b e^\alpha_a \nabla_\mu \xi_\alpha = \frac{1}{2} e^\mu_b e^\alpha_a (\pounds_\xi g_{\mu\alpha} + (d\boldsymbol{\xi})_{\mu\alpha}) = \mathrm{Y}_{ba} + \mathrm{F}_{ba}$. Using Lemma 4.22 applied to $\mathfrak{Z} = [\xi, k^{(\nu)}]$ and simplifying one then has

$$[\xi, k^{(\nu)}]^\alpha = \Big(\chi + 2\pounds_n(\mathfrak{w}(n)) - 2\mathfrak{S}(n,n)\Big)\xi^\alpha + \Big(W^a - 2P^{ab}\mathfrak{S}_{bc}n^c + \Sigma(\nu,\xi,\xi) n^a\Big) e^\alpha_a.$$

Identity (4.75) follows from this after applying Proposition B.5 to $T = g$ and $\zeta = [\xi, k^{(\nu)}]$. □

Now we are ready to prove that $\mathfrak{S}$ satisfies a transport equation along $n$.

**Proposition 4.24.** *Let $\{\mathcal{H}, \boldsymbol{\gamma}, \boldsymbol{\ell}, \ell^{(2)}, \mathbf{Y}, \mathbf{Z}^{(2)}\}$ be extended null hypersurface data $(\Phi, \xi)$-embedded in $(\mathcal{M}, g)$ and let $\eta \in \mathfrak{X}(\mathcal{M})$ and $\Sigma := \pounds_\eta \nabla$. Then,*

$$\begin{aligned}
2\pounds_n \mathfrak{S}_{ab} - 2\mathfrak{S}(n,n)\mathrm{Y}_{ab} - \pounds_\Xi \gamma_{ab} - 2\ell_{(a}\overset{\circ}{\nabla}_{b)}(\mathfrak{S}(n,n)) - 4P^{cd}\mathrm{U}_{c(a}\mathfrak{S}_{b)d} \\
+4(\mathrm{r}-\mathrm{s})_{(a}\mathfrak{S}_{b)c}n^c + (2\kappa + \operatorname{tr}_P \mathbf{U})\mathfrak{S}_{ab} + (\operatorname{tr}_P \mathfrak{S})\mathrm{U}_{ab} = \mathcal{I}_{ab},
\end{aligned} \qquad (4.78)$$

*where*

$$
\begin{aligned}
\mathcal{I}_{ab} = {} & (\pounds_\eta R)_{ab} + \frac{1}{2}\overset{\circ}{\nabla}_a\overset{\circ}{\nabla}_b\big(\operatorname{tr}_P \mathcal{K} + 2\mathfrak{w}(n)\big) + \frac{1}{2}\pounds_n\big(\operatorname{tr}_P \mathcal{K} + 2\mathfrak{w}(n)\big)\mathrm{Y}_{ab} - P^{cd}\overset{\circ}{\nabla}_c\Sigma_{dab} + 2V^c{}_{(a}\Sigma_{d|b)c}n^d \\
& + \left(P^{cd}\overset{\circ}{\nabla}_c\mathfrak{w}_d + \left(\operatorname{tr}_P \mathbf{Y} - \frac{1}{2}n(\ell^{(2)})\right)\mathfrak{w}(n) + (\operatorname{tr}_P \mathbf{U} + \kappa)\,\mathfrak{p} - 2P(\mathfrak{w},\mathbf{r}) - P^{cd}V^f{}_c\mathcal{K}_{df} + \frac{1}{2}n(\mathfrak{p})\right)\mathrm{U}_{ab} \\
& + \pounds_n\left(\overset{\circ}{\nabla}_{(a}\mathfrak{w}_{b)} + \mathfrak{p}\mathrm{U}_{ab}\right) - \mathcal{K}(n,n)\mathrm{Z}^{(2)}_{ab} - \frac{1}{2}\overset{\circ}{\nabla}_{(a}\ell^{(2)}\overset{\circ}{\nabla}_{b)}(\mathcal{K}(n,n)) - \pounds_X\mathrm{Y}_{ab} - \big(\chi + \pounds_n(\mathfrak{w}(n))\big)\mathrm{Y}_{ab} \\
& - \ell_{(a}\overset{\circ}{\nabla}_{b)}(\chi + 2\pounds_n(\mathfrak{w}(n))) - \frac{1}{2}\pounds_W\gamma_{ab} + 2P^{cd}(\mathrm{r}+\mathrm{s})_c\Sigma_{dab} - (\operatorname{tr}_P \mathbf{Y} - n(\ell^{(2)}))\Sigma_{cab}n^c \\
& - 2P^{cd}\mathrm{Y}_{c(a|}\Sigma_{de|b)}n^e - 2\mathrm{s}_{(a}\overset{\circ}{\nabla}_{b)}(\mathfrak{w}(n)) - 2(\mathrm{r}-\mathrm{s})_{(a}V^d{}_{b)}\mathcal{K}_{cd}n^c \\
& - 2P^{cd}\mathrm{U}_{c(a}\left(\overset{\circ}{\nabla}_{b)}\mathfrak{w}_d + \mathfrak{w}(n)\mathrm{Y}_{b)d} + \mathfrak{p}\mathrm{U}_{b)d} - V^c{}_{b)}\mathcal{K}_{cd}\right),
\end{aligned}
\tag{4.79}
$$

*and $\Xi$, $X$, $\chi$ and $W$ are defined as in Lemma 4.23.*

*Proof.* The proof is based on identity (4.16), which after pulled back to $\mathcal{H}$ will be shown to provide the transport equation (4.78) with a right hand side that depends only on extended hypersurface data and $\mathcal{K}_{ab}$, $\mathfrak{w}_a$, $\mathfrak{p}$, but no transverse derivatives of $\mathcal{K}_{\alpha\beta}$ on the hypersurface. We start by rewriting the divergence $\nabla_\mu\Sigma^\mu{}_{\alpha\beta}$ at points on $\mathcal{H}$ using (2.14) for $g^{\mu\rho}$ and replacing covariant derivatives by Lie derivatives,

$$
\begin{aligned}
\nabla_\mu\Sigma^\mu{}_{\alpha\beta} &= g^{\mu\rho}\nabla_\mu\Sigma_{\rho\alpha\beta} \\
&= \left(P^{cd}e^\mu_c e^\rho_d + \xi^\mu\nu^\rho + \xi^\rho\nu^\mu\right)\nabla_\mu\Sigma_{\rho\alpha\beta} \\
&= P^{cd}e^\mu_c e^\rho_d\nabla_\mu\Sigma_{\rho\alpha\beta} + \nu^\rho\left(\pounds_\xi\Sigma_{\rho\alpha\beta} - \Sigma_{\sigma\alpha\beta}\nabla_\rho\xi^\sigma - 2\Sigma_{\rho\sigma(\alpha}\nabla_{\beta)}\xi^\sigma\right) \\
&\quad + \xi^\rho\left(\pounds_\nu\Sigma_{\rho\alpha\beta} - \Sigma_{\sigma\alpha\beta}\nabla_\rho\nu^\sigma - 2\Sigma_{\rho\sigma(\alpha}\nabla_{\beta)}\nu^\sigma\right) \\
&= P^{cd}e^\mu_c e^\rho_d\nabla_\mu\Sigma_{\rho\alpha\beta} + \nu^\rho\pounds_\xi\Sigma_{\rho\alpha\beta} + \xi^\rho\pounds_\nu\Sigma_{\rho\alpha\beta} \\
&\quad - \Sigma_{\sigma\alpha\beta}\left(\nu^\rho\nabla_\rho\xi^\sigma + \xi^\rho\nabla_\rho\nu^\sigma\right) - 2\Sigma_{\rho\sigma(\alpha}\left(\nu^\rho\nabla_{\beta)}\xi^\sigma + \xi^\rho\nabla_{\beta)}\nu^\sigma\right).
\end{aligned}
$$

Since the result is independent of the extension of $\nu$ and $\xi$ off $\Phi(\mathcal{H})$ we assume without loss of generality $\nabla_\xi\xi = 0$ and $\pounds_\xi\nu = 0$, so the pullback of the divergence of $\Sigma$ on $\mathcal{H}$ takes the form

$$
\begin{aligned}
e^\alpha_a e^\beta_b\nabla_\mu\Sigma^\mu{}_{\alpha\beta} = {} & P^{cd}e^\mu_c e^\rho_d e^\alpha_a e^\beta_b\nabla_\mu\Sigma_{\rho\alpha\beta} + e^\alpha_a e^\beta_b\pounds_\xi(\nu^\rho\Sigma_{\rho\alpha\beta}) + e^\alpha_a e^\beta_b\pounds_\nu(\xi^\rho\Sigma_{\rho\alpha\beta}) \\
& - 2e^\alpha_a e^\beta_b\Sigma_{\sigma\alpha\beta}\nu^\rho\nabla_\rho\xi^\sigma - 2e^\alpha_a e^\beta_b\Sigma_{\rho\sigma(\alpha}\left(\nu^\rho\nabla_{\beta)}\xi^\sigma + \xi^\rho\nabla_{\beta)}\nu^\sigma\right).
\end{aligned}
\tag{4.80}
$$

We now proceed with the explicit computation of each term. The first term in the RHS of (4.80) can be evaluated using (B.1) applied to $T = \Sigma$ and (4.34), namely

$$
\begin{aligned}
P^{cd}(\nabla\Sigma)_{cdab} = {} & P^{cd}\overset{\circ}{\nabla}_c\Sigma_{dab} + (\operatorname{tr}_P \mathbf{Y})\Sigma_{eab}n^e + 2P^{cd}\mathrm{Y}_{c(a|}\Sigma_{de|b)}n^e \\
& + (\operatorname{tr}_P \mathbf{U})\mathfrak{S}_{ab} + 2P^{cd}\mathrm{U}_{c(a|}({}^{(2)}\Sigma)_{d|b)} \\
= {} & P^{cd}\overset{\circ}{\nabla}_c\Sigma_{dab} + (\operatorname{tr}_P \mathbf{Y})\Sigma_{eab}n^e + 2P^{cd}\mathrm{Y}_{c(a|}\Sigma_{de|b)}n^e + (\operatorname{tr}_P \mathbf{U})\mathfrak{S}_{ab} - 2P^{cd}\mathrm{U}_{c(a}\mathfrak{S}_{b)d} \\
& + 2P^{cd}\mathrm{U}_{c(a}\left(\overset{\circ}{\nabla}_{b)}\mathfrak{w}_d + \mathfrak{w}(n)\mathrm{Y}_{b)d} + \mathfrak{p}\mathrm{U}_{b)d} - (\mathrm{r}-\mathrm{s})_{b)}\mathfrak{w}_d - V^e{}_{b)}\mathcal{K}_{de}\right),
\end{aligned}
\tag{4.81}
$$

where in the second line we inserted (4.34) with $n^{(2)} = 0$. The third term in (4.80) is directly

$$e_a^\alpha e_b^\beta \pounds_\nu (\xi^\rho \Sigma_{\rho\alpha\beta}) = \pounds_n \mathfrak{S}_{ab}. \tag{4.82}$$

For the three terms in the second line of (4.80) we use (2.58) and (2.68), namely

$$\nu^\rho \nabla_\rho \xi^\sigma = -\kappa \xi^\sigma + \left( P^{cd}(\mathrm{r}+\mathrm{s})_c + \frac{1}{2} n(\ell^{(2)}) n^d \right) e_d^\sigma$$

and

$$\nabla_{e_a} \nu^\sigma = \left( P^{cd} \mathrm{U}_{bd} - (\mathrm{r}-\mathrm{s})_b n^c \right) e_c^\sigma.$$

Hence,

$$-2 e_a^\alpha e_b^\beta \Sigma_{\sigma\alpha\beta} \nu^\rho \nabla_\rho \xi^\sigma = 2\kappa \mathfrak{S}_{ab} - 2P^{cd}(\mathrm{r}+\mathrm{s})_c \Sigma_{dab} - n(\ell^{(2)}) \Sigma_{cab} n^c, \tag{4.83}$$

$$-2 e_a^\alpha e_b^\beta \Sigma_{\rho\sigma(\alpha} \xi^\rho \nabla_{\beta)} \nu^\sigma = -2 \mathfrak{S}_{c(a} \left( P^{cd} \mathrm{U}_{b)d} - (\mathrm{r}-\mathrm{s})_{b)} n^c \right), \tag{4.84}$$

$$\begin{aligned} -2 e_a^\alpha e_b^\beta \Sigma_{\rho\sigma(\alpha} \nu^\rho \nabla_{\beta)} \xi^\sigma &\overset{(2.52)}{=} -2n^c ({}^{(2)}\Sigma)_{c(a} (\mathrm{r}-\mathrm{s})_{b)} - 2n^c P^{de} \Sigma_{cd(a} (\mathrm{Y}+\mathrm{F})_{b)e} - n^c n^d \Sigma_{cd(a} \overset{\circ}{\nabla}_{b)} \ell^{(2)} \\ &\overset{(4.35)}{=} 2n^c \mathfrak{S}_{c(a} (\mathrm{r}-\mathrm{s})_{b)} - 2n^c P^{de} \Sigma_{cd(a} (\mathrm{Y}+\mathrm{F})_{b)e} - n^c n^d \Sigma_{cd(a} \overset{\circ}{\nabla}_{b)} \ell^{(2)} \\ &\quad - 2(\mathrm{r}-\mathrm{s})_{(a} \left( \overset{\circ}{\nabla}_{b)} (\boldsymbol{\varpi}(n)) - P^{cd} \mathrm{U}_{b)d} \boldsymbol{\varpi}_c - V^d{}_{b)} \mathcal{K}_{cd} n^c \right). \end{aligned} \tag{4.85}$$

In order to compute the remaining term of (4.80), namely $e_a^\alpha e_b^\beta \pounds_\xi \left( \nu^\rho \Sigma_{\rho\alpha\beta} \right)$, we make use of the identity (4.23) in Lemma 4.4 applied to $\zeta = \nu$,

$$\pounds_\xi \left( \nu^\rho \Sigma_{\rho\alpha\beta} \right) = \frac{1}{2} \pounds_\xi \left( \pounds_{k^{(\nu)}} g_{\alpha\beta} - \pounds_\nu \mathcal{K}_{\alpha\beta} \right) = \frac{1}{2} \pounds_{k^{(\nu)}} \pounds_\xi g_{\alpha\beta} - \frac{1}{2} \pounds_\nu \pounds_\xi \mathcal{K}_{\alpha\beta} + \frac{1}{2} \pounds_{[\xi, k^{(\nu)}]} g_{\alpha\beta}. \tag{4.86}$$

The pullbacks of the first and third terms in the right-hand side have been already computed in Lemma 4.23. The pullback of the second term is, by (4.33),

$$-\frac{1}{2} e_a^\alpha e_b^\beta \pounds_\nu \pounds_\xi \mathcal{K}_{\alpha\beta} = \pounds_n \mathfrak{S}_{ab} - \pounds_n \left( \overset{\circ}{\nabla}_{(a} \boldsymbol{\varpi}_{b)} + \boldsymbol{\varpi}(n) \mathrm{Y}_{ab} + \mathfrak{p} \mathrm{U}_{ab} \right)$$

Combining the three,

$$\begin{aligned} e_a^\alpha e_b^\beta \pounds_\xi (\nu^\rho \Sigma_{\rho\alpha\beta}) &= \mathcal{K}(n,n) \mathrm{Z}^{(2)}_{ab} + \frac{1}{2} \overset{\circ}{\nabla}_{(a} \ell^{(2)} \overset{\circ}{\nabla}_{b)} (\mathcal{K}(n,n)) + \pounds_X \mathrm{Y}_{ab} + 2\mathrm{r}_{(a} \overset{\circ}{\nabla}_{b)} (\boldsymbol{\varpi}(n)) + \pounds_n \mathfrak{S}_{ab} \\ &\quad - \pounds_n \left( \overset{\circ}{\nabla}_{(a} \boldsymbol{\varpi}_{b)} + \mathfrak{p} \mathrm{U}_{ab} \right) + \left( \chi + \pounds_n (\boldsymbol{\varpi}(n)) - 2\mathfrak{S}(n,n) \right) \mathrm{Y}_{ab} \\ &\quad + \ell_{(a} \overset{\circ}{\nabla}_{b)} \left( \chi + 2\pounds_n (\boldsymbol{\varpi}(n)) - 2\mathfrak{S}(n,n) \right) + \frac{1}{2} \pounds_W \gamma_{ab} - \pounds_\Xi \gamma_{ab} + \Sigma(\nu,\xi,\xi) \mathrm{U}_{ab}. \end{aligned} \tag{4.87}$$

With this, all the terms in the RHS of (4.80) have been evaluated. Inserting the corresponding expressions (4.81), (4.82), (4.83), (4.84), (4.85) and (4.87) into (4.80) the pullback of the divergence term is, finally,

$$
\begin{aligned}
e^\alpha_a e^\beta_b \nabla_\mu \Sigma^\mu{}_{\alpha\beta} = {} & P^{cd}\overset{\circ}{\nabla}_c \Sigma_{dab} + (\operatorname{tr}_P \mathbf{Y} - n(\ell^{(2)}))\Sigma_{cab} n^c + 2P^{cd}\mathrm{Y}_{c(a|}\Sigma_{de|b)} n^e + (\operatorname{tr}_P \mathbf{U} + 2\kappa)\mathfrak{S}_{ab} \\
& - 4P^{cd}\mathrm{U}_{c(a}\mathfrak{S}_{b)d} + 2P^{cd}\mathrm{U}_{c(a}\Big(\overset{\circ}{\nabla}_{b)}\mathfrak{w}_d + \mathfrak{w}(n)\mathrm{Y}_{b)d} + \mathfrak{p}\mathrm{U}_{b)d} - V^e{}_{b)}\mathcal{K}_{de}\Big) + 2\pounds_n \mathfrak{S}_{ab} \\
& - \pounds_n\left(\overset{\circ}{\nabla}_{(a}\mathfrak{w}_{b)} + \mathfrak{p}\mathrm{U}_{ab}\right) + \mathcal{K}(n,n)\mathrm{Z}^{(2)}_{ab} + \frac{1}{2}\overset{\circ}{\nabla}_{(a}\ell^{(2)}\overset{\circ}{\nabla}_{b)}(\mathcal{K}(n,n)) + \pounds_X \mathrm{Y}_{ab} + \frac{1}{2}\pounds_W \gamma_{ab} \\
& - \pounds_\Xi \gamma_{ab} - 2\ell_{(a}\overset{\circ}{\nabla}_{b)}(\mathfrak{S}(n,n)) + \Sigma(\nu,\xi,\xi)\mathrm{U}_{ab} + \Big(\chi + \pounds_n(\mathfrak{w}(n)) - 2\mathfrak{S}(n,n)\Big)\mathrm{Y}_{ab} \\
& + \ell_{(a}\overset{\circ}{\nabla}_{b)}\Big(\chi + 2\pounds_n(\mathfrak{w}(n))\Big) - 2P^{cd}(\mathrm{r}+\mathrm{s})_c\Sigma_{dab} + 4(\mathrm{r}-\mathrm{s})_{(a}\mathfrak{S}_{b)c}n^c \\
& - 2n^c P^{de}\Sigma_{cd(a}(\mathrm{Y}+\mathrm{F})_{b)e} - n^c n^d \Sigma_{cd(a}\overset{\circ}{\nabla}_{b)}\ell^{(2)} + 2\mathrm{s}_{(a}\overset{\circ}{\nabla}_{b)}\mathfrak{w}(n) + 2(\mathrm{r}-\mathrm{s})_{(a}V^d{}_{b)}\mathcal{K}_{cd}n^c.
\end{aligned}
$$

Now we consider the pullback of the RHS of (4.16). Using Proposition B.1 [Eq. (B.8)] applied to $f = \operatorname{tr}_g \mathcal{K}$ and taking into account that $\operatorname{tr}_g \mathcal{K} \stackrel{\mathcal{H}}{=} \operatorname{tr}_P \mathcal{K} + 2\mathfrak{w}(n)$ (cf. (2.14)),

$$
\begin{aligned}
e^\alpha_a e^\beta_b \left(\pounds_\eta R_{\alpha\beta} + \frac{1}{2}\nabla_\alpha\nabla_\beta \operatorname{tr}_g \mathcal{K}\right) = {} & (\pounds_\eta R)_{ab} + \frac{1}{2}\overset{\circ}{\nabla}_a\overset{\circ}{\nabla}_b\Big(\operatorname{tr}_P \mathcal{K} + 2\mathfrak{w}(n)\Big) \\
& + \frac{1}{2}\pounds_n(\operatorname{tr}_P \mathcal{K} + 2\mathfrak{w}(n))\mathrm{Y}_{ab} + \frac{1}{2}\xi(\operatorname{tr}_g \mathcal{K})\mathrm{U}_{ab}.
\end{aligned}
$$

Let us compute $\xi(\operatorname{tr}_g \mathcal{K})$. From the decomposition of $g^{\alpha\beta}$ in (2.14) together with ${}^{(1,2)}\nabla\mathcal{K} = {}^{(1,3)}\Sigma + {}^{(2,3)}\Sigma$ (see (4.9)),

$$
\begin{aligned}
\pounds_\xi(\operatorname{tr}_g \mathcal{K}) = g^{\alpha\beta}\xi^\mu\nabla_\mu \mathcal{K}_{\alpha\beta} &= P^{ab}({}^{(1)}\nabla\mathcal{K})_{ab} + 2({}^{(1,2)}\nabla\mathcal{K})_a n^a \\
&= P^{ab}({}^{(1)}\nabla\mathcal{K})_{ab} + 2\Sigma(\xi,\nu,\xi) + 2\Sigma(\nu,\xi,\xi).
\end{aligned}
$$

Inserting (4.30) and (4.53) (with $n^{(2)} = \beta = a_\parallel = 0$), but keeping the term[2] $\Sigma(\nu,\xi,\xi)$ gives

$$
\begin{aligned}
\pounds_\xi(\operatorname{tr}_g \mathcal{K}) = {} & 2P^{ab}\overset{\circ}{\nabla}_a\mathfrak{w}_b + (2\operatorname{tr}_P \mathbf{Y} - n(\ell^{(2)}))\mathfrak{w}(n) + 2(\operatorname{tr}_P \mathbf{U} + \kappa)\mathfrak{p} \\
& - 4P^{ab}\mathrm{r}_a\mathfrak{w}_b - 2P^{ab}V^c{}_a\mathcal{K}_{bc} - 2\operatorname{tr}_P \mathfrak{S} + n(\mathfrak{p}) + 2\Sigma(\nu,\xi,\xi),
\end{aligned}
$$

and thus, the pullback of the RHS of (4.16) is

$$
\begin{aligned}
\mathrm{RHS}_{ab} = {} & (\pounds_\eta R)_{ab} + \frac{1}{2}\overset{\circ}{\nabla}_a\overset{\circ}{\nabla}_b\Big(\operatorname{tr}_P \mathcal{K} + 2\mathfrak{w}(n)\Big) + \frac{1}{2}\pounds_n\Big(\operatorname{tr}_P \mathcal{K} + 2\mathfrak{w}(n)\Big)\mathrm{Y}_{ab} \\
& + \Bigg(P^{cd}\overset{\circ}{\nabla}_c\mathfrak{w}_d + \left(\operatorname{tr}_P \mathbf{Y} - \frac{1}{2}n(\ell^{(2)})\right)\mathfrak{w}(n) + (\operatorname{tr}_P \mathbf{U} + \kappa)\mathfrak{p} - 2P^{cd}\mathrm{r}_c\mathfrak{w}_d \\
& \qquad - P^{cd}V^f{}_c\mathcal{K}_{df} - \operatorname{tr}_P \mathfrak{S} + \frac{1}{2}n(\mathfrak{p}) + \Sigma(\nu,\xi,\xi)\Bigg)\mathrm{U}_{ab}.
\end{aligned}
$$

Identity (4.78) follows by equating the LHS and the RHS and rearranging terms. Observe that the only term that is not expressible in terms of extended hypersurface data and $\mathcal{K}|_{\Phi(\mathcal{H})}$ (namely $\Sigma(\nu,\xi,\xi)$) cancels out. □

In the next lemma, that will be needed in Corollary 4.26 below, we compute the contraction of $\pounds_{\mathfrak{z}}\gamma_{ab}$ with $n$ when $\mathfrak{z}$ is either of the form $\mathfrak{z}^b = P^{bc}\theta_c$ or $\mathfrak{z} = fn$.

2 As it depends on transverse derivatives of $\mathcal{K}_{\alpha\beta}$ because of the term $\mathfrak{w}^{(2)}(n)$ in (4.52).

**Lemma 4.25.** *Let $\{\mathcal{H},\boldsymbol{\gamma},\boldsymbol{\ell},\ell^{(2)}\}$ be null metric hypersurface data, $\boldsymbol{\theta}$ any one-form and $f$ any smooth function. Then,*

$$n^b \pounds_{P(\boldsymbol{\theta},\cdot)}\gamma_{ab} = \pounds_n\theta_a - \pounds_n(\boldsymbol{\theta}(n))\ell_a - 2\Big(\boldsymbol{\theta}(n)\mathrm{s}_a + P^{bc}\mathrm{U}_{ab}\theta_c\Big), \tag{4.88}$$

$$n^b \pounds_{fn}\gamma_{ab} = 0. \tag{4.89}$$

*Proof.* We start by noting that $\boldsymbol{\gamma}(n,\cdot)=0$ immediately implies for any vector field $\mathfrak{Z}$

$$n^b\pounds_{\mathfrak{Z}}\gamma_{ab} = -\gamma_{ab}\pounds_{\mathfrak{Z}}n^b = \gamma_{ab}\pounds_n\mathfrak{Z}^b.$$

When $\mathfrak{Z}$ is of the form $\mathfrak{Z}^b = P^{bc}\theta_c$ this becomes

$$\begin{aligned}
n^b \pounds_{P(\boldsymbol{\theta},\cdot)}\gamma_{ab} &= \gamma_{ab}\theta_c\pounds_n P^{bc} + \gamma_{ab}P^{bc}\pounds_n\theta_c \\
&\overset{(2.6)}{=} \gamma_{ab}\theta_c\pounds_n P^{bc} + \pounds_n\theta_a - \pounds_n(\theta(n))\ell_a \\
&\overset{(2.50)}{=} -2\gamma_{ab}P^{bd}\Big(n^c\theta_c\mathrm{s}_d + P^{cf}\mathrm{U}_{df}\theta_c\Big) + \pounds_n\theta_a - \pounds_n(\theta(n))\ell_a,
\end{aligned}$$

which after using again $\gamma_{ab}P^{bd} = \delta^d_a - n^d\ell_a$, $\mathbf{s}(n)=0$ and $\mathbf{U}(n,\cdot)=0$ gives (4.88). When $\mathfrak{Z} = fn$ then $n^b\pounds_{fn}\gamma_{ab} = \gamma_{ab}\pounds_n(fn^b) = 0$, which is (4.89). □

In the following corollary we show that when $\mathcal{K}|_{\Phi(\mathcal{H})}$ is pure trace, identity (4.78) admits again a hierarchical structure and can be rewritten as transport equations for $\operatorname{tr}_P\mathcal{K}$, all components of $\mathfrak{S}_{ab}n^b$ except $\mathfrak{S}(n,n)$ and $\operatorname{tr}_P\mathfrak{S}$. Before proving this we recall that given a 2-covariant, symmetric tensor field $T$, it can be decomposed uniquely as (see (2.24))

$$T_{ab} = \frac{\operatorname{tr}_P T}{\mathfrak{n}-1}\gamma_{ab} + 2\ell_{(a}T_{b)c}n^c + T(n,n)\left(\frac{\ell^{(2)}}{n-1}\gamma_{ab} - \ell_a\ell_b\right) + \widehat{T}_{ab}, \tag{4.90}$$

where $\widehat{T}$ is a symmetric tensor that satisfies $P^{ab}\widehat{T}_{ab}=0$ and $\widehat{T}_{ab}n^a=0$. It follows that when $T=\mu\gamma_{ab}$ then $\widehat{T}_{ab}=0$ and $T_{ab}n^b=0$. For the first item in the next corollary we only need to impose these two last conditions on $\mathcal{K}_{ab}$ while for items 2. and 3. we also need restrictions on the transverse components of $\mathcal{K}_{\mu\nu}$. The conditions needed here are well adapted to the ones appearing in Lemma 4.20. This is what allows one to close the argument in Theorem 4.27 below.

**Corollary 4.26.** *Assume $\mathcal{K}_{ab}n^b=0$, $\widehat{\mathcal{K}}=0$ and $R_{\alpha\beta}=\lambda g_{\alpha\beta}$ in a neighbourhood*[3] *of $\Phi(\mathcal{H})$ and let $\mu\in\mathbb{R}$ satisfying $\lambda\mu=0$. Then,*

1. $\left(\pounds^{(2)}_n + \left(\dfrac{2\operatorname{tr}_P\mathbf{U}}{\mathfrak{n}-1} - \kappa\right)\pounds_n\right)\operatorname{tr}_P\mathcal{K} = 2(\operatorname{tr}_P\mathbf{U})\mathfrak{S}(n,n).$

2. *If, moreover, $\mathcal{K}_{ab}=\mu\gamma_{ab}$ and $\boldsymbol{\mathfrak{w}}(n)=\mu$, one has*

$$\pounds_n(\mathfrak{S}_{ab}n^b) - \overset{\circ}{\nabla}_a(\mathfrak{S}(n,n)) + (\operatorname{tr}_P\mathbf{U})\mathfrak{S}_{ab}n^b = 0.$$

3 In fact it suffices that this equality holds up to first derivatives on $\Phi(\mathcal{H})$.

*3. If* $\mathcal{K}_{ab} = \mu\gamma_{ab}$ *and in addition* $\boldsymbol{\mathfrak{w}} = \mu\boldsymbol{\ell}$ *and* $\mathfrak{S}(n,\cdot) = 0$*, then*

$$2\pounds_n(\operatorname{tr}_P \mathfrak{S}) + 2(\kappa + \operatorname{tr}_P \mathbf{U}) \operatorname{tr}_P \mathfrak{S} = \Big((\mathfrak{p} - \mu\ell^{(2)}) \operatorname{tr}_P \mathbf{U} + \pounds_n(\mathfrak{p} - \mu\ell^{(2)}) + (\mathfrak{p} - \mu\ell^{(2)})\kappa\Big) \operatorname{tr}_P \mathbf{U} \\ + (\mathfrak{p} - \mu\ell^{(2)})\pounds_n(\operatorname{tr}_P \mathbf{U}).$$

***Remark:*** *Contracting the expression in item 2. with* $n^a$ *gives* $(\operatorname{tr}_P \mathbf{U})\mathfrak{S}(n,n) = 0$*, which implies* $\mathfrak{S}(n,n) = 0$ *at the points where* $\operatorname{tr}_P \mathbf{U} \neq 0$*. The same conclusion follows from item 1. provided* $\mathcal{K}_{ab} = \mu\gamma_{ab}$*. However, if* $\operatorname{tr}_P \mathbf{U}$ *vanishes on open sets of* $\mathcal{H}$ *one cannot use these equations to obtain the value of* $\mathfrak{S}(n,n)$*. This is why we do not assume* $\mathcal{K}_{ab} = \mu\gamma_{ab}$ *from the beginning.*

*Proof.* We start by noting that the quantities $X^a$ and $\chi$ in Lemma 4.23 vanish identically and that $\mathcal{K}_{ab} = \frac{1}{\mathfrak{n}-1}(\operatorname{tr}_P \mathcal{K})\gamma_{ab}$. Inserting this into (4.28) yields

$$\begin{aligned}\Sigma_{abc} &= \frac{1}{2(\mathfrak{n}-1)}\Big(2\gamma_{a(c}\overset{\circ}{\nabla}_{b)}(\operatorname{tr}_P \mathcal{K}) - \gamma_{bc}\overset{\circ}{\nabla}_a(\operatorname{tr}_P \mathcal{K})\Big) + \frac{\operatorname{tr}_P \mathcal{K}}{2(\mathfrak{n}-1)}\Big(2\overset{\circ}{\nabla}_{(b}\gamma_{c)a} - \overset{\circ}{\nabla}_a\gamma_{bc}\Big) + \boldsymbol{\mathfrak{w}}_a \mathrm{U}_{bc} \\ &\overset{(2.38)}{=} \frac{1}{2(\mathfrak{n}-1)}\Big(2\gamma_{a(c}\overset{\circ}{\nabla}_{b)}(\operatorname{tr}_P \mathcal{K}) - \gamma_{bc}\overset{\circ}{\nabla}_a(\operatorname{tr}_P \mathcal{K})\Big) + \left(\boldsymbol{\mathfrak{w}}_a - \frac{\operatorname{tr}_P \mathcal{K}}{\mathfrak{n}-1}\ell_a\right)\mathrm{U}_{bc}. \end{aligned} \tag{4.91}$$

The following contractions are immediate

$$\Sigma_{abc}n^b = \frac{\pounds_n(\operatorname{tr}_P \mathcal{K})}{2(\mathfrak{n}-1)}\gamma_{ac}, \qquad \Sigma_{abc}n^a n^b = 0, \qquad \Sigma_{abc}n^b n^c = 0, \tag{4.92}$$

and hence

$$P^{cd}n^a n^b \overset{\circ}{\nabla}_c \Sigma_{dab} = -2P^{cd}\Sigma_{dab}n^a \overset{\circ}{\nabla}_c n^b \overset{(2.42)}{=} -2P^{cd}P^{bf}\mathrm{U}_{fc}\Sigma_{dab}n^a = -\frac{\operatorname{tr}_P \mathbf{U}}{\mathfrak{n}-1}\pounds_n(\operatorname{tr}_P \mathcal{K}), \tag{4.93}$$

where in the last equality we used $P^{cd}P^{bf}\gamma_{db} = P^{cf} + \ell^{(2)}n^c n^f$, which is a consequence of (2.5)-(2.6). Many terms in the contraction of $\mathcal{I}_{ab}$ with $n^a n^b$ vanish identically. Using $(\pounds_\eta R)_{ab} = \lambda\mathcal{K}_{ab}$, $n^a \overset{\circ}{\nabla}_a n^b = 0$ and inserting (4.93) then gives

$$\mathcal{I}(n,n) = \frac{1}{2}\left(\pounds_n^{(2)} + \left(\frac{2\operatorname{tr}_P \mathbf{U}}{\mathfrak{n}-1} - \kappa\right)\pounds_n\right)\operatorname{tr}_P \mathcal{K}. \tag{4.94}$$

The contraction of the LHS of (4.78) with $n^a n^b$ simplifies to $(\operatorname{tr}_P \mathbf{U})\mathfrak{S}(n,n)$, and hence item 1. of the corollary follows.

To prove the second item we observe that now $\operatorname{tr}_P \mathcal{K} = (\mathfrak{n}-1)\mu$, so (4.91) becomes $\Sigma_{abc} = (\boldsymbol{\mathfrak{w}}_a - \mu\ell_a)\mathrm{U}_{bc}$ and hence $\Sigma_{abc}$ contracted with $n$ in any of its indices vanishes. Moreover, the vector $W^a$ (see Lemma 4.23) simplifies to

$$W^a = P^{ab}\Big(\pounds_n(\boldsymbol{\mathfrak{w}}_b - \mu\ell_b) - 2P^{cd}\mathrm{U}_{bc}(\boldsymbol{\mathfrak{w}}_d - \mu\ell_d)\Big) + \frac{1}{2}\pounds_n(\mathfrak{p} - \mu\ell^{(2)})n^a,$$

where we used $2\mathbf{s} = \pounds_n \boldsymbol{\ell}$ (see (2.22)) and $P^{cd}\mathrm{U}_{bc}\ell_d = 0$ (because of (2.5) and (2.23)). These two relations will be used throughout the rest of the proof without further notice. Then, since the term $\pounds_\eta R_{ab}$ in (4.79) is zero because $\lambda\mu = 0$,

$$
\begin{aligned}
\mathcal{I}_{ab} = {} & -P^{cd}\overset{\circ}{\nabla}_c\Big((\mathfrak{w}_d - \mu\ell_d)\mathrm{U}_{ab}\Big) + \bigg(P^{cd}\overset{\circ}{\nabla}_c\mathfrak{w}_d + \Big(\operatorname{tr}_P \mathbf{Y} - \frac{1}{2}n(\ell^{(2)})\Big)\mu + (\operatorname{tr}_P \mathbf{U} + \kappa)\mathfrak{p} \\
& -2P(\mathfrak{w}, \mathbf{r}) - \mu(\delta^c_f - n^c\ell_f)V^f{}_c + \frac{1}{2}n(\mathfrak{p})\bigg)\mathrm{U}_{ab} + \pounds_n\Big(\overset{\circ}{\nabla}_{(a}\mathfrak{w}_{b)} + \mathfrak{p}\mathrm{U}_{ab}\Big) - \frac{1}{2}\pounds_W\gamma_{ab} \\
& + 2P^{cd}(\mathrm{r}+\mathrm{s})_c(\mathfrak{w}_d - \mu\ell_d)\mathrm{U}_{ab} - 2P^{cd}\mathrm{U}_{c(a}\Big(\overset{\circ}{\nabla}_{b)}\mathfrak{w}_d + \mu\mathrm{Y}_{b)d} + \mathfrak{p}\mathrm{U}_{b)d} - \mu V^c{}_{b)}\gamma_{cd}\Big),
\end{aligned}
$$

where $V^a{}_b$ is as in (2.54), which after using $V^c{}_c = \operatorname{tr}_P \mathbf{Y} + \frac{1}{2}n(\ell^{(2)})$, $n^c\ell_f V^f{}_c = \kappa\ell^{(2)} + \frac{1}{2}n(\ell^{(2)})$ (see (2.57)) and $V^c{}_b\gamma_{cd} = \mathrm{Y}_{bd} + \mathrm{F}_{bd} - (\mathrm{r}-\mathrm{s})_b\ell_d$ (see (2.56)) simplifies to

$$
\begin{aligned}
\mathcal{I}_{ab} = {} & -P^{cd}\overset{\circ}{\nabla}_c\Big((\mathfrak{w}_d - \mu\ell_d)\mathrm{U}_{ab}\Big) + \Big(P^{cd}\overset{\circ}{\nabla}_c\mathfrak{w}_d + (\operatorname{tr}_P \mathbf{U})\mathfrak{p} + \kappa(\mathfrak{p} + \mu\ell^{(2)}) - 2P(\mathfrak{w}, \mathbf{r}) \\
& + \frac{1}{2}n(\mathfrak{p} - \mu\ell^{(2)})\Big)\mathrm{U}_{ab} + \pounds_n\Big(\overset{\circ}{\nabla}_{(a}\mathfrak{w}_{b)} + \mathfrak{p}\mathrm{U}_{ab}\Big) - \frac{1}{2}\pounds_W\gamma_{ab} + 2P^{cd}(\mathrm{r}+\mathrm{s})_c(\mathfrak{w}_d - \mu\ell_d)\mathrm{U}_{ab} \\
& - 2P^{cd}\mathrm{U}_{c(a}\Big(\overset{\circ}{\nabla}_{b)}\mathfrak{w}_d - \mu\mathrm{F}_{b)d} + \mathfrak{p}\mathrm{U}_{b)d}\Big).
\end{aligned}
\tag{4.95}
$$

The contraction of $\mathcal{I}_{ab}$ with $n^b$ then gives, after using (2.42) and $\mathrm{U}_{ab}n^b = 0$ in the first equality and identities (2.45) and (2.46) in Lemma 2.5 applied to $\theta_c = \mathfrak{w}_c$ (with $n^{(2)} = 0$) in the second,

$$
\begin{aligned}
\mathcal{I}_{ab}n^b & = P^{cd}P^{bf}\mathrm{U}_{fc}\mathrm{U}_{ab}(\mathfrak{w}_d - \mu\ell_d) + \pounds_n(n^b\overset{\circ}{\nabla}_{(a}\mathfrak{w}_{b)}) - \frac{1}{2}n^b\pounds_W\gamma_{ab} - P^{cd}\mathrm{U}_{ca}\Big(n^b\overset{\circ}{\nabla}_b\mathfrak{w}_d - \mu\mathrm{s}_d\Big) \\
& = 2P^{cd}P^{bf}\mathrm{U}_{fc}\mathrm{U}_{ab}(\mathfrak{w}_d - \mu\ell_d) + \frac{1}{2}\pounds_n\Big(\pounds_n(\mathfrak{w}_a - \mu\ell_a) - 2P^{cd}\mathrm{U}_{ac}(\mathfrak{w}_d - \mu\ell_d)\Big) - \frac{1}{2}n^b\pounds_W\gamma_{ab} \\
& \quad - P^{cd}\mathrm{U}_{ca}\pounds_n(\mathfrak{w}_d - \mu\ell_d).
\end{aligned}
\tag{4.96}
$$

From Lemma 4.25 and the fact that $\pounds_n(\mathfrak{w}_c - \mu\ell_c) - 2P^{fd}\mathrm{U}_{cf}(\mathfrak{w}_d - \mu\ell_d)$ contracted with $n^c$ is identically zero, the term $n^b\pounds_W\gamma_{ab}$ is

$$
n^b\pounds_W\gamma_{ab} = \Big(\delta^c_a\pounds_n - 2P^{bc}\mathrm{U}_{ab}\Big)\Big(\pounds_n(\mathfrak{w}_c - \mu\ell_c) - 2P^{fd}\mathrm{U}_{cf}(\mathfrak{w}_d - \mu\ell_d)\Big). \tag{4.97}
$$

This implies that all the terms in the RHS of (4.96) cancel each other and we arrive at $\mathcal{I}_{ab}n^b = 0$. Contracting the LHS of (4.78) with $n^b$ and using Lemma 4.25 [Eq. (4.88)] applied to $\theta_b = \mathfrak{S}_{bc}n^c$ gives $\pounds_n(\mathfrak{S}_{ab}n^b) - \overset{\circ}{\nabla}_a(\mathfrak{S}(n,n)) + (\operatorname{tr}_P \mathbf{U})\mathfrak{S}_{ab}n^b$, so item 2. follows.

Finally, to prove item 3. we note that now $\Sigma_{abc} = 0$ and $W^a = \dfrac{1}{2}n(\mathfrak{p} - \mu\ell^{(2)})n^a$, so $\pounds_W\gamma_{ab} = n(\mathfrak{p} - \mu\ell^{(2)})\mathrm{U}_{ab}$. In addition, $\mathfrak{w} = \mu\boldsymbol{\ell}$ implies $\overset{\circ}{\nabla}_a\mathfrak{w}_b \overset{(2.39)}{=} \mu(\mathrm{F}_{ab} - \ell^{(2)}\mathrm{U}_{ab})$, so (4.95) becomes

$$
\begin{aligned}
\mathcal{I}_{ab} = {} & \Big((\mathfrak{p} - \mu\ell^{(2)})\operatorname{tr}_P \mathbf{U} + (\mathfrak{p} - \mu\ell^{(2)})\kappa + \frac{1}{2}n(\mathfrak{p} - \mu\ell^{(2)})\Big)\mathrm{U}_{ab} + \pounds_n\Big((\mathfrak{p} - \mu\ell^{(2)})\mathrm{U}_{ab}\Big) \\
& - \frac{1}{2}n(\mathfrak{p} - \mu\ell^{(2)})\mathrm{U}_{ab} - 2(\mathfrak{p} - \mu\ell^{(2)})P^{cd}\mathrm{U}_{c(a}\mathrm{U}_{b)d}.
\end{aligned}
$$

Taking the trace with respect to $P^{ab}$ and using (2.51) in Lemma 2.6 with $T_{ab} = \mathcal{I}_{ab}$,

$$P^{ab}\mathcal{I}_{ab} = \Big( (\mathfrak{p} - \mu\ell^{(2)}) \operatorname{tr}_P \mathbf{U} + (\mathfrak{p} - \mu\ell^{(2)})\kappa + n(\mathfrak{p} - \mu\ell^{(2)}) \Big) \operatorname{tr}_P \mathbf{U} + (\mathfrak{p} - \mu\ell^{(2)})\pounds_n(\operatorname{tr}_P \mathbf{U}).$$

Contracting the LHS of (4.78) with $P^{ab}$ and using again identity (2.51) now with $T_{ab} = \mathfrak{S}_{ab}$ gives $2\pounds_n(\operatorname{tr}_P \mathfrak{S}) + 2(\kappa + \operatorname{tr}_P \mathbf{U}) \operatorname{tr}_P \mathfrak{S}$, which proves item 3. of the corollary. □

We can summarize all the results of this section in the following theorem.

**Theorem 4.27.** *Let $\{\mathcal{H}, \gamma, \boldsymbol{\ell}, \ell^{(2)}, \mathbf{Y}\}$ be null hypersurface data $(\Phi, \xi)$-embedded in $(\mathcal{M}, g)$ with $R_{\alpha\beta} = \lambda g_{\alpha\beta}$ in a neighbourhood[4] of $\Phi(\mathcal{H})$ and let $\eta$ be a vector field on $\mathcal{M}$ satisfying $\mathfrak{Q} \stackrel{\mathcal{H}}{=} 0$, $\mathfrak{S}(n,n) = 0$, $\mathcal{K}_{ab}n^b = 0$ and $\widehat{\mathcal{K}} = 0$. Assume that $\mathcal{H}$ admits a cross-section and let $\mu$ be a constant satisfying $\lambda\mu = 0$. Then,*

1. *If $P^{ab}\mathcal{K}_{ab} = (\mathfrak{n} - 1)\mu$ and $\pounds_n(P^{ab}\mathcal{K}_{ab}) = 0$ on a cross-section of $\mathcal{H}$, then $\mathcal{K}_{ab} = \mu\gamma_{ab}$ everywhere on $\mathcal{H}$. Moreover, if $\mathfrak{w}(n) = \mu$ on a cross-section, then $\mathfrak{w}(n) = \mu$ everywhere.*

2. *If in addition to item 1. one has $\mathfrak{w}_c = \mu\ell_c$ and $\mathfrak{S}_{bc}n^b = 0$ on a cross-section of $\mathcal{H}$, then $\mathfrak{w}_c = \mu\ell_c$ and $\mathfrak{S}_{bc}n^b = 0$ everywhere on $\mathcal{H}$.*

3. *Moreover, if in addition to items 1. and 2. one has $\mathfrak{p} = \mu\ell^{(2)}$ and $\operatorname{tr}_P \mathfrak{S} = 0$ on a cross-section of $\mathcal{H}$, then $\mathfrak{p} = \mu\ell^{(2)}$ and $\operatorname{tr}_P \mathfrak{S} = 0$ everywhere on $\mathcal{H}$.*

*Proof.* From the first equation in Corollary 4.26, $\operatorname{tr}_P \mathcal{K}$ satisfies a second order homogeneous transport equation along $n$. So, if $\operatorname{tr}_P \mathcal{K} = (\mathfrak{n} - 1)\mu$ and $\pounds_n(\operatorname{tr}_P \mathcal{K}) = 0$ on a cross-section of $\mathcal{H}$ it follows that $\operatorname{tr}_P \mathcal{K} = (\mathfrak{n} - 1)\mu$ everywhere on $\mathcal{H}$, and hence by decomposition (4.90) we conclude $\mathcal{K} = \mu\gamma$ everywhere. As a consequence of item 1. in Lemma 4.20, if $\mathfrak{w}(n) = \mu$ on a cross-section, then $\mathfrak{w}(n) = \mu$ everywhere. This proves item 1. of the theorem. The second item follows at once from item 2. in Lemma 4.20 and item 2. in Corollary 4.26, because $\mathfrak{w}_c - \mu\ell_c$ and $\mathfrak{S}_{bc}n^b$ satisfy a homogeneous system of transport equations and they vanish initially. Similarly, item 3. follows from the third item in Lemma 4.20 and item 3. in Corollary 4.26, since $\mathfrak{p} - \mu\ell^{(2)}$ and $\operatorname{tr}_P \mathfrak{S}$ also satisfy a homogeneous system and also vanish initially. □

## 4.6 CHARACTERISTIC HOMOTHETIC KID PROBLEM

In this section we present the homothetic (including Killing) KID problem for two characteristic hypersurfaces in the language of hypersurface data. In particular, we want to show that the initial data for the KID problem is at the same level as the characteristic data, meaning that all the extra restrictions one needs to impose can be written solely in terms of double null data (i.e. at the abstract level). The existence problem for Killing vectors in the characteristic case has already been addressed in [74] (see the review in Section 4.1). As we discussed there, in that work the authors assume the characteristic data to be embedded in a vacuum spacetime and find necessary and sufficient conditions that guarantee the existence of an ambient Killing vector (Theorem 4.2). As already mentioned,

4 As before, this condition is only needed up to first derivatives on $\Phi(\mathcal{H})$.

these conditions (cf. (4.5)-(4.6)) are written in a particular coordinate system and involve the ambient Levi-Civita connection and Riemann tensor, so in principle it is not clear whether they are at the level of the initial data or not. In this work we generalize the KID equations to the homothetic case and we show that, both in the homothetic and Killing cases, they can be written in terms of abstract data at the hypersurfaces.

Let $\{\mathcal{D}, \underline{\mathcal{D}}, \sigma, \phi\}$ be double null data satisfying the constraint equations $\boldsymbol{\mathcal{R}} = \lambda\gamma$ and $\underline{\boldsymbol{\mathcal{R}}} = \lambda\underline{\gamma}$. Consider two pairs $(\bar{\eta}, C) \in \mathfrak{X}(\mathcal{H}) \times \mathcal{F}(\mathcal{H})$ and $(\underline{\bar{\eta}}, \underline{C}) \in \mathfrak{X}(\underline{\mathcal{H}}) \times \mathcal{F}(\underline{\mathcal{H}})$ which on the respective boundaries $\partial\mathcal{H}$ and $\partial\underline{\mathcal{H}}$ are related by the $\partial$-isometry $\Psi$, namely $\Psi((\bar{\eta}, C)) \stackrel{\partial\underline{\mathcal{H}}}{=} (\underline{\bar{\eta}}, \underline{C})$. By Theorem 3.23 there exists a $\lambda$-vacuum spacetime $(\mathcal{M}, g)$ where $\{\mathcal{D}, \underline{\mathcal{D}}, \sigma, \phi\}$ is embedded (with respective embeddings $\Phi$ and $\underline{\Phi}$), and the abstract condition $\Psi((\bar{\eta}, C)) \stackrel{\partial\underline{\mathcal{H}}}{=} (\underline{\bar{\eta}}, \underline{C})$ guarantees that the vector field $\eta|_{\Phi(\mathcal{H})} := C\xi + \Phi_\star\bar{\eta}$ agrees with $\underline{\eta}|_{\underline{\Phi}(\underline{\mathcal{H}})} := \underline{C}\underline{\xi} + \underline{\Phi}_\star\underline{\bar{\eta}}$ on $\mathcal{S}$. Then, from the well-posedness of the wave equation with data on two intersecting characteristic hypersurfaces (see Theorem 3.2 in Section 3.1) one can construct a vector field $\eta$ on $(\mathcal{M}, g)$ by solving $\Box_g \eta = -\lambda\eta$ (i.e. $\mathfrak{Q} = 0$, see (4.14)) with initial data $\eta|_{\Phi(\mathcal{H})} = C\xi + \Phi_\star\bar{\eta}$ and $\eta|_{\underline{\Phi}(\underline{\mathcal{H}})} = \underline{C}\underline{\xi} + \underline{\Phi}_\star\underline{\bar{\eta}}$. The equation obtained from (4.21) after setting $\mathfrak{Q} = 0$ is homogeneous, so $\eta$ will be a homothety in $(\mathcal{M}, g)$ (i.e. satisfies $\mathcal{K} = \mu g$ on $\mathcal{M}$ for some $\mu \in \mathbb{R}$) if and only if one can guarantee that $\mathcal{K} = \mu g$ along $\Phi(\mathcal{H}) \cup \underline{\Phi}(\underline{\mathcal{H}})$. Thus, one must look for sufficient conditions on $\{\mathcal{D}, \underline{\mathcal{D}}, \sigma, \phi\}$ (i.e. at the abstract level) that a posteriori imply $\mathcal{K}|_{\Phi(\mathcal{H})} = \mu g|_{\Phi(\mathcal{H})}$ and $\mathcal{K}|_{\underline{\Phi}(\underline{\mathcal{H}})} = \mu g|_{\underline{\Phi}(\underline{\mathcal{H}})}$ once the data is embedded.

In Theorem 4.27 we have found sufficient conditions (they are also necessary) that the tensors $\mathcal{K}_{ab}$, $\mathfrak{w}_a$, $\mathfrak{p}$ and $\mathfrak{S}$ must satisfy on $\mathcal{H}$ and $\underline{\mathcal{H}}$ separately, and also on $\mathcal{S}$, in order to have $\mathcal{K} = \mu g$ at $\Phi(\mathcal{H}) \cup \underline{\Phi}(\underline{\mathcal{H}})$. On the full $\mathcal{H}$ the conditions are

$$1.\ \mathfrak{S}(n, n) = 0, \qquad 2.\ \mathcal{K}_{ab} n^b = 0, \qquad 3.\ \widehat{\mathcal{K}} = 0,$$

and similarly on $\underline{\mathcal{H}}$, while on $\mathcal{S}$ the conditions are

| | | | |
|---|---|---|---|
| 4. $P^{ab}\mathcal{K}_{ab} = (\mathfrak{n} - 1)\mu$, | 8. $\mathfrak{w}(n) = \mu$, | 12. $\mathfrak{S}_{ab} n^b = 0$, | |
| 5. $\underline{P}^{ab}\mathcal{K}_{ab} = (\mathfrak{n} - 1)\mu$, | 9. $\underline{\mathfrak{w}}(\underline{n}) = \mu$, | 13. $\underline{\mathfrak{S}}_{ab} \underline{n}^b = 0$, | 16. $\operatorname{tr}_P \mathfrak{S} = 0$, |
| 6. $\pounds_n(P^{ab}\mathcal{K}_{ab}) = 0$, | 10. $\mathfrak{w} = \mu\boldsymbol{\ell}$, | 14. $\mathfrak{p} = \mu\ell^{(2)}$, | 17. $\operatorname{tr}_{\underline{P}} \underline{\mathfrak{S}} = 0$. |
| 7. $\pounds_{\underline{n}}(\underline{P}^{ab}\underline{\mathcal{K}}_{ab}) = 0$, | 11. $\underline{\mathfrak{w}} = \mu\underline{\boldsymbol{\ell}}$, | 15. $\underline{\mathfrak{p}} = \mu\underline{\ell}^{(2)}$, | |

These conditions come from the analysis on $\Phi(\mathcal{H})$ and $\underline{\Phi}(\underline{\mathcal{H}})$ as separate null hypersurfaces. However, they are glued to each other across their boundaries via $\Psi$. This makes some of these conditions redundant. Our task now is to identify a complete subset of sufficient conditions and to prove that they can be written solely in terms of double null data. It is clear that conditions 1., 2., 3. and their underlined versions must necessarily be included in the set of sufficient conditions. However, as already mentioned many of the conditions on $\mathcal{S}$ are redundant because they are related to information coming from the other hypersurface. To get some intuition let us for the moment simplify the problem and assume a choice of riggings satisfying $\nu \stackrel{\mathcal{S}}{=} \sigma\underline{\xi}$ and $\underline{\nu} \stackrel{\mathcal{S}}{=} \sigma\xi$, which at the abstract level means $\ell^{(2)} \stackrel{\partial\mathcal{H}}{=} 0$, $\ell_A \stackrel{\partial\mathcal{H}}{=} 0$, $\underline{\ell}^{(2)} \stackrel{\partial\underline{\mathcal{H}}}{=} 0$ and $\underline{\ell}_A \stackrel{\partial\mathcal{H}}{=} 0$. With this choice of gauge $P^{ab} \stackrel{\mathcal{S}}{=} h^{AB}e^a_A e^b_B \stackrel{\mathcal{S}}{=} \underline{P}^{ab}$ (see (2.141)), so it is clear that condition 4. implies condition 5. Similarly, condition 8. implies condition 9., because $\mu \stackrel{\mathcal{S}}{=} \mathcal{K}(\xi,\nu) \stackrel{\mathcal{S}}{=} \mathcal{K}(\sigma^{-1}\underline{\nu},\sigma\underline{\xi}) \stackrel{\mathcal{S}}{=} \mathcal{K}(\underline{\nu},\underline{\xi})$, and conditions 10., 11., 14. and 15. are automatically fulfilled provided 2. and 8. hold. We discuss e.g. 15. with similar arguments applying to the other three cases. Condition 2. implies $\mathcal{K}(\nu,\nu)=0$ and since $\mathcal{K}(\nu,\nu) \stackrel{\mathcal{S}}{=} \mathcal{K}(\sigma^{-1}\underline{\xi},\sigma^{-1}\underline{\xi})$ one gets $\underline{\mathfrak{p}} \stackrel{\partial\mathcal{H}}{=} 0$, and by our choice of gauge $\underline{\ell}^{(2)} \stackrel{\partial\mathcal{H}}{=} 0$, so 15. follows. From (4.9) one can also guess that conditions 12. and 13. are not independent, so most likely only one of them needs to be imposed. That 16. and 17. are also redundant is less intuitive but we shall prove this below in an arbitrary gauge. In summary, the suggestion is that the complete subset of necessary conditions are 1., 2. and 3. on $\mathcal{H}$, their underlined versions on $\underline{\mathcal{H}}$, and 4., 6., 7., 8. and 12. on $\mathcal{S}$. We now take on the task of recasting these conditions in terms of double null data in an arbitrary gauge and of proving that they are indeed sufficient to guarantee $\mathcal{K}=\mu g$ on $\Phi(\mathcal{H})\cup\underline{\Phi}(\underline{\mathcal{H}})$. Let us start with the former. That 2., 3., 4., 6. and 7. are writeable in terms of double null data is clear from (4.25). In the following lemma we show that 1., 8. and 12. can also be written in terms of the data.

**Lemma 4.28.** *Let $\{\mathcal{D},\underline{\mathcal{D}},\sigma,\phi\}$ be double null data embedded in a $\lambda$-vacuum spacetime $(\mathcal{M},g)$ and let $\eta\in\mathfrak{X}(\mathcal{M})$. Define $C,\bar{\eta},\underline{C},\underline{\bar{\eta}}$ by $\eta|_{\mathcal{H}} = C\xi+\Phi_\star\bar{\eta}$ and $\eta|_{\underline{\mathcal{H}}} = \underline{C}\underline{\xi}+\underline{\Phi}_\star\underline{\bar{\eta}}$. Then,*

$$\mathfrak{S}(n,n) \stackrel{\mathcal{H}}{=} -C\left(\lambda + P^{bc}A_{bca}n^a + P(\mathbf{r}+\mathbf{s},\mathbf{r}+\mathbf{s}) - \frac{1}{2}\kappa n(\ell^{(2)}) - 2\pounds^{(2)}_n\ell^{(2)}\right) - n^a n^b \pounds_{\bar{\eta}}\mathrm{Y}_{ab}$$
$$- \left(\ell^{(2)}n(C) + (\pounds_{\bar{\eta}}\boldsymbol{\ell})(n)\right)\kappa + n(\ell^{(2)})n(C) + \ell^{(2)}\pounds^{(2)}_n C + \pounds_n((\pounds_{\bar{\eta}}\boldsymbol{\ell})(n)), \quad (4.98)$$

*with analogous expression on $\underline{\mathcal{H}}$, and*

$$\mathfrak{w}(n) \stackrel{\mathcal{S}}{=} \ell^{(2)}n(C) + \frac{1}{2}Cn(\ell^{(2)}) - (\mathbf{r}-\mathbf{s})(\bar{\eta}) + \boldsymbol{\ell}(\pounds_n\bar{\eta}) + \underline{\ell}^{(2)}\underline{n}(\underline{C}) + \frac{1}{2}\underline{C}\underline{n}(\underline{\ell}^{(2)}) - (\underline{\mathbf{r}}-\underline{\mathbf{s}})(\underline{\bar{\eta}})$$
$$+ \underline{\boldsymbol{\ell}}(\pounds_{\underline{n}}\underline{\bar{\eta}}) - \theta^a\left(\overset{\circ}{\nabla}_a C + C(\mathrm{r}-\mathrm{s})_a - \mathrm{U}_{ac}\bar{\eta}^c\right)$$
$$+ \sigma^{-1}\underline{\theta}^b\left(\underline{n}(\underline{C})\underline{\ell}_b + \underline{C}(\underline{\mathrm{r}}+\underline{\mathrm{s}})_b + \underline{\gamma}_{bc}\pounds_{\underline{n}}\underline{\bar{\eta}}^c + \underline{\mathrm{U}}_{ab}\underline{\bar{\eta}}^a\right), \quad (4.99)$$

$$\mathfrak{S}_{ab}n^b \stackrel{\mathcal{S}}{=} -C\Big(\lambda\ell_a + P^{bc}A_{bca} + P^{bc}(\mathrm{r}_b+\mathrm{s}_b)(\mathrm{Y}_{ac}+\mathrm{F}_{ac}) - \frac{1}{2}\kappa\overset{\circ}{\nabla}_a\ell^{(2)} - 2\overset{\circ}{\nabla}_a(n(\ell^{(2)}))$$
$$+ 2P^{bc}\mathrm{U}_{ab}\overset{\circ}{\nabla}_c\ell^{(2)}\Big) - n^b\pounds_{\bar{\eta}}\mathrm{Y}_{ab} + \left(\ell^{(2)}n(C) + (\pounds_{\bar{\eta}}\boldsymbol{\ell})(n)\right)(\mathrm{r}_a - \mathrm{s}_a)$$
$$+ \frac{1}{2}n(\ell^{(2)})\overset{\circ}{\nabla}_a C + \frac{1}{2}n(C)\overset{\circ}{\nabla}_a\ell^{(2)} + \ell^{(2)}\left(\overset{\circ}{\nabla}_a(n(C)) - P^{bc}\mathrm{U}_{ab}\overset{\circ}{\nabla}_c C\right)$$
$$+ \frac{1}{2}\pounds_n\pounds_{\bar{\eta}}\ell_a + \frac{1}{2}\overset{\circ}{\nabla}_a((\pounds_{\bar{\eta}}\boldsymbol{\ell})(n)) - P^{bc}\mathrm{U}_{ab}\pounds_{\bar{\eta}}\ell_c. \quad (4.100)$$

*Proof.* Relations (4.98) and (4.100) follow immediately from (4.73) and (4.72) after inserting $R_{\alpha\beta} = \lambda g_{\alpha\beta}$. To prove (4.99) recall that (see (3.16))

$$\nu \stackrel{\mathcal{S}}{=} \sigma(\underline{\xi} + \underline{\Phi}_\star\underline{\theta}), \qquad \underline{\nu} \stackrel{\mathcal{S}}{=} \sigma(\xi + \Phi_\star\theta), \quad (4.101)$$

with $\theta$ and $\underline{\theta}$ given by

$$\theta = -\frac{1}{2}(\ell^{(2)} - \ell^{(2)}_{\sharp})n - \ell^A e_A, \qquad \underline{\theta} = \frac{1}{2}(\underline{\ell}^{(2)} - \underline{\ell}^{(2)}_{\sharp})\underline{n} - \underline{\ell}^A \underline{e}_A,$$

from where it follows

$$\begin{aligned}\mathfrak{v}(n) &= \nu^\alpha \xi^\beta \nabla_\alpha \eta_\beta + \nu^\alpha \xi^\beta \nabla_\beta \eta_\alpha \\ &\stackrel{\mathcal{S}}{=} \nu^\alpha \xi^\beta \nabla_\alpha \eta_\beta + \underline{\xi}^\alpha \underline{\nu}^\beta \nabla_\beta \eta_\alpha - \nu^\alpha \theta^\beta \nabla_\beta \eta_\alpha + \sigma^{-1} \underline{\theta}^\alpha \underline{\nu}^\beta \nabla_\beta \eta_\alpha \\ &\stackrel{\mathcal{S}}{=} n^b ({}^{(2)}\nabla \boldsymbol{\eta})_b + \underline{n}^b ({}^{(2)}\nabla \boldsymbol{\eta})_b - \theta^a n^b (\nabla \boldsymbol{\eta})_{ab} + \sigma^{-1} \underline{n}^a \underline{\theta}^b (\nabla \boldsymbol{\eta})_{ab}. \end{aligned} \tag{4.102}$$

So we need to compute the tensors $(\nabla\boldsymbol{\eta})_{ab}$ and $({}^{(2)}\nabla\boldsymbol{\eta})_b$ and their contractions with $n$. For $(\nabla\boldsymbol{\eta})_{ab}$ we apply equation (B.1) in Proposition B.1 for $T = \boldsymbol{\eta}$ to get

$$\begin{aligned}(\nabla\boldsymbol{\eta})_{ab} &= \overset{\circ}{\nabla}_a \boldsymbol{\eta}_b + \mathrm{Y}_{ab} \boldsymbol{\eta}_c n^c + \mathrm{U}_{ab} \boldsymbol{\eta}(\xi) \\ &\stackrel{(4.24)}{=} \overset{\circ}{\nabla}_a \Big( C\ell_b + \gamma_{bc} \bar{\eta}^c \Big) + C\mathrm{Y}_{ab} + \Big( C\ell^{(2)} + \boldsymbol{\ell}(\bar{\eta}) \Big) \mathrm{U}_{ab} \\ &= \ell_b \overset{\circ}{\nabla}_a C + C(\mathrm{Y}_{ab} + \mathrm{F}_{ab}) - \ell_b \mathrm{U}_{ac} \bar{\eta}^c + \gamma_{bc} \overset{\circ}{\nabla}_a \bar{\eta}^c, \end{aligned}$$

where in the last line we used (2.38)-(2.39). The contractions with $n^a$ and $n^b$ follow after using $n^a \overset{\circ}{\nabla}_a \bar{\eta}^c = \pounds_n \bar{\eta}^c + \bar{\eta}^a \overset{\circ}{\nabla}_a n^c$ and (2.66),

$$n^a (\nabla\boldsymbol{\eta})_{ab} = n(C)\ell_b + C(\mathrm{r}+\mathrm{s})_b + \gamma_{bc} \pounds_n \bar{\eta}^c + \mathrm{U}_{ab} \bar{\eta}^a, \tag{4.103}$$

$$n^b (\nabla\boldsymbol{\eta})_{ab} = \overset{\circ}{\nabla}_a C + C(\mathrm{r}-\mathrm{s})_a - \mathrm{U}_{ac} \bar{\eta}^c. \tag{4.104}$$

To compute $({}^{(2)}\nabla\boldsymbol{\eta})_b$ we use equation (B.3) in Proposition B.1 for $T = \boldsymbol{\eta}$ and $j = 1$ together with (4.24) to get

$$\begin{aligned}({}^{(2)}\nabla\boldsymbol{\eta})_b &= \overset{\circ}{\nabla}_b \Big( C\ell^{(2)} + \boldsymbol{\ell}(\bar{\eta}) \Big) - \Big( C\ell^{(2)} + \boldsymbol{\ell}(\bar{\eta}) \Big)(\mathrm{r}-\mathrm{s})_b - V^c{}_b \Big( C\ell_c + \gamma_{cd} \bar{\eta}^d \Big) \\ &= \ell^{(2)} \overset{\circ}{\nabla}_b C + \frac{1}{2} C \overset{\circ}{\nabla}_b \ell^{(2)} + \overset{\circ}{\nabla}_b (\boldsymbol{\ell}(\bar{\eta})) - (\mathrm{Y}_{bd} + \mathrm{F}_{bd}) \bar{\eta}^d, \end{aligned}$$

where in the second equality we used (2.55)-(2.56). Contracting this with $n^b$ and using $\pounds_n(\boldsymbol{\ell}(\bar{\eta})) \stackrel{(2.22)}{=} 2\mathbf{s}(\bar{\eta}) + \ell_a \pounds_n \bar{\eta}^a$ we arrive at

$$n^b ({}^{(2)}\nabla\boldsymbol{\eta})_b = \ell^{(2)} n(C) + \frac{1}{2} C n(\ell^{(2)}) - (\mathbf{r}-\mathbf{s})(\bar{\eta}) + \boldsymbol{\ell}(\pounds_n \bar{\eta}). \tag{4.105}$$

Inserting (4.103)-(4.105) into (4.102) yields the result. □

**Remark 4.29.** *Note that in a gauge where* $\ell^{(2)} \stackrel{\partial\mathcal{H}}{=} 0$, $\ell_A \stackrel{\partial\mathcal{H}}{=} 0$, $\underline{\ell}^{(2)} \stackrel{\partial\underline{\mathcal{H}}}{=} 0$ *and* $\underline{\ell}_A \stackrel{\partial\underline{\mathcal{H}}}{=} 0$ *(and hence* $\nu \stackrel{\mathcal{S}}{=} \sigma\underline{\xi}$ *and* $\underline{\nu} \stackrel{\mathcal{S}}{=} \sigma\xi$*) the expression for* $\mathfrak{v}(n)$ *at* $\mathcal{S}$ *simplifies drastically. We keep the gauge arbitrary so that Theorem 4.31 below is written in a gauge-covariant form.*

From this lemma together with expression (4.25) it follows that conditions 1., 2., 3. on $\mathcal{H}$ can be written in terms of the data as

$$
\begin{aligned}
0 &\overset{\mathcal{H}}{=} -C\left(\lambda + P^{bc}A_{bca}n^a + P(\mathbf{r}+\mathbf{s},\mathbf{r}+\mathbf{s}) - \frac{1}{2}\kappa n(\ell^{(2)}) - 2\pounds_n^{(2)}\ell^{(2)}\right) - n^a n^b \pounds_{\bar\eta}\mathrm{Y}_{ab} \\
&\quad - \left(\ell^{(2)}n(C) + (\pounds_{\bar\eta}\boldsymbol{\ell})(n)\right)\kappa + n(\ell^{(2)})n(C) + \ell^{(2)}\pounds_n^{(2)}C + \pounds_n((\pounds_{\bar\eta}\boldsymbol{\ell})(n)), && (4.106)\\
0 &\overset{\mathcal{H}}{=} 2C\mathrm{r}_a + \overset{\circ}{\nabla}_a C + n(C)\ell_a + n^b\pounds_{\bar\eta}\gamma_{ab}, && (4.107)\\
0 &\overset{\mathcal{H}}{=} 2C\mathrm{Y}_{ab} + 2\ell_{(a}\overset{\circ}{\nabla}_{b)}C + \pounds_{\bar\eta}\gamma_{ab} - \frac{1}{\mathfrak{n}-1}\left(2C\operatorname{tr}_P\mathbf{Y} - 2\ell^{(2)}n(C) + P^{ab}\pounds_{\bar\eta}\gamma_{ab}\right)\gamma_{ab}, && (4.108)
\end{aligned}
$$

where in the last one we used $\widehat{\mathcal{K}} = \mathcal{K} - \frac{\operatorname{tr}_P\mathcal{K}}{\mathfrak{n}-1}\gamma$ because $\mathcal{K}_{ab}n^b = 0$. On $\underline{\mathcal{H}}$ the equations are analogous, while on $\partial\mathcal{H}$ and $\partial\underline{\mathcal{H}}$ conditions 4., 6., 7., 8. and 12. can be written as

$$
\begin{aligned}
(\mathfrak{n}-1)\mu &\overset{\partial\mathcal{H}}{=} 2C\operatorname{tr}_P\mathbf{Y} - 2\ell^{(2)}n(C) + P^{ab}\pounds_{\bar\eta}\gamma_{ab}, && (4.109)\\
0 &\overset{\partial\mathcal{H}}{=} \pounds_n\left(C\operatorname{tr}_P\mathbf{Y} - \ell^{(2)}n(C) + \frac{1}{2}P^{ab}\pounds_{\bar\eta}\gamma_{ab}\right), && (4.110)\\
0 &\overset{\partial\underline{\mathcal{H}}}{=} \pounds_{\underline{n}}\left(\underline{C}\operatorname{tr}_P\underline{\mathbf{Y}} - \underline{\ell}^{(2)}\underline{n}(\underline{C}) + \frac{1}{2}\underline{P}^{ab}\pounds_{\underline{\bar\eta}}\underline{\gamma}_{ab}\right), && (4.111)
\end{aligned}
$$

$$
\begin{aligned}
\mu &\overset{\partial\mathcal{H}}{=} \ell^{(2)}n(C) + \frac{1}{2}Cn(\ell^{(2)}) - (\mathbf{r}-\mathbf{s})(\bar\eta) + \boldsymbol{\ell}(\pounds_n\bar\eta) - \theta^a\left(\overset{\circ}{\nabla}_a C + C(\mathrm{r}-\mathrm{s})_a - \mathrm{U}_{ac}\bar\eta^c\right)\\
&\quad + \phi^\star\left(\underline{\ell}^{(2)}\underline{n}(\underline{C}) + \frac{1}{2}\underline{C}\underline{n}(\underline{\ell}^{(2)}) - (\underline{\mathbf{r}}-\underline{\mathbf{s}})(\underline{\bar\eta}) + \underline{\boldsymbol{\ell}}(\pounds_{\underline{n}}\underline{\bar\eta})\right)\\
&\quad + \phi^\star\left(\sigma^{-1}\underline{\theta}^b\left(\underline{n}(\underline{C})\underline{\ell}_b + \underline{C}(\underline{\mathrm{r}}+\underline{\mathrm{s}})_b + \underline{\gamma}_{bc}\pounds_{\underline{n}}\underline{\bar\eta}^c + \underline{\mathrm{U}}_{ab}\underline{\bar\eta}^a\right)\right), && (4.112)\\
0 &\overset{\partial\mathcal{H}}{=} -C\left(\lambda\ell_a + P^{bc}A_{bca} + P^{bc}(\mathrm{r}_b+\mathrm{s}_b)\Pi_{ac} - \frac{1}{2}\kappa\overset{\circ}{\nabla}_a\ell^{(2)} - 2\overset{\circ}{\nabla}_a(n(\ell^{(2)})) + 2P^{bc}\mathrm{U}_{ab}\overset{\circ}{\nabla}_c\ell^{(2)}\right)\\
&\quad - n^b\pounds_{\bar\eta}\mathrm{Y}_{ab} + \left(\ell^{(2)}n(C) + (\pounds_{\bar\eta}\boldsymbol{\ell})(n)\right)(\mathrm{r}_a - \mathrm{s}_a) + \frac{1}{2}n(\ell^{(2)})\overset{\circ}{\nabla}_a C + \frac{1}{2}n(C)\overset{\circ}{\nabla}_a\ell^{(2)}\\
&\quad + \ell^{(2)}\left(\overset{\circ}{\nabla}_a(n(C)) - P^{bc}\mathrm{U}_{ab}\overset{\circ}{\nabla}_c C\right) + \frac{1}{2}\pounds_n\pounds_{\bar\eta}\ell_a + \frac{1}{2}\overset{\circ}{\nabla}_a((\pounds_{\bar\eta}\boldsymbol{\ell})(n)) - P^{bc}\mathrm{U}_{ab}\pounds_{\bar\eta}\ell_c. && (4.113)
\end{aligned}
$$

Next we prove that whenever the data is embedded, these conditions imply the rest on $\mathcal{S}$, and thus that $\mathcal{K}$ is pure trace on $\Phi(\mathcal{H})\cup\underline{\Phi}(\underline{\mathcal{H}})$.

**Lemma 4.30.** *Let $\{\mathcal{D},\underline{\mathcal{D}},\sigma,\phi\}$ be double null data embedded in a $\lambda$-vacuum spacetime and satisfying* (4.106)-(4.113)*. Then, conditions 5., 9., 10., 11., 13., 14., 15., 16. and 17. hold on $\mathcal{S}$, and as a consequence, $\mathcal{K}_{\alpha\beta} = \mu g_{\alpha\beta}$ on $\Phi(\mathcal{H})\cup\underline{\Phi}(\underline{\mathcal{H}})$.*

*Proof.* Since the data satisfies (4.106)-(4.113) the embedded conditions 1., 2., 3., 4., 6., 7., 8. and 12. are true. Let us prove that they imply conditions 5., 9., 10., 11., 13., 14., 15., 16. and 17. on $\mathcal{S}$. We start with condition 5. From $P^{ab}\mathcal{K}_{ab} \overset{\mathcal{S}}{=} (\mathfrak{n}-1)\mu$, $\mathcal{K}_{ab}n^b = 0$ and using the decomposition (2.140) it follows $h^{AB}\mathcal{K}_{AB} = (\mathfrak{n}-1)\mu$ on $\mathcal{S}$, and hence condition $\underline{P}^{ab}\underline{\mathcal{K}}_{ab} = (\mathfrak{n}-1)\mu$ is automatically fulfilled on $\mathcal{S}$. To prove 9. note that conditions 1., 2., 3., 4. and 6. imply, by item 1. in Theorem 4.27, that $\mathcal{K}_{ab} = \mu\gamma_{ab}$ on $\mathcal{H}$ (and analogously $\underline{\mathcal{K}}_{ab} = \mu\underline{\gamma}_{ab}$ on $\underline{\mathcal{H}}$). Then, using (4.101) $\boldsymbol{\mathfrak{w}}(n) \overset{\mathcal{S}}{=} \mathcal{K}(\nu,\xi) \overset{\mathcal{S}}{=} \mathcal{K}(\underline{\xi} + \Phi_\star\underline{\theta},\underline{\nu}) - \mathcal{K}(\nu,\Phi_\star\theta) \overset{\mathcal{S}}{=} \mathcal{K}(\underline{\xi},\underline{\nu}) \overset{\mathcal{S}}{=} \underline{\boldsymbol{\mathfrak{w}}}(\underline{n})$.

So item 1. in Theorem 4.27 gives $\mathfrak{w}(n) = \mu$ (resp. $\underline{\mathfrak{w}}(\underline{n}) = \mu$) everywhere on $\mathcal{H}$ (resp. on $\underline{\mathcal{H}}$). Note also that $\mathfrak{w}_a = \mu \ell_a$ on $\mathcal{S}$ because $\mathfrak{w}(n) \stackrel{\mathcal{S}}{=} \mu = \mu \boldsymbol{\ell}(n)$ and

$$\mathfrak{w}(e_A) = \mathcal{K}(\xi, \Phi_\star e_A) \stackrel{\mathcal{S}}{=} \mathcal{K}\Big(\sigma^{-1}\underline{\nu}, \underline{\Phi}_\star \underline{e}_A\Big) - \mathcal{K}\Big(\Phi_\star \theta, \Phi_\star e_A\Big) \stackrel{\mathcal{S}}{=} \mu h_{BA}\ell^B \stackrel{\mathcal{S}}{=} \mu \ell_A.$$

Similarly $\underline{\mathfrak{w}}_a = \mu \underline{\ell}_a$ on $\mathcal{S}$, which proves items 10. and 11. To show 13. note that from the symmetry (4.9), (4.101) and (4.31) it follows that

$$\begin{aligned}
\underline{\mathfrak{S}}(\underline{n}, \underline{e}_A) = \Sigma(\underline{\xi}, \underline{\nu}, \underline{\Phi}_\star \underline{e}_A) &\stackrel{\mathcal{S}}{=} -\Sigma(\underline{\nu}, \underline{\xi}, \underline{\Phi}_\star \underline{e}_A) + (\nabla \mathcal{K})(\underline{\Phi}_\star \underline{e}_A, \underline{\xi}, \underline{\nu}) \\
&\stackrel{\mathcal{S}}{=} -\Sigma(\xi, \nu, \Phi_\star e_A) + \Sigma(\Phi_\star \theta, \nu, \Phi_\star e_A) + \Sigma(\underline{\nu}, \underline{\Phi}_\star \underline{\theta}, \underline{\Phi}_\star \underline{e}_A) \\
&\quad + ({}^{(2)}\nabla \underline{\mathcal{K}})_{ab} \underline{e}_A^a \underline{n}^b \\
&\stackrel{\mathcal{S}}{=} -\mathfrak{S}(n, e_A) + \Sigma(\theta, n, e_A) + \Sigma(\underline{n}, \underline{\theta}, \underline{e}_A) - \underline{P}^{bc} \underline{\mathrm{U}}_{ca} \underline{\mathfrak{w}}_b \underline{e}_A^a \stackrel{\mathcal{S}}{=} 0,
\end{aligned}$$

where in the third equality we used $\underline{P}^{bc} \underline{\mathrm{U}}_{ca} \underline{\mathfrak{w}}_b \stackrel{\partial\mathcal{H}}{=} \mu \underline{P}^{bc} \underline{\mathrm{U}}_{ca} \underline{\ell}_b \stackrel{\partial\mathcal{H}}{=} 0$, $\mathfrak{S}(n, e_A) \stackrel{\partial\mathcal{H}}{=} 0$ (because of 12.) and that $\underline{\Sigma}(\underline{n}, \underline{\theta}, \underline{e}_A) \stackrel{\partial\mathcal{H}}{=} 0$ and $\Sigma(\theta, n, e_A) \stackrel{\partial\mathcal{H}}{=} 0$ (see Rmk. 4.10). By item 2. of Theorem 4.27 we have that $\mathfrak{w} = \mu \boldsymbol{\ell}$ and $\mathfrak{S}(n, \cdot) = 0$ on $\mathcal{H}$, and $\underline{\mathfrak{w}} = \mu \underline{\boldsymbol{\ell}}$ and $\underline{\mathfrak{S}}(\underline{n}, \cdot) = 0$ on $\underline{\mathcal{H}}$.

From $\mathcal{K}(\underline{\nu}, \underline{\nu}) \stackrel{\mathcal{S}}{=} 0$ and (4.101) it follows that

$$0 = \sigma^{-2} \mathcal{K}(\underline{\nu}, \underline{\nu}) \stackrel{\mathcal{S}}{=} \mathcal{K}(\xi, \xi) + 2\mathcal{K}(\xi, \Phi_\star \theta) + \mathcal{K}(\Phi_\star \theta, \Phi_\star \theta) \stackrel{\mathcal{S}}{=} \mathfrak{p} + 2\mathfrak{w}(\theta) + \mathcal{K}(\theta, \theta),$$

which after inserting $\mathfrak{w} = \mu \boldsymbol{\ell}$, $\mathcal{K}_{ab} = \mu \gamma_{ab}$ and (3.17) gives

$$0 \stackrel{\partial\mathcal{H}}{=} \mathfrak{p} + 2\mu \boldsymbol{\ell}(\theta) + \mu \gamma(\theta, \theta) \stackrel{\partial\mathcal{H}}{=} \mathfrak{p} + \mu\Big( - (\ell^{(2)} - \ell^{(2)}_\sharp) - 2\ell^{(2)}_\sharp + \ell^{(2)}_\sharp\Big) \stackrel{\partial\mathcal{H}}{=} \mathfrak{p} - \mu \ell^{(2)},$$

so $\mathfrak{p} \stackrel{\partial\mathcal{H}}{=} \mu \ell^{(2)}$. Similarly $\underline{\mathfrak{p}} \stackrel{\partial\underline{\mathcal{H}}}{=} \mu \underline{\ell}^{(2)}$. Let us conclude by proving conditions 16. and 17. (we just prove 16. since 17. is analogous). Using again that $\mathfrak{w} = \mu \boldsymbol{\ell}$ and $\mathcal{K}_{ab} = \mu \gamma_{ab}$ it follows that

$$\begin{aligned}
({}^{(2)}\nabla \mathcal{K})_{ab} &\stackrel{(4.29)}{=} \mu\Big(\mathrm{F}_{ab} - \ell^{(2)} \mathrm{U}_{ab}\Big) + \mu \mathrm{Y}_{ab} + \mathfrak{p} \mathrm{U}_{ab} - \mu (\mathrm{r} - \mathrm{s})_a \ell_b - \mu V^c{}_a \gamma_{bc} \\
&\stackrel{(2.56)}{=} \mu(\mathrm{F}_{ab} + \mathrm{Y}_{ab}) + (\mathfrak{p} - \mu \ell^{(2)}) \mathrm{U}_{ab} - \mu(\mathrm{r} - \mathrm{s})_a \ell_b - \mu(\mathrm{F}_{ab} + \mathrm{Y}_{ab}) + \mu(\mathrm{r} - \mathrm{s})_a \ell_b \\
&= (\mathfrak{p} - \mu \ell^{(2)}) \mathrm{U}_{ab},
\end{aligned}$$

which in particular implies $({}^{(2)}\nabla \mathcal{K})_{ab} \stackrel{\partial\mathcal{H}}{=} 0$. Now, from (4.9), (4.101) and using $\Sigma_{abc}|_{\partial\mathcal{H}} = 0$ and $\underline{\Sigma}_{abc}|_{\partial\underline{\mathcal{H}}} = 0$ (see Remark 4.10) we get

$$\begin{aligned}
\Sigma(\xi, \Phi_\star e_A, \Phi_\star e_B) &\stackrel{\mathcal{S}}{=} -\Sigma(\Phi_\star e_A, \xi, \Phi_\star e_B) + (\nabla \mathcal{K})(\Phi_\star e_B, \xi, \Phi_\star e_A) \\
&\stackrel{\mathcal{S}}{=} -\Sigma(\underline{\Phi}_\star \underline{e}_A, \sigma^{-1}\underline{\nu}, \underline{\Phi}_\star \underline{e}_B) + \Sigma(\Phi_\star e_A, \Phi_\star \theta, \Phi_\star e_B) + ({}^{(2)}\nabla \mathcal{K})_{ab} e_A^a e_B^b \stackrel{\mathcal{S}}{=} 0.
\end{aligned}$$

This, combined with the decomposition (2.140) and $\mathfrak{S}(n, \cdot) \stackrel{\partial\mathcal{H}}{=} 0$ implies $\operatorname{tr}_P \mathfrak{S} \stackrel{\partial\mathcal{H}}{=} 0$, and hence by item 3. in Theorem 4.27 it follows that $\mathfrak{p} = \mu \ell^{(2)}$ and $\operatorname{tr}_P \mathfrak{S} = 0$ everywhere on $\mathcal{H}$ (and similarly $\underline{\mathfrak{p}} = \mu \underline{\ell}^{(2)}$ and $\operatorname{tr}_{\underline{P}} \underline{\mathfrak{S}} = 0$ on $\underline{\mathcal{H}}$). □

Thus, conditions (4.106)-(4.113) are everything one needs to impose to guarantee $\mathcal{K}_{\alpha\beta} = \mu g_{\alpha\beta}$ on $\Phi(\mathcal{H}) \cup \underline{\Phi}(\underline{\mathcal{H}})$, and hence $\pounds_\eta g = \mu g$ on $\mathcal{M}$. This concludes the proof of the main result of the section.

**Theorem 4.31.** *Let $\{\mathcal{D}, \underline{\mathcal{D}}, \sigma, \phi\}$ be double null data satisfying the constraint equations* (3.58)*. Let $(\bar{\eta}, C) \in \mathfrak{X}(\mathcal{H}) \times \mathcal{F}(\mathcal{H})$ and $(\underline{\bar{\eta}}, \underline{C}) \in \mathfrak{X}(\underline{\mathcal{H}}) \times \mathcal{F}(\underline{\mathcal{H}})$ satisfy $\Psi(\bar{\eta}, C) \overset{\partial\mathcal{H}}{=} (\underline{\bar{\eta}}, \underline{C})$. Assume conditions* (4.106)-(4.108) *on $\mathcal{H}$, the analogous ones in $\underline{\mathcal{H}}$, and conditions* (4.109)-(4.113) *on $\partial\mathcal{H}$ with $\mu \in \mathbb{R}$ satisfying $\lambda\mu = 0$. Then, there exists a $\lambda$-vacuum development $(\mathcal{M}, g)$ of $\{\mathcal{D}, \underline{\mathcal{D}}, \sigma, \phi\}$ and a vector field $\eta$ on $\mathcal{M}$ satisfying $\pounds_\eta g = \mu g$ such that $\eta|_{\Phi(\mathcal{H})} = C\xi + \bar{\eta}$ and $\eta|_{\Phi(\underline{\mathcal{H}})} = C\xi + \bar{\eta}$.*

**Remark 4.32.** *In my opinion this theorem is relevant for at least two reasons. Firstly because the homothetic KID equations are written solely in terms of the initial data for the characteristic problem (and hence detached from the spacetime). This allows one to decide whether an initial data set will give rise to an ambient spacetime admitting a homothetic/Killing vector before solving the Einstein equations. The second reason is that both the constraints and the KID equations are fully gauge and diffeomorphism covariant, which makes it possible to choose the gauge and the coordinates at will.*

As mentioned in the introduction, in [275, 276] the author analyzes general conditions under which the characteristic KID problem can be extended to include matter fields. However, these conditions are not expressed in terms of the initial data for the Einstein equations. An interesting open direction is to reformulate the necessary and sufficient conditions of [275, 276] directly in terms of double null data. I plan to address this in future work.

## 4.7 APPLICATIONS TO OTHER CAUCHY PROBLEMS

In this section I illustrate another application of the identities found in the previous sections, namely the homothetic KID problem in the context of smooth spacelike-characteristic initial data. Given a smooth achronal hypersurface $\mathcal{H}$ with normal $\nu$ embedded in a Lorentzian manifold $(\mathcal{M}, g)$, the equation $\Box_g F + L_1 F = f$, where $L_1$ is a a first order differential operator with smooth coefficients and $f$ is also smooth, admits a unique solution on the domain of dependence of $\mathcal{H}$ in $\mathcal{M}$ with initial data given by $(F|_{\mathcal{H}}, \nu(F)|_{\mathcal{H}})$ at the points where the hypersurface is spacelike and $F|_{\mathcal{H}}$ where it is null [172]. As far as we are aware, such a result is not yet known for arbitrary tensor fields, but it is believed to be true. Hence, one can also expect that the Cauchy problem for the Einstein equations with initial data posed on a smooth spacelike-null hypersurface is also well-posed. Addressing these issues is beyond the scope of this thesis, so I simply quote the following conjectures we made in [234].

**Conjecture 4.33.** *Let $\mathcal{H}$ be a smooth achronal hypersurface embedded in a Lorentzian manifold $(\mathcal{M}, g)$ with normal $\nu$. Let $L_1$ be a first order differential operator with smooth coefficients and $f$, $T_0$ and $T_1$ smooth $(p, q)$ tensor fields. Then there exists a unique $(p, q)$ smooth tensor field $T$ satisfying $\Box_g T + L_1 T = f$ on the domain of dependence of $\mathcal{H}$ in $\mathcal{M}$ with initial conditions $(T, \pounds_\nu T)|_{\mathcal{H}} = (T_0, T_1)$ at the points where $\mathcal{H}$ is spacelike and $T|_{\mathcal{H}} = T_0$ where it is null.*

**Conjecture 4.34.** *Let* $\{\mathcal{H}, \boldsymbol{\gamma}, \boldsymbol{\ell}, \ell^{(2)}, \mathbf{Y}\}$ *be hypersurface data satisfying* $n^{(2)} \leq 0$ *such that the tensor* $\mathcal{A}$ *defined in* (2.1) *is Lorentzian. Assume the constraint equations* (4.57)-(4.58) *are fulfilled on* $\mathcal{H}_S := \{n^{(2)} < 0\}$ *and* $\boldsymbol{\mathcal{R}} = \lambda\boldsymbol{\gamma}$ *on* $\mathcal{H}_0 := \{n^{(2)} = 0\}$*. Then there exists a* $\lambda$*-vacuum spacetime* $(\mathcal{M}, g)$*, embedding* $\Phi : \mathcal{H} \hookrightarrow \mathcal{M}$ *and rigging* $\xi$ *such that* $\{\mathcal{H}, \boldsymbol{\gamma}, \boldsymbol{\ell}, \ell^{(2)}, \mathbf{Y}\}$ *(possibly shrinking the data in the null part if necessary) is* $(\Phi, \xi)$*-embedded.*

Assuming these results to be true we can determine the necessary and sufficient conditions on $\{\mathcal{H}, \boldsymbol{\gamma}, \boldsymbol{\ell}, \ell^{(2)}, \mathbf{Y}\}$ for the existence of a homothetic vector field $\eta$ on $(\mathcal{M}, g)$. From the results of the previous sections it seems reasonable to impose $\mathcal{K}_{ab} = \mu\gamma_{ab}$ and $(\pounds_\xi\mathcal{K})_{ab} = 2\mu\mathrm{Y}_{ab}$ on $\mathcal{H}_S$ and also $\mathcal{K}_{ab}n^b = 0$, $\widehat{\mathcal{K}} = 0$ and $\mathfrak{S}(n,n) = 0$ on $\mathcal{H}_0$. As we show next, the remaining conditions needed on $\partial\mathcal{H}_S$ for the argument to work are automatically fulfilled.

**Theorem 4.35.** *Let* $\{\mathcal{H}, \boldsymbol{\gamma}, \boldsymbol{\ell}, \ell^{(2)}, \mathbf{Y}\}$ *be hypersurface data with the properties listed in Conjecture 4.34 and assume further that the boundary of* $\mathcal{H}_S$ *is a smooth cross-section of* $\mathcal{H}_0$*. Suppose the constraint equations* (4.57)-(4.58) *are satisfied on* $\mathcal{H}_S$ *and* $\boldsymbol{\mathcal{R}} = \lambda\boldsymbol{\gamma}$ *on* $\mathcal{H}_0$*. Assume that* $(C, \bar{\eta}) \in \mathcal{F}(\mathcal{H}) \times \mathfrak{X}(\mathcal{H})$ *satisfy* (4.61) *and* (4.63) *on* $\mathcal{H}_S$ *and* (4.106)-(4.108) *on* $\mathcal{H}_0$ *with* $\mu \in \mathbb{R}$ *such that* $\lambda\mu = 0$*. Then, assuming the validity of Conjectures 4.33 and 4.34, there exists a* $\lambda$*-vacuum spacetime* $(\mathcal{M}, g)$*, vector field* $\eta \in \mathfrak{X}(\mathcal{M})$*, embedding* $\Phi : \mathcal{H} \hookrightarrow \mathcal{M}$ *and rigging* $\xi$ *such that* $\{\mathcal{H}, \boldsymbol{\gamma}, \boldsymbol{\ell}, \ell^{(2)}, \mathbf{Y}\}$ *(possible shrunk if necessary) is* $(\Phi, \xi)$*-embedded on* $(\mathcal{M}, g)$*,* $\pounds_\eta g = \mu g$ *and* $\eta|_{\Phi(\mathcal{H})} = C\xi + \Phi_\star\bar{\eta}$*.*

*Proof.* By Conjecture 4.34 there exists a $\lambda$-vacuum spacetime $(\mathcal{M}, g)$, embedding $\Phi : \mathcal{H} \hookrightarrow \mathcal{M}$ and rigging $\xi$ such that $\{\mathcal{H}, \boldsymbol{\gamma}, \boldsymbol{\ell}, \ell^{(2)}, \mathbf{Y}\}$ is $(\Phi, \xi)$-embedded on $(\mathcal{M}, g)$. Let us extend $\xi$ to a neighbourhood of $\Phi(\mathcal{H})$ by $\nabla_\xi\xi = 0$ and construct a vector field $\eta \in \mathfrak{X}(\mathcal{M})$ by solving $\Box_g\eta = -\lambda\eta$ (i.e. $\mathfrak{Q} = 0$, see (4.14)) with initial data $\eta = C\xi + \Phi_\star\bar{\eta}$ on $\Phi(\mathcal{H})$ and (4.64) on $\Phi(\mathcal{H}_S)$. Under this condition, (4.21) becomes a homogeneous PDE. By the assumed validity of Conjecture 4.33 we only need to show that $\mathcal{K}_{\alpha\beta} = \mu g_{\alpha\beta}$ on $\Phi(\mathcal{H})$ and $\pounds_\xi\mathcal{K}_{\alpha\beta} = \mu\pounds_\xi g_{\alpha\beta}$ on $\Phi(\mathcal{H}_S)$. By the same argument as the one employed in Theorem 4.16 it follows $\mathcal{K}_{\alpha\beta} = \mu g_{\alpha\beta}$ and $\pounds_\xi\mathcal{K}_{\alpha\beta} = \mu\pounds_\xi g_{\alpha\beta}$ on $\Phi(\mathcal{H}_S)$. So, we only need to show that $\mathcal{K}_{\alpha\beta} = \mu g_{\alpha\beta}$ on the null region. If $\mathcal{H}_0$ has empty interior it follows by continuity. Otherwise, by Theorem 4.27 it suffices to prove that (i) $P^{ab}\mathcal{K}_{ab} = (\mathfrak{n}-1)\mu$, (ii) $\pounds_n(P^{ab}\mathcal{K}_{ab}) = 0$, (iii) $\mathfrak{w}_a = \mu\ell_a$, (iv) $\mathfrak{p} = \mu\ell^{(2)}$, (v) $\mathfrak{S}(n,\cdot) = 0$ and (vi) $\operatorname{tr}_P\mathfrak{S} = 0$ on $\partial\mathcal{H}_S$. Since $\mathcal{K}_{ab} = \mu\gamma_{ab}$, $\mathfrak{w}_a = \mu\ell_a$ and $\mathfrak{p} = \mu\ell^{(2)}$ on $\mathcal{H}_S$ and the hypersurface data is smooth it is in particular continuous and hence the three conditions also hold on the boundary $\partial\mathcal{H}_S$. So items (i), (iii) and (iv) are automatically fulfilled. Item (ii) follows after noting that again by continuity the vector $n$ cannot vanish in a neighbourhood of $\partial\mathcal{H}_S$, and since

$$\pounds_n(P^{ab}\mathcal{K}_{ab}) \stackrel{\mathcal{H}_S}{=} \mu\pounds_n(P^{ab}\gamma_{ab}) \stackrel{\mathcal{H}_S}{=} \mu\pounds_n(n^{(2)}\ell^{(2)}) \stackrel{\mathcal{H}_S}{=} \mu n^{(2)}n(\ell^{(2)}) + \mu\ell^{(2)}n(n^{(2)}),$$

it follows that $\pounds_n(P^{ab}\mathcal{K}_{ab}) \stackrel{\partial\mathcal{H}_S}{=} 0$, since $n^{(2)} \stackrel{\partial\mathcal{H}_S}{=} 0$ and $n(n^{(2)}) \stackrel{\partial\mathcal{H}_S}{=} 0$ (because $n^{(2)}$ is identically zero on $\mathcal{H}_0$). To prove items (v) and (vi) observe that since $\mathcal{K}_{\alpha\beta} = \mu g_{\alpha\beta}$ and $\pounds_\xi\mathcal{K}_{\alpha\beta} = \mu\pounds_\xi g_{\alpha\beta}$ on $\Phi(\mathcal{H}_S)$ then the full tensor $\Sigma$ vanishes on $\Phi(\mathcal{H}_S)$, so by continuity $\Sigma$ also vanishes on $\Phi(\partial\mathcal{H}_S)$, which in particular implies $\mathfrak{S}(n,\cdot) = 0$ and $\operatorname{tr}_P\mathfrak{S} = 0$ on $\partial\mathcal{H}_S$. Applying Theorem 4.27 it follows $\mathcal{K}_{\alpha\beta} = \mu g_{\alpha\beta}$ on $\Phi(\mathcal{H}_0)$, which proves the theorem. □

Another natural application of the results in this chapter is the spacelike-characteristic KID problem with corners. As mentioned in Section 3.1, the corresponding Cauchy problem is known to be well-posed (see e.g. [75, 91]). To analyze this situation from a detached viewpoint, however, one must first introduce an abstract notion of initial data, analogous to the construction in Section 3.2 for the characteristic case. In the spacelike-null setting, there are two possible ways of gluing the hypersurfaces, essentially because equation (2.123) admits two solutions (in contrast with the unique solution in the null-null case; see Remarks 2.24 and 2.25). Once the Cauchy problem is formulated in this detached framework, the experience acquired in this chapter suggests that the required KID equations reduce to (4.61) and (4.63) in the spacelike region and (4.106)-(4.108) in the null region, without any additional constraints on the intersection surface. Establishing this rigorously, however, requires substantial further work and lies beyond the scope of this thesis. This is an interesting direction for future research.

# 5

# TRANSVERSE EXPANSION AT NULL HYPERSURFACES

In this chapter we analyze the transverse expansion of the metric at a general null hypersurface. From a causal perspective, the absence of a second null hypersurface to pose data implies that extra information coming from the past can potentially influence the solution, hence spoiling uniqueness. However, in cases where the solution exhibits some symmetry, such as Killing vectors, initial data on a single null hypersurface may suffice to establish a unique solution to the Einstein equations. This topic has been extensively studied in recent years, for example when the hypersurface is a homothetic or a Killing horizon, as we review in Section 5.1. Motivated by these examples, in Section 5.2 we prove general identities that relate the transverse derivatives of the ambient Ricci tensor with the transverse expansion of the metric at a general null hypersurface. Our analysis is coordinate free, does not require any field equations and holds regardless the signature of the ambient metric, the dimension or the topology of the hypersurface. With these identities at hand, in Section 5.3 we prove general existence and uniqueness results concerning data at a single null hypersurface. In Section 5.4 we analyze the case where the ambient manifold admits a preferred vector field $\eta$ that is null and tangent to the hypersurface (e.g. a Killing or a homothetic vector field) and whose deformation tensor $\pounds_{\eta} g$ is known. Finally, we particularize these results to the Killing horizon case in Section 5.5. The results in this chapter have been published in [235, 236].

## 5.1 PRELIMINARIES

Let us begin by reviewing several known results where the transverse derivatives of the metric are obtained at a horizon. We begin in Subsection 5.1.1 with the homothetic horizon case, and then we move on to the non-degenerate and degenerate Killing horizon cases in Subsections 5.1.2 and 5.1.3, respectively. We emphasize that the following examples are not covered by the characteristic Cauchy problem presented in Chapter 3.

### 5.1.1 *Fefferman-Graham ambient space*

In a seminal paper [112] (later expanded into the monograph [113]), Fefferman and Graham developed a formalism to study conformal invariants from a differential geometry perspective. This formalism is also of significant importance in several physically motivated research areas such as holography (see [259] and references therein). As we shall see in Section 6.1, the Fefferman-Graham ambient metric has also been used to study the asymptotic Cauchy problem in higher dimensions.

The key idea is based on the observation that the lightcone in the $d$-dimensional Minkowski spacetime encodes the full conformal class of the sphere $\mathbb{S}^{d-2}$ [303]. In order to generalize this idea to general conformal manifolds $(S^{d-2},[h])$ of arbitrary signature, the authors first construct a null manifold (analogous to the lightcone in Minkowski) by lifting $(S^{d-2},[h])$, and then embed it as a homothetic horizon into a $d$-dimensional ambient manifold. Finally, by imposing Ricci flatness at such horizon, the authors construct the so-called ambient metric as a formal series. Let us now review in more detail this construction. The strategy we present differs slightly from the original one in order to simplify the exposition.

Let $(S,[h])$ be an $r := d-2$ dimensional smooth conformal manifold of signature $(p,q)$. Consider the product manifold $\mathbb{R}^+ \times S$ and denote by $t$ the coordinate in its first factor. Then, after picking up a representative $h \in [h]$ one introduces the vector field $T := t\partial_t$ and the degenerate bilinear form $\boldsymbol{\sigma} := t^2\pi^\star h$, where $\pi : \mathbb{R}^+ \times S \longrightarrow S$ is the projection onto the second factor. Note that $\pounds_T \boldsymbol{\sigma} = 2\boldsymbol{\sigma}$.

Now consider the space $\mathbb{R} \times \mathbb{R}^+ \times S$, and denote the coordinate in the first factor by $\rho$. The idea is to extend $T$ trivially off $\mathbb{R}^+ \times S$, and to construct a smooth metric $g$ of signature $(p+1, q+1)$ in a neighbourhood of $\rho = 0$, the so-called *ambient metric*, such that (i) $T$ is a homothety of $g$, (ii) the pullback of $g$ into $\mathbb{R}^+ \times S$ is $\boldsymbol{\sigma}$ and (iii) $g$ is Ricci-flat to infinite order at $\rho = 0$. As shown in [113], it is sufficient to consider ambient metrics written in *normal form*, i.e. such that at each point $(\rho = 0, t, x) \in \{0\} \times \mathbb{R}^+ \times S$ the metric is of the form

$$g = 2tdtd\rho + \boldsymbol{\sigma},$$

and the vector $\partial_\rho$ is geodesic. The reason is that every ambient metric can be diffeomorphically mapped into one that is written in normal form. Taking all these considerations into account, the line element in a neighbourhood of $\rho = 0$ can be written as

$$g = 2adt^2 + 2tdtd\rho + 2tb_A dtdx^A + t^2\mu, \tag{5.1}$$

where $a(\rho, x^A)$, $b_A(\rho, x^A)$ and $\mu(\rho, x^A)$ are to be determined as a series in $\rho$ by imposing Ricci flatness order by order at $\rho = 0$. We use the notation $\mu^{(m)}$, $a^{(m)}$, $b^{(m)}$ to denote the $m$-th term of the expansion, i.e. $\mu^{(m)} := \partial_\rho^{(m)}\mu|_{\rho=0}$, etc. By construction, one has $a^{(0)} = 0$, $b^{(0)} = 0$ and $\mu^{(0)} = h$.

The idea to obtain the expansion $\{a^{(m)}, b^{(m)}, \mu^{(m)}\}_{m\geq 1}$ is as follows. First, the $(t,t)$ and $(t,A)$ components of the ambient Ricci tensor at $\rho = 0$ are

$$R_{tt} \stackrel{\rho=0}{=} \frac{r}{t^2}(1 - a^{(1)}), \qquad R_{tA} \stackrel{\rho=0}{=} \frac{1}{2t}\Big(2\nabla^h_A a^{(1)} - r b^{(1)}_A\Big),$$

where $\nabla^h$ is the Levi-Civita connection of $h$. It is clear that only the choice $a^{(1)} = 1$ and $b^{(1)}_A = 0$ makes $R_{tt} \overset{\rho=0}{=} 0$ and $R_{tA} \overset{\rho=0}{=} 0$. After inserting them into the $(A, B)$ components of the ambient Ricci tensor at $\rho = 0$ one arrives at

$$R_{AB} \overset{\rho=0}{=} \left(1 - \frac{r}{2}\right) \mu^{(1)}_{AB} - \frac{r}{2} h^{CD} \mu^{(1)}_{CD} h_{AB} + R^h_{AB}, \tag{5.2}$$

where $R^h_{AB}$ is the Ricci tensor of $h$. Taking trace w.r.t. $h$,

$$h^{AB} R_{AB} \overset{\rho=0}{=} -(r-1) h^{AB} \mu^{(1)}_{AB} + R^h,$$

so there is a unique choice of $h^{AB}\mu^{(1)}_{AB}$ that makes $h^{AB} R_{AB} \overset{\rho=0}{=} 0$. Inserting it into (5.2) yields

$$R_{AB} \overset{\rho=0}{=} \left(1 - \frac{r}{2}\right) \mu^{(1)}_{AB} - \frac{R^h}{2(r-1)} h_{AB} + R^h_{AB},$$

so whenever $r > 2$ one can uniquely choose $\mu^{(1)}_{AB}$ to make $R_{AB} \overset{\rho=0}{=} 0$, and when $r = 2$ equation $R_{AB} \overset{\rho=0}{=} 0$ is automatically fulfilled because in dimension two every metric satisfies $R^h_{AB} = \frac{R^h}{2} h_{AB}$. Now suppose that $a^{(k)}$, $b^{(k)}$, $\mu^{(k)}$ $(1 \leq k \leq m)$ have already been determined. The $(t, t)$ component of the ambient Ricci tensor then reads

$$\partial^m_\rho R_{tt} \overset{\rho=0}{=} \frac{2m - r}{t^2} a^{(m+1)} + \mathcal{L}^{(m)},$$

where $\mathcal{L}^{(m)}$ gathers lower order terms, i.e. terms depending on $a^{(m)}$, $b^{(m)}$, $\mu^{(m)}$ and below. For every $m$ when $r$ is odd, and for every $m < \frac{r}{2}$ when $r$ is even, one can uniquely choose $a^{(m+1)}$ to make $\partial^m_\rho R_{tt} \overset{\rho=0}{=} 0$. Substituting it into the $(t, A)$ components ($\mathcal{L}^{(m)}_A$ are again lower order terms),

$$\partial^m_\rho R_{tA} \overset{\rho=0}{=} \frac{2m - r}{2t} b^{(m+1)}_A + \frac{1}{t} \nabla_A a^{(m+1)} + \mathcal{L}^{(m)}_A,$$

allows one to determine $b^{(m+1)}_A$ so that $\partial^m_\rho R_{tA} \overset{\rho=0}{=} 0$. Finally, the $(\rho, \rho)$ and $(A, B)$ components read (note the difference in the order of the transverse derivative in the left hand side)

$$\partial^{m-1}_\rho R_{\rho\rho} \overset{\rho=0}{=} -\frac{1}{2} h^{AB} \mu^{(m+1)}_{AB} + \widehat{\mathcal{L}}^{(m)} \tag{5.3}$$

and

$$\partial^m_\rho R_{AB} \overset{\rho=0}{=} \left(m + 1 - \frac{r}{2}\right) \mu^{(m+1)}_{AB} - \frac{1}{2} (h^{CD} \mu^{(m+1)}_{CD}) h_{AB} + \nabla_{(A} b^{(m+1)}_{B)} + 2a^{(m+1)} L^h_{AB} + \mathcal{L}^{(m)}_{AB}, \tag{5.4}$$

where $L^h_{AB}$ is the Schouten tensor of $h$,

$$L^h_{AB} = \frac{1}{r-2} \left( R^h_{AB} - \frac{R^h}{2(r-1)} h_{AB} \right),$$

and as before $\widehat{\mathcal{L}}^{(m)}$ and $\mathcal{L}^{(m)}_{AB}$ gather all the lower order terms, i.e. the terms depending on $a^{(m)}$, $b^{(m)}$, $\mu^{(m)}$ and below. From (5.3) one can always choose $h^{AB}\mu^{(m+1)}_{AB}$ so that $\partial^{m-1}_\rho R_{\rho\rho} \overset{\rho=0}{=} 0$ for every $m \geq 1$. The contracted Bianchi identity then implies that

$h^{AB}\partial_\rho^m R_{AB} \overset{\rho=0}{=} 0$, $\partial_\rho^{m-1}R_{\rho t} \overset{\rho=0}{=} 0$ and $\partial_\rho^{m-1}R_{\rho A} \overset{\rho=0}{=} 0$ are automatically fulfilled. As a consequence, the only non-trivial equation corresponds to the trace-free part of (5.4) w.r.t. $h$. Whenever $m < r/2-1$ (for $r$ even) and for all $m$ (for $r$ odd), one can uniquely choose $(\mu_{AB}^{(m+1)})^{TF}$ to make $(\partial_\rho^m R_{AB})^{TF} \overset{\rho=0}{=} 0$ ("TF" denotes the trace-free part w.r.t. $h$). So, for $r$ odd, the Einstein equations determine the full transverse expansion of the metric, $\{a^{(m)}, b^{(m)}, \mu^{(m)}\}_{m\geq 1}$, while for $r$ even the expansion is determined up to and including order $\{a^{(r/2)}, b_A^{(r/2)}, \mu_{AB}^{(r/2-1)}\}$, because only the trace of the coefficient $\mu_{AB}^{(r/2)}$ can be determined from the equations. Both in the odd and even cases, $a=\rho$ and $b=0$ hold up to the order they are determined, i.e. $a^{(m)} = \delta_1^m$ and $b_A^{(m)} = 0$ for all $m$ when $r$ is odd, and for all $m = 0, ..., r/2$ when $r$ is even. In general, the metrics constructed in this way are called *ambient metrics*.

As already mentioned, for $r$ even the determination of $\mu^{(r/2)}$ is obstructed due to the vanishing of the coefficient in front of $\mu_{AB}^{(r/2)}$ in (5.4) when $m = \frac{r}{2}-1$. Therefore, this equation does not fix the trace-free part of the term $\mu_{AB}^{(r/2)}$, which can therefore be freely specified in the form of a $(0,2)$ symmetric, traceless tensor $\Theta_{AB}$. In addition, $(\partial_\rho^{r/2-1}R_{AB})^{TF} \overset{\rho=0}{=} 0$ holds if and only if $(\mathcal{L}_{AB}^{(r/2-1)})^{TF}$ vanishes identically. This tensor (which in general does not vanish) only depends on the initial metric $h$ and it is called *obstruction tensor* $\mathcal{O}_{AB}^{\mathscr{F}\mathscr{G}}$. It has the following properties (indices $A, B, ...$ are raised and lowered with the metric $h$): (i) it is traceless $\mathcal{O}^{\mathscr{F}\mathscr{G}A}{}_A = 0$, (ii) divergence-free $\nabla_A^h \mathcal{O}^{\mathscr{F}\mathscr{G}A}{}_B = 0$, (iii) conformally covariant and (iv) it vanishes if $h$ is Einstein (but not only). As an example let us study the first non-trivial obstruction tensor, i.e. the one appearing for[1] $r=4$. Equation (5.4) for $m=1$ is

$$\partial_\rho R_{AB} \overset{\rho=0}{=} \left(2 - \frac{r}{2}\right)\mu_{AB}^{(2)} + 2B_{AB}^h - 2(r-4)L_A^{h\,C}L_{BC}^h,$$

where $B^h$ is the Bach tensor of $h$,

$$B_{AB}^h := W_A^{h\,C}{}_B{}^D L_{CD}^h + h^{CD}\nabla_C^h\nabla_D^h L_{AB}^h - h^{CD}\nabla_C^h\nabla_A^h L_{BD}^h, \tag{5.5}$$

and $W_{ABCD}^h$ is the Weyl of $h$. For $r \neq 4$, this equation determines the coefficient $\mu^{(2)}$, but for $r=4$ the trace-free part of $\mu^{(2)}$ remains undetermined and the ambient metric can only be Ricci flat provided the Bach tensor of $h$ (which is the obstruction tensor in dimension 4) vanishes.

The presence of an obstruction implies that for $r \geq 4$ even, it is not possible in general to have an ambient metric that is simultaneously more than $r/2-1$ times differentiable and satisfy $R_{\alpha\beta} \overset{\rho=0}{=} 0$ to infinite order. If one insists on forcing $R_{\alpha\beta} \overset{\rho=0}{=} 0$ to infinite order (or, more strongly, in a neighbourhood of the homothetic horizon) then one must include logarithmic terms into the expansion that compensate the presence of the obstruction tensor, but at the same time spoil the smoothness of the ambient metric at the horizon $\rho=0$.

1 As already discussed, for $r=2$ equation $R_{AB} \overset{\rho=0}{=} 0$ holds automatically, and hence there is no obstruction tensor when $r=2$.

When $r$ is odd, there is no obstruction tensor, but one can naturally consider solutions with expansions involving half-integral powers of $\rho$ which also have an indeterminacy at order $r/2$ and that satisfy $R_{\alpha\beta} \overset{\rho=0}{=} 0$ to infinite order. These ambient metrics are, in general, of class $\mathcal{C}^{\lfloor r/2 \rfloor -1}$ at $\rho = 0$. To sum up, ambient metrics satisfying $R_{\alpha\beta} \overset{\rho=0}{=} 0$ to infinite order present the following differentiability: for $r = 2$ they are smooth; when $r \geq 4$ is even they are, generically, no more than $r/2 - 1$ times differentiable; and for $r \geq 3$ odd they are smooth provided there are no half-integer powers, or $\lfloor r/2 \rfloor - 1$ times differentiable otherwise.

Given analytic data $(h, \Theta)$, standard convergence techniques establish existence of the ambient metric [34], even in the presence of a non-vanishing obstruction tensor [192]. In odd dimensions, the works [19, 20, 185] show existence of ambient metrics beyond the analytic setting. In a remarkable breakthrough, the authors of [285] study a class of metrics, called *self-similar*, defined as metrics admitting a homothetic horizon and solving *exactly* the Einstein equations on the hyperbolic region, i.e. where the homothety is timelike[2]. They work in double null coordinates in which the metric takes the form

$$g = -2\Delta^2 d\bar{u}d\bar{v} + \slashed{g}_{AB}\Big(dx^A - q^A d\bar{v}\Big)\Big(dx^B - q^B d\bar{v}\Big), \tag{5.6}$$

where $\Delta$, $q$ and $\slashed{g}$ are a function, a one-form and a Riemannian metric on the surfaces $S_{\bar{u},\bar{v}}$, respectively. The homothetic vector in these coordinates is $K = \bar{v}\partial_{\bar{v}} + \bar{u}\partial_{\bar{u}}$ and the homothetic horizon is at $\{\bar{u} = 0\}$. Their strategy is to define characteristic initial data on $\{\bar{u} = 0\} \cup \{\bar{v} = -1\}$ to construct self-similar solutions after applying a suitable scaling limit. On $\{\bar{u} = 0\}$ they specify the lapse $\Delta$, the shift $q$ and the conformal class subject to the normalization conditions $\Delta|_{\bar{u}=0} = 1$ and $q|_{\bar{u}=0} = 0$ (according to [285] the condition on $\Delta$ is not relevant, but the one on $q$ is essential to guarantee the regularity of $\Delta$). In principle, there are many ways to prescribe data on the hypersurface $\{\bar{v} = -1\}$, but after the limiting process involved in the construction of self-similar solutions the authors show that by prescribing solely a Riemannian metric $h$ and symmetric traceless $(0, 2)$-tensor $\Theta$, there exists a unique self-similar, *regular* solution such that (i) $\slashed{g}|_{S_{0,-1}} = h$, (ii) $\Theta$ agrees with the trace-free part of the coefficient $v^{r/2}$ of $\slashed{g}$ at $\bar{u} = 0$ and (iii) $\slashed{g}$ satisfies the normalization conditions $\Delta|_{\bar{u}=0} = 1$ and $q|_{\bar{u}=0} = 0$.

The notion of regularity can be understood as follows. For $r = 2$, $g$ is regular provided it is smooth. When $r \geq 3$ is odd, $g$ is regular provided it is smooth everywhere except at $\bar{u} = 0$, and there exists smooth tensors $\{\overset{(i)}{g}_{\alpha\beta}\}_{i=0}^{\frac{r-1}{2}}$ and $\widetilde{g}_{\alpha\beta}$ such that

$$g_{\alpha\beta} = \sum_{i=0}^{\frac{r-1}{2}} \overset{(i)}{g}_{\alpha\beta}\bar{u}^i + \widetilde{g}_{\alpha\beta}\bar{u}^{\frac{r}{2}} + O\Big(\bar{u}^{\frac{r+1}{2}}\Big).$$

2 In [152] a similar result showing existence in the elliptic region is shown, with the main difference that no free data at order $r/2$ needs to be prescribed in this case.

Finally, when $r \geq 4$ is even, $g$ is regular if it is smooth everywhere except at $\bar{u} = 0$, and there exists smooth tensors $\{g^{(i)}_{\alpha\beta}\}_{i=0}^{\frac{r}{2}}$ and $\widetilde{g}_{\alpha\beta}$ such that

$$g_{\alpha\beta} = \sum_{i=0}^{\frac{r}{2}-1} g^{(i)}_{\alpha\beta}\bar{u}^i + g^{(r/2)}_{\alpha\beta}\bar{u}^{r/2} + \widetilde{g}_{\alpha\beta}\bar{u}^{r/2}\log\bar{u} + O\Big(\bar{u}^{\frac{r}{2}+1}\log(\bar{u})\Big).$$

Observe that this definition of regularity is in complete agreement with the differentiability for Fefferman-Graham ambient metrics discussed above.

One natural question is whether an *exact ambient metric* (i.e. one solving $R_{\alpha\beta} = 0$ in a neighbourhood of the horizon) is self-similar in the sense of [285] and vice-versa. There is one specific situation where exact ambient metrics and self-similar metrics are in one-to-one correspondence, namely when the ambient metric is *straight*.

One says that an ambient metric is straight provided the homothety $T$ satisfies $d\boldsymbol{T} = 0$. Proposition 3.4 of [113] shows that this is equivalent to $\boldsymbol{T}$ being exact in a full neighbourhood of the horizon, and also equivalent to $a = 1$ and $b_A = 0$. In fact, the metric (5.1) with $a = 1$ and $b_A = 0$ satisfies $R_{tt} = R_{t\rho} = R_{tA} = 0$ exactly. As we already discussed, one of the results in [113] is that every ambient metric is straight up to the order determined by the initial metric $h$, i.e. to infinite order for $r$ odd (and no half integers are allowed) and up to and including order $r/2$ when $r$ is even. The question of whether the metric is straight to infinite order depends on the free data $\Theta$. When $r$ is even, the $(r/2-1)$ $\rho$-derivative of the $(\rho, A)$ components of the Einstein equations read

$$\partial_\rho^{r/2-1} R_{\rho A} \overset{\rho=0}{=} \frac{1}{2} b_A^{(r/2+1)} + \nabla^B (\mu^{(r/2)})^{TF}_{AB} - D_A, \tag{5.7}$$

where $D_A$ is a one-form that depends only on $h$. Hence, the condition $\nabla^B\Theta_{AB} = D_A$ is necessary for the ambient metric to be simultaneously Ricci flat and straight to infinite order. As proven in [113, Thm. 3.10], it is also sufficient. When $r$ is odd, the ambient metric is straight to infinite order if and only if $\nabla_B\Theta^{AB} = 0$ (see [113, Thm. 3.9]). One can combine these two conditions into a single one by just writing $\nabla_B\Theta^{AB} = D^A$, letting $D_A$ be identically zero when $r$ is odd.

In the context of self-similar metrics, Proposition B.7 of [285] shows that whenever the free data $\Theta$ satisfies $\nabla_B\Theta^{AB} = D^A$, the metric (5.6) has vanishing torsion one-form $\zeta$ and scalar $\omega$[3]. These are given in terms of $\Delta$ and $q$ by $\omega = -\frac{1}{2}\nabla_4(\log\Delta)$ and $\zeta_A = -\frac{1}{4}\Delta^{-1}e_4(q^B)\not{g}_{AB}$. Thus, the self-similar solutions that satisfy $\nabla^B\Theta_{AB} = D_A$ also have $\Delta = 1$ and $q = 0$ everywhere. The change of coordinates $t = \bar{v}$ and $\rho = \bar{u}\bar{v}^{-1}$ in (5.6) leads to

$$g = -2\rho dt^2 - 2t dt d\rho + \not{g}_{AB} dx^A dx^B,$$

3 In double null coordinates, the torsion $\zeta$ and the function $\omega$ are defined by $\zeta_A := \frac{1}{2}g(\nabla_A e_a, e_3)$ and $\omega := -\frac{1}{4}g(\nabla_4 e_3, e_4)$, where $e_3 := \Delta^{-1}\partial_{\bar{v}}$ and $e_3 := \Delta^{-1}(\partial_{\bar{u}} + q^A\partial_A)$ are null vectors satisfying $g(e_3, e_4) = -2$.

which after identifying $\not{g} = t^2\mu$ happens to be a straight, exact ambient metric. One can rephrase Theorem 1.3 of [285] in terms of straight, exact ambient metrics as follows.

**Theorem 5.1** ([285])**.** *Let $h$ be a Riemannian metric and $\Theta$ a symmetric, traceless tensor field on $S$ that satisfies $\nabla^B\Theta_{AB} = D_A$, where $D_A$ is an explicit one-form that only depends on $h$ when $r$ is even, and identically zero for $r$ odd. Then, there exists a unique straight, exact and regular ambient metric such that $g|_S = h$ and $\Theta$ agrees with the trace-free part of the coefficient $\bar{u}^{r/2}$ of $g$ at the horizon.*

### 5.1.2 *Non-degenerate Killing horizons*

Killing horizons are of an utmost importance in General Relativity. The celebrated no-hair theorem establishes that, under certain global assumptions, stationary, analytic, four-dimensional vacuum spacetimes must be isomeric to Kerr [154, 155] (see also [178, 179] for the previous works by Israel on Schwarzschild and Reissner-Nordstrom black holes). This theorem relies on proving that the event horizon is a Killing horizon and that there exists an extra Killing vector that is axisymmetric (see also [245]). As a consequence, the spacetime is isometric to Kerr [56, 284] (see also [68, 70]). While these results were originally proven for non-degenerate horizons, they were later generalized to the extremal case [15, 73, 117]. There exist analogous results for Einstein-Maxwell theory and the Kerr-Newman family [65, 239]. It is interesting to note that, in the context of extremal charged black holes, the uniqueness theorems allow for Majumdar-Papapetrou solutions [153, 210] that describe an arbitrary number of black holes in equilibrium. When the cosmological constant is present, no direct analogous results are known and only partial results are available [46, 77].

In four spacetime dimensions, the topology of the cross-sections are 2-spheres (see [79, 155]). This is no longer true in higher dimensions [110, 139, 207], where more topologies such as "black rings" with horizon topology $\mathbb{S}^2 \times \mathbb{S}^1$ are allowed [109, 273] (see also [108]). In fact, there are black rings and Myers-Perry solutions (which are the generalization of Kerr to higher dimensions) with the same mass and angular momentum, so these examples show that black hole uniqueness does not hold in higher than four dimensions. Further advances in the classification of higher dimensional black holes can be found in [167–169, 246].

It is also worth mentioning the work by Klainerman-Ionescu [175] and Alexakis-Ionescu-Klainerman [13] on the existence of Hawking's Killing field without assuming analyticity (see also [176]), and an analogous result by Petersen [269] and Petersen-Rácz [268] on compact Cauchy horizons with non-zero surface gravity. In this subsection and the following one, we discuss several approaches aimed at classifying vacuum spacetimes admitting data prescribed on a Killing horizon.

We begin by recalling some well-known properties of Killing horizons [150] that will be used throughout the remainder of this section. In particular, we first recall the precise definition of a Killing horizon. Given a spacetime $(M, g)$ with Killing $\eta$, a Killing horizon $\mathcal{H}$ is a smooth null hypersurface where $\eta|_{\mathcal{H}}$ is nowhere vanishing, null and tangent. The surface gravity of $\eta$ is the scalar field $\widetilde{\kappa}$ on $\mathcal{H}$ defined by $\nabla_\eta \eta \stackrel{\mathcal{H}}{=} \widetilde{\kappa}\eta$. We say that $\mathcal{H}$ is a non-degenerate (also non-extremal) Killing horizon provided $\widetilde{\kappa}$ nowhere vanishes. One says that $\mathcal{H}$ is degenerate (or extremal) if $\widetilde{\kappa} = 0$. In the non-degenerate case one can assume without loss of generality that $\widetilde{\kappa} > 0$ (by simply interchanging $\eta$ by $-\eta$). Recall also that the surface gravity is constant in many situations of physical interest, e.g. when the dominant energy condition holds [35, 323], when $\eta$ is integrable [288], for bifurcate horizons [135, 189, 288] or for multiple Killing horizons [227–229]. It is also well-known that Killing horizons are totally geodesic null hypersurfaces. Furthermore, it is possible to construct a one-form $\varpi$ on $\mathcal{H}$ such that [29, 135]

$$\nabla_X \eta \stackrel{\mathcal{H}}{=} \varpi(X)\eta \tag{5.8}$$

for every $X \in \mathfrak{X}(\mathcal{H})$.

The first attempt to characterize the spacetime geometry from data at a non-degenerate Killing horizon dates back to Moncrief [243] (see also [244]), who developed a technique to construct analytic, four-dimensional vacuum spacetimes with suitable data prescribed at a non-degenerate Killing horizon. The construction relies on a version of the Cauchy-Kowalewski theorem due to Fusaro [136], as we review next.

Consider the metric

$$g = e^{-2\phi}\Big( - N^2 dt^2 + h_{AB} dx^A dx^B \Big) + t^2 e^{2\phi}\Big(dx^3 + \beta_A dx^A\Big)^2,$$

defined on $T^3 \times \mathbb{R}^+$, where $t \in \mathbb{R}^+$ and $\{x^A, x^3\}$ with $A = 1, 2$ are periodic coordinates on $T^3$. Assume also that the coefficients of the metric only depend on $\{t, x^A\}$, $N > 0$ and that for each fixed $t$, $h_{AB}$ is a Riemannian metric. This metric is the most general one on $T^3 \times \mathbb{R}^+$ with $\partial_{x^3}$ as a spacelike Killing vector, $t = const.$ being spacelike hypersurfaces and zero shift. This metric degenerates at $t = 0$, but under a change of coordinates

$$t' := t^2, \qquad x^{3'} := 2x^3 - 2\log t, \qquad x^{A'} := x^A,$$

the transformed metric is smooth (and Lorentzian) in a neighbourhood $T^3 \times (-\varepsilon, \varepsilon)$ of $\mathcal{N} := \{t' = 0\}$ provided (i) the metric coefficients are smooth on $\mathcal{N}$, (ii) $N > 0$ and for each fixed $t'$, $h_{AB}$ is a Riemannian metric, and (iii) $(N^2 - e^{4\phi})/4t'$ is smooth on $\mathcal{N}$. Under these conditions, $\mathcal{N}$ is a non-degenerate Killing horizon w.r.t. $\partial_{x^{3'}}$, as one can directly show by checking $\Gamma^{3'}_{3'3'}|_{t'=0} \neq 0$.

Next, Moncrief decomposes the Einstein equations $R_{\alpha\beta} = 0$ into four "constraint equations" and six "evolution equations". In the original chart, the former are obtained from $R_{t\alpha} = 0$, while the latter correspond to $R_{33} = R_{3A} = R_{AB} = 0$. To get some intuition about what is happening, we only write down an informal version of such evolution equations,

$$\phi_{,tt} + \frac{1}{t}\phi_{,t} = \mathcal{D}^\phi, \qquad \beta_{A,tt} + \frac{3}{t}\beta_{A,t} = \mathcal{D}^\beta_A, \qquad h_{AB,tt} + \frac{1}{t}h_{AB,t} = \mathcal{D}^h_{AB}, \tag{5.9}$$

where $\mathcal{D}^\phi$, $\mathcal{D}^\beta_A$ and $\mathcal{D}^h_{AB}$ are differential operators acting on $\phi$, $\boldsymbol{\beta}$ and $h$ whose explicit form is not relevant for this discussion. Note that these six equations are singular at $t=0$. Since nothing depends on $x^3$, to rewrite the equations in the new chart it suffices to substitute $\partial_t \to 2\sqrt{t'}\partial_{t'}$, $\partial_{x^A} \to \partial_{x^{A'}}$ and $t \to \sqrt{t'}$.

In order to obtain solutions of the evolution equations (5.9) one immediately realizes that the classical version of the Cauchy-Kowalewski theorem [111] does not apply, due to the singularities $1/t$. One can also try to solve the transformed equations, that are of course free of singularities. However, they are hyperbolic in $t'>0$, parabolic at $t'=0$ and elliptic at $t'<0$, making it difficult to apply classical theorems too. Fortunately, there is a variant of the theorem due to Fusaro [136] that may be applied in this context. Fusaro's version of the theorem applies to systems of the form

$$\frac{\partial u_i}{\partial t} + \frac{k_i u_i}{t} = \sum_{j=1}^{N}\sum_{a=1}^{n} A^a_{ij}(u,t,x^b)\frac{\partial u_j}{\partial x^a} + B_i(u,t,x^b), \tag{5.10}$$

where the $k_i$ are constant different from $-1,-2$, etc. and $A^a_{ij}$ and $B_i$ are analytic. Fusaro's existence theorem establishes that equation (5.10) with analytic initial data $u_i(0,x^a)$ admits a unique analytic solution on a neighbourhood of $t=0$.

One of the main results in [243] is that one can transform the system (5.9) into the form (5.10) for which the theorem applies. More specifically, given analytic $\phi(0,x^A)$, $\beta_A(0,x^B)$ and $h_{AB}(0,x^C)$, there exists a unique analytic solution $(\phi,\boldsymbol{\beta},h)$ of (5.9). Moreover, the solution is even in $t$ (and hence also analytic in the variable $t'$ on a neighbourhood of $\mathcal{N}$). The last step is to show that the remaining Einstein equations (i.e., the "constraints") hold. For this one uses the equations of motion to derive "evolution equations" for the constraints to which Fusaro's theorem applies, so that the vanishing of the constraints at $t'=0$ (which turns out to be automatically true) implies its vanishing in a neighbourhood of this hypersurface. Moncrief's existence theorem can be then stated as follows.

**Theorem 5.2.** *Any analytic data $(\phi(0,x^A),\beta_A(0,x^B),h_{AB}(0,x^C))$ over $T^2$ (with $h$ a Riemannian metric) determines a unique analytic and vacuum spacetime in a neighbourhood of the (non-degenerate) Killing horizon.*

This theorem characterizes all possible 4-dimensional, analytic, vacuum spacetimes near a non-degenerate Killing horizon in terms of six coordinate dependent functions $(\phi(0,x^A),\beta_A(0,x^B),h_{AB}(0,x^C))$ at the horizon.

Since Theorem 5.2 relies on a particular coordinate system, this setup does not allow to determine whether two data give rise to the same spacetime (modulo isometries). In order to deal with this issue in [195, 196], K. Kröncke and O. Petersen formulate a more geometric notion of the initial data, as we describe next.

Let $(M,g)$ be a smooth spacetime with a Killing vector $\eta$ that admits a Killing horizon $\mathcal{H}\subset M$ with surface gravity $\widetilde{\kappa}$. Assume that $\widetilde{\kappa}$ is nowhere zero (e.g. $\widetilde{\kappa}>0$). One can show that the one-form $\boldsymbol{\varpi}$ defined in (5.8) satisfies $\pounds_\eta \boldsymbol{\varpi} = 0$ and that $\boldsymbol{\varpi}(\eta) = \widetilde{\kappa} > 0$, and hence the tensor $\boldsymbol{\sigma}$ defined on $\mathcal{H}$ by

$$\boldsymbol{\sigma} := g + \boldsymbol{\varpi} \otimes \boldsymbol{\varpi}$$

is a Riemannian metric on $\mathcal{H}$ that satisfies $\pounds_\eta \boldsymbol{\sigma} = 0$ and $\boldsymbol{\sigma}(\eta, \eta) = \widetilde{\kappa}^2$. The set $(\mathcal{H}, \boldsymbol{\sigma}, \eta)$ is called "induced non-degenerate Killing horizon data", and the main result in [195, 196] is that it uniquely characterizes the spacetime in a neighbourhood of the horizon. More specifically, the result is the following.

**Theorem 5.3.** *Let $(M_1, g_1)$ and $(M_2, g_2)$ be two spacetimes and $\iota_1 : \mathcal{H} \hookrightarrow M_1$ and $\iota_2 : \mathcal{H} \hookrightarrow M_2$ two embeddings of $\mathcal{H}$ as a non-degenerate Killing horizon with vector fields $\eta_1$ and $\eta_2$, respectively, and such that $(\mathcal{H}, \boldsymbol{\sigma}, \eta)$ is the induced non-degenerate Killing horizon data in both cases. Assume also that $M_1$ and $M_2$ are asymptotically vacuum spacetimes (i.e., their Ricci tensors and all their derivatives vanish at $\mathcal{H}$). Then, there exists neighbourhoods $U_1$ of $\mathcal{H}$ in $M_1$ and $U_2$ of $\mathcal{H}$ in $M_2$ and a diffeomorphism $\phi : U_1 \longrightarrow U_2$ such that $\phi \circ \iota_1 = \iota_2$, $\phi^\star \eta_2 = \eta_1$ and $\phi^\star g_2 - g_1$ vanishes to infinite order at $\mathcal{H}$.*

The strategy of the proof consists of picking up the unique null, geodesic, transversal vector $\partial_t$ at $\mathcal{H}$ such that $g(\partial_t, \eta) \stackrel{\mathcal{H}}{=} 1$, $[\partial_t, \eta] = 0$ and $g(\partial_t, X) = 0$ for all $X$ orthogonal to $\eta$ w.r.t. $\boldsymbol{\sigma}$ and to use the following well-known formula for the linearization of the Ricci tensor [38]

$$2\pounds_{\partial_t} \operatorname{Ric} = \square_L \pounds_{\partial_t} g + \pounds_\zeta g, \tag{5.11}$$

where $\square_L$ is the Lichnerowicz Laplacian [38, 205], defined over any symmetric tensor $T_{\alpha\beta}$ by

$$\square_L T_{\alpha\beta} := \square_g T_{\alpha\beta} + R^\mu{}_{(\alpha} T_{\beta)\mu} - R_{\alpha\mu\beta\nu} T^{\mu\nu},$$

and $\zeta$ is a certain vector field constructed from $g$ and $\pounds_t g$. After computing $m-1$ extra derivatives of (5.11) w.r.t. $\partial_t$ and using an induction argument, the authors show that the combination

$$(\pounds^{(m)}_{\partial_t} \operatorname{Ric})(X, Y)|_{\mathcal{H}} + (m+1)\widetilde{\kappa}(\pounds^{(m+1)}_{\partial_t} g)(X, Y)|_{\mathcal{H}} \tag{5.12}$$

for every $X, Y$ orthogonal to $\eta$ is uniquely determined by $(\mathcal{H}, \boldsymbol{\sigma}, \eta)$ for all $m \geq 0$. Hence, from the assumption that the Ricci tensor vanishes to infinite order, it follows that all the derivatives of the metric w.r.t. $\partial_t$ at $\mathcal{H}$ are uniquely given in terms of $(\mathcal{H}, \boldsymbol{\sigma}, \eta)$. As a consequence, two asymptotically vacuum spacetimes sharing the same induced data must have the same transverse expansion.

In addition, the authors in [195, 196] prove an existence result in terms of the data $(\mathcal{H}, \boldsymbol{\sigma}, \eta)$ which essentially consists of rewriting $\boldsymbol{\sigma}$ and $\eta$ in terms of $(\phi(0, x^A), \beta_A(0, x^B), h_{AB}(0, x^C))$ and applying Theorem 5.2. This existence result is therefore restricted to spacetime dimension four and to the analytic class.

The results in [195, 196] generalize the uniqueness part of [243] in the sense that the data $(\mathcal{H}, \sigma, \eta)$ does not depend on any coordinate system, and the topology and dimension of $\mathcal{H}$ are arbitrary. This allows one to decide whether two spacetimes constructed from different data by Theorem 5.2 are isometric or not. These results however still have room for improvement. Firstly, because they only hold for $\Lambda = 0$ vacuum; secondly, because no zeroes of $\eta$ are allowed[4]; thirdly, because the signature of the ambient manifold is restricted to be Lorentzian; and finally because the transverse vector $t$ if completely fixed from the beginning. One of the achievements of this chapter (specifically Section 5.5) is to generalize the results in [195, 196] in all these directions.

### 5.1.3 *Degenerate Killing horizons*

In this subsection we analyze the degenerate Killing horizon case, i.e. when the surface gravity is everywhere zero, $\widetilde{\kappa} = 0$. These horizons admit a so-called near horizon limit, as we now review [199].

Consider a spacetime metric $g$ written in Gaussian null coordinates (see Appendix C) as

$$g = 2dv \left( dr + r\beta_A dx^A + \frac{1}{2} r\phi dv \right) + h_{AB} dx^A dx^B, \tag{5.13}$$

where $\phi$, $\boldsymbol{\beta}$ and $h$ depend on $\{r, v, x^A\}$ and $h_{AB}$ defines a Riemannian metric. These coordinates are valid in a neighbourhood of any null hypersurface $\mathcal{N} = \{r = 0\}$ provided it admits a cross-section. If one assumes $\mathcal{N}$ to be a Killing horizon w.r.t. $K = \partial_v$, it then follows that $\phi$, $\boldsymbol{\beta}$ and $h$ depend only upon $\{r, x^A\}$. If one assumes further that $\mathcal{N}$ is degenerate, one has $\partial_r \phi|_{r=0} = 0$ and hence the metric (5.13) can be rewritten as

$$g = 2dv \left( dt + r\beta_A(r, x^B) dx^A + \frac{1}{2} r^2 F(r, x^A) dv \right) + h_{AB}(r, x^C) dx^A dx^B. \tag{5.14}$$

The near-horizon limit is defined as follows. Given any $\varepsilon > 0$, consider the diffeomorphism $\Phi(r, v, x^A) = (\varepsilon r, v/\varepsilon, x^A)$. This induces the following transformation of (5.14),

$$g = 2dv \left( dt + r\beta_A(\varepsilon r, x^B) dx^A + \frac{1}{2} r^2 F(\varepsilon r, x^A) dv \right) + h_{AB}(\varepsilon r, x^C) dx^A dx^B.$$

The near-horizon limit is obtained after putting $\varepsilon = 0$, i.e.

$$g = 2dv \left( dt + r\beta_A(x^B) dx^A + \frac{1}{2} r^2 F(x^A) dv \right) + h_{AB}(x^C) dx^A dx^B. \tag{5.15}$$

Note that now all the dependence on $r$ is explicit. Hence, the geometry is completely characterized by a function $F$, a one-form $\boldsymbol{\beta}$ and a metric $h$ at any cross section of $\mathcal{N}$. It is important to emphasize that this limit relies heavily on the assumption $g_{vv} = O(r^2)$. No

4 It is well-known [288] that $\lambda$-vacuum Killing horizons can always been extended to a bifurcate Killing horizon.

such limit exists for non-degenerate horizons. When $g$ is assumed to be $\lambda$-vacuum, the near horizon data $(F, \boldsymbol{\beta}, h)$ satisfies the *near horizon equations*

$$R^h_{AB} = \frac{1}{2}\beta_A\beta_B - \nabla^h_{(A}\beta_{B)} + \lambda h_{AB}, \tag{5.16}$$

$$F = \frac{1}{2}\beta_A\beta^A - \frac{1}{2}\nabla^h_A\beta^A + \lambda, \tag{5.17}$$

which correspond respectively to the Einstein equation components $R_{AB} \stackrel{\mathcal{N}}{=} \lambda h_{AB}$ and $R_{vr} \stackrel{\mathcal{N}}{=} \lambda$. It turns out that the metric (5.15) is in fact a solution to the $\lambda$-vacuum equations if and only if (5.16)-(5.17) hold [199]. One key property of (5.16) that makes the analysis of degenerate horizons so different to the non-degenerate case is the absence of first order transverse derivatives of the metric (compare (5.16) with (5.12) for $m = 0$). We shall discuss this further in Section 5.5. The near horizon equation also appears in the context of extremal isolated horizons [28, 29, 31, 45, 202, 327], Kundt's metrics [183] and multiple Killing horizons [227]. It has been recently generalized to arbitrary topology and allowing for fixed points in [211].

Equation (5.16) has been intensively studied in the literature [199], mostly in the compact case. For instance, it is known that four dimensional, static, vacuum solutions satisfy $\boldsymbol{\beta} = 0$ [76]. In higher dimensions and $\Lambda = 0$ the same conclusion holds, but for $\Lambda \neq 0$ there are examples with $\boldsymbol{\beta} \neq 0$ that solve (5.16) (see e.g. [33, 328]). Concerning the non-static case, defined by the condition $d\boldsymbol{\beta} \neq 0$, the general solution to (5.16) with spherical topology $\mathbb{S}^2$ admitting an axial Killing vector $K$ (that commutes with $\boldsymbol{\beta}$, i.e. $[K, \beta] = 0$) is the one that corresponds to extremal Kerr [199, 202]. A remarkable recent result [95] proves that whenever $\boldsymbol{\beta}$ is not exact and satisfies (5.16), $h$ admits an axial Killing $K$ that commutes with $\boldsymbol{\beta}$, and as a consequence extremal Kerr is the unique solution to (5.16) on a topological sphere $\mathbb{S}^2$ (see [86] for the generalization to the Einstein-Maxwell case). In other topologies, the only solution to (5.16) is the trivial one [94].

This uniqueness result can be thought of as an intrinsic version of the rigidity theorem for degenerate horizons, meaning that the zeroth order geometry at the horizon is uniquely determined to be that of Kerr. It is then natural to ask whether the higher order derivatives of the metric at the horizon agree with those of Kerr. In general, the higher order Einstein equations lead to complicated elliptic equations for the higher order derivatives of the metric. In the case of Kerr, it is known that the first order of these equations admits a unique solution [203], meaning that "Kerr is unique up to first order".

While the case of Kerr is still too complicated at higher orders due to the presence of a non-vanishing one-form $\boldsymbol{\beta}$, the most reasonable metric to try and prove uniqueness in the degenerate case is Schwarzschild-(A)de Sitter (precisely because $\boldsymbol{\beta} = 0$). This task has been successfully carried out by D. Katona and J. Lucietti in [188], as we describe next (see also [187] for the generalization to electrovacuum).

Recall first that, for $\Lambda > 0$, the $d$-dimensional Schwarzschild-de Sitter (SdS) metric is

$$g_{SdS} = -\left(1 - \frac{2m}{r^{d-3}} - \frac{\Lambda r^2}{d-1}\right) dt^2 + \left(1 - \frac{2m}{r^{d-3}} - \frac{\Lambda r^2}{d-1}\right)^{-1} dr^2 + r^2 \boldsymbol{\gamma}_{\mathbb{S}^{d-2}}, \tag{5.18}$$

where $\boldsymbol{\gamma}_{\mathbb{S}^{d-2}}$ is the unit metric on $\mathbb{S}^{d-2}$ and $0 \leq m \leq m_{max}$ is the mass parameter, with

$$m_{max} := \frac{R^{d-3}}{d-1}, \qquad R^2 := \frac{d-3}{\Lambda}. \tag{5.19}$$

For $m = 0$ this is just de Sitter spacetime. For $0 < m < m_{max}$ one has non-extremal SdS, that admits a cosmological horizon at $r = r_0^+ > R$ and a event horizon at $r = r_0^- < R$, where $r_0^+$ and $r_0^-$ are the two solutions of $g_{tt} = 0$. Finally, for $m = m_{max}$ one obtains extremal Schwarzschild-de Sitter, where the cosmological and the event horizon merge into a single degenerate horizon at $r_0^+ = r_0^- = R$. It is more convenient to write down the metric (5.18) in advanced Eddington-Finkelstein coordinates,

$$g_{SdS} = -\left(1 - \frac{2m}{r^{d-3}} - \frac{\Lambda r^2}{d-1}\right) dv^2 + 2dvdr + r^2 \boldsymbol{\gamma}_{\mathbb{S}^{d-2}}, \tag{5.20}$$

where the static Killing is $\partial_v$. In terms of $r_0$,

$$g_{SdS} = -\left(1 - \left(1 - \frac{d-3}{d-1}\frac{r_0^2}{R^2}\right)\left(\frac{r_0}{r}\right)^{d-3} - \frac{d-3}{d-1}\left(\frac{r}{R}\right)^2\right) dv^2 + 2dvdr + r^2 \boldsymbol{\gamma}_{\mathbb{S}^{d-2}}.$$

The extremal Schwarzschild-de Sitter metric is then ($r_0 = R$)

$$g_{SdSE} = -\left(1 - \frac{2}{d-1}\left(\frac{R}{r}\right)^{d-3} - \frac{d-3}{d-1}\left(\frac{r}{R}\right)^2\right) dv^2 + 2dvdr + r^2 \boldsymbol{\gamma}_{\mathbb{S}^{d-2}}. \tag{5.21}$$

This metric has near horizon data $F = \Lambda$, $\boldsymbol{\beta} = 0$ and $h = R^2 \boldsymbol{\gamma}_{\mathbb{S}^{d-2}}$. Its near horizon limit is called Nariai [248], and is obtained after setting $r \to R + \varepsilon r$ and $v \to v/\varepsilon$ and taking the limit $\varepsilon \to 0$. The result is

$$g_N = \frac{d-3}{R^2} r^2 dv^2 + 2dvdr + R^2 \boldsymbol{\gamma}_{\mathbb{S}^{d-2}} = \Lambda r^2 dv^2 + 2dvdr + R^2 \boldsymbol{\gamma}_{\mathbb{S}^{d-2}}, \tag{5.22}$$

which is the vacuum solution $dS_2 \times \mathbb{S}^{d-2}_R$, where $dS_2$ is the two-dimensional de Sitter space of constant scalar curvature $\Lambda$ and $\mathbb{S}^{d-2}_R$ the spherical metric of constant radius $R$ (recall that $\Lambda$ and $R$ are related by (5.19)). Of course, the corresponding near horizon data is the same as that of its parent spacetime. Apart from the degenerate Killing vector $\partial_v$, this metric admits another Killing vector $v\partial_v - r\partial_r$ that is null and tangent to the horizon $r = 0$ and whose surface gravity is different from zero. We will analyze this further in Remark 5.65.

Having summarized the SdS and Nariai metrics, we can proceed with the description of the results in [188]. The works [33] and [76] already mentioned imply that the most general solution of (5.16)-(5.17) for a $d \geq 4$ dimensional $\Lambda$-vacuum spacetime $(M, g)$ with a static degenerate Killing horizon with maximally symmetric[5], compact cross-sections corresponds to $\boldsymbol{\beta}=0$, $F=\Lambda$ and $k=\frac{\Lambda}{d-3}$, which is the near horizon data of Schwarzschild-de Sitter (and Nariai). The fundamental contribution of [188] is proving that the only higher order derivatives of the metric that are compatible with $\lambda$-vacuum and this near horizon data are precisely those corresponding to Schwarzschild-de Sitter (and Nariai). In order to show this, the authors write down the Einstein equations order by order in Gaussian null coordinates and use them to find the coefficients $F(r)$, $\beta_A(r)$ and $h_{AB}(r)$ to all orders, which turn out to be explicitly given in terms of a constant $C\in\mathbb{R}$ by[6]

$$\begin{gathered} h^{(0)}_{AB}=h_{AB}, \qquad h^{(1)}_{AB}=Ch_{AB}, \qquad h^{(2)}_{AB}=\frac{C^2}{2}h_{AB}, \qquad h^{(m)}_{AB}=0 \quad \forall m\geq 3, \\ \beta^{(m)}_A=0, \qquad F^{(m)}=\frac{\Lambda}{d-1}\left(\delta^m_0+(-1)^m\frac{(d-2+m)!}{(d-3)!}\left(\frac{C}{2}\right)^m\right)\binom{m+2}{2}^{-1} \qquad \forall m\geq 0. \end{gathered} \tag{5.23}$$

The key difference with respect to the non-degenerate case is that the equations that fix the tensors $h^{(m)}_{AB}$ are not algebraic, but PDEs. Crucially, for $\Lambda>0$ and maximally symmetric cross-sections, each of these PDEs admit a unique solution. After resumming the Taylor series, they show that these functions agree with those of Schwarzschild-de Sitter (when $C\neq 0$) or Nariai (when $C=0$). Their main result is the following.

**Theorem 5.4** ([188])**.** *Let $(M,g)$ be an analytic, $d\geq 4$ dimensional, $\Lambda>0$ vacuum spacetime that contains a static degenerate Killing horizon with maximally symmetric compact cross-sections. Then, $(M,g)$ is locally isometric either to extremal Schwarzschild-de Sitter or Nariai.*

As already mentioned, this result has also been generalized to the Einstein-Maxwell equations [187].

## 5.2 TRANSVERSE DERIVATIVES AT A NULL HYPERSURFACE

As we learned in Sections 2.2 and 2.3, given embedded null hypersurface data $\{\mathcal{H},\boldsymbol{\gamma},\boldsymbol{\ell},\ell^{(2)},\mathbf{Y}\}$ the fully tangential components of the ambient Ricci tensor at $\mathcal{H}$ can be written in terms of the data as (cf. (2.95))

$$\begin{aligned} \mathcal{R}_{ab}=&\ \overset{\circ}{R}_{(ab)}-2\pounds_n\mathrm{Y}_{ab}-(2\kappa+\operatorname{tr}_P\mathbf{U})\mathrm{Y}_{ab}+\overset{\circ}{\nabla}_{(a}(\mathrm{s}_{b)}+2\mathrm{r}_{b)}) \\ &-2\mathrm{r}_a\mathrm{r}_b+4\mathrm{r}_{(a}\mathrm{s}_{b)}-\mathrm{s}_a\mathrm{s}_b-(\operatorname{tr}_P\mathbf{Y})\mathrm{U}_{ab}+2P^{cd}\mathrm{U}_{d(a}(2\mathrm{Y}_{b)c}+\mathrm{F}_{b)c}). \end{aligned} \tag{5.24}$$

The contractions of $\mathcal{R}_{ab}$ with $P$ and $n$ are (see (2.97), (2.98) and (2.99))

$$\mathcal{R}_{ab}n^a=-\pounds_n(\mathrm{r}_b-\mathrm{s}_b)-\overset{\circ}{\nabla}_b\kappa-(\operatorname{tr}_P\mathbf{U})(\mathrm{r}_b-\mathrm{s}_b)-\overset{\circ}{\nabla}_b(\operatorname{tr}_P\mathbf{U})+P^{cd}\overset{\circ}{\nabla}_c\mathrm{U}_{bd}-2P^{cd}\mathrm{U}_{bd}s_c, \tag{5.25}$$

5 That is, $R^h_{ABCD}=k(h_{AC}h_{BD}-h_{AD}h_{BC})$ for some constant $k$.
6 We employ the notation $h^{(k)}_{AB}:=\pounds^{(k)}_{\partial_r}h_{AB}$, and similarly for $\beta^{(k)}_A$ and $F^{(k)}$.

$$\begin{aligned}\mathcal{R}_{ab}n^a n^b &= -n(\operatorname{tr}_P \mathbf{U}) + (\operatorname{tr}_P \mathbf{U})\kappa - P^{ab}P^{cd}\mathrm{U}_{ac}\mathrm{U}_{bd}, & (5.26)\\ P^{ab}\mathcal{R}_{ab} &= \operatorname{tr}_P \mathring{R} - 2\pounds_n(\operatorname{tr}_P \mathbf{Y}) - 2(\kappa + \operatorname{tr}_P \mathbf{U})\operatorname{tr}_P \mathbf{Y} + \operatorname{div}_P(\mathbf{s}+2\mathbf{r}) - 2P(\mathbf{r},\mathbf{r})\\ &\quad - 4P(\mathbf{r},\mathbf{s}) - P(\mathbf{s},\mathbf{s}) + 2\kappa n(\ell^{(2)}). & (5.27)\end{aligned}$$

The transverse-tangent and fully transverse components of $R_{\mu\nu}$ require, in addition, second order transverse derivatives of the metric (cf. (2.118)-(2.119)),

$$\begin{aligned}\ddot{\mathcal{R}} &= -\operatorname{tr}_P \mathbf{Z}^{(2)} + \operatorname{tr}_P(\mathbf{\Pi}\cdot\mathbf{\Pi}) + P(\mathbf{r}-\mathbf{s}, d\ell^{(2)}), & (5.28)\\ \dot{\mathcal{R}}_c &= -P^{ab}A_{abc} + \mathrm{z}^{(2)}_c - P^{ab}\Pi_{da}\Pi_{cb}n^d + \frac{1}{2}\kappa\mathring{\nabla}_c\ell^{(2)} - \frac{1}{2}n(\ell^{(2)})(\mathrm{r}-\mathrm{s})_c, & (5.29)\end{aligned}$$

where we recall the notation $\dot{\mathcal{R}} := \Phi^\star(\operatorname{Ric}(\xi,\cdot))$ and $\ddot{\mathcal{R}} := \Phi^\star(\operatorname{Ric}(\xi,\xi))$. One property of the tensor $A$ is [217]

$$\begin{aligned}P^{ab}A_{abc}n^c &= \frac{1}{2}P^{ab}(A_{abc}+A_{bac})n^c = -\frac{1}{2}P^{ab}(A_{acb}+A_{bca})n^c\\ &= P^{ab}\left(\pounds_n \mathrm{Y}_{ab} - \mathring{\nabla}_a(\mathrm{r}+\mathrm{s})_b + \kappa \mathrm{Y}_{ab} + (\mathrm{r}-\mathrm{s})_a(\mathrm{r}-\mathrm{s})_b - \frac{1}{2}n(\ell^{(2)})\mathrm{U}_{ab} - P^{cd}\mathrm{Y}_{ac}\mathrm{U}_{bd}\right)\\ &\overset{(2.49)}{=} \pounds_n(\operatorname{tr}_P \mathbf{Y}) + P^{ab}P^{cd}\mathrm{U}_{ac}\mathrm{Y}_{bd} - \operatorname{div}_P(\mathbf{r}+\mathbf{s}) + (\operatorname{tr}_P \mathbf{Y} - n(\ell^{(2)}))\kappa\\ &\quad + P(\mathbf{r}+\mathbf{s},\mathbf{r}+\mathbf{s}) - \frac{1}{2}n(\ell^{(2)})\operatorname{tr}_P \mathbf{U}.\end{aligned}$$

Therefore, the contraction of (5.29) with $n^c$ is

$$\begin{aligned}\dot{\mathcal{R}}_a n^a &= -z^{(2)} - P^{ab}A_{abc}n^c - P(\mathbf{r}+\mathbf{s},\mathbf{r}+\mathbf{s}) + n(\ell^{(2)})\kappa\\ &= -z^{(2)} - \pounds_n(\operatorname{tr}_P \mathbf{Y}) + \operatorname{div}(\mathbf{r}+\mathbf{s}) + (2n(\ell^{(2)}) - \operatorname{tr}_P \mathbf{Y})\kappa + \frac{1}{2}(\operatorname{tr}_P \mathbf{U})n(\ell^{(2)})\\ &\quad - P^{ad}P^{bc}\mathrm{U}_{ab}\mathrm{Y}_{cd} - 4P(\mathbf{r},\mathbf{s}) - 2P(\mathbf{r},\mathbf{r}) - 2P(\mathbf{s},\mathbf{s}). & (5.30)\end{aligned}$$

Expressions (5.27) and (5.30) are all one needs to compute the ambient scalar curvature at $\mathcal{H}$. Indeed, using (2.14), $R \overset{\mathcal{H}}{=} \operatorname{tr}_P \mathcal{R} + 2\dot{\mathcal{R}}_a n^a$, and inserting (5.27) and (5.30) one concludes

$$\begin{aligned}R &= -2z^{(2)} - 4\pounds_n(\operatorname{tr}_P \mathbf{Y}) - 2(2\kappa + \operatorname{tr}_P \mathbf{U})\operatorname{tr}_P \mathbf{Y} + 3\operatorname{div}\mathbf{s} + 4\operatorname{div}\mathbf{r} - 5P(\mathbf{s},\mathbf{s})\\ &\quad - 6P(\mathbf{r},\mathbf{r}) - 12P(\mathbf{r},\mathbf{s}) - 2P^{ab}P^{cd}\mathrm{U}_{ac}\mathrm{Y}_{bd} + (\operatorname{tr}_P \mathbf{U} + 6\kappa)n(\ell^{(2)}) + \operatorname{tr}_P \mathring{R}.\end{aligned} \tag{5.31}$$

The aim of this section is to compute the derivatives $\pounds^{(m)}_\xi \operatorname{Ric}$ on $\mathcal{H}$ in terms of transverse derivatives of $g$ on $\mathcal{H}$ up to order $m+1$, i.e. making $\pounds^{(m+2)}_\xi g$ and $\pounds^{(m+1)}_\xi g$ explicit. In order to simplify the notation let us introduce the tensors

$$\mathbf{Y}^{(m)} := \frac{1}{2}\Phi^\star(\pounds^{(m)}_\xi g), \qquad \mathbf{r}^{(m)} := \mathbf{Y}^{(m)}(n,\cdot), \qquad \kappa^{(m)} := -\mathbf{Y}^{(m)}(n,n), \tag{5.32}$$

as well as

$$\mathcal{R}^{(m)} := \Phi^\star(\pounds^{(m-1)}_\xi \operatorname{Ric}), \quad \dot{\mathcal{R}}^{(m)} := \Phi^\star(\pounds^{(m-1)}_\xi \operatorname{Ric}(\xi,\cdot)), \quad \ddot{\mathcal{R}}^{(m)} := \Phi^\star(\pounds^{(m-1)}_\xi \operatorname{Ric}(\xi,\xi)). \tag{5.33}$$

Observe that $\mathbf{Y}^{(1)}$, $\mathbf{r}^{(1)}$ and $\kappa^{(1)}$ agree with $\mathbf{Y}$, $\mathbf{r}$ and $\kappa$, respectively. In what follows we refer to the collection of tensors $\{\mathbf{Y}, \mathbf{Y}^{(2)}, ...\}$ as the *transverse* or *asymptotic expansion.*

In the following lemma we recall a well-known identity for derivatives of products of any two objects $S$ and $T$. Note than when $S$ and $T$ are tensors, the expression also holds when contractions are allowed.

**Lemma 5.5.** *Let $S$ and $T$ be two objects, $S \circledast T$ any product of them and $\mathcal{D}$ any derivative operator. Then,*

$$\mathcal{D}^{(m)}\Big(S \circledast T\Big) = \sum_{i=0}^{m} \binom{m}{i} \Big(\mathcal{D}^{(i)} S\Big) \circledast \Big(\mathcal{D}^{(m-i)} T\Big). \tag{5.34}$$

*Proof.* For $m = 1$ (5.34) is a consequence of the Leibniz rule for derivatives of products. Let us assume

$$\mathcal{D}^{(m-1)}\Big(S \circledast T\Big) = \sum_{i=0}^{m-1} \binom{m-1}{i} \Big(\mathcal{D}^{(i)} S\Big) \circledast \Big(\mathcal{D}^{(m-1-i)} T\Big).$$

Applying $\mathcal{D}$ to both sides it follows

$$\begin{aligned}
\mathcal{D}^{(m)}\Big(S \circledast T\Big) &= \sum_{i=0}^{m-1} \binom{m-1}{i} \Big(\mathcal{D}^{(i+1)} S\Big) \circledast \Big(\mathcal{D}^{(m-1-i)} T\Big) + \sum_{i=0}^{m-1} \binom{m-1}{i} \Big(\mathcal{D}^{(i)} S\Big) \circledast \Big(\mathcal{D}^{(m-i)} T\Big) \\
&= \Big(\mathcal{D}^{(m)} S\Big) \circledast T + \sum_{i=0}^{m-2} \binom{m-1}{i} \Big(\mathcal{D}^{(i+1)} S\Big) \circledast \Big(\mathcal{D}^{(m-1-i)} T\Big) + S \circledast \Big(\mathcal{D}^{(m)} T\Big) \\
&\quad + \sum_{i=1}^{m-1} \binom{m-1}{i} \Big(\mathcal{D}^{(i)} S\Big) \circledast \Big(\mathcal{D}^{(m-i)} T\Big).
\end{aligned}$$

Then, replacing $i$ by $i-1$ in the second term,

$$\begin{aligned}
\mathcal{D}^{(m)}\Big(S \circledast T\Big) &= S \circledast \Big(\mathcal{D}^{(m)} T\Big) + \Big(\mathcal{D}^{(m)} S\Big) \circledast T \\
&\quad + \sum_{i=1}^{m-1} \left[\binom{m-1}{i-1} + \binom{m-1}{i}\right] \Big(\mathcal{D}^{(i)} S\Big) \circledast \Big(\mathcal{D}^{(m-i)} T\Big) \\
&= \binom{m}{0} S \circledast \Big(\mathcal{D}^{(m)} T\Big) + \sum_{i=1}^{m-1} \binom{m}{i} \Big(\mathcal{D}^{(i)} S\Big) \circledast \Big(\mathcal{D}^{(m-i)} T\Big) + \binom{m}{m} \Big(\mathcal{D}^{(m)} S\Big) \circledast T \\
&= \sum_{i=0}^{m} \binom{m}{i} \Big(\mathcal{D}^{(i)} S\Big) \circledast \Big(\mathcal{D}^{(m-i)} T\Big),
\end{aligned}$$

where in the second equality we use the well-known property $\binom{j}{k} + \binom{j}{k+1} = \binom{j+1}{k+1}$. Then, (5.34) follows by induction. □

Following the notation of Chapter 4, given any vector field $\xi$ we define its deformation tensor $\mathcal{K}[\xi] := \pounds_\xi g$ and the tensor $\Sigma[\xi] := \pounds_\xi \nabla$, which in abstract index notation is [329]

$$\Sigma[\xi]^\alpha{}_{\mu\nu} = \frac{1}{2} g^{\alpha\beta} \left(\nabla_\mu \mathcal{K}[\xi]_{\nu\beta} + \nabla_\nu \mathcal{K}[\xi]_{\mu\beta} - \nabla_\beta \mathcal{K}[\xi]_{\mu\nu}\right). \tag{5.35}$$

In order not to overload the notation in this section we will simply use the symbols $\Sigma$ and $\mathcal{K}$ for $\Sigma[\xi]$ and $\mathcal{K}[\xi]$, respectively. In later sections, we will come back to the notation $\mathcal{K}[\xi]$ because deformation tensors of more than one vector field will occur.

To compute $\pounds_{\xi}^{(m)}$ Ric up to order $m+1$ we use the following classical result [329] relating the Lie derivative of the curvature with the Lie derivative of the connection (see also (4.20))

$$\pounds_{\xi} R^{\mu}{}_{\alpha\nu\beta} = H^{\gamma\rho}_{\nu\beta} \nabla_{\gamma} \Sigma^{\mu}{}_{\alpha\rho}, \qquad H^{\gamma\rho}_{\nu\beta} := \delta^{\gamma}_{\nu}\delta^{\rho}_{\beta} - \delta^{\gamma}_{\beta}\delta^{\rho}_{\nu}. \tag{5.36}$$

For future convenience we define

$$\underset{\smile}{\Sigma}_{\nu\alpha\beta} := g_{\mu\nu}\Sigma^{\mu}{}_{\alpha\beta} = \frac{1}{2}\left(\nabla_{\alpha}\mathcal{K}_{\beta\nu} + \nabla_{\beta}\mathcal{K}_{\alpha\nu} - \nabla_{\nu}\mathcal{K}_{\alpha\beta}\right) =: F^{\rho\lambda\gamma}_{\nu\alpha\beta}\nabla_{\rho}\mathcal{K}_{\lambda\gamma}, \tag{5.37}$$

where we have introduced the tensor

$$F^{\rho\lambda\gamma}_{\nu\alpha\beta} := \frac{1}{2}\left(\delta^{\rho}_{\alpha}\delta^{\lambda}_{\beta}\delta^{\gamma}_{\nu} + \delta^{\rho}_{\beta}\delta^{\lambda}_{\alpha}\delta^{\gamma}_{\nu} - \delta^{\rho}_{\nu}\delta^{\lambda}_{\alpha}\delta^{\gamma}_{\beta}\right).$$

The hat in $\underset{\smile}{\Sigma}_{\nu\alpha\beta}$ is not really necessary because $\underset{\smile}{\Sigma}_{\nu\alpha\beta}$ is just $\Sigma^{\nu}{}_{\alpha\beta}$ with the index lowered. However, the distinction will be necessary later for the Lie derivative of $\underset{\smile}{\Sigma}_{\nu\alpha\beta}$ and $\Sigma^{\nu}{}_{\alpha\beta}$.

**Remark 5.6.** *The notations $H^{\gamma\rho}_{\nu\beta}$ and $F^{\rho\lambda\gamma}_{\nu\alpha\beta}$ are unambiguous since we shall never lower/raise their indices. We stick to this rule for any tensor written with indices on top of each other.*

The idea now is to apply the operator $\pounds_{\xi}^{(m-1)}$ to (5.36) and express the result by making $\pounds_{\xi}^{(m+1)}g$ and $\pounds_{\xi}^{(m+2)}g$ explicit. In order to do that we need to commute $\pounds_{\xi}^{(m-1)}$ and $\nabla$ when they act on a $(q,p)$ tensor $A^{\alpha_1\cdots\alpha_q}_{\beta_1\cdots\beta_p}$. We introduce the notation $A^{(m)} := \pounds_{\xi}^{(m-1)}A$, $m\geq 1$. The commutator is found explicitly in the following proposition.

**Proposition 5.7.** *Let $\xi\in\mathfrak{X}(\mathcal{M})$ and $m\geq 1$ be a integer. Then, given any $(p,q)$ tensor $A^{\alpha_1\cdots\alpha_q}_{\beta_1\cdots\beta_p}$ the following identity holds*

$$\begin{aligned}\pounds_{\xi}^{(m)}\nabla_{\gamma}A^{\alpha_1\cdots\alpha_q}_{\beta_1\cdots\beta_p} = \nabla_{\gamma}A^{(m+1)}{}^{\alpha_1\cdots\alpha_q}_{\beta_1\cdots\beta_p} + \sum_{k=0}^{m-1}\binom{m}{k+1}\Bigg(&\sum_{j=1}^{q}A^{(m-k)}{}^{\alpha_1\cdots\alpha_{j-1}\sigma\alpha_{j+1}\cdots\alpha_q}_{\beta_1\cdots\beta_p}\Sigma^{(k+1)}{}^{\alpha_j}{}_{\sigma\gamma}\\ &-\sum_{i=1}^{p}A^{(m-k)}{}^{\alpha_1\cdots\alpha_q}_{\beta_1\cdots\beta_{i-1}\sigma\beta_{i+1}\cdots\beta_p}\Sigma^{(k+1)}{}^{\sigma}{}_{\beta_i\gamma}\Bigg).\end{aligned}$$

*Proof.* The case $m=1$ is the classical identity (4.11)

$$\pounds_{\xi}\nabla_{\gamma}A^{\alpha_1\cdots\alpha_q}_{\beta_1\cdots\beta_p} = \nabla_{\gamma}\pounds_{\xi}A^{\alpha_1\cdots\alpha_q}_{\beta_1\cdots\beta_p} + \sum_{j=1}^{q}A^{\alpha_1\cdots\alpha_{j-1}\sigma\alpha_{j+1}\cdots\alpha_q}_{\beta_1\cdots\beta_p}\Sigma^{\alpha_j}{}_{\sigma\gamma} - \sum_{i=1}^{p}A^{\alpha_1\cdots\alpha_q}_{\beta_1\cdots\beta_{i-1}\sigma\beta_{i+1}\cdots\beta_p}\Sigma^{\sigma}{}_{\beta_i\gamma}.$$

We prove the result by induction, so let us assume that the claim is true up to some $m \geq 1$ and show that it is then true for $m+1$ also. We compute, using the induction hypothesis,

$$
\begin{aligned}
\pounds_\xi^{(m+1)} \nabla_\gamma A^{\alpha_1\cdots\alpha_q}_{\beta_1\cdots\beta_p} &= \pounds_\xi \left( \pounds_\xi^{(m)} \nabla_\gamma A^{\alpha_1\cdots\alpha_q}_{\beta_1\cdots\beta_p} \right) \\
&= \pounds_\xi \left( \nabla_\gamma A^{(m+1)\alpha_1\cdots\alpha_q}_{\beta_1\cdots\beta_p} + \sum_{k=0}^{m-1} \binom{m}{k+1} \left( \sum_{j=1}^{q} A^{(m-k)\alpha_1\cdots\sigma\alpha_q}_{\beta_1\cdots\beta_p} \Sigma^{(k+1)\alpha_j}{}_{\sigma\gamma} \right.\right. \\
&\qquad \left.\left. - \sum_{i=1}^{p} A^{(m-k)\alpha_1\cdots\alpha_q}_{\beta_1\cdots\sigma\cdots\beta_p} \Sigma^{(k+1)\sigma}{}_{\beta_i\gamma} \right)\right) \\
&= \nabla_\gamma \pounds_\xi A^{(m+1)\alpha_1\cdots\alpha_q}_{\beta_1\cdots\beta_p} + \sum_{j=1}^{q} A^{(m+1)\alpha_1\cdots\sigma\cdots\alpha_q}_{\beta_1\cdots\beta_p} \Sigma^{\alpha_j}{}_{\sigma\gamma} - \sum_{i=1}^{p} A^{(m+1)\alpha_1\cdots\alpha_q}_{\beta_1\cdots\beta_{i-1}\sigma\beta_{i+1}\cdots\beta_p} \Sigma^{\sigma}{}_{\beta_i\gamma} \\
&\quad + \sum_{k=0}^{m-1} \binom{m}{k+1} \left( \sum_{j=1}^{q} A^{(m-k+1)\alpha_1\cdots\sigma\cdots\alpha_q}_{\beta_1\cdots\beta_p} \Sigma^{(k+1)\alpha_j}{}_{\sigma\gamma} - \sum_{i=1}^{p} A^{(m-k+1)\alpha_1\cdots\alpha_q}_{\beta_1\cdots\sigma\cdots\beta_p} \Sigma^{(k+1)\sigma}{}_{\beta_i\gamma} \right) \\
&\quad + \sum_{k=0}^{m-1} \binom{m}{k+1} \left( \sum_{j=1}^{q} A^{(m-k)\alpha_1\cdots\sigma\cdots\alpha_q}_{\beta_1\cdots\beta_p} \Sigma^{(k+2)\alpha_j}{}_{\sigma\gamma} - \sum_{i=1}^{p} A^{(m-k)\alpha_1\cdots\alpha_q}_{\beta_1\cdots\sigma\cdots\beta_p} \Sigma^{(k+2)\sigma}{}_{\beta_i\gamma} \right).
\end{aligned}
$$

In the last term we rename $k$ as $k-1$ and split the second sum in two parts and the last sum also in two parts,

$$
\begin{aligned}
\pounds_\xi^{(m+1)} \nabla_\gamma A^{\alpha_1\cdots\alpha_q}_{\beta_1\cdots\beta_p} &= \nabla_\gamma \pounds_\xi A^{(m+1)\alpha_1\cdots\alpha_q}_{\beta_1\cdots\beta_p} + \left( 1 + \binom{m}{1} \right) \left( \sum_{j=1}^{q} A^{(m+1)\alpha_1\cdots\alpha_q}_{\beta_1\cdots\beta_p} \Sigma^{\alpha_j}{}_{\sigma\gamma} \right. \\
&\qquad \left. - \sum_{i=1}^{p} A^{(m+1)\alpha_1\cdots\alpha_q}_{\beta_1\cdots\beta_p} \Sigma^{\sigma}{}_{\beta_i\gamma} \right) \\
&+ \sum_{k=1}^{m-1} \left( \binom{m}{k+1} + \binom{m}{k} \right) \left( \sum_{j=1}^{q} A^{(m-k+1)\alpha_1\cdots\alpha_q}_{\beta_1\cdots\beta_p} \Sigma^{(k+1)\alpha_j}{}_{\sigma\gamma} - \sum_{i=1}^{p} A^{(m-k+1)\alpha_1\cdots\alpha_q}_{\beta_1\cdots\beta_p} \Sigma^{(k+1)\sigma}{}_{\beta_i\gamma} \right) \\
&+ \binom{m}{m} \left( \sum_{j=1}^{q} A^{\alpha_1\cdots\alpha_q}_{\beta_1\cdots\beta_p} \Sigma^{(m+1)\alpha_j}_{\sigma\gamma} - \sum_{i=1}^{p} A^{\alpha_1\cdots\alpha_q}_{\beta_1\cdots\beta_p} \Sigma^{(m+1)\sigma}{}_{\beta_i\gamma} \right).
\end{aligned}
$$

Using the binomial identity $\binom{m}{i+1} + \binom{m}{i} = \binom{m+1}{i+1}$, the proposition follows by induction. □

Once we know how to commute $\pounds_\xi^{(m-1)}$ and $\nabla$ when acting on an arbitrary tensor field, we can apply the result to equation (5.36) to compute the explicit expression of the tensor $\pounds_\xi^{(m)} R^\mu{}_{\alpha\nu\beta}$. This will be used below to compute the derivatives $\pounds_\xi^{(m)}$ Ric on a null hypersurface $\mathcal{H}$ in terms of transverse derivatives of $g$ on $\mathcal{H}$ up to order $m+1$.

**Proposition 5.8.** *Let $\xi \in \mathfrak{X}(\mathcal{M})$ and $m \geq 2$ an integer. Then,*

$$
\pounds_\xi^{(m)} R^\mu{}_{\alpha\nu\beta} = H^{\gamma\rho}_{\nu\beta} \left( \nabla_\gamma \Sigma^{(m)\mu}{}_{\alpha\rho} + \sum_{k=0}^{m-2} \binom{m}{k+1} \Sigma^{(m-k-1)\sigma}{}_{\alpha\rho} \Sigma^{(k+1)\mu}{}_{\sigma\gamma} \right) \tag{5.38}
$$

*and*

$$
\pounds_\xi^{(m)} R_{\alpha\beta} = H^{\gamma\rho}_{\mu\beta} \left( \nabla_\gamma \Sigma^{(m)\mu}{}_{\alpha\rho} + \sum_{k=0}^{m-2} \binom{m}{k+1} \Sigma^{(m-k-1)\sigma}{}_{\alpha\rho} \Sigma^{(k+1)\mu}{}_{\sigma\gamma} \right). \tag{5.39}
$$

*Proof.* Applying $\pounds_\xi^{(m-1)}$ to (5.36) and using Proposition 5.7,

$$\pounds_\xi^{(m)} R^\mu{}_{\alpha\nu\beta} = H^{\gamma\rho}_{\nu\beta}\Big(\nabla_\gamma \Sigma^{(m)\mu}{}_{\alpha\rho} + \sum_{k=0}^{m-2}\binom{m-1}{k+1}\Big(\Sigma^{(m-k-1)\sigma}{}_{\alpha\rho}\Sigma^{(k+1)\mu}{}_{\sigma\gamma} - \Sigma^{(m-k-1)\mu}{}_{\sigma\alpha}\Sigma^{(k+1)\sigma}{}_{\rho\gamma} - \Sigma^{(m-k-1)\mu}{}_{\rho\sigma}\Sigma^{(k+1)\sigma}{}_{\alpha\gamma}\Big)\Big).$$

Since $H^{\gamma\rho}_{\nu\beta}$ is antisymmetric in $\gamma, \rho$ its contraction with the third term vanishes. Renaming $k' = m-2-k$ in the last term the sum simplifies to

$$\begin{aligned}
&H^{\gamma\rho}_{\nu\beta}\sum_{k=0}^{m-2}\binom{m-1}{k+1}\Big(\Sigma^{(m-k-1)\sigma}{}_{\rho\alpha}\Sigma^{(k+1)\mu}{}_{\sigma\gamma} - \Sigma^{(m-k-1)\mu}{}_{\rho\sigma}\Sigma^{(k+1)\sigma}{}_{\alpha\gamma}\Big)\\
&= H^{\gamma\rho}_{\nu\beta}\left(\sum_{k=0}^{m-2}\binom{m-1}{k+1}\Sigma^{(m-k-1)\sigma}{}_{\rho\alpha}\Sigma^{(k+1)\mu}{}_{\sigma\gamma} - \sum_{k'=0}^{m-2}\binom{m-1}{m-1-k'}\Sigma^{(k'+1)\mu}{}_{\rho\sigma}\Sigma^{(m-k'-1)\sigma}{}_{\alpha\gamma}\right).
\end{aligned}$$

From the antisymmetry of $H^{\gamma\rho}_{\nu\beta}$, the symmetry of $\Sigma^{(k+1)\mu}{}_{\rho\sigma}$ and the combinatorial properties $\binom{m-1}{m-1-k'} = \binom{m}{k'}$ and $\binom{m-1}{k+1} + \binom{m-1}{k} = \binom{m}{k+1}$, (5.38) follows. Equation (5.39) is immediate since the Lie derivative and the trace commute. □

Identity (5.39) constitutes the exact relation between the $m$-th Lie derivative of the Ricci tensor with the Lie derivatives of the tensor $\Sigma$. Before restricting it to a null hypersurface $\mathcal{H}$ with rigging $\xi$ and computing the leading order terms, we shall establish a property that will play a key role in Section 5.5, namely that when the rigging $\xi$ is extended off $\Phi(\mathcal{H})$ by $\nabla_\xi \xi = 0$, the tensors $\pounds_\xi^{(m)} R_{\alpha\beta}$ on $\mathcal{H}$ are geometrical in the following sense.

**Definition 5.9.** *Let $\{\mathcal{H}, \boldsymbol{\gamma}, \boldsymbol{\ell}, \ell^{(2)}\}$ be null metric hypersurface data $(\Phi, \xi)$-embedded in $(\mathcal{M}, g)$. Let $\mathcal{T}$ and $T$ be $(0,p)$ tensor fields on $\mathcal{H}$ and $\mathcal{M}$, respectively.*

- *We say that $\mathcal{T}$ is $\mathcal{H}$-geometrical provided that it depends at most on null metric data $\{\boldsymbol{\gamma}, \boldsymbol{\ell}, \ell^{(2)}\}$ and on the tensors $\{\mathbf{Y}^{(k)}\}_{k\geq 1}$.*
- *We say that $T$ is geometrical provided that the pullback of arbitrary contractions of $T$ with $\xi$ (including no contraction) into $\mathcal{H}$ is $\mathcal{H}$-geometrical.*

In order to prove that $\pounds_\xi^{(m)} R_{\alpha\beta}$ is geometrical for any natural number $m$ we first explore some general properties of geometrical objects and also establish that several building-block tensors that appear in the argument are indeed geometrical. We start with the latter.

**Lemma 5.10.** *Let $\{\mathcal{H}, \boldsymbol{\gamma}, \boldsymbol{\ell}, \ell^{(2)}\}$ be null metric hypersurface data $(\Phi, \xi)$-embedded in $(\mathcal{M}, g)$ and extend $\xi$ off $\Phi(\mathcal{H})$ by $\nabla_\xi \xi = 0$. Then the tensor $\nabla_\alpha \xi_\beta$ is geometrical.*

*Proof.* Since $\xi^\alpha \nabla_\alpha \xi_\beta = 0$ it suffices to check that $\widehat{e}_a^\alpha \widehat{e}_b^\beta \nabla_\alpha \xi_\beta$ and $\widehat{e}_a^\alpha \xi^\beta \nabla_\alpha \xi_\beta$ only depend on hypersurface data. Using $\widehat{e}_a^\mu \nabla_\mu \xi^\beta \overset{\mathcal{H}}{=} (\mathrm{r}-\mathrm{s})_a \xi^\beta + P^{cd}\Pi_{ac}\widehat{e}_d^\beta + \frac{1}{2}\nu^\beta \overset{\circ}{\nabla}_a \ell^{(2)}$ (cf. (2.52)) and Definition 2.2 a straightforward computation shows that

$$\widehat{e}_a^\alpha \widehat{e}_b^\beta \nabla_\alpha \xi_\beta = \Pi_{ab}, \qquad \widehat{e}_a^\alpha \xi^\beta \nabla_\alpha \xi_\beta = \frac{1}{2}\overset{\circ}{\nabla}_a \ell^{(2)}.$$

Hence, $\nabla_\alpha \xi_\beta$ is geometrical. □

Next we show that when $\nabla_\xi \xi = 0$ the tensor $\mathcal{K}^{(m)}_{\mu\nu}$ is geometrical for all $m \geq 0$. As a preliminary step we first compute in full generality the contraction $\xi^\alpha \mathcal{K}^{(m)}_{\alpha\beta}$ on $\mathcal{H}$.

**Proposition 5.11.** *Let $\mathcal{H}$ be a null hypersurface $(\Phi, \xi)$-embedded in $(\mathcal{M}, g)$ and let $Z$ be a vector field along $\Phi(\mathcal{H})$. Extend $\xi$ arbitrarily off $\Phi(\mathcal{H})$ and define $a_\xi := \nabla_\xi \xi$. Then for every $m \in \mathbb{N} \cup \{0\}$,*

$$\mathcal{K}^{(m+1)}(\xi, \xi) \stackrel{\mathcal{H}}{=} 2 \sum_{i=0}^{m} \binom{m}{i} \Big(\mathcal{K}^{(m-i)}\Big)\Big(\pounds^{(i)}_\xi a_\xi, \xi\Big), \tag{5.40}$$

$$\mathcal{K}^{(m+1)}(\xi, Z) \stackrel{\mathcal{H}}{=} \frac{1}{2} Z\left(\mathcal{K}^{(m)}(\xi, \xi)\right) + \sum_{i=0}^{m} \binom{m}{i} \Big(\mathcal{K}^{(m-i)}\Big)\Big(\pounds^{(i)}_\xi a_\xi, Z\Big). \tag{5.41}$$

*Proof.* It suffices to prove (5.41) because (5.40) is then easily obtained by particularizing to $Z = \xi$ and expanding the directional derivative along $Z$ as a Lie derivative. Let us extend $Z$ off $\mathcal{H}$ by $\pounds_\xi Z = 0$ (at the end we prove that the result is independent of the extension). First,

$$(\pounds_\xi g)(\xi, Z) = g(\nabla_\xi \xi, Z) + g(\xi, \nabla_Z \xi) = g(a_\xi, Z) + \frac{1}{2} Z\left(g(\xi, \xi)\right).$$

Applying $\pounds^{(m)}_\xi$ to both sides,

$$(\pounds^{(m+1)}_\xi g)(\xi, Z) = \sum_{i=0}^{m} \binom{m}{i} \Big(\pounds^{(m-i)}_\xi g\Big)\Big(\pounds^{(i)}_\xi a_\xi, Z\Big) + \frac{1}{2} \xi^{(m)}\Big(Z\left(g(\xi, \xi)\right)\Big),$$

which becomes (5.41) after using the commutation property $\xi^{(m)} Z = Z \xi^{(m)}$ (which is an immediate consequence of $[\xi, Z] = 0$). □

**Corollary 5.12.** *Let $\{\mathcal{H}, \boldsymbol{\gamma}, \boldsymbol{\ell}, \ell^{(2)}\}$ be null metric hypersurface data $(\Phi, \xi)$-embedded in $(\mathcal{M}, g)$ and extend $\xi$ off $\Phi(\mathcal{H})$ by $\nabla_\xi \xi = 0$. Then,*

$$\mathcal{K}^{(1)}(\xi, e_a) \stackrel{\mathcal{H}}{=} \frac{1}{2} \overset{\circ}{\nabla}_a \ell^{(2)}, \qquad \mathcal{K}^{(1)}(\xi, \xi) = 0, \qquad \mathcal{K}^{(m)}(\xi, e_a) \stackrel{\mathcal{H}}{=} 0, \qquad \mathcal{K}^{(m)}(\xi, \xi) \stackrel{\mathcal{H}}{=} 0 \tag{5.42}$$

*for all $m \geq 2$.*

As a consequence of this corollary and (2.14),

$$e_b^\mu \mathcal{K}_\mu{}^\rho \stackrel{\mathcal{H}}{=} \left(2P^{cd} \mathrm{Y}_{bc} + \frac{1}{2} n^d \overset{\circ}{\nabla}_b \ell^{(2)}\right) e_d^\rho + 2 \mathrm{r}_b \xi^\rho, \qquad \xi^\mu \mathcal{K}_\mu{}^\rho \stackrel{\mathcal{H}}{=} \frac{1}{2} P^{ab} \overset{\circ}{\nabla}_a \ell^{(2)} e_b^\rho + \frac{1}{2} n(\ell^{(2)}) \xi^\rho, \tag{5.43}$$

and

$$\mathcal{K}^\mu{}_\mu \stackrel{\mathcal{H}}{=} 2 \operatorname{tr}_P \mathbf{Y} + n(\ell^{(2)}), \qquad \mathcal{K}^{(m)\mu}{}_\mu = 2 \operatorname{tr}_P \mathbf{Y}^{(m)} \quad \forall m \geq 2. \tag{5.44}$$

In particular,

$$\nu^\mu \mathcal{K}_\mu{}^\rho \stackrel{\mathcal{H}}{=} \left(2P^{cd} \mathrm{r}_c + \frac{1}{2} n(\ell^{(2)}) n^d\right) e_d^\rho - 2\kappa \xi^\rho. \tag{5.45}$$

**Lemma 5.13.** *Let $\{\mathcal{H}, \boldsymbol{\gamma}, \boldsymbol{\ell}, \ell^{(2)}\}$ be null metric hypersurface data $(\Phi, \xi)$-embedded in $(\mathcal{M}, g)$ and extend $\xi$ off $\Phi(\mathcal{H})$ by $\nabla_\xi \xi = 0$. Then the tensor $\mathcal{K}^{(m)}_{\mu\nu}$ is geometrical for all $m \geq 0$.*

*Proof.* The tensor $\mathcal{K}^{(0)}_{\mu\nu} := g_{\mu\nu}$ is clearly geometrical by the definition of embedded metric hypersurface data. By Corollary 5.12 and recalling $\Phi^\star \mathcal{K}^{(m)} = 2\mathbf{Y}^{(m)}$, the tensor $\mathcal{K}^{(m)}_{\mu\nu}$ for $m \geq 1$ is geometrical as well. □

**Lemma 5.14.** *Let $\{\mathcal{H}, \boldsymbol{\gamma}, \boldsymbol{\ell}, \ell^{(2)}\}$ be null metric hypersurface data $(\Phi, \xi)$-embedded in $(\mathcal{M}, g)$ and extend $\xi$ off $\Phi(\mathcal{H})$ by $\nabla_\xi \xi = 0$. Let $T$ and $S$ be any two geometrical objects. Then, the following properties are true.*

1. *$T \otimes S$ is geometrical.*
2. *Any trace of $T$ w.r.t. $g$ is geometrical.*
3. *Any trace of $T$ w.r.t. $\pounds^{(m)}_\xi g^{\mu\nu}$ is geometrical for any $m \geq 1$.*

*As a consequence, any trace of $T \otimes S$ w.r.t. $g$ or $\pounds^{(m)}_\xi g^{\mu\nu}$ is also geometrical.*

*Proof.* The first property is obvious. The second one follows from the expression of $g^{\alpha\beta}$ in (2.14). We prove the third property by induction. First observe that for $m \geq 1$

$$\pounds^{(m)}_\xi (g^{\alpha\mu} g_{\mu\beta}) = 0 \qquad \Longrightarrow \qquad \pounds^{(m)}_\xi g^{\alpha\beta} = -g^{\beta\nu} \sum_{i=1}^{m} \binom{m}{i} (\pounds^{(m-i)}_\xi g^{\alpha\mu}) \pounds^{(i)}_\xi g_{\mu\nu}. \tag{5.46}$$

Particularizing for $m = 1$ gives $\pounds_\xi g^{\alpha\beta} = -g^{\beta\nu} g^{\alpha\nu} \mathcal{K}_{\mu\nu}$. By Lemma 5.13 and items 1. and 2. of this lemma, item 3. follows for $m = 1$. Assume it is true up to some integer $m-1$. Expression (5.46) together with Lemma 5.13 then show that it is also true for $m$. □

**Lemma 5.15.** *Let $\{\mathcal{H}, \boldsymbol{\gamma}, \boldsymbol{\ell}, \ell^{(2)}\}$ be null metric hypersurface data $(\Phi, \xi)$-embedded in $(\mathcal{M}, g)$ and extend $\xi$ off $\Phi(\mathcal{H})$ by $\nabla_\xi \xi = 0$. Then the tensor $\nabla_\rho \mathcal{K}^{(m)}_{\mu\nu}$ is geometrical for all $m \geq 0$.*

*Proof.* The argument relies on the following well-known relation between $\nabla$ and $\pounds_\xi$ acting on a (0,2) symmetric tensor $T$

$$\nabla_\xi T_{\lambda\gamma} = \pounds_\xi T_{\lambda\gamma} - 2T_{\mu(\lambda} \nabla_{\gamma)} \xi^\mu. \tag{5.47}$$

Particularizing to $T = \mathcal{K}^{(m)}$ gives

$$\xi^\rho \nabla_\rho \mathcal{K}^{(m)}_{\mu\nu} = (\pounds_\xi \mathcal{K}^{(m)})_{\mu\nu} - 2\mathcal{K}^{(m)}_{\rho(\mu} \nabla_{\nu)} \xi^\rho = \mathcal{K}^{(m+1)}_{\mu\nu} - 2g^{\rho\varepsilon} \mathcal{K}^{(m)}_{\rho(\mu} \nabla_{\nu)} \xi_\varepsilon.$$

The first term is geometrical by Lemma 5.13, and the second one is also geometrical as a consequence of Lemmas 5.10, 5.13 and 5.14. □

Next we show that the tensor $\underset{\smile}{\Sigma}^{(m)}$ is geometrical.

**Proposition 5.16.** *Let $\{\mathcal{H}, \boldsymbol{\gamma}, \boldsymbol{\ell}, \ell^{(2)}\}$ be null metric hypersurface data $(\Phi, \xi)$-embedded in $(\mathcal{M}, g)$ and extend $\xi$ off $\Phi(\mathcal{H})$ by $\nabla_\xi \xi = 0$. Then, the tensors $\underset{\smile}{\Sigma}^{(m)}{}_{\nu\alpha\beta}$ and $\Sigma^{(m)}{}_{\nu\alpha\beta}$ are geometrical for all $m \geq 1$.*

*Proof.* Recalling definition (5.37), namely $\underset{\smile}{\Sigma}_{\nu\alpha\beta} := g_{\nu\mu}\Sigma^{\mu}{}_{\alpha\beta}$, and applying identity (5.34) gives

$$\Sigma^{(m)}{}_{\nu\alpha\beta} = g_{\mu\nu}\Sigma^{(m)\mu}{}_{\alpha\beta} = g_{\mu\nu}\pounds_{\xi}^{(m-1)}(g^{\mu\rho}\underset{\smile}{\Sigma}_{\rho\alpha\beta}) = g_{\mu\nu}\sum_{i=0}^{m-1}\binom{m-1}{i}(\pounds_{\xi}^{(i)}g^{\mu\rho})\underset{\smile}{\Sigma}^{(m-i)}{}_{\rho\alpha\beta}. \tag{5.48}$$

So, by Lemma 5.14 if $\underset{\smile}{\Sigma}^{(k)}{}_{\nu\alpha\beta}$ is geometrical for all $k \leq m$, then $\Sigma^{(m)}{}_{\nu\alpha\beta}$ is geometrical as well. So let us prove that $\underset{\smile}{\Sigma}^{(m)}{}_{\nu\alpha\beta}$ is geometrical for all $m \geq 1$. We establish this by an induction argument. From $\underset{\smile}{\Sigma}_{\nu\alpha\beta} = F^{\rho\lambda\gamma}_{\nu\alpha\beta}\nabla_{\rho}\mathcal{K}_{\lambda\gamma}$ and Lemma 5.15 it follows that the tensor $\underset{\smile}{\Sigma}_{\nu\alpha\beta}$ is geometrical. Assume $\underset{\smile}{\Sigma}^{(k)}{}_{\nu\alpha\beta}$ (and hence also $\Sigma^{(k)}{}_{\nu\alpha\beta}$) is geometrical for all $k \leq m$. Applying Proposition 5.7 to $A = \mathcal{K}$ and taking into account $\pounds_{\xi}F = 0$ it follows

$$\begin{aligned}\underset{\smile}{\Sigma}^{(m+1)}_{\nu\alpha\beta} = \pounds_{\xi}^{(m)}\underset{\smile}{\Sigma}_{\nu\alpha\beta} &= F^{\rho\gamma\lambda}_{\nu\alpha\beta}\nabla_{\rho}\mathcal{K}^{(m+1)}_{\lambda\gamma} - 2\sum_{k=0}^{m-1}\binom{m}{k+1}F^{\rho\gamma\lambda}_{\nu\alpha\beta}\mathcal{K}^{(m-k)}_{\sigma(\lambda}\Sigma^{(k+1)\sigma}{}_{\gamma)\rho}\\ &= F^{\rho\gamma\lambda}_{\nu\alpha\beta}\nabla_{\rho}\mathcal{K}^{(m+1)}_{\lambda\gamma} - 2\sum_{k=0}^{m-1}\binom{m}{k+1}F^{\rho\gamma\lambda}_{\nu\alpha\beta}g^{\sigma\varepsilon}\mathcal{K}^{(m-k)}_{\sigma(\lambda}\Sigma^{(k+1)}{}_{\varepsilon|\gamma)\rho}.\end{aligned}$$

Using Lemmas 5.14 and 5.15 and the fact that every $\Sigma^{(k)}{}_{\nu\alpha\beta}$ for $k \leq m$ is geometrical, we conclude that $\underset{\smile}{\Sigma}^{(m+1)}_{\nu\alpha\beta}$ (and thus $\Sigma^{(m+1)}_{\nu\alpha\beta}$) is geometrical as well. □

That $R_{\alpha\beta}$ is geometrical when $\nabla_{\xi}\xi = 0$ is immediate from (2.95), (2.118) and (2.119) because in this case $\mathbf{Z}^{(2)} = \mathbf{Y}^{(2)}$ (see (2.113)). In order to prove that $\pounds_{\xi}^{(m)}R_{\alpha\beta}$ is geometrical for every $m \geq 1$ when $\nabla_{\xi}\xi = 0$ it is convenient to first rewrite (5.39) making the tensor $H$ explicit, namely

$$\begin{aligned}\pounds_{\xi}^{(m)}R_{\alpha\beta} = \nabla_{\mu}\Sigma^{(m)\mu}{}_{\alpha\beta} - \nabla_{\beta}\Sigma^{(m)\mu}{}_{\alpha\mu} + \sum_{k=0}^{m-2}\binom{m}{k+1}g^{\sigma\rho}g^{\mu\nu}\Sigma^{(m-k-1)}{}_{\rho\alpha\beta}\Sigma^{(k+1)}{}_{\nu\sigma\mu}\\ - \sum_{k=0}^{m-2}\binom{m}{k+1}g^{\sigma\rho}g^{\mu\nu}\Sigma^{(m-k-1)}{}_{\rho\alpha\mu}\Sigma^{(k+1)}{}_{\nu\sigma\beta}.\end{aligned} \tag{5.49}$$

By Lemma 5.14 and Proposition 5.16 all the terms in the two sums are geometrical, so it suffices to show that $\nabla_{\mu}\Sigma^{(m)\mu}{}_{\alpha\beta}$ and $\nabla_{\beta}\Sigma^{(m)\mu}{}_{\alpha\mu}$ are also geometrical. In order to do that we shall contract both tensors with two tangent vectors $\widehat{e}^{\alpha}_{a}\widehat{e}^{\beta}_{b}$, one tangent and one transverse $\widehat{e}^{\alpha}_{a}\xi^{\beta}$, and two transverse vectors $\xi^{\alpha}\xi^{\beta}$ and check that in all cases the result only depends on metric data and the expansion $\{\mathbf{Y}^{(k)}\}$. These computations rely on general expressions for the pullback of ambient tensor fields into an arbitrary null hypersurface $\mathcal{H}$ that are computed in full generality in Appendix B for general embedded hypersurfaces and have already been used in Chapter 4. For the sake of clarity we repeat once more the notation employed in Appendix B.

**Notation 5.17.** *Given a $(0,p)$ tensor field $T_{\alpha_1\cdots\alpha_p}$ on $\mathcal{M}$ we use the standard notation $T_{a_1\cdots a_p}$ to denote its pullback to $\mathcal{H}$. Moreover, we introduce the notation ${}^{(i)}T_{\alpha_1\cdots\alpha_{p-1}}$ for $\xi^{\mu}T_{\alpha_1\cdots\alpha_{i-1}\mu\alpha_i\cdots\alpha_{p-1}}$ and ${}^{(i)}T_{a_1\cdots a_{p-1}}$ for the pullback of ${}^{(i)}T_{\alpha_1\cdots\alpha_{p-1}}$ to $\mathcal{H}$. In addition, we use ${}^{(i,j)}T_{a_1\cdots a_{p-2}}$ for the pullback to $\mathcal{H}$ of the tensor obtained by first the contraction of $T$ with $\xi$ in the $j$-th slot and then in the $i$-th slot of the resulting $(0,p-1)$ tensor, i.e. ${}^{(i,j)}T =$*

*$^{(i)}\big(^{(j)}T\big)$. This notation requires care with the order of indices (note that $^{(1,1)}T = {}^{(1,2)}T$ and $^{(1,3)}T = {}^{(2,1)}T \neq {}^{(1,2)}T$).*

Let start by analyzing $\nabla_\mu \Sigma^{(m)\mu}{}_{\alpha\beta}$. Firstly, from Proposition B.3 and the fact that $\Sigma^{(m)}{}_{\nu\alpha\beta}$ is geometrical, it follows that

$$\widehat{e}_a^\alpha \widehat{e}_b^\beta \nabla_\mu \Sigma^{(m)\mu}{}_{\alpha\beta} = n^c (\pounds_\xi \Sigma^{(m)})_{cab} + \mathcal{H}\text{-geometrical terms} = n^c (\Sigma^{(m+1)})_{cab} + \mathcal{H}\text{-geo. terms},$$

and hence by Proposition 5.16 $\widehat{e}_a^\alpha \widehat{e}_b^\beta \nabla_\mu \Sigma^{(m)\mu}{}_{\alpha\beta}$ only depends on metric data and $\{\mathbf{Y}^{(k)}\}$. Secondly, its contraction with $\xi^\alpha \widehat{e}_b^\beta$ yields

$$\xi^\alpha \widehat{e}_b^\beta \nabla_\mu \Sigma^{(m)\mu}{}_{\alpha\beta} = g^{\mu\nu} \widehat{e}_b^\beta \nabla_\mu \big({}^{(2)}\Sigma^{(m)}\big)_{\nu\beta} - g^{\mu\nu} g^{\rho\alpha} \widehat{e}_b^\beta \Sigma^{(m)}{}_{\nu\alpha\beta} \nabla_\mu \xi_\rho.$$

By Lemma 5.14 and Proposition 5.16 the second term is $\mathcal{H}$-geometrical. Moreover, by Proposition B.3,

$$g^{\mu\nu} \widehat{e}_b^\beta \nabla_\mu \big({}^{(2)}\Sigma^{(m)}\big)_{\nu\beta} = n^c \Big(\pounds_\xi \big({}^{(2)}\Sigma^{(m)}\big)\Big)_{cb} + \mathcal{H}\text{-geo. terms} = \big({}^{(2)}\Sigma^{(m+1)}\big)_{cb}\, n^c + \mathcal{H}\text{-geo. terms},$$

where we used $\pounds_\xi \big({}^{(i)}T^{(m)}\big) = {}^{(i)}T^{(m+1)}$ since obviously $\pounds_\xi \xi = 0$. Again Prop. 5.16 shows that $\xi^\alpha \widehat{e}_b^\beta \nabla_\mu \Sigma^{(m)\mu}{}_{\alpha\beta}$ (and by symmetry also $\xi^\beta \widehat{e}_b^\alpha \nabla_\mu \Sigma^{(m)\mu}{}_{\alpha\beta}$) only depends on metric data and $\{\mathbf{Y}^{(k)}\}$. Finally, the contraction with $\xi^\alpha \xi^\beta$ is

$$\xi^\alpha \xi^\beta \nabla_\mu \Sigma^{(m)\mu}{}_{\alpha\beta} = g^{\mu\nu} \nabla_\mu \big({}^{(2,3)}\Sigma^{(m)}\big)_\nu - g^{\mu\nu} g^{\beta\rho} \big({}^{(2)}\Sigma^{(m)}\big)_{\nu\beta} \nabla_\mu \xi_\rho - g^{\mu\nu} g^{\rho\alpha} \big({}^{(3)}\Sigma^{(m)}\big)_{\nu\alpha} \nabla_\mu \xi_\rho.$$

By Prop. B.3 the first term is

$$g^{\mu\nu} \nabla_\mu \big({}^{(2,3)}\Sigma^{(m)}\big)_\nu = \big({}^{(2,3)}\Sigma^{(m+1)}\big)_c n^c + \mathcal{H}\text{-geo. terms},$$

so by Prop. 5.16 it is $\mathcal{H}$-geometrical. The second and third terms are $\mathcal{H}$-geometrical as well by Lemmas 5.10 and 5.14 and Prop. 5.16. Hence $\nabla_\mu \Sigma^{(m)\mu}{}_{\alpha\beta}$ is geometrical. For the tensor $\nabla_\beta \Sigma^{(m)\mu}{}_{\alpha\mu}$ we introduce $T^{(m)}{}_\alpha := \Sigma^{(m)\mu}{}_{\alpha\mu}$, which is geometrical by Lemma 5.14. The contractions $\widehat{e}_a^\alpha \widehat{e}_b^\beta \nabla_\beta T_\alpha$, $\widehat{e}_a^\alpha \xi^\beta \nabla_\beta T_\alpha$ and $\xi^\alpha \widehat{e}_b^\beta \nabla_\beta T_\alpha$ are automatically $\mathcal{H}$-geometrical after using identities (B.1)-(B.3) with $n^{(2)} = 0$ in Appendix B and $\pounds_\xi T^{(m)} = T^{(m+1)}$. Finally, $\xi^\alpha \xi^\beta \nabla_\beta T^{(m)}{}_\alpha = \pounds_\xi \big({}^{(1)}T^{(m)}\big) = {}^{(1)}T^{(m+1)}$ is also $\mathcal{H}$-geometrical, and hence $\nabla_\beta \Sigma^{(m)\mu}{}_{\alpha\mu}$ is geometrical. Thus, the following result has been proved.

**Proposition 5.18.** *Let $\{\mathcal{H}, \boldsymbol{\gamma}, \boldsymbol{\ell}, \ell^{(2)}\}$ be null metric hypersurface data $(\Phi, \xi)$-embedded in $(\mathcal{M}, g)$ and extend $\xi$ off $\Phi(\mathcal{H})$ by $\nabla_\xi \xi = 0$. Then, the tensors $\pounds_\xi^{(m)} R_{\alpha\beta}$ are geometrical for every $m \geq 0$.*

Once we have shown that the tensors $\pounds_\xi^{(m)} R_{\alpha\beta}$ are geometrical our next task is to compute the leading order terms of $\pounds_\xi^{(m)} R_{\alpha\beta}$ on an arbitrary null hypersurface $\mathcal{H}$. As we shall see, it turns out that the leading order terms of $\xi^\alpha \pounds_\xi^{(m)} R_{\alpha\beta}$ involve $m+2$ transverse derivatives of the metric, whereas the fully tangential components of $\pounds_\xi^{(m)} R_{\alpha\beta}$ depend at most on $m+1$ transverse derivatives. In order to write down equalities only to the leading order it is useful introduce the following notation.

**Notation 5.19.** *Let $(\mathcal{M}, g)$ be a semi-Riemannian manifold and $T, S$ be two tensor fields involving $g$ and its derivatives. The notation $T \overset{[m]}{=} S$ means that the tensor $T - S$ does not depend on derivatives of $g$ of order $m$ or higher.*

The only terms that have a chance to carry $m+1$ and $m+2$ derivatives of the metric in identity (5.49) are the ones of the form $\nabla_\gamma \Sigma^{(m)\mu}{}_{\alpha\beta}$. Thus, with the notation above,

$$\pounds_\xi^{(m)} R_{\alpha\beta} \overset{[m+1]}{=} \nabla_\mu \Sigma^{(m)\mu}{}_{\alpha\beta} - \nabla_\beta \Sigma^{(m)\mu}{}_{\mu\alpha}, \qquad m \geq 1. \tag{5.50}$$

Observe that equation (5.34) together with $\pounds_\xi g^{\mu\nu} = -\mathcal{K}^{\mu\nu}$ implies

$$\Sigma^{(m)\mu}{}_{\alpha\beta} = \pounds_\xi^{(m-1)} \left( g^{\mu\nu} \underset{\smile}{\Sigma}_{\nu\alpha\beta} \right) \overset{[m]}{=} g^{\mu\nu} \underset{\smile}{\Sigma}^{(m)}_{\nu\alpha\beta} - (m-1) \mathcal{K}^{\mu\nu} \underset{\smile}{\Sigma}^{(m-1)}_{\nu\alpha\beta}. \tag{5.51}$$

Lowering the index $\mu$ gives

$$\Sigma^{(m)}_{\mu\alpha\beta} \overset{[m]}{=} \underset{\smile}{\Sigma}^{(m)}_{\mu\alpha\beta} - (m-1) g^{\nu\rho} \mathcal{K}_{\mu\nu} \underset{\smile}{\Sigma}^{(m-1)}_{\rho\alpha\beta}, \qquad \Sigma^{(m)}_{\mu\alpha\beta} \overset{[m+1]}{=} \underset{\smile}{\Sigma}^{(m)}_{\mu\alpha\beta}, \tag{5.52}$$

and applying $\pounds_\xi$,

$$\pounds_\xi \Sigma^{(m)}_{\mu\alpha\beta} \overset{[m+1]}{=} \underset{\smile}{\Sigma}^{(m+1)}_{\mu\alpha\beta} - (m-1) g^{\nu\rho} \mathcal{K}_{\mu\nu} \underset{\smile}{\Sigma}^{(m)}_{\rho\alpha\beta}, \qquad \pounds_\xi \Sigma^{(m)}_{\mu\alpha\beta} \overset{[m+2]}{=} \underset{\smile}{\Sigma}^{(m+1)}_{\mu\alpha\beta}. \tag{5.53}$$

**Lemma 5.20.** *Let $\xi \in \mathfrak{X}(\mathcal{M})$ and $m \geq 1$ an integer. Then,*

$$\underset{\smile}{\Sigma}^{(m)}_{\nu\alpha\beta} \overset{[m]}{=} F^{\rho\lambda\gamma}_{\nu\alpha\beta} \nabla_\rho \mathcal{K}^{(m)}_{\lambda\gamma} - g^{\sigma\varepsilon} \mathcal{K}_{\sigma\nu} F^{\rho\lambda\gamma}_{\varepsilon\alpha\beta} \nabla_\rho \mathcal{K}^{(m-1)}_{\lambda\gamma}, \tag{5.54}$$

$$\Sigma^{(m)\mu}{}_{\alpha\beta} \overset{[m]}{=} g^{\mu\nu} F^{\rho\lambda\gamma}_{\nu\alpha\beta} \nabla_\rho \mathcal{K}^{(m)}_{\lambda\gamma} - m \mathcal{K}^{\mu\nu} F^{\rho\lambda\gamma}_{\nu\alpha\beta} \nabla_\rho \mathcal{K}^{(m-1)}_{\lambda\gamma}. \tag{5.55}$$

*Proof.* By Proposition 5.7 and (5.37), together with $\pounds_\xi F^{\rho\lambda\gamma}_{\nu\alpha\beta} = 0$, it follows

$$\pounds_\xi^{(m-1)} \underset{\smile}{\Sigma}_{\nu\alpha\beta} = F^{\rho\lambda\gamma}_{\nu\alpha\beta} \nabla_\rho \mathcal{K}^{(m)}_{\lambda\gamma} - \sum_{k=0}^{m-2} \binom{m-1}{k+1} \Big( \mathcal{K}^{(m-1-k)}_{\sigma(\beta} \Sigma^{(k+1)\sigma}{}_{\nu)\alpha} + \mathcal{K}^{(m-1-k)}_{\sigma(\alpha} \Sigma^{(k+1)\sigma}{}_{\nu)\beta} - \mathcal{K}^{(m-1-k)}_{\sigma(\alpha} \Sigma^{(k+1)\sigma}{}_{\beta)\nu} \Big),$$

which simplifies to

$$\pounds_\xi^{(m-1)} \underset{\smile}{\Sigma}_{\nu\alpha\beta} = F^{\rho\lambda\gamma}_{\nu\alpha\beta} \nabla_\rho \mathcal{K}^{(m)}_{\lambda\gamma} - \sum_{k=0}^{m-2} \binom{m-1}{k+1} \mathcal{K}^{(m-1-k)}_{\sigma\nu} \Sigma^{(k+1)\sigma}{}_{\alpha\beta}$$

because of the symmetries of $\mathcal{K}^{(i)}_{\alpha\beta}$ and $\Sigma^{(i)\sigma}{}_{\mu\nu}$. Hence,

$$\underset{\smile}{\Sigma}^{(m)}_{\nu\alpha\beta} \overset{[m]}{=} F^{\rho\lambda\gamma}_{\nu\alpha\beta} \nabla_\rho \mathcal{K}^{(m)}_{\lambda\gamma} - \mathcal{K}_{\sigma\nu} \Sigma^{(m-1)\sigma}{}_{\alpha\beta}, \tag{5.56}$$

$$\underset{\smile}{\Sigma}^{(m-1)}_{\nu\alpha\beta} \overset{[m]}{=} F^{\rho\lambda\gamma}_{\nu\alpha\beta} \nabla_\rho \mathcal{K}^{(m-1)}_{\lambda\gamma}. \tag{5.57}$$

Combining (5.51) and (5.57) it follows $\Sigma^{(m-1)\sigma}{}_{\alpha\beta} \stackrel{[m]}{=} g^{\sigma\nu} \underset{\smile}{\Sigma}^{(m-1)}_{\nu\alpha\beta} \stackrel{[m]}{=} g^{\sigma\varepsilon} F^{\rho\lambda\gamma}_{\varepsilon\alpha\beta} \nabla_\rho \mathcal{K}^{(m-1)}_{\lambda\gamma}$, which inserted into (5.56) gives

$$\underset{\smile}{\Sigma}^{(m)}_{\nu\alpha\beta} \stackrel{[m]}{=} F^{\rho\lambda\gamma}_{\nu\alpha\beta} \nabla_\rho \mathcal{K}^{(m)}_{\lambda\gamma} - g^{\sigma\varepsilon} \mathcal{K}_{\sigma\nu} F^{\rho\lambda\gamma}_{\varepsilon\alpha\beta} \nabla_\rho \mathcal{K}^{(m-1)}_{\lambda\gamma}.$$

This establishes (5.54). Introducing (5.57) and (5.54) into (5.51) gives (5.55). □

We quote the following immediate consequence for future reference.

**Corollary 5.21.** *Let $\xi \in \mathfrak{X}(\mathcal{M})$ and $m \geq 0$ an integer. Then,*

$$\underset{\smile}{\Sigma}^{(m)}_{\nu\alpha\beta} \stackrel{[m+1]}{=} F^{\rho\lambda\gamma}_{\nu\alpha\beta} \nabla_\rho \mathcal{K}^{(m)}_{\lambda\gamma}, \qquad (5.58) \qquad\qquad \Sigma^{(m)\mu}{}_{\alpha\beta} \stackrel{[m+1]}{=} g^{\mu\nu} F^{\rho\lambda\gamma}_{\nu\alpha\beta} \nabla_\rho \mathcal{K}^{(m)}_{\lambda\gamma}. \qquad (5.59)$$

Before computing $\mathcal{R}^{(m)}$, $\dot{\mathcal{R}}^{(m)}$ and $\ddot{\mathcal{R}}^{(m)}$ on $\mathcal{H}$ it is important to make the following observation based on Proposition 5.11.

**Remark 5.22.** *By Proposition 5.11 the derivatives $\mathcal{K}^{(k)}(\xi, \cdot)$ are given in terms of transverse derivatives of $a_\xi := \nabla_\xi \xi$ on $\Phi(\mathcal{H})$, metric hypersurface data as well as on the tensors $\{\mathbf{Y}, ..., \mathbf{Y}^{(k-1)}\}$. Hence, when computing $\mathcal{R}^{(m)}$, $\dot{\mathcal{R}}^{(m)}$ and $\ddot{\mathcal{R}}^{(m)}$, terms of the form $\mathcal{K}^{(k)}(\xi, \cdot)$ can always be replaced by lower order terms, i.e. terms that depend on metric hypersurface data and $\{\mathbf{Y}, ..., \mathbf{Y}^{(k-1)}\}$.*

**Notation 5.23.** *Let $(\mathcal{M}, g)$ be a semi-Riemannian manifold, $\mathcal{H}$ an embedded hypersurface and $T, S$ two tensors involving $g$ and its derivatives. We introduce the notation $T \stackrel{(m)}{=} S$ to denote that the tensor $(T - S)|_{\mathcal{H}}$ does not depend on **transverse** derivatives of $g$ at $\mathcal{H}$ of order $m$ or higher.*

In Lemma 5.24 and Proposition 5.25 we compute the pullback of $\underset{\smile}{\Sigma}^{(m)}$ to the hypersurface as well as several contractions of the tensors $\Sigma^{(m)}$ and $\underset{\smile}{\Sigma}^{(m)}$ that we shall use below. The computation relies again on the general identities for the pullback of ambient tensor fields to arbitrary hypersurfaces computed in Appendix B.

**Lemma 5.24.** *Let $\mathcal{H}$ be a null hypersurface $(\Phi, \xi)$-embedded in a semi-Riemannian manifold $(\mathcal{M}, g)$. Extend $\xi$ arbitrarily off $\Phi(\mathcal{H})$ and let $\{e_a\}$ be a local basis on $\mathcal{H}$ and $\widehat{e}_a := \Phi_\star e_a$. Then,*

$$\xi^\lambda \nabla_\rho \mathcal{K}^{(m)}_{\lambda\gamma} \stackrel{(m+1)}{=} 0, \qquad (5.60) \qquad \widehat{e}^\rho_a \widehat{e}^\lambda_b \widehat{e}^\gamma_c \nabla_\rho \mathcal{K}^{(m+1)}_{\lambda\gamma} \stackrel{(m+1)}{=} 2\overset{\circ}{\nabla}_a \mathrm{Y}^{(m+1)}_{bc} + 4\mathrm{Y}_{a(b} \mathrm{r}^{(m+1)}_{c)}, \qquad (5.61)$$

$$\xi^\gamma \widehat{e}^\rho_a \widehat{e}^\lambda_b \nabla_\rho \mathcal{K}^{(m)}_{\lambda\gamma} \stackrel{(m+1)}{=} 0, \qquad (5.62) \qquad \widehat{e}^\rho_a \widehat{e}^\lambda_b \widehat{e}^\gamma_c \nabla_\rho \mathcal{K}^{(m)}_{\lambda\gamma} \stackrel{(m+1)}{=} 0, \qquad (5.63)$$

$$\widehat{e}^\gamma_a \widehat{e}^\lambda_b \xi^\rho \nabla_\rho \mathcal{K}^{(m)}_{\lambda\gamma} \stackrel{(m+1)}{=} 2\mathrm{Y}^{(m+1)}_{ab}, \qquad (5.64) \qquad \widehat{e}^\lambda_a \widehat{e}^\gamma_b g^{\sigma\rho} \nabla_\rho \mathcal{K}^{(m)}_{\lambda\gamma} \stackrel{(m+1)}{=} 2\nu^\sigma \mathrm{Y}^{(m+1)}_{ab}, \qquad (5.65)$$

$$g^{\lambda\gamma} \widehat{e}^\rho_a \nabla_\rho \mathcal{K}^{(m)}_{\lambda\gamma} \stackrel{(m+1)}{=} 0, \qquad (5.66) \qquad \widehat{e}^\rho_a \widehat{e}^\lambda_b g^{\sigma\gamma} \nabla_\rho \mathcal{K}^{(m)}_{\lambda\gamma} \stackrel{(m+1)}{=} 0. \qquad (5.67)$$

*Proof.* Equations (5.62) and (5.63) are particular cases of (5.60) and (5.61), respectively. Moreover, the validity of (5.60) and (5.61) implies at once (5.66) and (5.67) because $g^{\sigma\gamma}$ can be decomposed using (2.14). Hence it suffices to prove the equations in the first and third lines. In order to prove (5.60) we contract $\xi^\lambda \nabla_\rho \mathcal{K}^{(m)}_{\lambda\gamma}$ with (i) $\xi^\rho \xi^\gamma$, (ii) $\xi^\rho \widehat{e}^\gamma_c$, (iii) $\widehat{e}^\rho_c \xi^\gamma$ and

(iv) $\widehat{e}_c^\rho \widehat{e}_d^\gamma$ and show that all of them are at most of order $m$. Let us start with (i). Applying (5.47) for $T = \mathcal{K}^{(m)}$ gives

$$\xi^\rho \xi^\gamma \xi^\lambda \nabla_\rho \mathcal{K}^{(m)}_{\lambda\gamma} = \xi^\lambda \xi^\gamma \Big( \pounds_\xi \mathcal{K}^{(m)}_{\lambda\gamma} - 2\mathcal{K}^{(m)}_{\mu(\lambda}\nabla_{\gamma)}\xi^\mu \Big) \overset{(m+1)}{=} \xi^\lambda \xi^\gamma \mathcal{K}^{(m+1)}_{\lambda\gamma} \overset{(m+1)}{=} 0,$$

where in the second equality we used that $\mathcal{K}^{(m)}$ is of order $m$ and in the last one Remark 5.22. For (ii) we contract $\xi^\lambda \nabla_\rho \mathcal{K}^{(m)}_{\lambda\gamma}$ with $\xi^\rho \widehat{e}_c^\gamma$ and use again that $\mathcal{K}^{(m)}$ is of order $m$,

$$\widehat{e}_c^\gamma \xi^\rho \xi^\lambda \nabla_\rho \mathcal{K}^{(m)}_{\lambda\gamma} = \widehat{e}_c^\gamma \xi^\rho \nabla_\rho ({}^{(1)}\mathcal{K}^{(m)})_\gamma - \widehat{e}_c^\gamma \xi^\rho \mathcal{K}^{(m)}_{\lambda\gamma} \nabla_\rho \xi^\lambda \overset{(m+1)}{=} \widehat{e}_c^\gamma \xi^\rho \nabla_\rho ({}^{(1)}\mathcal{K}^{(m)})_\gamma.$$

By Remark 5.22 the tensor ${}^{(1)}\mathcal{K}^{(m)}$ is of order $m-1$. Applying (B.2) for $T = {}^{(1)}\mathcal{K}^{(m)}$ all terms in the right hand side are at most of order $m$, so $\widehat{e}_c^\gamma \xi^\rho \xi^\lambda \nabla_\rho \mathcal{K}^{(m)}_{\lambda\gamma} \overset{(m+1)}{=} 0$. For item (iii) we contract $\xi^\lambda \nabla_\rho \mathcal{K}^{(m)}_{\lambda\gamma}$ with $\widehat{e}_c^\rho \xi^\gamma$, namely

$$\widehat{e}_c^\rho \xi^\gamma \xi^\lambda \nabla_\rho \mathcal{K}^{(m)}_{\lambda\gamma} = \widehat{e}_c^\rho \nabla_\rho \Big( \mathcal{K}^{(m)}(\xi,\xi) \Big) - 2\xi^\gamma \mathcal{K}^{(m)}_{\lambda\gamma} \nabla_\rho \xi^\lambda \overset{(m+1)}{=} 0,$$

because ${}^{(1)}\mathcal{K}^{(m)}$ is at most of order $m-1$. In order to prove (iv), i.e. $\widehat{e}_c^\rho \widehat{e}_d^\gamma \xi^\lambda \nabla_\rho \mathcal{K}^{(m)}_{\lambda\gamma} \overset{(m+1)}{=} 0$, we use identity (B.3) with $T = \mathcal{K}^{(m)}$. Since all the terms in the right hand side are at most of order $m$, $\widehat{e}_c^\rho \widehat{e}_d^\gamma \xi^\lambda \nabla_\rho \mathcal{K}^{(m)}_{\lambda\gamma} \overset{(m+1)}{=} 0$. This establishes (5.60). Equation (5.61) follows from (B.1) with $T = \mathcal{K}^{(m+1)}$ after recalling that ${}^{(i)}\mathcal{K}^{(m+1)}$ is at most of order $m$ and that $\Phi^\star \mathcal{K}^{(m+1)} = 2\mathbf{Y}^{(m+1)}$. It only remains to prove (5.64) and (5.65). For the first one we insert (2.14) into $\widehat{e}_a^\lambda \widehat{e}_b^\gamma g^{\sigma\rho} \nabla_\rho \mathcal{K}^{(m)}_{\lambda\gamma}$ and get

$$\widehat{e}_a^\lambda \widehat{e}_b^\gamma g^{\sigma\rho} \nabla_\rho \mathcal{K}^{(m)}_{\lambda\gamma} = \widehat{e}_a^\lambda \widehat{e}_b^\gamma \Big( P^{cd} \widehat{e}_c^\sigma \widehat{e}_d^\rho + \xi^\sigma \widehat{e}_c^\rho n^c + \xi^\rho \nu^\sigma \Big) \nabla_\rho \mathcal{K}^{(m)}_{\lambda\gamma} \overset{(m+1)}{=} \widehat{e}_a^\lambda \widehat{e}_b^\gamma \xi^\rho \nu^\sigma \nabla_\rho \mathcal{K}^{(m)}_{\lambda\gamma},$$

where equation (5.63) has been used in the first and second terms. This expression shows that (5.65) follows from (5.64). To establish the latter we use (B.2) with $T = \mathcal{K}^{(m)}$ because only the first term in the right hand side is of order $m+1$. □

**Proposition 5.25.** *Let $\mathcal{H}$ be a null hypersurface $(\Phi,\xi)$-embedded in a semi-Riemannian manifold $(\mathcal{M}, g)$ and extend $\xi$ arbitrarily off $\Phi(\mathcal{H})$. Then,*

$$\underset{\smile}{\Sigma}{}^{(m+1)}_{cab} \overset{(m+1)}{=} \overset{\circ}{\nabla}_a \mathrm{Y}^{(m+1)}_{bc} + \overset{\circ}{\nabla}_b \mathrm{Y}^{(m+1)}_{ac} - \overset{\circ}{\nabla}_c \mathrm{Y}^{(m+1)}_{ab} + 2\mathrm{r}^{(m+1)}_c \mathrm{Y}_{ab} + 2\mathrm{r}_c \mathrm{Y}^{(m+1)}_{ab},$$

$$\underset{\smile}{\Sigma}{}^{(m)}_{abc} \overset{(m+1)}{=} 0, \quad ({}^{(1)}\underset{\smile}{\Sigma}{}^{(m)})_{ab} \overset{(m+1)}{=} -\mathrm{Y}^{(m+1)}_{ab}, \quad ({}^{(2)}\underset{\smile}{\Sigma}{}^{(m)})_{ab} = ({}^{(3)}\underset{\smile}{\Sigma}{}^{(m)})_{ab} \overset{(m+1)}{=} \mathrm{Y}^{(m+1)}_{ab},$$

$$\xi^\alpha \xi^\beta \Sigma^{(m)\mu}{}_{\alpha\beta} \overset{(m+1)}{=} 0, \qquad \Sigma^{(m)\mu}{}_{\mu a} \overset{(m+1)}{=} 0, \qquad \xi^\alpha \Sigma^{(m)\mu}{}_{\mu\alpha} \overset{(m+1)}{=} \operatorname{tr}_P \mathbf{Y}^{(m+1)}.$$

*Proof.* Consider equation (5.54) for $m+1$ and contract it with $\widehat{e}_c^\nu \widehat{e}_a^\alpha \widehat{e}_b^\beta$, namely

$$\underset{\smile}{\Sigma}{}^{(m+1)}_{cab} \overset{[m+1]}{=} \widehat{e}_c^\nu \widehat{e}_a^\alpha \widehat{e}_b^\beta F^{\rho\lambda\gamma}_{\nu\alpha\beta} \nabla_\rho \mathcal{K}^{(m+1)}_{\lambda\gamma} - \widehat{e}_c^\nu \widehat{e}_a^\alpha \widehat{e}_b^\beta g^{\sigma\varepsilon} \mathcal{K}_{\sigma\nu} F^{\rho\lambda\gamma}_{\varepsilon\alpha\beta} \nabla_\rho \mathcal{K}^{(m)}_{\lambda\gamma}.$$

Defining $F^{def}_{cab} := \frac{1}{2}\left(\delta^d_a\delta^e_b\delta^f_c + \delta^d_b\delta^e_a\delta^f_c - \delta^d_c\delta^e_a\delta^f_b\right)$ and using equation (5.61) the first term takes the form

$$\begin{aligned} F^{def}_{cab}\left(\nabla\mathcal{K}^{(m+1)}\right)_{def} &\overset{(m+1)}{=} 2F^{def}_{cab}\left(\overset{\circ}{\nabla}_d \mathrm{Y}^{(m+1)}_{ef} + 2\mathrm{Y}_{d(e}\mathrm{r}^{(m+1)}_{f)}\right) \\ &\overset{(m+1)}{=} \overset{\circ}{\nabla}_a \mathrm{Y}^{(m+1)}_{bc} + \overset{\circ}{\nabla}_b \mathrm{Y}^{(m+1)}_{ac} - \overset{\circ}{\nabla}_c \mathrm{Y}^{(m+1)}_{ab} + 2\mathrm{r}^{(m+1)}_c \mathrm{Y}_{ab}. \end{aligned}$$

Making $F^{\rho\lambda\gamma}_{\varepsilon\alpha\beta}$ explicit the second term is

$$\widehat{e}^\nu_c \widehat{e}^\alpha_a \widehat{e}^\beta_b g^{\sigma\varepsilon}\mathcal{K}_{\sigma\nu}F^{\rho\lambda\gamma}_{\varepsilon\alpha\beta}\nabla_\rho\mathcal{K}^{(m)}_{\lambda\gamma} = \frac{1}{2}\left(\widehat{e}^\rho_a\widehat{e}^\lambda_b g^{\sigma\gamma} + \widehat{e}^\lambda_a\widehat{e}^\rho_b g^{\sigma\gamma} - \widehat{e}^\lambda_a\widehat{e}^\gamma_b g^{\sigma\rho}\right)\mathcal{K}_{\sigma\nu}\widehat{e}^\nu_c\nabla_\rho\mathcal{K}^{(m)}_{\lambda\gamma}.$$

Applying (5.67) to the first and second terms and (5.65) in the last term,

$$\widehat{e}^\nu_c \widehat{e}^\alpha_a \widehat{e}^\beta_b g^{\sigma\varepsilon}\mathcal{K}_{\sigma\nu}F^{\rho\lambda\gamma}_{\varepsilon\alpha\beta}\nabla_\rho\mathcal{K}^{(m)}_{\lambda\gamma} \overset{(m+1)}{=} -\mathcal{K}_{\sigma\nu}\widehat{e}^\nu_c n^d \widehat{e}^\sigma_d \mathrm{Y}^{(m+1)}_{ab} \overset{(m+1)}{=} -2\mathrm{r}_c\mathrm{Y}^{(m+1)}_{ab},$$

because $\mathbf{\Phi}^\star\mathcal{K} = 2\mathbf{Y}$. Hence the equation of the first line is established. The equations of the second line follow from (5.58), namely

$$\underset{\smile}{\Sigma}^{(m)}_{\nu\alpha\beta} \overset{[m+1]}{=} F^{\rho\lambda\gamma}_{\nu\alpha\beta}\nabla_\rho\mathcal{K}^{(m)}_{\lambda\gamma}. \tag{5.68}$$

Contracting this with $\widehat{e}^\nu_c\widehat{e}^\alpha_a\widehat{e}^\beta_b$ and using (5.63) gives $\underset{\smile}{\Sigma}^{(m)}_{cab} \overset{(m+1)}{=} 0$, and contracting it with $\xi^\nu e^\alpha_a \widehat{e}^\beta_b$ gives

$$({}^{(1)}\underset{\smile}{\Sigma}^{(m)})_{ab} \overset{[m+1]}{=} \frac{1}{2}\left(\xi^\gamma\widehat{e}^\rho_a\widehat{e}^\lambda_b + \xi^\gamma\widehat{e}^\lambda_a\widehat{e}^\rho_b - \xi^\rho\widehat{e}^\lambda_a\widehat{e}^\gamma_b\right)\nabla_\rho\mathcal{K}^{(m)}_{\lambda\gamma}.$$

Applying (5.62) to the first and second terms and (5.64) to the last one yields $({}^{(1)}\underset{\smile}{\Sigma}^{(m)})_{ab} \overset{(m+1)}{=} -\mathrm{Y}^{(m+1)}_{ab}$. In a similar way, contracting (5.68) with $\widehat{e}^\nu_a\xi^\alpha\widehat{e}^\beta_b$ gives

$$({}^{(2)}\underset{\smile}{\Sigma}^{(m)})_{ab} \overset{[m+1]}{=} \frac{1}{2}\left(\xi^\rho\widehat{e}^\gamma_a\widehat{e}^\lambda_b + \xi^\lambda\widehat{e}^\gamma_a\widehat{e}^\rho_b - \xi^\lambda\widehat{e}^\rho_a\widehat{e}^\gamma_b\right)\nabla_\rho\mathcal{K}^{(m)}_{\lambda\gamma}.$$

By (5.62) the only term that contributes is the first one, which after using (5.64) is $({}^{(2)}\underset{\smile}{\Sigma}^{(m)})_{ab} \overset{(m+1)}{=} \mathrm{Y}^{(m+1)}_{ab}$. By the symmetry of $\underset{\smile}{\Sigma}^{(m)}$, $({}^{(3)}\underset{\smile}{\Sigma}^{(m)})_{ab} \overset{(m+1)}{=} \mathrm{Y}^{(m+1)}_{ab}$. To prove the third line consider (5.59), namely

$$\Sigma^{(m)\mu}{}_{\alpha\beta} \overset{[m+1]}{=} g^{\mu\nu}F^{\rho\lambda\gamma}_{\nu\alpha\beta}\nabla_\rho\mathcal{K}^{(m)}_{\lambda\gamma}. \tag{5.69}$$

The contraction with $\xi^\alpha\xi^\beta$,

$$\xi^\alpha\xi^\beta\Sigma^{(m)\mu}{}_{\alpha\beta} \overset{[m+1]}{=} \frac{1}{2}\left(2\xi^\lambda\xi^\rho g^{\mu\gamma} - \xi^\gamma\xi^\lambda g^{\mu\rho}\right)\nabla_\rho\mathcal{K}^{(m)}_{\lambda\gamma}, \tag{5.70}$$

yields $\xi^\alpha\xi^\beta\Sigma^{(m)\mu}{}_{\alpha\beta} \overset{(m+1)}{=} 0$ after using (5.60). Taking trace in $\mu,\alpha$ in (5.69) gives

$$\Sigma^{(m)\mu}{}_{\mu\beta} \overset{[m+1]}{=} g^{\mu\nu}F^{\rho\lambda\gamma}_{\nu\mu\beta}\nabla_\rho\mathcal{K}^{(m)}_{\lambda\gamma} = \frac{1}{2}\left(g^{\rho\gamma}\delta^\lambda_\beta + g^{\lambda\gamma}\delta^\rho_\beta - g^{\lambda\rho}\delta^\gamma_\beta\right)\nabla_\rho\mathcal{K}^{(m)}_{\lambda\gamma} = \frac{1}{2}g^{\lambda\gamma}\nabla_\beta\mathcal{K}^{(m)}_{\lambda\gamma}, \tag{5.71}$$

where in the last equality we used the symmetry of $\mathcal{K}^{(m)}$. Contracting with $\widehat{e}_a^\beta$ and using equation (5.66) yields $\widehat{e}_a^\beta \Sigma^{(m)\mu}{}_{\mu\beta} \overset{(m+1)}{=} 0$. Finally, from (5.47) with $T = \mathcal{K}^{(m)}$,

$$\nabla_\xi \mathcal{K}^{(m)} \overset{[m+1]}{=} \mathcal{K}^{(m+1)}, \tag{5.72}$$

so contracting (5.71) with $\xi^\beta$ gives $\xi^\beta \Sigma^{(m)\mu}{}_{\mu\beta} \overset{(m+1)}{=} \frac{1}{2} g^{\lambda\gamma} \mathcal{K}^{(m+1)}_{\lambda\gamma}$. Using (2.13) and the fact that ${}^{(1)}\mathcal{K}^{(m+1)} \overset{(m+1)}{=} 0$, the last equation of the third line follows. □

We are now ready to compute the tensors $\mathcal{R}^{(m)}$, $\dot{\mathcal{R}}^{(m)}$ and $\ddot{\mathcal{R}}^{(m)}$ on $\mathcal{H}$ by contracting the identity (5.50) with $\xi^\alpha \xi^\beta$, $\widehat{e}_a^\alpha \xi^\beta$ and $\widehat{e}_a^\alpha \widehat{e}_b^\beta$, respectively.

**Proposition 5.26.** *Let $\mathcal{H}$ be a null hypersurface $(\Phi, \xi)$-embedded in $(\mathcal{M}, g)$ and extended $\xi$ arbitrarily off $\Phi(\mathcal{H})$. Then for every $m \geq 0$,*

$$\ddot{\mathcal{R}}^{(m+1)} \overset{(m+2)}{=} -\operatorname{tr}_P \mathbf{Y}^{(m+2)}. \tag{5.73}$$

*Proof.* The case $m = 0$ has already been established in (5.28). In order to prove the identity for $m \geq 1$ we contract the general expression (5.50) with $\xi^\alpha \xi^\beta$ so that

$$\begin{aligned}
\xi^\alpha \xi^\beta \pounds_\xi^{(m)} \operatorname{Ric}_{\alpha\beta} &\overset{[m+2]}{=} \xi^\alpha \xi^\beta \nabla_\mu \Sigma^{(m)\mu}{}_{\alpha\beta} - \xi^\alpha \xi^\beta \nabla_\beta \Sigma^{(m)\mu}{}_{\mu\alpha} \\
&\overset{[m+2]}{=} \nabla_\mu \Big( \xi^\alpha \xi^\beta \Sigma^{(m)\mu}{}_{\alpha\beta} \Big) - \nabla_\xi \Big( \xi^\alpha \Sigma^{(m)\mu}{}_{\mu\alpha} \Big),
\end{aligned}$$

where in the second line we used that $\Sigma^{(m)}$ involves up to $m+1$ derivatives of $g$. We start by computing the first term. For the same reason as before, by (5.70) it follows

$$\begin{aligned}
\nabla_\mu \Big( \xi^\alpha \xi^\beta \Sigma^{(m)\mu}{}_{\alpha\beta} \Big) &\overset{[m+2]}{=} \xi^\lambda \xi^\rho g^{\mu\gamma} \nabla_\mu \nabla_\rho \mathcal{K}^{(m)}_{\lambda\gamma} - \frac{1}{2} \xi^\gamma \xi^\lambda g^{\mu\rho} \nabla_\mu \nabla_\rho \mathcal{K}^{(m)}_{\lambda\gamma} \\
&\overset{[m+2]}{=} \xi^\lambda g^{\mu\gamma} \nabla_\mu \nabla_\xi \mathcal{K}^{(m)}_{\lambda\gamma} - \xi^\lambda g^{\mu\gamma} \nabla_\rho \mathcal{K}^{(m)}_{\lambda\gamma} \nabla_\mu \xi^\rho - \frac{1}{2} \xi^\gamma \xi^\lambda g^{\mu\rho} \nabla_\mu \nabla_\rho \mathcal{K}^{(m)}_{\lambda\gamma}.
\end{aligned}$$

From the expression of $g^{\mu\rho}$ in (2.14) and the fact that $\nabla \mathcal{K}^{(m)}$ is at most order $m+1$, the only term that has a chance of carrying $m+2$ transverse derivatives of $g$ is the first one. Using (5.72) and recalling (5.60),

$$\xi^\lambda g^{\mu\gamma} \nabla_\mu \nabla_\xi \mathcal{K}^{(m)}_{\lambda\gamma} \overset{[m+2]}{=} \xi^\lambda g^{\mu\gamma} \nabla_\mu \mathcal{K}^{(m+1)}_{\lambda\gamma} \overset{(m+2)}{=} 0.$$

We conclude by computing the last term. From (5.71) it follows

$$\nabla_\xi \Big( \xi^\alpha \Sigma^{(m)\mu}{}_{\mu\alpha} \Big) \overset{[m+2]}{=} \frac{1}{2} g^{\lambda\gamma} \nabla_\xi \Big( \xi^\mu \nabla_\mu \mathcal{K}^{(m)}_{\lambda\gamma} \Big) \overset{(m+2)}{=} \frac{1}{2} g^{\lambda\gamma} \mathcal{K}^{(m+2)}_{\lambda\gamma} \overset{(m+2)}{=} \operatorname{tr}_P \mathbf{Y}^{(m+2)},$$

where in the second equality we used $\nabla_\xi \mathcal{K}^{(m)} \overset{[m+1]}{=} \mathcal{K}^{(m+1)}$ and $\nabla_\xi \mathcal{K}^{(m+1)} \overset{[m+2]}{=} \mathcal{K}^{(m+2)}$ (see (5.72)) and in the third one we inserted (2.14) and used the fact that ${}^{(1)}\mathcal{K}^{(m+2)}$ is at most of order $m+1$ (see Remark 5.22). □

**Proposition 5.27.** *Let $\mathcal{H}$ be a null hypersurface $(\Phi,\xi)$-embedded in $(\mathcal{M},g)$ and extend $\xi$ arbitrarily off $\Phi(\mathcal{H})$. Then for any $m\geq 0$,*

$$\dot{\mathcal{R}}_a^{(m+1)} \overset{(m+2)}{=} \mathrm{r}_a^{(m+2)}. \tag{5.74}$$

*Proof.* The case $m=0$ is (5.29). From (5.50) and the fact that $\Sigma^{(m)}$ is at most of order $m+1$,

$$\begin{aligned}\widehat{e}_a^\alpha\xi^\beta\pounds_\xi^{(m)}R_{\alpha\beta} &\overset{[m+1]}{=} \widehat{e}_a^\alpha\xi^\beta\nabla_\mu\Sigma^{(m)\mu}{}_{\alpha\beta}-\widehat{e}_a^\alpha\xi^\beta\nabla_\beta\Sigma^{(m)\mu}{}_{\mu\alpha}\\ &\overset{[m+1]}{=} \widehat{e}_a^\alpha\nabla_\mu\Big(\xi^\beta\Sigma^{(m)\mu}{}_{\alpha\beta}\Big)-\widehat{e}_a^\alpha\Sigma^{(m)\mu}{}_{\alpha\beta}\nabla_\mu\xi^\beta-\widehat{e}_a^\alpha\nabla_\xi\Big(\Sigma^{(m)\mu}{}_{\mu\alpha}\Big),\end{aligned}$$

so the only terms capable of containing $m+2$ transverse derivatives are the first and third ones. By Proposition B.3 applied to ${}^{(3)}\Sigma^{(m)\mu}{}_\alpha$ and $\pounds_\xi\xi^\beta=0$ it follows

$$\widehat{e}_a^\alpha\nabla_\mu\Big(\xi^\beta\Sigma^{(m)\mu}{}_{\alpha\beta}\Big)\overset{(m+2)}{=} n^b\widehat{e}_a^\alpha\widehat{e}_b^\mu\xi^\beta\pounds_\xi\Sigma^{(m)}_{\mu\alpha\beta}$$

because the rest of the terms are of (transverse) order $m+1$ or below. Using the second equation in (5.53) and Proposition 5.25 we obtain

$$\widehat{e}_a^\alpha\nabla_\mu\Big(\xi^\beta\Sigma^{(m)\mu}{}_{\alpha\beta}\Big)\overset{(m+2)}{=} n^b\widehat{e}_a^\alpha\widehat{e}_b^\mu\xi^\beta\underset{\smile}{\Sigma}^{(m+1)}_{\mu\alpha\beta}\overset{(m+2)}{=}\mathrm{r}_a^{(m+2)}.$$

Finally, by (B.2) with $T=\Sigma^{(m)\mu}{}_{\mu\alpha}$ it follows

$$\widehat{e}_a^\alpha\nabla_\xi\Big(\Sigma^{(m)\mu}{}_{\mu\alpha}\Big)\overset{(m+2)}{=}\widehat{e}_a^\alpha\pounds_\xi\Sigma^{(m)\mu}{}_{\mu\alpha}\overset{(m+2)}{=}\Sigma^{(m+1)\mu}{}_{\mu a}\overset{(m+2)}{=}0,$$

where the last equality we used again Proposition 5.25. □

**Proposition 5.28.** *Let $\mathcal{H}$ be a null hypersurface $(\Phi,\xi)$-embedded in $(\mathcal{M},g)$ and extended $\xi$ arbitrarily off $\Phi(\mathcal{H})$. Let $m\geq 1$ be an integer. Then,*

$$\begin{aligned}\mathcal{R}_{ab}^{(m+1)} \overset{(m+1)}{=}& -2\pounds_n\mathrm{Y}_{ab}^{(m+1)}-\big(2(m+1)\kappa+\mathrm{tr}_P\,\mathbf{U}\big)\,\mathrm{Y}_{ab}^{(m+1)}-(\mathrm{tr}_P\,\mathbf{Y}^{(m+1)})\mathrm{U}_{ab}\\ &+4P^{cd}\mathrm{U}_{c(a}\mathrm{Y}_{b)d}^{(m+1)}+4(\mathrm{s}-\mathrm{r})_{(a}\mathrm{r}_{b)}^{(m+1)}+2\overset{\circ}{\nabla}_{(a}\mathrm{r}_{b)}^{(m+1)}-2\kappa^{(m+1)}\mathrm{Y}_{ab}.\end{aligned}\tag{5.75}$$

*Proof.* By the general formula (5.50) we have

$$\mathcal{R}_{ab}^{(m+1)} \overset{[m+1]}{=} \widehat{e}_a^\alpha\widehat{e}_b^\beta\nabla_\mu\Sigma^{(m)\mu}{}_{\alpha\beta}-\widehat{e}_a^\alpha\widehat{e}_b^\beta\nabla_\beta\Sigma^{(m)\mu}{}_{\mu\alpha}, \tag{5.76}$$

so we only need to compute each of these terms. For the first one we use Proposition B.3 applied to $\Sigma^{(m)}$, namely

$$\begin{aligned}(\operatorname{div}\Sigma^{(m)})_{ab} \stackrel{(m+1)}{=} & P^{bc}\overset{\circ}{\nabla}_b\Sigma^{(m)}_{cab} + n^c(\pounds_\xi\Sigma^{(m)})_{cab} + n^c\overset{\circ}{\nabla}_c({}^{(1)}\Sigma^{(m)})_{ab} + (2\kappa + \operatorname{tr}_P \mathbf{U})({}^{(1)}\Sigma^{(m)})_{ab}\\ & + (\operatorname{tr}_P \mathbf{Y} - n(\ell^{(2)}))n^c\Sigma^{(m)}_{cab} - 2P^{dc}(\mathrm{r}+\mathrm{s})_d\Sigma^{(m)}_{cab} + 2P^{dc}\mathrm{Y}_{d(a|}\Sigma^{(m)}_{c|b)f}n^f\\ & + 2P^{dc}\mathrm{U}_{d(a|}({}^{(2)}\Sigma^{(m)})_{c|b)} - 2(\mathrm{r}-\mathrm{s})_{(a|}({}^{(2)}\Sigma^{(m)})_{c|b)}n^c\\ & - 2V^c{}_{(a|}n^d\Sigma^{(m)}_{dc|b)} + 2\mathrm{r}_{(a|}({}^{(1)}\Sigma^{(m)})_{c|b)}n^c.\end{aligned}$$

From the second equation in (5.52), $\Sigma^{(m)}_{\alpha\beta\gamma} \stackrel{(m+1)}{=} \underset{\smile}{\Sigma}^{(m)}_{\alpha\beta\gamma}$, and thus by Proposition 5.25 it follows $\Sigma^{(m)}_{abc} \stackrel{(m+1)}{=} 0$, $({}^{(1)}\Sigma^{(m)})_{ab} \stackrel{(m+1)}{=} -\mathrm{Y}^{(m+1)}_{ab}$ and $({}^{(2)}\Sigma^{(m)})_{ab} \stackrel{(m+1)}{=} \mathrm{Y}^{(m+1)}_{ab}$. Hence the expression for $(\operatorname{div}\Sigma^{(m)})_{ab}$ simplifies to

$$\begin{aligned}(\operatorname{div}\Sigma^{(m)})_{ab} \stackrel{(m+1)}{=} & n^c(\pounds_\xi\Sigma^{(m)})_{cab} - n^c\overset{\circ}{\nabla}_c\mathrm{Y}^{(m+1)}_{ab} - (2\kappa + \operatorname{tr}_P\mathbf{U})\mathrm{Y}^{(m+1)}_{ab}\\ & + 2P^{dc}\mathrm{U}_{d(a|}\mathrm{Y}^{(m+1)}_{c|b)} - 2(\mathrm{r}-\mathrm{s})_{(a|}\mathrm{Y}^{(m+1)}_{c|b)}n^c - 2\mathrm{r}_{(a|}\mathrm{Y}^{(m+1)}_{c|b)}n^c.\end{aligned} \tag{5.77}$$

In order to compute the term $n^c(\pounds_\xi\Sigma^{(m)})_{cab}$ we contract the first equation in (5.53) with $\widehat{e}^\mu_c\widehat{e}^\alpha_a\widehat{e}^\beta_b$, which gives

$$(\pounds_\xi\Sigma^{(m)})_{cab} \stackrel{[m+1]}{=} \widehat{e}^\mu_c\widehat{e}^\alpha_a\widehat{e}^\beta_b\underset{\smile}{\Sigma}^{(m+1)}_{\mu\alpha\beta} - (m-1)\widehat{e}^\mu_c\widehat{e}^\alpha_a\widehat{e}^\beta_b g^{\nu\rho}\mathcal{K}_{\mu\nu}\underset{\smile}{\Sigma}^{(m)}_{\rho\alpha\beta}.$$

The first term is given by the first line in Proposition 5.25, and, for the second one, using (2.14) as well as $\underset{\smile}{\Sigma}^{(m)}_{abc} \stackrel{(m+1)}{=} 0$ and ${}^{(1)}\underset{\smile}{\Sigma}^{(m)}_{ab} \stackrel{(m+1)}{=} -\mathrm{Y}^{(m+1)}_{ab}$ one gets

$$\widehat{e}^\mu_c\widehat{e}^\alpha_a\widehat{e}^\beta_b g^{\nu\rho}\mathcal{K}_{\mu\nu}\underset{\smile}{\Sigma}^{(m)}_{\rho\alpha\beta} \stackrel{(m+1)}{=} -2\mathrm{r}_c\mathrm{Y}^{(m+1)}_{ab}.$$

Combining everything it follows

$$(\pounds_\xi\Sigma^{(m)})_{cab} \stackrel{(m+1)}{=} \overset{\circ}{\nabla}_a\mathrm{Y}^{(m+1)}_{bc} + \overset{\circ}{\nabla}_b\mathrm{Y}^{(m+1)}_{ac} - \overset{\circ}{\nabla}_c\mathrm{Y}^{(m+1)}_{ab} + 2\mathrm{r}^{(m+1)}_c\mathrm{Y}_{ab} + 2m\mathrm{r}_c\mathrm{Y}^{(m+1)}_{ab},$$

and after contracting with $n^c$ equation (5.77) becomes

$$\begin{aligned}(\operatorname{div}\Sigma^{(m)})_{ab} \stackrel{(m+1)}{=} & 2n^c\overset{\circ}{\nabla}_{(a}\mathrm{Y}^{(m+1)}_{b)c} - 2n^c\overset{\circ}{\nabla}_c\mathrm{Y}^{(m+1)}_{ab} - 2\kappa^{(m+1)}\mathrm{Y}_{ab} - (2(m+1)\kappa + \operatorname{tr}_P\mathbf{U})\mathrm{Y}^{(m+1)}_{ab}\\ & + 2P^{dc}\mathrm{U}_{d(a}\mathrm{Y}^{(m+1)}_{b)c} + 2(\mathrm{s}-2\mathrm{r})_{(a}\mathrm{r}^{(m+1)}_{b)}.\end{aligned}$$

Using

$$\begin{aligned}2n^c\overset{\circ}{\nabla}_{(a}\mathrm{Y}^{(m+1)}_{b)c} - 2n^c\overset{\circ}{\nabla}_c\mathrm{Y}^{(m+1)}_{ab} &= 2\overset{\circ}{\nabla}_{(a}\mathrm{r}^{(m+1)}_{b)} - 2\mathrm{Y}^{(m+1)}_{c(a}\overset{\circ}{\nabla}_{b)}n^c - 2\pounds_n\mathrm{Y}^{(m+1)}_{ab} + 4\mathrm{Y}^{(m+1)}_{c(a}\overset{\circ}{\nabla}_{b)}n^c\\ &= 2\overset{\circ}{\nabla}_{(a}\mathrm{r}^{(m+1)}_{b)} - 2\pounds_n\mathrm{Y}^{(m+1)}_{ab} + 2\mathrm{Y}^{(m+1)}_{c(a}\overset{\circ}{\nabla}_{b)}n^c\\ &= 2\overset{\circ}{\nabla}_{(a}\mathrm{r}^{(m+1)}_{b)} - 2\pounds_n\mathrm{Y}^{(m+1)}_{ab} + 2\mathrm{Y}^{(m+1)}_{c(a}\mathrm{U}_{b)d}P^{cd} + 2\mathrm{r}^{(m+1)}_{(a}\mathrm{s}_{b)},\end{aligned}$$

where the third equality follows from (2.66), the expression for $(\operatorname{div}\Sigma^{(m)})_{ab}$ is finally

$$\begin{aligned}(\operatorname{div}\Sigma^{(m)})_{ab} \overset{(m+1)}{=} & -2\pounds_n \mathrm{Y}^{(m+1)}_{ab} - (2(m+1)\kappa + \operatorname{tr}_P \mathbf{U})\mathrm{Y}^{(m+1)}_{ab} + 4P^{cd}\mathrm{Y}^{(m+1)}_{c(a}\mathrm{U}_{b)d} \\ & + 4(\mathrm{s}-\mathrm{r})_{(a}\mathrm{r}^{(m+1)}_{b)} + 2\overset{\circ}{\nabla}_{(a}\mathrm{r}^{(m+1)}_{b)} - 2\kappa^{(m+1)}\mathrm{Y}_{ab}.\end{aligned} \tag{5.78}$$

To compute the second term in (5.76) we use equation (B.1) for $T = \Sigma^{(m)\mu}{}_{\mu\alpha}$. Since by Proposition 5.25 $\Sigma^{(m)\mu}{}_{\mu a} \overset{(m+1)}{=} 0$ and $\xi^\alpha\Sigma^{(m)\mu}{}_{\mu\alpha} \overset{(m+1)}{=} \operatorname{tr}_P \mathbf{Y}^{(m+1)}$, we find that $\widehat{e}^\alpha_a \widehat{e}^\beta_b \nabla_\beta \Sigma^{(m)\mu}{}_{\mu\alpha} = (\operatorname{tr}_P \mathbf{Y}^{(m+1)})\mathrm{U}_{ab}$. Inserting this and (5.78) into (5.76) proves the Proposition. □

Note that (5.75) for $m = 0$ does not recover (5.24) due to the different coefficients in front of the term $\mathrm{r}_{(a}\mathrm{r}^{(m+1)}_{b)}$. We want to make clear that this is not an error. In fact, there is a simple heuristic argument that helps understanding why this is so. Consider a function $f(t)$ and construct $g = f^2$. Denoting by $g^{(m)}$ and $f^{(m)}$ the $m$-th derivative w.r.t. $t$, it follows from Newton's binomial formula that, for $m \geq 1$,

$$g^{(m)} = 2ff^{(m)} + \text{lower order derivatives.}$$

Clearly, this formula does not recover the case $m = 0$ by a factor of 2 (precisely like (5.75) and (5.24)).

Proposition 5.28 is interesting because it allows one to determine the evolution of the transverse expansion of the metric along the null generator in any null hypersurface. Moreover, no assumptions concerning neither the topology and dimension of $\mathcal{H}$, nor the ambient fields equations have been made. We next compute two contractions of (5.75) as well as its trace w.r.t. $P$.

**Corollary 5.29.** *Let $\mathcal{H}$ be a null hypersurface $(\Phi,\xi)$-embedded in $(\mathcal{M}, g)$ and extend $\xi$ arbitrarily off $\Phi(\mathcal{H})$. Let* Ric *be the Ricci tensor of $g$. Then,*

$$\mathcal{R}^{(m+1)}_{ab} n^b \overset{(m+1)}{=} -\pounds_n \mathrm{r}^{(m+1)}_a - (2m\kappa + \operatorname{tr}_P \mathbf{U})\mathrm{r}^{(m+1)}_a - \overset{\circ}{\nabla}_a \kappa^{(m+1)}, \tag{5.79}$$

$$\mathcal{R}^{(m+1)}_{ab} n^a n^b \overset{(m+1)}{=} (2m\kappa + \operatorname{tr}_P \mathbf{U})\kappa^{(m+1)}, \tag{5.80}$$

$$\begin{aligned}P^{ab}\mathcal{R}^{(m+1)}_{ab} \overset{(m+1)}{=} & -2\pounds_n(\operatorname{tr}_P \mathbf{Y}^{(m+1)}) - 2((m+1)\kappa + \operatorname{tr}_P \mathbf{U}) \operatorname{tr}_P \mathbf{Y}^{(m+1)} \\ & + 2\kappa^{(m+1)}(n(\ell^{(2)}) - \operatorname{tr}_P \mathbf{Y}) - 4P(\mathbf{r}+\mathbf{s}, \mathbf{r}^{(m+1)}) + 2\operatorname{div}_P \mathbf{r}^{(m+1)}.\end{aligned} \tag{5.81}$$

*Proof.* To prove (5.79) we contract (5.75) with $n^b$ and use (2.46) with $n^{(2)} = 0$, namely

$$2n^b\overset{\circ}{\nabla}_{(a}\mathrm{r}^{(m+1)}_{b)} = \pounds_n \mathrm{r}^{(m+1)}_a - \overset{\circ}{\nabla}_a\kappa^{(m+1)} + 2\kappa^{(m+1)}\mathrm{s}_a - 2P^{bc}\mathrm{U}_{ac}\mathrm{r}^{(m+1)}_b.$$

Contracting (5.79) with $n^a$, (5.80) follows at once. For the last one apply $P^{ab}$ to (5.75) and use (2.51) with $\mathbf{T} = \mathbf{Y}^{(m+1)}$. □

Later on we shall need to use the "complete" identities (5.73), (5.74) and (5.75), i.e. including a term that gathers all the lower order terms. By Proposition 5.18 these terms are $\mathcal{H}$-geometrical when $\nabla_\xi \xi = 0$.

**Corollary 5.30.** *Let $\{\mathcal{H}, \gamma, \boldsymbol{\ell}, \ell^{(2)}\}$ be null metric hypersurface data $(\Phi, \xi)$-embedded in $(\mathcal{M}, g)$ and extended $\xi$ arbitrarily off $\Phi(\mathcal{H})$. Let $m \geq 1$ be an integer. Then,*

$$\ddot{\mathcal{R}}^{(m)} = -\operatorname{tr}_P \mathbf{Y}^{(m+1)} + \mathcal{O}^{(m)}(\mathbf{Y}^{\leq m}), \quad (5.82) \qquad \dot{\mathcal{R}}^{(m)}_a = \mathrm{r}^{(m+1)}_a + \mathcal{O}^{(m)}_a(\mathbf{Y}^{\leq m}), \quad (5.83)$$

$$\begin{aligned}\mathcal{R}^{(m+1)}_{ab} &= -2\pounds_n \mathrm{Y}^{(m+1)}_{ab} - (2(m+1)\kappa + \operatorname{tr}_P \mathbf{U})\mathrm{Y}^{(m+1)}_{ab} - (\operatorname{tr}_P \mathbf{Y}^{(m+1)})\mathrm{U}_{ab} \\ &\quad + 4P^{cd}\mathrm{U}_{c(a}\mathrm{Y}^{(m+1)}_{b)d} + 4(\mathrm{s}-\mathrm{r})_{(a}\mathrm{r}^{(m+1)}_{b)} + 2\overset{\circ}{\nabla}_{(a}\mathrm{r}^{(m+1)}_{b)} \\ &\quad - 2\kappa^{(m+1)}\mathrm{Y}_{ab} + \mathcal{O}^{(m)}_{ab}(\mathbf{Y}^{\leq m}),\end{aligned} \tag{5.84}$$

*where $\mathcal{O}^{(m)}$, $\mathcal{O}^{(m)}_a$ and $\mathcal{O}^{(m)}_{ab}$ are, respectively, a scalar, a one-form and a (0,2) symmetric tensor on $\mathcal{H}$ with the property that when $\nabla_\xi \xi = 0$ they only depend on null metric data $\{\gamma, \boldsymbol{\ell}, \ell^{(2)}\}$ and on the tensors $\{\mathbf{Y}, ..., \mathbf{Y}^{(m)}\}$. Moreover,*

$$\mathcal{R}^{(m+1)}_{ab} n^b = -\pounds_n \mathrm{r}^{(m+1)}_a - (2m\kappa + \operatorname{tr}_P \mathbf{U})\mathrm{r}^{(m+1)}_a - \overset{\circ}{\nabla}_a \kappa^{(m+1)} + \mathcal{O}^{(m)}_{ab} n^b, \tag{5.85}$$

$$\begin{aligned}P^{ab}\mathcal{R}^{(m+1)}_{ab} &= -2\pounds_n(\operatorname{tr}_P \mathbf{Y}^{(m+1)}) - 2\left((m+1)\kappa + \operatorname{tr}_P \mathbf{U}\right)\operatorname{tr}_P \mathbf{Y}^{(m+1)} \\ &\quad + 2\kappa^{(m+1)}(n(\ell^{(2)}) - \operatorname{tr}_P \mathbf{Y}) - 4P(\mathbf{r}+\mathbf{s}, \mathbf{r}^{(m+1)}) + 2\operatorname{div}_P \mathbf{r}^{(m+1)} + P^{ab}\mathcal{O}^{(m)}_{ab}.\end{aligned} \tag{5.86}$$

As one can see, the scalar $\operatorname{tr}_P \mathbf{Y}^{(m)}$ and the one-form $\mathbf{r}^{(m)}$ appear both in identities (5.82), (5.83) and in (5.85), (5.86). It is straightforward to combine them and obtain

$$\mathcal{R}^{(m+1)}_{ab} n^b + \pounds_n \dot{\mathcal{R}}^{(m)}_a + (2m\kappa + \operatorname{tr}_P \mathbf{U})\dot{\mathcal{R}}^{(m)}_a - \overset{\circ}{\nabla}_a(\dot{\mathcal{R}}^{(m)}_b n^b) = \mathrm{T}^{(m)}_a, \tag{5.87}$$

and

$$\begin{gathered}P^{ab}\mathcal{R}^{(m+1)}_{ab} - 2\pounds_n \ddot{\mathcal{R}}^{(m)} - 2((m+1)\kappa + \operatorname{tr}_P \mathbf{U})\ddot{\mathcal{R}}^{(m)} \\ +2(n(\ell^{(2)}) - \operatorname{tr}_P \mathbf{Y})\dot{\mathcal{R}}^{(m)}_a n^a + 4P(\mathbf{r}+\mathbf{s}, \dot{\mathcal{R}}^{(m)}) - 2\operatorname{div}_P \dot{\mathcal{R}}^{(m)} = \mathrm{T}^{(m)},\end{gathered} \tag{5.88}$$

where $\mathrm{T}^{(m)}_a$ and $\mathrm{T}^{(m)}$ are, respectively, a one-form and a scalar function that depend only on metric data and $\{\mathbf{Y}, ..., \mathbf{Y}^{(m)}\}$ as well as in the way the rigging has been extended (this dependence drops out completely when $\xi$ is extended geodesically). Identities (5.87)-(5.88) are a manifestation of the (ambient) second contracted Bianchi identity order by order on $\Phi(\mathcal{H})$. The dependence of $\mathrm{T}^{(m)}_a$ and $\mathrm{T}^{(m)}$ in terms of the tensors $\mathcal{O}^{(m)}$, $\mathcal{O}^{(m)}_a$ and $\mathcal{O}^{(m)}_{ab}$ can be easily read of from equations (5.87)-(5.88). We do not write its explicit form since it will not be needed.

**Remark 5.31.** *With the notation introduced in 5.23 it is clear from* (5.82), (5.83), (5.84) *and* (6.55) *that* $\mathcal{R}^{(m)}_{ab} \overset{(m+1)}{=} 0$, $\dot{\mathcal{R}}^{(m)}_a \overset{(m+2)}{=} 0$, $\ddot{\mathcal{R}}^{(m)} \overset{(m+2)}{=} 0$ *and* $R^{(m)} \overset{(m+2)}{=} 0$.

For later use we need to compute the "gauge" transformations of the tensors $\mathcal{O}^{(m)}(\mathbf{Y}^{\leq m})$, $\mathcal{O}^{(m)}_a(\mathbf{Y}^{\leq m})$ and $\mathcal{O}^{(m)}_{ab}(\mathbf{Y}^{\leq m})$ under the change $\xi' = z\xi$ with $\xi(z) = 0$. In order not to overload the notation we drop the arguments "$\mathbf{Y}^{\leq m}$" and "$\mathbf{Y}'^{\leq m}$". When a prime is placed on the quantities $\mathcal{O}^{(m)\prime}$, $\mathcal{O}^{(m)\prime}_a$ and $\mathcal{O}^{(m)\prime}_{ab}$ it means that the object has been constructed with $\xi'$ according to (5.82)-(5.84).

**Proposition 5.32.** *Let $\{\mathcal{H},\boldsymbol{\gamma},\boldsymbol{\ell},\ell^{(2)}\}$ be null metric hypersurface data $(\Phi,\xi)$-embedded in $(\mathcal{M},g)$ and extend $\xi$ off $\Phi(\mathcal{H})$ by $\nabla_\xi\xi=0$. Assume $\ell^{(2)}=0$ and let $z\in\mathcal{F}^\star(\mathcal{M})$ satisfying $\xi(z)=0$ and define $\xi':=z\xi$. Then, for every $m\geq 1$,*

$$\mathcal{O}^{(m)\prime} = z^{m+1}\mathcal{O}^{(m)}, \tag{5.89}$$

$$\mathcal{O}^{(m)\prime}_a = z^m\mathcal{O}^{(m)}_a - (m-1)z^{m-1}\left(\operatorname{tr}_P \mathbf{Y}^{(m)} - \mathcal{O}^{(m-1)}\right)\overset{\circ}{\nabla}_a z, \tag{5.90}$$

$$\mathcal{O}^{(m)\prime}_{ab} = z^m\mathcal{O}^{(m)}_{ab} + 2mz^{m-1}\mathcal{O}^{(m)}_{(a}\overset{\circ}{\nabla}_{b)}z - m(m-1)z^{m-2}\left(\operatorname{tr}_P \mathbf{Y}^{(m)} - \mathcal{O}^{(m-1)}\right)\overset{\circ}{\nabla}_a z\overset{\circ}{\nabla}_b z. \tag{5.91}$$

*Proof.* The idea of the proof is to compute the transformations of the expansion $\{\mathbf{Y}^{(m)}\}$ and of the tensors $\ddot{\mathcal{R}}^{(m)}$, $\dot{\mathcal{R}}^{(m)}_a$ and $\mathcal{R}^{(m+1)}_{ab}$ and insert them into (5.82)-(5.84) to obtain the expressions for $\mathcal{O}^{(m)\prime}$, $\mathcal{O}^{(m)\prime}_a$ and $\mathcal{O}^{(m)\prime}_{ab}$. Under the rescaling $\xi'=z\xi$ the tensors $n$, $P$, $\mathbf{U}$, $\mathbf{s}-\mathbf{r}$ and the metric connection $\overset{\circ}{\nabla}$ transform as (see (2.34), (2.37), (2.33), (2.76) and (2.59), respectively)

$$n' = z^{-1}n,\qquad \mathbf{U}'=z^{-1}\mathbf{U},\qquad P'=P,\qquad \mathbf{s}'-\mathbf{r}'=\mathbf{s}-\mathbf{r}-z^{-1}dz, \tag{5.92}$$

$$\overset{\circ}{\nabla}{}' = \overset{\circ}{\nabla} + z^{-1}n\otimes\boldsymbol{\ell}\otimes_s dz. \tag{5.93}$$

In order to compute the transformations of $\{\mathbf{Y}^{(m)}\}$ and of $\{\ddot{\mathcal{R}}^{(m)},\dot{\mathcal{R}}^{(m)}_a,\mathcal{R}^{(m+1)}_{ab}\}$ we first prove the following identity valid for every symmetric $(0,2)$ tensor field $T$,

$$\begin{aligned}\pounds^{(m)}_{z\xi}T_{\alpha\beta} &= z^m\pounds^{(m)}_{\xi}T_{\alpha\beta} + 2mz^{m-1}\xi^\mu\pounds^{(m-1)}_{\xi}T_{\mu(\alpha}\nabla_{\beta)}z\\ &\quad + m(m-1)z^{m-2}\xi^\mu\xi^\nu\pounds^{(m-2)}_{\xi}T_{\mu\nu}\nabla_\alpha z\nabla_\beta z.\end{aligned} \tag{5.94}$$

For $m=1$ it is clearly true because it reduces to the standard identity

$$\pounds_{z\xi}T_{\alpha\beta} = z\pounds_\xi T_{\alpha\beta} + 2\xi^\mu T_{\mu(\alpha}\nabla_{\beta)}z.$$

Let us assume it holds for $m\geq 2$. Then, applying $\pounds_{z\xi}$ to (5.94) and using $\xi(z)=0$ gives

$$\begin{aligned}\pounds^{(m+1)}_{z\xi}T_{\alpha\beta} &= z\pounds_\xi\pounds^{(m)}_{z\xi}T_{\alpha\beta} + 2\xi^\mu\pounds^{(m)}_{z\xi}T_{\mu(\alpha}\nabla_{\beta)}z\\ &= z^{m+1}\pounds^{(m+1)}_{\xi}T_{\alpha\beta} + 2mz^m\xi^\mu\pounds^{(m)}_{\xi}T_{\mu(\alpha}\nabla_{\beta)}z\\ &\quad + m(m-1)z^{m-1}\xi^\mu\xi^\nu\pounds^{(m-1)}_{\xi}T_{\mu\nu}\nabla_\alpha z\nabla_\beta z\\ &\quad + 2z^m\xi^\mu\pounds^{(m)}_{\xi}T_{\mu(\alpha}\nabla_{\beta)}z + 2mz^{m-1}\xi^\mu\xi^\nu\pounds^{(m-1)}_{\xi}T_{\mu\nu}\nabla_\alpha z\nabla_\beta z\\ &= z^{m+1}\pounds^{(m+1)}_{\xi}T_{\alpha\beta} + 2(m+1)z^m\xi^\mu\pounds^{(m)}_{\xi}T_{\mu(\alpha}\nabla_{\beta)}z\\ &\quad + (m+1)mz^{m-1}\xi^\mu\xi^\nu\pounds^{(m-1)}_{\xi}T_{\mu\nu}\nabla_\alpha z\nabla_\beta z.\end{aligned}$$

Hence (5.94) follows by induction. Applying (5.94) to $T = g$ and using Corollary 5.12 with $\ell^{(2)} = 0$ gives

$$\mathcal{K}' = z\mathcal{K} + 2dz \otimes_s g(\xi, \cdot), \qquad \mathcal{K}^{(k)\prime} = z^k \mathcal{K}^{(k)} \quad (k \geq 2),$$

which becomes

$$\mathbf{Y}' = z\mathbf{Y} + dz \otimes_s \boldsymbol{\ell}, \qquad \mathbf{Y}^{(k)\prime} = z^k \mathbf{Y}^{(k)} \quad (k \geq 2) \tag{5.95}$$

after pulling it back to $\mathcal{H}$. Note that the transformation of $\mathbf{Y}$ is precisely (2.30) with $V = 0$, as it must be. Consequently,

$$\mathbf{r}' = \mathbf{r} + \frac{1}{2z}\left(dz + n(z)\boldsymbol{\ell}\right), \qquad \mathbf{r}^{(k)\prime} = z^{k-1}\mathbf{r}^{(k)}, \tag{5.96}$$

$$\kappa_n' = z^{-1}\kappa - z^{-2}n(z), \qquad \kappa^{(k)\prime} = z^{k-2}\kappa^{(k)}. \tag{5.97}$$

To calculate the transformation of $\{\ddot{\mathcal{R}}^{(m)}, \dot{\mathcal{R}}^{(m)}_a, \mathcal{R}^{(m+1)}_{ab}\}$ we particularize (5.94) to $T_{\alpha\beta} = R_{\alpha\beta}$ and contract it with $\xi'^\alpha \xi'^\beta$, $e^\alpha_a \xi'^\beta$ and $e^\alpha_a e^\beta_b$, which gives

$$\ddot{\mathcal{R}}^{(m)\prime} = z^{m+1}\ddot{\mathcal{R}}^{(m)}, \tag{5.98}$$

$$\dot{\mathcal{R}}^{(m)\prime}_a = z^m \dot{\mathcal{R}}^{(m)}_a + (m-1)z^{m-1}\ddot{\mathcal{R}}^{(m-1)}\overset{\circ}{\nabla}_a z, \tag{5.99}$$

$$\mathcal{R}^{(m+1)\prime}_{ab} = z^m \mathcal{R}^{(m+1)}_{ab} + 2mz^{m-1}\dot{\mathcal{R}}^{(m)}_{(a}\overset{\circ}{\nabla}_{b)}z + m(m-1)z^{m-2}\ddot{\mathcal{R}}^{(m-1)}\overset{\circ}{\nabla}_a z \overset{\circ}{\nabla}_b z. \tag{5.100}$$

In order to compute the transformed tensors $\mathcal{O}^{(m)\prime}$, $\mathcal{O}^{(m)\prime}_a$ and $\mathcal{O}^{(m)\prime}_{ab}$ one possibility is to insert the transformations we have just computed into (5.82)-(5.84), which gives (5.89)-(5.91) after a somewhat long computation. Fortunately there is a quicker way to deduce (5.89)-(5.91) without computing them explicitly. Let us rewrite (5.82)-(5.84) formally as

$$\ddot{\mathcal{R}}^{(m)} = \mathcal{B}^{(m+1)} + \mathcal{O}^{(m)}, \qquad \dot{\mathcal{R}}^{(m)}_a = \mathcal{B}^{(m+1)}_a + \mathcal{O}^{(m)}_a, \qquad \mathcal{R}^{(m+1)}_{ab} = \mathcal{B}^{(m+1)}_{ab} + \mathcal{O}^{(m)}_{ab}, \tag{5.101}$$

where $\mathcal{B}^{(m+1)}$, $\mathcal{B}^{(m+1)}_a$ and $\mathcal{B}^{(m+1)}_{ab}$ depend on $\mathbf{Y}^{(m+1)}$. Applying the rescaling $\xi' = z\xi$ gives

$$\ddot{\mathcal{R}}^{(m)\prime} = \mathcal{B}^{(m+1)\prime} + \mathcal{O}^{(m)\prime}, \qquad \dot{\mathcal{R}}^{(m)\prime}_a = \mathcal{B}^{(m+1)\prime}_a + \mathcal{O}^{(m)\prime}_a, \qquad \mathcal{R}^{(m+1)\prime}_{ab} = \mathcal{B}^{(m+1)\prime}_{ab} + \mathcal{O}^{(m)\prime}_{ab}.$$

Inserting (5.98)-(5.100) in the left hand sides and replacing (5.101) we get identities that must be true for every $z$ and every collection $\{\mathbf{Y}^{(m)}\}$. In particular, terms involving *only* $\mathbf{Y}^{(m+1)}$ must cancel each other. This also means that the terms that *do not* depend on $\mathbf{Y}^{(m+1)}$ must also cancel each other. The quantities $\mathcal{B}^{(m+1)}$, $\mathcal{B}^{(m+1)}_a$ and $\mathcal{B}^{(m+1)}_{ab}$ and their primed counterparts are homogeneous in $\mathbf{Y}^{(m+1)}$ (i.e. they vanish identically when $\mathbf{Y}^{(m+1)}$ vanishes identically). Thus, the tensors $\mathcal{O}^{(m)\prime}$, $\mathcal{O}^{(m)\prime}_a$ and $\mathcal{O}^{(m)\prime}_{ab}$ must agree with those terms in (5.98)-(5.100) involving only $\{\mathbf{Y}^{k\leq m}\}$. Using (5.82) gives

$$\mathcal{O}^{(m)\prime} = z^{m+1}\mathcal{O}^{(m)},$$

$$\mathcal{O}^{(m)\prime}_a = z^m \mathcal{O}^{(m)}_a - (m-1)z^{m-1}\left(\operatorname{tr}_P \mathbf{Y}^{(m)} - \mathcal{O}^{(m-1)}\right)\overset{\circ}{\nabla}_a z,$$

$$\mathcal{O}^{(m)\prime}_{ab} = z^m \mathcal{O}^{(m)}_{ab} + 2mz^{m-1}\mathcal{O}^{(m)}_{(a}\overset{\circ}{\nabla}_{b)}z - m(m-1)z^{m-2}\left(\operatorname{tr}_P \mathbf{Y}^{(m)} - \mathcal{O}^{(m-1)}\right)\overset{\circ}{\nabla}_a z \overset{\circ}{\nabla}_b z,$$

which proves (5.89)-(5.91). □

**Remark 5.33.** *As shown in Proposition 5.18 the objects $\mathcal{O}^{(m)}$, $\mathcal{O}^{(m)}_a$ and $\mathcal{O}^{(m)}_{ab}$ are universal, which implies that the quantities $\mathcal{O}^{(m)\prime}$, $\mathcal{O}^{(m)\prime}_a$ and $\mathcal{O}^{(m)\prime}_{ab}$ constructed with $\xi' = z\xi$ (extended geodesically) according to* (5.82)-(5.84) *are exactly*

$$\mathcal{O}^{(m)\prime} = \mathcal{O}^{(m)}\Big(\gamma', \boldsymbol{\ell}', \ell^{(2)\prime}, \mathbf{Y}'^{(\leq m)}\Big), \qquad \mathcal{O}^{(m)\prime}_a = \mathcal{O}^{(m)}_a\Big(\gamma', \boldsymbol{\ell}', \ell^{(2)\prime}, \mathbf{Y}'^{(\leq m)}\Big),$$
$$\mathcal{O}^{(m)\prime}_{ab} = \mathcal{O}^{(m)}_{ab}\Big(\gamma', \boldsymbol{\ell}', \ell^{(2)\prime}, \mathbf{Y}'^{(\leq m)}\Big),$$

*where in the universal functions on the right hand sides all the quantities associated to metric data (such as $\overset{\circ}{\nabla}$, $\mathbf{U}$, etc.) and the expansion ($\mathbf{Y}$, $\mathbf{Y}^{(2)}$, etc.) are replaced by their primed counterparts according to (cf.* (2.27)-(2.29) *and* (5.95)*)*

$$\gamma' = \gamma, \qquad \boldsymbol{\ell}' = z\boldsymbol{\ell}, \qquad \ell^{(2)\prime} = z^2\ell^{(2)}, \qquad \mathbf{Y}' = z\mathbf{Y} + \boldsymbol{\ell}\otimes_s dz, \qquad \mathbf{Y}'^{(k)} = z^k\mathbf{Y}^{(k)} \quad (k \geq 2).$$

## 5.3 EXISTENCE AND UNIQUENESS RESULTS

In the first part of this section we use the identities obtained in Section 5.2 to establish a general uniqueness result. In particular, we are interested in studying the sufficient conditions for two metric data $\{\mathcal{H}, \gamma, \boldsymbol{\ell}, \ell^{(2)}\}$ and $\{\mathcal{H}', \gamma', \boldsymbol{\ell}', \ell^{(2)\prime}\}$ along with their corresponding transverse expansions to be embedded in ambient manifolds $(\mathcal{M}, g)$ and $(\mathcal{M}', g')$ *asymptotically isometric at the hypersurface*, i.e. such that the ambient metrics $g$ and $g'$ at the two null hypersurfaces $\Phi(\mathcal{H})$ and $\Phi'(\mathcal{H}')$ agree at all orders. We begin by recalling some notation from Section 3.4.

**Notation 5.34.** *Let $\{\mathcal{H}', \gamma', \boldsymbol{\ell}', \ell^{(2)\prime}\}$ be null metric hypersurface data and $\chi : \mathcal{H} \longrightarrow \mathcal{H}'$ a diffeomorphism. We define*

$$\chi^\star\{\mathcal{H}', \gamma', \boldsymbol{\ell}', \ell^{(2)\prime}\} := \{\mathcal{H}, \chi^\star\gamma', \chi^\star\boldsymbol{\ell}', \chi^\star\ell^{(2)\prime}\}. \tag{5.102}$$

**Remark 5.35.** *As we mentioned in Section 3.4, it is immediate to check that if $\{\mathcal{H}', \gamma', \boldsymbol{\ell}', \ell^{(2)\prime}\}$ is null metric hypersurface data and $\chi : \mathcal{H} \longrightarrow \mathcal{H}'$ is a diffeomorphism, then $\chi^\star\{\mathcal{H}', \gamma', \boldsymbol{\ell}', \ell^{(2)\prime}\}$ is also null metric hypersurface data.*

In order to compare the metrics at $\Phi(\mathcal{H})$ and $\Phi'(\mathcal{H}')$ we need to construct suitable neighbourhoods around $\Phi(\mathcal{H})$ and $\Phi'(\mathcal{H}')$ and to map them via a diffeomorphism, as we accomplish next.

**Proposition 5.36.** *Let $\Phi : \mathcal{H} \hookrightarrow \mathcal{M}$ and $\Phi' : \mathcal{H}' \hookrightarrow \mathcal{M}'$ be two embedded hypersurfaces in ambient manifolds $(\mathcal{M}, g)$ and $(\mathcal{M}', g')$ and let $\xi$, $\xi'$ be respectively riggings of $\Phi(\mathcal{H})$, $\Phi'(\mathcal{H}')$ extended geodesically. Assume that there exists a diffeomorphism $\chi : \mathcal{H} \longrightarrow \mathcal{H}'$. Then, there exist open neighbourhoods $\mathcal{U} \subset \mathcal{M}$ and $\mathcal{U}' \subset \mathcal{M}'$ of $\Phi(\mathcal{H})$ and $\Phi'(\mathcal{H}')$ and a unique diffeomorphism $\Psi : \mathcal{U} \longrightarrow \mathcal{U}'$ satisfying $\Psi_\star\xi = \xi'$ and $\Phi' \circ \chi = \Psi \circ \Phi$.*

*Proof.* Pick a neighbourhood $\mathcal{U} \subset \mathcal{M}$ of $\Phi(\mathcal{H})$ small enough so that the integral curves of $\xi$ do not intersect each other and intersect $\Phi(\mathcal{H})$ precisely once. Then, given a point $q \in \mathcal{U}$ there exist a unique $p \in \mathcal{H}$ such that the integral curve $\sigma(\tau)$ of $\xi$ through $p$ reaches $q$ at a finite $\tau_q$, i.e. $\sigma(\tau_q) = q$. Now consider the point $p' := \chi(p) \in \mathcal{H}'$, a neighbourhood $\mathcal{U}'$ of $\Phi'(\mathcal{H}')$ and the integral curve $\sigma'$ of $\xi'$ through $p'$. By shrinking $\mathcal{U}$ and $\mathcal{U}'$ if necessary so that all the integral curves $\sigma'$ intersect $\Phi'(\mathcal{H}')$ exactly once and that the point $\Psi(q)$ is contained in $\mathcal{U}'$, one can construct the diffeomorphism $\Psi(q) := \sigma'(\tau_q)$, that by construction satisfies $\Psi_\star \xi = \xi'$ and $\Phi' \circ \chi = \Psi \circ \Phi$. □

Finally, once the neighbourhoods around $\Phi(\mathcal{H})$ and $\Phi'(\mathcal{H}')$ are constructed and mapped one to another, we prove that when the metric data and the asymptotic expansions at $\Phi(\mathcal{H})$ and $\Phi'(\mathcal{H}')$ agree, then $(\mathcal{M}, g)$ and $(\mathcal{M}', g')$ are isometric to infinite order (i.e. satisfy relation (5.103) below).

**Proposition 5.37.** *Let $\{\mathcal{H}, \boldsymbol{\gamma}, \boldsymbol{\ell}, \ell^{(2)}\}$ (respectively $\{\mathcal{H}', \boldsymbol{\gamma}', \boldsymbol{\ell}', \ell^{(2)\prime}\}$) be null metric hypersurface data $(\Phi, \xi)$-embedded in $(\mathcal{M}, g)$ (resp. $(\Phi', \xi')$-embedded in $(\mathcal{M}', g')$) with $\xi$ and $\xi'$ extended geodesically. Assume that there exists a diffeomorphism $\chi : \mathcal{H} \longrightarrow \mathcal{H}'$ such that $\chi^\star\{\mathcal{H}', \boldsymbol{\gamma}', \boldsymbol{\ell}', \ell^{(2)\prime}\} = \{\mathcal{H}, \boldsymbol{\gamma}, \boldsymbol{\ell}, \ell^{(2)}\}$ and $\chi^\star \mathbf{Y}^{(k)\prime} = \mathbf{Y}^{(k)}$ for every $k \geq 1$. Then, there exist neighbourhoods $\mathcal{U} \subset \mathcal{M}$ and $\mathcal{U}' \subset \mathcal{M}'$ of $\Phi(\mathcal{H})$ and $\Phi'(\mathcal{H}')$ and a diffeomorphism $\Psi : \mathcal{U} \longrightarrow \mathcal{U}'$ as in Prop. 5.36 such that*

$$\Psi^\star \pounds^{(i)}_{\xi'} g' \stackrel{\mathcal{H}}{=} \pounds^{(i)}_{\xi} g \tag{5.103}$$

*for every $i \in \mathbb{N} \cup \{0\}$.*

*Proof.* After appropriate choices of $\mathcal{U}$ and $\mathcal{U}'$, Proposition 5.36 ensures the existence of a unique diffeomorphism $\Psi : \mathcal{U} \longrightarrow \mathcal{U}'$ satisfying $\Psi_\star \xi = \xi'$ and $\Phi' \circ \chi = \Psi \circ \Phi$. Let us prove that $\Psi^\star \pounds^{(i)}_{\xi'} g' \stackrel{\mathcal{H}}{=} \pounds^{(i)}_{\xi} g$ for every $i \geq 0$. The case $i = 0$ is immediate because $\boldsymbol{\gamma} = \chi^\star \boldsymbol{\gamma}'$, $\boldsymbol{\ell} = \chi^\star \boldsymbol{\ell}'$ and $\ell^{(2)} = \chi^\star \ell^{(2)\prime}$. Proving the case $i \geq 1$ amounts to show $\Psi^\star(\pounds^{(i)}_{\xi'} g')(\xi', \cdot) = (\pounds^{(i)}_{\xi} g)(\xi, \cdot)$ and $\chi^\star \mathbf{Y}^{(i)\prime} = \mathbf{Y}^{(i)}$. The latter is part of the hypothesis, and the former is a direct consequence of Corollary 5.12 because $a_\xi = 0$, $a_{\xi'} = 0$ and $\ell^{(2)} = \chi^\star \ell^{(2)\prime}$. □

Next, we move on to the question of existence. In particular, we want to show that by prescribing null metric hypersurface data $\{\mathcal{H}, \boldsymbol{\gamma}, \boldsymbol{\ell}, \ell^{(2)}\}$ together with the full transverse expansion it is possible to construct a smooth semi-Riemannian manifold $(\mathcal{M}, g)$ where the data is embedded in the following sense: (i) the metric hypersurface data is $(\Phi, \xi)$-embedded and (ii) the given expansion agrees with the pullback of $\mathcal{K}^{(k)}$ for all $k \geq 1$. At this point it is crucial to distinguish the notion of expansion as abstract data from the concept of expansion as the pullback of $\frac{1}{2}\pounds^{(k)}_{\xi} g$, $k \geq 1$. For the former we will employ the notation $\mathbb{Y}^{(k)}$, and for the latter we continue using $\mathbf{Y}^{(k)}$. As usual, we introduce the tensors $\mathbb{r}^{(k)} := \mathbb{Y}^{(k)}(n, \cdot)$ and $\mathbb{k}^{(k)} := -\mathbb{r}^{(k)}(n)$.

The construction of $(\mathcal{M}, g)$ that we present does not require any field equations. If one is interested in $(\mathcal{M}, g)$ solving field equations such as vacuum, then the prescribed transverse expansion needs to satisfy suitable restrictions. In Subsection 5.3.1 we study the $\Lambda$-vacuum case and find necessary *and sufficient* conditions on the transverse expansion that guarantees that the $(\mathcal{M}, g)$ constructed in Theorem 5.40 below satisfies the equations at $\Phi(\mathcal{H})$ to infinite order. Before proving Theorem 5.40 we review a result due to Borel that we shall need, see [146, Lemma 2.5].

**Lemma 5.38** (Borel). *Let $\{F_n(x)\}_{n\geq 0}$ be a sequence of smooth functions defined on a given neighbourhood of* 0 *in* $\mathbb{R}^n$. *Then there exist a smooth function $F(r,x)$ defined on a neighbourhood of* 0 *in* $\mathbb{R}\times\mathbb{R}^n$ *such that*

$$\left.\frac{\partial^k F}{\partial r^k}\right|_{(0,x)} = F_k(x) \qquad \textit{for all } k\geq 0. \tag{5.104}$$

*Proof.* We sketch the argument provided in [146] since it will help us in the proof of Theorem 5.40. Choose any smooth function $\rho:\mathbb{R}\longrightarrow\mathbb{R}$ satisfying

$$\rho(t)=\begin{cases}1 & |r|\leq \frac{1}{2}\\ 0 & |r|\geq 1\end{cases}$$

and set

$$F(r,x)=\sum_{k=0}^{\infty}\frac{r^k}{k!}\rho(\mu_k r)F_k(x), \tag{5.105}$$

where $\{\mu_k\}_{k\geq 0}$ is an increasing and unbounded sequence of real values suitably chosen to make $F(r,x)$ smooth. By construction $F(r,x)$ satisfies (5.104). □

Borel's lemma will be also needed in Chapter 6 in the following specific form.

**Lemma 5.39** (Borel). *Let $\mathcal{M}$ be a smooth manifold, $\mathcal{H}\hookrightarrow\mathcal{M}$ a embedded smooth hypersurface with rigging $\xi$ and $\{\sigma^{(k)}\}_{k\geq 0}$ a collection of functions on $\mathcal{H}$. Then, there exists a function $\Omega$ in a neighbourhood of $\mathcal{H}$ in $\mathcal{M}$ such that $\pounds_\xi^{(k)}\Omega|_{\mathcal{H}}=\sigma^{(k)}$ for every $k\geq 0$.*

**Theorem 5.40.** *Let $\{\mathcal{H},\boldsymbol{\gamma},\boldsymbol{\ell},\ell^{(2)}\}$ be null metric hypersurface data and $\{\mathbb{Y}^{(k)}\}_{k\geq 1}$ a sequence of $(0,2)$ symmetric tensor fields on $\mathcal{H}$. Then there exists a semi-Riemannian manifold $(\mathcal{M},g)$, an embedding $\Phi:\mathcal{H}\hookrightarrow\mathcal{M}$ and a rigging vector $\xi$ satisfying $a_\xi=\nabla_\xi\xi=0$ on $\mathcal{M}$ such that (i) $\{\mathcal{H},\boldsymbol{\gamma},\boldsymbol{\ell},\ell^{(2)}\}$ is null metric hypersurface data $(\Phi,\xi)$-embedded in $(\mathcal{M},g)$ and (ii) $\{\mathbb{Y}^{(k)}\}_{k\geq 1}$ is the transverse expansion of $g$ at $\Phi(\mathcal{H})$, i.e. $\mathbb{Y}^{(k)}=\mathbf{Y}^{(k)}:=\frac{1}{2}\Phi^\star(\pounds_\xi^{(k)}g)$ for every $k\geq 1$.*

*Proof.* Define $\widetilde{\mathcal{M}}:=(-\varepsilon,\varepsilon)\times\mathcal{H}$ and let us construct a local coordinate system on $\widetilde{\mathcal{M}}$ around any point of $\{0\}\times\mathcal{H}$ as follows. Let $r$ be a coordinate in the first factor of $\widetilde{\mathcal{M}}$. Pick any $p\in\mathcal{H}$ and choose any local section $\mathcal{S}$ of $\mathcal{H}$ containing $p$ with (local) coordinates $\{x^A\}$. Let $u$ be a local function on $\mathcal{H}$ satisfying $u|_{\mathcal{S}}=0$ and $n(u)=1$ defined in a neighbourhood $\mathcal{V}\subset\mathcal{H}$ and extend $\{x^A\}$ to $\mathcal{V}$ by means of $n(x^A)=0$. Then $\{u,x^A\}$ are coordinates on $\mathcal{V}$ (note that $n|_{\mathcal{V}}=\partial_u$, so in these coordinates $\ell_u=\boldsymbol{\ell}(\partial_u)=\boldsymbol{\ell}(n)=1$). Finally, we extend trivially $\{u,x^A\}$ to $\widetilde{\mathcal{U}}:=(-\varepsilon,\varepsilon)\times\mathcal{V}$ so that $\{r,u,x^A\}$ are coordinates on $\widetilde{\mathcal{U}}$. In this coordinate system the natural embedding $\Phi_{\widetilde{\mathcal{U}}}:\mathcal{V}\hookrightarrow\widetilde{\mathcal{U}}$ is given by $\Phi_{\widetilde{\mathcal{U}}}(u,x^A)=(0,u,x^A)$.

The idea now is to construct smooth functions $f$, $f_A$, $h$, $h_A$ and $H_{AB}$ on a neighbourhood $\mathcal{U}\subset\widetilde{\mathcal{U}}$ of $p$ in terms of metric data and $\{\mathbb{Y}^{(k)}\}_{k\geq 1}$ using (5.105) so that items (i) and (ii) of the theorem follow. In order to do so we just fix the function $\rho$ and the sequence $\{\mu_k\}$ of the proof of Lemma 5.38 and define $f$ on $\mathcal{U}$ from the sequence $\{0, -2\mathbb{k}^{(i)}\}$, $f_A$ from $\{\ell_A, 0, 0, ...\}$, $h$ from $\{\ell^{(2)}, 0, 0, ...\}$, $h_A$ from $\{0, 2\mathbb{r}_A^{(i)}\}$ and $H_{AB}$ from $\{\gamma_{AB}, 2\mathbb{Y}_{AB}^{(i)}\}$. This guarantees that

$$f|_{r=0} = 0, \qquad f_A|_{r=0} = \ell_A, \qquad h|_{r=0} = \ell^{(2)}, \qquad h_A|_{r=0} = 0, \qquad H_{AB}|_{r=0} = \gamma_{AB}, \tag{5.106}$$

as well as

$$\partial_r^{(k)} f|_{r=0} = -2\mathbb{k}^{(k)}, \qquad \partial_r^{(k)} f_A|_{r=0} = 0, \qquad \partial_r^{(k)} h|_{r=0} = 0, \qquad \partial_r^{(k)} h_A|_{r=0} = 2\mathbb{r}_A^{(k)}, \\ \partial_r^{(k)} H_{AB}|_{r=0} = 2\mathbb{Y}_{AB}^{(k)}, \tag{5.107}$$

for every $k \geq 1$. Now we define the following smooth tensor on $\mathcal{U}$

$$g_{\mathcal{U}} = 2drdu + hdr^2 + 2f_A dr dx^A + 2h_A du dx^A + f du^2 + H_{AB} dx^A dx^B. \tag{5.108}$$

By (5.106) it follows $\Phi_{\mathcal{U}}^{\star} g_{\mathcal{U}} = \gamma_{AB} dx^A dx^B$, $\Phi_{\mathcal{U}}^{\star}(g_{\mathcal{U}}(\partial_r, \cdot)) = \boldsymbol{\ell}$ and $\Phi_{\mathcal{U}}^{\star}(g_{\mathcal{U}}(\partial_r, \partial_r)) = \ell^{(2)}$, so from Definition 2.1 the determinant of $g$ at $\Phi(\mathcal{H})$ coincides with the determinant of the tensor $\mathcal{A}$, and hence it vanishes nowhere on $\Phi(\mathcal{H})$. Since this is an open condition, by shrinking $\mathcal{U}$ if necessary $g_{\mathcal{U}}$ is a smooth semi-Riemannian metric on $\mathcal{U}$. Moreover, $\{\mathcal{V}, \boldsymbol{\gamma}, \boldsymbol{\ell}, \ell^{(2)}\}$ is $(\Phi_{\mathcal{U}}, \partial_r)$-embedded null metric hypersurface data in $(\mathcal{U}, g_{\mathcal{U}})$ and from (5.107) it is clear that the (embedded) transverse expansion of $g_{\mathcal{U}}$ at $\Phi(\mathcal{V})$ agrees with $\{\mathbb{Y}^{(k)}\}_{k\geq 1}$, i.e. $\mathbb{Y}^{(k)} = \mathbf{Y}^{(k)}$ for all $k \geq 1$.

Define $\mathcal{M}$ to be the union of all the $\mathcal{U}$'s. In order to finish the proof we need to check that two different metrics $g_{\mathcal{U}}$ and $\widehat{g}_{\widehat{\mathcal{U}}}$ constructed in this way agree on $\mathcal{U} \cap \widehat{\mathcal{U}}$ (i.e. that $g_{\mathcal{U}\cap\widehat{\mathcal{U}}}$ and $\widehat{g}_{\mathcal{U}\cap\widehat{\mathcal{U}}}$ are related by a change of coordinates), because this will then ensure that we have constructed a *bona fide* metric on $\mathcal{M}$. So let us consider two coordinate patches $\mathcal{V}$ and $\widehat{\mathcal{V}}$ on $\mathcal{H}$ with respective coordinates $\{u, x^A\}$ and $\{\widehat{u}, \widehat{x}^A\}$ such that $\mathcal{V} \cap \widehat{\mathcal{V}} \neq \emptyset$. Since in the intersection $n = \partial_u = \partial_{\widehat{u}}$ it follows that the most general transformation $\{u, x^A\} \longmapsto \{\widehat{u}, \widehat{x}^A\}$ must be of the form

$$\begin{cases} \widehat{u} = u + \phi(x^A), \\ \widehat{x}^A = \widehat{x}^A(x^B), \end{cases}$$

where $\phi(x^A)$ is arbitrary and $\widehat{x}^A(x^B)$ must be invertible (to define a change of coordinates). Under this transformation, $\ell^{(2)}$, $\ell_A$ and $\{\mathbb{k}^{(i)}, \mathbb{r}_A^{(i)}, \mathbb{Y}_{AB}^{(i)}\}_{i\geq 1}$ change by

$$\ell^{(2)} = \widehat{\ell}^{(2)}, \qquad \ell_A = \frac{\partial \widehat{x}^B}{\partial x^A}\widehat{\ell}_B + \frac{\partial \phi}{\partial x^A}, \qquad \mathbb{k}^{(i)} = \widehat{\mathbb{k}}^{(i)}, \qquad \mathbb{r}_A^{(i)} = \frac{\partial \widehat{x}^B}{\partial x^A}\widehat{\mathbb{r}}_B^{(i)} - \frac{\partial \phi}{\partial x^A}\widehat{\mathbb{k}}^{(i)}, \\ \mathbb{Y}_{AB}^{(i)} = \frac{\partial \widehat{x}^C}{\partial x^A}\frac{\partial \widehat{x}^D}{\partial x^B}\widehat{\mathbb{Y}}_{CD}^{(i)} + \frac{\partial \phi}{\partial x^A}\frac{\partial \widehat{x}^C}{\partial x^B}\widehat{\mathbb{r}}_C^{(i)} + \frac{\partial \widehat{x}^C}{\partial x^A}\frac{\partial \phi}{\partial x^B}\widehat{\mathbb{r}}_C^{(i)} - \frac{\partial \phi}{\partial x^A}\frac{\partial \phi}{\partial x^B}\widehat{\mathbb{k}}^{(i)}. \tag{5.109}$$

Inserting these expressions into (5.105) and taking into account that we used the same $\rho$ to construct all the functions, it follows

$$
\begin{aligned}
h = \widehat{h}, \qquad f_A = \frac{\partial \widehat{x}^B}{\partial x^A}\widehat{f}_B + \frac{\partial \phi}{\partial x^A}, \qquad f = \widehat{f}, \qquad h_A = \frac{\partial \widehat{x}^B}{\partial x^A}\widehat{h}_A + \frac{\partial \phi}{\partial x^A} f, \\
H_{AB} = \frac{\partial \widehat{x}^C}{\partial x^A}\frac{\partial \widehat{x}^D}{\partial x^B}\widehat{H}_{CD} + \frac{\partial \phi}{\partial x^A}\frac{\partial \widehat{x}^C}{\partial x^B}\widehat{f}_C + \frac{\partial \widehat{x}^C}{\partial x^A}\frac{\partial \phi}{\partial x^B}\widehat{f}_C + \frac{\partial \phi}{\partial x^A}\frac{\partial \phi}{\partial x^B} f.
\end{aligned}
\tag{5.110}
$$

Using these expressions, the coordinate change $\{r, u, x^A\} \longmapsto \{r, \widehat{u}, \widehat{x}^A\}$ applied to the metric $g_{\mathcal{U}\cap\widehat{\mathcal{U}}}$ gives precisely the metric $\widehat{g}_{\mathcal{U}\cap\widehat{\mathcal{U}}}$. So, the two tensorial objects are just different coordinate expressions of a single tensor field. This proves the existence of a metric $g$ on $\mathcal{M}$. □

**Remark 5.41.** *It is worth mentioning that Borel's theorem does not provide unique functions $f$, $f_A$, $h$, $h_A$ and $H_{AB}$, and thus the manifold $(\mathcal{M}, g)$ in Theorem 5.40 is not unique. This is due to the fact that we can choose any $\rho$ and any sequence $\{\mu_k\}$ fulfilling the conditions required in the proof of Lemma 5.38.*

### 5.3.1 $\Lambda$-*vacuum case*

The identities of Section 5.2 allow us to write down the necessary and sufficient conditions (i.e. the *higher order constraint equations*) on the sequence $\{\mathbb{Y}^{(k)}\}_{k\geq 1}$ in order for the ambient space to be $\Lambda$-vacuum to infinite order on $\mathcal{H}$. A similar analysis could be done for other field equations.

**Theorem 5.42.** *Let $\{\mathcal{H}, \boldsymbol{\gamma}, \boldsymbol{\ell}, \ell^{(2)}\}$ be null metric hypersurface data and $\{\mathbb{Y}^{(k)}\}_{k\geq 1}$ a sequence of $(0,2)$ symmetric tensors satisfying*

$$
\begin{aligned}
\lambda\gamma_{ab} &= \overset{\circ}{R}_{(ab)} - 2\pounds_n \mathbb{Y}_{ab} - (2\mathbb{k} + \operatorname{tr}_P \mathbf{U})\mathbb{Y}_{ab} + \overset{\circ}{\nabla}_{(a}(\mathrm{s}_{b)} + 2\mathbb{r}_{b)}) - 2\mathbb{r}_a\mathbb{r}_b \\
&\quad + 4\mathbb{r}_{(a}\mathrm{s}_{b)} - \mathrm{s}_a\mathrm{s}_b - (\operatorname{tr}_P \mathbb{Y})\mathrm{U}_{ab} + 2P^{cd}\mathrm{U}_{d(a}(2\mathbb{Y}_{b)c} + \mathrm{F}_{b)c}), \qquad (5.111) \\
\lambda\ell^{(2)}\delta_1^m &= -\operatorname{tr}_P \mathbb{Y}^{(m+1)} + \mathcal{O}^{(m)}(\mathbb{Y}^{\leq m}), \qquad (5.112) \\
\lambda\left(\delta_1^m \ell_a + \frac{1}{2}\delta_2^m \overset{\circ}{\nabla}_a \ell^{(2)}\right) &= \mathbb{r}_a^{(m+1)} + \mathcal{O}^{(m)}(\mathbb{Y}^{\leq m})_a, \qquad (5.113) \\
2\lambda\mathbb{Y}_{ab}^{(m)} &= -2\pounds_n \mathbb{Y}_{ab}^{(m+1)} - (2(m+1)\mathbb{k} + \operatorname{tr}_P \mathbf{U})\,\mathbb{Y}_{ab}^{(m+1)} - (\operatorname{tr}_P \mathbb{Y}^{(m+1)})\mathrm{U}_{ab} \\
&\quad + 4P^{cd}\mathrm{U}_{c(a}\mathbb{Y}_{b)d}^{(m+1)} + 4(\mathrm{s} - \mathbb{r})_{(a}\mathbb{r}_{b)}^{(m+1)} + 2\overset{\circ}{\nabla}_{(a}\mathbb{r}_{b)}^{(m+1)} \\
&\quad - 2\mathbb{k}^{(m+1)}\mathbb{Y}_{ab} + \mathcal{O}_{ab}^{(m)}(\mathbb{Y}^{\leq m}), \qquad (5.114)
\end{aligned}
$$

*for every $m \geq 1$, where the tensors $\mathcal{O}^{(m)}$, $\mathcal{O}_a^{(m)}$ and $\mathcal{O}_{ab}^{(m)}$ are the same as in Corollary 5.30 with the $\mathbf{Y}^{(k)}$'s replaced by the $\mathbb{Y}^{(k)}$'s. Then, the semi-Riemannian manifold $(\mathcal{M}, g)$ constructed from $\{\mathbb{Y}^{(k)}\}$ as in Theorem 5.40 solves the $\Lambda$-vacuum equations to infinite order on $\Phi(\mathcal{H})$, i.e. $R_{\mu\nu}^{(i)} \stackrel{\mathcal{H}}{=} \lambda\mathcal{K}_{\mu\nu}^{(i-1)}$ for every $i \geq 1$.*

*Proof.* By equations (2.95), (5.82) and (5.83) as well as (5.111), (5.112) and (5.113) for $m = 1$ together with the fact that the hypersurface data is $(\Phi, \xi)$-embedded (see Def. 2.2) it follows that $R_{\mu\nu} \stackrel{\mathcal{H}}{=} \lambda g_{\mu\nu}$. Moreover, by Corollary 5.12, $\mathcal{K}(\xi, \xi) = 0$, $\Phi^\star(\mathcal{K}(\xi, \cdot)) = \frac{1}{2}d\ell^{(2)}$ and $\mathcal{K}^{(i)}_{\alpha\beta}\xi^\alpha = 0$ for every $i \geq 2$. Now, (5.112) and (5.113) combined with (5.82) and (5.83) gives $R^{(2)}_{\alpha\beta}\xi^\alpha\xi^\beta = 0$, $R^{(2)}_{\alpha\beta}e^\alpha_a\xi^\beta = \frac{1}{2}\lambda\overset{\circ}{\nabla}_a\ell^{(2)}$ and $R^{(i+1)}_{\alpha\beta}\xi^\alpha = 0$ for $i \geq 2$ after taking into account $\mathbb{Y}^{(k)} = \mathbf{Y}^{(k)}$ for every $k \geq 1$. Thus, $R^{(i)}_{\alpha\beta}\xi^\alpha = \lambda\mathcal{K}^{(i-1)}_{\alpha\beta}\xi^\alpha$ for every $i \geq 1$. Finally, combining equations (5.114) and (5.84) and using again $\mathbb{Y}^{(k)} = \mathbf{Y}^{(k)}$ for $k \geq 1$ gives $\mathcal{R}^{(i)}_{ab} = 2\lambda\mathrm{Y}^{(i-1)}_{ab} = \lambda\mathcal{K}^{(i-1)}_{ab}$ for every $i \geq 1$, which finishes the proof. □

It is worth emphasizing that in Theorem 5.42 the topology of the null hypersurface does not need to be a product. This is particularly relevant because there are examples of null hypersurfaces whose topology is not a product, such as certain compact Cauchy horizons (see e.g. [134]). In these cases the constraint equations of Theorem 5.42 may or may not have solutions. However, when $\mathcal{H}$ admits a cross-section (and hence its topology is $\mathbb{R} \times \mathcal{S}$), these constraint equations can always be integrated given initial conditions at a section. Note however that due to (5.85)-(5.86) there are more constraints than independent components of $\{\mathbb{Y}^{(m)}\}$ to integrate. In the following proposition we show that some of the constraints are redundant.

**Proposition 5.43.** *Let $(\mathcal{M}, g)$ be an $(\mathfrak{n}+1)$-dimensional semi-Riemannian manifold and $\Phi : \mathcal{H} \hookrightarrow \mathcal{M}$ an embedded null hypersurface with rigging $\xi$ extended off $\Phi(\mathcal{H})$ arbitrarily. Fix a natural number $m \geq 1$ and assume the Ricci tensor of $g$ satisfies $R^{(i)}_{\mu\nu} \stackrel{\mathcal{H}}{=} \lambda\mathcal{K}^{(i-1)}_{\mu\nu}$ for every $i = 1, ..., m$. Then,*

$$\mathcal{R}^{(m+1)}_{ab} n^a \stackrel{\mathcal{H}}{=} \lambda\mathcal{K}^{(m)}_{ab} n^a, \qquad P^{ab}\mathcal{R}^{(m+1)}_{ab} \stackrel{\mathcal{H}}{=} \lambda P^{ab}\mathcal{K}^{(m)}_{ab}. \tag{5.115}$$

*Proof.* The idea is to prove the following identity

$$\nu^\alpha R^{(m+1)}_{\alpha\beta} - \frac{1}{2}g^{\mu\alpha}R^{(m+1)}_{\mu\alpha}\nu_\beta \stackrel{\mathcal{H}}{=} \lambda\left(\mathcal{K}^{(m)}_{\alpha\beta}\nu^\alpha - \frac{1}{2}g^{\mu\alpha}\mathcal{K}^{(m)}_{\mu\alpha}\nu_\beta\right), \tag{5.116}$$

since then its contraction with a tangent vector $\widehat{e}^\beta_b$ is (recall $\nu_\beta$ is normal to $\mathcal{H}$)

$$\widehat{e}^\beta_b \nu^\alpha R^{(m+1)}_{\alpha\beta} \stackrel{\mathcal{H}}{=} \lambda\mathcal{K}^{(m)}_{\alpha\beta}\nu^\alpha\widehat{e}^\beta_b,$$

which proves the first equation in (5.115), and its contraction with $\xi^\beta$ gives, after inserting (2.14),

$$\xi^\beta\nu^\alpha R^{(m+1)}_{\alpha\beta} - \xi^\mu\nu^\alpha R^{(m+1)}_{\mu\alpha} - \frac{1}{2}P^{ab}\widehat{e}^\mu_a\widehat{e}^\alpha_b R^{(m+1)}_{\mu\alpha} \stackrel{\mathcal{H}}{=} \lambda\left(\xi^\beta\nu^\alpha\mathcal{K}^{(m)}_{\alpha\beta} - \xi^\mu\nu^\alpha\mathcal{K}^{(m)}_{\mu\alpha} - \frac{1}{2}P^{ab}\widehat{e}^\mu_a\widehat{e}^\alpha_b\mathcal{K}^{(m)}_{\mu\alpha}\right),$$

so $P^{ab}\widehat{e}^\mu_a\widehat{e}^\alpha_b R^{(m+1)}_{\mu\alpha} \stackrel{\mathcal{H}}{=} \lambda P^{ab}\widehat{e}^\mu_a\widehat{e}^\alpha_b\mathcal{K}^{(m)}_{\mu\alpha}$ and hence the second equation in (5.115) is also established. So let us prove (5.116). By hypothesis $\pounds^{(i-1)}_\xi R_{\mu\nu} \stackrel{\mathcal{H}}{=} \lambda\pounds^{(i-1)}_\xi g_{\mu\nu}$ for every $i = 1, ..., m$, so for any $j \in \{0, ..., m-1\}$ it follows (cf. (5.34))

$$\begin{aligned}\pounds^{(j)}_\xi(R^\alpha{}_\beta) = \pounds^{(j)}_\xi(R_{\mu\beta}g^{\mu\alpha}) = \sum_{k=0}^{j}\binom{j}{k}\pounds^{(k)}_\xi R_{\mu\beta}\pounds^{(j-k)}_\xi g^{\mu\alpha} \stackrel{\mathcal{H}}{=} \lambda\sum_{k=0}^{j}\binom{j}{k}\pounds^{(k)}_\xi g_{\mu\beta}\pounds^{(j-k)}_\xi g^{\mu\alpha}\\ = \lambda\pounds^{(j)}_\xi\delta^\alpha_\beta,\end{aligned}$$

and thus

$$\pounds_\xi^{(j)}(R^\alpha{}_\beta) \stackrel{\mathcal{H}}{=} \begin{cases} \lambda\delta^\alpha_\beta & j=0, \\ 0 & 1\le j\le m-1, \end{cases} \qquad \pounds_\xi^{(j)} R \stackrel{\mathcal{H}}{=} \begin{cases} \lambda(\mathfrak{n}+1) & j=0, \\ 0 & 1\le j\le m-1. \end{cases} \tag{5.117}$$

Therefore the derivatives of the Einstein tensor $G_{\mu\nu} = R_{\mu\nu} - \frac{1}{2}Rg_{\mu\nu}$ satisfy

$$\pounds_\xi^{(j)}(G^\alpha{}_\beta) \stackrel{\mathcal{H}}{=} \begin{cases} \dfrac{1-\mathfrak{n}}{2}\lambda\delta^\alpha_\beta & j=0, \\ 0 & 1\le j\le m-1. \end{cases} \tag{5.118}$$

Applying $\pounds_\xi^{(m-1)}$ to the contracted Bianchi identity $\nabla_\alpha G^\alpha{}_\beta = 0$ and using Proposition 5.7 gives

$$\begin{aligned} 0 &= \pounds_\xi^{(m-1)}\nabla_\alpha G^\alpha{}_\beta \\ &= \nabla_\alpha(\pounds_\xi^{(m-1)}G^\alpha{}_\beta) + \sum_{k=0}^{m-2}(\pounds_\xi^{(m-2-k)}G^\sigma{}_\beta)\Sigma^{(k+1)\alpha}{}_{\alpha\sigma} - \sum_{k=0}^{m-2}(\pounds_\xi^{(m-2-k)}G^\alpha{}_\sigma)\Sigma^{(k+1)\sigma}{}_{\alpha\beta} \\ &\stackrel{\mathcal{H}}{=} \nabla_\alpha(\pounds_\xi^{(m-1)}G^\alpha{}_\beta) + \frac{1-\mathfrak{n}}{2}\lambda\delta^\sigma_\beta\Sigma^{(m-1)\alpha}{}_{\alpha\sigma} - \frac{1-\mathfrak{n}}{2}\lambda\delta^\alpha_\sigma\Sigma^{(m-1)\sigma}{}_{\alpha\beta} \\ &\stackrel{\mathcal{H}}{=} \nabla_\alpha(\pounds_\xi^{(m-1)}G^\alpha{}_\beta), \end{aligned}$$

where in the third equality we used (5.118). We now compute the trace in this equation by inserting $\delta^\alpha_\mu = \widehat{e}^\alpha_a\theta^a_\mu + \xi^\alpha\nu_\mu$ and get

$$\nabla_\alpha(\pounds_\xi^{(m-1)}G^\alpha{}_\beta) \stackrel{\mathcal{H}}{=} \left(\widehat{e}^\alpha_a\theta^a_\mu + \xi^\alpha\nu_\mu\right)\nabla_\alpha(\pounds_\xi^{(m-1)}G^\mu{}_\beta) \stackrel{\mathcal{H}}{=} 0.$$

The first term vanishes because we can take tangential derivatives of (5.118), and therefore $\xi^\alpha\nu_\mu\nabla_\alpha(\pounds_\xi^{(m-1)}G^\mu{}_\beta) \stackrel{\mathcal{H}}{=} 0$. Using again (5.118) together with the well-known identity

$$\nabla_\xi T_{\lambda\gamma} = \pounds_\xi T_{\lambda\gamma} - 2T_{\mu(\lambda}\nabla_{\gamma)}\xi^\mu,$$

valid for every $(0,2)$ symmetric tensor $T$, gives

$$\nu_\mu(\pounds_\xi^{(m)}G^\mu{}_\beta) \stackrel{\mathcal{H}}{=} 0. \tag{5.119}$$

Moreover, equations (5.34) and (5.118) yield

$$g_{\mu\alpha}(\pounds_\xi^{(m)}G^\mu{}_\beta) = \pounds_\xi^{(m)}G_{\alpha\beta} - \sum_{j=0}^{m-1}\binom{m}{j}(\pounds_\xi^{(j)}G^\mu{}_\beta)\mathcal{K}^{(m-j)}_{\mu\alpha} \stackrel{\mathcal{H}}{=} \pounds_\xi^{(m)}G_{\alpha\beta} - \frac{1-\mathfrak{n}}{2}\lambda\mathcal{K}^{(m)}_{\alpha\beta}.$$

Contracting with $\nu^\alpha$ and using (5.119) gives $\nu^\alpha\pounds_\xi^{(m)}G_{\alpha\beta} \stackrel{\mathcal{H}}{=} \dfrac{1-\mathfrak{n}}{2}\lambda\nu^\alpha\mathcal{K}^{(m)}_{\alpha\beta}$. Then, inserting $G_{\alpha\beta} = R_{\alpha\beta} - \dfrac{1}{2}Rg_{\alpha\beta}$ and using the second expression in (5.117) we conclude

$$\nu^\alpha\pounds_\xi^{(m)}R_{\alpha\beta} - \frac{1}{2}g_{\alpha\beta}\nu^\alpha\pounds_\xi^{(m)}R - \frac{\mathfrak{n}+1}{2}\lambda\mathcal{K}^{(m)}_{\alpha\beta}\nu^\alpha \stackrel{\mathcal{H}}{=} \frac{1-\mathfrak{n}}{2}\lambda\nu^\alpha\mathcal{K}^{(m)}_{\alpha\beta}. \tag{5.120}$$

The last step is to relate $\pounds_{\xi}^{(m)} R$ to the term of $\pounds_{\xi}^{(m)} R_{\alpha\beta}$. In order to do so we use again (5.34) and (5.117) to write

$$g^{\mu\alpha}\pounds_{\xi}^{(m)} R_{\mu\alpha} = g^{\mu\alpha}\pounds_{\xi}^{(m)}(R^{\nu}{}_{\alpha} g_{\nu\mu}) = g^{\mu\alpha} g_{\nu\mu}\pounds_{\xi}^{(m)}(R^{\nu}{}_{\alpha}) + \lambda g^{\mu\alpha}\delta^{\nu}_{\alpha}\mathcal{K}^{(m)}_{\nu\mu} = \pounds_{\xi}^{(m)} R + \lambda g^{\mu\nu}\mathcal{K}^{(m)}_{\mu\nu}.$$

It is immediate to combine this and (5.120) to get (5.116), which proves the proposition. □

**Remark 5.44.** *Some of the higher order constraints that $\{\mathbb{Y}^{(k)}\}$ need to satisfy in Theorem 5.42 are redundant. By Proposition 5.43 the components $n^b\mathcal{R}^{(i)}_{ab} = 2\lambda n^b \mathrm{Y}^{(i-1)}_{ab}$ and $P^{ab}\mathcal{R}^{(i)}_{ab} = 2\lambda P^{ab}\mathrm{Y}^{(i-1)}_{ab}$ of the equation $\mathcal{R}^{(i)}_{ab} = 2\lambda \mathrm{Y}^{(i-1)}_{ab}$ for $i \geq 2$ are automatically satisfied. Hence, one only needs to satisfy the constraint equations* (5.112), (5.113) *and the part of* (5.114) *that lies in the kernel of the energy-momentum map* (2.25).

Now we have all the ingredients to restate Theorem 5.42 when $\mathcal{H}$ admits cross sections.

**Theorem 5.45.** *Let $\{\mathcal{H},\boldsymbol{\gamma},\boldsymbol{\ell},\ell^{(2)}\}$ be null metric hypersurface data admitting a cross-section $\iota : \mathcal{S} \hookrightarrow \mathcal{H}$ with metric $h := \iota^{\star}\boldsymbol{\gamma}$ and assume there exists a smooth function $f$ on $\mathcal{H}$ satisfying*

$$-\pounds_n(\operatorname{tr}_P \mathbf{U}) + (\operatorname{tr}_P \mathbf{U}) f - P^{ab}P^{cd}\mathrm{U}_{ac}\mathrm{U}_{bd} = 0. \tag{5.121}$$

*Let $\chi$ and $\boldsymbol{\beta}$ be a function and a one-form on $\mathcal{S}$, respectively, and $\{\mathrm{S}^{(k)}_{AB}\}_{k\geq 1}$ a sequence of symmetric tensors on $\mathcal{S}$ traceless w.r.t. $h$. Then, there exists a semi-Riemannian manifold $(\mathcal{M}, g)$, an embedding $\boldsymbol{\Phi} : \mathcal{H} \hookrightarrow \mathcal{M}$ and a rigging $\xi$ such that (i) $\{\mathcal{H},\boldsymbol{\gamma},\boldsymbol{\ell},\ell^{(2)}\}$ is null metric hypersurface data $(\boldsymbol{\Phi},\xi)$-embedded in $(\mathcal{M}, g)$, (ii) $(\mathcal{M}, g)$ solves the $\Lambda$-vacuum equations to infinite order on $\boldsymbol{\Phi}(\mathcal{H})$ and (iii) its asymptotic expansion $\{\mathbf{Y}^{(k)}\}_{k\geq 1}$ satisfies $\operatorname{tr}_P \mathbf{Y}|_{\mathcal{S}} = \chi$, $\iota^{\star}\mathbf{r} = \boldsymbol{\beta}$, $\kappa = f$ and $tf\,(\iota^{\star}\mathbf{Y}^{(k)}) = \mathbf{S}^{(k)}$ for every $k \geq 1$, where tf stands for the trace-free part of a tensor w.r.t. $h$.*

*Proof.* The strategy is to construct explicitly the expansion $\{\mathbb{Y}^{(k)}\}_{k\geq 1}$ in order to satisfy the hypothesis of Theorem 5.42 and such that the initial conditions on $\mathcal{S}$ satisfy (iii). We first recall that from decomposition (2.140) it follows that given a $(0,2)$ tensor $\mathrm{T}_{ab}$ on $\mathcal{H}$,

$$\operatorname{tr}_P \mathbf{T} = h^{AB}\mathrm{T}_{AB} - \mathbf{T}(n, \iota_{\star}\ell^{\sharp}) - \mathbf{T}(\iota_{\star}\ell^{\sharp}, n) - (\ell^{(2)} - \ell^{(2)}_{\sharp})\mathbf{T}(n,n). \tag{5.122}$$

The construction of a tensor $\mathbb{Y}$ that satisfies equation (5.111) and $\operatorname{tr}_P \mathbb{Y}|_{\mathcal{S}} = \chi$, $\iota^{\star}\mathbb{r} = \boldsymbol{\beta}$, $\mathbb{k} = f$ and $(\iota^{\star}\mathbb{Y})^{\mathrm{tf}} = \mathbf{S}^{(1)}$ has already been made in Theorem 2.11, so here we only need to worry about the higher orders. For them, we follow a similar strategy, but with some important differences, particularly on how $\boldsymbol{c}^{(m)}$ and $t^{(m)}$ are constructed. Let us start by discussing the second order. Now, we define a scalar $t^{(2)}$ and a one-form $\boldsymbol{c}^{(2)}$ by means of (cf. (5.112)-(5.113))

$$\lambda\ell^{(2)} = -t^{(2)} + \mathcal{O}^{(1)}(\mathbb{Y}), \tag{5.123}$$

$$\lambda\ell_a = c^{(2)}_a + \mathcal{O}^{(1)}(\mathbb{Y})_a, \tag{5.124}$$

and the tensor field $\mathbb{Y}^{(2)}$ by integrating (cf. (5.114))

$$\begin{aligned} 2\lambda\mathbb{Y}_{ab} = {}& -2\pounds_n\mathbb{Y}^{(2)}_{ab} - (4\mathbb{k} + \operatorname{tr}_P \mathbf{U})\,\mathbb{Y}^{(2)}_{ab} - t^{(2)}\mathrm{U}_{ab} + 4P^{cd}\mathrm{U}_{c(a}\mathbb{Y}^{(2)}_{b)d} \\ & + 4(\mathrm{s} - \mathbb{r})_{(a}c^{(2)}_{b)} + 2\overset{\circ}{\nabla}_{(a}c^{(2)}_{b)} - 2c^{(2)}\mathbb{Y}_{ab} + \mathcal{O}^{(1)}_{ab}(\mathbb{Y}), \end{aligned} \tag{5.125}$$

where $\mathfrak{c}^{(2)} := -\boldsymbol{c}^{(2)}(n)$, with initial conditions $\iota^\star(\mathbb{Y}^{(2)}(n,\cdot)) = \iota^\star(\mathbb{Y}^{(2)}(\cdot,n)) = \iota^\star(\boldsymbol{c}^{(2)})$, $\mathbb{Y}^{(2)}(n,n) = -\mathfrak{c}^{(2)}$ on $\mathcal{S}$ and (see (5.122))

$$\iota^\star \mathbb{Y}^{(2)}_{AB} = \mathrm{S}^{(2)}_{AB} + \frac{1}{\mathfrak{n}-1}\Big(t^{(2)} + 2\boldsymbol{c}^{(2)}(\iota_\star \ell^\sharp) - (\ell^{(2)} - \ell^{(2)}_\sharp)\mathfrak{c}^{(2)}\Big) h_{AB}. \tag{5.126}$$

The equations that $\mathbb{r}^{(2)} := \mathbb{Y}^{(2)}(n,\cdot)$ and $\operatorname{tr}_P \mathbb{Y}^{(2)}$ satisfy are obtained, respectively, after contracting (5.125) with $n^b$ and $P^{ab}$ and using (2.96), (2.46), (2.66) and (2.50),

$$\begin{aligned}
2\lambda \mathbb{r}_a &= -2\pounds_n \mathbb{r}^{(2)}_a - (4\mathbb{k} + \operatorname{tr}_P \mathbf{U})\mathbb{r}^{(2)}_a + 2P^{cd}\mathrm{U}_{ca}\mathbb{r}^{(2)}_d + 2\mathbb{k} c^{(2)}_a + \pounds_n c^{(2)}_a - \overset{\circ}{\nabla}_a \mathfrak{c}^{(2)} \\
&\quad - 2P^{cd}\mathrm{U}_{ca} c^{(2)}_d + \mathcal{O}^{(1)}_{ab}(\mathbb{Y}) n^b, & (5.127)\\
2\lambda \operatorname{tr}_P \mathbb{Y} &= -2\pounds_n(\operatorname{tr}_P \mathbb{Y}^{(2)}) - 8P(\mathbf{s}, \mathbb{r}^{(2)}) + 2\mathbb{k}^{(2)} n(\ell^{(2)}) - (4\mathbb{k} + \operatorname{tr}_P \mathbf{U}) \operatorname{tr}_P \mathbb{Y}^{(2)} \\
&\quad - t^{(2)} \operatorname{tr}_P \mathbf{U} + 4P(\mathbf{s} - \mathbb{r}, \boldsymbol{c}^{(2)}) + 2\operatorname{div}_P \boldsymbol{c}^{(2)} - 2\mathfrak{c}^{(2)} \operatorname{tr}_P \mathbb{Y} + \operatorname{tr}_P \mathcal{O}^{(1)}(\mathbb{Y}). & (5.128)
\end{aligned}$$

Another contraction of (5.127) with $n^a$ yields

$$-2\lambda \mathbb{k} = 2\pounds_n(\mathbb{k}^{(2)} - \mathfrak{c}^{(2)}) + (4\mathbb{k} + \operatorname{tr}_P \mathbf{U})\mathbb{k}^{(2)} - 2\mathbb{k}\mathfrak{c}^{(2)} + \mathcal{O}^{(1)}_{ab}(\mathbb{Y}) n^a n^b. \tag{5.129}$$

Let us now check that $c^{(2)}_a = \mathbb{r}^{(2)}_a$ and $t^{(2)} = \operatorname{tr}_P \mathbb{Y}^{(2)}$ everywhere. Define the tensor field

$$\mathrm{T}^{(2)}_{ab} := \frac{t^{(2)} - \mathfrak{c}^{(2)}\ell^{(2)}}{\mathfrak{n}-1}\gamma_{ab} + 2c^{(2)}_{(a}\ell_{b)} + \mathfrak{c}^{(2)}\ell_a\ell_b.$$

It is a matter of simple computation to check that

$$P^{ab}\mathrm{T}^{(2)}_{ab} = t^{(2)}, \qquad \mathrm{T}^{(2)}_{ab} n^a = \mathrm{T}^{(2)}_{ba} n^a = c^{(2)}_b. \tag{5.130}$$

We now construct an auxiliary semi-Riemannian manifold $(\widetilde{\mathcal{M}}, \widetilde{g})$ using Theorem 5.40 with the sequence $\{\mathbb{Y}, \mathbf{T}^{(2)}, 0, 0, ...\}$ and denote its transverse expansion by $\widetilde{\mathbf{Y}}^{(k)} := \frac{1}{2}\widetilde{\mathbf{\Phi}}^\star(\pounds^{(k)}_\xi \widetilde{g})$. We shall use this space to relate the contractions $\operatorname{tr}_P \mathbb{Y}^{(2)}$ and $\mathbb{r}^{(2)}$ with $t^{(2)}$ and $\boldsymbol{c}^{(2)}$. It is important to note that these relations hold true independently of the auxiliary space we construct.

Since $\widetilde{\mathbf{Y}} = \mathbb{Y}$ and $\widetilde{\mathbf{Y}}^{(2)} = \mathbf{T}^{(2)}$, it is clear from Def. 2.2 and equations (2.95), (5.82), (5.83) for $m = 1$ as well as (5.111), (5.123), (5.124) that the Ricci tensor of $\widetilde{g}$ satisfies $\widetilde{R}_{\mu\nu} = \lambda \widetilde{g}_{\mu\nu}$ on $\mathbf{\Phi}(\mathcal{H})$. Hence, by Proposition 5.43 it then follows $n^b\widetilde{\mathcal{R}}^{(2)}_{ab} = \lambda n^b \mathcal{K}_{ab}$ as well as $P^{ab}\widetilde{\mathcal{R}}^{(2)}_{ab} = \lambda P^{ab}\mathcal{K}_{ab}$. Using the fact that $\mathcal{K}_{ab} = \mathbf{\Phi}^\star(\pounds_\xi \widetilde{g})_{ab} = 2\widetilde{\mathrm{Y}}_{ab} = 2\mathbb{Y}_{ab}$ together with (5.130), equations (5.85)-(5.86) read

$$\begin{aligned}
2\lambda \mathbb{r}_a &= -\pounds_n c^{(2)}_a - (2\mathbb{k} + \operatorname{tr}_P \mathbf{U})c^{(2)}_a - \overset{\circ}{\nabla}_a \mathfrak{c}^{(2)} + \mathcal{O}^{(1)}_{ab}(\mathbb{Y}) n^b, & (5.131)\\
2\lambda \operatorname{tr}_P \mathbb{Y} &= -2\pounds_n t^{(2)} - 2\,(2\mathbb{k} + \operatorname{tr}_P \mathbf{U})\, t^{(2)} + 2\mathfrak{c}^{(2)}(n(\ell^{(2)}) - \operatorname{tr}_P \mathbb{Y}) \\
&\quad - 4P(\mathbb{r} + \mathbf{s}, \boldsymbol{c}^{(2)}) + 2\operatorname{div}_P \boldsymbol{c}^{(2)} + \operatorname{tr}_P \mathcal{O}^{(1)}(\mathbb{Y}). & (5.132)
\end{aligned}$$

The contraction of (5.131) with $n^a$ gives

$$-2\lambda\Bbbk = (2\Bbbk + \operatorname{tr}_P \mathbf{U})\mathfrak{c}^{(2)} + \mathcal{O}^{(1)}_{ab}(\mathbb{Y})n^a n^b.$$

Subtracting it from (5.129) yields

$$2\pounds_n(\Bbbk^{(2)} - \mathfrak{c}^{(2)}) + (4\Bbbk + \operatorname{tr}_P \mathbf{U})(\Bbbk^{(2)} - \mathfrak{c}^{(2)}) = 0. \tag{5.133}$$

Since $\Bbbk^{(2)}$ initially agrees with $\mathfrak{c}^{(2)}$ it follows that $\Bbbk^{(2)} = \mathfrak{c}^{(2)}$ everywhere on $\mathcal{H}$. Next, subtracting (5.131) from (5.127)

$$2\pounds_n(\mathbb{r}^{(2)}_a - c^{(2)}_a) + (4\Bbbk + \operatorname{tr}_P \mathbf{U})(\mathbb{r}^{(2)}_a - c^{(2)}_a) - 2P^{cd}\mathrm{U}_{ca}(\mathbb{r}^{(2)}_d - c^{(2)}_d) = 0, \tag{5.134}$$

which is a homogeneous transport equation for $\mathbb{r}^{(2)}_a - c^{(2)}_a$, and since both tensors agree on $\mathcal{S}$, they agree everywhere. Finally, subtracting (5.132) from (5.128) and using $\mathbb{r}^{(2)} = \boldsymbol{c}^{(2)}$ and $\Bbbk^{(2)} = \mathfrak{c}^{(2)}$ gives

$$2\pounds_n(\operatorname{tr}_P \mathbb{Y}^{(2)} - t^{(2)}) + (4\Bbbk + \operatorname{tr}_P \mathbf{U})(\operatorname{tr}_P \mathbb{Y}^{(2)} - t^{(2)}) = 0. \tag{5.135}$$

Observe that the initial condition (5.126) together with $c^{(2)}_a = \mathbb{r}^{(2)}_a$ imply that $\operatorname{tr}_P \mathbb{Y}^{(2)} = t^{(2)}$ on $\mathcal{S}$ (see (5.122)), and thus by (5.135) they also coincide on all $\mathcal{H}$. We have shown that the tensor $\mathbb{Y}^{(2)}$ satisfy equations (5.112)-(5.114) for $m = 1$.

In order to construct the rest of the $\mathbb{Y}^{(k)}$'s we proceed analogously. Let us assume we have already constructed the sequence $\{\mathbb{Y}, ..., \mathbb{Y}^{(k)}\}$ for some integer $k \geq 2$ such that equations (5.111)-(5.114) hold for every $m \leq k-1$. We define the scalar $t^{(k+1)}$ and the one-form $\boldsymbol{c}^{(k+1)}$ by (cf. (5.112)-(5.113))

$$0 = -t^{(k+1)} + \mathcal{O}^{(k)}(\mathbb{Y}^{\leq k}), \tag{5.136}$$

$$\frac{\lambda}{2}\delta^k_2 \overset{\circ}{\nabla}_a \ell^{(2)} = c^{(k+1)}_a + \mathcal{O}^{(k)}_a(\mathbb{Y}^{\leq k}), \tag{5.137}$$

and the tensor $\mathbb{Y}^{(k+1)}$ by integrating (cf. (5.114))

$$\begin{aligned} 2\lambda \mathbb{Y}^{(k)}_{ab} &= -2\pounds_n \mathbb{Y}^{(k+1)}_{ab} - (2(k+1)\Bbbk + \operatorname{tr}_P \mathbf{U})\,\mathbb{Y}^{(k+1)}_{ab} - t^{(k+1)}\mathrm{U}_{ab} + 4P^{cd}\mathrm{U}_{c(a}\mathbb{Y}^{(k+1)}_{b)d} \\ &\quad + 4(\mathrm{s} - \mathbb{r})_{(a}c^{(k+1)}_{b)} + 2\overset{\circ}{\nabla}_{(a}c^{(k+1)}_{b)} - 2\mathfrak{c}^{(k+1)}\mathbb{Y}_{ab} + \mathcal{O}^{(k)}_{ab}(\mathbb{Y}^{\leq k}), \end{aligned} \tag{5.138}$$

where $\mathfrak{c}^{(k+1)} := -\boldsymbol{c}^{(k+1)}(n)$, with initial conditions given by $\iota^\star(\mathbb{Y}^{(k+1)}(n,\cdot)) = \iota^\star(\mathbb{Y}^{(k+1)}(\cdot,n)) = \iota^\star(\boldsymbol{c}^{(k+1)})$, $\mathbb{Y}^{(k+1)}(n,n) = -\mathfrak{c}^{(k+1)}$ on $\mathcal{S}$ and

$$\iota^\star \mathbb{Y}^{(k+1)}_{AB} = \mathrm{S}^{(k+1)}_{AB} + \frac{1}{\mathfrak{n}-1}\Big(t^{(k+1)} + 2\boldsymbol{c}^{(k+1)}(\iota_\star \ell^\sharp) - (\ell^{(2)} - \ell^{(2)}_\sharp)\mathfrak{c}^{(k+1)}\Big)h_{AB}. \tag{5.139}$$

As in the case $k = 1$ we define the tensor field

$$\mathrm{T}^{(k+1)}_{ab} := \frac{t^{(k+1)} - \mathfrak{c}^{(k+1)}\ell^{(2)}}{\mathfrak{n}-1}\gamma_{ab} + 2c^{(k+1)}_{(a}\ell_{b)} + \mathfrak{c}^{(2)}\ell_a\ell_b$$

which satisfies $P^{ab}\mathrm{T}^{(k+1)}_{ab} = t^{(k+1)}$ and $\mathrm{T}^{(k+1)}_{ab}n^a = \mathrm{T}^{(k+1)}_{ba}n^a = c^{(k+1)}_b$. Then we construct an auxiliary semi-Riemannian manifold $(\widetilde{\mathcal{M}}_k, \widetilde{g}_k)$ using Theorem 5.40 with $\{\mathbb{Y}, ..., \mathbb{Y}^{(k)}, \mathbf{T}^{(k+1)}, 0, ...\}$ and denote its transverse expansion by $\widetilde{\mathbf{Y}}^{(i)} := \frac{1}{2}\widetilde{\Phi}^\star(\pounds^{(i)}_\xi \widetilde{g}_k)$, $i \geq 1$. Since $\mathbb{Y}^{(i)} = \widetilde{\mathbf{Y}}^{(i)}$ for every $i = 1, ..., k$ and $\mathbf{T}^{(k+1)} = \widetilde{\mathbf{Y}}^{(k+1)}$, the same argument as in the case $k = 1$ proves that the Ricci tensor of $\widetilde{g}_k$ satisfies $\pounds^{(i)}_\xi \widetilde{R}_{\mu\nu} = \lambda \mathcal{K}^{(i)}_{\mu\nu}$ for every $i = 0, ..., k-1$. Hence, by Proposition 5.43 it then follows $n^b\widetilde{\mathcal{R}}^{(k+1)}_{ab} = \lambda n^b \mathcal{K}^{(k)}_{ab}$ and $P^{ab}\widetilde{\mathcal{R}}^{(k+1)}_{ab} = \lambda P^{ab}\mathcal{K}^{(k)}_{ab}$, and therefore (cf. (5.85)-(5.86))

$$2\lambda \mathbb{r}^{(k)}_a = -\pounds_n c^{(k+1)}_a - (2\mathbb{k} + \operatorname{tr}_P \mathbf{U})c^{(k+1)}_a - \overset{\circ}{\nabla}_a \mathfrak{c}^{(k+1)} + \mathcal{O}^{(k)}_{ab}(\mathbb{Y}^{\leq k})n^b, \tag{5.140}$$

$$\begin{aligned} 2\lambda \operatorname{tr}_P \mathbb{Y}^{(k)} &= -2\pounds_n t^{(k+1)} - 2\left(2\mathbb{k} + \operatorname{tr}_P \mathbf{U}\right) t^{(k+1)} + 2\mathfrak{c}^{(k+1)}(n(\ell^{(2)}) - \operatorname{tr}_P \mathbb{Y}) \\ &\quad - 4P(\mathbb{r} + \mathbf{s}, \boldsymbol{c}^{(k+1)}) + 2\operatorname{div}_P \boldsymbol{c}^{(k+1)} + \operatorname{tr}_P \mathcal{O}^{(k)}(\mathbb{Y}^{\leq k}). \end{aligned} \tag{5.141}$$

Contracting (5.138) with $n^b$ and $P^{ab}$ and combining the resulting equations with (5.140)-(5.141) leads to transport equations for $\mathbb{k}^{(k+1)} - \mathfrak{c}^{(k+1)}$, $\mathbb{r}^{(k+1)} - \boldsymbol{c}^{(k+1)}$ and $\operatorname{tr}_P \mathbb{Y}^{(k+1)} - t^{(k+1)}$ analogous to (5.133)-(5.135). Since the initial conditions in the three cases agree, then $\mathbb{k}^{(k+1)} = \mathfrak{c}^{(k+1)}$, $\mathbb{r}^{(k+1)} = \boldsymbol{c}^{(k+1)}$ and $\operatorname{tr}_P \mathbb{Y}^{(k+1)} = t^{(k+1)}$ everywhere on $\mathcal{H}$. Recall that this conclusion is independent on the auxiliary space $(\widetilde{\mathcal{M}}_k, \widetilde{g}_k)$ we have constructed. This proves that $\mathbb{Y}^{(k+1)}$ satisfies (5.114) for $m = k$, and therefore by induction the whole sequence $\{\mathbb{Y}^{(k)}\}$ constructed in this way satisfies the equations (5.111)-(5.114). Moreover $\operatorname{tr}_P \mathbb{Y}|_{\mathcal{S}} = \chi$, $\iota^\star \mathbb{r} = \boldsymbol{\beta}$, $\mathbb{k} = f$ and $\operatorname{tf}(\iota^\star \mathbb{Y}^{(k)}) = \mathbf{S}^{(k)}$ for every $k \geq 1$ (see (5.122) and (5.139)). Applying Theorem 5.42 the result is established. □

**Remark 5.46.** *The redundancy of equations* (5.123)-(5.124) *with* (5.127)-(5.128) *(and similarly for higher orders) is a manifestation of the (ambient) contracted Bianchi identity. When the ambient manifold is given, this identity ensures that the objects coming from* (5.127)-(5.128) *and the ones coming from* (5.123)-(5.124) *agree (see* (5.87)-(5.88)*). However, in Theorem 5.45 the ambient manifold is not given a priori, but it is being constructed, and thus analysing the compatibility between* (5.127)-(5.128) *and* (5.123)-(5.124) *is a crucial point in the proof of the Theorem.*

**Remark 5.47.** *The existence of a function $f$ satisfying* (5.121) *is automatic whenever* $\operatorname{tr}_P \mathbf{U} \neq 0$ *everywhere or* $\mathbf{U} = 0$ *everywhere. It is only when* $\operatorname{tr}_P \mathbf{U}$ *vanishes on a proper subset that* (5.121) *imposes restrictions on the data.*

Next we write down a by-product of the proof of Theorem 5.45 that will be important in Theorem 5.82 to show that the full set of Einstein equations is fulfilled order by order.

**Lemma 5.48.** *Let $(\mathcal{M}, g)$ be a semi-Riemannian manifold and $\Phi : \mathcal{H} \hookrightarrow \mathcal{M}$ an embedded null hypersurface with rigging $\xi$ extended off $\Phi(\mathcal{H})$ geodesically. Fix a natural number $m \geq 1$ and assume the Ricci tensor of $g$ satisfies $\mathcal{R}_{ab} \overset{\mathcal{H}}{=} \lambda \gamma_{ab}$ (when $m = 1$) and $\mathcal{R}^{(i)}_{ab} \overset{\mathcal{H}}{=} \lambda \mathcal{K}^{(i-1)}_{ab}$ and $\xi^\mu R^{(i-1)}_{\mu\nu} \overset{\mathcal{H}}{=} \lambda \xi^\mu \mathcal{K}^{(i-2)}_{\mu\nu}$ for $i = 1, ..., m$ (when $m \geq 2$). Let $t^{(m+1)}$ and $\boldsymbol{c}^{(m+1)}$ be defined by*

$$t^{(m+1)} := \mathcal{O}^{(m)}(\mathbf{Y}^{\leq m}) - \delta^m_1 \lambda \ell^{(2)}, \qquad c^{(m+1)}_a := \lambda\left(\delta^m_1 \ell_a + \frac{1}{2}\delta^m_2 \overset{\circ}{\nabla}_a \ell^{(2)}\right) - \mathcal{O}^{(m)}_a(\mathbf{Y}^{\leq m})$$

*and* $\mathfrak{c}^{(m)} := -\boldsymbol{c}^{(m)}(n)$*, where* $\mathcal{O}^{(m)}$ *and* $\mathcal{O}^{(m)}_a$ *are the same as in* (5.82)-(5.83)*. Then the following equations hold*

$$\begin{aligned}
2\lambda \mathrm{r}^{(m)}_a &= -\pounds_n c^{(m+1)}_a - (2m\kappa + \operatorname{tr}_P \mathbf{U}) c^{(m+1)}_a - \overset{\circ}{\nabla}_a \mathfrak{c}^{(m+1)} + \mathcal{O}^{(m)}_{ab} n^b, \\
2\lambda \operatorname{tr}_P \mathbf{Y}^{(m)} &= -2\pounds_n t^{(m+1)} - 2\left((m+1)\kappa + \operatorname{tr}_P \mathbf{U}\right) t^{(m+1)} + 2\mathfrak{c}^{(m+1)}(n(\ell^{(2)}) - \operatorname{tr}_P \mathbf{Y}) \\
&\quad - 4P(\mathbf{r}+\mathbf{s}, \boldsymbol{c}^{(m+1)}) + 2\operatorname{div}_P \boldsymbol{c}^{(m+1)} + P^{ab}\mathcal{O}^{(m)}_{ab}.
\end{aligned}$$

We conclude this section by showing an explicit application of our existence results that connects with one of the examples studied in Section 5.1.

**Example 5.49.** *Consider the degenerate Killing horizon case studied in Subsection 5.1.3. In Gaussian null coordinates* $\{r, v, x^A\}$ *(see Appendix C) the metric takes the form* (5.14)

$$g = r^2 F dv^2 + 2dvdr + 2r\beta_A dx^A dv + h_{AB} dx^A dx^B,$$

*and* $\eta := \partial_v$ *is the Killing field. Choosing* $\xi := \partial_r$ *as the rigging (which is geodesic, null and orthogonal to the cross-sections* $r = const.$*,* $v = const.$*), it is easy to see from* (5.23)

$$\begin{aligned}
&\kappa = 0, \qquad \kappa^{(2)} = -\Lambda, \qquad \kappa^{(m+2)} = -\frac{\Lambda}{\mathfrak{n}-1}(-C)^m \frac{(\mathfrak{n}-2+m)!}{(\mathfrak{n}-3)!}, \qquad \mathrm{r}^{(m)}_A = 0, \\
&\mathrm{Y}_{AB} = \frac{(\mathfrak{n}-3)C}{\Lambda} h_{AB}, \qquad \mathrm{Y}^{(2)}_{AB} = \frac{(\mathfrak{n}-3)C^2}{\Lambda} h_{AB}, \qquad \mathrm{Y}^{(m+1)}_{AB} = 0 \quad (m \geq 1).
\end{aligned} \tag{5.142}$$

*As already discussed, when* $C = 0$ *this expansion agrees with that of Nariai, and when* $C \neq 0$ *one can absorb* $C$ *by rescaling the coordinates* $r$ *and* $v$ *and the expansion* (5.142) *agrees with that of extremal Schwarzschild-de Sitter spacetime (Theorem 5.4). In this particular case the transverse expansion* (5.142) *is manifestly convergent and thus one can employ Theorem 5.40 with* $\rho = 1$ *(see* (5.105)*) to construct the spacetime* $(\mathcal{M}, g)$*, which clearly is either Schwarzschild-de Sitter (when* $C \neq 0$*) or Nariai (when* $C = 0$*). In both cases,* $(\mathcal{M}, g)$ *solves exactly the* $\lambda$*-vacuum equations.*

*There could be other situations (not necessarily Killing horizons) where one can find explicitly the transverse expansion but it may not converge (e.g. Fefferman-Graham metrics when the initial conformal class is not real analytic [113] and the series does not converge). In these cases, Theorem 5.40 still ensures the existence of a smooth (and not analytic) ambient manifold* $(\mathcal{M}, g)$*. This space may or may not be a true solution (note that now all the freedom in selecting the function* $\rho$ *and the constants* $\{\mu_k\}$ *kicks in) of the* $\Lambda$*-vacuum equations in a neighbourhood of the hypersurface. However, if the transverse expansion has been constructed by imposing the Einstein equations* $R_{vr} = \Lambda$*,* $R_{AB} = \Lambda h_{AB}$*,* $R^{(m)}_{rA} = 0$*,* $R^{(m+1)}_{rv} = 0$ *and* $R^{(m+1)}_{AB} = 2\Lambda \mathrm{Y}^{(m)}_{AB}$ *for all* $m \geq 1$ *(which is the case of Theorem 5.4), Proposition 5.43 guarantees that* $(\mathcal{M}, g)$ *solves the* $\Lambda$*-vacuum equations to infinite order on* $\mathcal{H}$*.*

## 5.4 GENERALIZED MASTER EQUATIONS

In some situations one has a privileged vector field $\eta$ on $(\mathcal{M}, g)$ whose deformation tensor $\mathcal{K}[\eta] := \pounds_\eta g$ is known. This is in general a very valuable information that one may want to incorporate into the identities of Section 5.2. For instance, in the Ph.D. thesis of M. Manzano [211] it is shown that the expression of the constraint tensor in (2.95) can be rewritten so that the dependence on the tensor $\mathbf{Y}$ is algebraic instead of via a transport equation. The corresponding identity was called the *generalized master equation.* The aim of this section is to extend the same idea to the higher order derivatives of the Ricci tensor, i.e. to combine identity (5.84) with information on $\mathcal{K}[\eta]$ so that the dependence on $\mathbf{Y}^{(m+1)}$ becomes algebraic. Let us start by reviewing some results from [211] relevant for this discussion.

**Lemma 5.50.** *Let $\{\mathcal{H}, \boldsymbol{\gamma}, \boldsymbol{\ell}, \ell^{(2)}\}$ be null metric hypersurface data $(\Phi, \xi)$-embedded in $(\mathcal{M}, g)$ and let $\eta \in \mathfrak{X}(\mathcal{M})$ be such that $\eta|_{\Phi(\mathcal{H})}$ is tangent to $\Phi(\mathcal{H})$. Denote by $\bar{\eta}$ the vector field on $\mathcal{H}$ satisfying $\Phi_\star(\bar{\eta}) = \eta|_{\Phi(\mathcal{H})}$. Then,*

$$[\eta, \xi] \stackrel{\mathcal{H}}{=} A_\eta \xi + \Phi_\star(X_\eta), \tag{5.143}$$

*where*

$$A_\eta := -\mathcal{K}[\eta](\xi, \nu) + (\pounds_{\bar{\eta}} \boldsymbol{\ell})(n), \tag{5.144}$$

*and the components of $X_\eta$ in any basis $\{e_a\}$ of $\mathcal{H}$ are given by*

$$X_\eta^a := \frac{1}{2}\mathcal{K}[\eta]\Big(\xi, n^a \xi - 2\theta^a\Big) + \frac{1}{2} n^a \bar{\eta}(\ell^{(2)}) + P^{ab} \pounds_{\bar{\eta}} \ell_b, \tag{5.145}$$

*where recall $\theta^a = P^{ab}\widehat{e}_b + n^a \xi$ (see (2.12)).*

From now on we assume $\eta|_{\mathcal{H}}$ to be null and tangent to $\Phi(\mathcal{H})$, so there must exist a function $\alpha \in \mathcal{F}(\mathcal{H})$ such that $\eta|_{\mathcal{H}} = \alpha\nu$ (in principle we do not assume $\alpha \neq 0$, thus allowing $\eta$ to have zeros). The gauge behaviour of the scalars $\alpha$ and $\mathcal{K}[\eta](\xi, \nu)$ are as follows.

**Lemma 5.51.** *Let $(z, V)$ be a gauge element and denote with a prime the gauge-transformed objects. Then,*

$$\alpha' = z\alpha, \qquad \mathcal{K}[\eta](\xi', \nu') = \mathcal{K}[\eta](\xi, \nu). \tag{5.146}$$

*Proof.* The first one follows from (2.34) and $\eta$ being gauge invariant. For the second one we use $\xi' = z(\xi + V)$ and $\nu' = z^{-1}\nu$, so

$$\mathcal{K}[\eta](\xi', \nu') = \mathcal{K}[\eta](\xi, \nu) + \mathcal{K}[\eta](V, \nu) = \mathcal{K}[\eta](\xi, \nu),$$

since $\Phi^\star \mathcal{K}[\eta] = \pounds_{\alpha n} \boldsymbol{\gamma} = 2\alpha \mathbf{U}$ and $\mathbf{U}(n, \cdot) = 0$. □

From this it is immediate to check that the transformation of $n(\alpha)$ and that of $A_\eta$ are then given by

$$n'(\alpha') = n(\alpha) + n(\log|z|)\alpha, \qquad (5.147) \qquad A'_\eta = A_\eta + n(\log|z|)\alpha. \qquad (5.148)$$

In particular, $n(\alpha)$ is gauge-invariant at the points where $\alpha = 0$.

Following [211], let us introduce the scalar function $\kappa_\star$ by means of

$$\kappa_\star := n(\alpha) + \alpha\kappa. \tag{5.149}$$

From (2.148) and Lemma 5.51 it follows that $\kappa_\star$ is gauge invariant. This function extends the standard notion of surface gravity. Indeed, at the points where the vector field $\eta$ does not vanish, its surface gravity $\widetilde{\kappa}$ is the scalar on $\mathcal{H}$ defined by

$$\nabla_\eta \eta \stackrel{\mathcal{H}}{=} \widetilde{\kappa}\eta. \tag{5.150}$$

Inserting $\eta|_{\mathcal{H}} = \alpha\nu$ into (5.150) and using (2.41) it follows that $\widetilde{\kappa} = \kappa_\star$ on the subset of $\mathcal{H}$ where $\bar{\eta} \neq 0$. Note however that $\kappa_\star$ is well defined and smooth everywhere on $\mathcal{H}$.

Using the identity $\pounds_{[X,Y]} = [\pounds_X, \pounds_Y]$ applied to the metric $g$ it follows

$$\pounds_\eta \pounds_\xi g = \pounds_\xi \pounds_\eta g - \pounds_{[\xi,\eta]} g. \tag{5.151}$$

Introducing (5.143) and pulling back this equation into $\mathcal{H}$ gives [211]

$$\pounds_{\bar{\eta}} \mathbf{Y} = A_\eta \mathbf{Y} + \boldsymbol{\ell} \otimes_s dA_\eta + \frac{1}{2}\pounds_{X_\eta}\gamma + \frac{1}{2}\Phi^\star(\pounds_\xi \mathcal{K}[\eta]). \tag{5.152}$$

Inserting $\bar{\eta} = \alpha n$ and recalling (2.22) one has $\pounds_{\bar{\eta}}\boldsymbol{\ell} = \pounds_{\alpha n}\boldsymbol{\ell} = 2\alpha \mathbf{s} + d\alpha$, so equations (5.144), (5.145) and (5.152) become

$$A_\eta = -\mathcal{K}[\eta](\xi,\nu) + n(\alpha), \tag{5.153}$$

$$X^a_\eta = \frac{1}{2}\mathcal{K}[\eta]\Big(\xi, n^a\xi - 2\theta^a\Big) + \frac{1}{2}\alpha n(\ell^{(2)})n^a + 2\alpha P^{ab}\mathrm{s}_b + P^{ab}\overset{\circ}{\nabla}_b\alpha, \tag{5.154}$$

$$\alpha \pounds_n \mathbf{Y} = A_\eta \mathbf{Y} - 2d\alpha \otimes_s \mathbf{r} + \boldsymbol{\ell} \otimes_s dA_\eta + \frac{1}{2}\pounds_{X_\eta}\gamma + \frac{1}{2}\Phi^\star(\pounds_\xi\mathcal{K}[\eta]). \tag{5.155}$$

The *generalized master equation* introduced in [211] relates the constraint tensor $\mathcal{R}_{ab}$, the metric hypersurface data and information on the deformation of $\eta$ codified via the tensorial quantities $\mathfrak{w}, \mathfrak{p}, \mathfrak{q}$ and $\mathfrak{I}$ defined by

$$\mathfrak{w} := \mathcal{K}[\eta](\xi,\nu), \qquad \mathfrak{p} := \mathcal{K}[\eta](\xi,\xi), \qquad \mathfrak{q} := \Phi^\star(\mathcal{K}[\eta](\xi,\cdot)), \qquad \mathfrak{I} := \frac{1}{2}\Phi^\star(\pounds_\xi\mathcal{K}[\eta]). \tag{5.156}$$

The equation is obtained by inserting (5.155) into (2.95) and using (5.156),

$$\begin{aligned} \alpha\mathcal{R}_{ab} = &-(2\kappa_\star + \alpha \operatorname{tr}_P \mathbf{U} - 2\mathfrak{w})\mathrm{Y}_{ab} + 2\alpha P^{cd}\mathrm{U}_{d(a}(2\mathrm{Y}_{b)c} + \mathrm{F}_{b)c}) - 2\mathfrak{I}_{ab} \\ &+ (\mathfrak{p} - \alpha(\operatorname{tr}_P \mathbf{Y}) - \alpha n(\ell^{(2)}))\mathrm{U}_{ab} - 2\alpha\overset{\circ}{\nabla}_{(a}(\mathrm{s}-\mathrm{r})_{b)} - 4(\mathrm{s}-\mathrm{r})_{(a}\overset{\circ}{\nabla}_{b)}\alpha \\ &- 2\alpha(\mathrm{s}-\mathrm{r})_a(\mathrm{s}-\mathrm{r})_b - 2\overset{\circ}{\nabla}_a\overset{\circ}{\nabla}_b\alpha - \alpha\overset{\circ}{\nabla}_{(a}\mathrm{s}_{b)} + \alpha\mathrm{s}_a\mathrm{s}_b + 2\overset{\circ}{\nabla}_{(a}\mathfrak{q}_{b)} + \alpha\overset{\circ}{R}_{(ab)}. \end{aligned} \tag{5.157}$$

As already mentioned before, the dependence on $\mathbf{Y}$ is now purely algebraic[7]. The contraction of (5.157) with $n$ is [211]

$$\begin{aligned}\alpha\mathcal{R}_{ab}n^b = (\mathfrak{w} - \alpha\,\mathrm{tr}_P\,\mathbf{U})\mathrm{r}_a - \overset{\circ}{\nabla}_a\kappa_\star - \mathfrak{I}_{ab}n^b + P^{cd}\overset{\circ}{\nabla}_c(\alpha\mathrm{U}_{ad}) + \alpha(\mathrm{tr}_P\,\mathbf{U})\mathrm{s}_a \\ - \alpha\overset{\circ}{\nabla}_a\,\mathrm{tr}_P\,\mathbf{U} + \frac{1}{2}\left(\overset{\circ}{\nabla}_n\mathfrak{q}_a + \overset{\circ}{\nabla}_a\mathfrak{w} - \mathfrak{w}\mathrm{s}_a - P^{bc}\mathrm{U}_{ca}\mathfrak{q}_b\right),\end{aligned} \tag{5.158}$$

and contracting again with $n$,

$$n(\kappa_\star) = n(\mathfrak{w}) + \kappa\mathfrak{w} + \mathfrak{I}(n,n). \tag{5.159}$$

In terms of (5.156), the function $A_\eta$ and the vector $X_\eta$ in (5.153)-(5.154) can be written as

$$A_\eta = n(\alpha) - \mathfrak{w}, \qquad X^a_\eta = \frac{1}{2}\Big(\alpha n(\ell^{(2)}) - \mathfrak{p}\Big)n^a + P^{ab}\Big(2\alpha\mathrm{s}_b + \overset{\circ}{\nabla}_b\alpha - \mathfrak{q}_b\Big). \tag{5.160}$$

Hence, relation (5.143) becomes, after using (5.149),

$$\pounds_\xi\eta^\mu \overset{\mathcal{H}}{=} (\mathfrak{w} + \alpha\kappa - \kappa_\star)\xi^\mu + \frac{1}{2}(\mathfrak{p} - \alpha n(\ell^{(2)}))\nu^\mu + P^{ab}\Big(\mathfrak{q}_b - 2\alpha\mathrm{s}_b - \overset{\circ}{\nabla}_b\alpha\Big)\widehat{e}^\mu_a. \tag{5.161}$$

The idea now is to repeat this process with the identities (5.84), i.e. to replace the derivative $\pounds_n\mathbf{Y}^{(m+1)}$ by derivatives of $\mathcal{K}[\eta]$, $\mathbf{Y}^{(m+1)}$ itself and lower order terms. As usual, let us define $\mathcal{K}[\eta]^{(m)} := \pounds_\xi^{(m-1)}\mathcal{K}[\eta]$ and

$$\mathfrak{w}^{(m-1)} := \mathcal{K}[\eta]^{(m)}(\xi,\nu), \quad \mathfrak{p}^{(m-1)} := \mathcal{K}[\eta]^{(m)}(\xi,\xi), \quad \mathfrak{q}^{(m-1)} := \Phi^\star\Big(\mathcal{K}[\eta]^{(m)}(\xi,\cdot)\Big), \tag{5.162}$$

$$\mathfrak{I}^{(m)} := \frac{1}{2}\Phi^\star\Big(\mathcal{K}[\eta]^{(m+1)}\Big). \tag{5.163}$$

Observe that $\mathfrak{w}^{(0)} = \mathfrak{w}$, $\mathfrak{p}^{(0)} = \mathfrak{p}$, $\mathfrak{q}^{(0)} = \mathfrak{q}$ and $\mathfrak{I}^{(1)} = \mathfrak{I}$. The rule of thumb is that the number inside the parenthesis denotes the number of transverse derivatives applied to $\mathcal{K}[\eta]$.

We start by finding the equations analogous to (5.152) and (5.155) but of "order $m$".

**Proposition 5.52.** *Let $\{\mathcal{H},\boldsymbol{\gamma},\boldsymbol{\ell},\ell^{(2)}\}$ be null metric hypersurface data $(\Phi,\xi)$-embedded in $(\mathcal{M},g)$ and extend $\xi$ off $\mathcal{H}$ arbitrarily. Let $\eta$ be a vector field on $(\mathcal{M},g)$ satisfying that $\eta|_{\Phi(\mathcal{H})} = \alpha\nu$ for some $\alpha \in \mathcal{F}(\mathcal{H})$. Then, for any integer $m \geq 2$*

$$\pounds_{\bar\eta}\mathbf{Y}^{(m)} = \mathfrak{I}^{(m)} + m(n(\alpha) - \mathfrak{w})\mathbf{Y}^{(m)} + \mathcal{P}^{(m)}, \tag{5.164}$$

*where $\mathcal{P}^{(m)}$ is a tensor that depends on $\{\mathbf{Y},...,\mathbf{Y}^{(m-1)}\}$ and $\{\pounds_\xi\eta,...,\pounds_\xi^{(m)}\eta\}\big|_{\mathcal{H}}$ and it is given explicitly by*

$$\mathcal{P}^{(m)} := m\pounds_{X_\eta}\mathbf{Y}^{(m-1)} - \frac{1}{2}\sum_{i=2}^{m}\binom{m}{i}\Phi^\star\Big(\pounds_{\pounds_\xi^{(i)}\eta}\pounds_\xi^{(m-i)}g\Big), \tag{5.165}$$

7 Actually, it is not algebraic due to the presence of $\overset{\circ}{\nabla}$-derivatives of $\mathbf{r} = \mathbf{Y}(n,\cdot)$ and contractions of $\mathbf{Y}$. However, $\mathbf{r}$ does appear algebraically on identity (5.158), thus reaffirming the hierarchical structure of the constraint tensor.

*and* $X^a_\eta = \frac{1}{2}\Big(\alpha n(\ell^{(2)}) - \mathfrak{p}\Big)n^a + P^{ab}\Big(2\alpha \mathrm{s}_b + \overset{\circ}{\nabla}_b\alpha - \mathfrak{q}_b\Big)$. *As a consequence,*

$$\alpha \pounds_n \mathbf{Y}^{(m)} = \mathfrak{I}^{(m)} + m(n(\alpha) - \mathfrak{w})\mathbf{Y}^{(m)} - 2d\alpha \otimes_s \mathbf{r}^{(m)} + \mathcal{P}^{(m)}. \tag{5.166}$$

*Proof.* We first show by induction the following relation

$$\pounds_\eta \pounds_\xi^{(m)} g = \pounds_\xi^{(m)} \pounds_\eta g - \sum_{i=1}^{m} \binom{m}{i} \pounds_{\pounds_\xi^{(i)}\eta} \pounds_\xi^{(m-i)} g. \tag{5.167}$$

For $m = 1$ it holds (cf. (5.151)). Let us assume (5.167) is true up to some $m \geq 1$ and show that it is then true for $m+1$ also. We compute

$$\begin{aligned}
\pounds_\eta \pounds_\xi^{(m+1)} g &= \pounds_\eta \pounds_\xi \pounds_\xi^{(m)} g \\
&= \pounds_\xi \pounds_\eta \pounds_\xi^{(m)} g + \pounds_{[\eta,\xi]} \pounds_\xi^{(m)} g \\
&= \pounds_\xi^{(m+1)} \pounds_\eta g - \sum_{i=1}^{m} \binom{m}{i} \pounds_\xi \pounds_{\pounds_\xi^{(i)}\eta} \pounds_\xi^{(m-i)} g - \pounds_{\pounds_\xi \eta} \pounds_\xi^{(m)} g \\
&= \pounds_\xi^{(m+1)} \pounds_\eta g - \sum_{i=1}^{m} \binom{m}{i} \pounds_{\pounds_\xi^{(i)}\eta} \pounds_\xi^{(m+1-i)} g - \sum_{i=1}^{m} \binom{m}{i} \pounds_{\pounds_\xi^{(i+1)}\eta} \pounds_\xi^{(m-i)} g - \pounds_{\pounds_\xi \eta} \pounds_\xi^{(m)} g \\
&= \pounds_\xi^{(m+1)} \pounds_\eta g - \sum_{i=1}^{m} \binom{m}{i} \pounds_{\pounds_\xi^{(i)}\eta} \pounds_\xi^{(m+1-i)} g - \sum_{i=0}^{m} \binom{m}{i} \pounds_{\pounds_\xi^{(i+1)}\eta} \pounds_\xi^{(m-i)} g,
\end{aligned}$$

where in the third line we introduced the induction hypothesis. Renaming $i \longmapsto i-1$ in the second sum and using the binomial identity $\binom{m}{i-1} + \binom{m}{i} = \binom{m+1}{i}$ it follows

$$\begin{aligned}
\pounds_\eta \pounds_\xi^{(m+1)} g &= \pounds_\xi^{(m+1)} \pounds_\eta g - \sum_{i=1}^{m} \binom{m+1}{i} \pounds_{\pounds_\xi^{(i)}\eta} \pounds_\xi^{(m+1-i)} g - \pounds_{\pounds_\xi^{(m+1)}\eta} g \\
&= \pounds_\xi^{(m+1)} \pounds_\eta g - \sum_{i=1}^{m+1} \binom{m+1}{i} \pounds_{\pounds_\xi^{(i)}\eta} \pounds_\xi^{(m+1-i)} g.
\end{aligned}$$

This establishes (5.167) for $m \geq 1$. The first term in the sum can be computed by means of (5.143). Pulling (5.167) back into $\mathcal{H}$ and using definition (5.163) gives

$$2\pounds_{\bar{\eta}} \mathbf{Y}^{(m)} = 2\mathfrak{I}^{(m)} + 2mA_\eta \mathbf{Y}^{(m)} + 2m\pounds_{X_\eta} \mathbf{Y}^{(m-1)} - \sum_{i=2}^{m} \binom{m}{i} \Phi^\star\Big(\pounds_{\pounds_\xi^{(i)}\eta} \pounds_\xi^{(m-i)} g\Big).$$

Relation (5.164) follows after using (5.160). □

An immediate consequence of Proposition 5.52 is the following.

**Corollary 5.53.** *Let* $\{\mathcal{H}, \boldsymbol{\gamma}, \boldsymbol{\ell}, \ell^{(2)}\}$ *be null metric hypersurface data* $(\Phi, \xi)$*-embedded in* $(\mathcal{M}, g)$ *and extend* $\xi$ *off* $\mathcal{H}$ *arbitrarily. Let* $\eta$ *be a vector field on* $(\mathcal{M}, g)$ *satisfying that* $\eta|_{\Phi(\mathcal{H})} = \alpha\nu$ *for some function* $\alpha \in \mathcal{F}(\mathcal{H})$*. Then, for any integer* $m \geq 1$

$$\pounds_{\bar{\eta}} \mathbf{Y}^{(m)} \overset{(m)}{=} \mathfrak{I}^{(m)} + m(n(\alpha) - \mathfrak{w})\mathbf{Y}^{(m)}. \tag{5.168}$$

*As a consequence,*

$$\alpha \pounds_n \mathbf{Y}^{(m)} \stackrel{(m)}{=} \mathfrak{J}^{(m)} + m(n(\alpha) - \mathfrak{w})\mathbf{Y}^{(m)} - 2d\alpha \otimes_s \mathbf{r}^{(m)}. \tag{5.169}$$

Inserting (5.166) into (5.84) we arrive at the main result of this section, namely the *generalized master equation of order* $m > 1$,

$$\begin{aligned} \alpha \mathcal{R}^{(m+1)}_{ab} =& - \left(2(m+1)(\kappa_\star - \mathfrak{w}) + \alpha \operatorname{tr}_P \mathbf{U}\right) \mathrm{Y}^{(m+1)}_{ab} - 2\mathfrak{J}^{(m+1)}_{ab} \\ & - \alpha(\operatorname{tr}_P \mathbf{Y}^{(m+1)})\mathrm{U}_{ab} + 4\alpha P^{cd}\mathrm{U}_{c(a}\mathrm{Y}^{(m+1)}_{b)d} + 4\alpha(\mathrm{s}-\mathrm{r})_{(a}\mathrm{r}^{(m+1)}_{b)} \\ & + 4\mathrm{r}^{(m+1)}_{(a}\overset{\circ}{\nabla}_{b)}\alpha + 2\alpha\overset{\circ}{\nabla}_{(a}\mathrm{r}^{(m+1)}_{b)} - 2\alpha\kappa^{(m+1)}\mathrm{Y}_{ab} + \alpha\mathcal{O}^{(m)}_{ab} + \mathcal{P}^{(m)}_{ab}, \end{aligned} \tag{5.170}$$

where recall that when $\nabla_\xi \xi = 0$ the tensor $\mathcal{O}^{(m)}$ depends *only* on metric hypersurface data and $\{\mathbf{Y}, ..., \mathbf{Y}^{(m)}\}$ (see Corollary 5.30). The tensor $\mathcal{P}^{(m)}$ also depends on $\{\mathbf{Y}, ..., \mathbf{Y}^{(m)}\}$ and in addition on the vectors $\pounds_\xi \eta, ..., \pounds^{(m+1)}_\xi \eta$ on $\Phi(\mathcal{H})$. Its key property is that it vanishes when $X_\eta = 0$ and $\pounds^{(i)}_\xi \eta \stackrel{\mathcal{H}}{=} 0$ for all $i = 2, ..., m+1$. Finally we prove an interesting property of the vector field $\pounds_\xi \eta$ that will play a key role in the next section.

**Lemma 5.54.** *Let* $\{\mathcal{H}, \boldsymbol{\gamma}, \boldsymbol{\ell}, \ell^{(2)}\}$ *be null metric hypersurface data* $(\Phi, \xi)$*-embedded in* $(\mathcal{M}, g)$ *and extend* $\xi$ *off* $\Phi(\mathcal{H})$ *by* $\nabla_\xi \xi = 0$*. Let* $\eta \in \mathfrak{X}(\mathcal{M})$ *be such that* $\eta|_{\Phi(\mathcal{H})}$ *is null and tangent to* $\Phi(\mathcal{H})$ *and such that its deformation tensor vanishes to all orders on* $\Phi(\mathcal{H})$*. If* $\pounds_\xi \eta \stackrel{\mathcal{H}}{=} h\xi$ *for some function* $h \in \mathcal{F}(\mathcal{H})$*, then* $\pounds^{(k)}_\xi \eta \stackrel{\mathcal{H}}{=} 0$ *for any integer* $k \geq 2$*.*

*Proof.* Let $m \geq 1$. By Proposition 5.7 it follows

$$\pounds^{(m)}_\xi \nabla_\gamma \xi^\alpha = \nabla_\gamma \pounds^{(m)}_\xi \xi^\beta + \sum_{k=0}^{m-1} \binom{m}{k+1} (\pounds^{(m-k-1)}_\xi \xi^\sigma) \Sigma[\xi]^{(k+1)\alpha}{}_{\sigma\gamma} = \xi^\sigma \Sigma[\xi]^{(m)\alpha}{}_{\sigma\gamma}. \tag{5.171}$$

Moreover, equation (5.34) and the fact that $\nabla_\xi \xi = 0$ imply

$$0 = \pounds^{(m)}_\xi \nabla_\xi \xi^\alpha = \xi^\gamma \pounds^{(m)}_\xi \nabla_\gamma \xi^\alpha. \tag{5.172}$$

Combining (5.171) and (5.172) gives

$$\xi^\gamma \xi^\sigma \Sigma[\xi]^{(i)\alpha}{}_{\gamma\sigma} \stackrel{\mathcal{H}}{=} 0 \qquad \forall i \geq 0. \tag{5.173}$$

The definition of $\Sigma[\eta]$ entails the general identity

$$-\nabla_\xi \pounds_\eta \xi = \Sigma[\eta](\xi, \xi) + \nabla_{\pounds_\eta \xi} \xi - \pounds_\eta (\nabla_\xi \xi). \tag{5.174}$$

This together with $\nabla_\xi \xi = 0$ gives

$$\pounds^{(2)}_\xi \eta = -\pounds_\xi \pounds_\eta \xi = -\nabla_\xi \pounds_\eta \xi + \nabla_{\pounds_\eta \xi} \xi = \Sigma[\eta](\xi, \xi) - 2\nabla_{\pounds_\xi \eta} \xi. \tag{5.175}$$

Since $\pounds_\xi \eta \stackrel{\mathcal{H}}{=} h\xi$ and $\Sigma[\eta]$ vanishes on $\Phi(\mathcal{H})$, it follows $\pounds_\xi^{(2)}\eta \stackrel{\mathcal{H}}{=} 0$. Applying $\pounds_\xi^{(m)}$ to (5.175) and using (5.34) and (5.171) gives

$$\pounds_\xi^{(m+2)}\eta = \Sigma[\eta]^{(m+1)}(\xi,\xi) - 2\sum_{k=0}^{m}\binom{m}{k}(\pounds_\xi^{(k+1)}\eta^\gamma)\xi^\sigma\Sigma[\xi]^{(m-k)}{}_{\sigma\gamma}. \tag{5.176}$$

We prove the statement by induction. Let $m \geq 1$, assume $\pounds_\xi^{(i)}\eta \stackrel{\mathcal{H}}{=} 0$ for all $i = 2, ..., m+1$ and let us prove that $\pounds_\xi^{(m+2)}\eta \stackrel{\mathcal{H}}{=} 0$. Since $\Sigma[\eta]^{(k)}(\xi,\xi) \stackrel{\mathcal{H}}{=} 0$ for all $k \geq 1$ equation (5.176) on $\Phi(\mathcal{H})$ reads

$$\pounds_\xi^{(m+2)}\eta \stackrel{\mathcal{H}}{=} -2(\pounds_\xi\eta^\gamma)\xi^\sigma\Sigma[\xi]^{(m)}{}_{\sigma\gamma} \stackrel{\mathcal{H}}{=} -2h\xi^\gamma\xi^\sigma\Sigma[\xi]^{(m)}{}_{\sigma\gamma} \stackrel{\mathcal{H}}{=} 0,$$

where in the last equality we used (5.173). □

**Remark 5.55.** *It is noteworthy that Lemma 5.54 is insensitive to the extension of $h$ off $\Phi(\mathcal{H})$. At first sight one could think that this is not possible and that counterexamples are easily constructed. For instance, assume that $\pounds_\xi\eta$ has the form $h\xi$ not just on $\mathcal{H}$ but everywhere. Clearly the lemma can only hold true if $\xi(h) = 0$. So, the conditions of the lemma must somehow ensure that this property holds. And indeed this can be proved, as we show next. Assume $\pounds_\xi\eta = h\xi$ everywhere, $\nabla_\xi\xi = 0$ and that the deformation tensor of $\eta$ vanishes. Then, equation (5.174) reads*

$$-\nabla_\xi\pounds_\eta\xi = \Sigma[\eta](\xi,\xi) + \nabla_{\pounds_\eta\xi}\xi - \pounds_\eta(\nabla_\xi\xi) = -h\nabla_\xi\xi = 0 \qquad \Longrightarrow \qquad \xi(h)\xi = 0.$$

*Since $\xi$ is a rigging of $\Phi(\mathcal{H})$ there exists a neighbourhood $\mathcal{U} \subset \mathcal{M}$ of $\Phi(\mathcal{H})$ in which $\xi \neq 0$, and hence $\xi(h) = 0$ on $\mathcal{U}$.*

When the deformation tensor of $\eta$ vanishes in a neighbourhood of $\Phi(\mathcal{H})$ (and not only to infinite order on $\Phi(\mathcal{H})$) and $h$ is assumed to be constant along $\xi$, then in fact $\pounds_\xi\eta = h\xi$ holds in a neighbourhood of $\Phi(\mathcal{H})$, as we prove next.

**Lemma 5.56.** *Let $(\mathcal{M}, g)$ admit a Killing vector $\eta$ and consider an embedded hypersurface $\Phi : \mathcal{H} \hookrightarrow \mathcal{M}$ with rigging $\xi$. Extend $\xi$ off $\Phi(\mathcal{H})$ by means of $\nabla_\xi\xi = 0$ and assume $\pounds_\xi\eta \stackrel{\mathcal{H}}{=} h\xi$ with $\xi(h) = 0$. Then, $\pounds_\xi\eta = h\xi$ in a neighbourhood of $\Phi(\mathcal{H})$.*

*Proof.* Define $\zeta := \pounds_\xi\eta - h\xi$. Equation (5.174) together with the fact that $\xi$ is geodesic yields

$$\nabla_\xi\zeta + \nabla_\zeta\xi - \Sigma[\eta](\xi,\xi) = 0.$$

Since $\eta$ is a Killing the last term vanishes and then we have a linear homogeneous transport equation for $\zeta$. Since $\zeta \stackrel{\mathcal{H}}{=} 0$ we conclude $\zeta = 0$ in a neighbourhood of $\Phi(\mathcal{H})$. □

## 5.5 ABSTRACT KILLING HORIZON DATA

In this section we study the case when $(\mathcal{M}, g)$ admits a Killing vector $\eta$ (and therefore $\mathcal{K}[\eta] = \pounds_\eta g = 0$). First, we review some properties of Killing horizons and particularize the identities of Section 5.4 to the present case. This will lead to a natural definition of "abstract Killing horizon data" as well as its embedded counterpart. Other abstract definitions of a Killing horizon (such as abstract Killing horizons of order zero ($AKH_0$) and one ($AKH_1$)) can be found in [215]. Then, we will prove that, fixing the extension of the rigging to being geodesic, the transverse expansion at any non-degenerate Killing horizon is uniquely determined in terms of its abstract Killing horizon data and the ambient Ricci tensor to infinite order on the horizon. Moreover, when the Ricci tensor fulfills a so-called hierarchical dependence (this will be defined precisely below), the transverse expansion only depends on abstract Killing horizon data. We subsequently apply this result to characterize $\Lambda$-vacuum manifolds near non-degenerate horizons. Finally, we move on to the question of existence and prove that given AKH data there exists an ambient manifold solving the $\Lambda$-vacuum equations to infinite order at the horizon and that such manifold admits a Killing vector which is null and tangent at the horizon.

As we have already recalled in Subsection 5.1.2, a Killing horizon of $\eta$ is an embedded hypersurface $\Phi : \mathcal{H} \hookrightarrow \mathcal{M}$ to which $\eta$ is tangent, null and nowhere zero. Its surface gravity $\widetilde{\kappa}$ is the scalar function on $\mathcal{H}$ satisfying $\nabla_\eta \eta = \widetilde{\kappa}\eta$ on $\Phi(\mathcal{H})$. In Subsection 5.1.2 we recalled the standard definitions of non-degenerate and degenerate horizons[8]. In this section, however, we prefer to say that the Killing horizon is non-degenerate when its surface gravity satisfies $\widetilde{\kappa} \neq 0$ at some point, and that it is degenerate when $\widetilde{\kappa} = 0$ everywhere. With this weaker definition a Killing horizon is either degenerate or non-degenerate. As we already indicated in Subsection 5.1.2, in many relevant cases of interest (e.g. $\Lambda$-vacuum spacetimes) both definitions turn out to be equivalent. In this section we also relax the condition that $\eta$ is nowhere zero at the horizon and allow the possibility of $\eta$ vanishing on a set with empty interior.

A well-known property of Killing horizons is that they are totally geodesic, i.e. $\mathbf{U} = 0$, which follows at once from $\Phi^\star \mathcal{K}[\eta] = 2\alpha \mathbf{U}$. As a consequence, given any $X \in \mathfrak{X}(\mathcal{H})$, equations (2.41) with $n^{(2)} = 0$ and (2.66) give

$$\nabla_{\Phi_\star X}\eta = \nabla_{\Phi_\star X}(\alpha\nu) = \alpha\nabla_{\Phi_\star X}\nu + X(\alpha)\nu = (\alpha(\mathbf{s}-\mathbf{r}) + d\alpha)(X)\nu. \tag{5.177}$$

In view of relation (5.177) it is useful to introduce the one-form

$$\boldsymbol{\tau} := \alpha(\mathbf{s}-\mathbf{r}) + d\alpha, \tag{5.178}$$

which is related to the one-form $\boldsymbol{\varpi}$ defined in (5.8) at the points where $\alpha \neq 0$ by means of $\boldsymbol{\tau} = \alpha\boldsymbol{\varpi}$. Note however that $\boldsymbol{\tau}$ is well-defined and smooth everywhere on $\mathcal{H}$. Let us now discuss some of the basic properties of $\boldsymbol{\tau}$.

**Proposition 5.57.** *Let $\boldsymbol{\tau}$ be as in (5.178). Then,*

*1.* $\mathcal{G}_{(z,V)}\boldsymbol{\tau} = z\boldsymbol{\tau}$, *2.* $\boldsymbol{\tau}(n) = \kappa_\star$, *3.* $\alpha\pounds_n\boldsymbol{\tau} = n(\alpha)\boldsymbol{\tau} - \boldsymbol{\tau}(n)d\alpha$.

8 A non-degenerate horizon was defined as one for which $\widetilde{\kappa}$ nowhere vanishes, and a degenerate horizon as one in which $\widetilde{\kappa} = 0$.

*Proof.* The gauge transformation of $\boldsymbol{\tau}$ follows from $\mathbf{r}' - \mathbf{s}' = \mathbf{r} - \mathbf{s} + d\log|z|$ (see (2.76)) and Lemma 5.51. Using (5.149) and $\mathbf{r}(n) = -\kappa$ the second property is immediate. For the third one we use that $\eta$ is a Killing vector, so $\pounds_\eta \nabla = 0$ and thus $0 = \pounds_\eta \nabla_X \eta - \nabla_X \pounds_\eta \eta - \nabla_{\pounds_\eta X}\eta$ for every $X \in \mathfrak{X}(\mathcal{M})$. When $X$ is tangent to $\mathcal{H}$ the vector $\pounds_\eta X$ on $\mathcal{H}$ is also tangent (because $\eta$ is tangent as well). Then,

$$0 = \pounds_\eta \nabla_X \eta - \nabla_{\pounds_\eta X}\eta = \pounds_{\bar\eta}(\boldsymbol{\tau}(X))\nu + \boldsymbol{\tau}(X)\pounds_\eta \nu - \boldsymbol{\tau}(\pounds_{\bar\eta} X)\nu = \Big(\pounds_{\bar\eta}\boldsymbol{\tau} - n(\alpha)\boldsymbol{\tau}\Big)(X)\nu,$$

where we used $\pounds_\eta \nu = \pounds_{\alpha\nu}\nu = -\nu(\alpha)\nu = -n(\alpha)\nu$. Then $\pounds_{\bar\eta}\boldsymbol{\tau} = n(\alpha)\boldsymbol{\tau}$, so item 3. follows after using $\pounds_{\bar\eta}\boldsymbol{\tau} = \pounds_{\alpha n}\boldsymbol{\tau} = \alpha\pounds_n \boldsymbol{\tau} + \boldsymbol{\tau}(n)d\alpha$. □

In the context of embedded horizons, equations (2.95), (2.97) and (2.98) simplify to

$$\mathcal{R}_{ab} = \overset{\circ}{R}_{(ab)} - 2\pounds_n \mathrm{Y}_{ab} - 2\kappa \mathrm{Y}_{ab} + \overset{\circ}{\nabla}_{(a}(\mathrm{s}_{b)} + 2\mathrm{r}_{b)}) - 2\mathrm{r}_a\mathrm{r}_b + 4\mathrm{r}_{(a}\mathrm{s}_{b)} - \mathrm{s}_a\mathrm{s}_b, \tag{5.179}$$

$$\mathcal{R}_{ab}n^a = \pounds_n(\mathrm{s}_b - \mathrm{r}_b) - \overset{\circ}{\nabla}_b \kappa, \tag{5.180}$$

$$\mathcal{R}_{ab}n^a n^b = 0, \tag{5.181}$$

while, using that the tensors (5.156) all vanish, equations (5.157)-(5.158) become

$$\begin{aligned} \alpha\mathcal{R}_{ab} = &-2\kappa_\star \mathrm{Y}_{ab} - 2\alpha\overset{\circ}{\nabla}_{(a}(\mathrm{s}-\mathrm{r})_{b)} - 4(\mathrm{s}-\mathrm{r})_{(a}\overset{\circ}{\nabla}_{b)}\alpha - 2\alpha(\mathrm{s}-\mathrm{r})_a(\mathrm{s}-\mathrm{r})_b \\ &- 2\overset{\circ}{\nabla}_a\overset{\circ}{\nabla}_b\alpha - \alpha\overset{\circ}{\nabla}_{(a}\mathrm{s}_{b)} + \alpha\mathrm{s}_a\mathrm{s}_b + \alpha\overset{\circ}{R}_{(ab)}, \end{aligned} \tag{5.182}$$

$$\alpha\mathcal{R}_{ab}n^b = -\overset{\circ}{\nabla}_a\kappa_\star. \tag{5.183}$$

Equations (5.179)-(5.183) have been obtained and discussed by M. Manzano in [211] where among other things the following results are obtained. Equation (5.159) implies $n(\kappa_\star) = 0$, so the surface gravity is constant along the null generator of the horizon. Hence, $\pounds_n d\kappa_\star = d(n(\kappa_\star)) = 0$ as well, so $d\kappa_\star$ is also Lie-constant along the null generators. For bifurcate horizons (5.183) implies $d\kappa_\star = 0$ on the bifurcation surface (and hence everywhere), which recovers the well-known fact that $\kappa_\star$ is constant for bifurcate horizons [189]. One can also recover easily the result that when $d\kappa_\star \neq 0$ geodesics terminate in a curvature singularity [288].

Particularizing (5.170) for $m > 1$ to the case $\mathbf{U} = 0$ and $\mathcal{K}[\eta] = 0$ and taking into account the definition (5.178),

$$\begin{aligned} \alpha\mathcal{R}^{(m+1)}_{ab} = &-2(m+1)\kappa_\star \mathrm{Y}^{(m+1)}_{ab} + 4\tau_{(a}\mathrm{r}^{(m+1)}_{b)} + 2\alpha\overset{\circ}{\nabla}_{(a}\mathrm{r}^{(m+1)}_{b)} \\ &- 2\alpha\kappa^{(m+1)}\mathrm{Y}_{ab} + \alpha\mathcal{O}^{(m)}_{ab} + \mathcal{P}^{(m)}_{ab}. \end{aligned} \tag{5.184}$$

In order to construct the abstract notion of "Killing horizon data" we want to find the minimum amount of data on the horizon that allows for the determination of the full transverse expansion. As we will see below, when $\kappa_\star \neq 0$ everywhere, the required information involves the null metric hypersurface data $\{\mathcal{H}, \boldsymbol{\gamma}, \boldsymbol{\ell}, \ell^{(2)}\}$, the one-form $\boldsymbol{\tau}$ and the function $\alpha$. The abstract conditions we need to incorporate into the definition must be such that, once the data is embedded, the corresponding horizon satisfies (i) $\Phi^{\star}(\pounds_{\eta} g) = 0$, (ii) the one-form $\boldsymbol{\tau}$ satisfies item 3. of Proposition 5.57, (iii) the set of zeros of $\alpha$ has empty interior and (iv) that one-form $\alpha^{-1}(\boldsymbol{\tau} - d\alpha)$ extends smoothly to all $\mathcal{H}$. Item (iii) is necessary because Killing vectors that vanish on any hypersurface are necessarily identically zero. In terms of $\eta$ this means that $\alpha$ vanishing on any open subset of $\mathcal{H}$ would only be compatible with $\eta$ being identically vanishing. Item (iv) is necessary because for actual Killing horizons, the following equality holds (cf. (5.178))

$$\alpha^{-1}(\boldsymbol{\tau} - d\alpha) = \mathbf{s} - \mathbf{r},$$

and the right hand side is smooth everywhere on $\mathcal{H}$. Note that the condition that the zeroes of $\alpha$ have empty interior means, in particular, that the extension of $\alpha^{-1}(\boldsymbol{\tau} - d\alpha)$ is necessarily unique. Note also than when $\alpha$ has no zeroes, condition (iv) is automatically satisfied. As we explained at the beginning of this section, condition (i) gives $\mathbf{U} = 0$. This discussion motivates the following definition.

**Definition 5.58.** *We say* $\{\mathcal{H}, \boldsymbol{\gamma}, \boldsymbol{\ell}, \ell^{(2)}, \boldsymbol{\tau}, \alpha\}$ *is* ***abstract Killing horizon data*** *(AKH data) provided that (i)* $\{\mathcal{H}, \boldsymbol{\gamma}, \boldsymbol{\ell}, \ell^{(2)}\}$ *is null metric hypersurface data satisfying* $\mathbf{U} = 0$*, (ii)* $\alpha$ *is a smooth function such that the set* $\{\alpha = 0\}$ *has empty interior and (iii)* $\boldsymbol{\tau}$ *is a one-form such that* $\alpha \pounds_n \boldsymbol{\tau} = n(\alpha)\boldsymbol{\tau} - \boldsymbol{\tau}(n) d\alpha$ *and the one-form* $\alpha^{-1}(\boldsymbol{\tau} - d\alpha)$ *extends smoothly to all* $\mathcal{H}$*.*

In a recent work [216] the authors analyze a more general definition of abstract horizon in which neither the deformation tensor $\mathcal{K}[\eta]$ nor the so-called non-isolation $\mathbf{\Pi}^{\eta}$ tensor need to vanish.

**Remark 5.59.** *It is worth comparing Definition 5.58 with the notions of abstract Killing horizons of order zero (AKH$_0$) and one (AKH$_1$) introduced in [215]. The main difference is that the definitions in [215] involve full hypersurface data (i.e. involve the tensor* $\mathbf{Y}$*) while the definition above makes no reference to* $\mathbf{Y}$*. In fact, we want to use Definition 5.58 in combination with the field equations to construct* $\mathbf{Y}$ *in such a way that the data corresponds to a Killing horizon. This is why we have added the term "data" in the definition of "AKH data".*

Next we extend the notion of gauge transformation to the context of AKH data motivated by Lemma 5.51 and Proposition 5.57.

**Definition 5.60.** *Let* $\mathscr{K} = \{\mathcal{H}, \boldsymbol{\gamma}, \boldsymbol{\ell}, \ell^{(2)}, \boldsymbol{\tau}, \alpha\}$ *be AKH data and* $(z, V)$ *gauge parameters. We define the gauge-transformed data by* $\mathcal{G}_{(z,V)}\mathscr{K} := \{\mathcal{H}, \mathcal{G}_{(z,V)}\boldsymbol{\gamma}, \mathcal{G}_{(z,V)}\boldsymbol{\ell}, \mathcal{G}_{(z,V)}\ell^{(2)}, z\boldsymbol{\tau}, z\alpha\}$*, where* $\{\mathcal{G}_{(z,V)}\boldsymbol{\gamma}, \mathcal{G}_{(z,V)}\boldsymbol{\ell}, \mathcal{G}_{(z,V)}\ell^{(2)}\}$ *are given by* (2.27)-(2.29)*.*

The condition $\mathbf{U} = 0$ of Definition 5.58 only guarantees that the pullback of the deformation tensor vanishes on $\mathcal{H}$. To capture the full information about the deformation tensor we need to restrict ourselves to the embedded case.

**Definition 5.61.** *Let* $\mathscr{K} = \{\mathcal{H}, \boldsymbol{\gamma}, \boldsymbol{\ell}, \ell^{(2)}, \boldsymbol{\tau}, \alpha\}$ *be AKH data and define* $\bar{\eta} := \alpha n$*. We say that* $\mathscr{K}$ *is* $(\Phi, \xi)$*-**embedded** in* $(\mathcal{M}, g)$ *if (i)* $\{\mathcal{H}, \boldsymbol{\gamma}, \boldsymbol{\ell}, \ell^{(2)}\}$ *is* $(\Phi, \xi)$*-embedded in* $(\mathcal{M}, g)$ *as in Def. 2.2 and (ii)* $\nabla_{\Phi_{\star} X} \Phi_{\star} \bar{\eta} = \boldsymbol{\tau}(X) \nu$ *for every* $X \in \mathfrak{X}(\mathcal{H})$*. Moreover, we say that* $\mathscr{K}$ *is* *an $(\Phi,\xi)$-**embedded Killing horizon data** (EKH data) if, additionally, (iii) there exist an extension $\eta$ of $\Phi_\star\bar{\eta}$ such that its deformation tensor $\mathcal{K}[\eta] := \pounds_\eta g$ vanishes to all orders at $\Phi(\mathcal{H})$.*

**Remark 5.62.** *It is easy to check that if $\mathscr{K} = \{\mathcal{H},\boldsymbol{\gamma},\boldsymbol{\ell},\ell^{(2)},\boldsymbol{\tau},\alpha\}$ is $(\Phi,\xi)$-embedded in $(\mathcal{M},g)$, then $\mathcal{G}_{(z,V)}\mathscr{K}$ is $(\Phi,\xi')$-embedded in $(\mathcal{M},g)$ with $\xi' = z(\xi+\Phi_\star V)$. Moreover, since the property of $\mathcal{K}[\eta]$ vanishing to all orders on $\Phi(\mathcal{H})$ is independent of $\xi|_{\Phi(\mathcal{H})}$ and its extension off $\Phi(\mathcal{H})$, it follows that if $\mathscr{K}$ is $(\Phi,\xi)$-EKH data, then $\mathcal{G}_{(z,V)}\mathscr{K}$ is $(\Phi,\xi')$-EKH data.*

**Remark 5.63.** *The definition of AKH data allows us to define a smooth one-form $\mathbb{r} := \mathbf{s} - \alpha^{-1}(\boldsymbol{\tau} - d\alpha)$ and a scalar $\mathbb{k} := \alpha^{-1}(\boldsymbol{\tau}(n) - n(\alpha))$. When the data happens to be embedded, the one-form $\mathbb{r}$ and the function $\mathbb{k}$ agree with $\mathbf{r}$ and $\kappa$, respectively. This is because from equation (5.177) and condition (ii) of Definition 5.61 it follows $\alpha(\mathbb{r}-\mathbf{r}) = 0$, and since the interior of the zeroes of $\alpha$ is empty, then $\mathbb{r} = \mathbf{r}$. Hence in the embedded case we shall not distinguish between $\mathbb{r}$ and $\mathbf{r}$ any more. Observe also that this fact together with (5.149) imply that the surface gravity $\kappa_\star$ of the hypersurface is given by $\kappa_\star = n(\alpha) + \alpha\kappa = \boldsymbol{\tau}(n)$.*

In the following two examples we recall the Schwarzschild-de Sitter and Nariai spacetimes introduced in Subsection 5.1.3, and we compute the induced EKH data at their horizons in both cases.

**Example 5.64.** *As reviewed in Subsection 5.1.3, the d-dimensional Schwarzschild-de Sitter metric is given by (5.20)*

$$g = -\left(1 - \frac{2M}{r^{d-3}} - \frac{\Lambda}{d-1}r^2\right)dv^2 + 2dvdr + r^2\gamma_{\mathbb{S}^{d-2}},$$

*where $\gamma_{\mathbb{S}^{d-2}}$ is the $d-2$ dimensional spherical metric. When $M$ and $\Lambda$ are both positive, and $M$ sufficiently small, $g_{vv}=0$ admits precisely two positive solutions $r_0^+ \geq R := \sqrt{\frac{d-3}{\Lambda}} \geq r_0^-$, and since*

$$\partial_r g_{vv}|_{r=r_0^\pm} = -\frac{2M(d-3)}{(r_0^\pm)^{d-2}} + \frac{2\Lambda}{d-1}r_0^\pm = \Lambda r_0^\pm - \frac{d-3}{r_0^\pm},$$

*it follows that when $r_0^+ > R > r_0^-$ the two null hypersurfaces $\mathcal{H}^\pm := \{r = r_0^\pm\}$ are non-degenerate Killing horizons with Killing $\eta = \partial_v$ ($\mathcal{H}^+$ is the cosmological horizon and $\mathcal{H}^-$ is the event horizon). When $r_0^+ = r_0^- = R$, $\mathcal{H}^+ = \mathcal{H}^-$ is a degenerate Killing horizon, and the metric becomes (5.21).*

*Let us compute the induced hypersurface data of any of the two non-degenerate horizons (we use $r_0$ to denote at once $r_0^+$ and $r_0^-$). Choosing $\xi = \partial_r$ as the rigging (observe $\nabla_\xi\xi = 0$) one obtains*

$$\boldsymbol{\gamma} = r_0^2\gamma_{\mathbb{S}^{d-2}}, \qquad \boldsymbol{\ell} = dv, \qquad \ell^{(2)} = 0, \qquad \mathbf{Y} = \left(-\frac{(d-3)M}{r_0^{d-2}} + \frac{\Lambda r_0}{d-1}\right)dv^2 + r_0\gamma_{\mathbb{S}^{d-2}}.$$

*It follows at once that* $\boldsymbol{\ell}(\eta) = 1$, $\mathbf{s} = \frac{1}{2}d\boldsymbol{\ell}(\eta, \cdot) = 0$, *so by* (5.178) *with* $\alpha = 1$

$$\boldsymbol{\tau} = -\mathbf{r} = -\mathbf{Y}(\eta, \cdot) = \left(\frac{(d-3)M}{r_0^{d-2}} - \frac{\Lambda r_0}{d-1}\right)dv,$$

*and thus* $\kappa_\star = \boldsymbol{\tau}(\eta) = \frac{(d-3)M}{r_0^{d-2}} - \frac{\Lambda r_0}{d-1} = \frac{d-3}{2r_0} - \frac{\Lambda r_0}{2}$. *When* $r_0 \neq R$ *one has* $\kappa_\star \neq 0$. *In this case* $\mathscr{K}_{SdS} := \{\mathcal{H}, \boldsymbol{\gamma}, \boldsymbol{\ell}, \ell^{(2)}, \boldsymbol{\tau}, \alpha = 1\}$ *is EKH in Schwarzschild-de Sitter spacetime with* $\kappa_\star \neq 0$ *everywhere.*

**Example 5.65.** *The Nariai metric [248] is obtained by taking the product manifold of the two-dimensional de Sitter spacetime* $dS_2$ *of constant curvature* $\Lambda$ *and the* $d-2$ *dimensional sphere* $\mathbb{S}^{d-2}$ *of radius* $\sqrt{\frac{d-3}{\Lambda}}$. *As we also reviewed in Subsection 5.1.3, the Nariai metric can be obtained as the near horizon limit of extremal Schwarzschild-de Sitter (see* (5.22)*), and apart from the degenerate Killing field inherited from extremal Schwarzschild-de Sitter, it admits another Killing vector for which* $r = 0$ *is a non-degenerate horizon (such horizon is therefore called a double Killing horizon (see [227, 228]). In coordinates adapted to the non-degenerate Killing, Nariai metric is given by*

$$g_N = r(1 + \Lambda r)dv^2 + 2dvdr + \frac{d-3}{\Lambda}\gamma_{\mathbb{S}^{d-2}},$$

*and the horizon is located at* $\mathcal{H} = \{r = 0\}$ *(note that* $\mathcal{H} \simeq \mathbb{R} \times \mathbb{S}^{d-2}$*). Let us perform a similar analysis as the one of the previous example. In these coordinates, the Killing vector is* $\eta = \partial_v$, *and we choose* $\xi = \partial_r$ *as the rigging vector (observe that again* $\nabla_\xi \xi = 0$*). The induced hypersurface data is*

$$\boldsymbol{\gamma} = R^2 \gamma_{\mathbb{S}^{d-2}}, \qquad \boldsymbol{\ell} = du, \qquad \ell^{(2)} = 0, \qquad \mathbf{Y} = \frac{1}{2}dv^2,$$

*and since* $\boldsymbol{\ell}(\eta) = 1$ *and* $\mathbf{s} = 0$ *it follows from* (5.178) *with* $\alpha = 1$ *that* $\boldsymbol{\tau} = \frac{1}{2}dv$ *and* $\kappa_\star = \boldsymbol{\tau}(n) = \frac{1}{2}$. *Therefore,* $\mathscr{K}_N := \{\mathcal{H}, \boldsymbol{\gamma}, \boldsymbol{\ell}, \ell^{(2)}, \boldsymbol{\tau}, \alpha = 1\}$ *is EKH in Nariai spacetime with* $\kappa_\star \neq 0$ *everywhere.*

### 5.5.1 *Uniqueness results*

In Proposition 5.68 we find necessary conditions for two AKH data to be embeddable in isometric semi-Riemannian manifolds. The idea is to use this information to then define a notion of isometry at the AKH data level. First we extend Notation 5.34 to the AKH case.

**Notation 5.66.** *Let* $\mathscr{K}' = \{\mathcal{H}', \boldsymbol{\gamma}', \boldsymbol{\ell}', \ell^{(2)\prime}, \boldsymbol{\omega}', \alpha'\}$ *be AKH data and* $\chi : \mathcal{H} \longrightarrow \mathcal{H}'$ *a diffeomorphism. We define*

$$\chi^\star \mathscr{K}' = \chi^\star\{\mathcal{H}', \boldsymbol{\gamma}', \boldsymbol{\ell}', \ell^{(2)\prime}, \boldsymbol{\tau}', \alpha'\} := \{\mathcal{H}, \chi^\star\boldsymbol{\gamma}', \chi^\star\boldsymbol{\ell}', \chi^\star\ell^{(2)\prime}, \chi^\star\boldsymbol{\tau}', \chi^\star\alpha'\}. \tag{5.185}$$

**Lemma 5.67.** *Let* $\mathscr{K}' = \{\mathcal{H}', \boldsymbol{\gamma}', \boldsymbol{\ell}', \ell^{(2)\prime}, \boldsymbol{\omega}', \alpha'\}$ *be AKH data and* $\chi : \mathcal{H} \longrightarrow \mathcal{H}'$ *a diffeomorphism. Then* $\chi^\star\mathscr{K}'$ *is AKH data.*

*Proof.* Firstly, from Remark 5.35 it follows that $\{\mathcal{H},\boldsymbol{\gamma},\boldsymbol{\ell},\ell^{(2)}\} := \chi^\star\{\mathcal{H}',\boldsymbol{\gamma}',\boldsymbol{\ell}',\ell^{(2)\prime}\}$ is null metric hypersurface data. Since $\chi^\star\pounds_{\chi_\star n}\boldsymbol{\gamma}' = \pounds_n\boldsymbol{\gamma}$ it also follows $\mathbf{U}=0$. Moreover, since $\chi$ is a diffeomorphism, $\alpha = \chi^\star\alpha'$ is smooth and has empty interior. Finally,

$$0 = \chi^\star\Big(\alpha'\pounds_{\chi_\star n}\boldsymbol{\tau}' - (\chi_\star n)(\alpha')\boldsymbol{\tau}' + \boldsymbol{\tau}'(\chi_\star n)d\alpha'\Big) = \alpha\pounds_n\boldsymbol{\tau} - n(\alpha)\boldsymbol{\tau} + \boldsymbol{\tau}(n)d\alpha$$

and $\chi^\star(\alpha'^{-1}(\boldsymbol{\tau}'-d\alpha')) = \alpha^{-1}(\boldsymbol{\tau}-d\alpha)$ extends smoothly to all $\mathcal{H}$. Hence $\{\mathcal{H},\boldsymbol{\gamma},\boldsymbol{\ell},\ell^{(2)},\boldsymbol{\tau},\alpha\}$ is AKH data. □

**Proposition 5.68.** *Let $\mathscr{K} = \{\mathcal{H},\boldsymbol{\gamma},\boldsymbol{\ell},\ell^{(2)},\boldsymbol{\omega},\alpha\}$ be AKH data $(\Phi,\xi)$-embedded in $(\mathcal{M},g)$ and let $\mathscr{K}' = \{\mathcal{H}',\boldsymbol{\gamma}',\boldsymbol{\ell}',\ell^{(2)\prime},\boldsymbol{\omega}',\alpha'\}$ be AKH data $(\Phi',\xi')$-embedded in $(\mathcal{M}',g')$. Let $\bar{\eta} := \alpha n$ and $\bar{\eta}' := \alpha' n'$. Assume there exists an isometry $\varphi : (\mathcal{M},g) \longrightarrow (\mathcal{M}',g')$ such that $\varphi(\Phi(\mathcal{H})) = \Phi'(\mathcal{H}')$ and $\varphi_\star\Phi_\star\bar{\eta} = \Phi'_\star\bar{\eta}'$. Then, there exist a diffeomorphism $\chi : \mathcal{H} \longrightarrow \mathcal{H}'$ and gauge parameters $(z,V)$ such that*

$$\chi^\star\{\mathcal{H}',\boldsymbol{\gamma}',\boldsymbol{\ell}',\ell^{(2)\prime},\boldsymbol{\tau}',\alpha'\} = \mathcal{G}_{(z,V)}\{\mathcal{H},\boldsymbol{\gamma},\boldsymbol{\ell},\ell^{(2)},\boldsymbol{\tau},\alpha\}. \tag{5.186}$$

*Proof.* First we prove that the vector field $\varphi^\star\xi'$ is everywhere transverse to $\Phi(\mathcal{H})$. Observe that since the vector field $\nu'$ is null and normal to $\Phi'(\mathcal{H}')$ and $\varphi$ is an isometry, the vector $\varphi^\star\nu'$ is also null and normal to $\Phi(\mathcal{H})$, so in particular it is proportional to $\nu$ with non-zero proportionality factor. Then, from $g'(\xi',\nu')=1$ it follows

$$1 = (\varphi^\star g')(\varphi^\star\xi',\varphi^\star\nu') = g(\varphi^\star\xi',\varphi^\star\nu') \qquad \Longrightarrow \qquad g(\varphi^\star\xi,\nu)\neq 0.$$

This proves that $\varphi^\star\xi'$ is everywhere transverse to $\Phi(\mathcal{H})$, so there must exist a function $z\in\mathcal{F}^\star(\mathcal{H})$ and a vector field $V\in\mathfrak{X}(\mathcal{H})$ such that $\varphi^\star\xi' = z(\xi+\Phi_\star V)$. Since $\Phi(\mathcal{H})$ and $\Phi'(\mathcal{H}')$ are diffeomorphic via $\varphi$ and both $\Phi$ and $\Phi'$ are embeddings, there exists a diffeomorphism $\chi$ making the following diagram commutative.

$$\begin{array}{ccc} \mathcal{H} & \xrightarrow{\ \chi\ } & \mathcal{H}' \\ \big\downarrow{\scriptstyle\Phi} & & \big\downarrow{\scriptstyle\Phi'} \\ \mathcal{M} & \xrightarrow{\ \varphi\ } & \mathcal{M}' \end{array}$$

Then,

$$\begin{aligned}
\chi^\star\boldsymbol{\gamma}' &= \chi^\star\Phi'^\star g' = \Phi^\star\varphi^\star g' = \Phi^\star g\\
&= \boldsymbol{\gamma},\\
\chi^\star\boldsymbol{\ell}' &= \chi^\star\Phi'^\star(g'(\xi',\cdot)) = \Phi^\star\varphi^\star(g'(\xi',\cdot)) = \Phi^\star(g(z(\xi+\Phi_\star V),\cdot))\\
&= z(\boldsymbol{\ell}+\boldsymbol{\gamma}(V,\cdot))\\
\chi^\star\ell^{(2)\prime} &= \chi^\star\Phi'^\star(g'(\xi',\xi')) = \Phi^\star\varphi^\star(g'(\xi',\xi')) = \Phi^\star(g(\varphi^\star\xi',\varphi^\star\xi'))\\
&= z^2(\ell^{(2)} + 2\boldsymbol{\ell}(V) + \boldsymbol{\gamma}(V,V)).
\end{aligned}$$

Hence $\chi^\star\{\mathcal{H}',\gamma',\boldsymbol{\ell}',\ell^{(2)}{}'\} = \mathcal{G}_{(z,V)}\{\mathcal{H},\gamma,\boldsymbol{\ell},\ell^{(2)}\}$ (see Def. 2.2), so $\chi^\star\nu' = z^{-1}\nu$ and $\chi^\star\alpha' = z\alpha$ because $\varphi_\star\Phi_\star\bar{\eta} = \Phi'_\star\bar{\eta}'$. Finally, from condition (ii) in Definition 5.61 together with $\varphi_\star\Phi_\star\bar{\eta} = \Phi'_\star\bar{\eta}'$ and the fact that $\varphi$ is an isometry, it follows

$$\varphi^\star\Big(\nabla'_{\Phi'_\star X'}\Phi'_\star\bar{\eta}'\Big) = \nabla_{\Phi_\star\chi^\star X'}\Phi_\star\bar{\eta} \qquad \Longrightarrow \qquad \chi^\star\boldsymbol{\tau}'\otimes\chi^\star\nu' = \boldsymbol{\tau}\otimes\nu \qquad \Longleftrightarrow \qquad \chi^\star\boldsymbol{\tau}' = z\boldsymbol{\tau}.$$

Taking into account Definition 5.60, (5.186) is established. □

This proposition motivates the following natural notion of isometry in the context of AKH data.

**Definition 5.69.** *Let $\mathscr{K} = \{\mathcal{H},\gamma,\boldsymbol{\ell},\ell^{(2)},\boldsymbol{\tau},\alpha\}$ and $\mathscr{K}' = \{\mathcal{H}',\gamma',\boldsymbol{\ell}',\ell^{(2)}{}',\boldsymbol{\tau}',\alpha'\}$ be two AKH data. We say $\mathscr{K}$ and $\mathscr{K}'$ are isometric provided that there exists a diffeomorphism $\chi:\mathcal{H}\longrightarrow\mathcal{H}'$ and gauge parameters $(z,V)$ such that*

$$\chi^\star\{\mathcal{H}',\gamma',\boldsymbol{\ell}',\ell^{(2)}{}',\boldsymbol{\tau}',\alpha'\} = \mathcal{G}_{(z,V)}\{\mathcal{H},\gamma,\boldsymbol{\ell},\ell^{(2)},\boldsymbol{\tau},\alpha\}. \tag{5.187}$$

Lemma 5.67 guarantees that Definition 5.69 is well defined.

Our next aim is to show that given EKH data $\mathscr{K}$ satisfying $\boldsymbol{\tau}(n)\neq 0$ everywhere, the full asymptotic expansion $\{\mathbf{Y}^{(k)}\}_{k\geq 1}$ is uniquely determined in terms of $\mathscr{K}$ and the set $\{R^{(m)}_{\alpha\beta}\}_{m\geq 1}$. As we shall see, to prove the uniqueness part of such statement we need to be able to extend the rigging vector $\xi$ such that the tensor $\pounds^{(m)}_\xi R_{\alpha\beta}$ is geometrical (in the sense of Definition 5.9) for every $m\geq 0$. By identities (5.82), (5.83) and (5.184), the extension of $\xi$ must be such that the tensors $\mathcal{O}^{(m)}$, $\mathcal{O}^{(m)}_a$, $\mathcal{O}^{(m)}_{ab}$ and $\mathcal{P}^{(m)}_{ab}$ are $\mathcal{H}$-geometrical for every $m\geq 0$. In Section 5.2 we have proved that by extending $\xi$ off $\Phi(\mathcal{H})$ by means of $\nabla_\xi\xi = 0$ the tensors $\mathcal{O}^{(m)}$, $\mathcal{O}^{(m)}_a$ and $\mathcal{O}^{(m)}_{ab}$ are $\mathcal{H}$-geometrical. However, this is not sufficient to guarantee that $\mathcal{P}^{(m)}_{ab}$ is $\mathcal{H}$-geometrical, because this tensor also depends on $X_\eta$ and $\pounds^{(i)}_\xi\eta\big|_{\mathcal{H}}$ for $i\geq 2$ (see the comment below equation (5.170)). A natural way to make this dependence disappear is to ensure that $X_\eta = 0$ and $\pounds^{(i)}_\xi\eta \stackrel{\mathcal{H}}{=} 0$ for every $i\geq 2$. Our strategy is as follows. In Lemma 5.70 we show that given AKH data embedded on an ambient manifold with rigging $\xi$, one can always choose the gauge such that $\pounds_\xi\eta$ is proportional to $\xi$ on $\Phi(\mathcal{H})$. With this choice $X_\eta$ automatically vanishes (cf. (5.143)). By combining this result with Lemma 5.54 we will be able to prove that $\pounds^{(i)}_\xi\eta \stackrel{\mathcal{H}}{=} 0$ for every $i\geq 2$ as well.

Particularizing equation (5.161) to the Killing horizon case, namely $\mathcal{K}[\eta]=0$,

$$\pounds_\xi\eta^\mu \stackrel{\mathcal{H}}{=} (\alpha\kappa-\kappa_\star)\xi^\mu - \frac{\alpha}{2}n(\ell^{(2)})\nu^\mu - P^{ab}\Big(2\alpha\mathrm{s}_b + \overset{\circ}{\nabla}_b\alpha\Big)\widehat{e}^\mu_a. \tag{5.188}$$

In the following lemma I quote a result from [235] where we show that there exists a choice of gauge in which $\pounds_\xi\eta \stackrel{\mathcal{H}}{=} (\alpha\kappa-\kappa_\star)\xi$.

**Lemma 5.70.** *Let $\mathscr{K} = \{\mathcal{H},\gamma,\boldsymbol{\ell},\ell^{(2)},\boldsymbol{\tau},\alpha\}$ be AKH data $(\Phi,\xi)$-embedded in $(\mathcal{M},g)$ satisfying $\boldsymbol{\tau}(n)\neq 0$ everywhere on $\mathcal{H}$. Assume there exists an extension $\eta$ of $\bar{\eta} := \alpha n$ off $\Phi(\mathcal{H})$* *satisfying* $\mathcal{K}[\eta](\xi,\cdot)=0$ *on* $\Phi(\mathcal{H})$. *Then there exists a family of gauges satisfying* $\boldsymbol{\ell}=\kappa_\star^{-1}\boldsymbol{\tau}$ *and* $\ell^{(2)}=0$. *Moreover, any element of the family satisfies*

$$\pounds_\xi\eta \stackrel{\mathcal{H}}{=} (\alpha\kappa-\kappa_\star)\xi, \tag{5.189}$$

*and the whole family can be generated from any element by the action of the subgroup of transformations* $\{\mathcal{G}_{(z,0)}\}$ *(i.e. this subgroup acts transitively on the family). Any element of this family will be said to be written in an "*$\eta$*-gauge".*

*Proof.* A vector $V\in\mathfrak{X}(\mathcal{H})$ is defined uniquely in terms of the one-form $\boldsymbol{w}:=\boldsymbol{\gamma}(V,\cdot)$ and the scalar function $f:=\boldsymbol{\ell}(V)$. From (2.6) and (2.10) it follows $P^{ab}w_aw_b=P^{ab}\gamma_{ac}\gamma_{bd}V^cV^d=(\delta^b_c-n^b\ell_c)\gamma_{bd}V^cV^d=\gamma_{cd}V^cV^d$, so in terms of $f$ and $\boldsymbol{w}$ the gauge transformations of $\boldsymbol{\ell}$ and $\ell^{(2)}$ in (2.28)-(2.29) read (as usual we use primes for the gauge transformed quantities)

$$\boldsymbol{\ell}'=z(\boldsymbol{\ell}+\boldsymbol{w}), \tag{5.190}$$

$$\ell^{(2)\prime}=z^2\Big(\ell^{(2)}+2f+P(\boldsymbol{w},\boldsymbol{w})\Big). \tag{5.191}$$

From the transformations of $\boldsymbol{\tau}$ and $\kappa_\star$, namely $\boldsymbol{\tau}'=z\boldsymbol{\tau}$ and $\kappa_\star'=\kappa_\star$, it is straightforward to check that by choosing

$$\boldsymbol{w}:=\kappa_\star^{-1}\boldsymbol{\tau}-\boldsymbol{\ell},\qquad\text{and}\qquad f:=-\ell^{(2)}-\frac{1}{2}\kappa_\star^{-2}P(\boldsymbol{\tau},\boldsymbol{\tau}), \tag{5.192}$$

the gauge-transformed data satisfies (i) $\boldsymbol{\ell}'=\kappa_\star^{-1}\boldsymbol{\tau}'$ and (ii) $\ell^{(2)\prime}=0$. Moreover, by the transformations (5.190)-(5.191) and those of $\boldsymbol{\tau}$ and $\kappa_\star$, it is clear that any additional transformation $\mathcal{G}_{(z',V')}$ will preserve properties (i) and (ii) if and only if $V'=0$. Thus, the whole family is generated by applying $\mathcal{G}_{(z,0)}$ (with $z\in\mathcal{F}^\star(\mathcal{H})$ arbitrary) to any element of the family. To prove that in this class of gauges expression (5.188) simplifies to (5.189) it suffices to show that $P^{ab}(2\alpha s_b+\overset{\circ}{\nabla}_b\alpha)=0$. Writing item 3. of Proposition 5.57 in terms of $\kappa_\star\boldsymbol{\tau}=\boldsymbol{\ell}$ gives

$$\begin{aligned}\alpha\pounds_n\boldsymbol{\tau}=\alpha n(\kappa_\star)\boldsymbol{\ell}+\alpha\kappa_\star\pounds_n\boldsymbol{\ell}=n(\alpha)\kappa_\star\boldsymbol{\ell}-\kappa_\star d\alpha\qquad&\Longrightarrow\\ \Longrightarrow\qquad\kappa_\star\Big(\alpha\pounds_n\boldsymbol{\ell}+d\alpha\Big)=\Big(\kappa_\star n(\alpha)-\alpha n(\kappa_\star)\Big)\boldsymbol{\ell}.&\end{aligned}$$

Using $\pounds_n\boldsymbol{\ell}=2\mathbf{s}$ (cf. (2.22)) it follows that the one-form $2\alpha\mathbf{s}+d\alpha$ is proportional to $\boldsymbol{\ell}$, and since $\ell^{(2)}=0$ then $P^{ab}(2\alpha s_b+\overset{\circ}{\nabla}_b\alpha)=0$ as a consequence of (2.5). □

Observe that the EKH data in Examples 5.64 and 5.65 are written in an $\eta$-gauge. By combining Lemmas 5.54 and 5.70 and Remark 5.62 we arrive at the following.

**Corollary 5.71.** *Let* $\{\mathcal{H},\boldsymbol{\gamma},\boldsymbol{\ell},\ell^{(2)},\boldsymbol{\tau},\alpha\}$ *be* $(\Phi,\xi)$*-EKH in* $(\mathcal{M},g)$. *Then there exists a choice of* $\xi$ *on* $\Phi(\mathcal{H})$ *in which* $\pounds_\xi\eta\stackrel{\mathcal{H}}{=}(\alpha\kappa-\kappa_\star)\xi$. *Such gauge is unique once* $n$ *is fixed. Moreover, extending* $\xi$ *off* $\Phi(\mathcal{H})$ *by* $\nabla_\xi\xi=0$, *then* $\pounds_\xi^{(k)}\eta\stackrel{\mathcal{H}}{=}0$ *for all* $k\geq 2$. *In particular, all the tensors* $\mathcal{P}^{(m)}$ *in* (5.165) *vanish.*

We are ready to show one of the main results of this section, namely that the full transverse expansion of an EKH data satisfying $\kappa_\star\neq 0$ everywhere is uniquely determined in terms of AKH data and the collection $\{R^{(m)}_{\alpha\beta}\}_{m\geq 1}$.

**Theorem 5.72.** *Let* $\mathscr{K} = \{\mathcal{H}, \boldsymbol{\gamma}, \boldsymbol{\ell}, \ell^{(2)}, \boldsymbol{\tau}, \alpha\}$ *be* $(\Phi, \xi)$*-EKH in any* $(\mathcal{M}, g)$*. Assume* $\boldsymbol{\tau}(n) \neq 0$ *everywhere on* $\mathcal{H}$*. Then, when the data is written in an* $\eta$*-gauge and* $\xi$ *is extended geodesically, the full transverse expansion* $\{\mathbf{Y}^{(m)}\}_{m\geq 1}$ *only depends on* $\mathscr{K}$ *and* $\{R^{(m)}_{\alpha\beta}\}_{m\geq 1}$*.*

*Proof.* Let us write the data in any of the $\eta$-gauges of Lemma 5.70, which fixes the vector $\xi$ up to a multiplicative non-vanishing function $\xi \longmapsto z\xi$, and extend $\xi$ off $\Phi(\mathcal{H})$ by means of $\nabla_\xi \xi = 0$. We want to prove that the tensors $\{\mathbf{Y}^{(k)}\}_{k\geq 1}$ only depend on $\mathscr{K}$ and $\{R^{(m)}_{\alpha\beta}\}_{m\geq 1}$, and thus they are insensitive to the particular manifold they are embedded in, provided their tensors $R^{(m)}_{\alpha\beta}$ agree on $\Phi(\mathcal{H})$. With this choice of gauge and extension of $\xi$ the tensors $\mathcal{O}^{(m)}$, $\mathcal{O}^{(m)}_a$ and $\mathcal{O}^{(m)}_{ab}$ only depend on metric data and $\{\mathbf{Y}, ..., \mathbf{Y}^{(m)}\}$ (see Corollary 5.30) and the tensor $\mathcal{P}^{(m)}$ vanishes for every $m \geq 2$ (see Corollary 5.71). These two facts will be used repeatedly throughout the proof. By Remark 5.63 we can identify the one-form $\mathbb{r} := \mathbf{s} - \alpha^{-1}(\boldsymbol{\tau} - d\alpha)$ with $\mathbf{r} := \mathbf{Y}(n, \cdot)$ and the scalar $\boldsymbol{\tau}(n)$ with $\kappa_\star$. Therefore, equation (5.182) can be rewritten as

$$\alpha \mathcal{R}_{ab} = -2\kappa_\star \mathrm{Y}_{ab} + C_{ab}, \tag{5.193}$$

where $C_{ab}$ is a tensor that only depends on AKH data $\{\boldsymbol{\gamma}, \boldsymbol{\ell}, \ell^{(2)}, \boldsymbol{\tau}, \alpha\}$. This proves that when $\kappa_\star \neq 0$ everywhere on $\mathcal{H}$, the full tensor $\mathbf{Y}$ is determined in terms of AKH data and the tensor $\mathcal{R}_{ab}$. Therefore, since in equations (5.82)-(5.83) for $m = 0$ the lower order terms only involve the tensor $\mathbf{Y}$ and metric data, it follows that the scalar $\operatorname{tr}_P \mathbf{Y}^{(2)}$ and the one-form $\mathbf{r}^{(2)}$ only depend on AKH data and the tensor $R_{\alpha\beta}$ on $\mathcal{H}$. Hence, equation (5.184) for $m = 1$ reads

$$\alpha \mathcal{R}^{(2)}_{ab} = -4\kappa_\star \mathrm{Y}^{(2)}_{ab} + C^{(2)}_{ab}, \tag{5.194}$$

where $C^{(2)}_{ab}$ only depends on AKH data and $R_{\alpha\beta}$. When $\kappa_\star \neq 0$ everywhere this shows that the tensor $\mathrm{Y}^{(2)}_{ab}$ is uniquely determined from AKH data and the tensors $R_{\alpha\beta}$ and $\mathcal{R}^{(2)}_{ab}$ on $\mathcal{H}$. Iterating this process by means of equations (5.82), (5.83) and (5.184) one obtains the full transverse expansion $\{\mathbf{Y}^{(k)}\}_{k\geq 1}$, and by Corollaries 5.71 and 5.30 this expansion only depends on AKH data and $\{R^{(m)}_{\alpha\beta}\}_{m\geq 1}$, and not on the particular $(\mathcal{M}, g)$ where $\mathscr{K}$ is embedded. □

This theorem shows that the asymptotic expansion of an EKH data satisfying $\kappa_\star \neq 0$ everywhere **only** depends on the abstract data $\mathscr{K}$ and the tensors $\{R^{(m)}_{\alpha\beta}\}_{m\geq 1}$, and thus it is insensitive to the particular $(\mathcal{M}, g)$ they may be embedded in. The collection $\{R^{(m)}_{\alpha\beta}\}_{m\geq 1}$ can be thought at least in two different ways. One possibility is to provide each $R^{(m)}_{\alpha\beta}$ as prescribed data on the null hypersurface, e.g. by some external matter field. Another option is to provide $R^{(m)}_{\alpha\beta}$ as a functional relation between the abstract data $\mathscr{K}$ and the transverse expansion $\{\mathbf{Y}^{(1)}, ..., \mathbf{Y}^{(m)}\}$. The simplest example of the second viewpoint is a manifold $(\mathcal{M}, g)$ satisfying the $\lambda$-vacuum equations $R^{(m+1)}_{\alpha\beta} = \lambda \mathcal{K}^{(m)}_{\alpha\beta}$. Then, $\mathcal{R} = \lambda \boldsymbol{\gamma}$, $\dot{\mathcal{R}} = \lambda \boldsymbol{\ell}$, $\ddot{\mathcal{R}} = \lambda \ell^{(2)}$ and $\mathcal{R}^{(m+1)} = 2\lambda \mathbf{Y}^{(m)}$, $\dot{\mathcal{R}}^{(m+1)} = \lambda \Phi^\star(\mathcal{K}^{(m)}(\xi, \cdot))$ and $\ddot{\mathcal{R}}^{(m+1)} = \lambda \Phi^\star(\mathcal{K}^{(m)}(\xi, \xi))$ for every $m \geq 1$. Since by Remark 5.22 the tensor $\mathcal{K}^{(m)}(\xi, \cdot)$ on $\Phi(\mathcal{H})$ depends on $\{\mathbf{Y}^{(1)}, ..., \mathbf{Y}^{(m-1)}\}$, one concludes that the tensor $R^{(m+1)}_{\alpha\beta}$ on $\Phi(\mathcal{H})$ depends at most on AKH data and $\{\mathbf{Y}, ..., \mathbf{Y}^{(m)}\}$. In general, when the functional relations are such that each $R^{(m)}_{\alpha\beta}$ depends on low enough transverse derivatives of the metric, i.e. such that the LHS in equations (5.82)-(5.84) depend on derivatives that we already have under control, the proof of The-orem 5.72 shows that the transverse expansion only depends on the abstract data. Let us make this property precise.

**Definition 5.73.** *Let $\mathscr{K} = \{\mathcal{H}, \boldsymbol{\gamma}, \boldsymbol{\ell}, \ell^{(2)}, \boldsymbol{\tau}, \alpha\}$ be AKH data $(\Phi, \xi)$-embedded in $(\mathcal{M}, g)$, and extend $\xi$ off $\Phi(\mathcal{H})$ by $\nabla_\xi \xi = 0$. We say that the Ricci tensor of $g$ is **hierarchical** on $\Phi(\mathcal{H})$ provided that $\mathcal{R}^{(1)}$ only depends on $\mathscr{K}$ and, for every $m \geq 1$,*

*(i) $\mathcal{R}^{(m+1)}$ depends (at most) on $\mathscr{K}$, $\{\mathbf{Y}, ..., \mathbf{Y}^{(m)}\}$, $\operatorname{tr}_P \mathbf{Y}^{(m+1)}$ and $\mathbf{r}^{(m+1)}$.*

*(ii) $\dot{\mathcal{R}}^{(m)}$ and $\ddot{\mathcal{R}}^{(m)}$ depend (at most) on $\mathscr{K}$ and $\{\mathbf{Y}, ..., \mathbf{Y}^{(m)}\}$.*

*When a Ricci tensor is hierarchical on $\Phi(\mathcal{H})$ we shall refer to its particular dependence stated in (i) and (ii) by its "hierarchical dependence".*

Recall that when $\xi$ is extended geodesically the tensor $R^{(m)}_{\alpha\beta}$ is geometrical for every $m \geq 1$ (cf. Proposition 5.18). As noted above, the canonical example is the $\Lambda$-vacuum equations, since in this case the tensors $\mathcal{R}^{(m+1)}$, $\dot{\mathcal{R}}^{(m)}$ and $\ddot{\mathcal{R}}^{(m)}$ only depend on AKH data and $\{\mathbf{Y}^{(1)}, ..., \mathbf{Y}^{(m)}\}$. In fact, $\mathcal{R}^{(1)}_{ab} = \lambda\gamma_{ab}$, $\dot{\mathcal{R}}_a = \lambda\ell_a$, $\ddot{\mathcal{R}} = \lambda\ell^{(2)}$, $\dot{\mathcal{R}}^{(2)}_a = \frac{\lambda}{2}\overset{\circ}{\nabla}_a\ell^{(2)}$, $\ddot{\mathcal{R}}^{(2)} = 0$ and $\mathcal{R}^{(m+1)} = 2\lambda\mathbf{Y}^{(m)}$, $\dot{\mathcal{R}}^{(m)}_a = 0$ and $\ddot{\mathcal{R}}^{(m)} = 0$ for every $m \geq 2$ (see Corollary 5.12). An immediate consequence of the proof of Theorem 5.72 and Definition 5.73 is the following.

**Theorem 5.74.** *Let $\mathscr{K} = \{\mathcal{H}, \boldsymbol{\gamma}, \boldsymbol{\ell}, \ell^{(2)}, \boldsymbol{\tau}, \alpha\}$ be $(\Phi, \xi)$-EKH in any $(\mathcal{M}, g)$. Assume $\boldsymbol{\tau}(n) \neq 0$ everywhere on $\mathcal{H}$ and that the field equations satisfied by $g$ are such that its Ricci tensor is hierarchical on $\Phi(\mathcal{H})$. Then, when the data is written in an $\eta$-gauge and $\xi$ is extended geodesically, the full transverse expansion $\{\mathbf{Y}^{(m)}\}_{m\geq 1}$ only depends on $\mathscr{K}$.*

This result extends Theorem 5.3 in all the directions mentioned in the last paragraph of Subsection 5.1.2. Indeed, Theorem 5.74 is more general firstly because we are allowing zeroes of $\eta$ on $\mathcal{H}$ (while Killing horizons by definition only include points where $\eta$ is non-zero), secondly because our hierarchical property includes more field equations besides vacuum with vanishing $\Lambda$, and finally because we have extended the result to arbitrary signature (compatible with the presence of null hypersurfaces). We recover the result in [195] simply by imposing that $R_{\alpha\beta}$ is zero to all orders on the horizon. As an interesting corollary of Theorem 5.74 we extend the geometric uniqueness of the transverse expansion to the case of asymptotic $\Lambda$-vacuum spacetimes.

**Corollary 5.75.** *Let $\mathscr{K} = \{\mathcal{H}, \boldsymbol{\gamma}, \boldsymbol{\ell}, \ell^{(2)}, \boldsymbol{\tau}, \alpha\}$ be $(\Phi, \xi)$-EKH in any $(\mathcal{M}, g)$ satisfying the $\Lambda$-vacuum equations to infinite order on $\mathcal{H}$. Assume $\boldsymbol{\tau}(n) \neq 0$ at least at one point $p \in \mathcal{H}$ and that $\mathscr{K}$ is written in an $\eta$-gauge and with $\xi$ extended geodesically off $\Phi(\mathcal{H})$. Then the full transverse expansion $\{\mathbf{Y}^{(m)}\}_{m\geq 1}$ is uniquely determined in terms of $\mathscr{K}$.*

*Proof.* Since $(\mathcal{M}, g)$ satisfies the $\Lambda$-vacuum equations to infinite order on $\mathcal{H}$ it follows $\mathcal{R}_{ab}n^b = 0$, and thus from (5.183) one has $d\kappa_\star = 0$, so $\kappa_\star = \boldsymbol{\tau}(n)$ is constant on $\mathcal{H}$. Since $\kappa_\star \neq 0$ at least at one point and $\mathcal{H}$ is assumed to be connected, we conclude $\kappa_\star \neq 0$ everywhere on $\mathcal{H}$, and hence Theorem 5.74 applies. □

Let us perform a detailed comparison between our AKH data and the "non-degenerate Killing horizon data" reviewed in Subsection 5.1.2. In order to compare both objects at the same footing we restrict our AKH data to $\alpha \neq 0$, $\boldsymbol{\gamma}$ semi-positive definite, $\boldsymbol{\tau}(n) \neq 0$ at some point and vacuum, which in particular implies $\boldsymbol{\tau}(n)$ must be a nonzero constant (see the proof of Corollary 5.75). Let us see the equivalence between AKH data and "non-degenerate Killing horizon data". In one direction, given $\{\mathcal{H}, \boldsymbol{\gamma}, \boldsymbol{\ell}, \ell^{(2)}, \boldsymbol{\tau}, \alpha\}$ the vector $\eta$ and the metric $\boldsymbol{\sigma}$ can be defined as follows:

$$\eta := \alpha n, \qquad \boldsymbol{\sigma} := \boldsymbol{\gamma} + \alpha^{-2} \boldsymbol{\tau} \otimes \boldsymbol{\tau}.$$

Clearly $\eta$ is nowhere vanishing and $\boldsymbol{\sigma}(\eta, \eta) = (\boldsymbol{\tau}(n))^2$ is a non-vanishing constant, so $\boldsymbol{\sigma}$ is a gauge-invariant Riemannian metric on $\mathcal{H}$ (see Lemma 5.51 and Proposition 5.57). Moreover, since $\pounds_\eta \boldsymbol{\tau} = n(\alpha)\boldsymbol{\tau}$, it follows

$$\pounds_\eta \boldsymbol{\sigma} = 2\alpha \mathbf{U} + \pounds_\eta(\alpha^{-2}) \boldsymbol{\tau} \otimes \boldsymbol{\tau} + 2\alpha^{-2} \pounds_\eta \boldsymbol{\tau} \otimes_s \boldsymbol{\tau} = \Big( - 2\alpha^{-2} n(\alpha) + 2\alpha^{-2} n(\alpha) \Big) \boldsymbol{\tau} \otimes \boldsymbol{\tau} = 0.$$

Conversely, given a "non-degenerate Killing horizon data" $(\mathcal{H}, \boldsymbol{\sigma}, \eta)$ as in [195] we define

$$\alpha := 1, \qquad \boldsymbol{\tau} := \frac{1}{\sqrt{\boldsymbol{\sigma}(\eta,\eta)}} \boldsymbol{\sigma}(\eta, \cdot), \qquad \boldsymbol{\gamma} := \boldsymbol{\sigma} - \frac{1}{\boldsymbol{\sigma}(\eta,\eta)} \boldsymbol{\sigma}(\eta, \cdot) \otimes \boldsymbol{\sigma}(\eta, \cdot)$$

$$\ell^{(2)} := 0, \qquad \boldsymbol{\ell} := \frac{1}{\boldsymbol{\sigma}(\eta,\eta)} \boldsymbol{\sigma}(\eta, \cdot).$$

Since $\boldsymbol{\ell}(\eta) = 1$ and $\boldsymbol{\gamma}(\eta, \cdot) = 0$ it follows at once that $\{\boldsymbol{\gamma}, \boldsymbol{\ell}, \ell^{(2)}\}$ is null metric hypersurface data and $n = \eta$. Moreover, from $\pounds_\eta \boldsymbol{\sigma} = 0$ one gets $\mathbf{U} = \frac{1}{2} \pounds_n \boldsymbol{\gamma} = 0$ and $\alpha \pounds_n \boldsymbol{\tau} = 0 = n(\alpha)\boldsymbol{\tau} - \boldsymbol{\tau}(n) d\alpha$. In addition $\boldsymbol{\tau}(n)$ is a non-zero constant and the one-form $\alpha^{-1}(\boldsymbol{\tau} - d\alpha)$ extends smoothly to all $\mathcal{H}$ (because $\alpha = 1$). Hence $\{\mathcal{H}, \boldsymbol{\gamma}, \boldsymbol{\ell}, \ell^{(2)}, \boldsymbol{\tau}, \alpha\}$ fulfills the conditions of Definition 5.58 with $\alpha \neq 0$ and $\boldsymbol{\tau}(n) = \textit{const.} \neq 0$.

Another key property of the approach followed in [195] is that the choice of the rigging $\xi$ is such that $\pounds_\xi \eta = 0$ (see the properties of the vector $\partial_t$ mentioned right after Theorem 5.3). This is possible only when $\alpha$ is nowhere vanishing. Indeed, when $\alpha \neq 0$ one can exploit the freedom in Lemma 5.70 to set $\alpha = 1$, and thus $\kappa_\star = \kappa$ (cf. (5.149)) and $\pounds_\xi \eta \stackrel{\mathcal{H}}{=} 0$ (see (5.189)). Lemma 5.56 then shows that $\pounds_\xi \eta = 0$ whenever $\eta$ is a Killing. However, if $\bar{\eta}$ admits zeroes on $\mathcal{H}$ this choice of gauge is not possible any more because the properties $\alpha \neq 0$ or $\alpha = 0$ at a point are gauge invariant, so if $\alpha = 0$ in some gauge it is impossible to make $\alpha = 1$ by a gauge transformation.

In Proposition 5.68 we proved that two AKH data $\mathscr{K}$ and $\mathscr{K}'$ embedded in isometric manifolds $(\mathcal{M}, g)$ and $(\mathcal{M}', g')$ are necessarily isometric (in the sense of Definition 5.69). Our next aim is to prove a kind of converse, namely that isometric EKH data $\mathscr{K}$ and $\mathscr{K}'$, both satisfying $\boldsymbol{\tau}(n) \neq 0$, imply that the respective manifolds they are embedded in are isometric to infinite order. The proof involves two steps. Firstly, we need to construct and map to each other suitable neighbourhoods of $\Phi(\mathcal{H})$ and $\Phi'(\mathcal{H}')$, and secondly we prove that within an appropriate gauge and extension of $\xi$, the two asymptotic expansions agree. The first task was accomplished in Proposition 5.36, and the second one is a consequence of Proposition 5.37, as we show next.

**Theorem 5.76.** *Let* $\mathscr{K} = \{\mathcal{H}, \boldsymbol{\gamma}, \boldsymbol{\ell}, \ell^{(2)}, \boldsymbol{\tau}, \alpha\}$ *and* $\mathscr{K}' = \{\mathcal{H}', \boldsymbol{\gamma}', \boldsymbol{\ell}', \ell^{(2)\prime}, \boldsymbol{\tau}', \alpha'\}$ *be two isometric EKH in respective ambient manifolds* $(\mathcal{M}, g)$ *and* $(\mathcal{M}', g')$ *and satisfying* $\boldsymbol{\tau}(n) \neq 0$ *everywhere on* $\mathcal{H}$*. Assume the field equations on* $(\mathcal{M}, g)$ *and* $(\mathcal{M}', g')$ *are such that their Ricci tensors satisfy the same hierarchical dependence on* $\Phi(\mathcal{H})$ *and* $\Phi'(\mathcal{H}')$*, respectively. Let* $\mathscr{K}'$ *be written in an* $\eta$*-gauge and* $\mathscr{K}$ *in the gauge in which* $\mathscr{K} = \chi^\star \mathscr{K}'$ *and extend the riggings geodesically. Then, there exist neighbourhoods* $\mathcal{U} \subset \mathcal{M}$ *and* $\mathcal{U}' \subset \mathcal{M}'$ *of* $\Phi(\mathcal{H})$ *and* $\Phi'(\mathcal{H}')$ *and a diffeomorphism* $\Psi : \mathcal{U} \longrightarrow \mathcal{U}'$ *such that*

$$\Psi^\star \pounds^{(i)}_{\xi'} g' \stackrel{\mathcal{H}}{=} \pounds^{(i)}_{\xi} g \tag{5.195}$$

*for every* $i \in \mathbb{N} \cup \{0\}$*.*

*Proof.* Since $\boldsymbol{\gamma} = \chi^\star \boldsymbol{\gamma}'$, $\boldsymbol{\ell} = \chi^\star \boldsymbol{\ell}'$ and $\ell^{(2)} = \chi^\star \ell^{(2)\prime}$, by Proposition 5.37 it suffices to prove $\chi^\star \mathbf{Y}^{(i)\prime} = \mathbf{Y}^{(i)}$ for every $i \geq 1$. Observe that since $\mathscr{K}'$ is written in an $\eta$-gauge and $\mathscr{K} = \chi^\star \mathscr{K}'$, then $\boldsymbol{\ell} = \chi^\star \boldsymbol{\ell}' = \chi^\star(\kappa_\star'^{-1} \boldsymbol{\tau}') = \kappa_\star^{-1} \boldsymbol{\tau}$ and $\ell^{(2)} = \chi^\star \ell^{(2)\prime} = 0$, so $\mathscr{K}$ is also written in an $\eta$-gauge. Equation (5.182) and its primed version read

$$-2\kappa_\star \mathrm{Y}_{ab} + C_{ab} = \alpha \mathcal{R}_{ab}, \qquad -2\kappa_\star' \mathrm{Y}'_{ab} + C_{ab} = \alpha' \mathcal{R}'_{ab},$$

where $C_{ab}$ and $C'_{ab}$ only depend on AKH data (see Corollaries 5.71 and 5.30) and thus agree. Since the dependence of $\mathcal{R}_{ab}$ on $\mathscr{K}$ is the same as the dependence of $\mathcal{R}'_{ab}$ on $\mathscr{K}'$, taking the pullback $\chi^\star$ of the second equation and subtracting the first one gives $\chi^\star \mathbf{Y}' = \mathbf{Y}$. A similar argument applied to (5.82)-(5.83) for $m = 0$ proves $\chi^\star(\mathrm{tr}_{P'} \mathbf{Y}^{(2)\prime}) = \mathrm{tr}_P \mathbf{Y}^{(2)}$ and $\chi^\star \mathbf{r}^{(2)\prime} = \mathbf{r}^{(2)}$. By iterating this process one shows that $\chi^\star \mathbf{Y}^{(i)\prime} = \mathbf{Y}^{(i)}$ for every $i \geq 1$. □

An example of two manifolds satisfying the same hierarchical dependence is two spacetimes $(\mathcal{M}, g)$ and $(\mathcal{M}', g')$ such that $R_{\alpha\beta} = \lambda g_{\alpha\beta}$ and $R'_{\alpha\beta} = \lambda g'_{\alpha\beta}$. This leads to the following immediate corollary.

**Corollary 5.77.** *Let* $\mathscr{K} = \{\mathcal{H}, \boldsymbol{\gamma}, \boldsymbol{\ell}, \ell^{(2)}, \boldsymbol{\tau}, \alpha\}$ *and* $\mathscr{K}' = \{\mathcal{H}', \boldsymbol{\gamma}', \boldsymbol{\ell}', \ell^{(2)\prime}, \boldsymbol{\tau}', \alpha'\}$ *be two isometric EKH in respective* $\Lambda$*-vacuum ambient manifolds* $(\mathcal{M}, g)$ *and* $(\mathcal{M}', g')$ *with the same cosmological constant. Assume* $\boldsymbol{\tau}(n) \neq 0$ *everywhere on* $\mathcal{H}$*. Let* $\mathscr{K}'$ *be written in an* $\eta$*-gauge and* $\mathscr{K}$ *in the gauge in which* $\mathscr{K} = \chi^\star \mathscr{K}'$ *and extend the riggings geodesically. Then, there exist neighbourhoods* $\mathcal{U} \subset \mathcal{M}$ *and* $\mathcal{U}' \subset \mathcal{M}'$ *of* $\Phi(\mathcal{H})$ *and* $\Phi'(\mathcal{H}')$ *and a diffeomorphism* $\Psi : \mathcal{U} \longrightarrow \mathcal{U}'$ *such that*

$$\Psi^\star \pounds^{(i)}_{\xi'} g' \stackrel{\mathcal{H}}{=} \pounds^{(i)}_{\xi} g \tag{5.196}$$

*for every* $i \in \mathbb{N} \cup \{0\}$*.*

A consequence of this corollary is that every analytic $\Lambda$-vacuum manifold admitting a non-degenerate Killing horizon (possibly with bifurcation surfaces) is characterized near the horizon by its AKH data. As we shall see in Subsection 5.5.2, every AKH data also gives rise to a spacetime which is $\Lambda$-vacuum to infinite order, irrespective of any assumption of analyticity neither on the data nor on the constructed spacetime.

We conclude this subsection by proving a uniqueness result for non-extremal Schwarzschild-de Sitter spacetime that complements the main result of [188] (Theorem 5.4), where the extremal case was considered. In Theorem 5.80 we improve our previous result [235, Prop. 6.21] by removing a hypothesis of analiticity, since this property turns out to follow from the assumption that the one-form $\boldsymbol{\tau}$ is integrable (by Theorem 5.79 below). We begin by finding sufficient conditions for $\boldsymbol{\tau}$ to be integrable.

**Proposition 5.78.** *Let $(\mathcal{M}, g)$ be a $d \geq 4$-dimensional spacetime admitting a non-degenerate Killing horizon $\Phi : \mathcal{H} \hookrightarrow \mathcal{M}$ with spherical cross-sections. Let $\eta$ be the corresponding Killing vector and let $\boldsymbol{\tau}$ be defined by* (5.178). *Assume that $\kappa_\star = \boldsymbol{\tau}(n)$ is a nonzero constant and that $\boldsymbol{\tau} \wedge d\boldsymbol{\tau} = 0$. Then, $\boldsymbol{\tau}$ is exact.*

*Proof.* Since on the horizon the Killing field is nowhere vanishing we can define $n := \eta$ and thus $\alpha = 1$. From item 3. of Proposition 5.57 it follows $\pounds_n \boldsymbol{\tau} = 0$. Let us pick any section $\mathcal{S} \subset \mathcal{H}$ and solve the differential equation $n(u) = 1$ with initial condition $u|_{\mathcal{S}} = 0$. This gives a global function $u \in \mathcal{F}(\mathcal{H})$ that defines a foliation $\{\mathcal{S}_u\}_{u \in \mathbb{R}}$. Let us decompose the cotangent space by $T^\star\mathcal{H} = \operatorname{span}\{du\} \oplus \langle n \rangle^\perp$, where as usual $\langle n \rangle^\perp$ is the set of covectors that annihilate $n$. Therefore the one-form $\boldsymbol{\tau}$ can be uniquely decomposed as $\boldsymbol{\tau} = A du + \boldsymbol{b}$ with $A \in \mathcal{F}(\mathcal{H})$ and $\boldsymbol{b} \in T^\star\mathcal{H}$ satisfying $\boldsymbol{b}(n) = 0$. Condition $\boldsymbol{\tau}(n) = \kappa_\star$ gives $A = \kappa_\star$, and condition $\boldsymbol{\tau} \wedge d\boldsymbol{\tau} = 0$ gives

$$(\kappa_\star du + \boldsymbol{b}) \wedge d\boldsymbol{b} = 0 \qquad \Longrightarrow \qquad \kappa_\star du \wedge d\boldsymbol{b} + \boldsymbol{b} \wedge d\boldsymbol{b} = 0. \tag{5.197}$$

Since $\pounds_n \boldsymbol{\tau} = \pounds_n(\kappa_\star du + \boldsymbol{b}) = \pounds_n \boldsymbol{b} = 0$ then from the Cartan identity one gets $d\boldsymbol{b}(n, \cdot) = \pounds_n \boldsymbol{b} - d(\boldsymbol{b}(n)) = 0$. As a consequence, the contraction of (5.197) with $n$ yields $\kappa_\star d\boldsymbol{b} = 0$. Since $\kappa_\star$ is a nonzero constant, the one-form $\boldsymbol{b}$ is closed, and hence exact because $\mathbb{R} \times \mathbb{S}^{d-2}$ is simply connected for $d \geq 4$. This proves that $\boldsymbol{\tau} = \kappa_\star du + \boldsymbol{b}$ is also exact. □

Whenever the spacetime of Proposition 5.78 is $\lambda$-vacuum and the horizon $\mathcal{H}$ includes a bifurcation surface $\mathcal{S}$, from the fact that $d\boldsymbol{\tau}|_{\mathcal{S}} = 0$ one has (cf. (5.178)) $0 \stackrel{\mathcal{S}}{=} d\alpha \wedge (\mathbf{r} - \mathbf{s})$, and since $d\alpha|_{\mathcal{S}} \neq 0$ (as otherwise $\kappa_\star = 0$, see (5.149)) we conclude that $\mathbf{r} - \mathbf{s} \stackrel{\mathcal{S}}{=} 0$. Recalling the results of Section 2.4 (see Remark 2.38) and taking into account $\Phi^\star \mathcal{K}[\eta] = 2\alpha \mathbf{U} = 0$ it follows that the torsion one-form $\daleth_{n\theta}$ vanishes at $\mathcal{S}$. Finally, by the main theorem in [72] the spacetime $(\mathcal{M}, g)$ is necessarily static, and consequently also analytic [66]. This proves the following.

**Theorem 5.79.** *Let $(\mathcal{M}, g)$ be a $\lambda$-vacuum $d \geq 4$-dimensional spacetime admitting a non-degenerate Killing horizon $\Phi : \mathcal{H} \hookrightarrow \mathcal{M}$ with spherical cross-sections that includes the bifurcation surface. Assume that the one-form $\boldsymbol{\tau}$ defined in* (5.178) *satisfies $\boldsymbol{\tau} \wedge d\boldsymbol{\tau} = 0$. Then, $(\mathcal{M}, g)$ is static and analytic.*

We are now ready to prove the following uniqueness result.

**Theorem 5.80.** *Let $(\mathcal{M}, g)$ be a $d \geq 4$-dimensional smooth spacetime satisfying the vacuum Einstein equations with cosmological constant $\Lambda > 0$ and admitting a non-degenerate Killing horizon $\Phi : \mathcal{H} \hookrightarrow \mathcal{M}$ with spherical cross-sections that includes the bifurcation surface. Let*

*$\eta$ be the corresponding Killing vector with surface gravity $\kappa_\star \neq 0$ (necessarily constant). Assume $\Phi^\star g = r_0^2 \gamma_{\mathbb{S}^{d-2}}$ for some $r_0^2 > 0$ and $\boldsymbol{\tau} \wedge d\boldsymbol{\tau} = 0$, where $\boldsymbol{\tau}$ is defined in* (5.178)*. Let $M := \frac{1}{2} r_0^{d-3} - \frac{\Lambda}{2(d-1)} r_0^{d-1}$.*

1. *If $r_0^2 > \frac{d-3}{\Lambda}$, $(\mathcal{M}, g)$ is isometric to non-extremal Schwarzschild-de Sitter spacetime with mass $M$ in a neighbourhood of its* ***cosmological horizon****.*

2. *If $r_0^2 < \frac{d-3}{\Lambda}$, $(\mathcal{M}, g)$ is isometric to non-extremal Schwarzschild-de Sitter spacetime with mass $M$ in a neighbourhood of its* ***event horizon****.*

3. *If $r_0^2 = \frac{d-3}{\Lambda}$, $(\mathcal{M}, g)$ is isometric to* ***Nariai*** *spacetime in a neighbourhood of its horizon.*

*Proof.* We denote all the objects referring to Schwarzschild-de Sitter described in Example 5.64 with the label "SdS", and those referring to Nariai (Example 5.65) with the label "N". We start by proving items 1. and 2. Let us scale the Killing vector $\eta$ by the constant $(\kappa_\star)_{SdS}/\kappa_\star$, where $(\kappa_\star)_{SdS} := \frac{d-3}{2r_0} - \frac{\Lambda r_0}{2} \neq 0$. Then, the surface gravity of (the re-scaled) $\eta$ is precisely $(\kappa_\star)_{SdS}$. Since on the horizon the Killing is nowhere vanishing we can define $n := \eta$ and thus $\alpha = 1$ (this removes the gauge function $z$). Following the notation of the proof of Proposition 5.78 we know that $\boldsymbol{\tau}$ is of the form $\boldsymbol{\tau} = \kappa_\star d(u + \kappa_\star^{-1} B)$, where $B$ is such that $\boldsymbol{b} = dB$. Let us define $u' := u + \kappa_\star^{-1} B$ (hence $\boldsymbol{\tau} = \kappa_\star du'$). Since $\boldsymbol{b}(n) = 0$ then $n(B) = 0$, so $n(u') = n(u) = 1$. Consider the new foliation defined by $u'$, $\{\mathcal{S}_{u'}\}_{u' \in \mathbb{R}}$. The remaining gauge freedom can be fixed by requiring $\xi$ to be null and orthogonal to this foliation, which in terms of the abstract data implies $\ell^{(2)} = 0$ and $\boldsymbol{\ell} = du'$. Then, the tuple

$$\mathscr{K} := \{\mathcal{H}, \boldsymbol{\gamma} = r_0^2 \gamma_{\mathbb{S}^{d-2}}, \boldsymbol{\ell} = du', \ell^{(2)} = 0, \boldsymbol{\tau} = \kappa_\star du', \alpha = 1\}$$

fulfills all the conditions of Definition 5.58, so $\mathscr{K}$ is AKH data satisfying $\boldsymbol{\tau}(n) = \kappa_\star \neq 0$ everywhere on $\mathcal{H}$. Moreover, $\mathscr{K}$ is written in an $\eta$-gauge. The last step is to construct the diffeomorphism $\chi : \mathcal{H} \longrightarrow \mathcal{H}_{SdS}$. Choose any isometry $\phi$ that maps the section $\{u' = 0\}$ of $\mathcal{H}$ with the section $\{v = 0\}$ of $\mathcal{H}_{SdS}$. Consider any point $p \in \mathcal{H}$ and let $\sigma(u')$ be the integral curve of $n$ through $p$. This curve intersects $\mathcal{S}_{u'=0}$ at a single point $q$. Now consider the integral curve $\sigma_{SdS}(v)$ of $n_{SdS}$ through $\phi(q)$. We define $\chi(p) := \sigma_{SdS}(u'(p))$. That $\chi$ is a diffeomorphism follows from the product topology of $\mathcal{H}$ and $\mathcal{H}_{SdS}$ and because $n$ and $n_{SdS}$ are both smooth and globally defined. Since $\phi$ is an isometry and the integral curves of $n$ and $n_{SdS}$ are identified it follows $\chi^\star \gamma_{SdS} = \boldsymbol{\gamma}$. With the definition $M := \frac{1}{2} r_0^{d-3} - \frac{\Lambda}{2(d-1)} r_0^{d-1}$ it is clear that $\mathscr{K}$ is isometric (as in Definition 5.69) either to $\mathscr{K}^+_{SdS}$ or $\mathscr{K}^-_{SdS}$. In order to distinguish the two cases recall that the equation $1 - \frac{2M}{{r_0}^{d-3}} - \frac{\Lambda}{d-1} {r_0}^2 = 0$ admits exactly two solutions ${r_0}^+ > \sqrt{\frac{d-3}{\Lambda}} > {r_0}^-$. Therefore, when $r_0^2 > \frac{d-3}{\Lambda}$ then $r_0 = r_0^+$ (and we are in the cosmological horizon case) and when $r_0^2 < \frac{d-3}{\Lambda}$ then $r_0 = r_0^-$ (which is the event horizon case). Items 1. and 2. follow after using Corollary 5.77 and the fact that $(\mathcal{M}, g)$ is analytic (Theorem 5.79).

The proof of item 3. is analogous. Following the same steps as before one finds that

$$\mathscr{K} := \left\{\mathcal{H}, \boldsymbol{\gamma} = \frac{d-3}{\Lambda} \gamma_{\mathbb{S}^{d-2}}, \boldsymbol{\ell} = du', \ell^{(2)} = 0, \boldsymbol{\tau} = \kappa_\star du', \alpha = 1\right\}$$

is AKH data satisfying $\boldsymbol{\tau}(n) = \kappa_\star \neq 0$ everywhere on $\mathcal{H}$, and that it is also written in an $\eta$-gauge. The construction of the diffeomorphism $\chi : \mathcal{H} \longrightarrow \mathcal{H}_N$ is also analogous, and then $\mathscr{K}$ is isometric to $\mathscr{K}_N$ in the sense of Def. 5.69. Finally, applying Corollary 5.77 and Theorem 5.79, item 3. also follows. □

This theorem is interesting because it provides a complete characterization of non-extremal Schwarzschild-de Sitter and Nariai spacetimes solely from properties of their non-degenerate horizons. In this respect, it contrasts with the classical uniqueness theorems (see the review of Subsection 5.1.2), which rely on global assumptions about the spacetime. Analogous results can be established for $\lambda = 0$ and $\lambda < 0$. From the AKH perspective, it is worth noting that the limit $r_0^2 \to \frac{d-3}{\lambda}$ yields to Nariai spacetime rather than to the extremal Schwarzschild-de Sitter solution.

Remarkably, our result does not require the spacetime to be neither analytic nor static a priori. This is one of the main differences with respect to the extremal case considered in [188] (see Theorem 5.4), where the spacetime is assumed to be analytic in order for the series to converge. It would be interesting to study whether the analyticity condition can be dropped in the extremal case, and also to see if a similar result holds for Kerr (both in the non-extremal and the extremal cases). Another difference with the extremal case reviewed in Subsection 5.1.3 is the way the transverse expansion is constrained order by order, because for non-degenerate horizons the equations that allow us to obtain recursively the asymptotic expansion are all algebraic, while in the degenerate case, once $\mathbf{r}^{(m+1)}$ has been replaced in (5.184) using (5.83), the leading order in identity (5.184) is $\mathbf{Y}^{(m)}$ because $\kappa_\star = 0$. The dependence between $\mathcal{R}^{(m)}$ and $\mathbf{Y}^{(m)}$ is not algebraic any more, but via a partial differential equation on the cross-sections of $\mathcal{H}$.

Finally, when the horizon $\mathcal{H}$ does not include the bifurcation surface $\mathcal{S}$, the only difference is that one cannot employ Theorem 5.79 to show analiticity, and instead the isometries in items 1., 2. and 3. of Theorem 5.80 need to be replaced by "isometry to infinite order". This is precisely how [235, Prop. 6.21] is formulated.

### 5.5.2 *Existence result*

In the last part of this section we analyze the question of whether an AKH data gives rise to a $\lambda$-vacuum manifold to infinite order admitting a Killing horizon with the given data. A similar analysis can be applied to other field equations. The idea of the theorem is to construct explicitly the transverse expansion $\{\mathbb{Y}^{(m)}\}$ from identities (5.182) and (5.184) where we replace the Ricci tensor by $\lambda g$ at all orders. A crucial point of the argument will rely on using that the transverse expansion constructed in this way has an appropriate behavior under gauge transformations of the AKH data. The following lemma is devoted to establishing that.

**Lemma 5.81.** *Let $\mathscr{K} = \{\mathcal{H}, \boldsymbol{\gamma}, \boldsymbol{\ell}, \ell^{(2)}, \boldsymbol{\tau}, \alpha\}$ be AKH data satisfying that $\boldsymbol{\tau}(n) \neq 0$ is constant. Assume $\mathscr{K}$ is written in an $\eta$-gauge and define the tensors $\boldsymbol{c} := \mathbf{s} + \alpha^{-1}(d\alpha - \boldsymbol{\tau})$, $\mathfrak{c} := -\boldsymbol{c}(n)$,*

$$\begin{aligned}\mathbb{Y}_{ab} := \frac{1}{2\kappa_\star}\Big(&\alpha \mathring{R}_{(ab)} + \alpha \mathrm{s}_a\mathrm{s}_b - \alpha\mathring{\nabla}_{(a}\mathrm{s}_{b)} - 2\mathring{\nabla}_a\mathring{\nabla}_b\alpha - 2\alpha(\mathrm{s}-c)_a(\mathrm{s}-c)_b \\ &-4(\mathrm{s}-c)_{(a}\mathring{\nabla}_{b)}\alpha - 2\alpha\mathring{\nabla}_{(a}(\mathrm{s}-c)_{b)} - \alpha\lambda\gamma_{ab}\Big),\end{aligned} \tag{5.198}$$

*and for $k \geq 2$*

$$\begin{aligned}\mathbb{Y}^{(k)}_{ab} := \frac{1}{2k\kappa_\star}\Big(&\alpha\mathcal{O}^{(k-1)}(\mathbb{Y}^{\leq k-1})_{ab} - 2\alpha\mathfrak{c}^{(k)}\mathbb{Y}_{ab} + 2\alpha\mathring{\nabla}_{(a}c^{(k)}_{b)} + 4c^{(k)}_{(a}\mathring{\nabla}_{b)}\alpha \\ &+4\alpha(\mathrm{s}-c)_{(a}c^{(k)}_{b)} - 2\alpha\lambda\mathbb{Y}^{(k-1)}_{ab}\Big),\end{aligned} \tag{5.199}$$

*where*

$$c^{(k)}_a := \lambda\left(\delta^k_2\ell_a + \frac{1}{2}\delta^k_3\mathring{\nabla}_a\ell^{(2)}\right) - \mathcal{O}^{(k-1)}(\mathbb{Y}^{\leq k-1})_a, \qquad \mathfrak{c}^{(k)} := -\boldsymbol{c}^{(k)}(n). \tag{5.200}$$

*Define the collection $\{\mathbb{Y}^{(m)\prime}\}$ exactly in the same way as above but with all the terms in the right hand sides expressed in the gauge $\mathscr{K}' = \mathcal{G}_{(z,0)}\mathscr{K}$ with $z \in \mathcal{F}^\star(\mathcal{H})$ arbitrary. Then,*

$$\mathbb{Y}'_{ab} = z\mathbb{Y}_{ab} + \boldsymbol{\ell}\otimes_s dz, \qquad \mathbb{Y}^{(2)\prime}_{ab} = z^2\mathbb{Y}^{(2)}_{ab}. \tag{5.201}$$

*Let $N \geq 2$ be an integer. If in addition $\operatorname{tr}_P \mathbb{Y}^{(j)} = \mathcal{O}^{(j-1)}(\mathbb{Y}^{\leq j-1})$ for all $j = 2, ..., N$, then*

$$\mathbb{Y}^{(k)\prime}_{ab} = z^k\mathbb{Y}^{(k)}_{ab} \quad \forall k = 3, ..., N+1. \tag{5.202}$$

*Proof.* By definition,

$$\begin{aligned}\mathbb{Y}'_{ab} := \frac{1}{2\kappa_\star}\Big(&\alpha'\mathring{R}'_{(ab)} + \alpha'\mathrm{s}'_a\mathrm{s}'_b - \alpha'\mathring{\nabla}'_{(a}\mathrm{s}'_{b)} - 2\mathring{\nabla}'_a\mathring{\nabla}'_b\alpha' - 2\alpha'(\mathrm{s}'-c')_a(\mathrm{s}'-c')_b \\ &-4(\mathrm{s}'-c')_{(a}\mathring{\nabla}'_{b)}\alpha' - 2\alpha'\mathring{\nabla}'_{(a}(\mathrm{s}'-c')_{b)} - \alpha'\lambda\gamma'_{ab}\Big)\end{aligned} \tag{5.203}$$

and

$$\begin{aligned}\mathbb{Y}^{(k)\prime}_{ab} := \frac{1}{2k\kappa_\star}\Big(&\alpha'\mathcal{O}^{(k-1)}_{ab}(\mathbb{Y}'^{\leq k-1}) - 2\alpha'\mathfrak{c}^{(k)\prime}\mathbb{Y}'_{ab} + 2\alpha'\mathring{\nabla}'_{(a}c^{(k)\prime}_{b)} + 4c^{(k)\prime}_{(a}\mathring{\nabla}'_{b)}\alpha' \\ &+4\alpha'(\mathrm{s}'-c')_{(a}c^{(k)\prime}_{b)} - 2\alpha'\lambda\mathbb{Y}^{(k-1)\prime}_{ab}\Big),\end{aligned} \tag{5.204}$$

where

$$c^{(k)\prime}_a := \lambda\left(\delta^k_2\ell'_a + \frac{1}{2}\delta^k_3\mathring{\nabla}'_a\ell^{(2)\prime}\right) - \mathcal{O}^{(k-1)}_a(\mathbb{Y}'^{\leq k-1}), \qquad \mathfrak{c}^{(k)\prime} := -\boldsymbol{c}^{(k)\prime}(n'). \tag{5.205}$$

Recall that by Remark 5.33 there is no possible confusion in denoting $\mathcal{O}^{(k-1)}_{ab}(\mathbb{Y}'^{\leq k-1})$ and $\mathcal{O}^{(k-1)}_{a}(\mathbb{Y}'^{\leq k-1})$ by $\mathcal{O}^{(k-1)\prime}_{ab}$ and $\mathcal{O}^{(k-1)\prime}_{a}$, respectively. Moreover, under the hypothesis $\operatorname{tr}_P \mathbb{Y}^{(j)} = \mathcal{O}^{(j-1)}(\mathbb{Y}^{\leq j-1})$ for all $j = 2, ..., N$ the transformations of Proposition 5.32 become

$$\mathcal{O}^{(k-1)\prime}_{ab} = \mathcal{O}^{(k-1)}_{ab}(\mathbb{Y}'^{\leq k-1}) = z^{k-1}\mathcal{O}^{(k-1)}_{ab}(\mathbb{Y}^{\leq k-1}) + 2(k-1)z^{k-2}\mathcal{O}^{(k-1)}_{(a}(\mathbb{Y}^{\leq k-1})\overset{\circ}{\nabla}_{b)}z, \tag{5.206}$$

$$\mathcal{O}^{(k-1)\prime}_{a} = \mathcal{O}^{(k-1)}_{a}(\mathbb{Y}'^{\leq k-1}) = z^{k-1}\mathcal{O}^{(k-1)}_{a}(\mathbb{Y}^{\leq k-1}). \tag{5.207}$$

Let us begin by showing that $\mathbb{Y}'_{ab} = z\mathbb{Y}_{ab} + \boldsymbol{\ell} \otimes_s dz$. We divide the computation into several pieces. Firstly, using $\alpha' = z\alpha$ and Corollary 2.9 the terms $\alpha \overset{\circ}{R}_{(ab)} + \alpha \mathrm{s}_a \mathrm{s}_b - \alpha \overset{\circ}{\nabla}_{(a}\mathrm{s}_{b)}$ transform as (recall $\mathbf{U} = 0$)

$$\alpha' \overset{\circ}{R}{}'_{(ab)} + \alpha' \mathrm{s}'_a \mathrm{s}'_b - \alpha' \overset{\circ}{\nabla}{}'_{(a}\mathrm{s}'_{b)} = z\left(\alpha \overset{\circ}{R}_{(ab)} + \alpha \mathrm{s}_a \mathrm{s}_b - \alpha \overset{\circ}{\nabla}_{(a}\mathrm{s}_{b)}\right).$$

Secondly, from $\overset{\circ}{\nabla}{}' = \overset{\circ}{\nabla} + z^{-1} n \otimes \boldsymbol{\ell} \otimes_s dz$ (see (2.59) with $V = 0$) it follows

$$\begin{aligned}
\overset{\circ}{\nabla}{}'_a \overset{\circ}{\nabla}{}'_b \alpha' &= \overset{\circ}{\nabla}{}'_a \overset{\circ}{\nabla}_b (z\alpha) \\
&= \overset{\circ}{\nabla}_a z \overset{\circ}{\nabla}_b \alpha + z \overset{\circ}{\nabla}{}'_a \overset{\circ}{\nabla}_b \alpha + \overset{\circ}{\nabla}_a \alpha \overset{\circ}{\nabla}_b z + \alpha \overset{\circ}{\nabla}{}'_a \overset{\circ}{\nabla}_b z \\
&= 2\overset{\circ}{\nabla}_{(a} z \overset{\circ}{\nabla}_{b)} \alpha + z \overset{\circ}{\nabla}_a \overset{\circ}{\nabla}_b \alpha + \alpha \overset{\circ}{\nabla}_a \overset{\circ}{\nabla}_b z - \Big(n(\alpha) + z^{-1}\alpha n(z)\Big)\ell_{(a}\overset{\circ}{\nabla}_{b)} z.
\end{aligned}$$

From $\boldsymbol{\tau}' = z\boldsymbol{\tau}$ and $\alpha' = z\alpha$, the transformation of $\mathbf{s} - \boldsymbol{c}$ is

$$\mathbf{s}' - \boldsymbol{c}' = \mathbf{s} - \boldsymbol{c} - z^{-1} dz, \tag{5.208}$$

from where it follows

$$(\mathrm{s}' - c')_a (\mathrm{s}' - c')_b = (\mathrm{s} - c)_a (\mathrm{s} - c)_b - 2z^{-1} (\mathrm{s} - c)_{(a} \overset{\circ}{\nabla}_{b)} z + z^{-2} \overset{\circ}{\nabla}_a z \overset{\circ}{\nabla}_b z$$

and

$$\begin{aligned}
\overset{\circ}{\nabla}{}'_{(a} (\mathrm{s}' - c')_{b)} &= \overset{\circ}{\nabla}_{(a}\Big(\mathrm{s} - c - z^{-1} dz\Big)_{b)} + z^{-1}\Big(\boldsymbol{c} - \mathbf{s} + z^{-1} dz\Big)(n)\ell_{(a}\overset{\circ}{\nabla}_{b)} z \\
&= \overset{\circ}{\nabla}_{(a}(\mathrm{s} - c)_{b)} - z^{-1}\overset{\circ}{\nabla}_a \overset{\circ}{\nabla}_b z + z^{-2}\overset{\circ}{\nabla}_a z \overset{\circ}{\nabla}_b z + z^{-1}\Big(z^{-1} n(z) - \mathfrak{c}\Big)\ell_{(a}\overset{\circ}{\nabla}_{b)} z.
\end{aligned}$$

Combining all the terms in the right hand side of (5.203),

$$\mathbb{Y}'_{ab} = z\mathbb{Y}_{ab} + \frac{1}{2\kappa_\star}\Big(2n(\alpha) + 2\alpha c\Big)\ell_{(a}\overset{\circ}{\nabla}_{b)} z = z\mathbb{Y}_{ab} + \ell_{(a}\overset{\circ}{\nabla}_{b)} z,$$

where we used $\kappa_\star = \boldsymbol{\tau}(n) = \alpha\mathfrak{c} + n(\alpha)$. Hence the first equation in (5.201) follows. Next we compute the transformation of $\mathbb{Y}^{(2)}$. Writing (5.204) for $k = 2$

$$\mathbb{Y}^{(2)\prime}_{ab} := \frac{1}{4\kappa_\star}\left(\alpha' \mathcal{O}^{(1)\prime}_{ab} - 2\alpha' \mathfrak{c}^{(2)\prime} \mathbb{Y}'_{ab} + 2\alpha' \overset{\circ}{\nabla}{}'_{(a} c^{(2)\prime}_{b)} + 4 c^{(2)\prime}_{(a} \overset{\circ}{\nabla}{}'_{b)} \alpha' + 4\alpha' (\mathrm{s}' - c')_{(a} c^{(2)\prime}_{b)} - 2\alpha' \lambda \mathbb{Y}'_{ab}\right),$$

where by (5.205) $c_a^{(2)\prime} := \lambda\ell'_a - \mathcal{O}_a^{(1)\prime} = z(\lambda\ell_a - \mathcal{O}_a^{(1)}) = zc_a^{(2)}$ (see (2.28) and (5.207)) and hence $\mathfrak{c}^{(2)\prime} = \mathfrak{c}^{(2)}$. Inserting $\alpha' = z\alpha$, (5.206), $\mathbb{Y}'_{ab} = z\mathbb{Y}_{ab} + \ell_{(a}\overset{\circ}{\nabla}_{b)}z$, (5.93) and (5.208) it can be easily checked that $\mathbb{Y}_{ab}^{(2)\prime} = z^2\mathbb{Y}_{ab}^{(2)}$. For arbitrary $3 \leq k \leq N+1$ the process is analogous. Indeed, since $c_a^{(k)\prime} := -\mathcal{O}_a^{(k-1)\prime}$ (recall that $\ell^{(2)} = 0$ in an $\eta$-gauge) equation (5.207) gives $c_a^{(k)\prime} = z^{k-1}c_a^{(k)}$ and $\mathfrak{c}^{(k)\prime} = z^{k-2}\mathfrak{c}^{(k)}$. Inserting the transformations of $\alpha$, $\mathcal{O}_{ab}^{(k-1)}$ (5.206), $\mathrm{\mathbf{s}} - \boldsymbol{c}$ (5.208) and (5.235) into (5.204) it follows $\mathbb{Y}^{(k)\prime} = z^k\mathbb{Y}^{(k)}$. □

Now we are ready to show that every AKH data with $\boldsymbol{\tau}(n) \neq 0$ constant gives rise to an ambient manifold solving the $\Lambda$-vacuum equations to infinite order and admitting a Killing field for which the initial data is a non-degenerate horizon. Our result is entirely covariant and does not make any assumptions regarding the dimension or topology of the null hypersurface.

**Theorem 5.82.** *Let $\widetilde{\mathscr{K}} = \{\mathcal{H}, \widetilde{\gamma}, \widetilde{\boldsymbol{\ell}}, \widetilde{\ell}^{(2)}, \widetilde{\boldsymbol{\tau}}, \widetilde{\alpha}\}$ be AKH data satisfying that $\widetilde{\boldsymbol{\tau}}(\widetilde{n})$ is a non-zero constant. Then, there exists a smooth semi-Riemannian manifold $(\mathcal{M}, g)$, an embedding $\Phi : \mathcal{H} \hookrightarrow \mathcal{M}$, a rigging $\zeta$ of $\Phi(\mathcal{H})$ and a smooth extension $\eta$ of $\bar{\eta} := \widetilde{\alpha}\widetilde{n}$ off $\Phi(\mathcal{H})$ such that (i) $\widetilde{\mathscr{K}}$ is $(\Phi, \zeta)$-embedded in $(\mathcal{M}, g)$, (ii) $\eta$ is a Killing vector of $g$ and (iii) $(\mathcal{M}, g)$ satisfies the $\Lambda$-vacuum equations to all orders on $\Phi(\mathcal{H})$.*

*Proof.* The proof is divided into three main parts. Firstly, we define a suitable sequence of $(0,2)$ symmetric tensors $\{\mathbb{Y}^{(k)}\}$ on $\mathcal{H}$ and we construct the ambient manifold $(\mathcal{M}, g)$ as in Theorem 5.40. Secondly, we show that the ambient Ricci tensor of $(\mathcal{M}, g)$ satisfies the $\Lambda$-vacuum equations to infinite order on $\mathcal{H}$, i.e. $R_{\mu\nu}^{(i)} \overset{\mathcal{H}}{=} \lambda\mathcal{K}_{\mu\nu}^{(i-1)}$ for every $i \geq 1$. Finally, we construct the Killing $\eta$ by extending $\bar{\eta} = \alpha n$ suitably.

1. Construction of the ambient space

The idea of the first part of the proof is to construct the sequence $\{\mathbb{Y}^{(k)}\}$ using identities (5.182) and (5.184), where the Ricci tensor is replaced by $\lambda g$ at all orders. Observe that since $\kappa_\star := \widetilde{\boldsymbol{\tau}}(\widetilde{n}) \neq 0$ is constant it follows from (5.183) that $\boldsymbol{\mathcal{R}}(\widetilde{n}, \cdot) = 0$.

Let us transform the AKH data $\widetilde{\mathscr{K}}$ to the $\eta$-gauge constructed by the transformation $\mathcal{G}_{(1,V)}$ with $V$ defined by (compare with (5.192))

$$\widetilde{\gamma}(V, \cdot) = \kappa_\star^{-1}\widetilde{\boldsymbol{\tau}} - \widetilde{\boldsymbol{\ell}}, \qquad \text{and} \qquad \widetilde{\boldsymbol{\ell}}(V) = -\widetilde{\ell}^{(2)} - \frac{1}{2}\kappa_\star^{-2}\widetilde{P}(\widetilde{\boldsymbol{\tau}}, \widetilde{\boldsymbol{\tau}}). \tag{5.209}$$

We denote the data in this $\eta$-gauge without the tilde and define the one-form $\boldsymbol{c} := \mathrm{\mathbf{s}} + \alpha^{-1}(d\alpha - \boldsymbol{\tau})$, the scalar $\mathfrak{c} := -\boldsymbol{c}(n) = \alpha^{-1}(\boldsymbol{\tau}(n) - n(\alpha))$ (recall that by Remark 5.63 both objects are smooth) and the tensor $\mathbb{Y}$ by (cf. (5.182))

$$\begin{aligned}\mathbb{Y}_{ab} := \frac{1}{2\kappa_\star}\Big(&\alpha\overset{\circ}{R}_{(ab)} + \alpha \mathrm{s}_a\mathrm{s}_b - \alpha\overset{\circ}{\nabla}_{(a}\mathrm{s}_{b)} - 2\overset{\circ}{\nabla}_a\overset{\circ}{\nabla}_b\alpha - 2\alpha(\mathrm{s}-c)_a(\mathrm{s}-c)_b \\ &-4(\mathrm{s}-c)_{(a}\overset{\circ}{\nabla}_{b)}\alpha - 2\alpha\overset{\circ}{\nabla}_{(a}(\mathrm{s}-c)_{b)} - \alpha\lambda\gamma_{ab}\Big).\end{aligned} \tag{5.210}$$

We also define the one-form $c_a^{(2)}$ by (cf. (5.113))

$$c_a^{(2)} := \lambda \ell_a - \mathcal{O}^{(1)}(\mathbb{Y})_a. \tag{5.211}$$

Similarly, for arbitrary $m \geq 2$ we define (cf. (5.184))

$$\begin{aligned}\mathbb{Y}_{ab}^{(m)} := \frac{1}{2m\kappa_\star}\Big(&\alpha\mathcal{O}^{(m-1)}(\mathbb{Y}^{\leq m-1})_{ab} - 2\alpha\mathfrak{c}^{(m)}\mathbb{Y}_{ab} + 2\alpha\mathring{\nabla}_{(a}c_{b)}^{(m)} + 4c_{(a}^{(m)}\mathring{\nabla}_{b)}\alpha \\ &+4\alpha(\mathrm{s}-c)_{(a}c_{b)}^{(m)} - 2\alpha\lambda\mathbb{Y}_{ab}^{(m-1)}\Big)\end{aligned} \tag{5.212}$$

and $\boldsymbol{c}^{(m+1)}$ by (cf. (5.113) and recall $\ell^{(2)} = 0$)

$$c_a^{(m+1)} := -\mathcal{O}^{(m)}(\mathbb{Y}^{\leq m})_a. \tag{5.213}$$

By Theorem 5.40 there exists an ambient space $(\mathcal{M}, g)$, an embedding $\Phi : \mathcal{H} \hookrightarrow \mathcal{M}$ and a rigging $\xi$ such that $\mathbb{Y}^{(k)} = \frac{1}{2}\Phi^\star(\pounds_\xi^{(k)} g)$ for every $k \geq 1$. By Remark 5.62, item (i) of the Theorem follows with $\zeta = \xi - V$, with $V$ given by (5.209). Our objective now is to prove that $(\mathcal{M}, g)$ satisfies the $\Lambda$-vacuum equations at infinite order on $\Phi(\mathcal{H})$ and that there exists an extension $\eta$ of $\bar{\eta} = \alpha n$ off $\Phi(\mathcal{H})$ such that $\pounds_\eta g = 0$ on $\mathcal{M}$.

2. $\Lambda$-vacuum equations

Now we concentrate in proving that the $\Lambda$-vacuum equations hold at infinite order on $\Phi(\mathcal{H})$, i.e. we want to show that for all $m \geq 1$ (we recall again that $\ell^{(2)} = 0$ in an $\eta$-gauge)

$$\mathcal{R}_{ab} = \lambda\gamma_{ab}, \qquad \ddot{\mathcal{R}}^{(m)} = 0, \qquad \dot{\mathcal{R}}_a^{(m)} = \delta_1^m \lambda \ell_a, \qquad \mathcal{R}_{ab}^{(m+1)} = 2\lambda \mathrm{Y}_{ab}^{(m)}. \tag{5.214}$$

By (2.95) and (5.82)-(5.84) this is equivalent to check that the embedded expansion $\{\mathbf{Y}^{(m)}\}$ satisfies the following equations (recall $\ell^{(2)} = 0$)

$$\begin{aligned}\lambda\gamma_{ab} = {}& \mathring{R}_{(ab)} - 2\pounds_n \mathrm{Y}_{ab} - (2\kappa + \operatorname{tr}_P \mathbf{U})\mathrm{Y}_{ab} + \mathring{\nabla}_{(a}(\mathrm{s}_{b)} + 2\mathrm{r}_{b)}) \\ & - 2\mathrm{r}_a\mathrm{r}_b + 4\mathrm{r}_{(a}\mathrm{s}_{b)} - \mathrm{s}_a\mathrm{s}_b - (\operatorname{tr}_P \mathbf{Y})\mathrm{U}_{ab} + 2P^{cd}\mathrm{U}_{d(a}(2\mathrm{Y}_{b)c} + \mathrm{F}_{b)c}),\end{aligned} \tag{5.215}$$

$$0 = -\operatorname{tr}_P \mathbf{Y}^{(m+1)} + \mathcal{O}^{(m)}(\mathbf{Y}^{\leq m}), \tag{5.216}$$

$$\lambda\ell_a\delta_1^m = \mathrm{r}_a^{(m+1)} + \mathcal{O}^{(m)}(\mathbf{Y}^{\leq m})_a, \tag{5.217}$$

$$\begin{aligned}2\lambda\mathrm{Y}_{ab}^{(m)} = {}& -2\pounds_n \mathrm{Y}_{ab}^{(m+1)} - \left(2(m+1)\kappa + \operatorname{tr}_P \mathbf{U}\right)\mathrm{Y}_{ab}^{(m+1)} - (\operatorname{tr}_P \mathbf{Y}^{(m+1)})\mathrm{U}_{ab} \\ & + 4P^{cd}\mathrm{U}_{c(a}\mathrm{Y}_{b)d}^{(m+1)} + 4(\mathrm{s}-\mathrm{r})_{(a}\mathrm{r}_{b)}^{(m+1)} + 2\mathring{\nabla}_{(a}\mathrm{r}_{b)}^{(m+1)} \\ & - 2\kappa^{(m+1)}\mathrm{Y}_{ab} + \mathcal{O}_{ab}^{(m)}(\mathbf{Y}^{\leq m}).\end{aligned} \tag{5.218}$$

There are two main difficulties at this point. Firstly, we shall check that the tensors $\{\boldsymbol{c}^{(m)}\}$ agree with $\{\mathbf{r}^{(m)}\}$ at all orders, and secondly we need to compute Lie derivatives along $n$ of the expansion $\{\mathbf{Y}^{(m)}\}$. Unfortunately we cannot do this directly in the $\eta$-gauge because we have no information on the Lie derivative along $n$ of the various quantities. The reason is that $\bar{\eta}$ and $n$ are different in this gauge as they are related by the proportionality function $\alpha$. The obvious thing to try is to go into a gauge where $\alpha = 1$. However, this requires some work because we are allowing $\alpha$ to have zeroes. This forces us to argue on a suitable open and dense subset and then come back to the original (and globally defined) $\eta$-gauge. Let us sketch the main lines of the argument:

1. We restrict to the subset $\{\alpha \neq 0\}$ and rescale the rigging by $\xi' = \alpha^{-1}\xi$. This rescaling induces a transformation on the embedded expansion $\{\mathbf{Y}^{(m)}\} \longmapsto \{\mathbf{Y}^{(m)\prime}\}$ (transformations (5.220)) as well as on the transverse derivatives of the ambient Ricci tensor $\{\ddot{\mathcal{R}}^{(m)}, \dot{\mathcal{R}}_a^{(m)}, \mathcal{R}_{ab}^{(m)}\} \longmapsto \{\ddot{\mathcal{R}}^{(m)\prime}, \dot{\mathcal{R}}_a^{(m)\prime}, \mathcal{R}_{ab}^{(m)\prime}\}$ in (5.223)-(5.225). At the same time, we transform $\mathscr{K}$ with gauge parameters $(z = \alpha^{-1}, 0)$ (for consistency we shall denote the transformed data by a prime, $\mathscr{K}'$).

2. We use (5.201) to write down $\mathbf{Y}'$ and $\mathbf{Y}^{(2)\prime}$ explicitly. This allows us to check that $\mathbf{r}' = \boldsymbol{c}'$, $\pounds_{n'}\mathbf{Y}' = \pounds_{n'}\mathbf{Y}^{(2)\prime} = 0$ and that $\mathcal{R}'_{ab} = \lambda\gamma'_{ab}$. Furthermore, applying Lemma 5.48 we will show $\mathbf{r}^{(2)\prime} = \boldsymbol{c}^{(2)\prime}$ and also $\dot{\mathcal{R}}_a^{(1)\prime} = \lambda\ell'_a$, $\ddot{\mathcal{R}}^{(1)\prime} = 0$ and $\mathcal{R}_{ab}^{(2)\prime} = 2\lambda\mathrm{Y}'_{ab}$.

3. In order to continue with the argument inductively we need to determine $\{\mathbf{Y}^{(k\geq 3)\prime}\}$ explicitly. To do that we rely on Lemma 5.81 that relates the unprimed quantities with the primed ones. This lemma has a key hypothesis, namely the validity of

   $$\mathrm{tr}_P\, \mathbb{Y}^{(j)} = \mathcal{O}^{(j-1)}(\mathbb{Y}^{\leq j-1}) \qquad \forall j = 2, ..., N.$$

   In order to check that this hypothesis holds we use the transformations (5.223)-(5.225) between the primed and unprimed components of the Ricci tensor. By a similar application of Lemma 5.48 as in the case $k = 2$ we will check that $\mathbf{r}^{(k)\prime} = \boldsymbol{c}^{(k)\prime}$, $\pounds_{n'}\mathbf{Y}^{(k)\prime} = 0$ and that $\ddot{\mathcal{R}}^{(k-1)\prime} = 0$, $\dot{\mathcal{R}}_a^{(k-1)\prime} = 0$ and $\mathcal{R}_{ab}^{(k)\prime} = 2\lambda\mathrm{Y}_{ab}^{(k-1)\prime}$.

4. Finally, we come back to the unprimed quantities via transformations (5.223)-(5.225) to show (5.215)-(5.218).

Let $\mathcal{H}_0$ be the subset of $\mathcal{H}$ consisting of the points where $\alpha \neq 0$ and let $\mathscr{K}_0$ the restriction of the AKH data $\mathscr{K}$ to $\mathcal{H}_0$. Observe that by definition of AKH data, the closure of $\mathcal{H}_0$ is $\mathcal{H}$. Let us extend $\alpha$ off $\Phi(\mathcal{H})$ by $\xi(\alpha) = 0$ and let $\mathcal{M}_0$ be the subset of $\mathcal{M}$ where $\alpha \neq 0$. Clearly the closure of $\mathcal{M}_0$ is the whole $\mathcal{M}$. Let us now define another transverse vector field $\xi'$ on $\mathcal{M}_0$ by $\xi' := \alpha^{-1}\xi$ (note that it is still geodesic). For such field the induced AKH data on $\mathcal{H}_0$, that we shall denote with a prime in accordance with the fact that the rigging is now $\xi'$, is written in the $\eta$-gauge satisfying $\alpha = 1$, i.e. $\mathscr{K}'_0 = \mathcal{G}_{(\alpha^{-1},0)}\mathscr{K}_0$. Such gauge will be called "adapted $\eta$-gauge" and it is unique (because the only gauge transformation of the form $\mathcal{G}_{(z,0)}$ that leaves $\alpha$ invariant is the identity). Let us explore some properties of the adapted $\eta$-gauge. Firstly, by item (iii) in Definition 5.58 the condition $0 = \pounds_{n'}\boldsymbol{\tau}' = \kappa_\star \pounds_{n'}\boldsymbol{\ell}'$ implies $\mathbf{s}' = 0$ (see (2.22)), and thus all the Lie derivatives along $n'$ of the metric data vanish, i.e.

$$\pounds_{n'}\boldsymbol{\gamma}' = 0, \qquad \pounds_{n'}\boldsymbol{\ell}' = 0, \qquad \pounds_{n'}\ell^{(2)\prime} = 0. \tag{5.219}$$

Secondly, $\mathfrak{c}' = \boldsymbol{\tau}'(n') = \kappa_\star$ is constant and $\boldsymbol{\tau}' = -\boldsymbol{c}'$, so $\pounds_{n'}\boldsymbol{c}' = 0$. And finally, the tensor defined by $\mathring{\Sigma}' := \pounds_{n'}\mathring{\nabla}'$ vanishes. Indeed, from (2.39) and $\pounds_{n'}\mathbf{F}' = \frac{1}{2}\pounds_{n'}d\boldsymbol{\ell}' = \frac{1}{2}d\pounds_{n'}\boldsymbol{\ell}' = 0$ as well as $\pounds_{n'}\boldsymbol{\ell}' = 0$ it follows (see Prop. 5.7 for $m = 1$)

$$0 = \pounds_{n'}\mathring{\nabla}'_a\ell'_b = \mathring{\nabla}'_a\pounds_{n'}\ell'_b - \mathring{\Sigma}'^c{}_{ab}\ell'_c = -\mathring{\Sigma}'^c{}_{ab}\ell'_c,$$

and from (2.38) one has

$$0 = \pounds_{n'}\mathring{\nabla}'_a\gamma'_{bc} = \mathring{\nabla}'_a\pounds_{n'}\gamma'_{bc} - 2\mathring{\Sigma}'^d{}_{a(b}\gamma'_{c)d} = -2\mathring{\Sigma}'^d{}_{a(b}\gamma'_{c)d}.$$

Hence,

$$0 = \pounds_{n'}\mathring{\nabla}'_a\gamma'_{bc} + \pounds_{n'}\mathring{\nabla}'_b\gamma'_{ca} - \pounds_{n'}\mathring{\nabla}'_c\gamma'_{ab} = 2\mathring{\Sigma}'^d{}_{ab}\gamma'_{cd}.$$

Conditions $\mathring{\Sigma}'^c{}_{ab}\ell'_c = 0$ and $\mathring{\Sigma}'^d{}_{ab}\gamma'_{cd} = 0$ automatically imply $\mathring{\Sigma}'^d{}_{ab} = 0$. Recalling identity (5.36) (which is valid for every vector and connection, and in particular for $n'$ and $\mathring{\nabla}'$) we conclude

$$\pounds_{n'}\mathring{R}'_{ab} = \mathring{\nabla}'_c\mathring{\Sigma}'^c{}_{ab} - \mathring{\nabla}'_b\mathring{\Sigma}'^d{}_{ad} = 0.$$

Under the transformation $\xi' = \alpha^{-1}\xi$ the transverse expansion $\{\mathbf{Y}^{(m)}\}$ transforms as (see (5.95)-(5.97) with $z = \alpha^{-1}$ and recall that $\xi(\alpha) = 0$)

$$\mathbf{Y}' = \alpha^{-1}\mathbf{Y} - \alpha^{-2}d\alpha \otimes_s \boldsymbol{\ell}, \qquad \mathbf{Y}^{(k)\prime} = \alpha^{-k}\mathbf{Y}^{(k)}, \tag{5.220}$$

$$\mathbf{r}' = \mathbf{r} - \frac{1}{2\alpha}\left(d\alpha + n(\alpha)\boldsymbol{\ell}\right), \qquad \mathbf{r}^{(k)\prime} = \alpha^{-k+1}\mathbf{r}^{(k)} \quad (k \geq 2), \tag{5.221}$$

$$\kappa' = \alpha\kappa + n(\alpha), \qquad \kappa^{(k)\prime} = \alpha^{-k+2}\kappa^{(k)}, \tag{5.222}$$

where the prime is to emphasize that $\mathbf{Y}^{(m)\prime} := \frac{1}{2}\Phi^\star(\pounds^{(m)}_{\xi'}g)$ on $\mathcal{H}_0$. Note that the transformation $\mathbf{Y} \longmapsto \mathbf{Y}'$ is analogous to the gauge transformation (2.30) with parameters $(\alpha^{-1}, 0)$, as it must be. In addition, under the change $\xi' = \alpha^{-1}\xi$ the derivatives of the Ricci tensor transform according to (5.98)-(5.100) with $z = \alpha^{-1}$

$$\ddot{\mathcal{R}}^{(m)\prime} = \alpha^{-m-1}\ddot{\mathcal{R}}^{(m)}, \tag{5.223}$$

$$\dot{\mathcal{R}}^{(m)\prime}_a = \alpha^{-m}\dot{\mathcal{R}}^{(m)}_a - (m-1)\alpha^{-m-1}\ddot{\mathcal{R}}^{(m-1)}\mathring{\nabla}_a\alpha, \tag{5.224}$$

$$\mathcal{R}^{(m+1)\prime}_{ab} = \alpha^{-m}\mathcal{R}^{(m+1)}_{ab} - 2m\alpha^{-m-1}\dot{\mathcal{R}}^{(m)}_{(a}\mathring{\nabla}_{b)}\alpha + m(m-1)\alpha^{-m-2}\ddot{\mathcal{R}}^{(m-1)}\mathring{\nabla}_a\alpha\,\mathring{\nabla}_b\alpha. \tag{5.225}$$

Our aim now is to prove that the abstract transverse expansion defined in (5.210) and (5.212) transform exactly as the embedded one in (5.220) under a gauge transformation with parameters $(z = \alpha^{-1}, V = 0)$, as this will imply that the expressions for the tensors $\{\mathbf{Y}^{(m)\prime}\}$ agree with those of $\{\mathbb{Y}^{(m)\prime}\}$ written in the adapted $\eta$-gauge. More specifically, we define the gauge-transformed tensors $\{\mathbb{Y}^{(m)\prime}\}$ by the expressions (5.203) and (5.204), where all the terms in the right hand sides are expressed in the gauge obtained by the transformation $\mathcal{G}_{(z,0)}$, and we want to show that, with these definitions, $\mathbb{Y}'_{ab}$ and $\mathbb{Y}_{ab}$ as well as $\mathbb{Y}^{(k)\prime}_{ab}$ and $\mathbb{Y}^{(k)}_{ab}$ are related by exactly the same expressions in (5.220) when $z$ is taken to be $z = \alpha^{-1}$, namely

$$\mathbb{Y}' = \alpha^{-1}\mathbb{Y} - \alpha^{-2} d\alpha \otimes_s \boldsymbol{\ell}, \qquad \mathbb{Y}^{(k)\prime} = \alpha^{-k}\mathbb{Y}^{(k)}, \tag{5.226}$$

$$\mathbb{r}' = \mathbb{r} - \frac{1}{2\alpha}\left(d\alpha + n(\alpha)\boldsymbol{\ell}\right), \qquad \mathbb{r}^{(k)\prime} = \alpha^{-k+1}\mathbb{r}^{(k)} \quad (k \geq 2), \tag{5.227}$$

$$\mathbb{k}' = \alpha\mathbb{k} + n(\alpha), \qquad \mathbb{k}^{(k)\prime} = \alpha^{-k+2}\mathbb{k}^{(k)}. \tag{5.228}$$

As proven in Lemma 5.81 the tensor $\mathbb{Y}'$ satisfies $\mathbb{Y}'_{ab} = z\mathbb{Y}_{ab} + \ell_{(a}\overset{\circ}{\nabla}_{b)}z$, so putting $z = \alpha^{-1}$ yields

$$\mathbb{Y}'_{ab} = \alpha^{-1}\mathbb{Y}_{ab} - \alpha^{-2}\ell_{(a}\overset{\circ}{\nabla}_{b)}\alpha,$$

which is the first relation in (5.226). This proves that the tensor $\mathbf{Y}'$ on $\mathcal{H}_0$ has the same expression as (5.210) but written in the adapted $\eta$-gauge, i.e. with $\alpha = 1$

$$\mathrm{Y}'_{ab} = \frac{1}{2\kappa_\star}\left(\overset{\circ}{R}{}'_{(ab)} - 2c'_a c'_b + 2\overset{\circ}{\nabla}{}'_{(a}c'_{b)} - \lambda\gamma'_{ab}\right). \tag{5.229}$$

By the properties of the adapted $\eta$-gauge described above it follows $\pounds_{n'}\mathbf{Y}' = 0$ on $\mathcal{H}_0$. Throughout the rest of the proof we will frequently use (2.46) with $\mathbf{s} = 0$, $\mathbf{U} = 0$ and $\pounds_n\boldsymbol{\theta} = 0$, i.e.

$$2n^b\overset{\circ}{\nabla}_{(a}\theta_{b)} = \overset{\circ}{\nabla}_a(\boldsymbol{\theta}(n)). \tag{5.230}$$

Contracting (5.229) with $n'^a$ and using $\mathfrak{c}' = \kappa_\star$, $\overset{\circ}{\nabla}{}'_a n'^b = 0$ (see (2.66)), $\overset{\circ}{R}{}'_{(ab)}n'^a = 0$ (see (2.96)) and $2n'^a\overset{\circ}{\nabla}{}'_{(a}c'_{b)} = -\overset{\circ}{\nabla}{}'_b\mathfrak{c}' = 0$ (see (5.230)) one also has $\mathbf{r}' = \boldsymbol{c}'$. In particular this implies that $\kappa' = \mathfrak{c}' = \kappa_\star$, so

$$\lambda\gamma'_{ab} = \overset{\circ}{R}{}'_{(ab)} - 2\pounds_{n'}\mathrm{Y}'_{ab} - 2\kappa_{\star n}{}'\mathrm{Y}'_{ab} + 2\overset{\circ}{\nabla}{}'_{(a}\mathrm{r}'_{b)} - 2\mathrm{r}'_a\mathrm{r}'_b. \tag{5.231}$$

From Lemma 5.81 the transformation of $\mathbb{Y}^{(2)}$ is $\mathbb{Y}^{(2)\prime}_{ab} = z^2\mathbb{Y}^{(2)}_{ab}$, which proves that the tensor $\mathbf{Y}^{(2)\prime}$ has the same expression as in (5.212) for $m = 2$ but with all the data written in the adapted $\eta$-gauge, namely

$$\mathrm{Y}^{(2)\prime}_{ab} = \frac{1}{4\kappa_\star}\left(\mathcal{O}^{(1)\prime}_{ab} - 2\mathfrak{c}^{(2)\prime}\mathrm{Y}'_{ab} + 2\overset{\circ}{\nabla}{}'_{(a}c^{(2)\prime}_{b)} - 4\mathrm{r}'_{(a}c^{(2)\prime}_{b)} - 2\lambda\mathrm{Y}'_{ab}\right), \tag{5.232}$$

where recall $c^{(2)\prime}_a := \lambda\ell'_a - \mathcal{O}^{(1)\prime}_a$. Since the tensors $\mathcal{O}^{(1)\prime}_{ab}$, $\mathcal{O}^{(1)\prime}_a$ and $\mathcal{O}^{(1)\prime}$ are constructed solely in terms of metric data and $\mathbf{Y}'$, it follows $\pounds_{n'}\mathcal{O}^{(1)\prime}_{ab} = 0$, $\pounds_{n'}\mathcal{O}^{(1)\prime}_a = 0$ and $\pounds_{n'}\mathcal{O}^{(1)\prime} = 0$, and thus $\pounds_{n'}\mathbf{Y}^{(2)\prime} = 0$ on $\mathcal{H}_0$ as well. Contracting (5.232) with $n'^b$, using $2n'^b\overset{\circ}{\nabla}{}'_{(a}c^{(2)\prime}_{b)} = -\overset{\circ}{\nabla}{}'_a\mathfrak{c}^{(2)}$ (see (5.230)) and Lemma 5.48 with $m = 1$, namely

$$2\lambda\mathrm{r}'_a = -2\kappa'_n c^{(2)\prime}_a - \overset{\circ}{\nabla}{}'_a c^{(2)\prime} + \mathcal{O}^{(1)\prime}_{ab}n'^b,$$

gives $\mathrm{r}^{(2)\prime}_a = c^{(2)\prime}_a := \lambda\ell'_a - \mathcal{O}^{(1)\prime}_a$. Contracting (5.232) with $P'^{ab}$ and using again Lemma 5.48 with $m = 1$ and $\ell^{(2)\prime} = 0$, i.e.

$$2\lambda\,\mathrm{tr}_{P'}\mathbf{Y}' = -4\kappa_\star\mathcal{O}^{(1)\prime} - 2\mathfrak{c}^{(2)\prime}\,\mathrm{tr}_{P'}\mathbf{Y}' - 4P'(\mathbf{r}', \boldsymbol{c}^{(2)\prime}) + 2\,\mathrm{div}_{P'}\boldsymbol{c}^{(2)\prime} + P'^{ab}\mathcal{O}^{(1)\prime}_{ab},$$

one shows that $P'^{ab}\mathrm{Y}^{(2)\prime}_{ab} = \mathcal{O}^{(1)\prime}$. By combining $\pounds_{n'}\mathbf{Y}^{(2)\prime} = 0$ and (5.232) it follows

$$2\lambda \mathrm{Y}'_{ab} = -2\pounds_{n'}\mathrm{Y}^{(2)\prime}_{ab} - 4\kappa'_n \mathrm{Y}^{(2)\prime}_{ab} - 4\mathrm{r}'_{(a}\mathrm{r}^{(2)\prime}_{b)} + 2\mathring{\nabla}'_{(a}\mathrm{r}^{(2)\prime}_{b)} - 2\kappa^{(2)\prime}\mathrm{Y}'_{ab} + \mathcal{O}^{(1)\prime}_{ab}.$$

So far we have shown that

$$\mathcal{R}^{(1)\prime}_{ab} = \lambda\gamma'_{ab}, \qquad \ddot{\mathcal{R}}^{(1)\prime} = 0, \qquad \dot{\mathcal{R}}^{(1)\prime}_a = \lambda\ell'_a, \qquad \mathcal{R}^{(2)\prime}_{ab} = 2\lambda \mathrm{Y}'_{ab}, \tag{5.233}$$

which by (5.220)-(5.225), $\boldsymbol{\ell} = \alpha\boldsymbol{\ell}'$ and the gauge invariance of the constraint tensor and that of $\boldsymbol{\gamma}$ is equivalent to

$$\mathcal{R}^{(1)}_{ab} = \lambda\gamma_{ab}, \qquad \ddot{\mathcal{R}}^{(1)} = 0, \qquad \dot{\mathcal{R}}^{(1)}_a = \lambda\ell_a, \qquad \mathcal{R}^{(2)}_{ab} = 2\lambda \mathrm{Y}_{ab}. \tag{5.234}$$

Let us show by induction that for every $m \geq 2$

$$\mathbb{Y}^{(k)\prime} = z^k \mathbb{Y}^{(k)}, \qquad \pounds_{n'}\mathbf{Y}^{(k)\prime} = 0 \quad \forall\, k = 2, ..., m \tag{5.235}$$

and also

$$\ddot{\mathcal{R}}^{(j)\prime} = 0, \qquad \dot{\mathcal{R}}^{(j)\prime}_a = \delta^j_1 \lambda\ell'_a, \qquad \mathcal{R}^{(j+1)\prime}_{ab} = 2\lambda \mathrm{Y}^{(j)\prime}_{ab} \quad \forall\, j = 1, ..., m-1. \tag{5.236}$$

For $m = 2$ it is clearly true because we have already shown $\mathbb{Y}^{(2)\prime} = z^2\mathbb{Y}^{(2)}$, $\pounds_{n'}\mathbf{Y}^{(2)\prime} = 0$ and (5.234). We assume (5.235)-(5.236) hold for some $m \geq 2$ and we prove that they also hold for $m+1$. By (5.220)-(5.225) equations (5.236) are equivalent to

$$\ddot{\mathcal{R}}^{(j)} = 0, \qquad \dot{\mathcal{R}}^{(j)}_a = \delta^j_1 \lambda\ell_a, \qquad \mathcal{R}^{(j+1)}_{ab} = 2\lambda \mathrm{Y}^{(j)}_{ab} \quad \forall\, j = 1, ..., m-1. \tag{5.237}$$

In particular, since $\ddot{\mathcal{R}}^{(j)} = 0$ it follows $\operatorname{tr}_P \mathbf{Y}^{(j+1)} = \mathcal{O}^{(j)}$ $\forall j = 1, ..., m-1$ and hence by Lemma 5.81 one concludes $\mathbb{Y}^{(m+1)\prime} = z^{m+1}\mathbb{Y}^{(m+1)}$, so the expression of the tensor $\mathbf{Y}^{(m+1)\prime}$ agrees with (5.212) in the adapted $\eta$-gauge, namely

$$\mathrm{Y}^{(m+1)\prime}_{ab} = \frac{1}{2(m+1)\kappa_\star}\left(\mathcal{O}^{(m)\prime}_{ab} - 2\mathfrak{c}^{(m+1)\prime}\mathrm{Y}'_{ab} + 2\mathring{\nabla}'_{(a}c^{(m+1)\prime}_{b)} - 4c'_{(a}c^{(m+1)\prime}_{b)} - 2\lambda\mathbb{Y}^{(m)\prime}_{ab}\right). \tag{5.238}$$

A similar argument as before based on Lemma 5.48 proves $\pounds_{n'}\mathbf{Y}^{(m+1)\prime} = 0$, $\mathrm{r}^{(m+1)\prime}_a = c^{(m+1)\prime}_a$ and $\operatorname{tr}_P \mathbf{Y}^{(m+1)\prime} = \mathcal{O}^{(m)\prime}$, so

$$0 = \mathrm{r}^{(m+1)\prime}_a + \mathcal{O}^{(m)\prime}_a, \qquad 0 = -\operatorname{tr}_P \mathbf{Y}^{(m+1)\prime} + \mathcal{O}^{(m)\prime}.$$

Moreover, combining $\pounds_{n'}\mathbf{Y}^{(m+1)\prime} = 0$ with (5.238) gives

$$\begin{aligned}2\lambda\mathrm{Y}^{(m)\prime}_{ab} = {} & -2\pounds_{n'}\mathbf{Y}^{(m+1)\prime}_{ab} - 2(m+1)\kappa'\mathrm{Y}^{(m+1)\prime}_{ab} - 2\kappa^{(m+1)\prime}\mathrm{Y}'_{ab} + 2\mathring{\nabla}'_{(a}\mathrm{r}^{(m+1)\prime}_{b)} - 4\mathrm{r}'_{(a}\mathrm{r}^{(m+1)\prime}_{b)} \\ & + \mathcal{O}^{(m)\prime}_{ab}.\end{aligned}$$

This proves

$$\ddot{\mathcal{R}}^{(m)\prime} = 0, \qquad \dot{\mathcal{R}}^{(m)\prime}_a = 0, \qquad \mathcal{R}^{(m+1)\prime}_{ab} = 2\lambda\mathrm{Y}^{(m)\prime}_{ab},$$

and since $\mathbb{Y}^{(m+1)\prime} = z^{m+1}\mathbb{Y}^{(m+1)}$ and $\pounds_{n'}\mathbf{Y}^{(m+1)\prime} = 0$ this closes the induction argument. Hence we have proved that for all $m \geq 1$

$$\pounds_{n'}\mathbf{Y}^{(m)\prime} = 0 \quad \forall m \geq 1, \tag{5.239}$$

and by (5.220)-(5.225),

$$\mathcal{R}_{ab} = \lambda\gamma_{ab}, \qquad \ddot{\mathcal{R}}^{(m)} = 0, \qquad \dot{\mathcal{R}}^{(m)}_a = \delta^m_1\lambda\ell_a, \qquad \mathcal{R}^{(m+1)}_{ab} = 2\lambda\mathrm{Y}^{(m)}_{ab}, \tag{5.240}$$

so equations (5.214) are established on $\mathcal{H}_0$, and by continuity on the full $\mathcal{H}$. Therefore $(\mathcal{M}, g)$ solves the $\Lambda$-vacuum equations to infinite order on $\mathcal{H}$. This proves item (iii) of the theorem.

3. Construction of the Killing field

In order to finish the proof it only remains to show that $(\mathcal{M}, g)$ admits a Killing that is null and tangent to $\Phi(\mathcal{H})$. Let us now extend $\overline{\eta} = \alpha n$ off $\Phi(\mathcal{H})$ by means of $\pounds_\xi \eta = \beta\xi$, where $\beta$ is the function on $\mathcal{M}$ that solves $\xi(\beta) = 0$ with initial condition $\beta = -n(\alpha)$ on $\Phi(\mathcal{H})$. Equation (5.189) (in fact combined with Lemma 5.56) suggests that this is indeed a natural candidate to be a Killing. Our aim is to prove that the deformation tensor of $\eta$ vanishes on $\mathcal{M}_0$, and thus by continuity on all $\mathcal{M}$. Recall that the construction of the metric $g$ in the proof of Theorem 5.40 (see (5.108)) consists of choosing local coordinates $\{r, u, x^A\}$, where $n = \partial_u$ and $\xi = \partial_r$, and then using Borel's lemma (see Lemma 5.38) to construct the coefficients $f_A$, $h_A$, $f$ and $H_{AB}$ in such a way that the transverse expansion at $\Phi(\mathcal{H})$ agrees with $\{\mathbb{Y}^{(k)}\}$. Each of the coefficients is given by a sum like (5.105), and from the properties of the sequence $\{\mu_k\}$ it follows that for any given $r > 0$ the sum has only finitely many terms. This will allows us to interchange the sum and a derivative when necessary. It is easy to check that the vector $\eta$ in these coordinates is given by $\eta = \alpha\partial_u + \beta r\partial_r = \alpha\partial_u - r\partial_u\alpha\,\partial_r$. In these coordinates the computation of $\pounds_\eta g$ is somewhat long and requires using the equations

$$\pounds_{\alpha n}\mathbf{Y} = n(\alpha)\mathbf{Y} + \alpha d(n(\alpha)) \otimes_s \boldsymbol{\ell}, \qquad \pounds_{\alpha n}\mathbf{Y}^{(k)} = kn(\alpha)\mathbf{Y}^{(k)} \quad (k \geq 2)$$

on $\mathcal{H}_0$, which follow after transforming $\pounds_{n'}\mathbf{Y}^{(m)\prime} = 0\ \forall m \geq 1$. Fortunately there exists a quicker strategy. If instead of working directly with the coordinate system $\{r, u, x^A\}$ one defines the new coordinates $\{r' := \alpha r, u' := \int \alpha^{-1}du, x^A\}$ on $\mathcal{M}_0$, it follows

$$\begin{aligned}
\partial_u &= \alpha^{-1}\partial_{u'} + r\partial_u\alpha\,\partial_{r'}, & du &= \alpha du', \\
\partial_r &= \alpha\partial_{r'}, & dr &= \alpha^{-1}dr' - r\partial_u\alpha\,du' - \alpha^{-1}r\partial_{x^A}\alpha\,dx^A,
\end{aligned}$$

and hence

$$\eta = \alpha\partial_u - r\partial_u\alpha\,\partial_r = \alpha\Big(\alpha^{-1}\partial_{u'} + r\partial_u\alpha\,\partial_{r'}\Big) - \alpha r\partial_u\alpha\,\partial_{r'} = \partial_{u'}.$$

Moreover, transforming the metric $g$ in (5.108) yields (recall $h=0$ because $\ell^{(2)}=0$)

$$g = 2dr'du' + 2f'_A dr'dx^A + 2h'_A du'dx^A + f'du'^2 + H'_{AB}dx^A dx^B, \tag{5.241}$$

where $f'_A$, $h'_A$, $f'$ and $H'_{AB}$ are given by

$$f'_A = \alpha^{-1} f_A, \qquad h'_A = \alpha h_A - r\partial_u \alpha \; f_A - r\partial_{x^A}\alpha,$$
$$f' = \alpha^2 f - 2\alpha r \partial_u \alpha, \qquad H'_{AB} = H_{AB} - 2\alpha^{-1} r f_{(A}\partial_{x^B)}\alpha.$$

By taking into account (5.106)-(5.107) this means that

$$f'|_{r'=0} = 0, \qquad f'_A|_{r'=0} = \alpha^{-1}\ell_A = \ell'_A, \qquad h'_A|_{r'=0} = 0, \qquad H'_{AB}|_{r'=0} = \gamma_{AB} = \gamma'_{AB},$$

as well as

$$\partial^{(k)}_{r'} f'|_{r'=0} = -2\alpha^{-k+2}\mathbb{k}^{(k)} - 2\delta^k_1 \partial_u \alpha = -2\mathbb{k}'^{(k)}, \qquad \partial^{(k)}_{r'} f'_A|_{r'=0} = 0,$$
$$\partial^{(k)}_{r'} h'_A|_{r'=0} = 2\alpha^{-k+1}\mathbb{r}^{(k)}_A - \delta^k_1 \alpha^{-1}\Big(n(\alpha)\ell_A + \partial_{x^A}(\alpha)\Big) = 2\mathbb{r}^{(k)\prime},$$
$$\partial^{(k)}_{r'} H'_{AB}|_{r'=0} = 2\alpha^{-k}\mathbb{Y}^{(k)}_{AB} - 2\delta^1_k \alpha^{-2}\ell_{(A}\partial_{x^B)}(\alpha) = 2\mathbb{Y}^{(k)\prime},$$

where the last equalities involving $\mathbb{k}^{(k)\prime}$, $\mathbb{r}^{(k)\prime}$ and $\mathbb{Y}^{(k)\prime}$ follow from (5.226)-(5.228). This proves that the coefficients $f'_A$, $h'_A$, $f'$ and $H'_{AB}$ are the same as the ones constructed as in Theorem 5.40 but in terms of $\{\mathbb{Y}^{(m)\prime}\}$ instead of $\{\mathbb{Y}^{(m)}\}$ (because $\xi' = \partial_{r'}$). Hence, after commuting $\pounds_\eta$ with the sum in (5.105), the computation of $\pounds_\eta g$ only requires calculating Lie derivatives of the metric data and of the expansion with respect to $\partial_{u'}|_{\mathcal{H}_0} = n'$, and thus all of them will automatically vanish (see (5.219) and (5.239)). This proves that $\eta$ is a Killing vector of $g$ on $\mathcal{M}_0$, and thus on $\mathcal{M}$. □

Note that when the function $\widetilde{\alpha}$ in the data of Theorem 5.82 has no zeroes the proof can be made much simpler because one can work directly in the adapted $\eta$-gauge from the beginning.

**Remark 5.83.** *Apart from the topology and dimension of the data, our existence theorem differs from Moncrief's result (as seen in Subsection 5.1.2) in that neither the AKH data nor the constructed spacetime are in general analytic. Of course, if we consider analytic AKH data and we restrict it to the subset where $\alpha \neq 0$, a direct application of Theorem 5.2 yields an analytic $\Lambda = 0$ vacuum spacetime.*

In the recent work [216] a similar existence result is proven (up first order transverse derivatives of the metric) for a broader class of abstract horizons in which neither the deformation nor the non-isolation tensors need to vanish.

# 6 ASYMPTOTIC EXPANSION AT NULL INFINITY

In this chapter we are interested in applying the techniques developed in Chapter 5 to analyze how the conformal Einstein equations restrict the geometry at null infinity. In Section 6.1 we begin with a brief historical review, including the Bondi-Sachs formalism and Penrose's conformal completion. As we will see, an important part of the literature focuses only on the four dimensional case and topology $\mathscr{I} \simeq \mathbb{R} \times \mathbb{S}^2$. In this chapter, however, we restrict neither the spacetime dimension nor the topology of $\mathscr{I}$ beyond the assumption that it admits a cross-section. In Section 6.2 we review the necessary tools of conformal geometry that will be needed in the rest of the chapter. In Section 6.3 we show that the Fefferman-Graham ambient metric reviewed in Subsection 5.1.1 admits a conformal completion with a null infinity, and furthermore we find a set of sufficient conditions that characterize the ambient metric from a conformal perspective. Later, in Section 6.4 we write down the conformal equations at null infinity order by order, and we use them in Section 6.5 to show uniqueness and existence theorems for asymptotic data at null infinity. Finally, in Section 6.6 we discuss the relation between one source of obstruction to the smoothness of $\mathscr{I}$ and the Fefferman-Graham obstruction tensor. The results of this chapter have been submitted for publication in [233, 237].

## 6.1 PRELIMINARIES

As already mentioned in the introductory chapter, studying gravitational radiation within the theory of General Relativity has always been a delicate task, as the metric that gives meaning to the fall-off conditions is also the dynamical variable. Since the controversial prediction of gravitational waves by Einstein and Rosen [105, 106, 286], there has been sustained interest in the study of gravitational radiation (see e.g. the seminal works of Pirani, Trautman, and Lichnerowicz [204, 206, 271, 315, 316]). In subsequent works [99, 184, 290, 291, 293], an invariant characterization of outgoing radiation based on the asymptotic fall-off behaviour of the Riemann tensor, known as the *peeling property*, was proposed.

The study of gravitational radiation received a decisive impulse in the 1960s through the work of the Bondi–Pirani group at King's College [44, 218–220, 270], in particular the analysis of axially symmetric radiating systems by Bondi, van der Burg, and Metzner [43], which was shortly afterwards generalized by Sachs [292]. Around the same time, Newman and Penrose developed the Newman–Penrose formalism, which they applied to the study of gravitational radiation [251, 260, 265]. A further important step was taken by Penrose [261, 263], who emphasized that these problems, including the definition of asymptotic flatness, radiation and the BMS group, are most naturally and effectively addressed within the framework of conformal geometry. This approach replaces the physical spacetime by an unphysical one where infinity is attached as a null boundary. We review these developments in Subsections 6.1.1 and 6.1.2 below.

In order to address the question of the existence of asymptotically flat spacetimes, one needs to set up appropriate initial value problems and prove existence results for the Einstein equations. In the conformal picture, this first requires rewriting the Einstein equations in the unphysical spacetime, and then establishing existence of solutions to these conformal field equations. These tasks were accomplished by Friedrich in the early 1980's in a series of seminal papers [122, 123]. His results provided the foundation for the formulation of initial value problems in four dimensions involving null infinity, such as the asymptotic characteristic problem [124, 186, 279] and the hyperboloidal one [21, 22, 125, 127], which we review later in Subsection 6.1.3.

The number of recent advances related to null infinity is remarkable. The structure of null infinity plays a central role in numerical relativity and in the extraction and interpretation of gravitational radiation [121]. It is also fundamental in the understanding of the gravitational memory effect [39, 40, 63, 119], the NP constants [49, 250, 253, 317] as well as in the study of asymptotic symmetries and generalizations of the BMS group [30, 53]. Moreover, null infinity has become a key arena in several developments in quantum gravity, including scattering amplitudes, the gravitational $S$-matrix, infrared issues, and soft theorems (see, for instance, [2–4, 51, 52, 92, 156, 157, 200] and references therein).

### 6.1.1 *Bondi-Sachs formalism*

In this subsection we review the Bondi-Sachs formalism to characterize gravitational radiation, based on the original references mentioned above and the reviews [14, 120, 209, 322].

Let $(M, g)$ be a four dimensional spacetime. The Bondi-Sachs coordinates are based on a family of outgoing null hypersurfaces $\mathcal{N}_u$, labelled by a coordinate $u = const.$ (hence $g^{uu} = 0$), and coordinates $\{r, x^A\}$ on each $\mathcal{N}_u$, where $r$ is the luminosity distance, meaning that

$$\partial_r \left(\det(r^{-2} g_{AB})\right) = 0,$$

and $\{x^A\}_{A=1}^2$ are angular coordinates constant along $\partial_r$, i.e. $g^{uA} = 0$. In this coordinate system, the metric takes the Bondi-Sachs form

$$g = -\frac{V}{r} e^{2\beta} du^2 - 2e^{2\beta} du dr + r^2 h_{AB} \left(dx^A - U^A du\right)\left(dx^B - U^B du\right), \tag{6.1}$$

where the determinant of $h_{AB}$ is that of the unit sphere in the coordinates $\{x^A\}$, and $V$, $\beta$ and $U^A$ are two scalar functions and a vector, respectively. To match with the Minkowski metric at large distances ($r \to \infty$), Sachs assumed that $g$ can be expanded in powers of $1/r$ and that

$$\lim_{r\to\infty} \beta = \lim_{r\to\infty} U^A = 0, \qquad \lim_{r\to\infty} \frac{V}{r} = 1, \qquad \lim_{r\to\infty} h_{AB} = q_{AB},$$

$q_{AB}$ being the unit sphere metric.

The metric components $V$, $\beta$, $U^A$ and $h_{AB}$ can be fixed order by order in powers of $1/r$ by imposing the Einstein equations $G_{\alpha\beta} = 0$ to infinite order at $r \to \infty$. The Bianchi identity $\nabla_\beta G^\beta{}_\alpha = 0$ implies that the main equations of interest are $G^u{}_\alpha = 0$ and $G_{AB} - \frac{G}{2} g_{AB} = 0$ (they entail 6 equations for the 6 components to be solved). The equation $G^u{}_r = 0$ is a first order differential equation with respect to $r$ for $\beta$, $G^u{}_A = 0$ are two second order radial equations determining $U^A$, $G^u{}_u = 0$ is a first order radial equation to determine $V$, and $G_{AB} - \frac{G}{2} g_{AB} = 0$ comprises a radial equation for $h_{AB}$. These equations are highly coupled, but present a hierarchical structure that allows one to integrate them in a sequential order. The prescribed free data is the following [289]

1. The metric $h_{AB}$ on a hypersurface $\mathcal{N}_{u_0}$, which has the following expansion

$$h_{AB}(u_0, r, x^D) = q_{AB} + \frac{c_{AB}(u_0, x^D)}{r} + \cdots .$$

2. The $1/r$ coefficient of $h_{AB}$ everywhere at infinity,

$$c_{AB}(u, x^D) := \lim_{r\to\infty} r(h_{AB}(u, r, x^D) - q_{AB}),$$

which encodes the gravitational radiation.

3. The mass aspect

$$M(u, x^A) := -\frac{1}{2} \lim_{r\to\infty} \Big( V(u, r, x^A) - r \Big)$$

at an initial sphere $u = u_0$ at infinity.

4. A covector $L_A(u_0, x^B)$ at an initial sphere $u = u_0$ at infinity,

$$L_A(u_0, x^B) := -\frac{1}{6} \lim_{r\to\infty} \Big( r^4 e^{-2\beta} h_{AC} \partial_r U^C - r D^B c_{AB} \Big),$$

which is known as the angular momentum aspect.

The derivative of the radiation field $c_{AB}$ along $u$ is called the news tensor,

$$N_{AB} = \frac{1}{2} \partial_u c_{AB}(u, x^D),$$

and is related with the mass aspect by means of

$$\partial_u M = \frac{1}{2} D_A D_B N^{AB} - N_{AB} N^{AB}.$$

Integrating over a sphere and defining $m(u) := \frac{1}{4\pi}\int_{\mathbb{S}^2} M(u,\theta,\phi)d\mu_{\mathbb{S}^2}$, one obtains the Bondi mass loss formula

$$\frac{d}{du}m(u) = -\frac{1}{4\pi}\int_{\mathbb{S}^2} |N|^2 d\mu_{\mathbb{S}^2}.$$

The positivity of the integrand shows that the presence of the news tensor makes the Bondi mass monotonically decreasing. Generalizations of this result to higher dimensions can be found in [145, 166, 170, 177, 312].

The Bondi-Sachs ansatz for the metric, along with the Einstein equations, imply the peeling property of the Weyl tensor [292]. In four dimensions, this property states that the Newman–Penrose components of the Weyl tensor satisfy a hierarchy of fall-off conditions,

$$\Psi_0 = \mathcal{O}(r^{-5}), \qquad \Psi_1 = \mathcal{O}(r^{-4}), \qquad \Psi_2 = \mathcal{O}(r^{-3}), \qquad \Psi_3 = \mathcal{O}(r^{-2}), \qquad \Psi_4 = \mathcal{O}(r^{-1}),$$

as $r \to \infty$. The leading $\mathcal{O}(r^{-1})$ term, encoded in $\Psi_4$, describes outgoing gravitational radiation, while the $\mathcal{O}(r^{-3})$ term, corresponding to $\Psi_2$, carries information about the Bondi mass aspect $M(u, x^A)$ (see, for instance, [322]). We emphasize that the peeling property relies crucially on the assumption that the metric coefficients admit asymptotic expansions in integer powers of $1/r$. In more general settings, such as polyhomogeneous expansions involving terms of the form $r^{-i}\log^j r$, the Weyl tensor typically fails to admit a smooth limit at null infinity, and the peeling hierarchy is violated (see, e.g., [71, 319, 320, 325]).

To conclude this review of the Bondi–Sachs formalism, we recall that the asymptotic symmetry group of the metric (6.1) is known as the Bondi–Metzner–Sachs (BMS) group. It is the semidirect product of the group of supertranslations [294], generated at null infinity by vector fields of the form $f\partial_u$, with the conformal group of the 2-sphere, which is isomorphic to the four-dimensional orthochronous Lorentz group. Unlike the subgroup of translations in the Poincaré group, the supertranslations in the BMS group form an infinite-dimensional Abelian group. The BMS group is closely related to the symmetry groups arising in the study of non-expanding horizons [309], isolated horizons [173], and weakly isolated horizons [32]. In higher dimensions, the specific definition of asymptotic flatness provided in [166] does not allow for an analogue of supertranslations, and the asymptotic isometry group at null infinity reduces to the Poincaré group [312].

### 6.1.2 *Penrose's conformal completion*

The study of asymptotics received a major impulse after Penrose's breakthrough, which showed that conformal infinity is better understood using the tools of conformal geometry. This approach allows one to study $\mathscr{I}$ not as a limit $r \to \infty$, but as a submanifold endowed with its own geometric structure. In this subsection we review the main ideas on this approach following the classic references [142, 261, 263] and the reviews [27, 120, 130–132, 322].

We start with the definition of asymptotic simplicity. A spacetime $(\widetilde{\mathcal{M}}, \widetilde{g})$ is called asymptotically simple if there exists another spacetime $(\mathcal{M}, g)$ such that

1. $\widetilde{\mathcal{M}}$ is an open submanifold of $\mathcal{M}$ with smooth boundary $\partial\widetilde{\mathcal{M}} =: \mathscr{I}$,
2. There exists a smooth function $\Omega$ on $\mathcal{M}$ such that $g = \Omega^2 \widetilde{g}$ on $\widetilde{\mathcal{M}}$, and $\Omega = 0$, $d\Omega \neq 0$ on $\mathscr{I}$.
3. Every null geodesic in $\widetilde{\mathcal{M}}$ acquires a future and past endpoint on $\mathscr{I}$.

A spacetime is called asymptotically flat if, in addition, it satisfies the vacuum equations in a neighbourhood of $\mathscr{I}$. Condition 1. guarantees that the "physical" spacetime $(\widetilde{\mathcal{M}}, \widetilde{g})$ is included in a larger "unphysical" spacetime $(\mathcal{M}, g)$, while condition 2. ensures that $\mathscr{I}$ is a smooth null hypersurface corresponding to the "infinity" of $(\widetilde{\mathcal{M}}, \widetilde{g})$. Condition 3. is meant to encode a completeness property, but in many cases it is definitely too restrictive. For instance, Schwarzschild spacetime is not asymptotically flat in this sense, as there exist null geodesics that do not reach $\mathscr{I}$. To avoid this, one can relax condition 3. and introduce the notion of weakly asymptotically simple spacetimes [264]. In essence, such spacetimes are required to be isometric to an asymptotically simple spacetime in a neighbourhood of $\mathscr{I}$. There are other possibilities to guarantee that $\mathscr{I}$ is complete in a certain way, see e.g. [143]. The completeness of $\mathscr{I}$ is a crucial notion for black holes.

Another key consequence of condition 3., together with the assumption of global hyperbolicity, is that the topology of $\mathscr{I}$ is $\mathbb{R} \times \mathbb{S}^2$ [254], and as a consequence the Weyl tensor vanishes at $\mathscr{I}$. The reason is that, due to the Bianchi identity, the only non-trivial components of ${C^\alpha}_{\beta\mu\nu}$ at $\mathscr{I}$ are those corresponding to a traceless divergence-free tensor on $\mathbb{S}^2$, and thus must vanish (see [116, 322] for the details). In fact, most of the definitions of asymptotic simplicity/flatness already assume either the topology $\mathbb{R} \times \mathbb{S}^2$ or the vanishing of the Weyl tensor at $\mathscr{I}$ from the outset, but it is important to emphasize that there are vacuum spacetimes with $\mathscr{I} \not\simeq \mathbb{R} \times \mathbb{S}^2$ such as e.g. $\mathscr{I} \simeq \mathbb{R} \times T^2$ [299]. As we will see later in Section 6.3, another example of vacuum spacetime with $\mathscr{I} \not\simeq \mathbb{R} \times \mathbb{S}^2$ is the Fefferman-Graham ambient metric. Note also that the non-existence of nontrivial TT tensors on the sphere is specific to two dimensions. In higher dimensions (or in four dimension with a topology other than $\mathbb{S}^2$) one cannot conclude in general that the full Weyl tensor vanishes from the fact that its non-trivial component is a TT tensor.

The common feature that all asymptotically flat spacetimes share is called the *universal structure*, which basically consists of a conformal class of degenerate metrics and null generators at $\mathscr{I}$. This data, however, does not contain the dynamical information of the spacetime. In four dimensions, the further ingredient (radiation) is encoded in the next leading order of the metric. There are many ways of characterizing radiation in the conformal picture. One possibility is to make use of the news tensor, which is defined by subtracting from the ambient Schouten tensor a sort of background, the so-called Geroch $\rho$-tensor [142] (which in spherical topology is uniquely defined). Another approach is by considering an equivalence class of connections on $\mathscr{I}$ (and their curvature), yielding an analogous definition of the news [26]. In any case, the vanishing of the news is equivalent to the vansihing of the radiative components of the rescaled Weyl tensor, and is then interpreted as the no-radiation condition (see [142] and also [114] for a novel conformally covariant definition of news). Other approaches to define radiation include the use of the Bel-Robinson tensor (commonly called superenergy) [115, 116], a tractor connection [90, 158, 160] or carrollian geometry [159, 280].

In higher dimensions, a definition of asymptotic flatness in even dimensions have been proposed using Bondi coordinates (see e.g. [166, 312], where the definition requires a specific rate in which the metric approaches the Minkowski one at infinity). The results in [171] point out that there are not smooth radiating odd dimensional spacetimes.

### 6.1.3 *Initial value problems*

Given an already compactified spacetime, solution of the Einstein equations, it is in principle straightforward to study the geometry inherited at its null infinity. Far more difficult, both conceptually and practically, is the converse problem, namely to analyze the existence of an asymptotically flat spacetime with prescribed boundary data. For that it becomes mandatory to set up an appropriate initial value problem and prove existence of solutions of the Einstein equations. A standard procedure to study this in the unphysical spacetime is to write down equations that are regular on the conformal spacetime (even at the points at infinity). Such equations were obtained by Friedrich in the early 1980s [122, 123]. We refer the reader to Subsection 6.2 where we rederive them.

A main equation in the Friedrich system is the Bianchi equation (see (6.24))

$$\nabla_\alpha \mathfrak{D}^\alpha{}_{\beta\mu\nu} = 0$$

for the rescaled Weyl tensor $\mathfrak{D} := \Omega^{3-d}\,\mathrm{Weyl}$. In four spacetime dimensions (and only in this case) these supply enough evolution equations to form a symmetric hyperbolic system. In higher dimensions, the contracted Bianchi equations alone do not provide enough evolution equations [132].

Just as for the standard Einstein equations described in Section 3.1, the Friedrich equations need to be transformed into partial differential equations for which standard theorems apply. This hyperbolic reduction requires first breaking the coordinate and conformal invariance of the equations, then extracting a proper subsystem of hyperbolic propagation equations, and finally making sure that the solution of the reduced system is indeed a solution of the full system (see [120, 322] for a complete review). The fact that the conformal equations can be cast into a hyperbolic system leads to the existence of smooth solutions to Friedrich's equations from suitable Cauchy data [126]. This was later extended to matter models such as Maxwell, Yang-Mills or scalar fields [128, 174].

The first result concerning the asymptotic characteristic problem is due to Friedrich in the analytic case [124]. It was later suggested by Rendall [279], and finally proven by Kannar [186], that this result can be extended to the smooth setting. In these results one typically has initial data posed on an ingoing null hypersurface $\mathcal{N}$ that intersects $\mathscr{I}$, and on a portion of $\mathscr{I}$ that lies in the future of $\mathcal{N}$. An informal version of such results is the following.

**Theorem 6.1** (Kannar [186])**.** *Given smooth initial data on $\mathcal{N} \cup \mathscr{I}$, there exists a smooth development of the data, solving the conformal Einstein equations, in a future neighbourhood of $\mathcal{N} \cap \mathscr{I}$.*

The initial data on $\mathscr{I}$ corresponds to the radiation field, those on $\mathcal{N}$ to the shear, and those on $\mathcal{N} \cap \mathscr{I}$ to the induced metric and the mass and angular momentum aspects. This is in total agreement with Sachs' analysis [289]. Later, following Luk's ideas [208], the existence region was enlarged to a full neighborhood of $\mathcal{N} \cup \mathscr{I}$ [163] (see also [330]).

Another asymptotic initial value problem is the hyperboloidal one, where the initial data is posed on a hyperboloidal hypersurface that intersects $\mathscr{I}$ transversely [22, 125, 127]. Generically, hyperboloidal initial data give rise to asymptotic expansions including logarithmic terms [21]. This is in accordance with the works already mentioned [71, 319, 320, 325].

The hyperboloidal problem has also been addressed in higher even dimensions by means of the Fefferman-Graham obstruction tensor $\mathcal{O}^{FG}$ [19, 20, 185]. In these works, the authors study the well-posedness of the equation $\mathcal{O}^{FG} = 0$, since the obstruction tensor necessarily vanishes for metrics conformal to an Einstein metric (see Subsection 5.1.1). This Cauchy problem admits many solutions that are not conformal to an Einstein metric, but by constraining appropriately the initial data one can show that the solution is indeed conformally Einstein.

## 6.2 QUASI-EINSTEIN MANIFOLDS

In this section we review the basic aspects of conformal completions of manifolds. The results are well-known. However, our presentation puts the emphasis on the notion of "quasi-Einstein" manifold. This follows [90, 230, 237]. We start by recalling the definition of the Schouten tensor for a $(d \geq 3)$-dimensional spacetime $(\mathcal{M}, g)$ in terms of the Ricci,

$$\mathrm{Sch}_g := \frac{1}{d-2}\left(\mathrm{Ric}_g - \frac{\mathrm{Scal}_g}{2(d-1)} g\right). \tag{6.2}$$

As in Subsection 5.1.1, we will employ the symbols $L_{\alpha\beta}$ and $L$ for the Schouten in abstract index notation and for its trace, respectively. One can easily check that the Schouten and Ricci scalars are related by $L = \frac{R}{2(d-1)}$. From the second Bianchi identity $\nabla_{(\delta} R_{\alpha\beta)\mu\nu} = 0$ it follows that

$$\nabla^\mu L_{\mu\alpha} - \nabla_\alpha L = 0, \quad (6.3) \qquad \nabla_\mu C^\mu{}_{\nu\alpha\beta} = (d-3)\left(\nabla_\alpha L_{\beta\nu} - \nabla_\beta L_{\alpha\nu}\right), \quad (6.4)$$

$C^{\mu}{}_{\nu\alpha\beta}$ being the Weyl tensor in index notation. The Weyl, Schouten and Ricci tensors transform as follows under a conformal rescaling $\widehat{g} = \omega^2 g$ [323],

$$\widehat{C}^{\alpha}{}_{\beta\mu\nu} = C^{\alpha}{}_{\beta\mu\nu}, \tag{6.5}$$

$$\widehat{L}_{\alpha\beta} = L_{\alpha\beta} - \frac{\nabla_\alpha \nabla_\beta \omega}{\omega} + \frac{2\nabla_\alpha \omega \nabla_\beta \omega}{\omega^2} - \frac{|\nabla\omega|^2_g}{2\omega^2} g_{\alpha\beta}, \tag{6.6}$$

$$\widehat{R}_{\alpha\beta} = R_{\alpha\beta} - \frac{(d-2)\nabla_\alpha \nabla_\beta \omega}{\omega} + \frac{2(d-2)\nabla_\alpha \omega \nabla_\beta \omega}{\omega^2} - \frac{(d-3)|\nabla\omega|^2_g + \omega \Box_g \omega}{\omega^2} g_{\alpha\beta}, \tag{6.7}$$

$$\widehat{\nabla} = \nabla + \omega^{-1}\left(2\,\mathrm{Id} \otimes_s d\omega - \nabla\omega \otimes g\right), \tag{6.8}$$

$$\widehat{L} = \omega^{-2}\left(L - \frac{\Box_g \omega}{\omega} - \frac{(d-4)|\nabla\omega|^2_g}{2\omega^2}\right). \tag{6.9}$$

In particular, from (6.6)

$$L^{tf}_{\alpha\beta} = \widetilde{L}^{tf}_{\alpha\beta} - \frac{m}{2}\left(\frac{\widetilde{\nabla}_\alpha \widetilde{\nabla}_\beta \omega}{\omega} - \frac{m+2}{2\omega^2}\widetilde{\nabla}_\alpha \omega \widetilde{\nabla}_\beta \omega\right)^{tf}, \tag{6.10}$$

where "tf" denotes the trace-free part w.r.t. $g$ (or w.r.t. $\widetilde{g}$, since "tf" is a conformally invariant operation). Let $(\mathcal{M}, [g])$ be a $d$-dimensional conformal structure of arbitrary pseudo-Riemannian signature. For each $g \in [g]$ we construct the differential operator

$$A_g(f) := (\mathrm{Hess}_g f + f\,\mathrm{Sch}_g)^{tf}, \qquad f \in \mathcal{F}(\mathcal{M}). \tag{6.11}$$

In the following lemma we check that $A_{\omega^2 g}(\omega f) = \omega A_g(f)$.

**Lemma 6.2.** *Let $\omega \in \mathcal{F}^\star(\mathcal{M})$ be everywhere positive. Then,*

$$A_{\omega^2 g}(\omega f) = \omega A_g(f). \tag{6.12}$$

*Proof.* First we compute the transformation of the trace-free part of $\mathrm{Hess}_g F$ for $F \in \mathcal{F}(\mathcal{M})$. From (6.8),

$$\widehat{\nabla}_\alpha \widehat{\nabla}_\beta F = \nabla_\alpha \nabla_\beta F - \omega^{-1}\left(2\nabla_{(\alpha}\omega \nabla_{\beta)} F - \langle \nabla\omega, \nabla F\rangle_g g_{\alpha\beta}\right), \tag{6.13}$$

so

$$\mathrm{Hess}^{tf}_{\widehat{g}} F = \mathrm{Hess}^{tf}_g F - 2\omega^{-1}\left(d\omega \otimes_s dF - \frac{\langle \nabla\omega, \nabla F\rangle_g}{d} g\right).$$

Putting $F = \omega f$ in the previous equation,

$$\mathrm{Hess}^{tf}_{\widehat{g}}(\omega f) = f\,\mathrm{Hess}^{tf}_g \omega + \omega\,\mathrm{Hess}^{tf}_g f - 2f\omega^{-1}\left(d\omega \otimes d\omega - d^{-1}|\nabla\omega|^2_g g\right). \tag{6.14}$$

On the other hand, from (6.6),

$$\mathrm{Sch}^{tf}_{\widehat{g}} = \mathrm{Sch}^{tf}_g - \omega^{-1}\,\mathrm{Hess}^{tf}_g \omega + 2\omega^{-2} d\omega \otimes d\omega - 2\omega^{-2} d^{-1}|\nabla\omega|^2 g. \tag{6.15}$$

Combining equations (6.14) and (6.15),

$$A_{\omega^2 g}(\omega f) = \omega\,\mathrm{Hess}^{tf}_g f + \omega f\,\mathrm{Sch}^{tf}_g = \omega A_g(f).$$

□

One immediate consequence of (6.12) is that

$$\Omega^{-2} g \in [g] \text{ is Einstein} \quad \Longleftrightarrow \quad (\operatorname{Hess}_g \Omega + \Omega \operatorname{Sch}_g)^{tf} = 0.$$

The equation $(\operatorname{Hess}_g \Omega + \Omega \operatorname{Sch}_g)^{tf} = 0$ can be rewritten in an equivalent way in terms of the scalar

$$q := \frac{\Box_g \Omega + \Omega L}{d} \tag{6.16}$$

as $\operatorname{Hess}_g \Omega + \Omega \operatorname{Sch}_g - qg = 0$. This discussion motivates the following definition.

**Definition 6.3.** *Let $(\mathcal{M}, g)$ be a pseudo-Riemannian manifold of dimension $d \geq 3$, $\Omega \in \mathcal{F}(\mathcal{M})$ a non-identically zero function and $\mathcal{T}$ a trace-free, two-covariant tensor field. We say that the four-tuple $(\mathcal{M}, g, \Omega, \mathcal{T})$ is a quasi-Einstein manifold provided that*

$$\operatorname{Hess}_g \Omega + \Omega \operatorname{Sch}_g - qg = \mathcal{T}. \tag{6.17}$$

*A quasi-Einstein manifold is called vacuum when $\mathcal{T} = 0$. Note that if $(\mathcal{M}, g, \Omega, \mathcal{T})$ is a quasi-Einstein manifold, then $(\mathcal{M}, \omega^2 g, \omega\Omega, \omega\mathcal{T})$ is also a quasi-Einstein manifold for every $\omega \in \mathcal{F}^\star(\mathcal{M})$.*

From the transformation $\widehat{\Omega} = \omega\Omega$ and using (6.8) and (6.9), the behaviour of $q$ under a conformal rescaling is

$$\widehat{q} = \omega^{-1} \left( q + \frac{\langle \nabla\Omega, \nabla\omega \rangle_g}{\omega} + \frac{\Omega |\nabla\omega|_g^2}{2\omega^2} \right). \tag{6.18}$$

Next we write down two direct consequences of equation (6.17).

**Lemma 6.4.** *Let $(\mathcal{M}, g, \Omega, \mathcal{T})$ be a quasi-Einstein manifold. Then,*

$$\nabla_\mu q = \frac{1}{d-1} \nabla_\rho \mathcal{T}^\rho{}_\mu - L^\rho{}_\mu \nabla_\rho \Omega, \tag{6.19}$$

*and*

$$C^\alpha{}_{\beta\mu\nu} \nabla_\alpha \Omega = 2\Omega \nabla_{[\mu} L_{\nu]\beta} - 2\nabla_{[\mu} \mathcal{T}_{\nu]\beta} + \frac{2}{d-1} g_{\beta[\mu} \nabla^\rho \mathcal{T}_{\nu]\rho} \tag{6.20}$$

$$= \frac{\Omega}{d-3} \nabla_\alpha C^\alpha{}_{\beta\mu\nu} - 2\nabla_{[\mu} \mathcal{T}_{\nu]\beta} + \frac{2}{d-1} g_{\beta[\mu} \nabla^\rho \mathcal{T}_{\nu]\rho}. \tag{6.21}$$

*Proof.* The Ricci identity for $\nabla_\beta \Omega$ is

$$2\nabla_{[\mu} \nabla_{\nu]} \nabla_\beta \Omega + R^\rho{}_{\beta\mu\nu} \nabla_\rho \Omega = 0,$$

which after inserting equation (6.17) and $C_{\alpha\beta\mu\nu} := R_{\alpha\beta\mu\nu} - 2(L_{\alpha[\mu} g_{\nu]\beta} + 2L_{\beta[\nu} g_{\mu]\alpha})$ yields

$$C^\rho{}_{\beta\mu\nu} \nabla_\rho \Omega + 2\nabla_{[\mu} q\, g_{\nu]\beta} - 2L_{\beta[\mu} \nabla_{\nu]} \Omega - 2\Omega \nabla_{[\mu} L_{\nu]\beta} + 2g_{\beta[\nu} L_{\mu]\rho} \nabla^\rho \Omega + 2\nabla_{[\mu} \mathcal{T}_{\nu]\beta} = 0. \tag{6.22}$$

Contracting the indices $\{\beta, \nu\}$ and using (6.3),

$$(d-1)\left(\nabla_\mu q + L_{\mu\rho}\nabla^\rho\Omega\right) - \nabla^\rho \mathcal{T}_{\rho\mu} = 0,$$

which is (6.19). Inserting this back into (6.22) gives (6.20). Equation (6.21) follows at once after using (6.4). □

As we show next, one can define a conformally invariant constant on each connected component of a quasi-Einstein manifold.

**Lemma 6.5.** *Let $(\mathcal{M}, g, \Omega, \mathcal{T})$ be a quasi-Einstein manifold satisfying*

$$\Omega\nabla_\beta \mathcal{T}^{\alpha\beta} - (d-1)\mathcal{T}^{\alpha\beta}\nabla_\beta\Omega = 0.$$

*Then, the quantity*

$$\lambda := 2q\Omega - |\nabla\Omega|^2_g \tag{6.23}$$

*is constant on each connected component of $\mathcal{M}$. Moreover, this constant is conformally invariant.*

*Proof.* Contracting (6.17) with $\nabla\Omega$ we get

$$\frac{1}{2}\nabla_\alpha\left(\nabla_\beta\Omega\nabla^\beta\Omega\right) + L_{\alpha\beta}\nabla^\beta\Omega - q\nabla_\alpha\Omega - \mathcal{T}_{\alpha\beta}\nabla^\beta\Omega = 0,$$

which upon inserting (6.19) gives

$$\nabla_\alpha\left(\nabla_\beta\Omega\nabla^\beta\Omega - 2q\Omega\right) + 2\left(\frac{\Omega}{d-1}\nabla^\beta\mathcal{T}_{\beta\alpha} - \mathcal{T}_{\alpha\beta}\nabla^\beta\Omega\right) = 0.$$

Under the conditions of the Lemma, $\nabla_\alpha\left(\nabla_\beta\Omega\nabla^\beta\Omega - 2q\Omega\right)$ vanishes, so the quantity $-\lambda = |\nabla\Omega|^2_g - 2q\Omega$ is constant along each connected component of $\mathcal{M}$. The transformation of $\lambda$ under conformal a rescaling is

$$\widehat{\lambda} = 2\widehat{q}\widehat{\Omega} - |\nabla\widehat{\Omega}|^2_{\omega^2 g} = 2\widehat{q}\omega\Omega - \omega^{-2}|\omega\nabla\Omega + \Omega\nabla\omega|^2_g = 2q\Omega - |\nabla\Omega|^2_g = \lambda,$$

where in the last step (6.18) has been introduced. □

The geometric interpretation of $\lambda$ is as follows. As we have already seen, given a vacuum quasi-Einstein manifold $(\mathcal{M}, g, \Omega, \mathcal{T} = 0)$, the metric $\widehat{g} := \Omega^{-2}g$ is Einstein. Then, plugging $\omega = \Omega^{-1}$ into (6.6) and using the quasi-Einstein equation (6.17),

$$\widehat{L}_{\alpha\beta} = L_{\alpha\beta} + \frac{\nabla_\alpha\nabla_\beta\Omega}{\Omega} - \frac{|\nabla\Omega|^2}{2\Omega^2}g_{\alpha\beta} = \frac{2q\Omega - |\nabla\Omega|^2_g}{2\Omega^2}g_{\alpha\beta} = \frac{\lambda}{2}\widehat{g}_{\alpha\beta}.$$

Then, ignoring a numerical factor, the quantity $\lambda$ is the cosmological constant associated to the Einstein representative of $(\mathcal{M}, [g])$. Thus, it makes sense to talk about $\lambda$-vacuum quasi-Einstein manifolds.

Recall that at the points where $\Omega \neq 0$ the rescaled Weyl tensor is defined as $\mathfrak{D}^{\alpha}{}_{\beta\mu\nu} := \Omega^{3-d}C^{\alpha}{}_{\beta\mu\nu}$. From (6.20),

$$\nabla_{\alpha}\mathfrak{D}^{\alpha}{}_{\beta\mu\nu} = 2(d-3)\Omega^{2-d}\left(\nabla_{[\mu}\mathcal{T}_{\nu]\beta} - \frac{1}{d-1}\nabla_{\alpha}\mathcal{T}^{\alpha}{}_{[\nu}g_{\mu]\beta}\right). \tag{6.24}$$

It then follows that for $\lambda$-vacuum quasi-Einstein manifolds, the rescaled Weyl tensor satisfies a regular PDE on the closure of $\{\Omega \neq 0\}$. As we show now, $\{\Omega \neq 0\}$ is dense on $\mathcal{M}$, so $\mathfrak{D}^{\alpha}{}_{\beta\mu\nu}$ satisfies a regular PDE everywhere on $\mathcal{M}$. This does not mean, however, that $\mathfrak{D}^{\alpha}{}_{\beta\mu\nu}$ extends regularly to $\{\Omega = 0\}$. The first part of the lemma is well-known, see e.g. [197]. The second is probably known to experts but we could not find an explicit proof in the literature.

**Lemma 6.6.** *Let $(\mathcal{M}, g, \Omega)$ a $\lambda$-vacuum quasi-Einstein manifold with constant $\lambda$ and assume $\{\Omega = 0\} \neq \emptyset$. Then,*

1. *If $\lambda \neq 0$, $\{\Omega = 0\}$ is an embedded hypersurface with non-zero normal given by $\nabla\Omega$. Furthermore, it is spacelike when $\lambda > 0$, and timelike when $\lambda < 0$.*
2. *If $\lambda = 0$, then except for a (maybe empty) collection of isolated points $\{p_i\}$, $\{\Omega = 0\}$ is an embedded null hypersurface with nowhere zero normal $\nabla\Omega$.*

*Proof.* A point $p \in \mathcal{M}$ is called singular provided that $\Omega|_p = 0$ and $\nabla\Omega|_p = 0$. Then, from equation (6.23) the set $\{\Omega = 0\}$ cannot admit any singular point in the case $\lambda \neq 0$, and therefore it is a smooth embedded hypersurface. The vector field $\nabla\Omega$ is normal to $\{\Omega = 0\}$ and again by equation (6.23) we see that $\{\Omega = 0\}$ is spacelike when $\lambda > 0$ and timelike when $\lambda < 0$.

For item 2. we first prove that singular points are necessarily isolated. Let $p \in \{\Omega = 0\}$ be a singular point and define $q_p := q(p)$. By (6.18) the transformation of $q_p$ is $\hat{q}_p = \omega(p)^{-1}q_p$, so being zero/non-zero is a conformally invariant statement. Let us prove that $q_p \neq 0$. Let $\{e_a\}$ be an orthonormal basis of $T_p\mathcal{M}$ and let $\gamma(v)$, $v \in \mathcal{I} \subseteq \mathbb{R}$, be an affinely parametrized geodesic starting at $p$. Let $\{e_a(v)\}$ the basis at $T_{\gamma(v)}\mathcal{M}$ obtained by parallel transport of $\{e_a\}$ along $\gamma(v)$. Define the following quantities

$$\tilde{\Omega}(v) := \Omega(\gamma(v)), \qquad \tilde{\Omega}_a(v) := \langle \nabla\Omega|_{\gamma(v)}, e_a(v)\rangle_g, \qquad \tilde{q}(v) := q(\gamma(v)).$$

If $\eta_{ab} := g(e_a, e_b)$ then $g(e_a(v), e_b(v)) = \eta_{ab}$ for all $v \in \mathcal{I}$. Observe that

$$\nabla\Omega|_{\gamma(v)} = \eta^{ab}\tilde{\Omega}_a(v)e_b(v), \tag{6.25}$$

where $\eta^{ab} := \eta_{ab}$. Let us establish the following equations,

$$\begin{aligned} \frac{d\tilde{\Omega}}{dv} &= \langle \dot{\gamma}, \nabla\Omega|_{\gamma(v)}\rangle_g = \eta^{ab}\tilde{\Omega}_a(v)\langle\dot{\gamma}(v), e_a(v)\rangle_g, \\ \frac{d\tilde{\Omega}_a}{dv} &= (\nabla_{\dot{\gamma}}\nabla_{e_a}\Omega)|_{\gamma} = (\operatorname{Hess}_g \Omega)|_{\gamma}(\dot{\gamma}, e_a) = \tilde{q}\langle\dot{\gamma}, e_a\rangle_g - \tilde{\Omega}\operatorname{Sch}_g(\dot{\gamma}, e_a), \\ \frac{d\tilde{q}}{dv} &= \nabla_{\dot{\gamma}}q|_{\gamma} = -\operatorname{Sch}_g(\dot{\gamma}, \nabla\Omega) = -\eta^{ab}\operatorname{Sch}_g(\dot{\gamma}, e_a)\tilde{\Omega}_b, \end{aligned} \tag{6.26}$$

where the first one follows from (6.25), the second from (6.17) and the last one from (6.19). We also compute for later use

$$\frac{d^2\tilde{\Omega}}{dv^2} = \nabla_{\dot{\gamma}}\nabla_{\dot{\gamma}}\Omega|_\gamma = \text{Hess}_g\,\Omega(\dot{\gamma},\dot{\gamma}) = \tilde{q}g(\dot{\gamma},\dot{\gamma}) - \tilde{\Omega}\,\text{Sch}_g(\dot{\gamma},\dot{\gamma}). \tag{6.27}$$

Equations (6.26) constitute a linear first order system of homogeneous ODE for $\{\tilde{\Omega},\tilde{\Omega}_a,\tilde{q}\}$. Since $p$ is a singular point we have $\tilde{\Omega}(0) = \tilde{\Omega}_a(0) = 0$. If, moreover, $q_p = 0$, then also $\tilde{q}(0) = 0$ and thus by uniqueness of the solution, all the points on $\gamma(v)$ would be singular with $q = 0$. Since $\gamma$ is arbitrary it follows that the set of singular points with $q = 0$ is open (and obviously also closed). Since $\mathcal{M}$ is connected, this set is the whole of $\mathcal{M}$, which contradicts the fact that $\Omega$ is not identically zero. Then, $q_p \neq 0$ at any singular point. Next we prove that singular points are isolated. Since $q_p \neq 0$ it follows from (6.17) that $\text{Hess}_g\,\Omega|_p \neq 0$ and thus $p$ is a non-degenerate critical point of $\Omega$. By Morse lemma [48] there exists a neighbourhood $\mathcal{U}$ of $p$ and coordinates $\{x^\alpha\}_{\alpha=1}^d$ with $x^\alpha(p) = 0$ such that $\Omega = -(x^1)^2 - \cdots - (x^\sigma)^2 + (x^{\sigma+1})^2 + \cdots + (x^d)^2$, the number $\sigma$ being the Morse index (i.e. the signature of the Hessian). Clearly the differential $d\Omega$ on $\mathcal{U}$ only vanishes at $p$, so $p$ is isolated and $\nabla\Omega$ is a non-zero normal to $(\{\Omega = 0\}\setminus\{p\})\cup\mathcal{U}$, and thus by (6.23) this set is a smooth null hypersurface (it may be empty, e.g. when $g$ is Riemannian). □

We then define $\mathscr{I}$ as the set of points where $\{\Omega = 0\}$ and $\nabla\Omega \neq 0$. Equations (6.17), (6.19) and (6.24) are known as conformal field equations (or Friedrich equations) [131, 132]. Their key property is that they are regular at $\mathscr{I}$.

Let $(\mathcal{M},g,\Omega)$ be a $d = \mathfrak{n}+1$ dimensional (vacuum) quasi-Einstein manifold with $\lambda = 0$. Consider an embedding $\Phi : \mathscr{I} \hookrightarrow \mathcal{M}$ and rigging $\xi$, and let $\{\mathscr{I},\boldsymbol{\gamma},\boldsymbol{\ell},\ell^{(2)}\}$ be the corresponding embedded null metric hypersurface data. Since $\nabla\Omega$ is non-vanishing, null and tangent to $\mathscr{I}$, there must exist a non-vanishing function $\sigma$ such that $\nabla\Omega = \sigma\nu$ (and hence $\pounds_\xi\Omega = \sigma$) on $\mathscr{I}$. Moreover, by pull-backing (6.17) to $\Omega = 0$ and noting that $e_a^\alpha e_b^\beta \nabla_\alpha\nabla_\beta\Omega = \frac{1}{2}e_a^\alpha e_b^\beta\left(\pounds_{\nabla\Omega}g\right)_{\alpha\beta} = \sigma\text{U}_{ab}$, it follows that $\mathbf{U} = q\sigma^{-1}\boldsymbol{\gamma}$, so $\mathscr{I}$ is totally umbilical. Furthermore, the conformal freedom $g' = \omega^2 g$ leaves a remnant conformal freedom on $\mathscr{I}$ of the form

$$\boldsymbol{\gamma}' = \omega^2\boldsymbol{\gamma}, \qquad \boldsymbol{\ell}' = \omega^2\boldsymbol{\ell}, \qquad \ell^{(2)\prime} = \omega^2\ell^{(2)}, \qquad P' = \omega^{-2}P, \qquad n' = \omega^{-2}n.$$

Since $\nabla'\Omega' = \omega^{-1}\nabla\Omega = \omega^{-1}\sigma\nu = \omega\sigma\nu'$ on $\mathscr{I}$, it follows that the function $\sigma$ scales by $\sigma' = \omega\sigma$ and $q|_{\mathscr{I}}$ by (cf. (6.18)) $q' \stackrel{\mathscr{I}}{=} \omega^{-1}q + \omega^{-2}\sigma\pounds_n\omega$. The functions $q$ and $\omega$ are obviously independent of the rigging, so $q|_{\mathscr{I}}$ and $\omega|_{\mathscr{I}}$ are gauge invariant quantities. One the other hand $\sigma = \pounds_\xi\Omega|_{\mathscr{I}}$ has gauge behavior

$$\mathcal{G}_{(z,V)}\sigma = z\sigma. \tag{6.28}$$

Observe also that $q|_{\mathscr{I}}$ and $\sigma|_{\mathscr{I}}$ are not independent, because from (6.23) $\nabla^\alpha\Omega\nabla_\alpha\nabla_\beta\Omega = \frac{1}{2}\nabla_\beta\Big(|\nabla\Omega|^2_g\Big) \stackrel{\mathscr{I}}{=} q\nabla_\beta\Omega$, so $q|_{\mathscr{I}}$ is the surface gravity of $\nabla\Omega = \sigma\nu$ at $\mathscr{I}$. Given that $\nabla_\nu\nu = \kappa\nu$ (cf. (2.41)) we also have

$$\nabla_{\alpha\nu}(\sigma\nu) = (\pounds_n\sigma + \sigma\kappa)\sigma\nu,$$

and hence $q \stackrel{\mathscr{I}}{=} \pounds_n\sigma + \sigma\kappa$. With this relation at hand, it is easy to check using (B.11) in Appendix B that the pullback of (6.16) to $\Omega = 0$ is automatically fulfilled. This motivates the following definition of "universal structure". As already reviewed in Section 6.1, this notion is commonly used in the literature (see e.g. [27, 82, 142, 160]). Here we are just adapting it to the context of metric hypersurface data.

**Definition 6.7.** *We say* $\{\mathscr{I},\gamma,\boldsymbol{\ell},\ell^{(2)},\sigma,\mathfrak{q}\}$ *is $\mathscr{I}$-**structure data** provided* $\{\mathscr{I},\gamma,\boldsymbol{\ell},\ell^{(2)}\}$ *is null metric hypersurface data,* $\sigma\in\mathcal{F}^\star(\mathscr{I})$*,* $\mathfrak{q}\in\mathcal{F}(\mathscr{I})$ *and* $\mathbf{U} = \mathfrak{q}\sigma^{-1}\gamma$*. Moreover, the gauge transformations of* $\sigma$ *and* $\mathfrak{q}$ *are* $\mathcal{G}_{(z,V)}\sigma := z\sigma$ *and* $\mathcal{G}_{(z,V)}\mathfrak{q} := \mathfrak{q}$*. Furthermore, we define the conformal transformation* $\mathcal{C}_\omega$ *of* $\{\mathscr{I},\gamma,\boldsymbol{\ell},\ell^{(2)},\sigma,\mathfrak{q}\}$*, where* $\omega\in\mathcal{F}^\star(\mathscr{I})$ *is a gauge-invariant function (called conformal factor), by*

$$\mathcal{C}_\omega\gamma = \omega^2\gamma, \qquad \mathcal{C}_\omega\boldsymbol{\ell} = \omega^2\boldsymbol{\ell}, \qquad \mathcal{C}_\omega\ell^{(2)} = \omega^2\ell^{(2)}, \qquad \mathcal{C}_\omega\sigma = \omega\sigma, \qquad \mathcal{C}_\omega\mathfrak{q} = \omega^{-1}\mathfrak{q} + \omega^{-2}\sigma\pounds_n\omega.$$

*It is straightforward to check that* $\mathcal{C}_\omega\circ\mathcal{G}_{(z,V)} = \mathcal{G}_{(z,V)}\circ\mathcal{C}_\omega$*.*

Note that given $\mathscr{I}$-structure data and $X\in\mathfrak{X}(\mathscr{I})$, the $\nabla$-derivative of $\nabla\Omega$ along $X$ at $\mathscr{I}$ is, by (2.41) and (2.66),

$$\nabla_X(\nabla\Omega) = \sigma P^{ab}\mathrm{U}_{bc}X^c e_a + \Big(\sigma(\mathbf{s}-\mathbf{r}) + d\sigma\Big)(X)n = \mathfrak{q}X + \Big(\sigma(\mathbf{s}-\mathbf{r}) + d\sigma - \mathfrak{q}\boldsymbol{\ell}\Big)(X)n. \tag{6.29}$$

Next we define the embedded version of the data to guarantee that $\sigma$ agrees with the proportionality function between $\nu$ and $\nabla\Omega$ at $\mathscr{I}$, and also that $\mathfrak{q}$ is the pullback of the scalar $q$ defined in (6.16).

**Definition 6.8.** *We say that* $\{\mathscr{I},\gamma,\boldsymbol{\ell},\ell^{(2)},\sigma,\mathfrak{q}\}$ *is* $(\Phi,\xi)$*-embedded in* $(\mathcal{M},g,\Omega)$ *provided that* $\{\mathscr{I},\gamma,\boldsymbol{\ell},\ell^{(2)}\}$ *is* $(\Phi,\xi)$*-embedded in* $(\mathcal{M},g)$*,* $\Omega = 0$ *in* $\Phi(\mathscr{I})$*, and in addition* $\nabla\Omega \stackrel{\mathscr{I}}{=} \sigma\nu$ *and* $\mathfrak{q} = \sigma\kappa + \pounds_n\sigma$*.*

From the conformal transformation of $\mathfrak{q}$ in Def. 6.7 it follows that whenever $\mathscr{I}$ admits cross-sections one can always find a conformal factor $\omega$ that satisfies $\omega^2\mathfrak{q} + \sigma\pounds_n\omega = 0$ and, as a consequence, $\mathcal{C}_\omega\mathfrak{q} = 0$. The remaining conformal freedom is the function $\omega$ at any cross-section. In any of such conformal gauges $\mathscr{I}$ is totally geodesic, $\mathbf{U} = 0$.

## 6.3 CONFORMAL CHARACTERIZATION OF THE AMBIENT METRIC

Before analyzing the geometry of null infinity in the general case (Section 6.4) using the tools of Chapter 5 and the identities of Section 6.2, in this section we consider the Fefferman–Graham ambient metric introduced in Subsection 5.1.1 and we show that every straight ambient metric admits a conformal completion with a well-defined null infinity. We start by considering a straight ambient metric

$$g = 2\rho dt^2 + 2tdtd\rho + t^2\mu \tag{6.30}$$

and the change of coordinates $\{t, \rho\} \longmapsto \{t, u := \rho t\}$, under which (6.30) takes the form

$$g = 2dtdu + t^2\mu, \tag{6.31}$$

where now $\mu$ is a series in powers of $\frac{u}{t}$. In these coordinates the homothetic horizon ($\rho = 0$) is placed at $u = 0$, and the infinity is reached when $t \to \infty$. Introducing the coordinate $v := t^{-1}$ the metric (6.31) becomes

$$g = v^{-2}(-2dudv + \mu), \tag{6.32}$$

where now $\mu$ is a series in powers of $uv$. Then, the metric

$$\widetilde{g} := v^2 g = -2dudv + \mu \tag{6.33}$$

is extendible beyond $\mathscr{I} := \{v = 0\}$, the (null) conformal infinity of the ambient metric. Note that if $g$ is Ricci flat, then $(\widetilde{g}, v)$ satisfies (6.17) with $\mathcal{T} = 0$. Following the notation of Subsection 5.1.1, for $r$ odd the tensor $\mu$ is given as a formal series by

$$\mu = \sum_{k=0}^{\frac{r-1}{2}} \frac{1}{k!}\mu^{(k)}(uv)^k + \widetilde{\mu}(uv)^{\frac{r}{2}} + O\Big((uv)^{\frac{r+1}{2}}\Big),$$

while for $r \geq 4$ even $\mu$ is given by

$$\mu = \sum_{k=0}^{\frac{r}{2}-1} \frac{1}{k!}\mu^{(k)}(uv)^k + \mu^{(r/2)}(uv)^{r/2} + \widetilde{\mu}(uv)^{\frac{r}{2}}\log(uv) + O\Big((uv)^{\frac{r}{2}+1}\log(uv)\Big).$$

As already said, the case $r = 2$ is special because no obstruction tensor appears and the ambient metric is always smooth.

Note that in these coordinates the vector $T$ is given by $T = u\partial_u - v\partial_v$. It is straightforward to check that $T$ is a Killing vector w.r.t. $\widetilde{g}$ and that the set $\{u = 0\} \cup \{v = 0\}$ is a (non-degenerate) bifurcate Killing horizon with bifurcation surface $\{u = v = 0\}$. Also note that the straightness condition implies that the one-form $\widetilde{\boldsymbol{T}} := \widetilde{g}(T, \cdot)$ is integrable in the conformal manifold $(\widetilde{M} := M \cup \mathscr{I}, \widetilde{g}, \Omega)$, i.e. $\widetilde{\boldsymbol{T}} \wedge d\widetilde{\boldsymbol{T}} = 0$. Being part of a Killing horizon, it follows that $\mathscr{I}$ is a totally geodesic null hypersurface with first fundamental form $h$. Moreover, the transverse derivatives of $\widetilde{g}$ at $\{v = 0\}$ are

$$\partial_v^m \widetilde{g} \stackrel{\mathscr{I}}{=} \mu^{(m)} u^m, \qquad 1 \leq m \leq \lceil r/2 - 1 \rceil,$$

while the transverse derivatives along $s := uv$ at $s = 0$ are given by[1]

$$\partial_s^m \widetilde{g} \overset{s=0}{=} \mu^{(m)}, \qquad 1 \leq m \leq \lceil r/2 - 1 \rceil. \tag{6.34}$$

This means that the free data $\Theta_{AB}$ agrees with the trace-free part of the coefficient in $s^{r/2}$.

Particularizing equation (6.23) to $\lambda = 0$ and taking into account that $|\widetilde{\nabla}\Omega|^2_{\widetilde{g}} = |\widetilde{\nabla}v|^2_{\widetilde{g}} = 0$ it follows that $q = 0$, so $\mathcal{Q}_{\alpha\beta} := r\left(\widetilde{\nabla}_\alpha \widetilde{\nabla}_\beta \Omega + \Omega \widetilde{L}_{\alpha\beta}\right) = \mathcal{Q}^{tf}_{\alpha\beta}$ and thus the quasi-Einstein equations can be written as $\mathcal{Q}_{\alpha\beta} = 0$. In other words, if $g$ is exactly Ricci flat, then $\widetilde{g}$ satisfies $\mathcal{Q}_{\alpha\beta} = 0$. Let us consider an even dimensional ambient metric constructed by solving $\partial_\rho^{(m)} R_{\alpha\beta} \overset{\rho=0}{=} 0$ up to an including order $m = r/2 - 2$. As we discussed in Subsection 5.1.1, the determination of the next coefficient of the expansion is (in general) obstructed, in the sense that $(\partial_\rho^{(r/2-1)} R_{AB})^{TF} \overset{\rho=0}{=} \mathcal{O}^{\mathscr{F}\mathscr{G}}_{AB} \neq 0$, where $\mathcal{O}^{\mathscr{F}\mathscr{G}}_{AB}$ is the Fefferman–Graham obstruction tensor of $h = \mu^{(0)}$. This obstruction can also be detected at $\mathscr{I}$, as we explain next.

First observe that $\partial_\rho^m R_{\alpha\beta} \overset{\rho=0}{=} 0 \quad \forall m \leq r/2 - 2$ implies $\partial_\rho^m L_{\alpha\beta} \overset{\rho=0}{=} 0$ and $\partial_\rho^m R \overset{\rho=0}{=} \partial_\rho^m L \overset{\rho=0}{=} 0 \quad \forall m \leq r/2 - 2$ ($R$ and $L$ are the Ricci and Schouten scalars of $g$, respectively), so (cf. (6.2))

$$\mathcal{O}^{\mathscr{F}\mathscr{G}}_{AB} = (\partial_\rho^{r/2-1} R_{AB}|_{\rho=0})^{TF} = \left( r\partial_\rho^{r/2-1} L_{AB}|_{\rho=0} + \frac{\partial_\rho^{r/2-1} R|_{\rho=0}}{2(r+1)} h_{AB} \right)^{TF},$$

and therefore

$$\mathcal{O}^{\mathscr{F}\mathscr{G}}_{AB} = r(\partial_\rho^{r/2-1} L_{AB}|_{\rho=0})^{TF}.$$

Recalling that $L^{tf}_{\alpha\beta} = \widetilde{L}^{tf}_{\alpha\beta} + \Omega^{-1}(\mathrm{Hess}_{\widetilde{g}}\Omega)^{tf}_{\alpha\beta} = \Omega^{-1} r^{-1} \mathcal{Q}_{\alpha\beta}$ (see (6.10)) it follows that

$$L_{\alpha\beta} - \frac{L}{r+2} g_{\alpha\beta} = \Omega^{-1} r^{-1} \mathcal{Q}_{\alpha\beta} \qquad \Longrightarrow \qquad L_{AB} - \frac{L}{r+2} t^2 \mu_{AB} = \Omega^{-1} r^{-1} \mathcal{Q}_{AB},$$

and then (note that $\partial_\rho \Omega = \partial_\rho t^{-1} = 0$)

$$\partial_\rho^{r/2-1} L_{AB}|_{\rho=0} - \frac{\partial_\rho^{r/2-1} L|_{\rho=0}}{r+2} t^2 h_{AB} = \Omega^{-1} r^{-1} \partial_\rho^{r/2-1} \mathcal{Q}_{AB}|_{\rho=0}.$$

Under the transformation $\{t = v^{-1}, \rho = uv\}$ this expression becomes (note that $\partial_\rho = v^{-1}\partial_u$)

$$(\partial_\rho^{r/2-1} L_{AB}|_{\rho=0})^{TF} = r^{-1} v^{-r/2} (\partial_u^{r/2-1} \mathcal{Q}_{AB}|_{u=0})^{TF} \qquad \Longrightarrow$$
$$\Longrightarrow \qquad \mathcal{O}^{\mathscr{F}\mathscr{G}}_{AB} = v^{-r/2} (\partial_u^{r/2-1} \mathcal{Q}_{AB}|_{u=0})^{TF}.$$

We are interested in obtaining an expression relating the obstruction tensor with transverse derivatives of the tensor $\mathcal{Q}$ at null infinity, i.e. derivatives of $\mathcal{Q}$ w.r.t. $\partial_v$ at $v = 0$. In other words, we want to "interchange" the roles of $u$ and $v$ in the formula for $\mathcal{O}^{\mathscr{F}\mathscr{G}}_{AB}$ we have just obtained. In order to do that, we exploit the fact that $uv = 0$ is a bifurcate Killing

1 We use $\partial_s$ as the derivative w.r.t. the product $uv$ for functions that only depend on $(uv, x^A)$.

horizon. Indeed, since $\mathcal{Q} = r(\mathrm{Hess}_{\widetilde{g}}\, v + v\, \mathrm{Sch}_{\widetilde{g}})$ and $T = u\partial_u - v\partial_v$ is a Killing vector of $\widetilde{g}$, one has $\pounds_T \mathcal{Q} = -\mathcal{Q}$ and hence $T(\mathcal{Q}_{AB}) = -\mathcal{Q}_{AB}$. This implies that $\mathcal{Q}_{AB}$ is of the form $\mathcal{Q}_{AB} = vT_{AB}(uv, x^C)$, and hence

$$\mathcal{O}^{\mathscr{F}\mathscr{G}}_{AB} = v^{-r/2+1}\Big((\partial_u^{r/2-1} T_{AB})|_{u=0}\Big)^{TF} = \Big(T^{(r/2-1)}_{AB}|_{uv=0}\Big)^{TF},$$

where $T^{(m)}$ denotes the $m$-th derivative w.r.t. $uv$ and we used that $T_{AB}(uv, x^C)|_{u=0} = T_{AB}(uv, x^C)|_{uv=0}$. Recall that $\partial_u^k = v^k \partial_s^k$ and $\partial_v^k = u^k \partial_s^k$ for every $k$, and note also that

$$k!\Big(T^{(k)}_{AB}|_{uv=0}\Big)^{TF} = \Big(\partial_u^k (u^k T^{(k)}_{AB})|_{v=0}\Big)^{TF} = \Big(\partial_u^k \Big(\partial_v^k T_{AB}\Big)|_{v=0}\Big)^{TF},$$

where we have used that $\partial_u^k (u^k T^{(k)}_{AB})|_{v=0} = k! T^{(k)}_{AB}|_{v=0}$. Note also that the horizon is totally geodesic and then $\partial_u$ and $\partial_v$ commute with $TF$. Then,

$$\left(\frac{r}{2} - 1\right)!\, \mathcal{O}^{\mathscr{F}\mathscr{G}}_{AB} = \partial_u^{r/2-1}\Big((\partial_v^{r/2-1} T_{AB})|_{v=0}\Big)^{TF},$$

and since $\Big(\partial_v^{r/2}(vT_{AB})\Big)|_{v=0} = \frac{r}{2}\Big(\partial_v^{r/2-1} T_{AB}\Big)|_{v=0}$, we conclude that

$$r\left(\frac{r}{2} - 1\right)!\, \mathcal{O}^{\mathscr{F}\mathscr{G}}_{AB} = 2\partial_u^{r/2-1}\Big((\partial_v^{r/2} \mathcal{Q}_{AB})|_{v=0}\Big)^{TF}. \tag{6.35}$$

This shows that the presence of a non-vanishing obstruction tensor can also be detected from the conformal infinity, as it makes the $r/2$-transverse derivative of the tensor $\mathcal{Q}$ at $\mathscr{I}$ to be different from zero.

Our next aim is to characterize the ambient metric from a conformal viewpoint. More specifically, we want to find a set of conformally covariant conditions on a conformal manifold $(\mathcal{M}, g, \Omega)$ that univocally lead to the ambient metric. As we have seen before, the conformal completion of the ambient metric exhibits two key properties, namely the existence of a bifurcate Killing horizon where one of the horizons is $\mathscr{I}$, and that the one-form $\widetilde{\boldsymbol{T}}$ is integrable. While the latter condition is already conformally invariant, the former is not. The obvious replacement is to ask for the existence of a bifurcate *conformal* Killing horizon. Additionally, it is clear that one should impose some extra condition relating the conformal Killing field (that we now denote by $\eta$ to avoid confusion with the previous section) and the function $\Omega$. In the conformal completion of the previous section $\Omega = v$ and $T = u\partial_u - v\partial_v$, so $T(\Omega) = -v$. The natural replacement for this condition is to ask that $\eta(\Omega) = (\psi - 1)\Omega$, where $\psi$ is the function such that $\pounds_\eta g = 2\psi g$. That this condition is conformally invariant follows because under a conformal rescaling $g' := \omega^2 g$ the function $\psi$ transforms by $\psi' = \psi + \eta(\log|\omega|)$, and hence $\eta(\omega\Omega) = \omega\eta(\Omega) + \omega\Omega\eta(\log|\omega|) = (\psi' - 1)\omega\Omega$. The main result of this section is that these three conditions fully characterize the Fefferman-Graham ambient metric (Theorem 6.16). We start by showing that the function $\psi$ necessarily vanishes at the bifurcation surface.

**Lemma 6.9.** *Let $\eta$ be a conformal Killing field and $\psi$ the function defined by $\pounds_\eta g = 2\psi g$. Assume $\eta$ admits a bifurcation surface $\mathcal{S}$. Then, $\psi|_{\mathcal{S}} = 0$.*

*Proof.* Let $V, W \in T\mathcal{S}$. Since $[\eta, V] \stackrel{\mathcal{S}}{=} 0$ it follows $\nabla_V \eta \stackrel{\mathcal{S}}{=} 0$, and then

$$0 \stackrel{\mathcal{S}}{=} V^\alpha W^\beta \nabla_{(\alpha}\eta_{\beta)} = \psi g(V, W).$$

Since $V, W$ are arbitrary it follows that $\psi \stackrel{\mathcal{S}}{=} 0$. □

Next we show that it is possible to fix the gauge such that $|\nabla\Omega|^2 = 0$.

**Lemma 6.10.** *Let $(\mathcal{M}, g, \Omega)$ be a conformal manifold with $\lambda = 0$ and $\mathcal{H}$ a hypersurface transverse to $\mathscr{I}$. Let $\omega_0$ be a non-vanishing function on $\mathcal{H}$. Then, there exists a unique $(\widetilde{g} = \omega^2 g, \widetilde{\Omega} = \omega\Omega)$ in a neighbourhood of $\mathscr{I} \cap \mathcal{H}$ such that $|\widetilde{\nabla}\widetilde{\Omega}|^2_{\widetilde{g}} = 0$ and $\widetilde{\Omega} \stackrel{\mathcal{H}}{=} \omega_0\Omega$.*

*Proof.* Let $(M, g, \Omega)$ be a conformal manifold, $\omega > 0$ a function, and define $\widetilde{g} := \omega^2 g$, $\widetilde{\Omega} := \omega\Omega$ and $F := |\nabla\Omega|^2_g$. Then,

$$\widetilde{F} := |\widetilde{\nabla}\widetilde{\Omega}|^2_{\widetilde{g}} = \Omega^2 \nabla_\alpha f \nabla^\alpha f + 2\Omega \nabla_\alpha f \nabla^\alpha \Omega + F,$$

where we defined $f := \log\omega$. We want to show that the equation $\widetilde{F} = 0$ admits a unique solution given the function $f$ in a hypersurface transverse to $\mathscr{I}$. Since $F|_{\Omega=0} = 0$, the function $\Omega^{-1}F$ has a good limit at $\Omega = 0$ and henceforth we can equivalently look for solutions to the following equation

$$\widetilde{F}' = \Omega\nabla_\alpha f \nabla^\alpha f + 2\nabla_\alpha f \nabla^\alpha \Omega + \Omega^{-1}F = 0. \tag{6.36}$$

In order to do that we use the method of characteristics (see [111]), which basically consists of rewriting the PDE as a first order system of ODEs for $f$ and its gradient $p^\alpha := \nabla^\alpha f$ along the so-called characteristic curves, with the aim of solving it from data at a hypersurface $\mathcal{H}$. In our specific setup, since $\nabla^\alpha\Omega$ is tangent to $\mathscr{I}$, it is also necessarily transverse to $\mathcal{H}$ (at least in a neighbourhood of $\mathcal{H} \cap \mathscr{I}$). Then, we can complete any local coordinate system $\{x^a\}$ on $\mathcal{H}$ to a local coordinate system $\{x^0, x^a\}$ on $M$ by extending $\{x^a\}$ trivially along $\nabla\Omega$, and solving $\nabla\Omega(x^0) = 1$, $x^0|_{\mathcal{H}} = 0$ (then, $\nabla\Omega = \partial_{x^0}$). Given a smooth function $f$ on $\mathcal{H}$, an initial condition for $p$ is called admissible provided that $p_a|_{\mathcal{H}} = df_a$ and $\widetilde{F}'(f, p|_{\mathcal{H}}) \stackrel{\mathcal{H}}{=} 0$. Note that, in general, a covector $p|_{\mathcal{H}}$ satisfying these conditions may not exist or may not be unique. It is only when the problem is non-characteristic, i.e. $\xi^\alpha \mathcal{D}_{p_\alpha}\widetilde{F}'|_{\mathcal{H}} \neq 0$ for any non-vanishing vector $\xi$ transverse to $\mathcal{H}$, when a unique solution for $p$ exists. In our specific setup, choosing $\xi = \nabla\Omega = \partial_{x^0}$, equation (6.36) becomes

$$\widetilde{F}' = \Omega p_\alpha p^\alpha + 2p_0 + \Omega^{-1}F = 0, \tag{6.37}$$

and then one can check that $\xi^\alpha \mathcal{D}_{p_\alpha}\widetilde{F}'|_{\mathcal{H}} = \mathcal{D}_{p_0}\widetilde{F}' = 2 + O(\Omega)$, so $\mathcal{D}_{p_0}\widetilde{F}' \neq 0$ in a neighbourhood of $\mathcal{H} \cap \mathscr{I}$. This proves that the problem is non-characteristic and therefore the equation $|\nabla\Omega|^2_g = 0$ admits a unique solution given the function $\omega$ on a hypersurface transverse to $\mathscr{I}$. □

Let $(M, g, \Omega)$ be a conformal manifold admitting a bifurcate conformal Killing horizon such that one of the horizons is $\mathscr{I}$. Let $\eta$ be the conformal field and $\psi$ the function satisfying $\pounds_\eta g = 2\psi g$. Assume also that $\eta(\Omega) = (\psi - 1)\Omega$. Now we show that one can restrict further the gauge to set $\psi = 0$.

**Proposition 6.11.** *Let $(\mathcal{M}, g, \Omega)$ be a conformal manifold admitting a bifurcate conformal Killing vector $\eta$ where one of the horizons is $\mathscr{I}$ and $\eta(\Omega) = (\psi - 1)\Omega$. Then, there exists a conformal gauge in which simultaneously $|\nabla\Omega|^2 = 0$ and $\pounds_\eta g = 0$ in a neighbourhood of the bifurcation surface. Moreover, the remaining gauge freedom is a function $\omega(x^A)$.*

*Proof.* We choose the transverse hypersurface $\mathcal{H}$ of the previous proposition as the conformal Killing horizon transverse to $\mathscr{I}$, so that $\eta$ is tangent to $\mathcal{H}$ and $\mathcal{S} := \mathcal{H} \cap \mathscr{I}$ is the bifurcation surface of $\eta$. The idea of the proof is to show that one can make $\psi$ to vanish at $\mathcal{H}$ by choosing the free function $\omega$ of the proposition to satisfy $\psi' = \psi + \eta(\log\omega) = 0$ at $\mathcal{H}$, and then to prove that condition $|\nabla\Omega|^2 = 0$ implies that $\psi$ vanishes everywhere. Recall by Lemma 6.9 that $\psi = 0$ at $\mathcal{H} \cap \mathscr{I}$, so in order to prove that the equation $\psi + \eta(\log\omega) = 0$ along $\mathcal{H}$ admits a solution we must show that $\eta|_{\mathcal{H}}$ has a zero of order one at $\mathcal{S}$. We choose double null coordinates $\{v, u, x^A\}$ adapted to $\mathscr{I} = \{v = 0\}$ and $\mathcal{H} = \{u = 0\}$, where the metric is given by

$$g = 2Gdudv + \gamma\!\!\!/_{AB}(dx^A + q^A dv)(dx^B + q^B dv).$$

Note that $n := \partial_v - q^A\partial_{x^A}$ is a null generator of $\mathcal{H}$ and by construction $\eta|_{\mathcal{H}} = Hn$ for some function $H$ defined on $\mathcal{H}$. We want to show that $H$ satisfies $H|_{v=0} = 0$ and $\partial_v H|_{v=0} \neq 0$.

Consider the equation $\eta^\alpha\nabla_\alpha\Omega = (\psi - 1)\Omega$. Applying $\nabla_\beta$ to both sides it follows that

$$\nabla_\beta\eta^\alpha\nabla_\alpha\Omega + \eta^\alpha\nabla_\beta\nabla_\alpha\Omega = \Omega\nabla_\beta\psi + (\psi - 1)\nabla_\beta\Omega.$$

Defining $F_{\alpha\beta} := \nabla_{[\alpha}\eta_{\beta]}$ and inserting $\nabla_\beta\eta^\alpha = \psi\delta^\alpha_\beta + F_\beta{}^\alpha$,

$$\psi\nabla_\beta\Omega + F_\beta{}^\alpha\nabla_\alpha\Omega + \eta^\alpha\nabla_\beta\nabla_\alpha\Omega = \Omega\nabla_\beta\psi + (\psi - 1)\nabla_\beta\Omega,$$

which evaluated at $\mathcal{S}$ gives $F_\beta{}^\alpha\nabla_\alpha\Omega = -\nabla_\beta\Omega$. Since $d\Omega = fdv$ on $v = 0$ with $f \neq 0$ it follows that $F_\beta{}^v = -\delta^v_\beta$, and hence taking $\beta = v$ and using that $g^{v\alpha} = G^{-1}\delta^\alpha_u$ we get $F_{vu} \stackrel{\mathcal{S}}{=} \partial_v\eta_u - \partial_u\eta_v \stackrel{\mathcal{S}}{=} 2\partial_v\eta_u \stackrel{\mathcal{S}}{=} -G \neq 0$, where we used $\partial_\alpha\eta_\beta + \partial_\beta\eta_\alpha \stackrel{\mathcal{S}}{=} 0$. Since $\boldsymbol{\eta}|_{\mathcal{H}} = g(\eta, \cdot)|_{\mathcal{H}} = Hdu$ we conclude $\partial_v H|_{\mathcal{S}} \neq 0$, so $\eta|_{\mathcal{H}}$ has a zero of order one at $\mathcal{S}$ and then equation $\psi + \eta(\log\omega) = 0$ admits a solution along $\mathcal{H}$.

Once we have guaranteed that (part of) the remaining freedom in Prop. 6.10 can be chosen so that $\psi = 0$ on $\mathcal{H}$, it remains to show that $\psi$ actually vanishes everywhere. Using that $\pounds_\eta(g^{\alpha\beta}) = -2\psi g^{\alpha\beta}$,

$$\begin{aligned}\pounds_\eta|\nabla\Omega|^2 &= -2\psi|\nabla\Omega|^2 + 2\nabla^\alpha\Omega\nabla_\alpha((\psi - 1)\Omega)\\ &= -2\psi|\nabla\Omega|^2 + 2\Omega\nabla^\alpha\Omega\nabla_\alpha\psi + 2(\psi - 1)|\nabla\Omega|^2\\ &= 2\Big(\Omega\nabla^\alpha\Omega\nabla_\alpha\psi - |\nabla\Omega|^2\Big).\end{aligned}$$

Then, in a gauge in which $|\nabla\Omega|^2 = 0$ one has $\nabla_\alpha\Omega\nabla^\alpha\psi = 0$, and since $\nabla_\alpha\Omega$ is transverse to $\mathcal{H}$, the condition $\psi \stackrel{\mathcal{H}}{=} 0$ extends to a neighbourhood of $\mathcal{S}$. □

An immediate corollary of this proposition is the following.

**Corollary 6.12.** *Let $f_1$ and $f_2$ be two functions satisfying $|\nabla f_1|^2 = |\nabla f_2|^2 = 0$ and $\eta(f_1) = -f_1$, $\eta(f_2) = -f_2$. Assume $f_1 \stackrel{\mathcal{S}}{=} f_2$. Then, $f_1 = f_2$.*

*Proof.* In Proposition 6.11 it is shown that the solution to $|\nabla f|^2 = 0$ and $\eta(f) = -f$ is characterized by a free function $\omega$ on $\mathcal{S}$. Since both $f_1$ and $f_2$ agree on $\mathcal{S}$ the function $\omega$ is identically 1, and then $f_1 = f_2$ everywhere. □

In this conformal gauge, from now on called *geodesic Killing gauge*, the metric $g$ admits a bifurcate Killing horizon, so it can be written in Rácz-Wald coordinates as [288]

$$\widetilde{g} = 2Gdu\Big(dv + v\beta_A dx^A\Big) + \mu_{AB}dx^A dx^B, \tag{6.38}$$

where $G$, $\boldsymbol{\beta}$ and $\mu$ are a function, a one-form and a metric on the codimension-two surfaces $S_{u,v}$ that depend only upon $\{s := uv, x^A\}$. In these coordinates the Killing field is given by $\eta = u\partial_u - v\partial_v$. We note that in the geodesic Killing gauge, the vector field grad$\Omega$ is geodesic (hence the name). This follows from $0 = \nabla_\beta(\nabla_\alpha\Omega\nabla^\alpha\Omega) = 2\nabla^\alpha\Omega\nabla_\alpha\nabla_\beta\Omega = 0$.

The Rácz–Wald (RW) construction allows for the freedom to choose any cross-section $\Sigma$ on $\{v = 0\}$ not intersecting the bifurcation surface $\mathcal{S}$ (see Section C.2 in Appendix C). From this surface, the coordinate $u$ is uniquely defined so that $u|_{\mathcal{S}} = 0$, $u|_\Sigma = 1$, and $\partial_u$ is geodesic. Given any conformal factor $\Omega$ in a geodesic Killing gauge, we fix the RW coordinate $u$ as follows. Choose the family of geodesic curves $\gamma(\tau)$ on $\{v = 0\}$ satisfying $\tau|_{\mathcal{S}} = 0$ and $\dot{\gamma}|_{\mathcal{S}} = \text{grad}\Omega|_{\mathcal{S}}$ (note that grad$\Omega$ is tangential to the hypersurface $\{v = 0\}$). We then choose $\Sigma := \{\tau = 1\}$ in the construction described above. Since $u|_{\mathcal{S}} = \tau|_{\mathcal{S}} = 0$, $u|_\Sigma = \tau|_\Sigma = 1$, and both $\partial_u$ and $\dot{\gamma}$ are geodesic, it follows that $u = \tau$, so grad $\Omega \stackrel{v=0}{=} \partial_u$. In particular, $d\Omega \stackrel{\mathcal{S}}{=} Gdv$. The resulting RW coordinates only admit an additional scaling freedom for the coordinate $v$ of the form $v = \bar{v}h(x^A)$. Next, we show that one can rescale the coordinate $v$ to set $\boldsymbol{\beta} = 0$ locally when $\eta$ is integrable.

**Lemma 6.13.** *Assume that the Killing $\eta = u\partial_u - v\partial_v$ is integrable w.r.t. the metric* (6.38)*. Then $\boldsymbol{\beta}$ is closed, and hence locally exact.*

*Proof.* It is easier to work away from $u = 0$ or $v = 0$ and define the one-form $\widetilde{\boldsymbol{\eta}} := (Guv)^{-1}\widetilde{g}(\eta, \cdot) = \boldsymbol{\beta} - d\log\left|\frac{u}{v}\right|$. Now, condition $\boldsymbol{\eta} \wedge d\boldsymbol{\eta} = 0$ is equivalent to $\widetilde{\boldsymbol{\eta}} \wedge d\widetilde{\boldsymbol{\eta}} = 0$, and hence

$$0 = \left(\boldsymbol{\beta} - d\log\left|\frac{u}{v}\right|\right) \wedge d\boldsymbol{\beta} = \boldsymbol{\beta} \wedge ds \wedge \dot{\boldsymbol{\beta}} + \boldsymbol{\beta} \wedge d\!\!\!/\boldsymbol{\beta} - d\log\left|\frac{u}{v}\right| \wedge ds \wedge \dot{\boldsymbol{\beta}} - d\log\left|\frac{u}{v}\right| \wedge d\!\!\!/\boldsymbol{\beta},$$

where the dot denotes derivative w.r.t. $s$ and $d\!\!\!/$ is the exterior derivative on the codimension two surfaces $S_{u,v}$. Since the four terms on the RHS are linearly independent, the last term shows that $d\!\!\!/\boldsymbol{\beta} = 0$ (note that although the computation has been done away from $u = 0$ or $v = 0$, by continuity the result is valid everywhere). □

**Corollary 6.14.** *Assume the Killing $\eta = u\partial_u - v\partial_v$ is integrable w.r.t. the metric* (6.38)*. Then, there exists a change of coordinates that respects the form of $\eta$ and such as the metric takes the form* (6.38) *with $\boldsymbol{\beta} = 0$, i.e.*

$$g = 2Gdudv + \mu. \tag{6.39}$$

*Proof.* Since $\boldsymbol{\eta}$ is integrable, by the previous lemma it is locally exact, i.e. there exists a function $f$ such that $\boldsymbol{\beta} = df$. Moreover, since $\dot{\boldsymbol{\beta}} = 0$ it follows that $f = f(x^A)$. Inserting this and $v = \bar{v}h(x^A)$ into (6.38) yields

$$g = 2Gdu\Big(\bar{v}dh + hd\bar{v} + h\bar{v}df\Big) + \mu.$$

By choosing $h$ such that $dh + hdf = 0$ and redefining $G$, the metric $g$ takes (locally) the form (6.39). Note that this change keeps the same form of $\eta$, since $v\partial_v = h\bar{v}h^{-1}\partial_{\bar{v}} = \bar{v}\partial_{\bar{v}}$. □

Note that this change of $v$ does not affect the coordinate $u$ and that the remaining freedom is scaling $v$ by a non-zero constant. In these new coordinates, we still have $d\Omega \stackrel{\mathcal{S}}{=} Gdv$. This means that $\Omega = vF$, where $F$ satisfies $F|_{\mathcal{S}} = G|_{\mathcal{S}}$. Since $\eta(\Omega) = -\Omega$ and $\eta(v) = -v$ it follows that $\eta(F) = 0$ and thus $F = F(uv, x^A)$.

Next, we define $\widehat{\Omega} := F^{-1}\Omega$ and $\widehat{g} := F^{-2}g$ (note that $\widehat{\Omega} = v$). We write the metric as $\widehat{g} = 2\widehat{G}dudv + \widehat{\mu}$ and note that $\widehat{G} = F^{-2}G$, and in particular $\widehat{G}|_{\mathcal{S}} = F^{-1}|_{\mathcal{S}}$ (this is the key reason for our specific choice of RW coordinates above). Using the conformal covariance of the equations, showing that $(g, \Omega)$ is quasi-Einstein is equivalent to showing that $(\widehat{g}, v)$ satisfy $\widehat{\mathcal{Q}} := r(\operatorname{Hess} v + v \operatorname{Sch}_{\widehat{g}}) - q\widehat{g} = 0$ (note that a priori the conformal factor $\widehat{\Omega}$ need not to satisfy $|\nabla\widehat{\Omega}|^2_{\widehat{g}} = 0$, i.e. we have momentarily abandoned the geodesic Killing gauge and the term $q\widehat{g}$ reappears in the conformal Killing equations). In Appendix C we have computed the components of the tensor $r(\operatorname{Hess} v + v \operatorname{Sch}_{\widehat{g}})$ for the metric (6.39). In particular, we are interested in the equations $\mathcal{Q}_{vv} = 0$ and $\mathcal{Q}_{vA} = 0$. Since $\widehat{g}_{vv} = \widehat{g}_{vA} = 0$, the $(v, v)$ and $(v, A)$ equations that $(\widehat{g}, v)$ satisfy are obtained by simply replacing $G$ by $\widehat{G}$ and $\mu$ by $\widehat{\mu}$. In particular, equation $\mathcal{Q}_{vv} = 0$ is

$$2\Big(s \operatorname{tr}_{\widehat{\mu}} \dot{\widehat{\mu}} - 2r\Big)\frac{\dot{\widehat{G}}}{\widehat{G}} + s\Big(|\dot{\widehat{\mu}}|^2 - 2 \operatorname{tr}_{\widehat{\mu}} \ddot{\widehat{\mu}}\Big) = 0,$$

which proves $\dot{\widehat{G}} = 0$ (at least in a neighbourhood of $s = 0$), and equation $\mathcal{Q}_{vA} = 0$ reads

$$\Big(s \operatorname{tr}_{\widehat{\mu}} \dot{\widehat{\mu}} - 2r\Big)\frac{\nabla_A \widehat{G}}{\widehat{G}} - 2s\Big(\nabla_A \operatorname{tr}_{\widehat{\mu}} \dot{\widehat{\mu}} - (\operatorname{div}_{\widehat{\mu}} \dot{\widehat{\mu}})_A\Big) = 0,$$

which evaluated at $s = 0$ implies $\widehat{G} = A \in \mathbb{R} \setminus \{0\}$ is constant on $s = 0$ (we assume $S$ is connected), and hence everywhere. It follows that $F|_S = A^{-1}$. Now, in the original geodesic Killing gauge we have that $\Omega$ satisfies $|\nabla\Omega|^2_g = 0$ and $\pounds_\eta \Omega = -\Omega$. The function $v$ has the same properties and satisfies $\Omega = A^{-1}v$ on $\mathcal{S}$, so by Corollary 6.12 one has $\Omega = A^{-1}v$ and $F = A^{-1}$ everywhere. Moreover, $G = \widehat{G}F^2 = A^{-1}$ because $\widehat{G} = A$. Performing a final constant rescaling of $v$ we finally arrive at

$$\Omega = v, \qquad g = 2dudv + \mu. \tag{6.40}$$

We emphasize that we have arrived at this expression starting with any $\Omega$ that belongs to the geodesic Killing gauge. The RW coordinates $\{u, v\}$ in the final metric (6.40) are fully determined in terms of $\Omega$. We call them *adapted Rácz-Wald coordinates.* A conformal change within the geodesic Killing gauge has a highly non-trivial effect in the coordinates $\{u, v\}$ and also on $\mu$. However, at $\mathcal{S}$, the effect is simple. Any other $\widetilde{\Omega}$ in this gauge is uniquely parametrized by a positive function $\omega$ on $\mathcal{S}$ by means of $\widetilde{\Omega}|_{\mathcal{S}} = \omega\Omega|_{\mathcal{S}}$, and then $\tilde{\mu}|_{\mathcal{S}} = \omega^2\mu|_{\mathcal{S}}$. So our construction keeps the full conformal freedom at $\mathcal{S}$. This is exactly the conformal freedom that exists in the original Fefferman-Graham construction.

Whenever $(g, \Omega)$ is quasi-Einstein one has that $\widetilde{g} = \Omega^{-2}g = v^{-2}(2dudv + \mu)$ is Ricci flat. Defining $\bar{u} := u$ and $\bar{v} := v^{-1}$ one finds

$$\widetilde{g} = -2d\bar{u}d\bar{v} + \bar{v}^2\mu.$$

Therefore, this metric is exactly Ricci flat, admits a homothetic horizon ($\bar{u} = 0$) and it is written in a double null coordinate system with $\Delta = 1$, $q = 0$ and $\not{g} = \bar{v}^2\mu$. This proves that the class of ambient metrics of Theorem 5.1 is in one-to-one correspondence with the class of quasi-Einstein manifolds $(M, g, \Omega)$ that we have considered in this section.

**Proposition 6.15.** *There exists a one-to-one correspondence between straight, exact and regular ambient metrics and quasi-Einstein manifolds $(\mathcal{M}, g, \Omega)$ with $\lambda = 0$ admitting an integrable conformal field $\pounds_\eta g = 2\psi g$ satisfying $\pounds_\eta\Omega = (\psi - 1)\Omega$ and such that $\eta$ admits a bifurcate horizon where one of the horizons is $\mathscr{I}$.*

This result relaxes the condition of $\eta$ being closed. As a consequence of the regularity in Theorem 5.1 and relation (6.34), the regularity of $g$ is as follows: For $r = 2$, $g$ is smooth. When $r \geq 3$ is odd, $g$ is smooth everywhere except at $s = 0$, and there exists smooth tensors $\{g^{(i)}_{\alpha\beta}\}_{i=0}^{\frac{r-1}{2}}$ and $\widetilde{g}_{\alpha\beta}$ such that

$$g_{\alpha\beta} = \sum_{i=0}^{\frac{r-1}{2}} g^{(i)}_{\alpha\beta}s^i + s^{\frac{r}{2}}\widetilde{g}_{\alpha\beta} + O\Big(s^{\frac{r+1}{2}}\Big).$$

Finally, when $r \geq 4$ is even, $g$ is smooth everywhere except at $s = 0$, and there exists smooth tensors $\{g^{(i)}_{\alpha\beta}\}_{i=0}^{\frac{r}{2}}$ and $\widetilde{g}_{\alpha\beta}$ such that

$$g_{\alpha\beta} = \sum_{i=0}^{\frac{r}{2}-1} g^{(i)}_{\alpha\beta}s^i + g^{(r/2)}_{\alpha\beta}s^{r/2} + \widetilde{g}_{\alpha\beta}s^{r/2}\log(s) + O\Big(s^{\frac{r}{2}+1}\log(s)\Big).$$

This allows us to identify the free data from the conformal picture as follows.

**Theorem 6.16.** *Let $(\mathcal{M}, g, \Omega)$ be a regular quasi-Einstein manifold of dimension $r+2$ that admits an integrable conformal Killing vector $\pounds_\eta g = 2\psi g$ and a bifurcate horizon where one of the horizons is $\mathscr{I} = \{\Omega = 0\} \simeq \mathcal{S} \times \mathbb{R}$ and such that $\pounds_\eta \Omega = (\psi - 1)\Omega$. Write $(g, \Omega)$ in a geodesic Killing gauge and let $h$ be the metric of the bifurcation surface $\mathcal{S}$ and $\Theta$ the trace-free part of the term $s^{r/2}$ in the expansion of $g$ in adapted Rácz-Wald coordinates. Then, $\Omega^{-2}g$ is the exact, straight and regular ambient metric with data $(\mathcal{S}, h, \Theta)$.*

It is also instructive to see how this free data appears by analysing the conformal Einstein equations order by order at $s = 0$, as a similar analysis will be carried out in a more general context in the subsequent sections. Let us consider the $(A, B)$ and $(v, u)$ components of the quasi-Einstein equations of the metric (6.39), that we write again here for completeness (see Appendix C)

$$\mathcal{Q}_{uv} = \frac{v}{4(r+1)}\Big(2R^{(h)} + (r-2)s|\dot{\mu}|^2 - 2(r-1)s\,\mathrm{tr}_\mu\,\ddot{\mu} + \big(s\,\mathrm{tr}_\mu\,\dot{\mu} - 2(r-1)\big)\,\mathrm{tr}_\mu\,\dot{\mu}\Big),$$

$$\begin{aligned}\mathcal{Q}_{AB} = -\frac{v}{2}\Bigg(&\big(r - 2 - s\,\mathrm{tr}_\mu\,\dot{\mu}\big)\dot{\mu}_{AB} - 2\Big(R^{(h)}_{AB} - s(\dot{\mu}\cdot\dot{\mu})_{AB} + s\ddot{\mu}_{AB}\Big)\\ &+ \frac{1}{2(r+1)}\Big(2R^{(h)} - 3s|\dot{\mu}|^2 + 4s\,\mathrm{tr}_\mu\,\ddot{\mu} + (4 + s\,\mathrm{tr}_\mu\,\dot{\mu})\,\mathrm{tr}_\mu\,\dot{\mu}\Big)h_{AB}\Bigg).\end{aligned}$$

In order to study how the equations fix the geometry order by order at $s = 0$ we start by solving $\mathcal{Q}_{uv} \overset{s=0}{=} 0$ for $\mathrm{tr}_\mu\,\dot{\mu}$, which gives $\mathrm{tr}_\mu\,\dot{\mu} = \frac{1}{r-1}R^{(h)}$. Inserting it into equation $\mathcal{Q}_{AB} \overset{s=0}{=} 0$ we obtain

$$(r-2)\dot{\mu}_{AB} - 2R^{(h)}_{AB} - \frac{R^{(h)}}{r-1}h_{AB} = 0.$$

For $r > 2$ this equation fixes the tensor $\dot{\mu}_{AB}$ at $s = 0$ to be twice the Schouten of $h_{AB}$, while for $r = 2$ this equation do not determine $\dot{\mu}_{AB}$ but holds automatically because in two dimensions $R^{(h)}_{AB} = \frac{1}{2}R^{(h)}h_{AB}$. Let us now analyse the equations order by order. To do that the strategy is to take $m$ derivatives of the equations w.r.t. $s$ and keep track of the leading order terms. Dropping irrelevant global factors the result is (the symbol $\propto$ means proportionality with a non-zero factor)

$$\begin{aligned}\partial_s^{(m)}(v^{-1}\mathcal{Q}_{uv}) &\overset{s=0}{\propto} \mathrm{tr}_h\,\mu^{(m+1)} + \text{l.o.t},\\ \partial_s^{(m)}(v^{-1}\mathcal{Q}_{AB}) &\overset{s=0}{\propto} (r-2-2m)\mu^{(m+1)}_{AB} + \frac{2m\,\mathrm{tr}_h\,\mu^{(m+1)}}{r+1}h_{AB} + \text{l.o.t},\end{aligned}$$

where $\mu^{(m)} := \partial_s^{(m)}\mu$ and "l.o.t" stands for lower order terms. For every $1 \leq m < \frac{r-2}{2}$ it is clear that the first equation determines the trace of $\mu^{(m+1)}_{AB}$, which inserted into the second gives the full $\mu^{(m+1)}_{AB}$. When $m = \frac{r-2}{2}$ the trace-free part of $\mu^{(r/2)}_{AB}$ cannot be determined from the equations, which makes the $\frac{r}{2}$-term of the expansion free. Once such free data $\Theta_{AB}$ has been specified, one can continue determining the rest of the expansion.

That the free data $\Theta_{AB}$ must satisfy a divergence condition follows from Proposition 6.15, as otherwise the ambient metric would not be straight. This can also be detected from the conformal viewpoint. Indeed, consider the $(u, A)$ components of the conformal equations, namely

$$(\operatorname{div}_\mu \mu)_A - \nabla_A \operatorname{tr}_\mu \mu = 0.$$

After taking $(r/2 - 1)$ derivatives along $\partial_s$ one arrives to

$$\operatorname{div}_h \Theta - \widetilde{D} \overset{s=0}{=} 0,$$

where $\widetilde{D}$ is a tensor that only depends on $h_{AB}$. By Theorem 6.15, this condition must be equivalent to (5.7) (i.e. $\widetilde{D} = D$).

## 6.4 TRANSVERSE EXPANSION AT NULL INFINITY

In the previous section we have studied in detail the conformal infinity of the Fefferman-Graham ambient metric. In particular, we have illustrated how the conformal equations constrain the geometry of its null infinity. The purpose of this section is to perform a similar analysis in a complete general context, i.e. to analyze how the quasi-Einstein equations fix the geometry at null infinity, or in other words how the transverse expansion of the metric is constrained when the conformal Einstein equations are imposed at $\mathscr{I}$ to infinite order. To do that we shall start by recalling some of the results developed in Section 5.2 (that we particularize to $\nabla_\xi \xi = 0$) and by extending some others. In particular, we will need to compute the tensor $\dot{\mathcal{R}}_a^{(m)}$ up to one order higher than in (5.83), i.e. up to an including order $m$. This is accomplished in Appendix D (see Proposition D.2). Our starting point will be the particularization of identities (5.82), (D.22) and (5.84) for $m \geq 2$ and $\nabla_\xi \xi = 0$

$$\ddot{\mathcal{R}}^{(m)} = -\operatorname{tr}_P \mathbf{Y}^{(m+1)} + \mathcal{O}^{(m)}, \tag{6.41}$$

$$\begin{aligned}\dot{\mathcal{R}}_a^{(m)} &= \mathrm{r}_a^{(m+1)} + P^{bc}\overset{\circ}{\nabla}_b \mathrm{Y}_{ac}^{(m)} - 2mP^{bc}\mathrm{r}_b \mathrm{Y}_{ac}^{(m)} - \overset{\circ}{\nabla}_a(\operatorname{tr}_P \mathbf{Y}^{(m)}) + (\operatorname{tr}_P \mathbf{Y}^{(m)})(\mathrm{r}_a - \mathrm{s}_a) \\ &\quad + (P^{bc}\mathrm{Y}_{ab} - 3V^c{}_a)\mathrm{r}_c^{(m)} + \left(\operatorname{tr}_P \mathbf{Y} - \frac{m}{2} n(\ell^{(2)})\right)\mathrm{r}_a^{(m)} + \mathcal{O}_a^{(m-1)},\end{aligned} \tag{6.42}$$

$$\begin{aligned}\mathcal{R}_{ab}^{(m)} &= -2\pounds_n \mathrm{Y}_{ab}^{(m)} - (2m\kappa + \operatorname{tr}_P \mathbf{U})\,\mathrm{Y}_{ab}^{(m)} - (\operatorname{tr}_P \mathbf{Y}^{(m)})\mathrm{U}_{ab} + 4P^{cd}\mathrm{U}_{c(a}\mathrm{Y}_{b)d}^{(m)} \\ &\quad + 4(\mathrm{s}-\mathrm{r})_{(a}\mathrm{r}_{b)}^{(m)} + 2\overset{\circ}{\nabla}_{(a}\mathrm{r}_{b)}^{(m)} - 2\kappa^{(m)}\mathrm{Y}_{ab} + \mathcal{O}_{ab}^{(m-1)},\end{aligned} \tag{6.43}$$

where $\mathcal{O}^{(m)}$, $\mathcal{O}_a^{(m)}$ and $\mathcal{O}_{ab}^{(m)}$ are, respectively, a scalar, a one-form and a (0,2) symmetric tensor depending only on metric data $\{\boldsymbol{\gamma}, \boldsymbol{\ell}, \ell^{(2)}\}$ and $\{\mathbf{Y}, ..., \mathbf{Y}^{(m)}\}$. Moreover (cf. (5.85), (5.86), (D.23))

$$\mathcal{R}_{ab}^{(m)} n^b = -\pounds_n \mathrm{r}_a^{(m)} - \Big(2(m-1)\kappa + \operatorname{tr}_P \mathbf{U}\Big)\mathrm{r}_a^{(m)} - \overset{\circ}{\nabla}_a \kappa^{(m)} + \mathcal{O}_{ab}^{(m-1)} n^b, \tag{6.44}$$

$$\begin{aligned}P^{ab}\mathcal{R}_{ab}^{(m)} &= -2\pounds_n(\operatorname{tr}_P \mathbf{Y}^{(m)}) - 2\,(m\kappa + \operatorname{tr}_P \mathbf{U})\operatorname{tr}_P \mathbf{Y}^{(m)} + 2\kappa^{(m)}(n(\ell^{(2)}) - \operatorname{tr}_P \mathbf{Y}) \\ &\quad - 4P(\mathbf{r} + \mathbf{s}, \mathbf{r}^{(m)}) + 2\operatorname{div}_P \mathbf{r}^{(m)} + P^{ab}\mathcal{O}_{ab}^{(m-1)},\end{aligned} \tag{6.45}$$

$$\begin{aligned}\dot{\mathcal{R}}_a^{(m)} n^a &= -\kappa^{(m+1)} + \operatorname{div}_P \mathbf{r}^{(m)} - P^{ab}P^{cd}\mathrm{U}_{bd}\mathrm{Y}_{ac}^{(m)} - \pounds_n(\operatorname{tr}_P \mathbf{Y}^{(m)}) - \kappa \operatorname{tr}_P \mathbf{Y}^{(m)} \\ &\quad - 2P\Big((m+1)\mathbf{r} + 2\mathbf{s}, \mathbf{r}^{(m)}\Big) - \left(\operatorname{tr}_P \mathbf{Y} - \frac{m+3}{2} n(\ell^{(2)})\right)\kappa^{(m)} + \mathcal{O}_a^{(m-1)} n^a.\end{aligned} \tag{6.46}$$

The corresponding expressions in the case $m = 1$, also for $\nabla_\xi \xi = 0$, are (cf. (5.24), (5.28), (5.29) and recall $\mathbf{Z}^{(2)} = \mathbf{Y}^{(2)}$ because $\nabla_\xi \xi = 0$, see (2.113))

$$\begin{aligned}\mathcal{R}_{ab} &= \mathring{R}_{(ab)} - 2\pounds_n \mathrm{Y}_{ab} - (2\kappa + \operatorname{tr}_P \mathbf{U})\mathrm{Y}_{ab} + \mathring{\nabla}_{(a}(\mathrm{s}_{b)} + 2\mathrm{r}_{b)}) \\ &\quad - 2\mathrm{r}_a\mathrm{r}_b + 4\mathrm{r}_{(a}\mathrm{s}_{b)} - \mathrm{s}_a\mathrm{s}_b - (\operatorname{tr}_P \mathbf{Y})\mathrm{U}_{ab} + 2P^{cd}\mathrm{U}_{d(a}(2\mathrm{Y}_{b)c} + \mathrm{F}_{b)c}), \qquad (6.47)\\ \ddot{\mathcal{R}} &= -P^{ab}\mathrm{Y}^{(2)}_{ab} + P^{ab}P^{cd}(\mathrm{Y}+\mathrm{F})_{ac}(\mathrm{Y}+\mathrm{F})_{bd} + P^{ab}(\mathrm{r}-\mathrm{s})_a\mathring{\nabla}_b\ell^{(2)}, \qquad (6.48)\\ \dot{\mathcal{R}}_c &= \mathrm{r}^{(2)}_c - P^{ab}A_{abc} - P^{ab}(\mathrm{r}+\mathrm{s})_a(\mathrm{Y}+\mathrm{F})_{cb} + \frac{1}{2}\kappa\mathring{\nabla}_c\ell^{(2)} - \frac{1}{2}n(\ell^{(2)})(\mathrm{r}-\mathrm{s})_c. \qquad (6.49)\end{aligned}$$

The contraction of (2.119) with $n^c$ is (cf. (5.30))

$$\begin{aligned}\dot{\mathcal{R}}_a n^a &= -\kappa^{(2)} - \pounds_n(\operatorname{tr}_P \mathbf{Y}) + \operatorname{div}(\mathbf{r}+\mathbf{s}) + (2n(\ell^{(2)}) - \operatorname{tr}_P \mathbf{Y})\kappa + \frac{1}{2}(\operatorname{tr}_P \mathbf{U})n(\ell^{(2)}) \\ &\quad - P^{ad}P^{bc}\mathrm{U}_{ab}\mathrm{Y}_{cd} - 4P(\mathbf{r},\mathbf{s}) - 2P(\mathbf{r},\mathbf{r}) - 2P(\mathbf{s},\mathbf{s}), \qquad (6.50)\end{aligned}$$

and contractions of $\mathcal{R}_{ab}$ with $P$ and $n$ are (cf. (5.25)-(5.27))

$$\mathcal{R}_{ab}n^a = -\pounds_n(\mathrm{r}_b - \mathrm{s}_b) - \mathring{\nabla}_b\kappa - (\operatorname{tr}_P \mathbf{U})(\mathrm{r}_b - \mathrm{s}_b) - \mathring{\nabla}_b(\operatorname{tr}_P \mathbf{U}) + P^{cd}\mathring{\nabla}_c\mathrm{U}_{bd} - 2P^{cd}\mathrm{U}_{bd}s_c, \qquad (6.51)$$

$$\begin{aligned}\mathcal{R}_{ab}n^a n^b &= -\pounds_n(\operatorname{tr}_P \mathbf{U}) + (\operatorname{tr}_P \mathbf{U})\kappa - P^{ab}P^{cd}\mathrm{U}_{ac}\mathrm{U}_{bd}, \qquad (6.52)\\ P^{ab}\mathcal{R}_{ab} &= \operatorname{tr}_P \mathring{R} - 2\pounds_n(\operatorname{tr}_P \mathbf{Y}) - 2(\kappa + \operatorname{tr}_P \mathbf{U})\operatorname{tr}_P \mathbf{Y} + \operatorname{div}_P(\mathbf{s}+2\mathbf{r}) - 2P(\mathbf{r},\mathbf{r}) \\ &\quad - 4P(\mathbf{r},\mathbf{s}) - P(\mathbf{s},\mathbf{s}) + 2\kappa n(\ell^{(2)}). \qquad (6.53)\end{aligned}$$

Finally, the scalar curvature $R$ reads (see (5.31))

$$\begin{aligned}R &= -2\kappa^{(2)} - 4\pounds_n(\operatorname{tr}_P \mathbf{Y}) - 2(2\kappa + \operatorname{tr}_P \mathbf{U})\operatorname{tr}_P \mathbf{Y} + 3\operatorname{div}\mathbf{s} + 4\operatorname{div}\mathbf{r} - 5P(\mathbf{s},\mathbf{s}) \\ &\quad - 6P(\mathbf{r},\mathbf{r}) - 12P(\mathbf{r},\mathbf{s}) - 2P^{ab}P^{cd}\mathrm{U}_{ac}\mathrm{Y}_{bd} + (\operatorname{tr}_P \mathbf{U} + 6\kappa)n(\ell^{(2)}) + \operatorname{tr}_P \mathring{R}. \end{aligned} \qquad (6.54)$$

In order to compute $R^{(m)}$ we apply $\pounds_\xi^{(m-1)}$ to $g^{\alpha\beta}R_{\alpha\beta}$ for $m \geq 2$ and using identity (5.34) and $\pounds_\xi g^{\alpha\beta} = -g^{\alpha\mu}g^{\beta\nu}\mathcal{K}_{\mu\nu}$ one obtains

$$\begin{aligned}R^{(m)} &\overset{[m]}{=} g^{\alpha\beta}R^{(m)}_{\alpha\beta} - (m-1)g^{\alpha\mu}g^{\beta\nu}\mathcal{K}_{\mu\nu}R^{(m-1)}_{\alpha\beta} \\ &\overset{[m]}{=} P^{ab}\mathcal{R}^{(m)}_{ab} + 2\dot{\mathcal{R}}^{(m)}_a n^a - (m-1)g^{\alpha\mu}g^{\beta\nu}\mathcal{K}_{\mu\nu}R^{(m-1)}_{\alpha\beta},\end{aligned}$$

where in the second equality we inserted (2.14). The first two terms are (6.45) and (6.46). Concerning the third one, we note that (6.43) implies that terms of the form $\mathbf{Y}^{(m)}$ and $\mathbf{Y}^{(m+1)}$ can only appear when the tensor $R^{(m-1)}_{\alpha\beta}$ is contracted with $\xi$ at least once, i.e. (cf. (2.14))

$$
\begin{aligned}
g^{\alpha\mu}g^{\beta\nu}\mathcal{K}_{\mu\nu}R^{(m-1)}_{\alpha\beta} &\overset{(m)}{=} \Big(2P^{ac}e^{\mu}_{c}\nu^{\nu}e^{\alpha}_{a}\xi^{\beta} + 2\xi^{\mu}\nu^{\nu}\nu^{\alpha}\xi^{\beta} + \nu^{\mu}\nu^{\nu}\xi^{\alpha}\xi^{\beta}\Big)\mathcal{K}_{\mu\nu}R^{(m-1)}_{\alpha\beta} \\
&\overset{(m)}{=} 4P^{ab}\mathrm{r}_a\dot{\mathcal{R}}^{(m-1)}_b + n(\ell^{(2)})\dot{\mathcal{R}}^{(m-1)}_a n^a - 2\kappa\ddot{\mathcal{R}}^{(m-1)} \\
&\overset{(m)}{=} 4P^{ab}\mathrm{r}_a\mathrm{r}^{(m)}_b - n(\ell^{(2)})\kappa^{(m)} + 2\kappa\,\mathrm{tr}_P\,\mathbf{Y}^{(m)},
\end{aligned}
$$

where in the second line we used $\mathcal{K}_{ab} = 2\mathrm{Y}_{ab}$ and $\mathcal{K}_{\mu\nu}\xi^{\mu}e^{\nu}_a = \frac{1}{2}\overset{\circ}{\nabla}_a\ell^{(2)}$ (see (5.42)), and in the third line (6.42), (6.46) and (6.41). Finally, inserting this, (6.45) and (6.46) into the expression of $R^{(m)}$, we arrive at

$$
\begin{aligned}
R^{(m)} \overset{(m)}{=}& -2\kappa^{(m+1)} - 4\pounds_n(\mathrm{tr}_P\,\mathbf{Y}^{(m)}) - 2(2m\kappa + \mathrm{tr}_P\,\mathbf{U})\,\mathrm{tr}_P\,\mathbf{Y}^{(m)} - 2P^{ab}P^{cd}\mathrm{U}_{ac}\mathrm{Y}^{(m)}_{bd} \\
&+ 4\,\mathrm{div}_P\,\mathbf{r}^{(m)} - 4P\Big((2m+1)\mathbf{r} + 3\mathbf{s}, \mathrm{r}^{(m)}_b\Big) + 2\Big((m+2)n(\ell^{(2)}) - 2\,\mathrm{tr}_P\,\mathbf{Y}\Big)\kappa^{(m)}.
\end{aligned}
\tag{6.55}
$$

An immediate consequence is

$$R^{(m)} \overset{(m+1)}{=} -2\kappa^{(m+1)}. \tag{6.56}$$

In order to analyze how the quasi-Einstein equations fix the geometry at null infinity, we shall use the following tensors ($\mathcal{Q}$ is the same tensor already defined in Section 6.3)

$$\mathcal{Q}_{\alpha\beta} := (\mathfrak{n}-1)\Big(\nabla_\alpha\nabla_\beta\Omega + \Omega L_{\alpha\beta}\Big) = (\mathfrak{n}-1)\nabla_\alpha\nabla_\beta\Omega + \Omega\left(R_{\alpha\beta} - \frac{R}{2\mathfrak{n}}g_{\alpha\beta}\right), \tag{6.57}$$

$$\mathcal{L}_\alpha := (\mathfrak{n}-1)L_{\alpha\beta}\nabla^\beta\Omega = R_{\alpha\beta}\nabla^\beta\Omega - \frac{R}{2\mathfrak{n}}\nabla_\alpha\Omega, \tag{6.58}$$

and the function $f := |\nabla\Omega|^2$. Note that these quantities are obviously related by

$$\mathcal{Q}_{\alpha\beta}\nabla^\beta\Omega = \frac{\mathfrak{n}-1}{2}\nabla_\alpha f + \Omega\mathcal{L}_\alpha. \tag{6.59}$$

Less immediate is the identity

$$\nabla_\mu(\mathrm{tr}_g\,\mathcal{Q}) + \mathfrak{n}\mathcal{L}_\mu = \nabla_\rho\mathcal{Q}^\rho{}_\mu, \tag{6.60}$$

which follows from the Bianchi identity (6.19) after replacing $q = \frac{\mathrm{tr}_g\,\mathcal{Q}}{\mathfrak{n}^2-1}$ and $(\mathfrak{n}-1)\mathcal{T} = \mathcal{Q} - \frac{\mathrm{tr}\,\mathcal{Q}}{\mathfrak{n}+1}g$. In accordance of the general notation of this thesis, we introduce the tensors

$$\mathcal{Q}^{(m)}_{\alpha\beta} := \pounds^{(m-1)}_\xi\mathcal{Q}_{\alpha\beta}, \qquad \mathcal{L}^{(m)}_\alpha := \pounds^{(m-1)}_\xi\mathcal{L}_\alpha, \qquad f^{(m)} := \pounds^{(m-1)}_\xi f$$

and

$$\mathcal{Q}^{(m)}_{ab} := (\Phi^\star\mathcal{Q}^{(m)})_{ab}, \qquad \dot{\mathcal{Q}}^{(m)}_a = (\Phi^\star\mathcal{Q}^{(m)}(\xi,\cdot))_a, \qquad \ddot{\mathcal{Q}}^{(m)} := \Phi^\star(\mathcal{Q}^{(m)}(\xi,\xi)), \tag{6.61}$$

$$\mathcal{L}^{(m)}_a := (\Phi^\star\mathcal{L}^{(m)})_a, \qquad \dot{\mathcal{L}}^{(m)} := \Phi^\star(\mathcal{L}^{(m)}(\xi)). \tag{6.62}$$

We also denote the transverse derivatives of $\Omega$ at $\mathscr{I}$ by $\sigma^{(k)} := \pounds^{(k)}_\xi\Omega|_{\mathscr{I}}$. Note that $\sigma^{(1)}$ agrees with the function $\sigma$ introduced in Section 6.2, so we shall use both symbols

indistinctly. As already mentioned, this function cannot vanish anywhere on $\mathscr{I}$.

The idea now is to impose the conformal equations to infinite order at $\mathscr{I}$ to see how $\{\mathbf{Y}^{(k)}\}_{k\geq 1}$ and $\{\sigma^{(k)}\}_{k\geq 1}$ are constrained. Clearly, this depends on how the conformal factor $\Omega$ has been fixed. One sensible choice commonly used in the literature is to require the transformed Ricci scalar to vanish, which amounts solving a wave equation of the form $\Box_g \Omega = \Omega F$ for some function $F$, see [131]. Another interesting possibility is to fix $\Omega$ by solving $|\nabla\Omega|^2 = 0$. As proven in Lemma 6.10, this conformal gauge always exists locally near $\mathscr{I}$ and depends on a free function on a hypersurface transverse to $\mathscr{I}$. In either of the two choices mentioned above, one can verify that $\mathscr{I}$ is totally geodesic, which, in terms of hypersurface data means $\mathbf{U} = 0$. However, in a generic conformal gauge, $\mathscr{I}$ is only totally umbilical. In this thesis we shall be mostly concerned with the choice $|\nabla\Omega|^2 = 0$, so in Proposition 6.18 below we write down the tensors $\mathcal{Q}^{(m)}_{ab}$, $\dot{\mathcal{Q}}^{(m)}_a$, $\ddot{\mathcal{Q}}^{(m)}$, $\mathcal{L}^{(m)}_a$, $\dot{\mathcal{L}}^{(m)}$, and $f^{(m)}$ under the assumption $\mathbf{U} = 0$. Each computation is performed up to the order that it will be needed. Given that other conformal choices are possible, in Appendix E we compute the expressions in full generality, i.e. without the assumption $\mathbf{U} = 0$. The formulae in Prop. 6.18 are simply their particularization to $\mathbf{U} = 0$. We also extend the meaning of $\overset{(m)}{=}$ in Notation 5.23 as follows.

**Notation 6.17.** *Let $(\mathcal{M}, g, \Omega)$ be a conformal manifold with null infinity $\mathscr{I}$ and $T$, $S$ two tensors on $\mathscr{I}$. We use $T \overset{(m)}{=} S$ to denote that $T - S$ does not depend on transverse derivatives of $g$ and $\Omega$ at $\mathscr{I}$ of order $m$ or higher.*

**Proposition 6.18.** *Let $\mathscr{I} = \{\Omega = 0\}$ be $(\Phi, \xi)$-embedded in $(\mathcal{M}, g)$ and extend $\xi$ off $\Phi(\mathscr{I})$ geodesically. Assume $\mathbf{U} = 0$. Then, for every $m \geq 2$,*

$$\begin{aligned}\mathcal{Q}^{(m+1)}_{ab} \overset{(m)}{=}& (\mathfrak{n} - 1 - 2m)\sigma^{(1)}\pounds_n \mathrm{Y}^{(m)}_{ab} + m\Big((\mathfrak{n}-1)(\pounds_n\sigma + \sigma\kappa) + (\mathfrak{n} - 1 - 2m)\sigma^{(1)}\kappa\Big)\mathrm{Y}^{(m)}_{ab} \\ &+ \frac{m\sigma^{(1)}}{\mathfrak{n}}\Big(\kappa^{(m+1)} + 2\pounds_n(\operatorname{tr}_P \mathbf{Y}^{(m)}) + 2m\kappa \operatorname{tr}_P \mathbf{Y}^{(m)}\Big)\gamma_{ab} \\ &+ \frac{m(m-1)}{2\mathfrak{n}}\sigma^{(2)}\kappa^{(m)}\gamma_{ab} + \widetilde{\mathscr{R}}^{(m)}_{ab},\end{aligned} \tag{6.63}$$

$$\dot{\mathcal{Q}}^{(m+1)}_a \overset{(m+1)}{=} (\mathfrak{n}-1)\Big(\mathring{\nabla}_a\sigma^{(m+1)} - \sigma^{(m+1)}(\mathrm{r}-\mathrm{s})_a\Big) - (\mathfrak{n} - 1 - m)\sigma^{(1)}\mathrm{r}^{(m+1)}_a + \frac{m\sigma^{(1)}}{\mathfrak{n}}\kappa^{(m+1)}\ell_a, \tag{6.64}$$

$$\ddot{\mathcal{Q}}^{(m+1)} \overset{(m+1)}{=} (\mathfrak{n}-1)\sigma^{(m+2)} - m\sigma^{(1)}\operatorname{tr}_P \mathbf{Y}^{(m+1)} + \frac{m\sigma^{(1)}\ell^{(2)}}{\mathfrak{n}}\kappa^{(m+1)}, \tag{6.65}$$

$$\mathcal{L}^{(m)}_a \overset{(m)}{=} -\sigma^{(1)}\pounds_n \mathrm{r}^{(m)}_a + (m-1)(\pounds_n\sigma^{(1)})\mathrm{r}^{(m)}_a - \sigma^{(1)}\mathring{\nabla}_a\kappa^{(m)} + \frac{(m-1)\kappa^{(m)}}{\mathfrak{n}}\mathring{\nabla}_a\sigma^{(1)}, \tag{6.66}$$

$$\begin{aligned}\mathfrak{n}\dot{\mathcal{L}}^{(m)} \overset{(m)}{=}& -(\mathfrak{n}-1)\sigma^{(1)}\kappa^{(m+1)} - (\mathfrak{n}-2)\sigma^{(1)}\pounds_n(\operatorname{tr}_P \mathbf{Y}^{(m)}) \\ &+ \Big((2m(1-\mathfrak{n}) + \mathfrak{n})\sigma^{(1)}\kappa - (m-1)\mathfrak{n}\pounds_n\sigma^{(1)}\Big)\operatorname{tr}_P \mathbf{Y}^{(m)} + \dot{\mathscr{R}}^{(m)}_{\mathcal{L}},\end{aligned} \tag{6.67}$$

$$f^{(m+1)} \overset{(m)}{=} 2\sigma^{(1)}(\pounds_n\sigma^{(m)} + \sigma^{(1)}\kappa^{(m)}) + 2m(\pounds_n\sigma^{(1)} + 2\sigma^{(1)}\kappa)\sigma^{(m)}, \tag{6.68}$$

*where* $\widetilde{\mathscr{R}}^{(m)}_{ab}$*,* $\dot{\mathscr{R}}^{(m)}_{\mathcal{L}}$ *are tensors that depend on* $\mathbf{r}^{(m)}$*,* $\sigma^{(m)}$ *and lower order terms and we do not write for simplicity (they can be easily read out by performing explicitly the calculations in their respective proofs).*

*Proof.* The first three expressions are the particularization to the case $\mathbf{U}=0$ of Proposition E.4, the next two of Proposition E.2 and the last one of Proposition E.5. □

The tensors $\mathcal{L}^{(1)}_{\mu}$, $\mathcal{Q}^{(1)}_{\alpha\beta}$, $\mathcal{Q}^{(2)}_{\alpha\beta}$ and $f^{(2)}$ are not covered by this proposition. Again under the assumption $\mathbf{U}=0$ they are given by (see (E.35)-(E.38), (E.20)-(E.21) and (E.40) in Appendix E)

$$\mathcal{Q}^{(1)}_{ab}=0,\qquad \dot{\mathcal{Q}}^{(1)}_{a}=\mathring{\nabla}_a\sigma^{(1)}-\sigma^{(1)}(\mathrm{r}-\mathrm{s})_a,\qquad \ddot{\mathcal{Q}}^{(1)}=\sigma^{(2)},\tag{6.69}$$

$$\begin{aligned}\mathcal{Q}^{(2)}_{ab}&=(\mathfrak{n}-3)\sigma^{(1)}\pounds_n\mathrm{Y}_{ab}+\Big((\mathfrak{n}-1)(\pounds_n\sigma+\sigma\kappa)+(\mathfrak{n}-3)\sigma^{(1)}\kappa\Big)\mathrm{Y}_{ab}+(\mathfrak{n}-1)\mathring{\nabla}_a\mathring{\nabla}_b\sigma^{(1)}\\&\quad-2\sigma^{(1)}(\mathfrak{n}-2)\mathring{\nabla}_{(a}\mathrm{r}_{b)}+\sigma^{(1)}\Big(\mathring{\nabla}_{(a}\mathrm{s}_{b)}-2\mathrm{r}_a\mathrm{r}_b+4\mathrm{r}_{(a}\mathrm{s}_{b)}-\mathrm{s}_a\mathrm{s}_b+\mathring{R}_{(ab)}\Big)-\frac{\sigma R}{2\mathfrak{n}}\gamma_{ab},\qquad(6.70)\\\dot{\mathcal{Q}}^{(2)}_{a}&=-(\mathfrak{n}-2)\sigma^{(1)}\mathrm{r}^{(2)}_a-(\mathfrak{n}-1)V_a{}^b\mathring{\nabla}_b\sigma^{(1)}-\frac{1}{2}\sigma^{(1)}(\mathfrak{n}-2)n(\ell^{(2)})(\mathrm{r}-\mathrm{s})_a+2\sigma^{(1)}(\mathfrak{n}-1)V^c{}_a\mathrm{r}_c\\&\quad-\sigma^{(1)}P^{bc}A_{bca}-\sigma^{(1)}P^{cb}(\mathrm{r}+\mathrm{s})_c(\mathrm{Y}_{ab}+\mathrm{F}_{ab})+\frac{1}{2}\sigma^{(1)}\kappa\mathring{\nabla}_a\ell^{(2)}-\frac{\sigma^{(1)}R}{2\mathfrak{n}}\ell_a,\qquad(6.71)\\\ddot{\mathcal{Q}}^{(2)}&=(\mathfrak{n}-1)\sigma^{(2)}-\sigma^{(1)}\operatorname{tr}_P\mathbf{Y}^{(2)}+\sigma^{(1)}P^{ab}P^{cd}(\mathrm{Y}+\mathrm{F})_{ac}(\mathrm{Y}+\mathrm{F})_{bd}+\sigma^{(1)}P(\mathbf{r}-\mathbf{s},d\ell^{(2)})\\&\quad-\frac{R}{2\mathfrak{n}}\sigma^{(1)}\ell^{(2)},\qquad(6.72)\end{aligned}$$

$$\begin{aligned}\frac{2\mathfrak{n}}{\sigma^{(1)}}\dot{\mathcal{L}}^{(1)}&=-2(\mathfrak{n}-1)\kappa^{(2)}-2(\mathfrak{n}-2)\pounds_n(\operatorname{tr}_P\mathbf{Y})-2(\mathfrak{n}-2)\kappa\operatorname{tr}_P\mathbf{Y}+2(\mathfrak{n}-2)\operatorname{div}\mathbf{r}\\&\quad+(2\mathfrak{n}-3)\operatorname{div}\mathbf{s}+2(2\mathfrak{n}-3)\kappa n(\ell^{(2)})-2(2\mathfrak{n}-3)P(\mathbf{r},\mathbf{r})-(4\mathfrak{n}-5)P(\mathbf{s},\mathbf{s})\\&\quad-4(2\mathfrak{n}-3)P(\mathbf{r},\mathbf{s})-\operatorname{tr}_P\mathring{R},\qquad(6.73)\\\frac{1}{\sigma^{(1)}}\mathcal{L}^{(1)}_a&=-\pounds_n(\mathrm{r}_b-\mathrm{s}_b)-\mathring{\nabla}_b\kappa,\qquad(6.74)\end{aligned}$$

where $R$ is explicitly given by (6.54). In addition (cf. (E.40))

$$f^{(2)}=2\sigma(\pounds_n\sigma+\sigma\kappa).\tag{6.75}$$

Although $f^{(3)}$ is covered in Proposition 6.18, we will need shortly its explicit expression (cf. (E.41))

$$\begin{aligned}f^{(3)}&=2\sigma^{(1)}(\sigma^{(1)}\kappa^{(2)}+\pounds_n\sigma^{(2)})+2\Big(2\sigma^{(2)}-\sigma^{(1)}n(\ell^{(2)})\Big)(\pounds_n\sigma^{(1)}+2\sigma^{(1)}\kappa)\\&\quad+8\sigma^{(1)}P(\mathbf{r},\sigma^{(1)}\mathbf{r}-d\sigma^{(1)})+2P^{ab}\mathring{\nabla}_a\sigma^{(1)}\mathring{\nabla}_b\sigma^{(1)}.\end{aligned}\tag{6.76}$$

One immediate consequence of (E.35) is that, for every embedded $\mathscr{I}$-structure data satisfying $\dot{\mathcal{Q}}_a=\mathfrak{q}\ell_a$, the $\nabla_X$ derivative of $\nabla\Omega$ at $\mathscr{I}$ is given by (see (6.29))

$$\nabla_X(\nabla\Omega)\stackrel{\mathscr{I}}{=}\mathfrak{q}X,$$

and therefore every $\mathscr{I}$-structure data written in a conformal gauge with $\mathfrak{q} = 0$ is in particular a weakly isolated horizon (see [32]).

In the conformal gauge in which $|\nabla\Omega|^2 = 0$, the "higher order conformal equations" that we must solve are $\mathcal{Q}^{(m)}_{ab} = 0$, $\dot{\mathcal{Q}}^{(m)}_a = 0$, $\ddot{\mathcal{Q}}^{(m)} = 0$, $\mathcal{L}^{(m)}_a = 0$, $\dot{\mathcal{L}}^{(m)} = 0$ and $f^{(m)} = 0$ for every $m \geq 1$. Observe that, for each $m$, there are $2(\mathfrak{n}+1)$ equations more than components of the metric to be fixed ($\mathbf{Y}^{(m)}$ and $\sigma^{(m)}$). This overdeterminacy is related to the identities (6.59) and (6.60). Taking $m$ transverse derivatives in (6.59) and applying (5.7) we arrive at

$$\sum_{k=0}^{m}\binom{m}{k}\mathcal{Q}^{(m+1-k)}_{\alpha\beta}\sum_{j=0}^{k}\binom{k}{j}(\pounds^{(j)}_{\xi}g^{\beta\mu})\nabla_\mu\pounds^{(k-j)}_{\xi}\Omega = \frac{\mathfrak{n}-1}{2}\nabla_\alpha\pounds^{(m)}_{\xi}f + \sum_{k=0}^{m}\binom{m}{k}(\pounds^{(k)}_{\xi}\Omega)\mathcal{L}^{(m-k+1)}_{\alpha}, \tag{6.77}$$

and applying $\pounds^{(m-1)}_{\xi}$ to (6.60) gives

$$\begin{aligned}\left(\nabla_\rho\mathcal{Q}^{(m)\rho}{}_{\mu} + \sum_{k=0}^{m-2}\binom{m-1}{k+1}\left(\mathcal{Q}^{(m-1-k)\sigma}{}_{\mu}\Sigma^{(k+1)\rho}{}_{\rho\sigma} - \mathcal{Q}^{(m-1-k)\rho}{}_{\sigma}\Sigma^{(k+1)\sigma}{}_{\rho\mu}\right)\right) = \\ = \nabla_\mu\left(\sum_{k=0}^{m-1}\binom{m-1}{k}(\pounds^{(k)}_{\xi}g^{\alpha\beta})\mathcal{Q}^{(m-k)}_{\alpha\beta}\right) + \mathfrak{n}\mathcal{L}^{(m)}_{\mu}.\end{aligned} \tag{6.78}$$

In the next two lemmas we prove some direct consequences of these identities. They will be essential in Section 6.5 to show, order by order, that the set of $2(\mathfrak{n}+1)$ redundant equations is automatically satisfied provided the remaining equations hold. This is analogous (though considerably more involved) to Proposition 5.43 in Chapter 5. Each lemma comes with its own general hypothesis, namely the validity of the quasi-Einstein equations up to a certain order, and is divided into several items with additional conditions. Although some items follow directly from others, each will be used separately below. For the sake of clarity, we therefore present them individually.

**Lemma 6.19.** *Fix $m \geq 1$ and assume $\mathcal{Q}^{(k)}_{\alpha\beta} = 0$ and $\mathcal{L}^{(k)}_{\alpha\beta} = 0$ for every $k = 1, ..., m-1$ whenever $m \geq 2$.*

1. *If $\mathcal{Q}^{(m)}_{ab}n^b = 0$, then*

$$\sigma^{(1)}\mathcal{Q}^{(m+1)}_{ab}n^a n^b + m(2\sigma^{(1)}\kappa + \pounds_n\sigma^{(1)})\dot{\mathcal{Q}}^{(m)}_a n^a = \frac{\mathfrak{n}-1}{2}\pounds_n f^{(m+1)} + m\sigma^{(1)}\mathcal{L}^{(m)}_a n^a. \tag{6.79}$$

2. *If $\mathcal{Q}^{(m)}_{ab} = 0$, then*

$$\sigma^{(1)}\mathcal{Q}^{(m+1)}_{ab}n^b + m(2\sigma^{(1)}\kappa + \pounds_n\sigma^{(1)})\dot{\mathcal{Q}}^{(m)}_a = \frac{\mathfrak{n}-1}{2}\overset{\circ}{\nabla}_a f^{(m+1)} + m\sigma^{(1)}\mathcal{L}^{(m)}_a \tag{6.80}$$

3. *If $\mathcal{Q}^{(m)}_{ab} = 0$ and $\dot{\mathcal{Q}}^{(m)}_a = 0$, then*

$$\sigma^{(1)}\dot{\mathcal{Q}}^{(m+1)}_a n^a + m(2\sigma^{(1)}\kappa + \pounds_n\sigma^{(1)})\ddot{\mathcal{Q}}^{(m)} = \frac{\mathfrak{n}-1}{2}f^{(m+2)} + m\sigma^{(1)}\dot{\mathcal{L}}^{(m)} \tag{6.81}$$

*and*

$$\sigma^{(1)}\mathcal{Q}^{(m+1)}_{ab}n^b = \frac{\mathfrak{n}-1}{2}\overset{\circ}{\nabla}_a f^{(m+1)} + m\sigma^{(1)}\mathcal{L}^{(m)}_a. \tag{6.82}$$

*Proof.* Directly from (6.77),

$$\mathcal{Q}^{(m+1)}_{\alpha\beta}\nabla^\beta\Omega + m\mathcal{Q}^{(m)}_{\alpha\beta}\Big((\pounds_\xi g^{\beta\mu})\nabla_\mu\Omega + g^{\beta\mu}\nabla_\mu\pounds_\xi\Omega\Big) \stackrel{\mathscr{I}}{=} \frac{\mathfrak{n}-1}{2}\nabla_\alpha\pounds^{(m)}_\xi f + m\sigma^{(1)}\mathcal{L}^{(m)}_\alpha,$$

which after inserting $\nabla^\beta\Omega \stackrel{\mathscr{I}}{=} \sigma^{(1)}\nu^\beta$, (5.45) and (2.14) becomes

$$\begin{aligned}\sigma^{(1)}\mathcal{Q}^{(m+1)}_{\alpha\beta}\nu^\beta + m\mathcal{Q}^{(m)}_{\alpha\beta}\Big(P^{bc}\big(\mathring{\nabla}_c\sigma^{(1)} - 2\sigma^{(1)}\mathrm{r}_c\big)e^\beta_b + \Big(\sigma^{(2)} - \frac{1}{2}\sigma^{(1)}n(\ell^{(2)})\Big)\nu^\beta\Big)\\ +m(2\sigma^{(1)}\kappa + \pounds_n\sigma^{(1)})\xi^\beta\mathcal{Q}^{(m)}_{\alpha\beta} \stackrel{\mathscr{I}}{=} \frac{\mathfrak{n}-1}{2}\nabla_\alpha\pounds^{(m)}_\xi f + m\sigma^{(1)}\mathcal{L}^{(m)}_\alpha.\end{aligned} \tag{6.83}$$

The contraction of this identity with $\nu^\alpha$ gives (6.79) at once after using that $\mathcal{Q}^{(m)}_{ab}n^b = 0$. To prove (6.80) one contracts (6.83) with $e^\alpha_a$ and uses the hypothesis $\mathcal{Q}^{(m)}_{ab} = 0$. Identity (6.82) is an immediate consequence of (6.80). Finally, to show (6.81) we contract (6.83) with $\xi^\alpha$ and use $\mathcal{Q}^{(m)}_{ab} = 0$ and $\dot{\mathcal{Q}}^{(m)}_a = 0$. □

**Lemma 6.20.** *Fix $m \geq 1$ and assume $\mathcal{Q}^{(k)}_{\alpha\beta} = 0$ for every $k = 1, ..., m-1$ whenever $m \geq 2$.*

1. *If $\mathcal{Q}^{(m)}_{ab}n^b = 0$, then*

$$\begin{aligned}\mathcal{Q}^{(m+1)}_{ab}n^a n^b = \pounds_n(\operatorname{tr}_P \mathcal{Q}^{(m)}) + \frac{\operatorname{tr}_P \mathbf{U}}{\mathfrak{n}-1}\operatorname{tr}_P \mathcal{Q}^{(m)} + P^{bc}P^{da}\mathrm{U}_{bd}\widehat{\mathcal{Q}}^{(m)}_{ac} + \pounds_n(\dot{\mathcal{Q}}^{(m)}_a n^a)\\ -(2\kappa + \operatorname{tr}_P \mathbf{U})\dot{\mathcal{Q}}^{(m)}_a n^a + \mathfrak{n}\pounds^{(m)}_a n^a.\end{aligned} \tag{6.84}$$

2. *If $\mathcal{Q}^{(m)}_{ab} = 0$, then*

$$\mathcal{Q}^{(m+1)}_{ab}n^b + \pounds_n\dot{\mathcal{Q}}^{(m)}_a + (2\kappa + \operatorname{tr}_P \mathbf{U})\dot{\mathcal{Q}}^{(m)}_a = 2\mathring{\nabla}_a(\dot{\mathcal{Q}}^{(m)}_b n^b) + \mathfrak{n}\mathcal{L}^{(m)}_a. \tag{6.85}$$

3. *If $\mathcal{Q}^{(m)}_{ab} = 0$ and $\dot{\mathcal{Q}}^{(m)}_a = 0$, then*

$$\pounds_n\ddot{\mathcal{Q}}^{(m)} + \Big((3-2m)\kappa + \operatorname{tr}_P \mathbf{U}\Big)\ddot{\mathcal{Q}}^{(m)} = \dot{\mathcal{Q}}^{(m+1)}_a n^a + \operatorname{tr}_P \mathcal{Q}^{(m+1)} + \mathfrak{n}\dot{\mathcal{L}}^{(m)}. \tag{6.86}$$

4. *Finally, if $\mathcal{Q}^{(m)}_{\alpha\beta} = 0$, then*

$$\dot{\mathcal{Q}}^{(m+1)}_a n^a + \operatorname{tr}_P \mathcal{Q}^{(m+1)} + \mathfrak{n}\dot{\mathcal{L}}^{(m)} = 0 \tag{6.87}$$

*and*

$$\mathcal{Q}^{(m+1)}_{ab}n^a n^b = \mathfrak{n}\mathcal{L}^{(m)}_a n^a. \tag{6.88}$$

*Proof.* Relations (6.87) and (6.88) are particularizations respectively of (6.86) and (6.85), so it suffices to prove (6.84)-(6.86). Our strategy is to write down (6.78) at $\mathscr{I}$ and then compute its contractions first with $e^\mu_c$ and then with $\xi^\mu$. The former will show (6.84)-(6.85), and the later will establish (6.86). With the assumption $\mathcal{Q}^{(k)}_{\alpha\beta} = 0$ for $k = 1, ..., m-1$, identity (6.78) reads

$$
\begin{aligned}
0 &\stackrel{\mathscr{I}}{=} -\nabla_\rho \mathcal{Q}^{(m)\rho}{}_\mu + \nabla_\mu \left( g^{\alpha\beta} \mathcal{Q}^{(m)}_{\alpha\beta} + (m-1)(\pounds_\xi g^{\alpha\beta}) \mathcal{Q}^{(m-1)}_{\alpha\beta} \right) + \mathfrak{n}\mathcal{L}^{(m)}_\mu \\
&\stackrel{\mathscr{I}}{=} -\nabla_\rho \mathcal{Q}^{(m)\rho}{}_\mu + g^{\alpha\beta} \nabla_\mu \mathcal{Q}^{(m)}_{\alpha\beta} - (m-1)\mathcal{K}^{\alpha\beta} \nabla_\mu \mathcal{Q}^{(m-1)}_{\alpha\beta} + \mathfrak{n}\mathcal{L}^{(m)}_\mu
\end{aligned}
\tag{6.89}
$$

where in the second line we used $\mathcal{Q}^{(m-1)}_{\alpha\beta} \stackrel{\mathscr{I}}{=} 0$ and $\pounds_\xi g^{\alpha\beta} = -\mathcal{K}^{\alpha\beta}$.

We now compute the contraction of (6.89) with $e^\mu_c$ under the assumption $\mathcal{Q}^{(m)}_{ab} n^b = 0$, i.e. (cf. (2.24)) $\mathcal{Q}^{(m)}_{ab} = \frac{\operatorname{tr}_P \mathcal{Q}^{(m)}}{\mathfrak{n}-1} \gamma_{ab} + \widehat{\mathcal{Q}}^{(m)}_{ab}$, from which (6.84) and (6.85) will then follow at once. This contraction is

$$
0 \stackrel{\mathscr{I}}{=} -(\operatorname{div} \mathcal{Q}^{(m)})_a + \overset{\circ}{\nabla}_a \left( \operatorname{tr}_P \mathcal{Q}^{(m)} + 2\dot{\mathcal{Q}}^{(m)}_b n^b \right) + \mathfrak{n}\mathcal{L}^{(m)}_a
\tag{6.90}
$$

because the tangential derivatives of $\mathcal{Q}^{(m-1)}_{\alpha\beta}$ vanish at $\mathscr{I}$. Only the first term requires further analysis. Applying Proposition B.3 to $\mathcal{H} = \mathscr{I}$ and using $\mathcal{Q}^{(m)}_{ab} n^b = 0$ we get

$$
\begin{aligned}
(\operatorname{div} \mathcal{Q}^{(m)})_a &= \mathcal{Q}^{(m+1)}_{ab} n^b + P^{bc} \overset{\circ}{\nabla}_b \mathcal{Q}^{(m)}_{ac} + n^b \overset{\circ}{\nabla}_b \dot{\mathcal{Q}}^{(m)}_a + (2\kappa + \operatorname{tr}_P \mathbf{U}) \dot{\mathcal{Q}}^{(m)}_a \\
&\quad - 2P^{bc} (\mathrm{r}+\mathrm{s})_b \mathcal{Q}^{(m)}_{ac} + P^{bc} \mathrm{U}_{ba} \dot{\mathcal{Q}}^{(m)}_c + \mathrm{s}_a \dot{\mathcal{Q}}^{(m)}_b n^b \\
&= \mathcal{Q}^{(m+1)}_{ab} n^b + P^{bc} \overset{\circ}{\nabla}_b \mathcal{Q}^{(m)}_{ac} + \pounds_n \dot{\mathcal{Q}}^{(m)}_a + (2\kappa + \operatorname{tr}_P \mathbf{U}) \dot{\mathcal{Q}}^{(m)}_a - 2P^{bc} (\mathrm{r}+\mathrm{s})_b \mathcal{Q}^{(m)}_{ac},
\end{aligned}
$$

where we used (cf. (2.66)) $\pounds_n \dot{\mathcal{Q}}^{(m)}_a = n^b \overset{\circ}{\nabla}_b \dot{\mathcal{Q}}^{(m)}_a + P^{bc} \mathrm{U}_{ba} \dot{\mathcal{Q}}^{(m)}_c + \mathrm{s}_a \dot{\mathcal{Q}}^{(m)}_b n^b$. Since $\mathcal{Q}^{(m)}_{ab} = \frac{\operatorname{tr}_P \mathcal{Q}^{(m)}}{\mathfrak{n}-1} \gamma_{ab} + \widehat{\mathcal{Q}}^{(m)}_{ab}$, the second and fifth terms can be elaborated further by means of

$$
\begin{aligned}
P^{bc} \overset{\circ}{\nabla}_b \mathcal{Q}^{(m)}_{ac} &\underset{(2.38)}{=} \frac{1}{\mathfrak{n}-1} \left( P^{bc} \gamma_{ac} \overset{\circ}{\nabla}_b (\operatorname{tr}_P \mathcal{Q}^{(m)}) - 2 \operatorname{tr}_P \mathcal{Q}^{(m)} P^{bc} \mathrm{U}_{b(a} \ell_{c)} \right) + P^{bc} \overset{\circ}{\nabla}_b \widehat{\mathcal{Q}}^{(m)}_{ac} \\
&\underset{(2.6),(2.5)}{=} \frac{1}{\mathfrak{n}-1} \left( \overset{\circ}{\nabla}_a (\operatorname{tr}_P \mathcal{Q}^{(m)}) - \left( \pounds_n (\operatorname{tr}_P \mathcal{Q}^{(m)}) + \operatorname{tr}_P \mathbf{U} \operatorname{tr}_P \mathcal{Q}^{(m)} \right) \ell_a \right) + P^{bc} \overset{\circ}{\nabla}_b \widehat{\mathcal{Q}}^{(m)}_{ac}
\end{aligned}
$$

and

$$
P^{bc} \mathcal{Q}^{(m)}_{ac} \stackrel{(2.6)}{=} \frac{\operatorname{tr}_P \mathcal{Q}^{(m)}}{\mathfrak{n}-1} (\delta^b_a - n^b \ell_a) + P^{bc} \widehat{\mathcal{Q}}^{(m)}_{ac}.
$$

Consequently,

$$
\begin{aligned}
(\operatorname{div} \mathcal{Q}^{(m)})_a &= \mathcal{Q}^{(m+1)}_{ab} n^b + \pounds_n \dot{\mathcal{Q}}^{(m)}_a + (2\kappa + \operatorname{tr}_P \mathbf{U}) \dot{\mathcal{Q}}^{(m)}_a + P^{bc} \overset{\circ}{\nabla}_b \widehat{\mathcal{Q}}^{(m)}_{ac} - 2P^{bc} (\mathrm{r}+\mathrm{s})_b \widehat{\mathcal{Q}}^{(m)}_{ac} \\
&\quad + \frac{1}{\mathfrak{n}-1} \left( \overset{\circ}{\nabla}_a \operatorname{tr}_P \mathcal{Q}^{(m)} - \pounds_n (\operatorname{tr}_P \mathcal{Q}^{(m)}) \ell_a - \left( (2\kappa + \operatorname{tr}_P \mathbf{U}) \ell_a + 2(\mathrm{r}+\mathrm{s})_a \right) \operatorname{tr}_P \mathcal{Q}^{(m)} \right).
\end{aligned}
$$

Inserting this into (6.90) and simplifying gives

$$
\begin{aligned}
0 &= -\mathcal{Q}^{(m+1)}_{ab} n^b - \pounds_n \dot{\mathcal{Q}}^{(m)}_a - (2\kappa + \operatorname{tr}_P \mathbf{U}) \dot{\mathcal{Q}}^{(m)}_a + \frac{1}{\mathfrak{n}-1} \left( (\mathfrak{n}-2) \overset{\circ}{\nabla}_a \operatorname{tr}_P \mathcal{Q}^{(m)} + \pounds_n (\operatorname{tr}_P \mathcal{Q}^{(m)}) \ell_a \right) \\
&\quad + \frac{1}{\mathfrak{n}-1} \left( (2\kappa + \operatorname{tr}_P \mathbf{U}) \ell_a + 2(\mathrm{r}+\mathrm{s})_a \right) \operatorname{tr}_P \mathcal{Q}^{(m)} - P^{bc} \overset{\circ}{\nabla}_b \widehat{\mathcal{Q}}^{(m)}_{ac} + 2P^{bc} (\mathrm{r}+\mathrm{s})_b \widehat{\mathcal{Q}}^{(m)}_{ac} \\
&\quad + 2\overset{\circ}{\nabla}_a (\dot{\mathcal{Q}}^{(m)}_b n^b) + \mathfrak{n}\mathcal{L}^{(m)}_a.
\end{aligned}
$$

A contraction with $n^a$ gives (6.84) after noting $\widehat{Q}^{(m)}_{ab} n^a = 0$ and $n^a \overset{\circ}{\nabla}_b \widehat{\mathcal{Q}}^{(m)}_{ac} = -P^{ad} \mathrm{U}_{db} \widehat{\mathcal{Q}}^{(m)}_{ac}$ (cf. (2.66)), while (6.85) is just this expression after setting $\operatorname{tr}_P \mathcal{Q}^{(m)} = 0$ and $\widehat{\mathcal{Q}}^{(m)}_{ab} = 0$.

To show (6.86) we now contract (6.89) with $\xi^\mu$ and use

$$\xi^\mu \nabla_\rho \mathcal{Q}^{(m)\rho}{}_\mu = \nabla_\rho \dot{\mathcal{Q}}^{(m)\rho} - \mathcal{Q}^{(m)\rho}{}_\mu \nabla_\rho \xi^\mu, \qquad \xi^\mu \nabla_\mu \mathcal{Q}^{(m)}_{\alpha\beta} = \mathcal{Q}^{(m+1)}_{\alpha\beta} - 2\mathcal{Q}^{(m)}_{\mu(\alpha} \nabla_{\beta)} \xi^\mu,$$

and $\xi^\mu \nabla_\mu \mathcal{Q}^{(m-1)}_{\alpha\beta} \stackrel{\mathscr{I}}{=} \mathcal{Q}^{(m)}_{\alpha\beta}$ (because $\mathcal{Q}^{(m-1)}_{\alpha\beta} = 0$) to get

$$0 \stackrel{\mathscr{I}}{=} -\nabla_\rho \dot{\mathcal{Q}}^{(m)\rho} - \mathcal{Q}^{(m)\rho}{}_\mu \nabla_\rho \xi^\mu + g^{\alpha\beta} \mathcal{Q}^{(m+1)}_{\alpha\beta} - (m-1) \mathcal{K}^{\alpha\beta} \mathcal{Q}^{(m)}_{\alpha\beta} + \mathfrak{n} \dot{\mathcal{L}}^{(m)}.$$

The first term is given by Prop. B.3, which under the assumption $\dot{\mathcal{Q}}^{(m)}_a = 0$ is $\nabla_\rho \dot{\mathcal{Q}}^{(m)\rho} = \dot{\mathcal{Q}}^{(m+1)}_a n^a + \pounds_n \ddot{\mathcal{Q}}^{(m)} + \left(2\kappa + \operatorname{tr}_P \mathbf{U}\right) \ddot{\mathcal{Q}}^{(m)}$. The second term is computed from (2.72) after noting that, after the assumptions of item 3., namely $\mathcal{Q}^{(m)}_{ab} = 0$ and $\dot{\mathcal{Q}}^{(m)}_a = 0$, only the term with two riggings survives, $\mathcal{Q}^{(m)\rho}{}_\mu \nabla_\rho \xi^\mu = -\kappa \ddot{\mathcal{Q}}^{(m)}$. The third term is simply $g^{\alpha\beta} \mathcal{Q}^{(m+1)}_{\alpha\beta} \underset{(2.13)}{=} \operatorname{tr}_P \mathcal{Q}^{(m+1)} + 2\dot{\mathcal{Q}}^{(m+1)}_a n^a$, and for the fourth term we use again that the only piece that does not vanish is the contraction with $\xi$ twice, obtaining

$$\mathcal{K}^{\alpha\beta} \mathcal{Q}^{(m)}_{\alpha\beta} = g^{\alpha\mu} \mathcal{K}_\mu{}^\beta \mathcal{Q}^{(m)}_{\alpha\beta} \underset{(2.13)}{=} \nu^\mu \xi^\alpha \mathcal{K}_\mu{}^\beta \mathcal{Q}^{(m)}_{\alpha\beta} \underset{(5.45)}{=} -2\kappa \xi^\alpha \xi^\beta \mathcal{Q}^{(m)}_{\alpha\beta} = -2\kappa \ddot{\mathcal{Q}}^{(m)}.$$

Adding up the four terms and $\mathfrak{n}\dot{\mathcal{L}}^{(m)}$, (6.86) is established. □

Combining (6.80) and (6.85) one has the following corollary.

**Corollary 6.21.** *Let $m \geq 1$ and assume $\mathcal{Q}^{(k)}_{\alpha\beta} = 0$ and $\mathcal{L}^{(k)}_\mu = 0$ for every $k = 1, ..., m-1$, as well as $f^{(m+1)} = 0$, $\dot{\mathcal{Q}}^{(m)}_a n^a = 0$, $\mathcal{Q}^{(m)}_{ab} n^a = 0$ and $\operatorname{tr}_P \mathcal{Q}^{(m)} = 0$. If $m = \mathfrak{n}$, then $\mathcal{Q}^{(\mathfrak{n}+1)}(n,n) = \mathfrak{n} \mathcal{L}^{(\mathfrak{n})}_a n^a$; and if $m \neq \mathfrak{n}$, then $\mathcal{Q}^{(m+1)}(n,n) = \mathcal{L}^{(m)}_a n^a = 0$.*

*Proof.* From (6.79) one has $\mathcal{Q}^{(m+1)}(n,n) = m\mathcal{L}^{(m)}_a n^a$, and from $\mathcal{Q}_{ab} = 0$ we have $\mathrm{U}_{ab} = 0$ (see (E.35)), so (6.84) gives $\mathcal{Q}^{(m+1)}(n,n) = \mathfrak{n} \mathcal{L}^{(m)}_a n^a$. When $\mathfrak{n} = m$, then $\mathcal{Q}^{(m+1)}(n,n) = \mathfrak{n} \mathcal{L}^{(m)}_a n^a$; and when $\mathfrak{n} \neq m$, $\mathcal{Q}^{(m+1)}(n,n) = \mathcal{L}^{(m)}_a n^a = 0$. □

As we will see in detail below, these results show that, for generic values of $m$, determining the transverse expansion at $\mathscr{I}$ reduces to analyzing, order by order, the equations $\ddot{\mathcal{Q}}^{(m)} = 0$, $f^{(m)} = 0$, $\dot{\mathcal{L}}^{(m)} = 0$ as well as $\widehat{\mathcal{L}}^{(m)}_a := \mathcal{L}^{(m)}_a - (\mathcal{L}^{(m)}_b n^b) \ell_a = 0$ (cf. (2.26)) and (cf. (2.24))

$$\widehat{\mathcal{Q}}^{(m)}_{ab} := \mathcal{Q}^{(m)}_{ab} - \frac{\operatorname{tr}_P \mathcal{Q}^{(m)} + \mathcal{Q}^{(m)}(n,n) \ell^{(2)}}{\mathfrak{n} - 1} \gamma_{ab} - 2\ell_{(a} \mathcal{Q}^{(m)}_{b)c} n^c - \mathcal{Q}^{(m)}(n,n) \ell_a \ell_b = 0.$$

The remaining equations will turn out to be automatically satisfied as a consequence of Lemmas 6.19 and 6.20, together with Corollary 6.21.

**Proposition 6.22.** *Let $(\mathcal{M}, g, \Omega)$ be an $(\mathfrak{n}+1)$-dimensional conformal manifold with null infinity $\mathscr{I}$ and assume $\mathscr{I}$ admits a cross-section $\Sigma$. Fix an integer $\ell \geq 1$ and assume that equations $\mathcal{L}^{(k)}_\mu = 0$, $\mathcal{Q}^{(k)}_{\mu\nu} = 0$ hold for every $k \leq \ell$, $f^{(k)} = 0$ for every $k \leq \ell + 2$, and also $\mathcal{L}^{(\mathfrak{n})}_a n^a = 0$ when $\ell = \mathfrak{n} - 1$ and $f^{(\mathfrak{n}+1)}|_\Sigma = 0$ when $\ell = \mathfrak{n} - 2$. Then,*

1. *$\dot{\mathcal{Q}}^{(\ell+1)}_a n^a = 0$, $\mathcal{Q}^{(\ell+1)}_{ab} n^b = 0$, $P^{ab}\mathcal{Q}^{(\ell+1)}_{ab} = 0$ and $\mathcal{Q}^{(\ell+2)}_{ab} n^a n^b = \mathcal{L}^{(\ell+1)}_a n^a = 0$.*

*Assume in addition that $\widehat{\mathcal{L}}^{(\ell+1)}_a = 0$, $\dot{\widehat{\mathcal{Q}}}^{(\ell+1)}_a|_\Sigma = 0$ and $\widehat{\mathcal{Q}}^{(\ell+1)}_{ab} = 0$. Then,*

2. *$\dot{\mathcal{Q}}^{(\ell+1)}_a = 0$ and $\mathcal{Q}^{(\ell+2)}_{ab} n^b = 0$.*

*If, moreover, we suppose $\ddot{\mathcal{Q}}^{(\ell+1)} = \dot{\mathcal{L}}^{(\ell+1)} = 0$ and, provided $\ell \neq \mathfrak{n}-2$, that $f^{(\ell+3)} = 0$. Then*

3. *$P^{ab}\mathcal{Q}^{(\ell+2)}_{ab} = 0$, $\dot{\mathcal{Q}}^{(\ell+2)}_a n^a = 0$, and if $\ell = \mathfrak{n}-2$ also $f^{(\mathfrak{n}+1)} = 0$ and $\mathcal{Q}^{(\mathfrak{n}+1)}_{ab} n^a n^b = 0$.*

*Proof.* In order to prove items 1., 2. and 3. we make use of the identities in Lemmas 6.19 and 6.20 particularized to $\mathbf{U} = 0$ (because $\mathcal{Q}_{ab} = 0$, see (E.35)). First we prove that the equations $\dot{\mathcal{Q}}^{(\ell+1)}_a n^a = 0$, $\mathcal{Q}^{(\ell+1)}_{ab} n^b = 0$ and $P^{ab}\mathcal{Q}^{(\ell+1)}_{ab} = 0$ hold under the assumptions $\mathcal{L}^{(k)}_\mu = 0$, $\mathcal{Q}^{(k)}_{\mu\nu} = 0$ for $k \leq \ell$, and $f^{(k)} = 0$ for $k \leq \ell+2$. Under the present hypothesis all the conditions of item 3. in Lemma 6.19 with $m = \ell$ are verified and equation (6.82) simplifies to $\mathcal{Q}^{(\ell+1)}_{ab} n^b = 0$. Similarly, $P^{ab}\mathcal{Q}^{(\ell+1)}_{ab} = 0$ is a consequence of (6.87) also for $m = \ell$. The remaining two claims in this item are now immediate by Corollary 6.21 for $m = \ell+1$ (note that when $\ell+1 = \mathfrak{n}$, $\mathcal{L}^{(\ell+1)}_a n^a = \mathcal{L}^{(\mathfrak{n})}_a n^a = 0$ holds by hypothesis).

Now we prove item 2. We first note that the result $\mathcal{L}^{(\ell+1)}_a n^a = 0$ combined with the hypothesis $\widehat{\mathcal{L}}^{(\ell+1)}_a = 0$ gives $\mathcal{L}^{(\ell+1)}_a = 0$, and the results in item 1. combined with the hypothesis $\widehat{\mathcal{Q}}^{(\ell+1)}_{ab} = 0$ gives $\mathcal{Q}^{(\ell+1)}_{ab} = 0$. Since by item 1. we also have $\dot{\mathcal{Q}}^{(\ell+1)}_a n^a = 0$, identities (6.80) and (6.85) for $m = \ell+1$ simplify to

$$
\begin{aligned}
\sigma^{(1)}\mathcal{Q}^{(\ell+2)}_{ab} n^b + (\ell+1)(2\sigma^{(1)}\kappa + \pounds_n \sigma^{(1)})\dot{\mathcal{Q}}^{(\ell+1)}_a &= 0, \\
\mathcal{Q}^{(\ell+2)}_{ab} n^b + \pounds_n \dot{\mathcal{Q}}^{(\ell+1)}_a + 2\kappa \dot{\mathcal{Q}}^{(\ell+1)}_a &= 0.
\end{aligned}
$$

Solving for $\mathcal{Q}^{(\ell+2)}_{ab} n^b$ in the second and inserting it into the first gives

$$
\sigma^{(1)}\pounds_n \dot{\mathcal{Q}}^{(\ell+1)}_a - \Big(2\ell\sigma^{(1)}\kappa + (\ell+1)\pounds_n\sigma^{(1)}\Big)\dot{\mathcal{Q}}^{(\ell+1)}_a = 0.
$$

This is a homogeneous transport equation for $\dot{\mathcal{Q}}^{(\ell+1)}_a$, and since $\dot{\mathcal{Q}}^{(\ell+1)}_a \stackrel{\Sigma}{=} 0$ (because $\dot{\widehat{\mathcal{Q}}}^{(\ell+1)}_a \stackrel{\Sigma}{=} 0$ and $\dot{\mathcal{Q}}^{(\ell+1)}_a n^a = 0$ everywhere, and in particular at $\Sigma$), one concludes that $\dot{\mathcal{Q}}^{(\ell+1)}_a = 0$, and consequently $\mathcal{Q}^{(\ell+2)}_{ab} n^b = 0$.

It only remains to prove item 3. which has the additional hypotheses $\ddot{\mathcal{Q}}^{(\ell+1)} = \dot{\mathcal{L}}^{(\ell+1)} = 0$ (which imply $\mathcal{Q}^{(\ell+1)}_{\alpha\beta} = 0$ and $\mathcal{L}^{(\ell+1)}_\mu = 0$) and, provided $\ell \neq \mathfrak{n}-2$, that $f^{(\ell+3)} = 0$. Then, by (6.81) for $m = \ell+1$ one has $\dot{\mathcal{Q}}^{(\ell+2)}_a n^a = 0$, and since $\mathcal{Q}^{(\ell+1)}_{\alpha\beta} = 0$, from item 4. in Lemma 6.20 for $m = \ell+1$ we get $\operatorname{tr}_P \mathcal{Q}^{(\ell+2)} = 0$, which proves the item for $\ell \neq \mathfrak{n}-2$. For the case $\ell = \mathfrak{n}-2$ we cannot use (6.81) to prove $\dot{\mathcal{Q}}^{(\mathfrak{n})}_a n^a = 0$ because we do not know yet that $f^{(\mathfrak{n}+1)} = 0$. Instead, the strategy is to consider the identities (6.81) and (6.87) for $m = \mathfrak{n}-1$ and the identities (6.79) and (6.84) for $m = \mathfrak{n}$ (recall that $\mathcal{Q}^{(\mathfrak{n}-1)}_{\alpha\beta} = 0$, $\mathcal{L}^{(\mathfrak{n}-1)}_{\mu} = 0$, $\mathcal{L}^{(\mathfrak{n})}_{a} n^a = 0$ and, by item 2., $\mathcal{Q}^{(\mathfrak{n})}_{ab} n^a = 0$),

$$\begin{aligned} \sigma^{(1)} \dot{\mathcal{Q}}^{(\mathfrak{n})}_{a} n^a - \frac{\mathfrak{n}-1}{2} f^{(\mathfrak{n}+1)} &= 0, \\ \dot{\mathcal{Q}}^{(\mathfrak{n})}_{a} n^a + \operatorname{tr}_P \mathcal{Q}^{(\mathfrak{n})} &= 0, \\ \sigma^{(1)} \mathcal{Q}^{(\mathfrak{n}+1)}_{ab} n^a n^b + \mathfrak{n}(2\sigma^{(1)}\kappa + \pounds_n \sigma^{(1)}) \dot{\mathcal{Q}}^{(\mathfrak{n})}_{a} n^a - \frac{\mathfrak{n}-1}{2} \pounds_n f^{(\mathfrak{n}+1)} &= 0, \\ \mathcal{Q}^{(\mathfrak{n}+1)}_{ab} n^a n^b - \pounds_n(\operatorname{tr}_P \mathcal{Q}^{(\mathfrak{n})}) - \pounds_n(\dot{\mathcal{Q}}^{(\mathfrak{n})}_{a} n^a) + 2\kappa \dot{\mathcal{Q}}^{(\mathfrak{n})}_{a} n^a &= 0. \end{aligned}$$

Combining the second and the fourth yields $\mathcal{Q}^{(\mathfrak{n}+1)}_{ab} n^a n^b + 2\kappa \dot{\mathcal{Q}}^{(\mathfrak{n})}_{a} n^a = 0$, which after inserted in the third and taking into account the first gives

$$\sigma^{(1)} \pounds_n f^{(\mathfrak{n}+1)} - \Big(2(\mathfrak{n}-1)\sigma^{(1)}\kappa + \mathfrak{n} \pounds_n \sigma^{(1)}\Big) f^{(\mathfrak{n}+1)} = 0.$$

Since $f^{(\mathfrak{n}+1)}|_{\Sigma} = 0$ we conclude $f^{(\mathfrak{n}+1)} = 0$, and as a consequence, $\dot{\mathcal{Q}}^{(\mathfrak{n})}_{a} n^a = 0$, $\operatorname{tr}_P \mathcal{Q}^{(\mathfrak{n})} = 0$ and $\mathcal{Q}^{(\mathfrak{n}+1)}_{ab} n^a n^b = 0$ also hold. □

This proposition shows that it suffices to analyse equations $\widehat{\mathcal{L}}^{(k)}_{a} = 0$, $\ddot{\mathcal{Q}}^{(k)} = 0$, $\dot{\mathcal{L}}^{(k)} = 0$, $\dot{\widehat{\mathcal{Q}}}^{(k)}_{a}|_{\Sigma} = 0$ and $\widehat{\mathcal{Q}}^{(k)}_{ab} = 0$ at all orders, as well as $f^{(k)} = 0$ (except for $k = \mathfrak{n}+1$), $f^{(\mathfrak{n}+1)}|_{\Sigma} = 0$ and $\mathcal{L}^{(\mathfrak{n})}_{a} n^a = 0$, since the rest of the conformal equations follow from them. Note that the number of equations agrees with the degrees of freedom to be fixed at each order (i.e. a symmetric tensor $\mathbf{Y}^{(k)}$ and a scalar function $\sigma^{(k)}$). Indeed, the reduced set of equations $\widehat{\mathcal{L}}^{(k)}_{a} = 0$, $\ddot{\mathcal{Q}}^{(k)} = 0$, $\dot{\mathcal{L}}^{(k)} = 0$, $f^{(k)} = 0$ and $\widehat{\mathcal{Q}}^{(k)}_{ab} = 0$ constitute exactly $(\mathfrak{n}-1) + 1 + 1 + 1 + \frac{(\mathfrak{n}+1)(\mathfrak{n}-2)}{2} = \frac{\mathfrak{n}(\mathfrak{n}+1)}{2} + 1$ equations, the same as the number of degrees of freedom of $\mathbf{Y}^{(k)}$ and $\sigma^{(k)}$. An immediate corollary of this proposition that will be useful in Theorem 6.30 is the following.

**Corollary 6.23.** *Let $(\mathcal{M}, g, \Omega)$ be an $(\mathfrak{n}+1)$-dimensional conformal manifold with null infinity $\mathscr{I}$ and assume $\mathscr{I}$ admits a cross-section $\Sigma$. Fix an integer $\ell \geq 1$ and assume that equations $\mathcal{L}^{(k)}_{\mu} = 0$, $\mathcal{Q}^{(k)}_{\mu\nu} = 0$ hold for every $k \leq \ell$, $f^{(k)} = 0$ for every $k \leq \ell + 2$, and also $\mathcal{L}^{(\mathfrak{n})}_{a} n^a = 0$ when $\ell = \mathfrak{n}-1$ and $f^{(\mathfrak{n}+1)}|_{\Sigma} = 0$ when $\ell = \mathfrak{n}-2$. Then,*

1. $\mathcal{L}^{(\ell+1)}_{a} n^a = 0$.

*Assume in addition that $\mathcal{L}^{(\ell+1)}_{a} = 0$, $\dot{\widehat{\mathcal{Q}}}^{(\ell+1)}_{a}|_{\Sigma} = 0$ and $\mathcal{Q}^{(\ell+1)}_{ab} = 0$. Then,*

2. $\mathcal{Q}^{(\ell+2)}_{ab} n^b = 0$ *and* $\dot{\mathcal{Q}}^{(\ell+1)}_{a} = 0$.

*If, moreover, we suppose $\ddot{\mathcal{Q}}^{(\ell+1)} = \dot{\mathcal{L}}^{(\ell+1)} = 0$ and, provided $\ell \neq \mathfrak{n}-2$, that $f^{(\ell+3)} = 0$. Then*

3. $P^{ab} \mathcal{Q}^{(\ell+2)}_{ab} = 0$.

Actually, one could have proven Corollary 6.23 directly with much less effort. However, presented this way it would have not allowed us to count the number of degrees of freedom to be fixed, and ultimately know which equations need to be imposed and which ones are automatically fulfilled form the others.

In the following remark we motivate how the reduced set of equations mentioned below Proposition 6.22 constrain the asymptotic expansion at null infinity. Later, in Section 6.5 we will use this motivated free data to characterize the full asymptotic expansion at null infinity.

**Remark 6.24.** *Consider a conformal manifold $(\mathcal{M}, g, \Omega)$ with null infinity $\mathscr{I}$ (assumed to admit a cross-section $\iota : \Sigma \hookrightarrow \mathscr{I}$) and denote as usual the embedded metric hypersurface data by $\{\boldsymbol{\gamma}, \boldsymbol{\ell}, \ell^{(2)}\}$, the asymptotic expansion by $\{\mathbf{Y}^{(k)}\}$, and the transverse derivatives of $\Omega$ at $\mathscr{I}$ by $\{\sigma^{(k)}\}$. The idea is to see how the reduced set of conformal equations described above Corollary 6.23 constrain $\{\boldsymbol{\gamma}, \boldsymbol{\ell}, \ell^{(2)}\}$, $\{\mathbf{Y}^{(k)}\}_{k\geq 1}$ and $\{\sigma^{(k)}\}_{k\geq 2}$ order by order (the quantity $\sigma^{(1)}$ is part of the data, see Definitions 6.7 and 6.8). Of course, the first restriction is $\mathbf{U} = \frac{1}{2}\pounds_n \boldsymbol{\gamma} = 0$. Equations $\mathcal{Q}^{(1)}_{\alpha\beta} = 0$ (see (6.69)) impose the following two extra restrictions on the data: (i) $\mathbf{r} = \mathbf{s} + d(\log|\sigma^{(1)}|)$ and (ii) $\sigma^{(2)} = 0$, so the full one-form $\mathbf{r}$ is determined in terms of metric data and $\sigma$. In particular, this means that the surface gravity of $\nabla\Omega$ vanishes, $\kappa_\star := \sigma^{(1)}\kappa + \pounds_n \sigma^{(1)} = 0$, so $f^{(2)} = 0$ and $\mathcal{L}^{(1)}_a = 0$ follow automatically (by (6.75) and (6.74)). Moreover, equations $\dot{\mathcal{L}}^{(1)} = 0$ (6.73) and $f^{(3)} = 0$ (6.76) constrain the pair $\{\kappa^{(2)}, \pounds_n(\operatorname{tr}_P \mathbf{Y})\}$ such that, given a function $\chi$ at $\Sigma$, there is a unique pair $\{\kappa^{(2)}, \operatorname{tr}_P \mathbf{Y}\}$ satisfying $\dot{\mathcal{L}}^{(1)} = f^{(3)} = 0$ and $\operatorname{tr}_P \mathbf{Y}|_\Sigma = \chi$. For $\mathfrak{n} \neq 3$, the remaining part of the tensor $\mathbf{Y}$ that is not yet fixed, namely $\widehat{\mathbf{Y}}$, becomes completely constrained by equation $\widehat{\mathcal{Q}}^{(2)}_{ab} = 0$ (which is equivalent to (6.70), because by item 1. in Prop. 6.22 for $\ell = 1$, $\mathcal{Q}^{(2)}_{ab} n^b = 0$ and $P^{ab}\mathcal{Q}^{(2)}_{ab} = 0$) in terms of $\widehat{\mathbf{Y}}|_\Sigma$. Showing this statement rigorously requires analyzing the compatibility between the equations, something we will accomplish in Theorem 6.30.*

*When $\mathfrak{n} = 3$, the equation $\widehat{\mathcal{Q}}^{(2)}_{ab} = 0$ does not constrain $\widehat{\mathbf{Y}}$ because the coefficients in front of $\pounds_n \mathbf{Y}$ and $\mathbf{Y}$ in (6.70) both vanish. In principle, this could mean that $\widehat{\mathcal{Q}}^{(2)}_{ab} = 0$ is a new restriction on the data $\{\boldsymbol{\gamma}, \boldsymbol{\ell}, \ell^{(2)}, \chi\}$. However, the tensor $\mathcal{Q}^{(2)}_{ab}$ in dimension $\mathfrak{n} = 3$ identically vanishes, so it imposes no new restrictions on the data. To see this it suffices to prove it in a gauge in which $\sigma^{(1)} = 1$ and $\ell^{(2)} = 0$ (by the transformations laws (6.28) and (2.29) it is easy to see that this gauge can always be achieved), because $\mathcal{Q}^{(2)}_{ab}$ being (or not) zero is a gauge invariant property provided $\mathcal{Q}^{(1)}_{\alpha\beta} = 0$ (this is a particular case of Lemma 6.32 below). So, in this gauge we also have $\mathbf{r} = \mathbf{s}$ and $\kappa = 0$. Under these assumptions, the values of $\kappa^{(2)}$ and $\pounds_n(\operatorname{tr}_P \mathbf{Y})$ that are obtained from $\dot{\mathcal{L}}^{(1)} = f^{(3)} = 0$ are $\kappa^{(2)} = -4P(\mathbf{s}, \mathbf{s})$ and $\pounds_n(\operatorname{tr}_P \mathbf{Y}) = -\frac{1}{2}\Big(\operatorname{tr}_P \mathring{R} - 5\operatorname{div}\mathbf{s} + 9P(\mathbf{s}, \mathbf{s})\Big)$, and therefore the scalar curvature at $\mathscr{I}$ is given by $R = 3(\operatorname{tr}_P \mathring{R} + P(\mathbf{s}, \mathbf{s}) - \operatorname{div}\mathbf{s})$ (see (6.54)). Inserting this into (6.70) for $\mathfrak{n} = 3$ gives*

$$\mathcal{Q}^{(2)}_{ab} = \mathring{R}_{(ab)} + \mathrm{s}_a \mathrm{s}_b - \mathring{\nabla}_{(a}\mathrm{s}_{b)} - \frac{1}{2}\left(\operatorname{tr}_P \mathring{R} + P(\mathbf{s}, \mathbf{s}) - \operatorname{div}\mathbf{s}\right)\gamma_{ab}.$$

*This tensor is manifestly traceless and its contraction with $n^a$ vanishes because $\gamma_{ab} n^a = 0$ and $\left(\mathring{R}_{(ab)} + \mathrm{s}_a \mathrm{s}_b - \mathring{\nabla}_{(a}\mathrm{s}_{b)}\right) n^a = 0$ (see (2.96) and recall $\mathbf{s}(n) = 0$). Moreover, its pullback to a cross section is [217] $\mathcal{Q}^{(2)}_{AB} = R^h_{AB} - \frac{R^h}{2} h_{AB}$, which is identically zero in dimension two. So we conclude that the equation $\mathcal{Q}^{(2)}_{ab} = 0$ is fulfilled when $\mathscr{I}$ is three-dimensional.*

*For the higher order derivatives the idea is similar but with some important differences. Let us assume that we already know how* $\{\mathbf{Y}^{(k)}, \sigma^{(k+1)}, \kappa^{(k+1)}\}$ *are fixed for every* $k = 1, ..., m-1$. *Then, equations* $\mathcal{L}_a^{(m)} = 0$ (6.66) *and* $\dot{\mathcal{Q}}_a^{(m)}|_\Sigma = 0$ (6.64) *constrain the remaining part of the one-form* $\mathbf{r}^{(m)}$ *provided* $m \neq \mathfrak{n}$, *while equations* $\ddot{\mathcal{Q}}^{(m)} = 0$, $\dot{\mathcal{L}}^{(m)} = 0$ *and* $f^{(m+2)} = 0$ *(cf.* (6.65), (6.67) *and* (6.68)*) read*

$$
\begin{aligned}
(\mathfrak{n}-1)\sigma^{(m+1)} - (m-1)\sigma^{(1)} \operatorname{tr}_P \mathbf{Y}^{(m)} &= L^m_{\ddot{\mathcal{Q}}}, \\
(\mathfrak{n}-1)\sigma^{(1)}\kappa^{(m+1)} + (\mathfrak{n}-2)\sigma^{(1)}\pounds_n(\operatorname{tr}_P \mathbf{Y}^{(m)}) + F_m \operatorname{tr}_P \mathbf{Y}^{(m)} &= L^m_{\dot{\mathcal{L}}}, \\
\sigma^{(1)}\pounds_n(\sigma^{(m+1)}) + (\sigma^{(1)})^2\kappa^{(m+1)} + (m+1)(\pounds_n\sigma^{(1)} + 2\sigma^{(1)}\kappa)\sigma^{(m+1)} &= L^{m+2}_f,
\end{aligned}
\tag{6.91}
$$

*where* $L^m_{\ddot{\mathcal{Q}}}$, $L^m_{\dot{\mathcal{L}}}$ *and* $L^{m+2}_f$ *gather the lower order terms that we already know how are constrained, and* $F_m := (2m(\mathfrak{n}-1) - \mathfrak{n})\sigma^{(1)}\kappa + (m-1)\mathfrak{n}\pounds_n\sigma^{(1)}$. *Given* $\sigma^{(m+1)}|_\Sigma$ *(this is a completely free function associated to the remaining conformal freedom present in Lemma 6.10), this system admits a unique solution for* $\{\sigma^{(m+1)}, \kappa^{(m+1)}, \operatorname{tr}_P \mathbf{Y}^{(m)}\}$ *provided* $m \neq \mathfrak{n}-1$. *Indeed, taking the Lie derivative of the first one along* $n$ *and combining it with the three equations in* (6.91) *gives (the precise combination can be read off from the RHS)*

$$
\begin{aligned}
&(\mathfrak{n}-m-1)(\sigma^{(1)})^2\pounds_n(\operatorname{tr}_P \mathbf{Y}^{(m)}) + \sigma^{(1)}G_m \operatorname{tr}_P \mathbf{Y}^{(m)} \\
&= \sigma^{(1)}L^m_{\dot{\mathcal{L}}} - (\mathfrak{n}-1)L^{m+2}_f + \sigma^{(1)}\pounds_n(L^m_{\ddot{\mathcal{Q}}}) + (m+1)(\pounds_n\sigma^{(1)} + 2\sigma^{(1)}\kappa)L^m_{\ddot{\mathcal{Q}}},
\end{aligned}
\tag{6.92}
$$

*where* $G_m := m(\mathfrak{n}-m-1)\sigma\kappa + (m-1)(\mathfrak{n}-m-2)\kappa = m(\mathfrak{n}-m-1)\sigma\kappa$ *(recall that the surface gravity* $\kappa_\star = \sigma\kappa + \pounds_n\sigma$ *has already been shown to be zero). This admits a unique solution for* $\operatorname{tr}_P \mathbf{Y}^{(m)}$ *with "initial" data determined by the first equation in* (6.91), *namely* $\operatorname{tr}_P \mathbf{Y}^{(m)}|_\Sigma = \frac{1}{(m-1)\sigma^{(1)}}\Big((\mathfrak{n}-1)\sigma^{(m+1)}|_\Sigma - L^m_{\ddot{\mathcal{Q}}}|_\Sigma\Big)$ *except when* $m = \mathfrak{n}-1$, *because then the coefficient in front of* $\pounds_n(\operatorname{tr}_P \mathbf{Y}^{(m)})$ *vanishes and* $G_{\mathfrak{n}-1} = 0$. *Finally, equation* $\widehat{\mathcal{Q}}^{(m+1)}_{ab} = 0$, *which is equivalent to* (6.63) *because* $\mathcal{Q}^{(m+1)}_{ab}n^b = 0$ *and* $P^{ab}\mathcal{Q}^{(m+1)}_{ab} = 0$ *by item 1. in Prop. 6.22 for* $\ell = m$, *constrains the remaining part of* $\mathbf{Y}^{(m)}$, *i.e.* $\widehat{\mathbf{Y}}^{(m)}$, *in terms of free data* $\widehat{\mathbf{Y}}^{(m)}|_\Sigma$ *provided* $2m \neq \mathfrak{n}-1$. *As before, this statement requires checking the compatibility with the rest of the equations, see Theorem 6.30 below. Let us now analyse the "special" cases* $m = \frac{\mathfrak{n}-1}{2}$, $m = \mathfrak{n}-1$ *and* $m = \mathfrak{n}$.

<u>*Case* $m = \frac{\mathfrak{n}-1}{2}$ *(*$\mathfrak{n}$ *even)*</u>

*In this case, the coefficients multiplying* $\pounds_n\mathbf{Y}^{(m)}$ *and* $\mathbf{Y}^{(m)}$ *in* $\mathcal{Q}^{(m+1)}_{ab}$ *(cf.* (6.63)*) vanish (recall that* $\kappa = 0$*), and therefore the equation* $\widehat{\mathcal{Q}}^{(m+1)}_{ab} = 0$ *does not constrain*[2] $\widehat{\mathbf{Y}}^{(m)}$ *in terms of* $\widehat{\mathbf{Y}}^{(m)}|_\Sigma$. *Instead, it becomes a potential constraint on the remaining free data, i.e. the metric hypersurface data* $\{\gamma, \boldsymbol{\ell}, \ell^{(2)}\}$, *the functions* $\operatorname{tr}_P \mathbf{Y}|_\Sigma$, $\{\sigma^{(k)}|_\Sigma\}_{k\leq m}$ *and the tensors* $\{\widehat{\mathbf{Y}}^{(k)}|_\Sigma\}_{k<m}$. *We denote the resulting tensor by* $\mathcal{O}^{\mathscr{I}}_{ab}$ *and refer to it as the "radiative obstruction tensor". It is symmetric (since* $\mathcal{Q}$ *is), and by item 1. in Prop. 6.22 it is also*

2 The absence of $\widehat{\mathbf{Y}}^{(\frac{\mathfrak{n}-1}{2})}$ in $\widehat{\mathcal{Q}}^{(\frac{\mathfrak{n}+1}{2})}_{ab}$ is reminiscent of the situation at a degenerate Killing horizon, where the tensor $\mathbf{Y}^{(m)}$ does not appear in $\mathcal{R}^{(m)}_{ab}$ for any $m \geq 1$ (see the discussion following Theorem 5.80). Further parallels between extreme horizons and null infinity have been recently highlighted in [6, 47].

*transverse, i.e.* $P^{ab}\mathcal{O}^{\mathscr{I}}_{ab} = 0$ *and* $\mathcal{O}^{\mathscr{I}}_{ab} n^a = 0$. *A more detailed analysis of* $\mathcal{O}^{\mathscr{I}}_{ab}$ *is presented in Section 6.6.*

Case $m = \mathfrak{n} - 1$

*As already discussed, when* $m = \mathfrak{n} - 1$ *the system* (6.91) *no longer determines the set of tensors* $\{\sigma^{(\mathfrak{n})}, \kappa^{(\mathfrak{n})}, \operatorname{tr}_P \mathbf{Y}^{(\mathfrak{n}-1)}\}$ *in terms of* $\sigma^{(\mathfrak{n})}|_\Sigma$, *as it does for the remaining values of* $m$. *Instead, these quantities become fully determined in terms of* $\sigma^{(\mathfrak{n})}|_\Sigma$ *and* $\kappa^{(\mathfrak{n})}|_\Sigma$ *through the equations* $\ddot{Q}^{(\mathfrak{n}-1)} = 0$, $\dot{\mathcal{L}}^{(\mathfrak{n}-1)} = 0$, $\mathcal{L}^{(\mathfrak{n})}_a n^a = 0$ *and* $f^{(\mathfrak{n}+1)}|_\Sigma = 0$ *(that is, we replace the third equation* $f^{(\mathfrak{n}+1)} = 0$ *in the system* (6.91) *by* $f^{(\mathfrak{n}+1)}|_\Sigma = 0$ *and* $\mathcal{L}^{(\mathfrak{n})}_a n^a = 0$*), as we discuss next. First, the equations* $\ddot{Q}^{(\mathfrak{n}-1)} = 0$ *and* $\dot{\mathcal{L}}^{(\mathfrak{n}-1)} = 0$ *(see* (6.65) *and* (6.67)*) take the form*

$$(\mathfrak{n}-1)\sigma^{(\mathfrak{n})} - (\mathfrak{n}-2)\sigma^{(1)} \operatorname{tr}_P \mathbf{Y}^{(\mathfrak{n}-1)} = L^{\mathfrak{n}-1}_{\ddot{Q}}, \tag{6.93}$$

$$(\mathfrak{n}-1)\sigma^{(1)}\kappa^{(\mathfrak{n})} + (\mathfrak{n}-2)\sigma^{(1)}\pounds_n(\operatorname{tr}_P \mathbf{Y}^{(\mathfrak{n}-1)}) + F \operatorname{tr}_P \mathbf{Y}^{(\mathfrak{n}-1)} = L^{\mathfrak{n}-1}_{\dot{\mathcal{L}}}. \tag{6.94}$$

*Substituting the expression for* $\kappa^{(\mathfrak{n})}$ *obtained from* (6.94) *into the equation* $\mathcal{L}^{(\mathfrak{n})}_a n^a = 0$ *(see* (C.4)*) yields*[3] *a second-order transport equation for* $\operatorname{tr}_P \mathbf{Y}^{(\mathfrak{n}-1)}$ *of the form*

$$\sigma \pounds^{(2)}_n(\operatorname{tr}_P \mathbf{Y}^{(\mathfrak{n}-1)}) + \widetilde{H}_1 \pounds_n(\operatorname{tr}_P \mathbf{Y}^{(\mathfrak{n}-1)}) + \widetilde{H}_2 \operatorname{tr}_P \mathbf{Y}^{(\mathfrak{n}-1)} + \textit{known terms} = 0 \tag{6.95}$$

*for some functions* $\widetilde{H}_i$, $i = 1, 2$. *Thus, given* $\sigma^{(\mathfrak{n})}|_\Sigma$ *and* $\kappa^{(\mathfrak{n})}|_\Sigma$, *one can then uniquely determine* $\operatorname{tr}_P \mathbf{Y}^{(\mathfrak{n}-1)}|_\Sigma$ *and* $\pounds_n\sigma^{(\mathfrak{n})}|_\Sigma$ *from* $\ddot{Q}^{(\mathfrak{n}-1)}|_\Sigma = 0$ *and* $f^{(\mathfrak{n}+1)}|_\Sigma = 0$ *(see the third equation in* (6.92) *which recall is just a rewriting of* (6.68)*), namely*

$$(\mathfrak{n}-1)\sigma^{(\mathfrak{n})}|_\Sigma - (\mathfrak{n}-2)\sigma^{(1)} \operatorname{tr}_P \mathbf{Y}^{(\mathfrak{n}-1)}|_\Sigma = L^{\mathfrak{n}-1}_{\ddot{Q}}|_\Sigma,$$

$$\sigma^{(1)}\pounds_n\sigma^{(\mathfrak{n})}|_\Sigma + (\sigma^{(1)})^2\kappa^{(\mathfrak{n})}|_\Sigma + \mathfrak{n}(\pounds_n\sigma^{(1)} + 2\sigma^{(1)}\kappa)\sigma^{(\mathfrak{n})}|_\Sigma = L^{\mathfrak{n}+1}_f|_\Sigma,$$

*and from* $\pounds_n\sigma^{(\mathfrak{n})}|_\Sigma$ *one gets* $\pounds_n(\mathbf{Y}^{(\mathfrak{n}-1)})|_\Sigma$ *using* (6.93). *Integrating* (6.95) *with this initial data* $\operatorname{tr}_P \mathbf{Y}^{(\mathfrak{n}-1)}|_\Sigma$ *and* $\pounds_n(\mathbf{Y}^{(\mathfrak{n}-1)})|_\Sigma$ *gives* $\operatorname{tr}_P \mathbf{Y}^{(\mathfrak{n}-1)}$ *everywhere. Once* $\operatorname{tr}_P \mathbf{Y}^{(\mathfrak{n}-1)}$ *is known, equations* (6.93) *and* (6.94) *determine* $\sigma^{(\mathfrak{n})}$ *and* $\kappa^{(\mathfrak{n})}$ *uniquely. Moreover, since the initial condition* $\pounds_n(\mathbf{Y}^{(\mathfrak{n}-1)})|_\Sigma$ *has been fitted so that* $f^{(\mathfrak{n}+1)}|_\Sigma = 0$, *Proposition 6.22 ensures* $f^{(\mathfrak{n}+1)} = 0$ *everywhere. In summary, we have shown that the equations* $\ddot{Q}^{(\mathfrak{n}-1)} = 0$, $\dot{\mathcal{L}}^{(\mathfrak{n}-1)} = 0$, $\mathcal{L}^{(\mathfrak{n})}_a n^a = 0$ *and* $f^{(\mathfrak{n}+1)} = 0$ *form a compatible system which determines uniquely* $\{\sigma^{(\mathfrak{n})}, \kappa^{(\mathfrak{n})}, \operatorname{tr}_P \mathbf{Y}^{(\mathfrak{n}-1)}\}$ *in terms of* $\{\sigma^{(\mathfrak{n})}|_\Sigma, \kappa^{(\mathfrak{n})}|_\Sigma\}$. *In particular the system* (6.91) *for* $m = \mathfrak{n} - 2$ *is compatible and hence equation* (6.92) *is satisfied. Since the LHS of this equations is identically zero when* $m = \mathfrak{n} - 1$, *it follows that the RHS is also identically zero, which means that there is no "obstruction" in this case.*

3 Equation (C.4) is written in a particular gauge, and therefore it will change under another choice of rigging $\xi$. Nevertheless, its second-order character is gauge invariant.

Case $m = \mathfrak{n}$

*For this value of $m$, the coefficient multiplying $\mathbf{r}^{(m+1)}$ in (6.64) vanishes, and as a consequence the equation $\dot{\mathcal{Q}}^{(\mathfrak{n})}|_{\Sigma} = 0$ does not impose any new constraint on the one-form $\mathbf{r}^{(\mathfrak{n})}$ on $\Sigma$. Instead, the tensor $\dot{\mathcal{Q}}^{(\mathfrak{n})}|_{\Sigma}$ only depends on $\sigma^{(\mathfrak{n})}$, $\kappa^{(\mathfrak{n})}$ and lower order terms, that have already been determined, so $\dot{\mathcal{Q}}^{(\mathfrak{n})}|_{\Sigma} = 0$ becomes a potential constraint on the free data. We denote the resulting tensor by $\mathcal{O}^{\Sigma}_{a}$ and refer to it as the "Coulombic obstruction". Note that by item 1. of Proposition 6.22 it satisfies $\mathcal{O}^{\Sigma}_{a} n^a = 0$. A more detailed analysis of this obstruction tensor is undertaken in Section 6.6.*

In summary, when $\mathfrak{n}$ is even, the data that can be freely prescribed at $\mathscr{I}$ consist of:

1. Metric hypersurface data $\{\boldsymbol{\gamma}, \boldsymbol{\ell}, \ell^{(2)}\}$ satisfying $\mathbf{U} = 0$,
2. The collection of functions $\{\sigma, \sigma^{(2)} = 0, \sigma^{(3)}, ...\}$ on $\Sigma$ ($\sigma$ is pure metric hypersurface gauge, cf. (6.28)),
3. The scalar $\operatorname{tr}_P \mathbf{Y}$ on $\Sigma$,
4. The family of tensors $\{\widehat{\mathrm{Y}}^{(1)}_{AB}, \widehat{\mathrm{Y}}^{(2)}_{AB}, ...\}$ on $\Sigma$, and
5. The full one-form $\mathbf{r}^{(\mathfrak{n})}$ on $\Sigma$.

When $\mathfrak{n}$ is odd, one must additionally prescribe the tensor $\widehat{\mathbf{Y}}^{(\frac{\mathfrak{n}-1}{2})}$ on $\mathscr{I}$. This extra data must of course be compatible with the $\widehat{\mathbf{Y}}^{(\frac{\mathfrak{n}-1}{2})}|_{\Sigma}$ already prescribed in point 4. One way of ensuring this is to just prescribe $\pounds_n \widehat{\mathbf{Y}}^{(\frac{\mathfrak{n}-1}{2})}$ and solve for $\widehat{\mathbf{Y}}^{(\frac{\mathfrak{n}-1}{2})}$ with the already given data at $\Sigma$.

The previous analysis shows that the number of degrees of freedom in the geometric and detached approach that we present agrees with the analysis in Bondi coordinates in dimension four [87, 289] and also in higher even dimensions [54, 280]. Furthermore, the fact that the tensor $\widehat{\mathbf{Y}}^{(\frac{\mathfrak{n}-1}{2})}$ is only freely specifiable in even dimensions enforces the idea already pointed out in [171] that there are not smooth radiating odd dimensional spacetimes, as we already discussed in Section 6.1.

**Remark 6.25.** *An alternative approach to analyzing how the transverse expansion $\{\mathbf{Y}^{(k)}\}_{k\geq 1}$ is constrained by the conformal equations is to exploit the identities (5.166) with $\eta = \nabla\Omega$. The two approaches are equivalent, since imposing $\mathcal{Q}_{\alpha\beta} = 0$ to infinite order is equivalent to requiring that the deformation tensor of $\nabla\Omega$, $\mathcal{K}[\nabla\Omega]_{\alpha\beta} = 2\nabla_\alpha\nabla_\beta\Omega$, equals $-2\Omega L_{\alpha\beta}$ to all orders.*

## 6.5 UNIQUENESS AND EXISTENCE RESULTS

In this section we prove that the free data described in Remark 6.24 fully characterizes the geometric structure at null infinity (Theorem 6.27), and reciprocally, that given such free data there exists a conformal spacetime realizing it (Theorem 6.30). It is convenient first to set up the following notions of asymptotic flatness of finite and infinite order.

**Definition 6.26.** *Let $(\mathcal{M}, g, \Omega)$ be a conformal manifold with null infinity $\mathscr{I}$. We say $(\mathcal{M}, g, \Omega)$ is k-asymptotically flat provided it satisfies the quasi-Einstein equation* (6.17) *with $\mathcal{T}^{(\ell)}_{\alpha\beta} \stackrel{\mathscr{I}}{=} 0$ for every $\ell \leq k$. When $k = \infty$ we simply say that $(\mathcal{M}, g, \Omega)$ is asymptotically flat.*

This definition is less restrictive than others usually made in the literature, in particular those discussed in Section 6.1, but it is well-adapted to the asymptotic expansion analysis we perform in this chapter. We are ready to establish our uniqueness theorem.

**Theorem 6.27.** *Let $(\mathcal{M}, g, \Omega)$, $(\mathcal{M}', g', \Omega')$ be two asymptotically flat manifolds with respective null infinities $\Phi : \mathscr{I} \hookrightarrow \mathcal{M}$, $\Phi' : \mathscr{I}' \hookrightarrow \mathcal{M}'$. Choose any conformal gauge on $(\mathcal{M}, g, \Omega)$ and $(\mathcal{M}', g', \Omega')$ such that $|\nabla\Omega|^2_g = 0$ and $|\nabla'\Omega'|^2_{g'} = 0$, and let $\xi$, $\xi'$ be respective riggings of $\Phi(\mathscr{I})$, $\Phi'(\mathscr{I}')$ extended geodesically. Suppose that $\mathscr{I}$ admits a cross section $\iota : \Sigma \hookrightarrow \mathscr{I}$ and that there exists a diffeomorphism $\chi : \mathscr{I} \longrightarrow \mathscr{I}'$. Let $\iota' := \chi \circ \iota$ so that $\Sigma' := \chi(\Sigma) \stackrel{\iota'}{\hookrightarrow} \mathscr{I}'$ and let $\Psi$ be the diffeomorphism of Prop. 5.36 constructed between neighbourhoods $\mathcal{U} \subset \mathcal{M}$, $\mathcal{U} \subset \mathcal{M}'$ of $\Phi(\Sigma)$, $\Phi'(\Sigma')$. Define the function $\omega := \Omega^{-1}\Psi^\star\Omega'$, which is smooth and nowhere zero on $\mathcal{U}$, and let $\overline{\mathbf{Y}}^{(k)} := \Phi^\star \pounds^{(k)}_{\xi}(\omega^2 g)$ and $\varpi := \Phi^\star\omega$. Finally, assume*

1. *$\chi^\star\{\boldsymbol{\gamma}', \boldsymbol{\ell}', \ell^{(2)\prime}\} = \{\varpi^2\boldsymbol{\gamma}, \varpi^2\boldsymbol{\ell}, \varpi^2\ell^{(2)}\}$,*
2. *$\chi^\star \operatorname{tr}_{P'} \mathbf{Y}'|_\Sigma = \varpi^{-2} \operatorname{tr}_P \overline{\mathbf{Y}}|_\Sigma$ and $\chi^\star\widehat{\mathbf{Y}}'^{(k)}|_\Sigma = \widehat{\overline{\mathbf{Y}}}^{(k)}|_\Sigma$ for every $k \geq 1$,*
3. *$\chi^\star\mathbf{r}'^{(\mathfrak{n})}|_\Sigma = \overline{\mathbf{r}}^{(\mathfrak{n})}|_\Sigma$,*
4. *and $\chi^\star\widehat{\mathbf{Y}}'^{(\frac{\mathfrak{n}-1}{2})} = \widehat{\overline{\mathbf{Y}}}^{(\frac{\mathfrak{n}-1}{2})}$ for $\mathfrak{n}$ odd.*

*Then, for all $k \geq 0$,*

$$\Psi^\star \pounds^{(k)}_{\xi'} g' \stackrel{\mathscr{I}}{=} \pounds^{(k)}_{\xi}(\omega^2 g).$$

*Proof.* Let $\overline{g} := \omega^2 g$ and $\overline{\Omega} := \omega\Omega$. Since $\Phi_\star\xi = \xi'$ we have $\Psi^\star\sigma^{(k)\prime} = \overline{\sigma}^{(k)}$ for every $k \geq 0$, and given that $\chi^\star\{\boldsymbol{\gamma}', \boldsymbol{\ell}', \ell^{(2)\prime}\} = \{\varpi^2\boldsymbol{\gamma}, \varpi^2\boldsymbol{\ell}, \varpi^2\ell^{(2)}\}$, by Proposition 5.37 it suffices to show that $\chi^\star\mathbf{Y}'^{(k)} = \overline{\mathbf{Y}}^{(k)}$ for all $k \geq 1$. This follows after noting that, except at the exceptional orders, the equations that constrain $\mathbf{Y}'^{(k)}$ and $\overline{\mathbf{Y}}^{(k)}$ at each order (see Remark 6.24) are identical and share the same initial conditions (by item 2.), which necessarily forces $\mathbf{Y}'^{(k)} = \overline{\mathbf{Y}}^{(k)}$ order by order. The first exception is related to the one-form $\mathbf{r}^{(\mathfrak{n})}$, which is constrained by the conformal equations only up to its value at one cross-section, but since by item 3. one has that $\chi^\star\mathbf{r}'^{(\mathfrak{n})} = \overline{\mathbf{r}}^{(\mathfrak{n})}$ at $\Sigma$, then $\chi^\star\mathbf{r}'^{(\mathfrak{n})} = \overline{\mathbf{r}}^{(\mathfrak{n})}$ everywhere. Additionally, when $\mathfrak{n}$ is even there is another exceptional case regarding $\widehat{\mathbf{Y}}^{(\frac{\mathfrak{n}-1}{2})}$ because the conformal equations do not constrain it at all. This is taken care by item 4., which imposes $\chi^\star\widehat{\mathbf{Y}}'^{(\frac{\mathfrak{n}-1}{2})} = \widehat{\overline{\mathbf{Y}}}^{(\frac{\mathfrak{n}-1}{2})}$. Hence the conformal equations along with the agreement on the "free data" in items 1.-4. completely constrain the transverse expansion, and then a direct application of Proposition 5.37 yields the result. □

**Remark 6.28.** *Note that the statement of the theorem already rules out all possible obstructions discussed in Remark 6.24, because $(\mathcal{M}, g, \Omega)$ and $(\mathcal{M}', g', \Omega')$ are assumed to be asymptotically flat (as in Def. 6.26). With comparable effort one can prove that $\Psi$ is an asymptotic isometry up to order $\frac{\mathfrak{n}-3}{2}$ for asymptotically flat metrics with differentiability $C^{\frac{\mathfrak{n}-3}{2}}$ and non-vanishing obstruction tensors; or an isometry to order $\frac{\mathfrak{n}-1}{2}$ for smooth metrics that are $\frac{\mathfrak{n}-1}{2}$-asymptotically flat.*

**Remark 6.29.** *It is worth connecting the uniqueness result in Theorem 6.27 with the results in Section 6.3, where recall we considered conformal manifolds* $(\mathcal{M}, g, \Omega)$ *arising from Fefferman-Graham ambient metrics. Equivalently, in Section 6.3 one deals with* $(\mathcal{M}, g, \Omega)$ *admitting a conformal Killing vector* $\pounds_\eta g = 2\psi g$ *with* $\mathscr{I}$ *as one of the horizons and satisfying* $\pounds_\eta(\Omega) = (\psi - 1)\Omega$. *As we proved there, it is always possible to choose the conformal gauge such that* $\psi = 0$, *and thus* $\eta$ *is Killing. We now connect this with the fact that, under the assumption that* $\eta$ *is a Killing vector and* $\eta \stackrel{\mathscr{I}}{=} \alpha\nu$ *for some function* $\alpha$, *each element of the transverse expansion* $\{\mathbf{Y}^{(k)}\}$ *satisfies a constraint given by* (5.166)

$$\alpha \pounds_n \mathbf{Y}^{(m)} = m \pounds_n \alpha \mathbf{Y}^{(m)} - 2d\alpha \otimes_s \mathbf{r}^{(m)} + \mathcal{P}^{(m)} \qquad \forall m \geq 1, \tag{6.96}$$

*where (cf.* (5.155) *and* (5.165)*)* $\mathcal{P}^{(1)} := \boldsymbol{\ell} \otimes_s d(\pounds_n \alpha) + \frac{1}{2}\pounds_{X_\eta}\boldsymbol{\gamma}$,

$$\mathcal{P}^{(m)} := m \pounds_{X_\eta} \mathbf{Y}^{(m-1)} - \frac{1}{2}\sum_{i=2}^{m} \binom{m}{i} \Phi^\star \left( \pounds_{\pounds_\xi^{(i)}\eta} \pounds_\xi^{(m-i)} g \right) \qquad m \geq 1, \tag{6.97}$$

*and (cf.* (5.154)*)* $X_\eta^a = \frac{1}{2}\alpha n(\ell^{(2)}) n^a + P^{ab}(2\alpha \mathrm{s}_b + \mathring{\nabla}_b \alpha)$. *As proven in Lemma 5.70, when* $\eta$ *is non-degenerate, i.e. its surface gravity is nowhere zero, one can always choose* $\xi|_{\mathscr{I}}$ *and extend it geodesically so that* $\mathcal{P}^{(1)} = \boldsymbol{\ell} \otimes_s d(\pounds_n \alpha)$ *and* $\mathcal{P}^{(m)} = 0$ *for every* $m \geq 2$ *(see also Corollary 5.71). Contracting* (6.96) *for* $m \geq 2$ *with* $n$ *once and twice in this gauge gives*

$$\alpha \pounds_n \mathbf{r}^{(m)} = (m-1)(\pounds_n \alpha)\mathbf{r}^{(m)} + \kappa^{(m)} d\alpha, \qquad \alpha \pounds_n \kappa^{(m)} = (m-2)(\pounds_n \alpha)\kappa^{(m)}.$$

*Taking* $\Sigma$ *to be the bifurcation surface, where* $\alpha = 0$ *and* $\pounds_n \alpha \neq 0$, *one concludes that the free data of item 2. in Theorem 6.27 must be trivial. Note that this statement is gauge invariant because the leading order of* $\mathbf{Y}'^{(k)}$ *under a gauge transformation is* $z^k \mathbf{Y}^{(k)}$. *Moreover, condition* $\eta(\Omega) = -\Omega$ *at all orders implies that the free functions* $\{\sigma^{(k)}\}_{k\geq 1}$ *at* $\Sigma$ *are also zero. In conclusion, the only remaining freedom is a traceless tensor* $\Psi_{AB} = \widehat{\mathrm{Y}}_{AB}^{(\frac{\mathfrak{n}-1}{2})}|_\Sigma$ *for* $\mathfrak{n}$ *odd, from where the full* $\mathbf{Y}^{(\frac{\mathfrak{n}-1}{2})}$ *is constrained by equation* (6.96). *This freedom agrees with the one we obtained by a different argument in Section 6.3, which related this tensor to the free data that appears in the Fefferman-Graham construction of the ambient metric.*

We now prove one of the main results of this chapter, namely that given $\mathscr{I}$-structure data (as in Definition 6.7) together with the free data described in Theorem 6.27, it is possible to construct a smooth asymptotically flat[4] spacetime $(\mathcal{M}, g, \Omega)$ provided the obstruction tensors vanish.

**Theorem 6.30.** *Let* $\{\mathscr{I}, \boldsymbol{\gamma}, \boldsymbol{\ell}, \ell^{(2)}, \sigma, \mathfrak{q} = 0\}$ *be an* $\mathfrak{n}$*-dimensional* $\mathscr{I}$*-structure data admitting a cross-section* $\iota : \Sigma \hookrightarrow \mathscr{I}$ *with induced metric* $h := \iota^\star \boldsymbol{\gamma}$. *Let*

*a.* $\chi$ *be a scalar function on* $\Sigma$,

*b.* $\{\sigma_\Sigma^{(k)}\}_{k\geq 3}$ *a collection of scalar functions on* $\Sigma$,

4 Again, in the sense of Def. 6.26, which as already noted is weaker than other notions in the literature as those reviewed in Section 6.1. Stronger conclusions such as e.g. $\Omega^{-2} g$ being vacuum in a full neighbourhood of $\mathscr{I}$ clearly cannot be proven with data given only at $\mathscr{I}$.

*c.* $\{\mathcal{Y}^{(k)}_{AB}\}_{k\geq 1}$ *a set of symmetric tensors on* $\Sigma$ *traceless w.r.t.* $h$*,*

*d.* $\boldsymbol{\beta}$ *and* $\mathfrak{m}$ *a one-form and a function on* $\Sigma$*, and*

*e. for* $\mathfrak{n}$ *odd, let* $\widehat{\mathbb{Y}}^{(\frac{\mathfrak{n}-1}{2})}_{ab}$ *be a symmetric transverse tensor on* $\mathscr{I}$ *satisfying* $\iota^\star\widehat{\mathbb{Y}}^{(\frac{\mathfrak{n}-1}{2})} = \mathcal{Y}^{(\frac{\mathfrak{n}-1}{2})}$*.*

*Assume the obstruction tensors* $\mathcal{O}^{\Sigma}_a$ *and* $\mathcal{O}^{\mathscr{I}}_{ab}$ *constructed from this data as described in Remark 6.24 both vanish. Then, there exists a smooth conformal manifold* $(\mathcal{M}, g, \Omega)$*, embedding* $\Phi : \mathscr{I} \hookrightarrow \mathcal{M}$ *and geodesic vector* $\xi$ *such that (i)* $(\mathcal{M}, g, \Omega)$ *satisfies* $|\nabla\Omega|^2_g = 0$ *and the quasi-Einstein equations to infinite order at* $\mathscr{I} = \{\Omega = 0\}$*, (ii)* $\{\mathscr{I}, \gamma, \boldsymbol{\ell}, \ell^{(2)}, \sigma, \mathfrak{q}\}$ *is* $(\Phi, \xi)$*-embedded in* $(\mathcal{M}, g)$ *in the sense of Def. 6.8, and (iii)* $\operatorname{tr}_P \mathbf{Y}|_\Sigma = \chi$*,* $\pounds^{(2)}_\xi \Omega|_\Sigma = 0$ *and* $\pounds^{(k)}_\xi \Omega|_\Sigma = \sigma^{(k)}_\Sigma$ *for every* $k \geq 3$*,* $(\iota^\star \mathbf{Y}^{(k)})^{tf} = \mathcal{Y}^{(k)}$ *for every* $k \geq 1$*,* $\iota^\star \mathbf{r}^{(\mathfrak{n})} = \boldsymbol{\beta}$ *and* $\kappa^{(\mathfrak{n})}|_\Sigma = \mathfrak{m}$*. For* $\mathfrak{n}$ *odd one has, in addition to items (i)-(iii) above, that* $\widehat{\mathrm{Y}}^{(\frac{\mathfrak{n}-1}{2})}_{ab} = \widehat{\mathbb{Y}}^{(\frac{\mathfrak{n}-1}{2})}_{ab}$*.*

***Remark:*** *Note that there is not loss of generality in assuming* $\mathfrak{q} = 0$ *from the beginning because by the comment right after Def. 6.8 a conformal gauge in which* $\mathfrak{q} = 0$ *can always be achieved.*

*Proof.* The strategy of the proof is similar to the one we followed in Theorem 5.45 with some important differences. The idea is to construct a spacetime $(\mathcal{M}, g)$ and a function $\Omega$ using Theorem 5.40 and Lemma 5.39 from a collection of "abstract" tensors $\{\mathbb{Y}^{(k)}\}_{k\geq 1}$ and functions $\{\sigma^{(k)}\}_{k\geq 1}$ on $\mathscr{I}$ so that $(\mathcal{M}, g, \Omega)$ satisfies the quasi-Einstein equations to infinite order at $\mathscr{I}$. To do that, we will construct each $\mathbb{Y}^{(k)}$ and $\sigma^{(k)}$ from the equations derived in Section 6.4 obtained after setting $\mathcal{Q}^{(k)}_{\alpha\beta} = 0$, $\mathcal{L}^{(k)}_\mu = 0$ and $f^{(k)} = 0$ for all $k$ and replacing each $\mathbf{Y}^{(k)}$ by $\mathbb{Y}^{(k)}$. To follow a consistent notation we introduce $\mathbb{r}^{(k)} := \mathbb{Y}^{(k)}(n, \cdot)$ and $\mathbb{k}^{(k)} := -\mathbb{Y}^{(k)}(n, n)$ to denote the would-be tensors $\mathbf{r}^{(k)}$ and $\kappa^{(k)}$, respectively.

1. The strategy

To construct the abstract expansion $\{\mathbb{Y}^{(k)}\}$ there arises the following difficulty. Let's say that we want to use the equation $\mathcal{Q}^{(k+1)}_{ab} = 0$ to construct $\mathbb{Y}^{(k)}$ from some initial condition on $\Sigma$. By looking at the right hand side of (6.63) (recall that the tensor $\widetilde{O}^{(k)}_{ab}$ collects additional terms depending on $\mathbb{r}^{(k)}$ and $\sigma^{(k)}$) one immediately realizes that this is not a transport equation, but a partial differential system due to the presence of $\mathring{\nabla}$-derivatives of $\mathbb{r}^{(k)}$ in $\widetilde{O}^{(k)}_{ab}$ and of $\operatorname{tr}_P \mathbb{Y}^{(k)}$. This is a complicated system and to the best of our knowledge there is no existence theorem available. In addition, there appears the scalar $\kappa^{(k+1)}$, which is one order higher. Thus, our strategy is to first construct a suitable one-form $\boldsymbol{c}^{(k)}$ and three scalar functions $t^{(k)}$, $\mathfrak{c}^{(k)}$ and $\mathfrak{c}^{(k+1)}$, and then rewrite the equation $\mathcal{Q}^{(k+1)}_{ab} = 0$ but replacing $\mathbb{k}^{(k)}$ by $\mathfrak{c}^{(k)}$, $\mathbb{r}^{(k)}$ by $\boldsymbol{c}^{(k)}$, $\operatorname{tr}_P \mathbb{Y}^{(k)}$ by $t^{(k)}$, and $\kappa^{(k+1)}$ by $\mathfrak{c}^{(k+1)}$. The resulting equation is now an actual transport equation for $\mathbb{Y}^{(k)}$, from which we can build $\mathbb{Y}^{(k)}$ given an initial condition at $\Sigma$. Later, we will need to close the argument by showing that $\boldsymbol{c}^{(k)}(n) = -\mathfrak{c}^{(k)}$, $\mathbb{Y}^{(k)}(n, \cdot) := \mathbb{r}^{(k)} = \boldsymbol{c}^{(k)}$, $\operatorname{tr}_P \mathbb{Y}^{(k)} = t^{(k)}$ and $\mathbb{Y}^{(k+1)}(n, n) = -\mathfrak{c}^{(k+1)}$ for every $k \geq 1$.

Now we provide the summary of the argument step by step: Suppose that for a given value[5] of $m \geq 1$ we have already constructed the collection $\{\mathbb{Y}^{(k)}, \sigma^{(k)}\}_{k\leq m}$ and also $\sigma^{(m+1)}$ and $\mathfrak{c}^{(m+1)}$ such that the equations $\mathcal{Q}^{(k)}_{\alpha\beta} = 0$, $\mathcal{L}^{(k)}_{\mu} = 0$ and $f^{(k+2)} = 0$ for all $k \leq m$, and $\mathcal{Q}^{(m+1)}_{ab} = 0$ are all satisfied with the replacements $\mathbf{Y}^{(k)} \to \mathbb{Y}^{(k)}$ for every $k \leq m$, and $\kappa^{(m+1)} \to \mathfrak{c}^{(m+1)}$. Then we construct $\mathbb{Y}^{(m+1)}$, $\sigma^{(m+2)}$ and $\mathfrak{c}^{(m+2)}$ by following these steps:

1. We define $\boldsymbol{c}^{(m+1)}|_{\Sigma}$ as the unique one-form that satisfies (i) $\boldsymbol{c}^{(m+1)}(n)|_{\Sigma} = -\mathfrak{c}^{(m+1)}|_{\Sigma}$ and (ii) the equation $\iota^{\star}\dot{\mathcal{Q}}^{(m+1)}_{a} = 0$ (see (6.64)) with the replacements $\mathbf{Y}^{(k)} \to \mathbb{Y}^{(k)}$ for every $k \leq m$, $\mathbf{r}^{(m+1)} \to \boldsymbol{c}^{(m+1)}$ and $\kappa^{(m+1)} \to \mathfrak{c}^{(m+1)}$.

2. Given $\boldsymbol{c}^{(m+1)}|_{\Sigma}$, we integrate the one-form $\boldsymbol{c}^{(m+1)}$ by solving the equation $\mathcal{L}^{(m+1)}_{a} = 0$ (6.66) with the replacements $\mathbf{Y}^{(k)} \to \mathbb{Y}^{(k)}$ for every $k \leq m$, $\mathbf{r}^{(m+1)} \to \boldsymbol{c}^{(m+1)}$ and $\kappa^{(m+1)} \to \mathfrak{c}^{(m+1)}$. We emphasize that our working hypothesis is the function $\mathfrak{c}^{(m+1)}$ has already been fixed at this stage.

3. Having determined $\boldsymbol{c}^{(m+1)}$, we build three functions $\sigma^{(m+2)}$, $t^{(m+1)}$ and $\mathfrak{c}^{(m+2)}$ by solving the system of equations $f^{(m+3)} = \dot{\mathcal{L}}^{(m+1)} = \ddot{\mathcal{Q}}^{(m+1)} = 0$ with the replacements $\mathbf{Y}^{(k)} \to \mathbb{Y}^{(k)}$ for every $k \leq m$, $\mathbf{r}^{(m+1)} \to \boldsymbol{c}^{(m+1)}$, $\kappa^{(m+1)} \to \mathfrak{c}^{(m+1)}$, $\operatorname{tr}_P \mathbf{Y}^{(m+1)} \to t^{(m+1)}$ and $\kappa^{(m+2)} \to \mathfrak{c}^{(m+2)}$, from an initial condition $\sigma^{(m+2)}|_{\Sigma} = \sigma^{(m+2)}_{\Sigma}$ (see (6.91) and the subsequent discussion on existence of solutions). The free function $\sigma^{(m+2)}_{\Sigma}$ simply encodes the residual conformal freedom, as described in Lemma 6.10.

4. Finally, with the tensors $\boldsymbol{c}^{(m+1)}$, $\sigma^{(m+2)}$, $t^{(m+1)}$ and $\mathfrak{c}^{(m+2)}$ just determined, we construct $\mathbb{Y}^{(m+1)}$ by integrating the equation $\mathcal{Q}^{(m+2)}_{ab} = 0$ with the replacements $\mathbf{Y}^{(k)} \to \mathbb{Y}^{(k)}$ for every $k \leq m$, $\mathbf{r}^{(m+1)} \to \boldsymbol{c}^{(m+1)}$, $\kappa^{(m+1)} \to \mathfrak{c}^{(m+1)}$, $\operatorname{tr}_P \mathbf{Y}^{(m+1)} \to t^{(m+1)}$, $\kappa^{(m+2)} \to \mathfrak{c}^{(m+2)}$ and $\mathbf{Y}^{(m+1)} \to \mathbb{Y}^{(m+1)}$, with the initial condition $\operatorname{tr}_P \mathbb{Y}^{(m+1)}|_{\Sigma} = t^{(m+1)}|_{\Sigma}$, $\mathbb{Y}^{(m+1)}(n,\cdot)|_{\Sigma} = \boldsymbol{c}^{(m+1)}|_{\Sigma}$ and $(\mathbb{Y}^{(m+1)}_{AB})^{tf}|_{\Sigma} = \mathcal{Y}^{(m+1)}_{AB}$.

To close the argument we need to show that the collection $\{\mathbb{Y}^{(k)}, \sigma^{(k)}\}_{k\leq m+1}$ and also the functions $\sigma^{(m+2)}$ and $\mathfrak{c}^{(m+2)}$ satisfy the equations $\mathcal{Q}^{(k)}_{\alpha\beta} = 0$, $\mathcal{L}^{(k)}_{\mu} = 0$ and $f^{(k+2)} = 0$ for all $k \leq m+1$, and also $\mathcal{Q}^{(m+2)}_{ab} = 0$ (each one with the replacements $\mathbf{Y}^{(k)} \to \mathbb{Y}^{(k)}$ for every $k \leq m+1$, and $\kappa^{(m+2)} \to \mathfrak{c}^{(m+2)}$). This is equivalent to proving three things: (1) That the scalars $\boldsymbol{c}^{(m+1)}(n)$ and $-\mathfrak{c}^{(m+1)}$ are the same; (2) That $\mathbb{r}^{(m+1)} := \mathbb{Y}^{(m+1)}(n,\cdot)$ agrees with $\boldsymbol{c}^{(m+1)}$; And (3) that $\operatorname{tr}_P \mathbb{Y}^{(m+1)}$ is the same as $t^{(m+1)}$. Finally, we will construct the conformal manifold $(\mathcal{M}, g, \Omega)$ from the collection $\{\mathbb{Y}^{(k)}, \sigma^{(k)}\}_{k\geq 1}$ using Theorem 5.40 and Lemma 5.39, and given that these tensors satisfy $\mathcal{Q}^{(k)}_{\alpha\beta} = 0$, $\mathcal{L}^{(k)}_{\mu} = 0$ and $f^{(k)} = 0$ (with the replacements $\mathbf{Y}^{(k)} \to \mathbb{Y}^{(k)}$) for every $k \geq 1$, it is clear that $(\mathcal{M}, g, \Omega)$ will also satisfy $\mathcal{Q}^{(k)}_{\alpha\beta} = 0$, $\mathcal{L}^{(k)}_{\mu} = 0$ and $f^{(k)} = 0$ to all orders because $\mathbf{Y}^{(k)} = \mathbb{Y}^{(k)}$ and $\pounds^{(k)}_{\xi}\Omega|_{\mathscr{I}} = \sigma^{(k)}$ for every $k \geq 1$.

5 Assume for the sake of the argument that $m+1$ is not one of the exceptional values $\mathfrak{n}$, $\mathfrak{n}-1$ or $\frac{\mathfrak{n}-1}{2}$. These will be dealt with separately.

2. The first order

The first order is special because, as already noted several times, the conformal equations have a different structural form at their lowest orders. This fact reflects itself in the way how the one-form $\boldsymbol{c}$ and the scalars $\mathfrak{c}$, $t$ and $\sigma^{(2)}$ will be built. Let us start by defining $\boldsymbol{c} := \mathbf{s} + d\log|\sigma|$ (cf. (6.69)), and the scalars[6] $\mathfrak{c} := -\pounds_n(\log|\sigma|)$, $\sigma^{(2)} = 0$ (see (6.69)) and $\mathfrak{c}^{(2)} := \mathfrak{c}n(\ell^{(2)}) - 4P(\boldsymbol{c},\mathbf{s}) - 2P(\mathring{\nabla}\log|\sigma|, \mathring{\nabla}\log|\sigma|)$ (which is obtained after equating (6.76) to zero with the replacements $\kappa^{(2)} \to \mathfrak{c}^{(2)}$, $\sigma^{(2)} \to 0$, $\mathbf{r} \to \boldsymbol{c}$ and $\kappa \to \mathfrak{c}$). We can now construct the function $t$ by integrating the ODE $\dot{\mathcal{L}}^{(1)} = 0$ (see (6.73)) with the replacements $\mathbf{r}$ by $\boldsymbol{c}$, $\kappa$ by $\mathfrak{c}$, $\kappa^{(2)}$ by $\mathfrak{c}^{(2)}$ and $\operatorname{tr}_P \mathbf{Y}$ by $t$, and with the initial condition $t|_\Sigma = \chi$. Next, if $\mathfrak{n} > 3$ we construct the tensor $\mathbb{Y}_{ab}$ by integrating the transport equation $\mathcal{Q}^{(2)}_{ab} = 0$ (cf. (6.70)) obtained after replacing $\mathbf{r}$ by $\boldsymbol{c}$, $\kappa$ by $\mathfrak{c}$, $\kappa^{(2)}$ by $\mathfrak{c}^{(2)}$ and $\operatorname{tr}_P \mathbf{Y}$ by $t$, with the initial conditions $\mathbb{Y}^{tf}_{AB}|_\Sigma = \mathcal{Y}_{AB}$, $\operatorname{tr}_P \mathbb{Y}|_\Sigma = t|_\Sigma$ and $\mathbb{r}|_\Sigma = \boldsymbol{c}|_\Sigma$, namely

$$
\begin{aligned}
0 = (\mathfrak{n}-3)\sigma\pounds_n\mathbb{Y}_{ab} + (\mathfrak{n}-3)\sigma\mathfrak{c}\mathbb{Y}_{ab} + (\mathfrak{n}-1)\mathring{\nabla}_a\mathring{\nabla}_b\sigma - 2\sigma(\mathfrak{n}-2)\mathring{\nabla}_{(a}c_{b)} \\
+ \sigma\Big(\mathring{\nabla}_{(a}\mathrm{s}_{b)} - 2c_ac_b + 4c_{(a}\mathrm{s}_{b)} - \mathrm{s}_a\mathrm{s}_b + \mathring{R}_{(ab)}\Big) - \frac{\sigma\mathfrak{R}}{2\mathfrak{n}}\gamma_{ab},
\end{aligned}
\tag{6.98}
$$

where $\mathfrak{R}$ is given by (6.54) with the corresponding replacements ($\operatorname{tr}_P \mathbf{Y} \to t$, $\kappa^{(2)} \to \mathfrak{c}^{(2)}$, $\mathbf{r} \to \boldsymbol{c}$, $\kappa \to \mathfrak{c}$). If $\mathfrak{n} = 3$ we simply define the symmetric tensor $\mathbb{Y}_{ab}$ by means of the decomposition (2.24) with the values $\mathbb{Y}_{ab}n^b = c_a$, $\operatorname{tr}_P \mathbb{Y} = t$ and its transverse part given by the data $\widehat{\mathbb{Y}}_{ab}$ prescribed in item (e) of the theorem. Recall that for $\mathfrak{n} = 3$ the equation $\mathcal{Q}^{(2)}_{ab} = 0$ holds automatically.

We now prove that in the case $\mathfrak{n} > 3$, $\mathbb{r} = \boldsymbol{c}$ and $\operatorname{tr}_P \mathbb{Y} = t$ everywhere (for $\mathfrak{n} = 3$ it holds by construction). Firstly, we contract equation (6.98) with $n$ and use identity (2.46) (for $n^{(2)} = 0$) twice (with $\theta_a = c_a$ and $\theta_a = \mathrm{s}_a$) and $\mathring{R}_{(ab)}n^b = \frac{1}{2}\pounds_n\mathrm{s}_a$ (see (2.96) with $\mathbf{U} = 0$) to get

$$
\begin{aligned}
0 = (\mathfrak{n}-3)\sigma\pounds_n\mathbb{r}_a + (\mathfrak{n}-3)\sigma\mathfrak{c}\mathbb{r}_a + (\mathfrak{n}-1)n^b\mathring{\nabla}_a\mathring{\nabla}_b\sigma - \sigma(\mathfrak{n}-2)(\pounds_n c_a - \mathring{\nabla}_a\mathfrak{c} + 2\mathfrak{c}\mathrm{s}_a) \\
+ \sigma\pounds_n\mathrm{s}_a + 2\sigma\mathfrak{c}(c_a - \mathrm{s}_a).
\end{aligned}
\tag{6.99}
$$

Contracting it again with $n$ and recalling the notation $\mathbb{k} := -\mathbb{r}(n) = -\mathbb{Y}(n,n)$,

$$
0 = -(\mathfrak{n}-3)\sigma\pounds_n\mathbb{k} - (\mathfrak{n}-3)\sigma\mathfrak{c}\,\mathbb{k} + (\mathfrak{n}-1)n^an^b\mathring{\nabla}_a\mathring{\nabla}_b\sigma + 2\sigma(\mathfrak{n}-2)\pounds_n\mathfrak{c} - 2\sigma\mathfrak{c}^2. \tag{6.100}
$$

Additionally, the contraction of (6.98) with $P^{ab}$ gives, after using (2.51),

$$
\begin{aligned}
0 = (\mathfrak{n}-3)\sigma\pounds_n(\operatorname{tr}_P \mathbb{Y}) + 4\sigma(\mathfrak{n}-3)P(\mathbb{r},\mathbf{s}) - \sigma(\mathfrak{n}-3)n(\ell^{(2)})\mathbb{k} + (\mathfrak{n}-3)\sigma\mathfrak{c}\operatorname{tr}_P \mathbb{Y} + (\mathfrak{n}-1)\mathring{\square}\sigma \\
- 2\sigma(\mathfrak{n}-2)\operatorname{div}_P \boldsymbol{c} + \sigma\Big(\operatorname{div}_P \mathbf{s} - 2P(\boldsymbol{c},\boldsymbol{c}) + 4P(\boldsymbol{c},\boldsymbol{s}) - P(\boldsymbol{s},\boldsymbol{s}) + \operatorname{tr}_P \mathring{R}\Big) - \frac{\mathfrak{n}-1}{2\mathfrak{n}}\sigma\mathfrak{R}.
\end{aligned}
\tag{6.101}
$$

6 Observe that by construction $\mathfrak{c} = -\boldsymbol{c}(n)$ so we do not need to worry about proving this at this order, in contrast with the rest of orders.

Next, we construct an auxiliary spacetime $(\mathcal{M}_1, g_1)$ using Theorem 5.40 from the sequence $\{\mathrm{T}^{(k)}_{ab}\}_{k\geq 1}$ defined by (cf. (2.24))

$$\mathrm{T}^{(1)}_{ab} := \frac{t - \mathfrak{c}\ell^{(2)}}{\mathfrak{n}-1}\gamma_{ab} + 2\widehat{c}_{(a}\ell_{b)} - \mathfrak{c}\ell_a\ell_b + \widehat{\mathbb{Y}}_{ab}, \qquad \mathrm{T}^{(2)}_{ab} := -\mathfrak{c}^{(2)}\ell_a\ell_b,$$

and $\mathrm{T}^{(k)}_{ab} = 0$ for every $k \geq 3$. It is immediate to check that $P^{ab}\mathrm{T}^{(1)}_{ab} = t$, $\mathrm{T}^{(1)}_{ab}n^a = \widehat{c}_b - \mathfrak{c}\ell_b = c_b$ and $\mathrm{T}^{(2)}_{ab}n^a n^b = -\mathfrak{c}^{(2)}$. We also construct a function $\Omega_1$ on $\mathcal{M}_1$ using Borel's Lemma 5.39 from the sequence $\{0, \sigma, 0, 0, ...\}$. Since $(\mathcal{M}_1, g_1, \Omega_1)$ satisfies $\mathcal{L}^{(1)}_\mu = 0$, $\mathcal{Q}^{(1)}_{\alpha\beta} = 0$ and $f^{(1)} = f^{(2)} = f^{(3)} = 0$ ($\dot{\mathcal{L}} = 0$, $\dot{\mathcal{Q}}_a = 0$, $\ddot{\mathcal{Q}} = 0$ and $f^{(3)} = 0$ hold by construction, and $\mathcal{Q}_{ab} = 0$, $\mathcal{L}_a = 0$ and $f^{(1)} = f^{(2)} = 0$ are automatically true, see (6.69), (6.74), (6.75)), the first item in Proposition 6.22 (with $\ell = 1$) ensures that it also satisfies $\mathcal{Q}^{(2)}_{ab}n^a = 0$ and $P^{ab}\mathcal{Q}^{(2)}_{ab} = 0$, which imply the following:

1. The scalar $\mathfrak{c} = -\mathbf{T}^{(1)}(n, n)$ satisfies the same equation as $\mathbb{k}$ (cf. (6.100)) with the replacement $\mathbb{k} \to \mathfrak{c}$, namely

$$0 = -(\mathfrak{n}-3)\sigma\pounds_n\mathfrak{c} - (\mathfrak{n}-1)\sigma\mathfrak{c}^2 + (\mathfrak{n}-1)n^a n^b\mathring{\nabla}_a\mathring{\nabla}_b\sigma + 2\sigma(\mathfrak{n}-2)\pounds_n\mathfrak{c}.$$

   Subtracting both equations one then arrives at

$$0 = (\mathfrak{n}-3)\sigma\pounds_n(\mathbb{k} - \mathfrak{c}) + (\mathfrak{n}-1)\sigma\mathfrak{c}(\mathbb{k} - \mathfrak{c}).$$

   This is a linear homogeneous ODE for $\mathbb{k} - \mathfrak{c}$, and since $\mathbb{k}$ and $\mathfrak{c}$ agree on $\Sigma$ (because of the initial data we have imposed when solving (6.98)) we conclude $\mathbb{k} = \mathfrak{c}$ everywhere.

2. Once $\mathbb{k} = \mathfrak{c}$ is known to be true, the one-form $\boldsymbol{c} = \mathbf{T}^{(1)}(n, \cdot)$ satisfies the same equation than $\mathbb{r}$ (6.99) with the replacement $\mathbb{r} \to \boldsymbol{c}$, namely

$$\begin{aligned}0 = (\mathfrak{n}-3)\sigma\pounds_n c_a + (\mathfrak{n}-3)\sigma\mathfrak{c}c_a + (\mathfrak{n}-1)n^b\mathring{\nabla}_a\mathring{\nabla}_b\sigma - \sigma(\mathfrak{n}-2)(\pounds_n c_a - \mathring{\nabla}_a\mathfrak{c} + 2\mathfrak{c}\mathrm{s}_a)\\ + \sigma\pounds_n\mathrm{s}_a + 2\sigma\mathfrak{c}(c_a - \mathrm{s}_a),\end{aligned}$$

   so subtracting it from (6.99) and using that $\mathbb{r}|_\Sigma = \boldsymbol{c}|_\Sigma$, it follows that $\boldsymbol{c} = \mathbb{r}$ everywhere.

3. Finally, once we have shown that $\boldsymbol{c} = \mathbb{r}$, the scalar $t = P^{ab}\mathrm{T}^{(1)}_{ab}$ satisfies the same equation than $\operatorname{tr}_P\mathbb{Y}$ (6.101) with the replacement $\operatorname{tr}_P\mathbb{Y} \to t$, namely

$$\begin{aligned}0 = (\mathfrak{n}-3)\sigma\pounds_n t + 4\sigma(\mathfrak{n}-3)P(\mathbb{r}, \mathbf{s}) - \sigma(\mathfrak{n}-3)n(\ell^{(2)})\mathbb{k} + (\mathfrak{n}-3)\sigma\mathfrak{c}t + (\mathfrak{n}-1)\mathring{\Box}\sigma\\ - 2\sigma(\mathfrak{n}-2)\operatorname{div}_P\boldsymbol{c} + \sigma\Big(\operatorname{div}_P\mathbf{s} - 2P(\boldsymbol{c}, \boldsymbol{c}) + 4P(\boldsymbol{c}, \boldsymbol{s}) - P(\boldsymbol{s}, \boldsymbol{s}) + \operatorname{tr}_P\mathring{R}\Big) - \frac{\mathfrak{n}-1}{2\mathfrak{n}}\sigma\mathfrak{R}.\end{aligned}$$

   Subtracting them and using that they coincide at $\Sigma$ we also conclude that $t = \operatorname{tr}_P\mathbb{Y}$ everywhere.

Note that at this point we have already constructed $\sigma^{(2)}$, $\mathfrak{c}^{(2)}$ and the full tensor $\mathbb{Y}_{ab}$. The construction guarantees that the equations $\ddot{Q}^{(1)} = 0$, $f^{(3)} = 0$, $\dot{\mathcal{Q}}^{(1)}_a = 0$, $\dot{\mathcal{L}}^{(1)} = 0$ and $\mathcal{Q}^{(2)}_{ab} = 0$ are fulfilled. Recall also that $\mathcal{Q}^{(1)}_{ab} = 0$ and $\mathcal{L}^{(1)}_a = 0$ hold automatically.

3. Higher order terms

For the higher order terms we apply a similar strategy. Fix $m \geq 1$ (suppose that $m+1$ is not one of the exceptional values, i.e. that $m+1 \neq \frac{\mathfrak{n}-1}{2}$, $m+1 \neq \mathfrak{n}$ and $m+1 \neq \mathfrak{n}-1$) and that we have already constructed $\{\mathbb{Y}^{(k)}, \sigma^{(k)}\}$ for all $k \leq m$ and also $\sigma^{(m+1)}$ and $\mathfrak{c}^{(m+1)}$, and that the equations $\mathcal{Q}^{(k)}_{\alpha\beta} = 0$, $\mathcal{L}^{(k)}_{\mu} = 0$, $f^{(k+2)} = 0$ (with the replacements $\{\mathbf{Y}^{(k)} \to \mathbb{Y}^{(k)}\}_{k\leq m}$ and $\kappa^{(m+1)} \to \mathfrak{c}^{(m+1)}$) hold for every $k \leq m$, and also $\mathcal{Q}^{(m+1)}_{ab} = 0$. Note that this is what we achieved in the first order. We now follow the steps 1.-4. that we explained in the strategy of the proof.

1. We construct uniquely $\boldsymbol{c}^{(m+1)}|_{\Sigma}$ from conditions (i) $\boldsymbol{c}^{(m+1)}(n)|_{\Sigma} = -\mathfrak{c}^{(m+1)}|_{\Sigma}$ and such that (ii) the equation $\iota^{\star}\dot{\mathcal{Q}}^{(m+1)}_{a} = 0$ with the replacements $\mathbf{Y}^{(k)} \to \mathbb{Y}^{(k)}$ for every $k \leq m$ and $\kappa^{(m+1)} \to \mathfrak{c}^{(m+1)}$ is fulfilled.

2. Given $\boldsymbol{c}^{(m+1)}|_{\Sigma}$, we integrate the one-form $\boldsymbol{c}^{(m+1)}$ by solving the equation $\mathcal{L}^{(m+1)}_{a} = 0$ (6.66) with the replacements $\mathbf{Y}^{(k)} \to \mathbb{Y}^{(k)}$ for every $k \leq m$, $\mathbf{r}^{(m+1)} \to \boldsymbol{c}^{(m+1)}$ and $\kappa^{(m+1)} \to \mathfrak{c}^{(m+1)}$, namely
$$-\sigma \pounds_n c^{(m+1)}_a + m\pounds_n \sigma c^{(m+1)}_a - \sigma \overset{\circ}{\nabla}_a \mathfrak{c}^{(m+1)} - \frac{m}{\mathfrak{n}}\mathfrak{c}^{(m+1)} \overset{\circ}{\nabla}_a \sigma = \text{lower order terms.} \tag{6.102}$$
To prove that $\boldsymbol{c}^{(m+1)}(n) = -\mathfrak{c}^{(m+1)}$ later we will need its contraction with $n$, which is
$$-\sigma \pounds_n (c^{(m+1)}_a n^a) + m(\pounds_n \sigma)\, c^{(m+1)}_a n^a - \sigma \pounds_n \mathfrak{c}^{(m+1)} - \frac{m}{\mathfrak{n}}\mathfrak{c}^{(m+1)} \pounds_n \sigma = \text{lower order terms.} \tag{6.103}$$

3. With the $\boldsymbol{c}^{(m+1)}$ constructed in 2., we build the three functions $\sigma^{(m+2)}$, $t^{(m+1)}$ and $\mathfrak{c}^{(m+2)}$ by solving the system of equations $f^{(m+3)} = \dot{\mathcal{L}}^{(m+1)} = \ddot{\mathcal{Q}}^{(m+1)} = 0$ (see (6.91)) with the replacements $\mathbf{Y}^{(k)} \to \mathbb{Y}^{(k)}$ for every $k \leq m$, $\kappa^{(m+1)} \to \mathfrak{c}^{(m+1)}$, $\mathbf{r}^{(m+1)} \to \boldsymbol{c}^{(m+1)}$, $\operatorname{tr}_P \mathbf{Y}^{(m+1)} \to t^{(m+1)}$ and $\kappa^{(m+2)} \to \mathfrak{c}^{(m+2)}$) from the initial condition $\sigma^{(m+2)}|_{\Sigma} = \sigma^{(m+2)}_{\Sigma}$.

4. Finally, we construct $\mathbb{Y}^{(m+1)}_{ab}$ by integrating the equation $\mathcal{Q}^{(m+2)}_{ab} = 0$ with the replacements $\mathbf{Y}^{(k)} \to \mathbb{Y}^{(k)}$ for every $k \leq m$, $\mathbf{r}^{(m+1)} \to \boldsymbol{c}^{(m+1)}$, $\kappa^{(m+1)} \to \mathfrak{c}^{(m+1)}$, $\operatorname{tr}_P \mathbf{Y}^{(m+1)} \to t^{(m+1)}$, $\kappa^{(m+2)} \to \mathfrak{c}^{(m+2)}$ and $\mathbf{Y}^{(m+1)} \to \mathbb{Y}^{(m+1)}$ with the initial conditions $\operatorname{tr}_P \mathbb{Y}^{(m+1)}|_{\Sigma} = t^{(m+1)}|_{\Sigma}$, $\mathbb{Y}^{(m+1)}(n,\cdot)|_{\Sigma} = \boldsymbol{c}^{(m+1)}|_{\Sigma}$ and $(\mathbb{Y}^{(m+1)}_{AB})^{tf}|_{\Sigma} = \mathcal{Y}^{(m+1)}_{AB}$.

Once the tensors $\{\mathbb{Y}^{(k)}, \sigma^{(k)}\}_{k\leq m+1}$ and the functions $\sigma^{(m+2)}$ and $\mathfrak{c}^{(m+2)}$ are built, we need to check that the equations $\mathcal{Q}^{(k)}_{\alpha\beta} = 0$, $\mathcal{L}^{(k)}_{\mu} = 0$ and $f^{(k+2)} = 0$ hold for all $k \leq m+1$, and also $\mathcal{Q}^{(m+2)}_{ab} = 0$ (each one with the replacements $\mathbf{Y}^{(k)} \to \mathbb{Y}^{(k)}$ for every $k \leq m+1$, and $\kappa^{(m+2)} \to \mathfrak{c}^{(m+2)}$). To do that it suffices to prove that (1) $\boldsymbol{c}^{(m+1)}(n) = -\mathfrak{c}^{(m+1)}$, (2) $\mathbb{Y}^{(m+1)}(n,\cdot) = \boldsymbol{c}^{(m+1)}$, and (3) $\operatorname{tr}_P \mathbb{Y}^{(m+1)} = t^{(m+1)}$.

To prove the three claims we shall construct an auxiliary spacetime $(\mathcal{M}_m, g_m)$ using Theorem 5.40 from the sequence $\{\mathbf{T}^{(k)}\}_{k\geq 1}$, with $\mathbf{T}^{(k)} = \mathbb{Y}^{(k)}$ for all $k \leq m$,

$$\mathrm{T}^{(m+1)}_{ab} := \frac{t^{(m+1)} - \mathfrak{c}^{(m+1)}\ell^{(2)}}{\mathfrak{n}-1}\gamma_{ab} + 2\widehat{c}^{(m+1)}_{(a}\ell_{b)} - \mathfrak{c}^{(m+1)}\ell_a\ell_b + \widehat{\mathbb{Y}}^{(m+1)}_{ab},$$
$$\mathrm{T}^{(m+2)}_{ab} := -\mathfrak{c}^{(m+2)}\ell_a\ell_b,$$

and $\mathrm{T}^{(k)}_{ab} = 0$ for every $k \geq m+3$. We also construct a function $\Omega_m$ on $\mathcal{M}_m$ from the sequence $\{0,\sigma,0,...,\sigma^{(m+2)},0,...\}$ using Borel's Lemma 5.39. Note that $P^{ab}\mathrm{T}^{(m+1)}_{ab} = t^{(m+1)}$, $\mathrm{T}^{(m+1)}_{ab}n^b = \widehat{c}^{(m+1)} - \mathfrak{c}^{(m+1)}\ell_a$, $\mathrm{T}^{(m+1)}_{ab}n^an^b = -\mathfrak{c}^{(m+1)}$ and $\mathrm{T}^{(m+2)}_{ab}n^an^b = -\mathfrak{c}^{(m+2)}$, but we do not yet know that $\boldsymbol{c}^{(m+1)}(n) = -\mathfrak{c}^{(m+1)}$. To establish this we note that by construction this spacetime satisfies the equations $\mathcal{Q}^{(k)}_{\alpha\beta} = 0$, $\mathcal{L}^{(k)}_{\mu} = 0$ for $k \leq m$ and $f^{(k)} = 0$ for $k \leq m+2$. So, item 1. of Corollary 6.23 for $\ell = m$ yields firstly $\mathcal{L}^{(m+1)}_a n^a = 0$, which by (6.66) takes the form

$$\sigma\pounds_n\mathfrak{c}^{(m+1)} - m(\pounds_n\sigma)\,\mathfrak{c}^{(m+1)} - \sigma\pounds_n\mathfrak{c}^{(m+1)} + \frac{m}{\mathfrak{n}}\mathfrak{c}^{(m+1)}\pounds_n\sigma = \text{lower order terms.}$$

The key point is that the lower order terms in this equation are, by construction, the same ones as in (6.103). Thus, subtracting both equations we get a homogeneous first order ODE for $c^{(m+1)}_a n^a + \mathfrak{c}^{(m+1)}$, and since $\boldsymbol{c}^{(m+1)}(n)|_\Sigma = -\mathfrak{c}^{(m+1)}|_\Sigma$ we conclude $\boldsymbol{c}^{(m+1)}(n) = -\mathfrak{c}^{(m+1)}$ everywhere.

Once we have shown $\boldsymbol{c}^{(m+1)}(n) = -\mathfrak{c}^{(m+1)}$ we also have $\mathrm{T}^{(m+1)}_{ab}n^b = \widehat{c}^{(m+1)} - \mathfrak{c}^{(m+1)}\ell_a = c^{(m+1)}_a$ and we are ready to prove the other two claims, namely that $\mathbb{Y}^{(m+1)}(n,\cdot) = \boldsymbol{c}^{(m+1)}$ and $\operatorname{tr}_P \mathbb{Y}^{(m+1)} = t^{(m+1)}$. First of all note that $(\mathcal{M}_m, g_m, \Omega_m)$ satisfies $\mathcal{L}^{(m+1)}_a = 0$, $\iota^\star\dot{\mathcal{Q}}^{(m+1)}_a = 0$, $\ddot{\mathcal{Q}}^{(m+1)} = \dot{\mathcal{L}}^{(m+1)} = f^{(m+3)} = 0$ (because these were the equations we used to obtain $\boldsymbol{c}^{(m+1)}$, $\sigma^{(m+2)}$, $t^{(m+1)}$ and $\mathfrak{c}^{(m+2)}$) and $\mathcal{Q}^{(m+1)}_{ab} = 0$ (by assumption). Then, items 2. and 3. in Corollary 6.23 (with $\ell = m$) imply that $(\mathcal{M}_m, g_m, \Omega_m)$ satisfies also $\mathcal{Q}^{(m+2)}_{ab}n^an^b = 0$, $\mathcal{Q}^{(m+2)}_{ab}n^a = 0$ and $P^{ab}\mathcal{Q}^{(m+2)}_{ab} = 0$. From (6.63) with $\mathbf{Y}^{(m+1)} \to \mathbf{T}^{(m+1)}$ we get (for the third one we use (2.49), and define for shortness $N^{\mathfrak{n}}_m := \mathfrak{n} - 3 - 2m \neq 0$)

$$\begin{aligned}
0 &= -N^{\mathfrak{n}}_m\sigma\pounds_n\mathfrak{c}^{(m+1)} - (m+1)N^{\mathfrak{n}}_m\sigma\Bbbk\mathfrak{c}^{(m+1)} + \widetilde{O}^{(m+1)}_{ab}[\boldsymbol{c}^{(m+1)},\mathfrak{c}^{(m+1)},\sigma^{(m+1)}]n^an^b + \text{l.o.t},\\
0 &= N^{\mathfrak{n}}_m\sigma\pounds_n\boldsymbol{c}^{(m+1)}_b + (m+1)N^{\mathfrak{n}}_m\sigma\Bbbk\boldsymbol{c}^{(m+1)}_b + \widetilde{O}^{(m+1)}_{ab}[\boldsymbol{c}^{(m+1)},\mathfrak{c}^{(m+1)},\sigma^{(m+1)}]n^a + \text{l.o.t},\\
0 &= N^{\mathfrak{n}}_m\sigma\Big(\pounds_n t^{(m+1)} + 4P(\boldsymbol{c}^{(m+1)},\mathbf{s}) - n(\ell^{(2)})\mathfrak{c}^{(m+1)}\Big) + (m+1)N^{\mathfrak{n}}_m\Bbbk t^{(m+1)}\\
&\quad + \frac{(m+1)(\mathfrak{n}-1)\sigma}{\mathfrak{n}}\Big(\mathfrak{c}^{(m+2)} + 2\pounds_n t^{(m+1)} + 2(m+1)\Bbbk t^{(m+1)}\Big)\\
&\quad + P^{ab}\widetilde{O}^{(m+1)}_{ab}[\boldsymbol{c}^{(m+1)},\mathfrak{c}^{(m+1)},\sigma^{(m+1)}] + \text{l.o.t.}
\end{aligned}$$

But the tensor $\mathbb{Y}^{(m+1)}$ is the solution of the equation $\mathcal{Q}^{(m+2)}_{ab} = 0$ with the replacements $\mathbf{Y}^{(k)} \to \mathbb{Y}^{(k)}$ for every $k \leq m$, $\mathbf{r}^{(m+1)} \to \boldsymbol{c}^{(m+1)}$, $\kappa^{(m+1)} \to \mathfrak{c}^{(m+1)}$, $\operatorname{tr}_P \mathbf{Y}^{(m+1)} \to t^{(m+1)}$, $\kappa^{(m+2)} \to \mathfrak{c}^{(m+2)}$ and $\mathbf{Y}^{(m+1)} \to \mathbb{Y}^{(m+1)}$. This means that the equations $\mathcal{Q}^{(m+2)}_{ab}n^an^b = 0$, $\mathcal{Q}^{(m+2)}_{ab}n^a = 0$ and $P^{ab}\mathcal{Q}^{(m+2)}_{ab} = 0$ (with the corresponding replacements) are also satisfied. Explicitly,

$$\begin{aligned}
0 &= -N^{\mathfrak{n}}_m \sigma \pounds_n \mathbb{k}^{(m+1)} - (m+1)N^{\mathfrak{n}}_m \sigma \mathbb{k}\mathbb{k}^{(m+1)} + \widetilde{O}^{(m+1)}_{ab}[\boldsymbol{c}^{(m+1)}, \mathfrak{c}^{(m+1)}, \sigma^{(m+1)}]n^a n^b + \text{l.o.t},\\
0 &= N^{\mathfrak{n}}_m \sigma \pounds_n \mathbb{r}^{(m+1)}_b + (m+1)N^{\mathfrak{n}}_m \sigma \mathbb{k}\mathbb{r}^{(m+1)}_b + \widetilde{O}^{(m+1)}_{ab}[\boldsymbol{c}^{(m+1)}, \mathfrak{c}^{(m+1)}, \sigma^{(m+1)}]n^a + \text{l.o.t},\\
0 &= N^{\mathfrak{n}}_m \sigma \Big(\pounds_n \Big(\operatorname{tr}_P \mathbb{Y}^{(m+1)}\Big) + 4P(\mathbb{r}^{(m+1)}, \mathbf{s}) - n(\ell^{(2)})\mathbb{k}^{(m+1)}\Big) + (m+1)N^{\mathfrak{n}}_m \sigma \mathbb{k} \operatorname{tr}_P \mathbb{Y}^{(m+1)}\\
&\quad + \frac{(m+1)(\mathfrak{n}-1)\sigma}{\mathfrak{n}} \Big(\mathfrak{c}^{(m+2)} + 2\pounds_n t^{(m+1)} + 2(m+1)\mathbb{k}t^{(m+1)}\Big)\\
&\quad + P^{ab}\widetilde{O}^{(m+1)}_{ab}[\boldsymbol{c}^{(m+1)}, \mathfrak{c}^{(m+1)}, \sigma^{(m+1)}] + \text{l.o.t.}
\end{aligned}$$

By subtracting both systems and recalling that $N^{\mathfrak{n}}_m \neq 0$ (because we have assumed $m+1 \neq \frac{\mathfrak{n}-1}{2}$) and that the lower order terms agree (by construction), one arrives at a homogeneous hierarchical system of ODEs for $\mathfrak{c}^{(m+1)} - \mathbb{k}^{(m+1)}$, $\boldsymbol{c}^{(m+1)} - \mathbb{r}^{(m+1)}$ and $t^{(m+1)} - \operatorname{tr}_P \mathbb{Y}^{(m+1)}$. Since these quantities vanish at $\Sigma$ (because of the initial conditions employed to build $\mathbb{Y}^{(m+1)}_{ab}$) one concludes that they vanish everywhere.

Summarizing, the tensors $\mathbb{Y}^{(m+1)}$, $\sigma^{(m+2)}$ and $\mathfrak{c}^{(m+2)}$ we have just constructed satisfy the equations $\mathcal{Q}^{(m+2)}_{ab} = 0$, $\mathcal{L}^{(m+1)}_a = 0$, $\dot{\mathcal{L}}^{(m+1)} = \ddot{\mathcal{Q}}^{(m+1)} = f^{(m+3)} = 0$ and $\dot{\hat{\mathcal{Q}}}^{(m+1)}_a|_\Sigma = 0$. An application of item 2. in Corollary 6.23 (with $\ell = m$) for $(\mathcal{M}_m, g_m, \Omega_m)$ shows that $\mathbb{Y}^{(m+1)}$ also satisfies $\dot{\mathcal{Q}}^{(m+1)}_a = 0$ because $\mathbb{Y}^{(m+1)}_{ab} = \mathrm{T}^{(m+1)}_{ab}$. Then, the equations $\mathcal{Q}^{(k)}_{\alpha\beta} = 0$, $\mathcal{L}^{(k)}_\mu = 0$ and $f^{(k+2)} = 0$ hold for all $k \leq m+1$, and also $\mathcal{Q}^{(m+2)}_{ab} = 0$. This closes the induction argument. To conclude the proof we only need to analyze the three exceptional cases.

- In the case $m+1 = \frac{\mathfrak{n}-1}{2}$ (when $\mathfrak{n}$ is odd) the only thing that changes is that the equation $\mathcal{Q}^{(\frac{\mathfrak{n}+1}{2})}_{ab} = 0$ cannot be employed to build $\mathbb{Y}^{(\frac{\mathfrak{n}-1}{2})}$. Instead, we construct it using decomposition (2.24) with $\mathbb{r}^{(\frac{\mathfrak{n}-1}{2})} := \boldsymbol{c}^{(\frac{\mathfrak{n}-1}{2})}$, $\operatorname{tr}_P \mathbb{Y}^{(\frac{\mathfrak{n}-1}{2})} := t^{(\frac{\mathfrak{n}-1}{2})}$ and the transverse part given by the free data $\widehat{\mathbb{Y}}^{(\frac{\mathfrak{n}-1}{2})}$. Note that equation $\mathcal{Q}^{(\frac{\mathfrak{n}+1}{2})}_{ab} = 0$ still holds by hypothesis because we have assumed $\mathcal{O}^{\mathscr{I}}_{ab} = 0$. Here obviously we do not need to prove that $\mathbb{Y}^{(m+1)}(n,\cdot) = \boldsymbol{c}^{(m+1)}$ and $\operatorname{tr}_P \mathbb{Y}^{(m+1)} = t^{(m+1)}$. The rest of the argument remains unchanged.

- When $m+1 = \mathfrak{n}-1$ the problem is that the system (6.91) does not determine $\{\sigma^{(\mathfrak{n})}, \mathfrak{c}^{(\mathfrak{n})}, t^{(\mathfrak{n}-1)}\}$, so instead we integrate the second-order transport equation (6.95) with the replacement $\operatorname{tr}_P \mathbf{Y}^{(\mathfrak{n}-1)} \to t^{(\mathfrak{n}-1)}$ with the initial conditions $t^{(\mathfrak{n}-1)}|_\Sigma$ and $\pounds_n t^{(\mathfrak{n}-1)}|_\Sigma$ determined from $\sigma^{(\mathfrak{n})}_\Sigma$ and $\mathfrak{c}^{(\mathfrak{n})}|_\Sigma := \mathfrak{m}$ using the equations $\ddot{\mathcal{Q}}^{(\mathfrak{n}-1)}|_\Sigma = 0$ and $f^{(\mathfrak{n}+1)}|_\Sigma = 0$, exactly as explained in Remark 6.24. The remainder of the argument proceeds identically.

- Finally, for $m+1 = \mathfrak{n}$ the issue is that the equation $\dot{\hat{\mathcal{Q}}}^{(\mathfrak{n})}_a|_\Sigma = 0$ cannot be imposed to obtain the value of the one-form $\boldsymbol{c}^{(\mathfrak{n})}$ at $\Sigma$, and instead we establish $\iota^\star \boldsymbol{c}^{(\mathfrak{n})} = \boldsymbol{\beta}^{(\mathfrak{n})}$ and $\boldsymbol{c}^{(\mathfrak{n})}(n)|_\Sigma = -\mathfrak{c}^{(\mathfrak{n})}|_\Sigma$ as initial condition. Note again that the equation $\dot{\hat{\mathcal{Q}}}^{(\mathfrak{n})}_a|_\Sigma = 0$ holds by hypothesis because we have assumed $\mathcal{O}^\Sigma_a = 0$. The argument continues in the same manner.

Once the full collections $\{\sigma^{(k)}\}_{k\geq 0}$ and $\{\mathbb{Y}^{(k)}\}_{k\geq 1}$ have been constructed, we use Borel's Lemma 5.39 and Theorem 5.40 to build a function $\Omega$ and a spacetime $(\mathcal{M}, g)$ that satisfies $f^{(k)} = 0$, $\mathcal{Q}^{(k)}_{\alpha\beta} = 0$ and $\mathcal{L}^{(k)}_{\alpha} = 0$ for all $k \geq 1$. Therefore, $(\mathcal{M}, g, \Omega)$ solves the conformal Einstein equations to infinite order at $\mathscr{I}$ and realizes the initial data. □

## 6.6 OBSTRUCTION TENSORS AT NULL INFINITY

The purpose of this section is to study the obstruction tensors at their lowest non-trivial orders, namely the Coulombian obstruction tensor $\mathcal{O}^{\Sigma}_{a}$ in spacetime dimension four ($\mathfrak{n} = 3$) and the radiative obstruction tensor $\mathcal{O}^{\mathscr{I}}_{ab}$ in spacetime dimension six ($\mathfrak{n} = 5$), because recall that $\mathcal{Q}^{(2)}_{ab}$ is identically zero in spacetime dimension four. First of all we put forward the precise definition of the obstruction tensors.

**Definition 6.31.** *Let $(\mathcal{M}, g, \Omega)$ be an $(\mathfrak{n}+1)$-dimensional conformal manifold with null infinity $\Phi : \mathscr{I} \hookrightarrow \mathcal{M}$ written in a conformal gauge satisfying $|\nabla\Omega|^2 = 0$. Let $\xi$ be a rigging extended off $\Phi(\mathscr{I})$ geodesically, $\mathcal{Q} := (\mathfrak{n}-1)\Big(\operatorname{Hess}\Omega + \Omega\operatorname{Sch}_g\Big)$ and $\iota : \Sigma \hookrightarrow \mathscr{I}$ a cross-section. We define the radiative obstruction tensor $\mathcal{O}^{\mathscr{I}}$ (for $\mathfrak{n}$ odd) and the Coulombian obstruction tensor $\mathcal{O}^{\Sigma}$ (for any $\mathfrak{n}$) by means of*

$$\mathcal{O}^{\mathscr{I}} := \Phi^{\star}\Big(\pounds_{\xi}^{(\frac{\mathfrak{n}-1}{2})}\mathcal{Q}\Big), \qquad \mathcal{O}^{\Sigma} := \iota^{\star}\Phi^{\star}\Big(\pounds_{\xi}^{(\mathfrak{n}-1)}\mathcal{Q}(\xi,\cdot)\Big).$$

As recalled in the introduction (Chapter 1) these obstructions have already appeared in the literature in specific situations with other names (see e.g. [280] where they are denoted as the "Coulombian" and "radiative" anomalies). A definition of the obstruction tensors should also exist in an arbitrary conformal gauge, but this is beyond the scope of this thesis.

### 6.6.1 *Coulombian obstruction in four dimensions*

As already indicated in the previous section, the factor $(\mathfrak{n}-m)$ multiplying $\mathbf{r}^{(m)}$ in equation $\dot{\mathcal{Q}}^{(m)}_{a} = 0$ leads to two important consequences when $m = \mathfrak{n}$. The first one is that the equation does not constrain the value of the one-form $\widehat{\mathbf{r}}^{(\mathfrak{n})}$ (which therefore becomes free data), and the second one is that if the reminder of the equation does not vanish, either the conformal spacetime is not smooth or does not satisfy the Einstein equations beyond order $m-1$. This reminder defines the Coulombian obstruction tensor $\mathcal{O}^{\Sigma}_{a}$. In spacetime dimension four, $\mathcal{O}^{\mathscr{I}}_{a} = \dot{\mathcal{Q}}^{(3)}_{a}$, and hence it depends on the $\mathscr{I}$-structure data, $\mathbf{Y}$, $\mathbf{Y}^{(2)}$, $\sigma^{(3)}$ and $\kappa^{(3)}$. These tensors are uniquely given in terms of the free data $\{\chi, \sigma^{(3)}_{\Sigma}, \mathfrak{m}, \mathcal{Y}^{(1)}_{AB}\}$ on $\Sigma$ and the radiation field $\widehat{\mathbb{Y}}_{ab}$ on $\mathscr{I}$ after solving the equations $\mathcal{Q}^{(1)}_{\alpha\beta} = \mathcal{Q}^{(2)}_{\alpha\beta} = 0$, $\mathcal{L}_{\mu} = \mathcal{L}^{(2)}_{\mu} = 0$, $f^{(2)} = f^{(3)} = 0$, $\mathcal{L}^{(3)}_{a} n^a = 0$ and $f^{(4)}|_{\Sigma} = 0$. Note also that by Prop. 6.22 the equations $\dot{\mathcal{Q}}^{(3)}_{a} n^a = 0$, $\mathcal{Q}^{(3)}_{ab} n^a = 0$, $P^{ab}\mathcal{Q}^{(3)}_{ab} = 0$ and $f^{(4)} = 0$ follow automatically, so in particular the obstruction tensor satisfies $\mathcal{O}^{\Sigma}_{a} n^a = 0$.

In order to find a necessary and sufficient condition for $\mathcal{O}^{\Sigma}_{a}$ to vanish, let us note that since $\mathcal{Q}_{\alpha\beta}$ vanishes up to an including order 2, it suffices to study the tensor $\dot{\mathcal{Q}}^{(3)}_{a}$ in any gauge, since its vanishing is a gauge-invariant statement. This is a consequence of the following simple observation.

**Lemma 6.32.** *Assume* $\mathcal{Q}^{(k)}_{\alpha\beta} = 0$ *for all* $k = 1, ..., m$. *Let* $\xi' = z(\xi + V)$, *with* $z$ *and* $V$ *extended arbitrarily off* $\mathscr{I}$. *Then,* $\mathcal{Q}^{(m+1)\prime}_{ab} := (\pounds^{(m)}_{\xi'}\mathcal{Q})_{ab} = z^m \mathcal{Q}^{(m+1)}_{ab}$.

*Proof.* The result is obtained at once by inserting $\xi' = z(\xi + V)$ into $\pounds^{(m)}_{\xi'}\mathcal{Q}_{\alpha\beta}$ and using $\mathcal{Q}^{(k)}_{\alpha\beta} = 0$ for all $k = 1, ..., m$. □

Hence, it is sufficient to analyze $\mathcal{O}^{\Sigma}_{a}$ in a gauge in which $\sigma^{(1)} = 1$, $\ell^{(2)} = 0$ and the pullback of $\boldsymbol{\ell}$ to the cross-sections of $\mathscr{I}$ vanishes, $\boldsymbol{\ell}_{\parallel} = 0$. This immediately implies $\mathbf{s} = 0$ (see Section 2.4) and hence $\mathbf{r} = \mathbf{s} + d\sigma^{(1)} = 0$. Moreover, the tensor $P$ at $\Sigma$ decomposes as (2.141) $P^{ab} = h^{AB} e^a_A e^b_B$, where $h^{AB}$ is the inverse metric of $h_{AB}$ and $\{e_A\}$ is a basis in $\Sigma$ with dual $\{\boldsymbol{\theta}^A\}$, and therefore $\delta^\alpha_\rho = e^\alpha_B \theta^B_\rho + \xi^\alpha \nu_\rho + \nu^\alpha \xi_\rho$. Since the tensor $\mathcal{Q}$ involves up to third derivatives of the metric and the quasi-Einstein equations to second order are imposed, it is to be expected that $\mathcal{Q}$ may have some relation to derivatives of the Weyl tensor. We pursue this idea by applying a transverse derivative to the identity (6.20) and evaluating the result at $\mathscr{I}$. Since $\mathcal{Q}$ vanishes up to order two, we may perform the substitution $\mathcal{T}_{\alpha\beta} = \frac{1}{\mathfrak{n}-1}\mathcal{Q}_{\alpha\beta} = \frac{\Omega^2}{2(\mathfrak{n}-1)}\mathcal{P}_{\alpha\beta}$ for some tensor $\mathcal{P}_{\alpha\beta}$ satisfying $\mathcal{Q}^{(3)} = \mathcal{P}^{(1)}$ at $\mathscr{I}$. Rewriting identity (6.20) (recall $d = \mathfrak{n} + 1$) in terms of $\mathcal{P}_{\alpha\beta}$ gives

$$\begin{aligned} C^\alpha{}_{\beta\mu\nu}\nabla_\alpha\Omega - \frac{\Omega}{\mathfrak{n}-2}\nabla_\alpha C^\alpha{}_{\beta\mu\nu} &= -\frac{\Omega}{\mathfrak{n}-1}\left(\Omega\nabla_{[\mu}\mathcal{P}_{\nu]\beta} + 2\mathcal{P}_{\beta[\nu}\nabla_{\mu]}\Omega\right) \\ &\quad + \frac{\Omega}{\mathfrak{n}(\mathfrak{n}-1)} g_{\beta[\mu}\left(\Omega\nabla^\rho\mathcal{P}_{\nu]\rho} + 2\mathcal{P}_{\rho[\nu}\nabla^\rho\Omega\right). \end{aligned}$$

Applying $\nabla_\xi$ (here it turns out to be more useful to take a covariant derivative along $\xi$ rather than a Lie derivative) and evaluating the result at $\Omega = 0$ gives (we use that $\pounds_\xi\Omega \stackrel{\mathscr{I}}{=} \sigma^{(1)} \stackrel{\mathscr{I}}{=} 1$ and hence $\nu_\mu \stackrel{\mathscr{I}}{=} \nabla_\mu\Omega$)

$$\nu_\alpha\xi^\rho\nabla_\rho C^\alpha{}_{\beta\mu\nu} + C^\alpha{}_{\beta\mu\nu}\xi^\rho\nabla_\rho\nabla_\alpha\Omega - \frac{1}{\mathfrak{n}-2}\nabla_\alpha C^\alpha{}_{\beta\mu\nu} \stackrel{\mathscr{I}}{=} -\frac{2}{\mathfrak{n}-1}\mathcal{P}_{\beta[\nu}\nu_{\mu]} + \frac{2}{\mathfrak{n}(\mathfrak{n}-1)}g_{\beta[\mu}\mathcal{P}_{\nu]\rho}\nu^\rho.$$

The second term in the left-hand side vanishes because $\xi^\rho\nabla_\rho\nabla_\alpha\Omega \stackrel{\mathscr{I}}{=} \frac{1}{\mathfrak{n}-1}\mathcal{Q}^{(1)}_{\rho\alpha}\xi^\rho - \Omega\xi^\rho L_{\rho\alpha} \stackrel{\mathscr{I}}{=} 0$, so we arrive at

$$\nu_\alpha\xi^\rho\nabla_\rho C^\alpha{}_{\beta\mu\nu} - \frac{1}{\mathfrak{n}-2}\nabla_\alpha C^\alpha{}_{\beta\mu\nu} \stackrel{\mathscr{I}}{=} -\frac{2}{\mathfrak{n}-1}\mathcal{P}_{\beta[\nu}\nu_{\mu]} + \frac{2}{\mathfrak{n}(\mathfrak{n}-1)}g_{\beta[\mu}\mathcal{P}_{\nu]\rho}\nu^\rho. \tag{6.104}$$

We now contract this equation with $\xi^\beta e^\mu_A \xi^\nu$ to make the tensor $\mathfrak{C}_{\alpha\beta} := \xi^\mu\xi^\nu C_{\alpha\mu\beta\nu}$ appear. The contraction of (6.104) with $\xi^\beta e^\mu_A\xi^\nu$ then gives, after using $e^\mu_A\nu_\mu = e^\mu_A\xi_\mu = \xi^\mu\xi_\mu = 0$, $\nu_\alpha\xi^\alpha = 1$ and $\nabla_\xi\xi = 0$,

$$\nu_\alpha e^\mu_A\xi^\rho\nabla_\rho\mathfrak{C}^\alpha{}_\mu - \frac{1}{\mathfrak{n}-2}\xi^\beta e^\mu_A\xi^\nu\nabla_\alpha C^\alpha{}_{\beta\mu\nu} \stackrel{\mathscr{I}}{=} \frac{1}{\mathfrak{n}-1}\xi^\beta e^\mu_A\mathcal{P}_{\beta\mu} \stackrel{\mathscr{I}}{=} \frac{1}{\mathfrak{n}-1}\dot{\mathcal{Q}}^{(3)}_A.$$

We still need to elaborate the second term in the left-hand side. Note that due to $\mathbf{r} = \mathbf{s} = 0$ and $\ell^{(2)} = 0$ we have from (2.72), (2.54) and (2.57) that $\nabla^\alpha \xi^\beta = P^{ac} V^b{}_c e^\alpha_a e^\beta_b = P^{ac} P^{bd} (\mathrm{Y}_{cd} + \mathrm{F}_{cd}) e^\alpha_a e^\beta_b = h^{AC} h^{BD} (\mathrm{Y}_{CD} + \mathrm{F}_{CD}) e^\alpha_A e^\beta_B$, and hence

$$\begin{aligned} \xi^\beta e^\mu_A \xi^\nu \nabla_\alpha C^\alpha{}_{\beta\mu\nu} &\stackrel{\mathscr{I}}{=} e^\mu_A \nabla_\alpha \mathfrak{C}^\alpha{}_\mu - e^\mu_A C_{\alpha\beta\mu\nu} \Big( \xi^\beta \nabla^\alpha \xi^\nu + \xi^\nu \nabla^\alpha \xi^\beta \Big) \\ &\stackrel{\mathscr{I}}{=} e^\mu_A \nabla_\alpha \mathfrak{C}^\alpha{}_\mu + \Big( {}^{(2)}C^B{}_A{}^C + {}^{(2)}C_A{}^{BC} \Big) (\mathrm{Y}_{BC} + \mathrm{F}_{BC}), \end{aligned}$$

where we recall the notation in Appendix B for ${}^{(2)}C_{\alpha\beta\mu} := \xi^\nu C_{\alpha\nu\beta\mu}$. We finally use $\delta^\alpha_\rho = e^\alpha_B \theta^B_\rho + \xi^\alpha \nu_\rho + \nu^\alpha \xi_\rho$ in the first term of the right-hand side to get

$$\begin{aligned} e^\mu_A \nabla_\alpha \mathfrak{C}^\alpha{}_\mu &\stackrel{\mathscr{I}}{=} e^\mu_A e^\alpha_B \theta^B_\rho \nabla_\alpha \mathfrak{C}^\rho{}_\mu + \nu_\rho e^\mu_A \xi^\alpha \nabla_\alpha \mathfrak{C}^\rho{}_\mu + e^\mu_A \nu^\alpha \xi_\rho \nabla_\alpha \mathfrak{C}^\rho{}_\mu \\ &\stackrel{\mathscr{I}}{=} \nabla^h_B \mathfrak{C}^B{}_A + \nu_\alpha e^\mu_A \xi^\rho \nabla_\rho \mathfrak{C}^\alpha{}_\mu, \end{aligned}$$

where in the first term we used $e^\mu_A e^\alpha_B \theta^B_\rho \nabla_\alpha \mathfrak{C}^\rho{}_\mu = \nabla^h_B \mathfrak{C}^B{}_A$ (because $\mathfrak{C}$ is orthogonal both to $\xi$ and $\nu$), and the third term vanishes because $\nu^\alpha \nabla_\alpha \xi^\rho \stackrel{\mathscr{I}}{=} 0$ (by (2.53) and (2.57)) and $\mathfrak{C}(\xi, \cdot) = 0$. Therefore, we conclude

$$(\mathfrak{n}-3) \nu_\alpha e^\mu_A \xi^\rho \nabla_\rho \mathfrak{C}^\alpha{}_\mu - \nabla^h_B \mathfrak{C}^B{}_A - \Big( {}^{(2)}C^B{}_A{}^C + {}^{(2)}C_A{}^{BC} \Big) (\mathrm{Y}_{BC} + \mathrm{F}_{BC}) \stackrel{\mathscr{I}}{=} \frac{\mathfrak{n}-2}{\mathfrak{n}-1} \dot{\mathcal{Q}}^{(3)}_A. \quad (6.105)$$

In four spacetime dimensions ($\mathfrak{n} = 3$) the first term vanishes and the tensor ${}^{(2)}C_{ABC}$ is zero, because the identity (6.20) entails $C_{\alpha\beta\mu\nu} \nu^\alpha \stackrel{\mathscr{I}}{=} 0$, so $C_{\alpha\beta\mu\nu}$ has Petrov type $N$ at $\mathscr{I}$, and all its non-vanishing components at $\mathscr{I}$ are encoded in $\mathfrak{C}_{\alpha\beta}$ (see [116, 257]). Since $\dot{\mathcal{Q}}^{(3)}_A = \mathcal{O}^\Sigma_A$, then a necessary and sufficient condition for the Coulombian obstruction tensor to vanish is $\mathfrak{C}_{AB}$ being divergence-free. As we already recalled in Section 6.1, when the cuts of $\mathscr{I}$ are 2-spheres (e.g. for asymptotically simple spacetimes [254]), the fact that there are no TT tensors on $\mathbb{S}^2$ [322] imply that $\mathcal{O}^\Sigma_a = 0$ if and only if $\mathfrak{C}_{AB} = 0$, and hence the full Weyl tensor vanishes at $\mathscr{I}$. For other topologies of $\mathscr{I}$, such as $\mathbb{R} \times T^2$ (see [299]), $\mathfrak{C}_{AB}$ being divergence-free does not imply $\mathfrak{C}_{AB} = 0$. One can then construct four dimensional, smooth, asymptotically flat spacetimes (in the sense of Def. 6.26) as a particular case of Theorem 6.16 with $\mathscr{I} \not\simeq \mathbb{R} \times \mathbb{S}^2$ whose Weyl tensor does not vanish at $\mathscr{I}$. Establishing existence in the stronger sense that the spacetime $\Omega^{-2} g$ is Ricci flat in a neighborhood of $\mathscr{I}$ is an interesting and open problem. To the best of our knowledge, the asymptotic characteristic problem has only been solved under the assumption that the Weyl tensor vanishes at $\mathscr{I}$ and in dimension 4 [163, 186], see Subsection 6.1.3 for a brief review.

Equation (6.105) is interesting also in higher dimensions, because if one is interested in using the Weyl as a variable to be determined iteratively from an expansion, (6.105) can be used to compute the second order term $\nu_\alpha e^\mu_A \xi^\rho \nabla_\rho \mathfrak{C}^\alpha{}_\mu$ in terms of $\mathfrak{C}$, ${}^{(2)}C_{ABC}$ and $\dot{\mathcal{Q}}^{(3)}_A$ provided $\mathfrak{n} \neq 3$ (recall that in higher dimension the tensor ${}^{(2)}C_{ABC}$ need not to vanish, see [257]). When $\mathfrak{n} = 3$ the Coulombian obstruction would manifest itself also in this approach.

### 6.6.2 *Radiative obstruction in six dimensions*

As already mentioned, for $\mathfrak{n}$ odd the equation $\mathcal{Q}_{ab}^{(\frac{\mathfrak{n}+1}{2})} = 0$ does not fix the full tensor $\mathbf{Y}^{(\frac{\mathfrak{n}-1}{2})}$. Furthermore, after having assumed that the previous orders are satisfied, the tensor $\mathcal{Q}_{ab}^{(\frac{\mathfrak{n}+1}{2})}$ turns out to only depend on null metric hypersurface data, $\chi$, $\{\sigma_{\Sigma}^{(k)}\}_{k\leq\frac{\mathfrak{n}+1}{2}}$ and $\{\mathcal{Y}_{AB}^{(k)}\}_{k\leq\frac{\mathfrak{n}-3}{2}}$. This defines the radiative obstruction tensor $\mathcal{O}_{ab}^{\mathscr{I}}$ whose vanishing determines whether $\mathcal{Q}_{ab}^{(\frac{\mathfrak{n}+1}{2})} = 0$ can be satisfied. This behaviour is reminiscent of the Fefferman-Graham obstruction tensor $\mathcal{O}^{\mathscr{F}\mathscr{G}}$ in the context of ambient metrics (see Subsection 5.1.1). Recall that for $\mathfrak{n} = 3$ the tensor $\mathcal{O}_{ab}^{\mathscr{I}}$ automatically vanishes, just like $\mathcal{O}^{\mathscr{F}\mathscr{G}}$, and that they appear at the same order. This suggests a strong connection between $\mathcal{O}_{ab}^{\mathscr{I}}$ and the FG obstruction tensor at the cross-sections of $\mathscr{I}$. Establishing this connection would require understanding in detail all the lower order terms arising in (6.63). This task is challenging and well beyond the scope of this chapter, so in this section we just analyze the first non-trivial case, namely $\mathfrak{n} = 5$ (i.e. spacetime dimension six), where $\mathcal{O}_{ab}^{\mathscr{I}} = \mathcal{Q}_{ab}^{(3)}$.

Assume $(\mathcal{M}, g, \Omega)$ is a six-dimensional conformal manifold with $\lambda = 0$ satisfying $\mathcal{Q}_{\alpha\beta}^{(1)} = \mathcal{Q}_{\alpha\beta}^{(2)} = 0$, $\mathcal{L}_{\mu}^{(1)} = \mathcal{L}_{\mu}^{(2)} = 0$ and $f^{(1)} = f^{(2)} = f^{(3)} = f^{(4)} = 0$. By item 1. in Prop. 6.22 we know that the tensor $\mathcal{Q}_{ab}^{(3)}$ satisfies $P^{ab}\mathcal{Q}_{ab}^{(3)} = 0$ and $\mathcal{Q}_{ab}^{(3)} n^b = 0$. In Lemma 6.32 we have established that under any change of rigging $\xi' = z(\xi + V)$ (with $z$ and $V$ extended arbitrarily off $\mathscr{I}$) one has $\mathcal{Q}_{ab}^{(3)\prime} = z^2 \mathcal{Q}_{ab}^{(3)}$. So, in order to analyze the obstruction tensor at $\mathscr{I}$ it suffices to compute $\mathcal{Q}_{ab}^{(3)}$ in a simple gauge. We choose, as in the previous subsection, the gauge in which $\sigma^{(1)} = 1$, $\ell^{(2)} = 0$ and the pullback of $\boldsymbol{\ell}$ to the cross-sections of $\mathscr{I}$ vanishes, $\boldsymbol{\ell}_{\parallel} = 0$. More specifically, we shall work in Gaussian null coordinates $\{r, t, x^A\}$ (see Appendix C) in which the metric in a neighbourhood of $\mathscr{I} = \{r = 0\}$ takes the form

$$g = 2dtdr + \phi dt^2 + 2\beta_A dx^A dt + \mu_{AB} dx^A dx^B,$$

where $\phi$ and $\boldsymbol{\beta}$ vanish at $r = 0$ and the rigging is $\xi = \partial_r$. As in Appendix C, we have $\gamma_{ab} = \delta_a^A \delta_b^B h_{AB}$, where $h_{AB} := \mu_{AB}|_{r=0}$, $\kappa^{(m)} = -\frac{1}{2}\dot{\phi}^{(m)}$, $\mathrm{r}_A^{(m)} = \frac{1}{2}\dot{\boldsymbol{\beta}}^{(m)}$, $\mathrm{Y}_{AB}^{(m)} = \frac{1}{2}\dot{\mu}_{AB}^{(m)}$, $P^{ab} = \mu^{AB}\delta_A^a \delta_B^b$ and $n^a = \delta_t^a$. In particular, we use a prime to denote derivative w.r.t. $t$ and a dot for derivative w.r.t. $r$.

Computing the quasi-Einstein equation (including all terms) by hand becomes intractable very quickly. Therefore, and since we need the full expression of the quasi-Einstein equation up to order $\mathcal{Q}_{ab}^{(3)}$, we have performed the computation with the aid of the `xAct` package [93] in `Mathematica`. The outcome of the computation has the following consequences. Firstly, equations $\mathcal{Q}_{AB}^{(1)} = 0$, $\dot{\mathcal{Q}}_A^{(1)} = 0$, $\dot{\mathcal{Q}}_t^{(1)} = 0$ and $\ddot{\mathcal{Q}}^{(1)} = 0$ at $\mathscr{I}$ imply $h'_{AB} = 0$, $\mathrm{r}_A = 0$, $\kappa = 0$ and $\sigma^{(2)} = 0$, while $\mathcal{Q}_{At}^{(1)} = \mathcal{Q}_{tt}^{(1)} = \mathcal{L}_A^{(1)} = \mathcal{L}_t^{(1)} = 0$ hold automatically. Next, equation $f^{(3)} = 0$ fixes $\kappa^{(2)} = 0$, which inserted into $\dot{\mathcal{L}}^{(1)} = 0$ gives $\pounds_n(P^{AB}\mathrm{Y}_{AB}) = -\frac{R^h}{4}$. One then checks that $\mu^{AB}\mathcal{Q}_{AB}^{(2)} = \mathcal{Q}_{tA}^{(2)} = \mathcal{Q}_{tt}^{(2)} = 0$ hold automatically, and from $\mathcal{Q}_{AB}^{(2)} = 0$ one obtains

$$\pounds_n \mathrm{Y}_{AB} = -L_{AB}^h, \qquad (6.106)$$

where $L^h_{AB}$ is the Schouten tensor of $h_{AB}$. One can also check that $\dot{\mathcal{Q}}^{(2)}_t = \mathcal{L}^{(2)}_t = 0$ hold automatically. Equation $\dot{\mathcal{Q}}_A = 0$ then gives $\mathrm{r}^{(2)}_A = \frac{1}{3}\left(\nabla_B \mathrm{Y}^B{}_A - \nabla_A \mathrm{Y}^B{}_B\right)$, where $\nabla$ is the Levi-Civita derivative of $h$, and $\mathcal{L}^{(2)}_A = \mathcal{Q}^{(3)}_{At} = \mathcal{Q}^{(3)}_{tt} = 0$ are automatically satisfied. Next, equations $\ddot{\mathcal{Q}}^{(2)} = 0$, $f^{(4)} = 0$ and $\dot{\mathcal{L}}^{(2)} = 0$ read, respectively,

$$
\begin{aligned}
4\sigma^{(3)} + \mathrm{Y}^{AB}\mathrm{Y}_{AB} - \operatorname{tr}_P \mathbf{Y}^{(2)} &= 0,\\
\pounds_n \sigma^{(3)} + \kappa^{(3)} &= 0,\\
6\pounds_n(\operatorname{tr}_P \mathbf{Y}^{(2)}) + 8\kappa^{(3)} + 6\mathrm{Y}^{AB}R_{AB} - R^h \operatorname{tr}_P \mathbf{Y} &= 0.
\end{aligned}
$$

Taking a derivative of the first equation w.r.t. $t$ and solving the system one obtains $\pounds_n\sigma^{(3)} = \phi^{(3)} = 0$ and $\pounds_n(\operatorname{tr}_P \mathbf{Y}^{(2)}) = -2h^{AB}h^{CD}L^h_{AC}\mathrm{Y}_{BD}$. One can now check that $P^{ab}\mathcal{Q}^{(3)}_{ab} = 0$ is automatically verified. Substituting all these expressions into the tensor $\mathcal{Q}^{(3)}_{AB}$ yields

$$
\begin{aligned}
\mathcal{Q}^{(3)}_{AB} = -\frac{1}{3}\Big(& 6\Box_h \mathrm{Y}_{AB} - 6\nabla^C\nabla_{(A}\mathrm{Y}_{B)C} + 6W^h_{ACBD}\mathrm{Y}^{CD} - 2\nabla_{(A}\nabla^C\mathrm{Y}_{B)C} + 2\nabla_A\nabla_B\mathrm{Y}^C{}_C\\
&- 2\left(\Box_h \mathrm{Y}^C{}_C - \nabla^C\nabla^D\mathrm{Y}_{CD}\right)h_{AB}\Big),
\end{aligned}
\tag{6.107}
$$

where $W^h_{ACBD}$ denotes the Weyl tensor of $h_{AB}$. Observe that the right-hand side is manifestly traceless. In accordance with the general results in the previous sections, equation $\mathcal{Q}^{(3)}_{AB} = 0$ does not determine the tensor $\mathrm{Y}^{(2)}_{AB}$ via a transport equation. Instead, the right-hand side of (6.107) defines a symmetric, traceless tensor

$$
\begin{aligned}
\mathcal{O}^{\mathscr{I}}_{AB} := -\frac{1}{3}\Big(& 6\Box_h \mathrm{Y}_{AB} - 6\nabla^C\nabla_{(A}\mathrm{Y}_{B)C} + 6W^h_{ACBD}\mathrm{Y}^{CD} - 2\nabla_{(A}\nabla^C\mathrm{Y}_{B)C} + 2\nabla_A\nabla_B\mathrm{Y}^C{}_C\\
&- 2\left(\Box_h \mathrm{Y}^C{}_C - \nabla^C\nabla^D\mathrm{Y}_{CD}\right)h_{AB}\Big),
\end{aligned}
\tag{6.108}
$$

constructed solely from $h_{AB}$ and $\mathrm{Y}_{AB}$. Taking the derivative of $\mathcal{O}^{\mathscr{I}}_{AB}$ along $\partial_t$ and using $\partial_t h_{AB} = 0$, (6.106), and the identity $\nabla_{(A}\nabla^C L^h_{B)C} = h^{CD}\nabla_A\nabla_B L^h_{CD}$ (which follows at once form the contracted Bianchi identity, see (6.3)), one finds

$$
\partial_t \mathcal{Q}^{(3)}_{AB} = 2\left(\Box_h L^h_{AB} - \nabla^C\nabla_{(A}L^h_{B)C} + W^h{}_A{}^C{}_B{}^D L^h_{CD}\right).
$$

The term between round brackets in the right-hand side is precisely the Bach tensor $B_{AB}$ of $h_{AB}$ (see (5.5)). Summarizing, we have obtained $\partial_t\mathcal{Q}^{(3)}_{AB} = 2B_{AB}$. Recall from Subsection 5.1.1 that the Bach tensor is precisely the Fefferman and Graham obstruction tensor in the case of conformal metrics of dimension four. This provides strong support for our expectation that $\mathcal{O}^{\mathscr{I}}$ is closely related to the FG obstruction tensor $\mathcal{O}^{\mathscr{F}\mathscr{G}}$ of the corresponding dimension.

We now give a different argument to show that $\partial_t\mathcal{Q}^{(3)}_{AB}$ must be proportional to the FG obstruction tensor. In Section 6.3 we proved that the Fefferman–Graham ambient metric associated to the conformal class $[h]$ admits a null infinity whose $\mathscr{I}$-structure is given precisely by $h$. Furthermore, we showed that the derivative of $\mathcal{Q}^{(3)}_{AB}$ along $\partial_t$ at $\mathscr{I}$ is proportional to the Fefferman–Graham obstruction tensor of $[h]$ (see (6.35)). Since the derivative $\partial_t \mathcal{Q}^{(3)}_{AB}$ of (6.107) only depends on $h_{AB}$, it must be the same for all spacetimes sharing the same $h_{AB}$ at $\mathscr{I}$. Thus, the only possibility is $\partial_t \mathcal{Q}^{(3)}_{AB}$ being proportional to the FG obstruction tensor in dimension four, i.e. the Bach tensor.

In summary, the vanishing of the Bach tensor of $h_{AB}$, together with the condition $\mathcal{O}^{\mathscr{I}}_{AB} = 0$ on a cross-section $\Sigma$ (which by (6.107) may be interpreted as a restriction on the free data $\mathrm{Y}_{AB}|_{\Sigma}$), guarantees that the full obstruction tensor $\mathcal{O}^{\mathscr{I}}$ vanishes everywhere on $\mathscr{I}$. Thus, the hypothesis $\mathcal{O}^{\mathscr{I}}_{ab} = 0$ in Theorem 6.16 for $\mathfrak{n} = 5$ can be relaxed to $B_{AB} = 0$ and $\mathcal{O}^{\mathscr{I}}_{AB}|_{\Sigma} = 0$. Note that the FG obstruction tensor is conformally covariant, and hence the condition $B_{AB} = 0$ does not depend on the conformal representative of the $\mathscr{I}$-structure.

Our results concerning a geometric characterization of conformal infinity for the Fefferman–Graham ambient metric in Section 6.3 together with preliminary analysis of the general case, suggest that a similar picture emerges in higher dimensions, namely that the FG obstruction tensor arises after taking a sufficient number of derivatives along $n$ on the radiative obstruction tensor. We therefore expect the following conjecture to be true.

**Conjecture 6.33.** *Let* $\{\mathscr{I}, \gamma, \boldsymbol{\ell}, \ell^{(2)}, \sigma, \mathfrak{q}\}$ *be* $\mathscr{I}$*-structure data of odd dimension* $\mathfrak{n} \geq 7$ *admitting a cross-section* $\iota : \Sigma \hookrightarrow \mathscr{I}$ *with induced metric* $h := \iota^\star \gamma$ *and let* $\mathcal{O}^{\mathscr{I}}_{ab}$ *be the radiative obstruction tensor. Denote by* $\mathcal{O}^{\mathscr{F}\mathscr{G}}$ *the Fefferman-Graham obstruction tensor of* $[h]$*. Then,*

$$\iota^\star \left( \mathcal{L}_n^{\left(\frac{\mathfrak{n}-3}{2}\right)} \mathcal{O}^{\mathscr{I}} \right) = C(\mathfrak{n})\, \mathcal{O}^{\mathscr{F}\mathscr{G}},$$

*where* $C(\mathfrak{n})$ *is a constant depending only on* $\mathfrak{n}$*.*

Comparing with the formula (6.35), the constant $C(\mathfrak{n})$ is expected to take the value $C(\mathfrak{n}) = \frac{\mathfrak{n}-1}{2} \left( \frac{\mathfrak{n}-3}{2} \right)!$. However, to remain conservative, we formulate the conjecture without fixing the precise value of $C(\mathfrak{n})$.

Recall from the review in Subsection 5.1.1 that the requirement that the Fefferman-Graham obstruction tensor vanishes is not particularly restrictive and is satisfied in many physically relevant situations, e.g. when $h$ is Einstein or conformally flat, among others. Establishing this conjecture would require a detailed analysis of the lower order terms appearing in (6.63). We intend to analyze this problem in future work.

# 7 CONCLUSIONS AND FUTURE WORK

In Chapter 2 we introduced several new developments within the hypersurface data formalism that play a central role throughout this thesis. These include the definition of the constraint tensor in the null setting in Section 2.2, the notion of extended hypersurface data introduced in Section 2.3, the analysis of non-degenerate submanifolds in Section 2.4, and the formulation of the harmonic gauge in Section 2.5. Promising directions for future research include investigating the relation between null hypersurface data and Carrollian geometry, two different but certainly related approaches to describing null hypersurfaces in a detached setting. Another idea is to explore possible extensions to conformal contexts, with potential links with tractor geometry [90]. This may offer further insight into the geometric structure of null infinity as well.

The results of Section 2.4 are subsequently employed in Chapter 3 to define the notion of double null data, thereby detaching the geometric concept of two transversely intersecting null hypersurfaces from any ambient spacetime. As shown in Section 3.3, double null data satisfying the abstract constraint equations always admits a $\Lambda$-vacuum development. Moreover, in Section 3.4 we establish necessary and sufficient conditions for two such data sets to give rise to isometric spacetimes. Taken together, these results yield a satisfactory geometrization of the characteristic Cauchy problem for the Einstein equations, placing it on the same conceptual footing as the standard spacelike Cauchy problem.

Several natural directions for further investigation emerge from this work. One such direction is the extension of the geometrization procedure along the lines suggested by Luk and others, which would require incorporating the Weyl tensor as an independent field variable and treating the Bianchi identities as additional constraint equations. Further avenues include a systematic analysis of the constraint equations on general hypersurfaces of mixed causal character and the derivation of geometric criteria for their embeddability into an ambient spacetime. Combined with the abstract gluing procedure required to define double null data, which we aim to generalize to hypersurface data of arbitrary causal character, we plan to investigate the conditions under which two hypersurface data can be glued together to form a “corner” and to study the differentiability properties of the resulting spacetime. In particular, it is of interest to determine when such gluings give rise to thin shells of energy, a problem that naturally connects to the active “matching” community [83, 214, 223, 238].

In Chapter 4, and in particular in Section 4.2, we developed a collection of general identities relating the deformation tensor $\mathcal{K} := \pounds_\eta g$ of an arbitrary vector field $\eta$ to the tensor $\Sigma := \pounds_\eta \nabla$ on an abstract hypersurface $\mathcal{H}$ of arbitrary causal character. These identities allowed us to derive explicit necessary conditions that an ambient vector field must satisfy on $\mathcal{H}$ in order to be a homothety. In Section 4.4 we recovered, within the hypersurface data formalism, the classical homothetic Killing initial data equations in the spacelike case [37, 84, 140, 242]. In Section 4.5 we identified two distinct types of conditions for null hypersurfaces, namely those that must be satisfied along the entire hypersurface and those that only need to hold on a single cross-section. Both sets of conditions were formulated in a fully detached and geometric manner. Applying this framework to the characteristic Cauchy problem, we extended in Section 4.6 the results of [74] by expressing the characteristic KID equations in a fully gauge-covariant form, placing them at the same level as the initial data for the characteristic problem (i.e. detached from the spacetime one wishes to construct). In this way, the characteristic KID problem is put on equal footing with its spacelike counterpart. Finally, the versatility of the formalism was illustrated in Section 4.7 through its application to the smooth spacelike–characteristic setting. We emphasize that results of this type, involving hypersurfaces of mixed causal character, rely crucially on a formalism capable of treating hypersurfaces of arbitrary causal character within a unified framework, and would not be possible in approaches restricted to a single causal type.

The identities established in Chapter 4 have several noteworthy consequences. Since they are formulated entirely in terms of abstract hypersurface data, they apply in a wide range of settings and are valid for hypersurfaces of arbitrary causal character. Moreover, they are fully covariant under both diffeomorphisms and gauge transformations. Beyond the specific applications developed in this thesis, these identities provide a systematic framework for analyzing homothetic Killing initial data in other well-posed Cauchy problems, showing that the existence of a homothetic Killing field can be characterized purely at the level of initial data. One potential application of our framework is the study of the KID equations for the spacelike–characteristic initial value problem with corners (the corresponding Cauchy problem is analyzed in [75, 91]). Preliminary analysis suggest that, in this setting, the KID equations reduce to (4.61), (4.63) on the spacelike portion and to (4.106)–(4.108) on the null portion, without the need to impose additional constraints at the intersection surface. A rigorous proof of this statement within a fully detached approach would, however, require an abstract gluing procedure for hypersurface data of arbitrary causal character. As indicated above, developing such a construction is left for future work. Nevertheless, this preliminary analysis further supports the conclusion that the homothetic KID problem is completely characterized at the level of abstract initial data. Furthermore, since the identities hold for an arbitrary vector field $\eta$, and independently of any specific field equations, they are applicable to a broader class of initial value problems beyond the $\Lambda$-vacuum Einstein equations $\mathrm{Ric} = \lambda g$.

Several interesting directions for future research arise from these results. One is the study of asymptotic characteristic (homothetic) KID conditions within the framework of the conformal Einstein field equations, which would require extending the notion of double null data to null infinity. Another appealing application is the construction of candidate Killing or homothetic vector fields by integrating the KID equations from data prescribed on a cross-section of $\mathcal{H}$, as well as to allow for multiple Killing fields along the lines of the concept of multiple Killing horizon [227–229]. These topics are natural continuations of the present work and will be pursued in future investigations.

In Chapter 5 we developed a collection of general identities relating the transverse expansion of the metric at a general null hypersurface to the geometry of the ambient spacetime in which the hypersurface is embedded. These identities allowed us to formulate sufficient conditions on null hypersurface data for the existence of an ambient spacetime satisfying the Einstein equations to infinite order on the hypersurface. We further particularized these results to the case in which the ambient spacetime admits a preferred vector field, such as a symmetry generator, leading to a new family of identities applicable, in particular, to settings including Killing and homothetic horizons. In the Killing horizon case, this analysis naturally led to the introduction of *abstract Killing horizon data*, which completely characterizes the transverse expansion at the hypersurface. For the $\Lambda$-vacuum Einstein equations, this yields a characterization of all analytic $\Lambda$-vacuum spacetimes in a neighbourhood of a non-degenerate Killing horizon. As a direct application, we obtained a local uniqueness theorem in the smooth setting for the non-extremal Schwarzschild–de Sitter spacetime. Finally, we showed that every abstract Killing horizon data set admits a development satisfying the $\Lambda$-vacuum equations to infinite order, with the prescribed data realized as a non-degenerate Killing horizon.

The study of transverse expansions at null hypersurfaces carried out in this thesis opens several promising avenues for future research. One such direction is the *characteristic gluing problem*, which seeks to glue two characteristic initial data sets across a transition region [23, 24]. While the original constructions addressed gluing only up to second order in the metric derivatives, more recent work has extended these results to third order and beyond [69, 297]. Existing approaches, however, rely heavily on specific coordinate systems, assume embedded data, and typically operate in perturbative regimes near highly symmetric backgrounds. A natural objective is to use the identities developed in Chapter 5 to formulate a fully geometric and coordinate-free version of the characteristic gluing problem, enabling one to determine whether two initial data sets are gluable without explicitly solving the Einstein equations.

Another central goal is to establish a local uniqueness result for the non-extremal Kerr spacetime, analogous to the one obtained in this thesis for Schwarzschild–de Sitter. As reviewed in Section 5.1, black-hole uniqueness has been a central topic in General Relativity since the foundational work of Hawking, Carter, Robinson, Israel, and others [56, 155, 284], and continues through recent advances by Alexakis, Ionescu, and Klainerman [13, 175]. While most existing uniqueness results rely on global assumptions such as asymptotic flatness, it is both natural and desirable to investigate uniqueness from a purely local perspective. The techniques developed in Chapter 5, which capture the full transverse expansion of the metric at a horizon, provide a promising framework for addressing this problem.

The extremal case is considerably more challenging. As discussed in Subsection 5.1.3, the static situation has been successfully analyzed in the analytic setting [187, 188], but in the general stationary context the resulting equations are significantly more involved, making uniqueness a difficult open problem. One possible approach is to analyze the equations order by order in the transverse expansion, with the aim of identifying structural patterns that could lead to a general uniqueness argument. For this purpose, it will be useful to implement symbolic computations, e.g. using `xAct` [93] or `SageManifold` [295], to systematically generate and study the equations. Another interesting strategy is to combine the constraint equations arising from the transverse expansion with the KID equations developed in Chapter 4, in order to identify a unique set of transverse data from which to solve an associated characteristic problem. Further directions include establishing local existence and uniqueness results for data prescribed on a homothetic horizon, thereby providing a geometric framework that encompasses and extends the constructions of Fefferman–Graham [112, 113] and Rodnianski–Shlapentokh-Rothman [285].

In Chapter 6 we began by showing that any straight ambient metric admits a conformal completion with a bifurcate Killing horizon with integrable Killing one-form, in which one branch corresponds to $\mathscr{I}$ and the other to the original homothetic horizon. This construction implies that the transverse expansion of the metric at null infinity can be directly related to that at the homothetic horizon. In particular, the free data in the Fefferman–Graham construction appears at the same order as gravitational radiation at $\mathscr{I}$. We then established a one-to-one correspondence between exact, straight ambient metrics and solutions of the conformal Einstein equations admitting a bifurcate conformal Killing horizon with integrable Killing one-form, whose conformal factor satisfies a suitable additional condition. Notably, this identifies the Fefferman–Graham ambient metric as the unique conformal spacetime with these properties. From a conformal-geometric viewpoint, this result characterizes the ambient metric while relaxing the requirement that the homothety be exact as a one-form.

We then turned to the general setting and analyzed how the conformal Einstein equations constrain the geometry at null infinity without imposing restrictions on the spacetime dimension, on fall-off conditions for the Weyl tensor, or on the topology of $\mathscr{I}$ beyond the existence of a foliation by cross-sections. This analysis led to the identification of a collection of free tensors on $\mathscr{I}$ that completely characterize asymptotically flat spacetimes within this framework. Moreover, we proved that, assuming the obstruction tensors vanish, any such choice of free data gives rise to a smooth asymptotically flat spacetime. We also established a relation between the radiative obstruction tensor in six dimensions and the Fefferman–Graham obstruction tensor, and we expect an analogous relation to hold in higher even dimensions, as we explicitly conjecture in Conjecture 6.33.

Several natural directions for future research emerge from these results. Concerning the Fefferman–Graham construction, it would be of interest to analyze the associated conformal freedom from a purely conformal perspective, namely to characterize which pairs of data $(\omega^2 h, \Theta')$ give rise to the same ambient metric as $(h, \Theta)$. This analysis could shed light on the general (and complicated) question of when two asymptotically flat spacetimes in arbitrary dimension with conformally related universal structures are equivalent, or in other words, how radiation behaves under conformal transformations. One possible approach to this problem is through tractor calculus [90], which is naturally adapted to conformally invariant structures and is closely related with the ambient metric construction itself. Reformulating the ideas of Chapter 6 within the tractor framework could therefore provide a more intrinsic understanding of the conformal freedom in the Fefferman–Graham construction. Such a reformulation would also facilitate a direct comparison with existing tractor-based treatments of null infinity, for instance those in [160], and may help clarify how radiation data transforms under conformal rescalings. Another interesting problem is to understand the compatibility between the existence of Fefferman–Graham ambient metrics in the elliptic region established in [152] and their existence in the hyperbolic region proven in [285], in particular determining under which conditions the corresponding solutions can be matched across the horizon.

With regard to the general framework developed in Chapter 6, an important open question is the physical interpretation of the free tensors $\mathfrak{m}$ and $\boldsymbol{\beta}$ appearing in Theorem 6.16, and their relation to higher-dimensional notions of Bondi mass and angular momentum, as developed in [145, 166, 170, 177, 312]. Addressing this problem will require the formulation of conformally covariant definitions of these quantities, together with appropriate higher-dimensional generalizations of the news tensor and of Geroch's $\rho$-tensor. We also plan to investigate the radiative obstruction tensor in arbitrary dimension and conformal gauge, possibly using tools such as `xAct` or `Sage` [93, 295], as well as its relation to the Fefferman–Graham obstruction tensor, hopefully turning Conjecture 6.33 into a theorem. As suggested above, one interesting approach to all these questions is through tractor calculus. In this context, it would be particularly interesting to connect our results with those of [280, 320] and related works, where logarithmic terms in the asymptotic expansions at infinity are permitted.

Further open problems include extending the notion of double null data introduced in Chapter 3 to the asymptotic characteristic problem, incorporating the new free data identified in Chapter 6. It would be interesting to determine whether this detached data set suffices to construct a conformal spacetime solving Einstein's equations in a neighbourhood of $\mathscr{I}$, once suitable existence theorems for the conformal equations are available, either in higher dimensions or in four dimensions with non-spherical topology. Finally, another prom-ising direction is the incorporation of Killing and homothetic initial data, making use of the general identities developed in Chapter 4, and the study of the corresponding asymptotic KID problem within this setting.

# A

# PROPERTIES OF THE CONSTRAINT TENSOR

In this appendix we prove that the tensors $A$ and $B$ as defined in (2.85) and (2.86) transform as

$$\mathcal{G}_{(z,V)} A_{abc} = z(A_{abc} + V^d B_{dabc}), \qquad \mathcal{G}_{(z,V)} B_{abcd} = B_{abcd}. \tag{A.1}$$

These are the expected transformations if one thinks the data as embedded, but we prove this statement in full generality.

**Proposition A.1.** *Let* $\mathcal{D} = \{\mathcal{H}, \boldsymbol{\gamma}, \boldsymbol{\ell}, \ell^{(2)}, \mathbf{Y}\}$ *be null hypersurface data and* $(z, V)$ *gauge parameters. Let* $A$ *and* $B$ *be the tensors defined in* (2.85) *and* (2.86)*, respectively. Then*

1. $\mathcal{G}_{(z,V)}(A) = z(A + i_V B)$,
2. $\mathcal{G}_{(z,V)}(B) = B$.

*Proof.* The statement to be proven is local, so we can suppose the existence of a local cross-section and assume that $\mathcal{D}$ is characteristic hypersurface data. By the composition law $(z, V) = (z, 0) \circ (1, V)$ (see (2.35)) it suffices to prove the result with $(z, 0)$ and $(1, V)$ independently. We start by writing $\mathcal{D}$ in a characteristic gauge (see Def. 2.43), we shall deal later with the general case. Then,

$$A_{bcd} = \ell_a \overline{R}^a{}_{bcd}, \tag{A.2}$$

$$B_{abcd} = \gamma_{af} \overline{R}^f{}_{bcd} - 2\overline{\nabla}_{[c}(\mathrm{K}_{d]b}\ell_a). \tag{A.3}$$

By the transformation laws (2.28), (2.37), (2.75) and (2.78), the expression of $B$ above is insensitive to the transformations $(z, 0)$, and thus we only need to show the invariance of $B$ under transformations of the form $(1, V)$. Denote with a prime the transformed data,

$$B'_{abcd} = \gamma_{af} \overline{R}'^f{}_{bcd} - 2\overline{\nabla}'_{[c}(\mathrm{K}'_{d]b}\ell'_a) + 2\ell'^{(2)}\mathrm{K}'_{a[d}\mathrm{K}'_{c]b}. \tag{A.4}$$

The first term is given by (2.78),

$$\gamma_{af} \overline{R}'^f{}_{bcd} = \gamma_{af} \left( \overline{R}^f{}_{bcd} + 2\overline{\nabla}_{[c}(V^f \mathrm{K}_{d]b}) + 2V^f V^g \mathrm{K}_{g[c}\mathrm{K}_{d]b} \right). \tag{A.5}$$

For the second one we apply (2.28), (2.37), as well as (2.75),

$$2\overline{\nabla}'_{[c}(\mathrm{K}'_{d]b}\ell'_a) = 2\overline{\nabla}_{[c}(\mathrm{K}_{d]b}\ell_a) + 2\overline{\nabla}_{[c}(\mathrm{K}_{d]b}\gamma_{af}V^f)$$
$$+ 2(\ell_a + \gamma_{af}V^f)V^g \mathrm{K}_{b[d}\mathrm{K}_{c]g} + 2\left(\boldsymbol{\ell}(V) + \boldsymbol{\gamma}(V,V)\right)\mathrm{K}_{a[d}\mathrm{K}_{c]b}.$$

Expanding the second term and inserting (2.62),

$$\begin{aligned} 2\overline{\nabla}'_{[c}(\mathrm{K}'_{d]b}\ell'_a) &= 2\overline{\nabla}_{[c}(\mathrm{K}_{d]b}\ell_a) + 2\gamma_{af}\overline{\nabla}_{[c}(V^f\mathrm{K}_{d]b}) \\ &\quad + 2\left(2\boldsymbol{\ell}(V) + \boldsymbol{\gamma}(V,V)\right)\mathrm{K}_{a[d}\mathrm{K}_{c]b} + 2\gamma_{af}V^fV^g\mathrm{K}_{g[c}\mathrm{K}_{d]b}. \end{aligned} \tag{A.6}$$

Introducing (A.5) and (A.6) into (A.4), and taking into account $\ell'^{(2)} = z^2\left(2\boldsymbol{\ell}(V) + \boldsymbol{\gamma}(V,V)\right)$,

$$B'_{abcd} = \gamma_{af}\overline{R}^f{}_{bcd} - 2\overline{\nabla}_{[c}(\mathrm{K}_{d]b}\ell_a), \tag{A.7}$$

so the gauge invariance of $B$ follows. Now we study the transformation of $A$ in the same way. In the primed gauge,

$$A'_{bcd} = \ell'_a\overline{R}'^a{}_{bcd} + 2\ell'^{(2)}\overline{\nabla}'_{[d}\mathrm{K}'_{c]b} + \mathrm{K}'_{b[c}\overline{\nabla}_{d]}\ell'^{(2)}, \tag{A.8}$$

while in a characteristic gauge $A$ takes the form (A.2). From (2.28), (2.78), (2.29), (2.75) and (2.37) it follows $\mathcal{G}_{(z,0)}(A) = zA$. We start focusing on the transformations of the form $(1,V)$. Firstly from (2.62),

$$\overline{\nabla}_a\left(\boldsymbol{\gamma}(V,V)\right) = 2\gamma_{cb}V^c\overline{\nabla}_aV^b - 2\boldsymbol{\ell}(V)\mathrm{K}_{ab}V^b. \tag{A.9}$$

Since $\ell'^{(2)} = 2\boldsymbol{\ell}(V) + \boldsymbol{\gamma}(V,V)$, the second term of (A.8) can be written as

$$2\ell'^{(2)}\overline{\nabla}'_{[d}\mathrm{K}'_{c]b} = 2\left(2\boldsymbol{\ell}(V) + \boldsymbol{\gamma}(V,V)\right)\left(\overline{\nabla}_{[d}\mathrm{K}_{c]b} - V^a\mathrm{K}_{b[d}\mathrm{K}_{c]a}\right),$$

while using (A.9) the third one is

$$\mathrm{K}'_{b[c}\overline{\nabla}'_{d]}\ell'^{(2)} = 2z\mathrm{K}_{b[c}\left(\gamma_{fg}V^f\overline{\nabla}_{d]}V^g - \boldsymbol{\ell}(V)\mathrm{K}_{d]f}V^f\right) + 2z\mathrm{K}_{b[c}\left(V^f\overline{\nabla}_{d]}\ell_f + \ell_f\overline{\nabla}_{d]}V^f\right).$$

For the first term in (A.8) we use (2.78) and (2.28). Putting everything together and simplifying,

$$\begin{aligned} z^{-1}A'_{bcd} &= \ell_a\overline{R}^a{}_{bcd} + \gamma_{af}V^f\overline{R}^a{}_{bcd} + 2\ell_a\overline{\nabla}_{[c}(V^a\mathrm{K}_{d]b}) \\ &\quad + 4\boldsymbol{\ell}(V)\overline{\nabla}_{[d}\mathrm{K}_{c]b} + 2V^f\mathrm{K}_{b[c}\overline{\nabla}_{d]}\ell_f + 2\ell_f\mathrm{K}_{b[c}\overline{\nabla}_{d]}V^f \\ &= \ell_a\overline{R}^a{}_{bcd} + \gamma_{af}V^f\overline{R}^a{}_{bcd} + 2\boldsymbol{\ell}(V)\overline{\nabla}_{[d}\mathrm{K}_{c]b} + 2V^f\mathrm{K}_{b[c}\overline{\nabla}_{d]}\ell_f, \end{aligned}$$

and the claim $z^{-1}A'_{bcd} = A_{bcd} + V^aB_{abcd}$ follows because $V^a\overline{\nabla}_{[d}(\mathrm{K}_{c]b}\ell_a) = \boldsymbol{\ell}(V)\overline{\nabla}_{[d}\mathrm{K}_{c]b} + V^f\mathrm{K}_{b[c}\overline{\nabla}_{d]}\ell_f$. To finish the proof we still need to show that the assumption that the initial gauge is characteristic does not spoil the generality of the argument. Let $\mathcal{D}$ be CHD and $\mathcal{D}'' = \mathcal{G}_{(z'',V'')}\mathcal{D}$. From Proposition 2.42 there always exist gauge parameters $(z,V)$ such that $\mathcal{D}' := \mathcal{G}_{(z,V)}\mathcal{D}$ is in a CG. Let $(z',V')$ be the parameters making the following diagram commutative

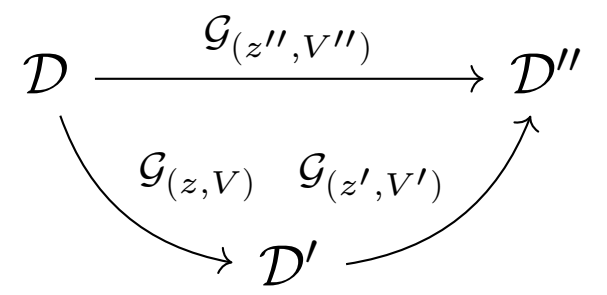

In other words, $(z', V') = (z'', V'') \circ (z, V)^{-1} = (z''z^{-1}, z(V'' - V))$, where we have made use of (2.35) and (2.36). Since $\mathcal{D}'$ is in a CG, and from the fact that a transformations of the form $\mathcal{G}_{(z',0)}$ keeps the gauge characteristic, we can write[1] $\mathcal{G}_{(z',V')}(A') = A'' = z'(A' + i_{V'}B')$ and $\mathcal{G}_{(z',V')}(B') = B'' = B'$, as well as $\mathcal{G}_{(z,V)^{-1}}(A') = A = z^{-1}(A' + i_{-zV}B)$ and $\mathcal{G}_{(z,V)^{-1}}(B') = B = B'$, after using (2.36) again. Then, since $A'' = z'(A' + i_{V'}B)$ it follows

$$A'' = z'' \left( z^{-1}A' + i_{V''}B' - i_V B' \right) = z'' \left( A + i_V B' + i_{V''}B' - i_V B' \right) = z'' \left( A + i_{V''}B \right),$$

and hence $\mathcal{G}_{(z'',V'')}(A) = z'' \left( A + i_{V''}B \right)$. The fact that $\mathcal{G}_{(z'',V'')}(B) = B$ is obvious. □

Next we study the symmetries of $A$ and $B$. When the data is embedded, these symmetries are the ones inherited from the curvature tensor of the ambient space. Here we establish them without assuming embeddedness of the data.

**Proposition A.2.** *The tensors $A$ and $B$ possess the following symmetries:*

1. $A_{bcd} = -A_{bdc}$,
2. $A_{bcd} + A_{cdb} + A_{dbc} = 0$,
3. $B_{abcd} + B_{acdb} + B_{adbc} = 0$,
4. $B_{abcd} = -B_{abdc} = -B_{bacd}$,
5. $B_{abcd} = B_{cdab}$.

*Proof.* The symmetries $A_{bcd} = -A_{bdc}$ and $B_{abcd} = -B_{abdc}$ are obvious, so item 1. and the first equality in item 4. are proven. Item 2. is a consequence of the first Bianchi identity for $\overline{R}$ and the fact that $\mathbf{K}$ is symmetric. Item 3. is analogous. In order to prove the second equality in item 4., namely $B_{abcd} + B_{bacd} = 0$, we first compute the symmetrization of the first term in (2.86) which is

$$\begin{aligned}
\gamma_{af}\overline{R}^f{}_{bcd} + \gamma_{bf}\overline{R}^f{}_{acd} &= \overline{\nabla}_d \overline{\nabla}_c \gamma_{ab} - \overline{\nabla}_c \overline{\nabla}_d \gamma_{ab} \\
&\overset{(2.62)}{=} -\overline{\nabla}_d \left( \mathrm{K}_{ca}\ell_b + \mathrm{K}_{cb}\ell_a \right) + \overline{\nabla}_c \left( \mathrm{K}_{da}\ell_b + \mathrm{K}_{db}\ell_a \right) \\
&= 2\overline{\nabla}_{[c}(\mathrm{K}_{d]a}\ell_b) + 2\overline{\nabla}_{[c}(\mathrm{K}_{d]b}\ell_a).
\end{aligned}$$

For any symmetric tensor it holds $\mathrm{K}_{b[d}\mathrm{K}_{c]a} + \mathrm{K}_{a[d}\mathrm{K}_{c]b} = 0$, so the symmetrization relative to the indices $(a, b)$ in the third term in (2.86) vanishes. Thus,

$$B_{abcd} + B_{bacd} = 2\overline{\nabla}_{[c}(\mathrm{K}_{d]a}\ell_b) + 2\overline{\nabla}_{[c}(\mathrm{K}_{d]b}\ell_a) - 2\overline{\nabla}_{[c}(\mathrm{K}_{d]b}\ell_a) - 2\overline{\nabla}_{[c}(\mathrm{K}_{d]a}\ell_b) = 0.$$

1 Actually, this transformation must be understood as first acting with $(z', 0)$ and then with $(1, V')$, as otherwise the intermediate gauge would not be characteristic and we could not apply the gauge transformations obtained in the first part of the proof. This choice is possible because $(z', V') = (1, V) \circ (z', 0)$, see (2.35)

Item 5. is consequence of items 3. and 4. Indeed, let $\mathring{B}_{abcd}$ be the cyclic permutation $\mathring{B}_{abcd} := B_{abcd} + B_{acdb} + B_{adbc}$, which by the third item vanishes. Then, taking into account item (4),

$$0 = \mathring{B}_{abcd} - \mathring{B}_{bcda} - \mathring{B}_{cdab} + \mathring{B}_{dcab} = B_{abcd} - B_{bacd} - B_{cdab} + B_{dcab} = 2B_{abcd} - 2B_{cdab}. \qquad \square$$

## A.1 SOME CONTRACTIONS OF THE TENSORS $A$ AND $B$

In Lemma A.4 below we compute the contractions of the tensors $A$ and $B$ that are needed in Section 2.4 to write down the constraint tensors in terms of the foliation tensors. Before we prove an intermediate result.

**Lemma A.3.** *Let $\mathcal{D} = \{\mathcal{H}, \boldsymbol{\gamma}, \boldsymbol{\ell}, \ell^{(2)}, \mathbf{Y}\}$ be null hypersurface data. Then,*

$$\ell_a \overline{R}^a{}_{bcd} = 2\overline{\nabla}_{[d}\Pi_{c]b} + 2\mathrm{K}_{b[d}\overline{\nabla}_{c]}\ell^{(2)} + 2\ell^{(2)}\overline{\nabla}_{[c}\mathrm{K}_{d]b} \tag{A.10}$$

$$\begin{aligned} n^c\ell_a \overline{R}^a{}_{bcd} = {} & \overline{\nabla}_d(\Pi_{cb}n^c) - \pounds_n \mathrm{Y}_{db} - (ds)_{db} + P^{ca}\mathrm{K}_{ab}\Pi_{dc} - \Pi_{dc}\Pi_{ba}n^c n^a \\ & + n(\ell^{(2)})\mathrm{K}_{db} + \ell^{(2)}\Big(\pounds_n \mathrm{K}_{db} - P^{ac}\mathrm{K}_{dc}\mathrm{K}_{ab}\Big). \end{aligned} \tag{A.11}$$

*Proof.* The first equation follows from the Ricci identity applied to $\ell_a$ and the fact that $\overline{\nabla}_a \ell_b = \Pi_{ab} - \ell^{(2)}\mathrm{K}_{ab}$ (see (2.63)),

$$\ell_a \overline{R}^a{}_{bcd} = \overline{\nabla}_d\overline{\nabla}_c\ell_b - \overline{\nabla}_c\overline{\nabla}_d\ell_b = 2\overline{\nabla}_{[d}\Pi_{c]b} + 2\mathrm{K}_{b[d}\overline{\nabla}_{c]}\ell^{(2)} + 2\ell^{(2)}\overline{\nabla}_{[c}\mathrm{K}_{d]b}.$$

For the second, we "integrate" by parts equation (A.10) and use $\mathbf{K}(n,\cdot) = 0$ to find

$$n^c \boldsymbol{\ell}_a \overline{R}^a{}_{bcd} = \overline{\nabla}_d\left(\Pi_{cb}n^c\right) - \Pi_{cb}\overline{\nabla}_d n^c - n^c\overline{\nabla}_c\Pi_{db} + \mathrm{K}_{db}\overline{\nabla}_n\ell^{(2)} + \ell^{(2)}\left(n^c\overline{\nabla}_c\mathrm{K}_{db} + \mathrm{K}_{cb}\overline{\nabla}_d n^c\right).$$

Inserting $\overline{\nabla}_d n^c = P^{ca}\mathrm{K}_{ad} - \Pi_{da}n^c n^a$ (see (2.67)) one obtains

$$\begin{aligned} n^c\ell_a \overline{R}^a{}_{bcd} = {} & \overline{\nabla}_d\left(\Pi_{cb}n^c\right) - n^c\overline{\nabla}_c\Pi_{db} - P^{ca}\mathrm{K}_{ad}\Pi_{cb} + \Pi_{cb}\Pi_{da}n^c n^a + n(\ell^{(2)})\mathrm{K}_{db} \\ & + \ell^{(2)}\left(n^c\overline{\nabla}_c\mathrm{K}_{db} + P^{ca}\mathrm{K}_{cb}\mathrm{K}_{ad}\right). \end{aligned} \tag{A.12}$$

Finally we express the derivatives $\overline{\nabla}_n$ in terms of Lie derivatives along $n$ using

$$\begin{aligned} n^c\overline{\nabla}_c T_{db} &= \pounds_n T_{db} - T_{cb}\overline{\nabla}_d n^c - T_{dc}\overline{\nabla}_b n^c \\ &\overset{(2.67)}{=} \pounds_n T_{db} - P^{ac}T_{cb}\mathrm{K}_{ad} - P^{ac}T_{dc}\mathrm{K}_{ab} + T_{cb}\Pi_{da}n^c n^a + T_{dc}\Pi_{ba}n^c n^a, \end{aligned}$$

valid for any tensor $T_{ab}$. Particularizing it for $T_{db} = \Pi_{db}$ first, and $T_{db} = \mathrm{K}_{db}$ second, it follows

$$n^c\overline{\nabla}_c\Pi_{db} + P^{ac}\mathrm{K}_{ad}\Pi_{cb} - \Pi_{cb}\Pi_{da}n^c n^a = \pounds_n\Pi_{db} - P^{ac}\Pi_{dc}\mathrm{K}_{ab} + \Pi_{dc}\Pi_{ba}n^c n^a$$

and (recall $\mathbf{K}(n,\cdot) = 0$)

$$n^c\overline{\nabla}_c\mathrm{K}_{db} + P^{ac}\mathrm{K}_{ad}\mathrm{K}_{cb} = \pounds_n\mathrm{K}_{db} - P^{ac}\mathrm{K}_{dc}\mathrm{K}_{ab}.$$

Inserting both into (A.12) and using Cartan formula applied to $\mathbf{F}$, namely (recall $\mathbf{s} := \mathbf{F}(n,\cdot)$ and $2d\mathbf{F} = d^2\boldsymbol{\ell} = 0$)

$$\pounds_n \mathbf{F} = d\iota_n \mathbf{F} + \iota_n d\mathbf{F} = d\iota_n \mathbf{F} = d\mathbf{s},$$

identity (A.11) follows. □

**Lemma A.4.** *Let $\mathcal{D}$ be null hypersurface data, $\mathcal{S}$ a non-degenerate submanifold, $\{e_A\}$ a basis of $T\mathcal{S}$ and let $A$ and $B$ the tensors defined in* (2.85) *and* (2.86). *Then,*

$$A_{bcd}n^b n^d = \pounds_n(\Pi_{cb}n^b) + \overline{\nabla}_c\kappa + P^{ab}\mathrm{K}_{cb}\Pi_{da}n^d, \tag{A.13}$$

$$B_{abcd}n^a P^{bd} = -(\mathrm{tr}_P\,\mathbf{K})\Pi_{cd}n^d + P^{bd}\mathrm{K}_{bc}\Pi_{da}n^a - \overline{\nabla}_c(\mathrm{tr}_P\,\mathbf{K}) + P^{ab}\overline{\nabla}_a\mathrm{K}_{bc}. \tag{A.14}$$

*Moreover, in a gauge in which $\ell^{(2)} = 0$ and $\boldsymbol{\ell}_{\parallel} = 0$ at a section $\mathcal{S} \hookrightarrow \mathcal{H}$ (which exists by Lemma 2.31),*

$$B_{cadb}P^{cd}e_A^a e_B^b \stackrel{\mathcal{S}}{=} R^h_{AB} - (\mathrm{tr}_h\,\mathbf{K}_{\parallel})\mathrm{Y}_{AB} - (\mathrm{tr}_h\,\mathbf{Y}_{\parallel})\mathrm{K}_{AB} + 2h^{CD}\mathrm{K}_{C(A}\mathrm{Y}_{B)D}, \tag{A.15}$$

$$\begin{aligned} A_{bca}n^c e_A^a e_B^b \stackrel{\mathcal{S}}{=} & -\pounds_n \mathrm{Y}_{AB} - \kappa \mathrm{Y}_{AB} + h^{CD}\Pi_{AC}\mathrm{K}_{BD} + \nabla^h_A(\mathrm{r}+\mathrm{s})_B \\ & - (\mathrm{r}-\mathrm{s})_A(\mathrm{r}-\mathrm{s})_B - (d\mathrm{s})_{AB} + \frac{1}{2}n(\ell^{(2)})\mathrm{K}_{AB}. \end{aligned} \tag{A.16}$$

*Proof.* Identity (A.13) follows from the expression of $A$ in (2.85) and Lemma A.3. Indeed, using $\mathbf{K}(n,\cdot) = 0$,

$$\begin{aligned} A_{bcd}n^b n^d &= \ell_a \overline{R}^a{}_{bcd}n^b n^d + \ell^{(2)} n^b n^d \overline{\nabla}_d \mathrm{K}_{cb} \\ &= n^b n^d \overline{\nabla}_d \Pi_{cb} - n^b n^d \overline{\nabla}_c \Pi_{db} \\ &= n^d \overline{\nabla}_d(\Pi_{cb}n^b) - \Pi_{cb}n^d\overline{\nabla}_d n^b - \overline{\nabla}_c(\mathbf{\Pi}(n,n)) + \Pi_{db}n^d\overline{\nabla}_c n^b + \Pi_{db}n^b\overline{\nabla}_c n^d \\ &= \pounds_n(\Pi_{cb}n^b) + \overline{\nabla}_c\kappa + P^{ab}\mathrm{K}_{ca}\Pi_{db}n^d, \end{aligned}$$

where in the fourth equality we used $n^d\overline{\nabla}_d(\Pi_{cb}n^b) = \pounds_n(\Pi_{cb}n^b) - \Pi_{db}n^b\overline{\nabla}_c n^d$ and (2.67). This establishes (A.13). Now, from (2.86) and taking into account $\gamma_{af}\overline{R}^f{}_{bcd}n^a = 0$ and $\mathrm{K}_{ab}n^a = 0$,

$$B_{abcd}n^a = 2\overline{\nabla}_{[d}\mathrm{K}_{c]b} - 2\ell_a \mathrm{K}_{b[c}\overline{\nabla}_{d]}n^a = 2\overline{\nabla}_{[d}\mathrm{K}_{c]b} + 2\mathrm{K}_{b[c}\Pi_{d]a}n^a,$$

where in both equalities we used $\boldsymbol{\ell}(n) = 1$ and in the second we inserted (2.67) and used $\ell_a P^{ab}\mathrm{K}_{bc} = -\ell^{(2)}\mathrm{K}_{bc}n^b = 0$ (see (2.5)). Contracting with $P^{bd}$ and using $\mathrm{K}_{bc}\overline{\nabla}_a P^{bc} = 0$, which is a direct consequence of (2.65) and $\mathbf{K}(n,\cdot) = 0$,

$$B_{abcd}n^a P^{bd} = P^{bd}\overline{\nabla}_d \mathrm{K}_{bc} - \overline{\nabla}_c(P^{bd}\mathrm{K}_{bd}) + P^{bd}\mathrm{K}_{bc}\Pi_{da}n^a - P^{bd}\mathrm{K}_{bd}\Pi_{cd}n^d,$$

and hence (A.14) follows.

Next we proceed with (A.15). From (2.86) and taking into account that $\ell^{(2)} \stackrel{\mathcal{S}}{=} 0$, $\ell^\sharp = 0$ and (2.141),

$$B_{acbd}P^{cd}e_A^a e_B^b = \left(\gamma_{cf}\overline{R}^f{}_{adb} + 2\overline{\nabla}_{[b}(\mathrm{K}_{d]a}\ell_c)\right) h^{CD} e_C^c e_D^d e_A^a e_B^b.$$

For the first term we employ Gauss identity (2.159),

$$\begin{aligned}\boldsymbol{\gamma}_{cf}\overline{R}^{f}{}_{adb}h^{CD}e_C^c e_D^d e_A^a e_B^b &= \boldsymbol{\gamma}\left(e_C, \overline{R}(e_D, e_B)e_A\right)h^{CD}\\ &= R^h_{AB} - (\operatorname{tr}_h \mathbf{K}_{\parallel})\mathrm{Y}_{AB} + h^{CD}\mathrm{Y}_{DA}\mathrm{K}_{BC},\end{aligned}$$

where $R^h_{AB}$ is the Ricci tensor of $h$. For the second term we use (2.63), which in this gauge is $\overline{\nabla}_a \ell_b \stackrel{\mathcal{S}}{=} \Pi_{ab}$, and the fact that $\Pi_{AB} \stackrel{\mathcal{S}}{=} \mathrm{Y}_{AB}$ (because $\boldsymbol{\ell}_{\parallel} = 0$ and thus $2i^{\star}\mathbf{F} = i^{\star}d\boldsymbol{\ell} = d\boldsymbol{\ell}_{\parallel} = 0$, see Remark 2.45). Then,

$$2h^{CD}e_C^c e_D^d e_A^a e_B^b \overline{\nabla}_{[b}(\mathrm{K}_{d]a}\ell_c) = 2h^{CD}e_C^c e_D^d e_A^a e_B^b \mathrm{K}_{a[d}\overline{\nabla}_{b]}\ell_c = -2h^{CD}\mathrm{K}_{A[B}\mathrm{Y}_{D]C},$$

where the first equality holds because $\boldsymbol{\ell}_{\parallel} = 0$. Combining the two terms, (A.15) follows. Finally we compute $A_{bca}n^c e_A^a e_B^b$. In the present gauge the quantity $A_{bca}n^c$ is simply (note that $\ell^{(2)}$ is not assumed to be zero off $\mathcal{S}$)

$$A_{bca}n^c = n^c\left(\ell_d \overline{R}^{d}{}_{bca} + \mathrm{K}_{b[c}\overline{\nabla}_{a]}\ell^{(2)}\right) = n^c \ell_d \overline{R}^{d}{}_{bca} - \frac{1}{2}n(\ell^{(2)})\mathrm{K}_{ba}, \tag{A.17}$$

where $\mathbf{K}(n,\cdot) = 0$ has been used. Inserting (A.11),

$$A_{bca}n^c = \overline{\nabla}_a(\Pi_{cb}n^c) - \pounds_n \mathrm{Y}_{ab} - (d\mathrm{s})_{ab} + P^{cd}\mathrm{K}_{db}\Pi_{ac} - \Pi_{ac}\Pi_{bd}n^c n^d + \frac{1}{2}n(\ell^{(2)})\mathrm{K}_{ab}.$$

Equation (A.16) follows after contracting with $e_A^a e_B^b$ and using (2.141), $\mathbf{\Pi}(n,\cdot) = \mathbf{r} + \mathbf{s}$, $\mathbf{\Pi}(\cdot, n) = \mathbf{r} - \mathbf{s}$ and that by decomposition (2.130) with $Q_{AB} \stackrel{\mathcal{S}}{=} \mathrm{Y}_{AB}$ (see Rmk. 2.45) one has

$$\left(\overline{\nabla}(\mathbf{r}+\mathbf{s})\right)_{AB} = \nabla^h_A(\mathrm{r}+\mathrm{s})_B + \mathbf{r}(n)\mathrm{Y}_{AB} = \nabla^h_A(\mathrm{r}+\mathrm{s})_B - \kappa \mathrm{Y}_{AB}.$$

□

# B

# PULLBACKS INTO ABSTRACT HYPERSURFACES

In this appendix we compute the pullback of several derivatives of ambient tensors into an embedded hypersurface that we shall need in the main text of this thesis. Given a $(0,p)$ tensor field $T_{\alpha_1\cdots\alpha_p}$ on $\mathcal{M}$ we use the standard notation $T_{a_1\cdots a_p}$ to denote the pullback of $T$ to $\mathcal{H}$. Moreover, we also use the notation ${}^{(i)}T_{\alpha_1\cdots\alpha_{p-1}}$ for $\xi^\mu T_{\alpha_1\cdots\alpha_{i-1}\mu\alpha_i\cdots\alpha_{p-1}}$ and ${}^{(i)}T_{a_1\cdots a_{p-1}}$ for the pullback of ${}^{(i)}T_{\alpha_1\cdots\alpha_{p-1}}$ to $\mathcal{H}$. In addition, we use ${}^{(i,j)}T_{a_1\cdots a_{p-2}}$ for the pullback to $\mathcal{H}$ of the tensor obtained by first the contraction of $T$ with $\xi$ in the $j$-th slot and then in the $i$-th slot of the resulting $(0,p-1)$ tensor, i.e. ${}^{(i,j)}T = {}^{(i)}({}^{(j)}T)$. This notation requires care with the order of indices. For instance, note that ${}^{(1,1)}T = {}^{(1,2)}T$ and ${}^{(1,3)}T = {}^{(2,1)}T \neq {}^{(1,2)}T$. The rest of the notation is as in Chapter 2. In particular, the tensor $V^b{}_a$ is defined in (2.54).

**Proposition B.1.** *Let $(\mathcal{M},g)$ be a semi-Riemannian manifold and $\Phi : \mathcal{H} \hookrightarrow \mathcal{M}$ a smooth hypersurface with rigging $\xi$. Let $T$ be a $(0,p)$-tensor on $\mathcal{M}$. Then,*

$$
\begin{aligned}
(\nabla T)_{ba_1\cdots a_p} &= \overset{\circ}{\nabla}_b T_{a_1\cdots a_p} + \sum_{i=1}^{p} \mathrm{Y}_{ba_i} T_{a_1\cdots a_{i-1}ca_{i+1}\cdots a_p} n^c \\
&\quad + \sum_{i=1}^{p} \Big(\mathrm{U}_{ba_i} + n^{(2)}\mathrm{Y}_{ba_i}\Big) ({}^{(i)}T)_{a_1\cdots a_{i-1}a_{i+1}\cdots a_p}, \qquad \text{(B.1)}\\
({}^{(1)}\nabla T)_{a_1\cdots a_p} &= (\pounds_\xi T)_{a_1\cdots a_p} - \sum_{i=1}^{p} \left(\mathrm{r}_{a_i} - \mathrm{s}_{a_i} + \frac{1}{2} n^{(2)} \overset{\circ}{\nabla}_{a_i} \ell^{(2)}\right) ({}^{(i)}T)_{a_1\cdots a_{i-1}a_{i+1}\cdots a_p} \\
&\quad - \sum_{i=1}^{p} V^c{}_{a_i} T_{a_1\cdots a_{i-1}ca_{i+1}\cdots a_p}, \qquad \text{(B.2)}
\end{aligned}
$$

$$
\begin{aligned}
({}^{(j+1)}\nabla T)_{ba_1\cdots a_{p-1}} &= \overset{\circ}{\nabla}_b ({}^{(j)}T)_{a_1\cdots a_{p-1}} + \sum_{i=1}^{p-1} \mathrm{Y}_{ba_i} ({}^{(j)}T)_{a_1\cdots a_{i-1}ca_{i+1}\cdots a_{p-1}} n^c \\
&\quad + \sum_{i=1}^{p-1} \Big(\mathrm{U}_{ba_i} + n^{(2)}\mathrm{Y}_{ba_i}\Big) ({}^{(i,j)}T)_{a_1\cdots a_{i-1}a_{i+1}\cdots a_{p-1}} \\
&\quad - \left(\mathrm{r}_b - \mathrm{s}_b + \frac{1}{2} n^{(2)} \overset{\circ}{\nabla}_b \ell^{(2)}\right) ({}^{(j)}T)_{a_1\cdots a_{p-1}} - V^c{}_b T_{a_1\cdots a_{j-1}ca_j\cdots a_{p-1}}. \qquad \text{(B.3)}
\end{aligned}
$$

*Proof.* Let $\{e_a\}$ be a local basis of $\mathcal{H}$ and $\widehat{e}_a := \Phi_\star(e_a)$. To prove the first identity we contract $\nabla_\beta T_{\alpha_1\cdots\alpha_p}$ with $\widehat{e}_b^\beta \widehat{e}_{a_1}^{\alpha_1} \cdots \widehat{e}_{a_p}^{\alpha_p}$ and use (2.41), namely

$$
\nabla_{\widehat{e}_b^\beta} \widehat{e}_a^\alpha = (\overset{\circ}{\nabla}_{e_b} e_a)^c \widehat{e}_c^\alpha - \mathrm{Y}_{ba} n^c \widehat{e}_c^\alpha - (\mathrm{U}_{ba} + n^{(2)}\mathrm{Y}_{ba})\xi^\alpha.
$$

For the second one we use the relation between $\pounds_\xi$ and $\nabla$

$$\xi^\mu \nabla_\mu T_{\alpha_1\cdots\alpha_p} = \pounds_\xi T_{\alpha_1\cdots\alpha_p} - \sum_{i=1}^{p} T_{\alpha_1\cdots\alpha_{i-1}\mu\alpha_{i+1}\cdots\alpha_p}\nabla_{\alpha_i}\xi^\mu, \tag{B.4}$$

contract it with $\widehat{e}^{\alpha_1}_{a_1}\cdots\widehat{e}^{\alpha_p}_{a_p}$ and employ (2.53). To prove the third one just contract the relation

$$\xi^\mu \nabla_\beta T_{\mu\alpha_1\cdots\alpha_{p-1}} = \nabla_\beta ({}^{(1)}T)_{\alpha_1\cdots\alpha_{p-1}} - T_{\mu\alpha_1\cdots\alpha_{p-1}}\nabla_\beta \xi^\mu.$$

with $\widehat{e}^{\beta}_{b}\widehat{e}^{\alpha_1}_{a_1}\cdots\widehat{e}^{\alpha_{p-1}}_{a_{p-1}}$ and use (B.1) and (2.53). □

In addition, we shall also need to relate contractions of $\nabla T$ with two or more $\xi$ with either Lie derivatives or with pullbacked quantities in $\mathcal{H}$.

**Proposition B.2.** *Let $(\mathcal{M},g)$ be a semi-Riemannian manifold, $\Phi:\mathcal{H}\hookrightarrow\mathcal{M}$ a smooth embedded hypersurface with rigging $\xi$ and let $T$ be a $(0,p)$-tensor on $\mathcal{M}$. Then,*

$$\begin{aligned} e^\mu_b \xi^{\alpha_1}\cdots\xi^{\alpha_p}\nabla_\mu T_{\alpha_1\cdots\alpha_p} \stackrel{\mathcal{H}}{=}{}& \overset{\circ}{\nabla}_b(T(\xi,\cdots,\xi)) - pT(\xi,\cdots,\xi)\left(\mathrm{r}_b - \mathrm{s}_b + \frac{1}{2}n^{(2)}\overset{\circ}{\nabla}_b \ell^{(2)}\right) \\ & - \sum_{i=1}^{p} V^c{}_b\left({}^{(1,\dots,i-1,i+1,\dots p)}T\right)_c, \end{aligned} \tag{B.5}$$

$$\xi^\mu\xi^{\alpha_1}\cdots\xi^{\alpha_p}\nabla_\mu T_{\alpha_1\cdots\alpha_p} \stackrel{\mathcal{H}}{=} (\pounds_\xi T)(\xi,\cdots,\xi) - p\beta T(\xi,\cdots,\xi) - \sum_{i=1}^{p} a^c_\parallel\left({}^{(1,\dots,i-1,i+1,\dots,p)}T\right)_c, \tag{B.6}$$

$$\begin{aligned} \xi^\mu\xi^\nu e^{\alpha_1}_{a_1}\cdots e^{\alpha_{p-1}}_{a_{p-1}}\nabla_\mu T_{\nu\alpha_1\cdots\alpha_{p-1}} \stackrel{\mathcal{H}}{=}{}& (\pounds_\xi({}^{(1)}T))_{a_1\cdots a_{p-1}} - \beta\,({}^{(1)}T)_{a_1\cdots a_{p-1}} - a^b_\parallel T_{ba_1\cdots a_{p-1}} \\ & - \sum_{i=1}^{p-1} V^b{}_{a_i}({}^{(1)}T)_{a_1\cdots a_{i-1}ba_{i+1}\cdots a_{p-1}} \\ & - \sum_{i=1}^{p-1}\left(\mathrm{r}_{a_i} - \mathrm{s}_{a_i} + \frac{1}{2}n^{(2)}\overset{\circ}{\nabla}_{a_i}\ell^{(2)}\right)({}^{(i,1)}T)_{a_1\cdots a_{i-1}a_{i+1}\cdots a_{p-1}}, \end{aligned} \tag{B.7}$$

*where $\nabla_\xi\xi \stackrel{\mathcal{H}}{=} \beta\xi + \Phi_\star a_\parallel$.*

*Proof.* The first identity follows at once from the expression of $\nabla_{e_c}\xi$ in (2.53), and the second and third identities are a consequence of (B.4). □

Next we compute the pullback of the divergence of a tensor. For our needs, we restrict the proof to the null case.

**Proposition B.3.** *Let $(\mathcal{M}, g)$ be a semi-Riemannian manifold and $\Phi : \mathcal{H} \hookrightarrow \mathcal{M}$ a smooth null hypersurface with rigging $\xi$. Let $T$ be a $(0, p+1)$-tensor on $\mathcal{M}$ and denote by $\operatorname{div} T$ the $p$-covariant tensor defined by $(\operatorname{div} T)_{\alpha_1 \cdots \alpha_p} := g^{\mu\nu} \nabla_\mu T_{\nu \alpha_1 \cdots \alpha_p}$. Then,*

$$\begin{aligned}
\Phi^\star (\operatorname{div} T)_{a_1 \cdots a_p} &= P^{bc} \overset{\circ}{\nabla}_b T_{c a_1 \cdots a_p} + n^b (\pounds_\xi T)_{b a_1 \cdots a_p} + n^c \overset{\circ}{\nabla}_c ({}^{(1)}T)_{a_1 \cdots a_p} + (2\kappa + \operatorname{tr}_P \mathbf{U}) ({}^{(1)}T)_{a_1 \cdots a_p} \\
&\quad + (\operatorname{tr}_P \mathbf{Y} - n(\ell^{(2)})) n^c T_{c a_1 \cdots a_p} - 2 P^{ac} (\mathrm{r} + \mathrm{s})_a T_{c a_1 \cdots a_p} \\
&\quad + \sum_{i=1}^p P^{bc} \mathrm{Y}_{b a_i} T_{c a_1 \cdots a_{i-1} d a_{i+1} \cdots a_p} n^d + \sum_{i=1}^p P^{bc} \mathrm{U}_{b a_i} ({}^{(i+1)}T)_{c a_1 \cdots a_{i-1} a_{i+1} \cdots a_p} \\
&\quad - \sum_{i=1}^p (\mathrm{r} - \mathrm{s})_{a_i} n^b \, ({}^{(i+1)}T)_{b a_1 \cdots a_{i-1} a_{i+1} \cdots a_p} - \sum_{i=1}^p V^c{}_{a_i} n^b T_{b a_1 \cdots a_{i-1} c a_{i+1} \cdots a_p} \\
&\quad + \sum_{i=1}^p \mathrm{r}_{a_i} ({}^{(1)}T)_{a_1 \cdots a_{i-1} c a_{i+1} \cdots a_p} n^c .
\end{aligned}$$

*Proof.* From (2.14),

$$\begin{aligned}
\widehat{e}^{\alpha_1}_{a_1} \cdots \widehat{e}^{\alpha_p}_{a_p} \nabla_\mu T^\mu{}_{\alpha_1 \cdots \alpha_p} &= \widehat{e}^{\alpha_1}_{a_1} \cdots \widehat{e}^{\alpha_p}_{a_p} g^{\mu\nu} \nabla_\mu T_{\nu \alpha_1 \cdots \alpha_p} \\
&= \widehat{e}^{\alpha_1}_{a_1} \cdots \widehat{e}^{\alpha_p}_{a_p} \left( P^{bc} \widehat{e}^\mu_b \widehat{e}^\nu_c + \xi^\mu n^b \widehat{e}^\nu_b + \xi^\nu n^b \widehat{e}^\mu_b \right) \nabla_\mu T_{\nu \alpha_1 \cdots \alpha_p}.
\end{aligned}$$

Using (B.1), the first term is

$$\begin{aligned}
P^{bc} \overset{\circ}{\nabla}_b T_{c a_1 \cdots a_p} + (\operatorname{tr}_P \mathbf{Y}) T_{d a_1 \cdots a_p} n^d + \sum_{i=1}^p P^{bc} \mathrm{Y}_{b a_i} T_{c a_1 \cdots a_{i-1} d a_{i+1} \cdots a_p} n^d \\
+ (\operatorname{tr}_P \mathbf{U}) ({}^{(1)}T)_{a_1 \cdots a_p} + \sum_{i=1}^p P^{bc} \mathrm{U}_{b a_i} ({}^{(i+1)}T)_{c a_1 \cdots a_{i-1} a_{i+1} \cdots a_p}.
\end{aligned}$$

From (B.2), $V^c{}_b n^b = P^{ac} (\mathrm{r} + \mathrm{s})_a + \frac{1}{2} n(\ell^{(2)}) n^c$ (see (2.57)) and $\mathbf{r}(n) = -\kappa$, $\mathbf{s}(n) = 0$ the second term becomes

$$\begin{aligned}
n^b (\pounds_\xi T)_{b a_1 \cdots a_p} + \kappa ({}^{(1)}T)_{a_1 \cdots a_p} - \sum_{i=1}^p (\mathrm{r} - \mathrm{s})_{a_i} n^b \, ({}^{(i+1)}T)_{b a_1 \cdots a_{i-1} a_{i+1} \cdots a_p} \\
- \left( P^{ac} (\mathrm{r} + \mathrm{s})_a + \frac{1}{2} n(\ell^{(2)}) n^c \right) T_{c a_1 \cdots a_p} - \sum_{i=1}^p V^c{}_{a_i} n^b T_{b a_1 \cdots a_{i-1} c a_{i+1} \cdots a_p}.
\end{aligned}$$

Finally, using (B.3) and $\mathbf{U}(n, \cdot) = 0$ the last term is

$$\begin{aligned}
n^c \overset{\circ}{\nabla}_c ({}^{(1)}T)_{a_1 \cdots a_p} + \sum_{i=1}^p \mathrm{r}_{a_i} ({}^{(1)}T)_{a_1 \cdots a_{i-1} c a_{i+1} \cdots a_p} n^c + \kappa ({}^{(1)}T)_{a_1 \cdots a_p} \\
- \left( P^{ac} (\mathrm{r} + \mathrm{s})_a + \frac{1}{2} n(\ell^{(2)}) n^c \right) T_{c a_1 \cdots a_p}.
\end{aligned}$$

Adding up the three terms the result follows. □

All the possible contractions of the ambient Hessian are computed next.

**Proposition B.4.** *Let $(\mathcal{M}, g)$ be a semi-Riemannian manifold and $\Phi : \mathcal{H} \hookrightarrow \mathcal{M}$ a smooth hypersurface with rigging $\xi$. Let $f$ be function on $\mathcal{M}$. Then,*

$$\Phi^\star \left(\operatorname{Hess} f\right)_{ab} = \overset{\circ}{\nabla}_a \overset{\circ}{\nabla}_b f + \pounds_n(f) \mathrm{Y}_{ab} + \pounds_\xi(f)\Big(\mathrm{U}_{ab} + n^{(2)}\mathrm{Y}_{ab}\Big), \tag{B.8}$$

$$\Phi^\star\Big((\operatorname{Hess} f)(\xi,\cdot)\Big)_a = \overset{\circ}{\nabla}_a\Big(\pounds_\xi(f)\Big) - \pounds_\xi(f)\left(\mathrm{r}_a - \mathrm{s}_a + \frac{1}{2}n^{(2)}\overset{\circ}{\nabla}_a \ell^{(2)}\right) - V^b{}_a \overset{\circ}{\nabla}_b f, \tag{B.9}$$

$$\Phi^\star\Big(\Big(\operatorname{Hess} f\Big)(\xi,\xi)\Big) = \pounds^{(2)}_\xi(f) - \beta \pounds_\xi(f) - \pounds_{a_\parallel}(f), \tag{B.10}$$

*where $\nabla_\xi \xi \overset{\mathcal{H}}{=} \beta\xi + \Phi_\star a_\parallel$. As a consequence,*

$$\begin{aligned}\Box_g f \overset{\mathcal{H}}{=} \overset{\circ}{\Box} f &+ (\operatorname{tr}_P \mathbf{Y} - n(\ell^{(2)}))\pounds_n(f) + (\operatorname{tr}_P \mathbf{U} + 2\kappa)\, \pounds_\xi(f) + 2\pounds_n(\pounds_\xi(f)) - 2P^{ab}(\mathrm{r}+\mathrm{s})_a \overset{\circ}{\nabla}_b f \\ &+ n^{(2)}\left((\operatorname{tr}_P \mathbf{Y} - n(\ell^{(2)}))\pounds_\xi(f) + \pounds^{(2)}_\xi(f) - \beta\pounds_\xi(f) - \pounds_{a_\parallel}(f)\right),\end{aligned} \tag{B.11}$$

*where $\overset{\circ}{\Box} := P^{ab}\overset{\circ}{\nabla}_a \overset{\circ}{\nabla}_b$.*

*Proof.* In order to prove (B.8) we contract the (ambient) Hessian $\nabla_\alpha \nabla_\beta f$ with $e^\alpha_a e^\beta_b$ and use (2.41),

$$\begin{aligned} e^\alpha_a e^\beta_b \nabla_\alpha \nabla_\beta f &= \nabla_{e_a}\nabla_{e_b} f - (\nabla_{e_a} e_b)(f) \\ &= \overset{\circ}{\nabla}_{e_a}\overset{\circ}{\nabla}_{e_b} f - (\overset{\circ}{\nabla}_{e_a} e_b)(f) + \pounds_n(f)\mathrm{Y}_{ab} + \pounds_\xi(f)\Big(\mathrm{U}_{ab} + n^{(2)}\mathrm{Y}_{ab}\Big) \\ &= \overset{\circ}{\nabla}_a \overset{\circ}{\nabla}_b f + \pounds_n(f)\mathrm{Y}_{ab} + \pounds_\xi(f)\Big(\mathrm{U}_{ab} + n^{(2)}\mathrm{Y}_{ab}\Big). \end{aligned}$$

To prove the second one we contract $\nabla_\alpha\nabla_\beta f$ with $e^\alpha_a \xi^\beta$ and use (2.53),

$$\begin{aligned} e^\alpha_a \xi^\beta \nabla_\alpha \nabla_\beta f = \nabla_{e_a}(\pounds_\xi(f)) - (\nabla_{e_a}\xi)^\beta \nabla_\beta f = \overset{\circ}{\nabla}_{e_a}(\pounds_\xi(f)) - \left(\mathrm{r}_a - \mathrm{s}_a + \frac{1}{2}n^{(2)}\overset{\circ}{\nabla}_a \ell^{(2)}\right)\pounds_\xi(f) \\ - e^c_a V^b{}_c \overset{\circ}{\nabla}_b f. \end{aligned}$$

Expression (B.10) is immediate because $\xi^\alpha \xi^\beta \nabla_\alpha \nabla_\beta f = \nabla_\xi \nabla_\xi f - \nabla_{\nabla_\xi \xi} f = \pounds^{(2)}_\xi(f) - \pounds_{\nabla_\xi \xi}(f)$. Finally, identity (B.11) follows from (see (2.13)) $g^{\alpha\beta}\nabla_\alpha\nabla_\beta f = P^{ab}e^\alpha_a e^\beta_b \nabla_\alpha \nabla_\beta f + 2\xi^\alpha \nu^\beta \nabla_\alpha \nabla_\beta f + n^{(2)}\xi^\alpha\xi^\beta\nabla_\alpha\nabla_\beta f$ after inserting (B.8)-(B.10) and using (2.57). □

Next we compute the pullback of the Lie derivative of an arbitrary covariant tensor field along any vector.

**Proposition B.5.** *Let $(\mathcal{M}, g)$ be a semi-Riemannian manifold, $\Phi : \mathcal{H} \hookrightarrow \mathcal{M}$ a smooth embedded hypersurface with rigging $\xi$ and $T$ a $(0,p)$ tensor field on $\mathcal{M}$. For any vector field $\zeta \in \mathfrak{X}(\mathcal{M})$ define $\zeta_t \in \mathcal{F}(\mathcal{H})$ and $\zeta_\parallel \in \mathfrak{X}(\mathcal{H})$ by $\zeta \overset{\mathcal{H}}{=} \zeta_t \xi + \Phi_\star \zeta_\parallel$. Then,*

$$\Phi^\star(\pounds_\zeta T)_{a_1\cdots a_p} = \zeta_t(\pounds_\xi T)_{a_1\cdots a_p} + \sum_{i=1}^p (\overset{\circ}{\nabla}_{a_i}\zeta_t)({}^{(i)}T)_{a_1\cdots a_{i-1}a_{i+1}\cdots a_p} + (\pounds_{\zeta_\parallel} T)_{a_1\cdots a_p}. \tag{B.12}$$

*Proof.* Inserting $\zeta \stackrel{\mathcal{H}}{=} \zeta_t \xi + \Phi_\star \zeta_\parallel$ in the Lie derivative $\pounds_\zeta T_{\alpha_1 \cdots \alpha_p}$,

$$\pounds_\zeta T_{\alpha_1\cdots\alpha_p} \stackrel{\mathcal{H}}{=} \pounds_{\zeta_t \xi + \Phi_\star \zeta_\parallel} T_{\alpha_1\cdots\alpha_p} \stackrel{\mathcal{H}}{=} \zeta_t (\pounds_\xi T)_{\alpha_1\cdots\alpha_p} + \sum_{i=1}^{p} (\nabla_{\alpha_i} \zeta_t) T_{\alpha_1\cdots\alpha_{i-1}\mu\alpha_{i+1}\cdots\alpha_p} \xi^\mu$$
$$+ (\pounds_{\Phi_\star \zeta_\parallel} T)_{\alpha_1\cdots\alpha_p}.$$

Identity (B.12) is just the pullback of this after using the property $\Phi^\star \pounds_{\Phi_\star \zeta_\parallel} T = \pounds_{\zeta_\parallel} \Phi^\star T$. □

C

# GAUSSIAN NULL AND RÁCZ-WALD COORDINATES

In this appendix we review the construction of the Gaussian null and Rácz-Wald coordinate systems needed in Chapter 6.

## C.1 GAUSSIAN NULL COORDINATES

We begin by recalling the construction of the Gaussian null coordinates. For the details we refer to [134, 245]. Let $(\mathcal{M}, g)$ be a $d$-dimensional spacetime, $\mathcal{H}$ a null hypersurface and $N$ a vector normal to $\mathcal{H}$ so that its integral curves are null generators of $\mathcal{H}$. Assume $\mathcal{H}$ admits a cross section. One can construct coordinates in a neighbourhood of $\mathcal{H}$ in $\mathcal{M}$ as follows.

Pick a cross section $\Sigma$ of $\mathcal{H}$ with (local) coordinates $\{x^A\}$, define $t$ by $N(t) = 1$ with $t|_\Sigma = 0$ and extend $\{x^A\}$ to $\mathcal{H}$ as constant along $N$. Then, $\{t, x^A\}$ are (local) coordinates on $\mathcal{H}$. Consider the unique null rigging $\xi$ of $\mathcal{H}$ satisfying $g(N, \xi) = 1$ and that $\xi$ is orthogonal to the submanifolds of $\mathcal{H}$ defined by $t = const$. Extend $\xi$ off $\mathcal{H}$ geodesically, i.e. $\nabla_\xi \xi = 0$, and let $r$ be the affine parameter along these geodesics. After extending $\{t, x^A\}$ as constants along $\xi$, $\{r, t, x^A\}$ becomes a coordinate system in a neighbourhood of $\mathcal{H}$ on $\mathcal{M}$, in which $\mathcal{H} = \{r = 0\}$ and the spacetime metric $g$ takes the following form

$$g = 2dtdr + r\beta_A dx^A dt + r\phi dt^2 + \mu_{AB} dx^A dx^B, \tag{C.1}$$

where $\phi$, $\beta_A$ and $\mu_{AB}$ are smooth functions of $\{r, t, x^A\}$ and $\mu_{AB}$ is a positive definite $(d-2) \times (d-2)$ matrix. The Ricci components in these coordinates can be found e.g. in [245]. Later we will need the $(t,t)$-component, which reads

$$\begin{aligned}
\mathcal{R}_{tt} = {} & \frac{1}{2}\Big(\phi\dot{\phi} + \beta^A\Big(D_A\phi - 2\beta'_A - \dot{\phi}\beta_A\Big)\Big)^{\bullet} + \frac{1}{4}\left(\phi' + \phi\dot{\phi} + \beta^A\Big(D_A\phi - 2\beta'_A - \dot{\phi}\beta_A\Big)\right)\mu^{AB}\dot{\mu}_{AB} \\
& + \frac{1}{2}D^A\left(2\beta'_A - D_A\phi + \dot{\phi}\beta_A\right) - \frac{1}{4}\dot{\phi}\mu^{AB}\mu'_{AB} - \frac{1}{2}\mu^{AB}\mu''_{AB} + \frac{1}{4}\mu^{AC}\mu^{BD}\mu'_{AB}\mu'_{CD} \\
& - \frac{1}{2}(\dot{\phi} - \beta^A\dot{\beta}_A)^2 - \mu^{AC}\mu^{BD}D_{[B}\beta_{A]}D_{[C}\beta_{D]} \\
& - \frac{1}{2}\dot{\beta}^C\Big(2D_C\phi + \phi\dot{\beta}_C - \beta^A\Big(4D_{[C}\beta_{A]} + \beta_A\dot{\beta}_C\Big) - 2\beta'_C\Big),
\end{aligned} \tag{C.2}$$

where the prime means derivative w.r.t. $t$, the dot denotes derivative w.r.t. $r$ and $D$ is the Levi-Civita connection on the cross-sections.

#### C.1.0.1 *Computation of* $\mathcal{L}^{(m+1)}_a n^a$

In Chapter 6 we need the contraction $\mathcal{L}^{(m+1)}_a n^a$ up to order $m$. This computation requires knowing $\mathcal{R}^{(m+1)}_{ab} n^a n^b$ up to order $m$ and turns out to be quite laborious using the general formalism of hypersurface data. To keep the computations to a reasonable length we have decided to perform the computation using a particular gauge and assuming that the hypersurface is totally geodesic and admits a foliation by cross-sections. The result presented here will be sufficient for the purposes of this thesis. We refer the reader to Notations 5.19 and 5.23 to recall the meaning of the symbols $[m]$ and $(m)$ above an equal sign.

**Proposition C.1.** *Let $\mathcal{H}$ be a null hypersurface embedded in $(\mathcal{M}, g)$ and assume the existence of a foliation $\{\mathcal{S}_t\}_{t\in\mathbb{R}}$ of $\mathcal{H}$ by cross-sections. Let $\xi$ be the (uniquely defined) rigging satisfying (i) $g(\xi,\xi) \stackrel{\mathcal{H}}{=} 0$ and (ii) $\Phi^\star \boldsymbol{\xi} = dt$. Suppose in addition that $\pounds_n \gamma = 0$. Then,*

$$\begin{aligned}\mathcal{R}^{(m+1)}_{ab} n^a n^b \stackrel{(m)}{=} & \, 2m\kappa\kappa^{(m+1)} - \pounds_n^{(2)}(\operatorname{tr}_P \mathbf{Y}^{(m)}) + \kappa \pounds_n (\operatorname{tr}_P \mathbf{Y}^{(m)}) \\ & + (2\kappa^2 - \pounds_n \kappa) m \operatorname{tr}_P \mathbf{Y}^{(m)} + \mathcal{O}_{\mathcal{R}},\end{aligned} \tag{C.3}$$

*where $\mathcal{O}_{\mathcal{R}}$ is a scalar that depends on $\mathbf{r}^{(m)}$ and lower order terms.*

*Proof.* We construct Gaussian null coordinates $\{r, t, x^A\}$ from $\mathcal{H} = \{r = 0\}$ in which the metric takes the form (C.1). In these coordinates, the rigging is $\xi = \partial_r$ and $\nu = \partial_t$. Denoting the derivative w.r.t. $r$ at $r = 0$ with a dot, it is easy to check that, at $r = 0$, $\kappa = -\frac{1}{2}\dot{\phi}$, $\mathrm{r}_A = \frac{1}{2}\dot{\boldsymbol{\beta}}$ and $\mathrm{Y}_{AB} = \frac{1}{2}\dot{\mu}_{AB}$. If we denote by $\dot{f}^{(m)}$ the $m$-th $t$-derivative at $r = 0$, one also cheeks easily that $\kappa^{(m)} = -\frac{1}{2}\dot{\phi}^{(m)}$, $\mathrm{r}^{(m)}_A = \frac{1}{2}\dot{\boldsymbol{\beta}}^{(m)}$ and $\mathrm{Y}^{(m)}_{AB} = \frac{1}{2}\dot{\mu}^{(m)}_{AB}$. We also note that $P^{ab} = \mu^{AB}\delta^a_A \delta^b_B$ and $n^a = \delta^a_t$. In these coordinates, the $(t,t)$ component of the Ricci tensor is given by (C.2), where recall that the prime means derivative w.r.t. $t$ and $D$ is the Levi-Civita connection of the induced metric in the codimension-two surfaces $t = const.$, $r = const.$ Obviously this connection depends on $t$ and $r$ (we do not reflect this dependence for notational simplicity). Note however that at $r = 0$, the connection is independent of $t$ because $\gamma_{AB}$ does not depend on $t$ (since $\pounds_n \gamma = 0$). So, when an expression is evaluated at $r = 0$ (e.g. all expressions with an $(m)$ on top of the equal sign), $D$ will mean the covariant derivative associated to the metric $\gamma_{AB}$. The strategy now is to take $m$ derivatives along $r$ and keep only the terms depending on $\dot{\phi}^{(m+1)}$, $\dot{\boldsymbol{\beta}}^{(m+1)}$, $\dot{\mu}^{(m+1)}$ and $\dot{\mu}^{(m)}$. This requires commuting $\partial_r$ with $D$ using a formula analogous to (5.7),

$$\begin{aligned}\left[\partial_r^{(m)}, D_A\right] T^{B_1\cdots B_q}{}_{C_1\cdots C_p} = \sum_{k=0}^{m-1} \binom{m}{k+1} \Bigg( & \sum_{j=1}^{q} \left(\dot{T}^{(m-k-1)}\right)^{B_1\cdots B_{j-1} D B_{j+1}\cdots B_q}{}_{C_1\cdots C_p} \Xi^{(k+1)B_j}{}_{DA} \\ & - \sum_{i=1}^{p} \left(\dot{T}^{(m-k-1)}\right)^{B_1\cdots B_q}{}_{C_1\cdots C_{i-1} D C_{i+1}\cdots C_p} \Xi^{(k+1)D}{}_{C_i A} \Bigg),\end{aligned}$$

where $\Xi^A{}_{BC} = D_{(B}\dot{\mu}_{C)}{}^A - \frac{1}{2}D^A\dot{\mu}_{BC}$. Then, $\Xi^{(k)A}{}_{BC}$ will depend on $\dot{\mu}^{(j)}$ with $j = 1, ..., k$. Consequently, for any tensor field $T$ satisfying $T|_{r=0} = 0$, the previous identity implies

$$\partial_r^{(m)} D_A T^{B_1\cdots B_q}{}_{C_1\cdots C_p} \stackrel{(m)}{=} D_A(\dot{T}^{(m)})^{B_1\cdots B_q}{}_{C_1\cdots C_p}.$$

This is the case, in particular, of $D_A\phi$, $\beta_A$, $\beta'_A$, $\mu'_{AB}$ and $\mu''_{AB}$. Using this rule and that $\mu'_{AB}(r=0) = \beta_A(r=0) = \phi(r=0) = 0$, the $m$-th $r$-derivative of $\mathcal{R}_{tt}$ at $r=0$ is

$$\mathcal{R}_{tt}^{(m+1)} = \dot{\mathcal{R}}_{tt}^{(m)} \stackrel{(m)}{=} \frac{1}{2}m\dot{\phi}\dot{\phi}^{(m+1)} + \frac{1}{4}m(\dot{\phi}' + \dot{\phi}^2)\mu^{AB}\dot{\mu}_{AB}^{(m)} - \frac{1}{4}\dot{\phi}\mu^{AB}\dot{\mu}_{AB}^{(m)\prime} - \frac{1}{2}\mu^{AB}\dot{\mu}_{AB}^{(m)\prime\prime}$$
$$+ \text{ terms depending on } \dot{\boldsymbol{\beta}}^{(m)} \text{ and } \dot{\phi}^{(m)},$$

which is (C.3) after replacing $\dot{\phi} = -2\kappa$, $\dot{\phi}^{(m+1)} = -2\kappa^{(m+1)}$, $\dot{\mu}_{AB}^{(m)} = 2\mathrm{Y}_{AB}^{(m)}$ and taking into account $\gamma'_{AB} = 0$. □

We now use Proposition C.1 to compute $\mathcal{L}_a^{(m+1)}n^a$ up to order $m$ under the same assumptions on $\mathscr{I}$ and the rigging.

**Proposition C.2.** *Assume $\mathscr{I}$ admits a foliation of cross-sections $\{\mathcal{S}_u\}_{u\in\mathbb{R}}$ and let $\xi$ be the rigging satisfying $g(\xi,\xi) \stackrel{\mathscr{I}}{=} 0$ and $\Phi^\star\boldsymbol{\xi} = du$. Suppose in addition that $\pounds_n\gamma = 0$. Then,*

$$\mathcal{L}_a^{(m+1)}n^a \stackrel{(m)}{=} H_1\kappa^{(m+1)} - \sigma^{(1)}\pounds_n^{(2)}(\operatorname{tr}_P \mathbf{Y}^{(m)}) + H_2\pounds_n(\operatorname{tr}_P \mathbf{Y}^{(m)}) + H_3 \operatorname{tr}_P \mathbf{Y}^{(m)} + \mathcal{O}_{\mathcal{L}}, \quad \text{(C.4)}$$

*where $\mathcal{O}_{\mathcal{L}}$ is a scalar that depends on $\mathbf{r}^{(m)}$ and lower order terms and whose explicit form is not relevant for the purposes of this thesis, and*

$$H_1 := -\frac{m(\mathfrak{n}-1)\pounds_n\sigma^{(1)}}{\mathfrak{n}}, \qquad H_2 := -(2m-1)\sigma^{(1)}\kappa - m(\mathfrak{n}-2)\pounds_n\sigma^{(1)},$$
$$H_3 := \frac{(2m-\mathfrak{n})m}{\mathfrak{n}}\pounds_n\sigma^{(1)} - m\sigma^{(1)}\pounds_n\kappa.$$

*Proof.* We begin by computing $\mathcal{L}_\alpha^{(m+1)}$ to order $[m]$. From (E.18) and Proposition E.1, the only terms that could depend on $\mathbf{Y}^{(m)}$ are $i = m$, $i = m-1$ ($j = 0,1$) and $i = m-2$ ($j = 0,1,2$), namely

$$\mathcal{L}_\alpha^{(m+1)} \stackrel{[m]}{=} (\mathfrak{n}-1)\Bigg( L_{\alpha\beta}^{(m+1)}\nabla^\beta\Omega + mL_{\alpha\beta}^{(m)}\left(g^{\beta\mu}\nabla_\mu\pounds_\xi\Omega + (\pounds_\xi g^{\beta\mu})\nabla_\mu\Omega\right)$$
$$+ \frac{m(m-1)}{2}L_{\alpha\beta}^{(m-1)}\left(g^{\beta\mu}\nabla_\mu\pounds_\xi^{(2)}\Omega + 2(\pounds_\xi g^{\beta\mu})\nabla_\mu\pounds_\xi\Omega + (\pounds_\xi^{(2)}g^{\beta\mu})\nabla_\mu\Omega\right)\Bigg)$$

Contracting with $\nu^\alpha$, evaluating at $\mathscr{I}$ and using as usual $\nabla^\beta\Omega \stackrel{\mathscr{I}}{=} \sigma^{(1)}\nu^\beta$ we get

$$\mathcal{L}_a^{(m+1)}n^a \stackrel{(m)}{=} (\mathfrak{n}-1)\left(\sigma^{(1)}L_{\alpha\beta}^{(m+1)}\nu^\alpha\nu^\beta + m\nu^\alpha L_{\alpha\beta}^{(m)}\left(g^{\beta\mu}\nabla_\mu\pounds_\xi\Omega + (\pounds_\xi g^{\beta\mu})\nabla_\mu\Omega\right)\right.$$
$$\left. + \frac{m(m-1)}{2}\nu^\alpha L_{\alpha\beta}^{(m-1)}\left(g^{\beta\mu}\nabla_\mu\pounds_\xi^{(2)}\Omega + 2(\pounds_\xi g^{\beta\mu})\nabla_\mu\pounds_\xi\Omega + (\pounds_\xi^{(2)}g^{\beta\mu})\nabla_\mu\Omega\right)\right).$$

It is easier to split the computation into three pieces, namely $\mathrm{I} = (\mathfrak{n}-1)\sigma^{(1)} L^{(m+1)}_{\alpha\beta}\nu^\alpha\nu^\beta$, $\mathrm{II} = (\mathfrak{n}-1)m\nu^\alpha L^{(m)}_{\alpha\beta}\left(g^{\beta\mu}\nabla_\mu \pounds_\xi\Omega + (\pounds_\xi g^{\beta\mu})\nabla_\mu\Omega\right)$ and

$$\mathrm{III} = \frac{m(m-1)(\mathfrak{n}-1)}{2}\nu^\alpha L^{(m-1)}_{\alpha\beta}\left(g^{\beta\mu}\nabla_\mu\pounds^{(2)}_\xi\Omega + 2(\pounds_\xi g^{\beta\mu})\nabla_\mu\pounds_\xi\Omega + (\pounds^{(2)}_\xi g^{\beta\mu})\nabla_\mu\Omega\right),$$

and compute each term separately. Those involving only $\mathbf{r}^{(m)}$ or $\kappa^{(m)}$ are not needed in explicit form. We shall gather all of them into the tensor $\mathcal{O}_{\mathcal{L}}$ in the statement of the proposition. For the calculation we shall write $\cdots$ to mean additional terms of this form. For the first term I we contract (E.2) (with $m+1$ instead of $m$) with $\nu^\alpha\nu^\beta$ and use $\mathcal{K}_{\alpha\beta}\nu^\alpha\nu^\beta = -2\kappa$ and $\mathcal{K}^{(2)}_{\alpha\beta}\nu^\alpha\nu^\beta = -2\kappa^{(2)}$ to get

$$\mathrm{I} \overset{(m)}{=} \sigma^{(1)}\left(\mathcal{R}^{(m+1)}_{ab}n^a n^b + \frac{1}{\mathfrak{n}}\left(m\kappa R^{(m)} + \frac{m(m-1)}{2}\kappa^{(2)}R^{(m-1)}\right)\right).$$

Inserting (C.3) as well as

$$R^{(m)} \overset{(m)}{=} -2\kappa^{(m+1)} - 4\pounds_n(\operatorname{tr}_P \mathbf{Y}^{(m)}) - 4m\kappa\operatorname{tr}_P\mathbf{Y}^{(m)} + \cdots, \qquad R^{(m-1)} \overset{(m)}{=} \cdots,$$

we arrive at

$$\begin{aligned}\mathrm{I} \overset{(m)}{=}& \frac{2m(\mathfrak{n}-1)\sigma^{(1)}\kappa}{\mathfrak{n}}\kappa^{(m+1)} - \sigma^{(1)}\pounds^{(2)}_n(\operatorname{tr}_P\mathbf{Y}^{(m)}) + \frac{\mathfrak{n}-4m}{\mathfrak{n}}\kappa\sigma^{(1)}\pounds_n(\operatorname{tr}_P\mathbf{Y}^{(m)})\\ &+ \left(\frac{2(\mathfrak{n}-2m)\kappa^2}{\mathfrak{n}} - \pounds_n\kappa\right)m\sigma^{(1)}\operatorname{tr}_P\mathbf{Y}^{(m)} + \cdots.\end{aligned}$$

To compute II we now use that (cf. (E.9)-(E.10))

$$L^{(m)}_{ab}n^a \overset{(m)}{=} \cdots \tag{C.5}$$

$$\overset{\bullet}{L}{}^{(m)}_{a}n^a \overset{(m)}{=} \frac{1}{\mathfrak{n}(\mathfrak{n}-1)}\left((1-\mathfrak{n})\kappa^{(m+1)} + (2-\mathfrak{n})\pounds_n(\operatorname{tr}_P\mathbf{Y}^{(m)}) + (2m-\mathfrak{n})\kappa\operatorname{tr}_P\mathbf{Y}^{(m)}\right) + \cdots. \tag{C.6}$$

Inserting these into the definition of II and using (2.13) and (E.43) gives

$$\begin{aligned}\mathrm{II} &= (\mathfrak{n}-1)m\nu^\alpha L^{(m)}_{\alpha\beta}\left(g^{\beta\mu}\nabla_\mu\pounds_\xi\Omega + (\pounds_\xi g^{\beta\mu})\nabla_\mu\Omega\right)\\ &\overset{(m)}{=} (\mathfrak{n}-1)m(\pounds_n\sigma^{(1)} + 2\sigma^{(1)}\kappa)\overset{\bullet}{L}{}^{(m)}_{a}n^a\\ &\overset{(m)}{=} \frac{m(\pounds_n\sigma^{(1)} + 2\sigma^{(1)}\kappa)}{\mathfrak{n}}\left((1-\mathfrak{n})\kappa^{(m+1)} + (2-\mathfrak{n})\pounds_n(\operatorname{tr}_P\mathbf{Y}^{(m)}) + (2m-\mathfrak{n})\kappa\operatorname{tr}_P\mathbf{Y}^{(m)}\right) + \cdots.\end{aligned}$$

Finally, it is clear from (C.5)-(C.6) that $\mathrm{III} = \cdots$. To show (C.4) we only need to add $\mathcal{L}^{(m+1)}_a n^a \overset{(m)}{=} \mathrm{I} + \mathrm{II} + \mathrm{III}$ and simplify. □

## C.2 RÁCZ-WALD COORDINATES

In this section we sketch the construction of the Rácz-Wald coordinates. We refer the reader to the original construction [288] for the details (see also Appendix D in [211] for a construction closer in spirit to the Gaussian null construction above). Consider a $d$-dimensional spacetime $(\mathcal{M}, g)$ possessing a Killing horizon $\mathcal{H}$ admitting cross-sections and such that orbits of the Killing $K$ are complete. Assume also that the surface gravity $\kappa$ of $K$ at $\mathcal{H}$ is a non-zero constant. Then, one constructs Gaussian null coordinates $\{r, t, x^A\}$ associated to $N = K$ where the Killing horizon is placed at $\mathcal{H} = \{r = 0\}$. Since $K$ is Killing, $\pounds_K \xi = 0$ and consequently $\pounds_\xi (K - \partial_t) = 0$, which proves that in these coordinates $K = \partial_t$ because $K|_{\mathcal{H}} = \partial_t|_{\mathcal{H}}$. Then, the metric is given by

$$g = -rf dt^2 + 2dtdr + 2rh_A dtdx^A + \mu_{AB} dx^A dx^B,$$

where $f$, $h_A$ and $\mu_{AB}$ are smooth functions only of $\{r, x^A\}$, $f(r = 0) = 2\kappa$ and $\mu_{AB}$ is a positive definite $(d-2) \times (d-2)$ matrix.

Next one constructs Kruskal-type coordinates $\{u, v\}$ as follows. By noting that $f$ is non-vanishing in a neighbourhood of $\mathcal{H}$ (and hence $1/f$ is smooth there) it follows that there exists a smooth function $g$ satisfying

$$\frac{1}{f} - \frac{1}{2\kappa} = rg.$$

One then defines the smooth functions

$$u := e^{\kappa t}, \qquad v := -e^{-\kappa t} r \exp\left(2\kappa \int_0^r g dr\right),$$

and

$$\varphi := UV = -r \exp\left(2\kappa \int_0^r g dr\right).$$

The Killing vector in these coordinates is $K = U\partial_U - V\partial_V$. Since $\partial_r \varphi \neq 0$ on $\mathcal{H}$ one can solve $r$ as a function $r = q(\varphi, x^A)$, and since $r = 0$ at $\varphi = 0$,

$$r = \varphi\psi(\varphi, x^A) = uv\psi(uv, x^A),$$

where $\psi$ is a smooth function. To express the metric in the coordinates $\{u, v, x^A\}$ we first compute

$$du = \kappa u dt, \qquad dv = -\kappa v dt + \frac{2\kappa v}{rf} dr + 2\kappa v \left(\int_0^r \partial_{x^A} g dr\right) dx^A$$

and define

$$G := \frac{rf}{\kappa^2 uv} = \frac{f\psi}{\kappa^2}.$$

Then,

$$G du dv = -rf dt^2 + 2dtdr + 2rf \left(\int_0^r \partial_{x^A} g dr\right) dtdx^A,$$

and hence

$$g = Gdudv + v\beta_A dudx^A + \mu_{AB}dx^A dx^B, \tag{C.7}$$

where

$$\beta_A := \frac{\psi}{\kappa}\left(h_A - f\int_0^r \partial_{x^A} g dr\right).$$

Originally the coordinate $u$ is defined in the region $u > 0$, but the metric (C.7) is extendible beyond $u = 0$. The Killing vector $K = u\partial_u - v\partial_v$ admits a bifurcate horizon $uv = 0$, and the bifurcation surface corresponds to $u = v = 0$. The coordinates $\{u, v, x^A\}$ where the metric takes the form (C.7) are called Rácz-Wald coordinates.

To conclude this appendix we write down explicitly the quasi-Einstein equations (see Chapter 6) associated to the metric (C.7) with $\boldsymbol{\beta} = 0$. More specifically, we compute the components of the tensor $\mathcal{Q} := r\Big(\operatorname{Hess}_g v + v\operatorname{Sch}_g\Big)$. The computation has been done using the `xAct` package of `Mathematica`. We introduce the variable $s := uv$ and we denote a derivative w.r.t. $s$ with a dot.

$$\begin{aligned}
\mathcal{Q}_{vv} &= \frac{u}{4}\left(2(s\operatorname{tr}_\mu\dot{\mu} - 2r)\frac{\dot{G}}{G} + s\Big(|\dot{\mu}|^2 - 2\operatorname{tr}_\mu\ddot{\mu}\Big)\right),\\
\mathcal{Q}_{uv} &= \frac{v}{4(r+1)}\Big(-4r\frac{\dot{G}}{G} - 2GR^{(\mu)} + 4rs\frac{\dot{G}^2}{G^2} - 4rs\frac{\ddot{G}}{G} + (r-2)s|\dot{\mu}|^2 - 2(r-1)s\operatorname{tr}_\mu\ddot{\mu}\\
&\qquad + \Big(s\operatorname{tr}_\mu\dot{\mu} - 2(r-1)\Big)\operatorname{tr}_\mu\dot{\mu} - 2(r-1)\Box_\mu G - G^{-1}|\nabla G|^2_\mu\Big),\\
\mathcal{Q}_{uu} &= \frac{v^3}{4}\left(|\dot{\mu}|^2 - 2\operatorname{tr}_\mu\ddot{\mu} + 2\frac{\dot{G}}{G}\operatorname{tr}_\mu\dot{\mu}\right),\\
\mathcal{Q}_{vA} &= \frac{1}{4}\left(\left(-2r + 2s\frac{\dot{G}}{G} + s\operatorname{tr}_\mu\dot{\mu}\right)\frac{\nabla_A G}{G} - 2s\left(\frac{\nabla_A\dot{G}}{G} + \Big(\nabla_A\operatorname{tr}_\mu\dot{\mu} - (\operatorname{div}_\mu\dot{\mu})_A\Big)\right)\right),\\
\mathcal{Q}_{uA} &= \frac{v^2}{4}\left(\left(2\frac{\dot{G}}{G} + \operatorname{tr}_\mu\dot{\mu}\right)\frac{\nabla_A G}{G} - 2\left(\frac{\nabla_A\dot{G}}{G} + \Big(\nabla_A\operatorname{tr}_\mu\dot{\mu} - (\operatorname{div}_\mu\dot{\mu})_A\Big)\right)\right),\\
\mathcal{Q}_{AB} &= \frac{v}{4G}\Bigg(\Big(2\Big(r - 2 - s\operatorname{tr}_\mu\dot{\mu}\Big)\dot{\mu}_{AB} + 2\Big(2GR^{(\mu)}_{AB} + 2s(\dot{\mu}\cdot\dot{\mu})_{AB} - 2s\ddot{\mu}_{AB} - 2\nabla_A\nabla_B G\\
&\qquad + G^{-1}\nabla_A G\nabla_B G\Big)\Big) + \frac{1}{r+1}\Bigg(4\frac{\dot{G}}{G} - 2GR^{(\mu)} + 4s\left(\frac{\ddot{G}}{G} - \frac{\dot{G}^2}{G^2}\right) - 3s|\dot{\mu}|^2 + 4s\operatorname{tr}_\mu\ddot{\mu}\\
&\qquad + (4 + s\operatorname{tr}_\mu\dot{\mu})\operatorname{tr}_\mu\dot{\mu} + 4\Box_\mu G - G^{-1}|\nabla G|^2\Bigg)\mu_{AB}\Bigg),
\end{aligned}$$

where $\operatorname{tr}_\mu$ is the trace w.r.t. $\mu$ and $(\dot{\mu}\cdot\dot{\mu})_{AB} := \mu^{CD}\dot{\mu}_{AC}\dot{\mu}_{BD}$. The particular case $G = -1$ of these equations will be also of relevance. Their expressions are as follows.

$$
\begin{aligned}
\mathcal{Q}_{vv} &= \frac{us}{4}\Big(|\dot{\mu}|^2 - 2\,\mathrm{tr}_\mu\,\ddot{\mu}\Big),\\
\mathcal{Q}_{uv} &= \frac{v}{4(r+1)}\bigg(2R^{(\mu)} + (r-2)s|\dot{\mu}|^2 - 2(r-1)s\,\mathrm{tr}_\mu\,\ddot{\mu} + \Big(s\,\mathrm{tr}_\mu\,\dot{\mu} - 2(r-1)\Big)\,\mathrm{tr}_\mu\,\dot{\mu}\bigg),\\
\mathcal{Q}_{uu} &= \frac{v^3}{4}\left(|\dot{\mu}|^2 - 2\,\mathrm{tr}_\mu\,\ddot{\mu}\right),\\
\mathcal{Q}_{vA} &= -\frac{s}{2}\left(\nabla_A\,\mathrm{tr}_\mu\,\dot{\mu} - (\mathrm{div}_\mu\,\dot{\mu})_A\right),\\
\mathcal{Q}_{uA} &= -\frac{v^2}{2}\left(\nabla_A\,\mathrm{tr}_\mu\,\dot{\mu} - (\mathrm{div}_\mu\,\dot{\mu})_A\right),\\
\mathcal{Q}_{AB} &= -\frac{v}{2}\Bigg(\Big(r-2-s\,\mathrm{tr}_\mu\,\dot{\mu}\Big)\dot{\mu}_{AB} - 2\Big(R^{(\mu)}_{AB} - s(\dot{\mu}\cdot\dot{\mu})_{AB} + s\ddot{\mu}_{AB}\Big)\\
&\qquad + \frac{1}{2(r+1)}\Big(2R^{(\mu)} - 3s|\dot{\mu}|^2 + 4s\,\mathrm{tr}_\mu\,\ddot{\mu} + (4+s\,\mathrm{tr}_\mu\,\dot{\mu})\,\mathrm{tr}_\mu\,\dot{\mu}\Big)\mu_{AB}\Bigg).
\end{aligned}
$$

# D

# AUXILIARY COMPUTATIONS

In this appendix we compute several additional contractions of the tensors $\underset{\smile}{\Sigma}^{(m)}$ and $\Sigma^{(m)}$, as well as the tensor $\dot{\mathcal{R}}_a^{(m)}$ to one order higher. These computations are placed here rather than in Section 5.2 because they are not required in Chapter 5, and used only in Chapter 6. We hope that presenting them here simplifies the reading of Chapter 5 and improves the flow of the exposition. Let us begin by recalling the following expressions computed in Chapter 5 (see Proposition 5.25)

$$\underset{\smile}{\Sigma}^{(m)}_{cab} \overset{(m)}{=} \overset{\circ}{\nabla}_a \mathrm{Y}^{(m)}_{bc} + \overset{\circ}{\nabla}_b \mathrm{Y}^{(m)}_{ac} - \overset{\circ}{\nabla}_c \mathrm{Y}^{(m)}_{ab} + 2\mathrm{r}^{(m)}_c \mathrm{Y}_{ab} + 2\mathrm{r}_c \mathrm{Y}^{(m)}_{ab}, \tag{D.1}$$

$$({}^{(3)}\underset{\smile}{\Sigma}^{(m)})_{ab} \overset{(m+1)}{=} -({}^{(1)}\underset{\smile}{\Sigma}^{(m)})_{ab} \overset{(m+1)}{=} \mathrm{Y}^{(m+1)}_{ab}, \tag{D.2}$$

$$\xi^\beta \Sigma^{(m)\mu}{}_{\mu\beta} \overset{(m+1)}{=} \operatorname{tr}_P \mathbf{Y}^{(m+1)}. \tag{D.3}$$

The next expressions also proved in Chapter 5 (formulas (5.52) and (5.54)) and valid for $m \geq 1$ will be also needed

$$\Sigma^{(m)}_{\mu\alpha\beta} \overset{[m]}{=} \underset{\smile}{\Sigma}^{(m)}_{\mu\alpha\beta} - (m-1)\mathcal{K}^\rho{}_\mu \underset{\smile}{\Sigma}^{(m-1)}_{\rho\alpha\beta}, \tag{D.4}$$

$$\Sigma^{(m)}_{\rho\alpha\beta} \overset{[m+1]}{=} \underset{\smile}{\Sigma}^{(m)}_{\rho\alpha\beta}, \tag{D.5}$$

$$\underset{\smile}{\Sigma}^{(m)}_{\mu\alpha\beta} \overset{[m]}{=} \nabla_{(\alpha}\mathcal{K}^{(m)}_{\beta)\mu} - \frac{1}{2}\nabla_\mu \mathcal{K}^{(m)}_{\alpha\beta} - \mathcal{K}^\varepsilon{}_\mu \nabla_{(\alpha}\mathcal{K}^{(m-1)}_{\beta)\varepsilon} + \frac{1}{2}\mathcal{K}^\varepsilon{}_\mu \nabla_\varepsilon \mathcal{K}^{(m-1)}_{\alpha\beta}. \tag{D.6}$$

Note that these three identities become exact for $m = 1$. Later in this appendix we will also need several contractions of $\xi^\beta \underset{\smile}{\Sigma}^{(m)}_{\mu\alpha\beta}$ that we compute next. First observe that the contraction of (D.6) with $\xi^\beta$ can be written as

$$\xi^\beta \underset{\smile}{\Sigma}^{(m)}_{\mu\alpha\beta} \overset{[m]}{=} \frac{1}{2}\xi^\beta \nabla_\beta \mathcal{K}^{(m)}_{\alpha\mu} + \xi^\beta \nabla_{[\alpha}\mathcal{K}^{(m)}_{\mu]\beta} - \frac{1}{2}\mathcal{K}^\varepsilon{}_\mu \xi^\beta \nabla_\beta \mathcal{K}^{(m-1)}_{\alpha\varepsilon} - \mathcal{K}^\varepsilon{}_\mu \xi^\beta \nabla_{[\alpha}\mathcal{K}^{(m-1)}_{\varepsilon]\beta}.$$

We now use the two identities (5.43) and get

$$\begin{aligned} e^\mu_c e^\alpha_a \xi^\beta \underset{\smile}{\Sigma}^{(m)}_{\mu\alpha\beta} \overset{(m)}{=}\ & \frac{1}{2}\Big({}^{(1)}\nabla\mathcal{K}^{(m)}\Big)_{ac} + \Big({}^{(2)}\nabla\mathcal{K}^{(m)}\Big)_{[ac]} - \mathrm{r}_c\Big({}^{(2,3)}\nabla\mathcal{K}^{(m-1)}\Big)_a \\ & + \Big(2P^{bd}\mathrm{Y}_{bc} + \frac{1}{2}n^d \overset{\circ}{\nabla}_c \ell^{(2)}\Big)\bigg(\Big({}^{(2)}\nabla\mathcal{K}^{(m-1)}\Big)_{[da]} - \frac{1}{2}\Big({}^{(1)}\nabla\mathcal{K}^{(m-1)}\Big)_{ad}\bigg), \end{aligned} \tag{D.7}$$

$$\begin{aligned} e^\mu_c \xi^\alpha \xi^\beta \underset{\smile}{\Sigma}^{(m)}_{\mu\alpha\beta} \overset{(m)}{=}\ & \Big({}^{(1,2)}\nabla\mathcal{K}^{(m)}\Big)_c - \frac{1}{2}\Big({}^{(2,3)}\nabla\mathcal{K}^{(m)}\Big)_c - \mathrm{r}_c\Big({}^{(1,2,3)}\nabla\mathcal{K}^{(m-1)}\Big) \\ & - \Big(2P^{bd}\mathrm{Y}_{bc} + \frac{1}{2}n^d \overset{\circ}{\nabla}_c \ell^{(2)}\Big)\bigg(\Big({}^{(1,2)}\nabla\mathcal{K}^{(m-1)}\Big)_d - \frac{1}{2}\Big({}^{(2,3)}\nabla\mathcal{K}^{(m-1)}\Big)_d\bigg) \end{aligned} \tag{D.8}$$

$$\xi^\mu e_a^\alpha \xi^\beta \underset{\smile}{\Sigma}^{(m)}_{\mu\alpha\beta} \stackrel{(m)}{=} \frac{1}{2}\Big({}^{(2,3)}\nabla\mathcal{K}^{(m)}\Big)_a + \frac{1}{2}P^{bc}\overset{\circ}{\nabla}_b\ell^{(2)}\Big({}^{(2)}\nabla\mathcal{K}^{(m-1)}\Big)_{[ca]} - \frac{1}{4}P^{bc}\overset{\circ}{\nabla}_b\ell^{(2)}\Big({}^{(1)}\nabla\mathcal{K}^{(m-1)}\Big)_{ac}$$
$$- \frac{1}{4}n(\ell^{(2)})\Big({}^{(2,3)}\nabla\mathcal{K}^{(m-1)}\Big)_a. \tag{D.9}$$

Taking trace in (D.4) in the indices $\mu, \alpha$ and using (D.6) one gets

$$\Sigma^{(m)\mu}{}_{\mu\beta} \stackrel{[m]}{=} \frac{1}{2}\left(\nabla_\beta \mathcal{K}^{(m)\mu}{}_\mu - m\mathcal{K}^{\mu\rho}\nabla_\beta\mathcal{K}^{(m-1)}_{\mu\rho}\right). \tag{D.10}$$

Finally, the $m-1$ Lie derivative of the Ricci tensor is for $m \geq 2$ (cf. (5.50))

$$\pounds_\xi^{(m-1)} R_{\alpha\beta} \stackrel{[m]}{=} \nabla_\mu \Sigma^{(m-1)\mu}{}_{\alpha\beta} - \nabla_\beta \Sigma^{(m-1)\mu}{}_{\mu\alpha}. \tag{D.11}$$

**Proposition D.1.** *Let $\{\mathcal{H}, \boldsymbol{\gamma}, \boldsymbol{\ell}, \ell^{(2)}\}$ be null metric hypersurface data $(\Phi, \xi)$-embedded in $(\mathcal{M}, g)$ and extend $\xi$ off $\Phi(\mathcal{H})$ by $\nabla_\xi \xi = 0$. Then, for any $m \geq 2$,*

$$\Sigma^{(m)}_{cab} \stackrel{(m)}{=} \overset{\circ}{\nabla}_a \mathrm{Y}^{(m)}_{bc} + \overset{\circ}{\nabla}_b \mathrm{Y}^{(m)}_{ac} - \overset{\circ}{\nabla}_c \mathrm{Y}^{(m)}_{ab} + 2\mathrm{r}^{(m)}_c \mathrm{Y}_{ab} + 2m\mathrm{r}_c \mathrm{Y}^{(m)}_{ab}, \tag{D.12}$$

$$e_a^\beta \Sigma^{(m)\mu}{}_{\mu\beta} \stackrel{(m)}{=} \overset{\circ}{\nabla}_a(\operatorname{tr}_P \mathbf{Y}^{(m)}), \quad \text{(D.13)} \qquad ({}^{(2,3)}\underset{\smile}{\Sigma}^{(m)})_a \stackrel{(m+1)}{=} ({}^{(1,3)}\underset{\smile}{\Sigma}^{(m)})_a \stackrel{(m+1)}{=} 0, \quad \text{(D.14)}$$

$$({}^{(3)}\underset{\smile}{\Sigma}^{(m)})_{ca} \stackrel{(m)}{=} \mathrm{Y}^{(m+1)}_{ac} - 2V^b{}_a \mathrm{Y}^{(m)}_{cb} - 2P^{bd}\mathrm{Y}_{bc}\mathrm{Y}^{(m)}_{ad} - \frac{1}{2}\mathrm{r}^{(m)}_a \overset{\circ}{\nabla}_c \ell^{(2)}, \tag{D.15}$$

$$({}^{(3)}\Sigma^{(m)})_{ca} \stackrel{(m)}{=} \mathrm{Y}^{(m+1)}_{ac} - 2V^b{}_a \mathrm{Y}^{(m)}_{cb} - 2mP^{bd}\mathrm{Y}_{bc}\mathrm{Y}^{(m)}_{ad} - \left(m - \frac{1}{2}\right)\mathrm{r}^{(m)}_a \overset{\circ}{\nabla}_c \ell^{(2)}. \tag{D.16}$$

*As a consequence,*

$$n^c \Sigma^{(m)}_{cab} \stackrel{(m)}{=} 2\overset{\circ}{\nabla}_{(a}\mathrm{r}^{(m)}_{b)} - \pounds_n \mathrm{Y}^{(m)}_{ab} - 2\kappa^{(m)}\mathrm{Y}_{ab} - 2m\kappa \mathrm{Y}^{(m)}_{ab}, \tag{D.17}$$

$$({}^{(3)}\underset{\smile}{\Sigma}^{(m)})_{ca} n^c \stackrel{(m)}{=} \mathrm{r}^{(m+1)}_a - 2V^b{}_a \mathrm{r}^{(m)}_b - 2P^{bc}\mathrm{r}_b \mathrm{Y}^{(m)}_{ac} - \frac{1}{2}n(\ell^{(2)})\mathrm{r}^{(m)}_a, \tag{D.18}$$

$$({}^{(3)}\Sigma^{(m)})_{ca} n^c \stackrel{(m)}{=} \mathrm{r}^{(m+1)}_a - 2V^b{}_a \mathrm{r}^{(m)}_b - 2mP^{bc}\mathrm{r}_b \mathrm{Y}^{(m)}_{ac} - \frac{m}{2}n(\ell^{(2)})\mathrm{r}^{(m)}_a. \tag{D.19}$$

*Proof.* Identity (D.12) will be a consequence of (D.1) and (D.4). The pullback of the second is, after taking into account (5.43), (D.2) and $\underset{\smile}{\Sigma}^{(m-1)}_{cab} \stackrel{(m)}{=} 0$ (by (D.1)),

$$\Sigma^{(m)}_{cab} \stackrel{(m)}{=} \underset{\smile}{\Sigma}^{(m)}_{cab} + 2(m-1)\mathrm{r}_c \mathrm{Y}^{(m)}_{ab}.$$

Replacing here $\underset{\smile}{\Sigma}^{(m)}_{cab}$ from (D.1) yields (D.12). The contraction of (D.12) with $n^c$ gives (D.17) after using

$$n^c\Big(\overset{\circ}{\nabla}_a \mathrm{Y}^{(m)}_{bc} + \overset{\circ}{\nabla}_b \mathrm{Y}^{(m)}_{ac} - \overset{\circ}{\nabla}_c \mathrm{Y}^{(m)}_{ab}\Big) = 2\overset{\circ}{\nabla}_{(a}\mathrm{r}^{(m)}_{b)} - \pounds_n \mathrm{Y}^{(m)}_{ab}.$$

To show (D.13) we simply contract (D.10) with $e_a^\beta$ and use (5.44) (recall that a tangential derivative of $\mathcal{K}^{(m-1)}$ is at most of transverse order $m-1$).

The remaining identities will rely on (D.7)-(D.9). Before applying them we need to determine the terms of the form $({}^{(i)}\nabla\mathcal{K}^{(m)})_{ab}$ and $({}^{(i,j)}\nabla\mathcal{K}^{(m)})_a$ for various values of $i, j$. For the former we use (B.2)-(B.3), and for the latter (B.5)-(B.7). Taking also into account (5.42) the result is

$$\left({}^{(1)}\nabla\mathcal{K}^{(m)}\right)_{ac} \overset{(m)}{=} \mathcal{K}^{(m+1)}_{ac} - 2V^b{}_{(a}\mathcal{K}^{(m)}_{c)b}, \qquad \left({}^{(2)}\nabla\mathcal{K}^{(m)}\right)_{[ac]} \overset{(m)}{=} V^b{}_{[c}\mathcal{K}^{(m)}_{a]b}, \tag{D.20}$$
$$\left({}^{(1,2)}\nabla\mathcal{K}^{(m)}\right)_c \overset{(m)}{=} 0, \qquad \left({}^{(2,3)}\nabla\mathcal{K}^{(m)}\right)_c \overset{(m)}{=} 0, \qquad \left({}^{(1,2,3)}\nabla\mathcal{K}^{(m-1)}\right) \overset{(m)}{=} 0.$$

Note that these expressions immediately imply

$$\left({}^{(1)}\nabla\mathcal{K}^{(m-1)}\right)_{ad} \overset{(m)}{=} \mathcal{K}^{(m)}_{ad}, \qquad \left({}^{(2)}\nabla\mathcal{K}^{(m-1)}\right)_{[da]} \overset{(m)}{=} 0,$$
$$\left({}^{(2,3)}\nabla\mathcal{K}^{(m-1)}\right)_a \overset{(m)}{=} \left({}^{(1,2)}\nabla\mathcal{K}^{(m-1)}\right)_c \overset{(m)}{=} 0. \tag{D.21}$$

With all these expressions at hand, identities (D.8) and (D.9) become, respectively,

$$({}^{(2,3)}\underset{\smile}{\Sigma}{}^{(m)})_c \overset{(m+1)}{=} 0, \qquad ({}^{(1,3)}\underset{\smile}{\Sigma}{}^{(m)})_c \overset{(m+1)}{=} 0,$$

which proves (D.14). Replacing (D.20)-(D.21) into (D.7) and substituting $\mathcal{K}^{(m)}_{ab} = 2\mathrm{Y}^{(m)}_{ab}$ gives

$$\begin{aligned}({}^{(3)}\underset{\smile}{\Sigma}{}^{(m)})_{ca} &\overset{(m)}{=} \frac{1}{2}\mathcal{K}^{(m+1)}_{ac} - V^b{}_{(a}\mathcal{K}^{(m)}_{c)b} + V^b{}_{[c}\mathcal{K}^{(m)}_{a]b} - \frac{1}{2}\left(2P^{bd}\mathrm{Y}_{bc} + \frac{1}{2}n^d\overset{\circ}{\nabla}_c\ell^{(2)}\right)\mathcal{K}^{(m)}_{ad}\\
&\overset{(m)}{=} \mathrm{Y}^{(m+1)}_{ac} - 2V^b{}_a\mathrm{Y}^{(m)}_{cb} - 2P^{bd}\mathrm{Y}_{bc}\mathrm{Y}^{(m)}_{ad} - \frac{1}{2}\mathrm{r}^{(m)}_a\overset{\circ}{\nabla}_c\ell^{(2)},\end{aligned}$$

which is (D.15). A contraction with $n^c$ gives (D.18). Now, identity (D.16) is a consequence of replacing (D.14), (D.15) and (5.43) into (D.4), namely

$$\begin{aligned}e^\mu_c e^\alpha_a \xi^\beta \Sigma^{(m)}_{\mu\alpha\beta} &\overset{(m)}{=} e^\mu_c e^\alpha_a \xi^\beta \underset{\smile}{\Sigma}{}^{(m)}_{\mu\alpha\beta} - (m-1)\mathcal{K}_\mu{}^\rho e^\mu_c e^\alpha_a \xi^\beta \underset{\smile}{\Sigma}{}^{(m-1)}_{\rho\alpha\beta}\\
&\overset{(m)}{=} ({}^{(3)}\underset{\smile}{\Sigma}{}^{(m)})_{ca} - (m-1)\left(2P^{bd}\mathrm{Y}_{cb} + \frac{1}{2}n^d\overset{\circ}{\nabla}_c\ell^{(2)}\right)({}^{(3)}\underset{\smile}{\Sigma}{}^{(m-1)})_{da} - 2\mathrm{r}_c({}^{(1,3)}\Sigma^{(m-1)})_a\\
&\overset{(m)}{=} \mathrm{Y}^{(m+1)}_{ac} - 2V^b{}_a\mathrm{Y}^{(m)}_{cb} - 2mP^{bd}\mathrm{Y}_{bc}\mathrm{Y}^{(m)}_{ad} - \frac{m}{2}\mathrm{r}^{(m)}_a\overset{\circ}{\nabla}_c\ell^{(2)}.\end{aligned}$$

Finally, (D.19) is its contraction with $n^c$. □

We already have all the necessary ingredients to compute $\dot{\mathcal{R}}^{(m)}_a$ up to order $m$.

**Proposition D.2.** *Let $\mathcal{H}$ be a null hypersurface $(\Phi,\xi)$-embedded in $(\mathcal{M},g)$ and extend $\xi$ off $\Phi(\mathcal{H})$ by $\nabla_\xi\xi = 0$. Then for any $m \geq 2$,*

$$\begin{aligned}\dot{\mathcal{R}}^{(m)}_a &\overset{(m)}{=} \mathrm{r}^{(m+1)}_a + P^{bc}\overset{\circ}{\nabla}_b\mathrm{Y}^{(m)}_{ac} - 2mP^{bc}\mathrm{r}_b\mathrm{Y}^{(m)}_{ac} - \overset{\circ}{\nabla}_a(\mathrm{tr}_P\,\mathbf{Y}^{(m)}) + (\mathrm{tr}_P\,\mathbf{Y}^{(m)})(\mathrm{r}_a - \mathrm{s}_a)\\
&\quad + (P^{bc}\mathrm{Y}_{ab} - 3V^c{}_a)\mathrm{r}^{(m)}_c + \left(\mathrm{tr}_P\,\mathbf{Y} - \frac{m}{2}n(\ell^{(2)})\right)\mathrm{r}^{(m)}_a,\end{aligned} \tag{D.22}$$

*and consequently*

$$\begin{aligned}\dot{\mathcal{R}}^{(m)}_a n^a \overset{(m)}{=} & -\kappa^{(m+1)} + P^{bc}\mathring{\nabla}_b \mathrm{r}^{(m)}_c - P^{bc}P^{ad}\mathrm{U}_{bd}\mathrm{Y}^{(m)}_{ac} - \mathcal{L}_n(\mathrm{tr}_P\,\mathbf{Y}^{(m)}) - \kappa\,\mathrm{tr}_P\,\mathbf{Y}^{(m)} \\ & - 2P^{bc}\Big((m+1)\mathrm{r}_b + 2\mathrm{s}_b\Big)\mathrm{r}^{(m)}_c + \left(\frac{m+3}{2}n(\ell^{(2)}) - \mathrm{tr}_P\,\mathbf{Y}\right)\kappa^{(m)}.\end{aligned} \tag{D.23}$$

*Proof.* From (D.11),

$$\begin{aligned}\widehat{e}^{\alpha}_a\xi^{\beta}\mathcal{L}^{(m-1)}_{\xi}R_{\alpha\beta} & \overset{[m]}{=} \widehat{e}^{\alpha}_a\xi^{\beta}\nabla_{\mu}\Sigma^{(m-1)\mu}{}_{\alpha\beta} - \widehat{e}^{\alpha}_a\xi^{\beta}\nabla_{\beta}\Sigma^{(m-1)\mu}{}_{\mu\alpha} \\ & \overset{[m]}{=} \widehat{e}^{\alpha}_a\nabla_{\mu}\Big(\xi^{\beta}\Sigma^{(m-1)\mu}{}_{\alpha\beta}\Big) - \widehat{e}^{\alpha}_a\Sigma^{(m-1)\mu}{}_{\alpha\beta}\nabla_{\mu}\xi^{\beta} - \widehat{e}^{\alpha}_a\xi^{\beta}\nabla_{\beta}\Big(\Sigma^{(m-1)\mu}{}_{\mu\alpha}\Big).\end{aligned} \tag{D.24}$$

We evaluate each term separately. For the first we apply Prop. B.3 to $T = {}^{(3)}\Sigma^{(m-1)}$ and take into account $({}^{(3)}\Sigma^{(m-1)})_{ab} \overset{(m)}{=} \mathrm{Y}^{(m)}_{ab}$ (by (D.2) and (D.5)) and ${}^{(2,3)}\Sigma^{(m-1)} \overset{(m)}{=} {}^{(1,3)}\Sigma^{(m-1)} \overset{(m)}{=} 0$ (by (D.14) and (D.5)). Thus,

$$\begin{aligned}\widehat{e}^{\alpha}_a\nabla_{\mu}\Big(\xi^{\beta}\Sigma^{(m-1)\mu}{}_{\alpha\beta}\Big) \overset{(m)}{=} & \; n^b\widehat{e}^{\alpha}_a\widehat{e}^{\mu}_b\xi^{\beta}\mathcal{L}_{\xi}\Sigma^{(m-1)}_{\mu\alpha\beta} + P^{bc}\mathring{\nabla}_b\mathrm{Y}^{(m)}_{ca} + (\mathrm{tr}_P\,\mathbf{Y} - n(\ell^{(2)}))\mathrm{r}^{(m)}_a \\ & - 2P^{bc}(\mathrm{r}+\mathrm{s})_b\mathrm{Y}^{(m)}_{ca} + P^{bc}\mathrm{Y}_{ba}\mathrm{r}^{(m)}_c - V^c{}_a\mathrm{r}^{(m)}_c.\end{aligned} \tag{D.25}$$

Only the first term needs further analysis. Note that

$$\mathcal{L}_{\xi}\Sigma^{(m-1)}_{\mu\alpha\beta} = \mathcal{L}_{\xi}\Big(g_{\mu\nu}\Sigma^{(m-1)\nu}{}_{\alpha\beta}\Big) = \mathcal{K}_{\mu\nu}\Sigma^{(m-1)\nu}{}_{\alpha\beta} + g_{\mu\nu}\Sigma^{(m)\nu}{}_{\alpha\beta} \underset{(\mathrm{D.4})}{\overset{[m]}{=}} \underset{\smile}{\Sigma}^{(m)}_{\mu\alpha\beta} - (m-2)\mathcal{K}_{\mu}{}^{\rho}\underset{\smile}{\Sigma}^{(m-1)}_{\rho\alpha\beta},$$

so

$$n^b e^{\alpha}_a e^{\mu}_b\xi^{\beta}\mathcal{L}_{\xi}\Sigma^{(m-1)}_{\mu\alpha\beta} \overset{[m]}{=} n^b e^{\alpha}_a e^{\mu}_b\xi^{\beta}\underset{\smile}{\Sigma}^{(m)}_{\mu\alpha\beta} - (m-2)n^b e^{\alpha}_a e^{\mu}_b\xi^{\beta}\mathcal{K}_{\mu}{}^{\rho}\underset{\smile}{\Sigma}^{(m-1)}_{\rho\alpha\beta}. \tag{D.26}$$

The first term in the right hand side is directly given by (D.18), and for the second one we use

$$\begin{aligned}n^b e^{\alpha}_a e^{\mu}_b\xi^{\beta}\mathcal{K}_{\mu}{}^{\rho}\underset{\smile}{\Sigma}^{(m-1)}_{\rho\alpha\beta} & \underset{(5.43)}{=} \left(2P^{cd}\mathrm{r}_c + \frac{1}{2}n(\ell^{(2)})n^d\right)({}^{(3)}\underset{\smile}{\Sigma}^{(m-1)})_{da} + 2\mathrm{r}_b({}^{(1,3)}\underset{\smile}{\Sigma}^{(m-1)})_a \\ & \underset{(\mathrm{D.2})}{\overset{(m)}{=}} 2P^{cd}\mathrm{r}_c\mathrm{Y}^{(m)}_{da} + \frac{1}{2}n(\ell^{(2)})\mathrm{r}^{(m)}_a,\end{aligned}$$

where in the second line we also used (D.14). Replacing this and (D.18) into (D.26) gives

$$n^b e^{\alpha}_a e^{\mu}_b\xi^{\beta}\mathcal{L}_{\xi}\Sigma^{(m-1)}_{\mu\alpha\beta} \overset{(m)}{=} \mathrm{r}^{(m+1)}_a - 2V^c{}_a\mathrm{r}^{(m)}_c - 2(m-1)P^{bc}\mathrm{r}_b\mathrm{Y}^{(m)}_{ca} - \frac{1}{2}(m-1)n(\ell^{(2)})\mathrm{r}^{(m)}_a,$$

and hence (D.25) takes the form

$$\begin{aligned}\widehat{e}^{\alpha}_a\nabla_{\mu}(\xi^{\beta}\Sigma^{(m-1)\mu}{}_{\alpha\beta}) \overset{(m)}{=} & \; \mathrm{r}^{(m+1)}_a - 3V^c{}_a\mathrm{r}^{(m)}_c - 2P^{bc}(m\mathrm{r}_b + \mathrm{s}_b)\mathrm{Y}^{(m)}_{ca} \\ & + \left(\mathrm{tr}_P\,\mathbf{Y} - \frac{m+1}{2}n(\ell^{(2)})\right)\mathrm{r}^{(m)}_a + P^{bc}\mathring{\nabla}_b\mathrm{Y}^{(m)}_{ca} + P^{bc}\mathrm{Y}_{ab}\mathrm{r}^{(m)}_c.\end{aligned} \tag{D.27}$$

For the second term of (D.24) we insert (2.72) and use that $e^\alpha_a \xi^\mu \xi^\beta \Sigma^{(m-1)}_{\mu\alpha\beta} \stackrel{(m)}{=} 0$ (by (D.14)) and $e^\mu_c e^\alpha_a e^\beta_b \Sigma^{(m-1)}_{\mu\alpha\beta} \stackrel{(m)}{=} 0$ (by (D.1)), obtaining

$$
\begin{aligned}
-g^{\mu\rho}\widehat{e}^{\alpha}_a \Sigma^{(m-1)}_{\mu\alpha\beta}\nabla_\rho \xi^\beta &\stackrel{(m)}{=} -P^{cd}(\mathrm{r}-\mathrm{s})_c ({}^{(3)}\Sigma^{(m-1)})_{da} - V^b{}_c n^c ({}^{(1)}\Sigma^{(m-1)})_{ab} \\
&\underset{(\mathrm{D}.2)}{\stackrel{(m)}{=}} 2P^{bc}\mathrm{s}_b \mathrm{Y}^{(m)}_{ca} + \frac{1}{2} n(\ell^{(2)}) \mathrm{r}^{(m)}_a,
\end{aligned}
\tag{D.28}
$$

where in the second equality we also inserted (2.57). Finally, the last term in (D.24) is

$$
\begin{aligned}
-\widehat{e}^{\alpha}_a \xi^\beta \nabla_\beta (\Sigma^{(m-1)\mu}{}_{\mu\alpha}) &= -e^\alpha_a \pounds_\xi \Sigma^{(m-1)\mu}{}_{\mu\alpha} + e^\alpha_a \Sigma^{(m-1)\mu}{}_{\mu\beta} \nabla_\alpha \xi^\beta \\
&= -e^\alpha_a \Sigma^{(m)\mu}{}_{\mu\alpha} + e^\alpha_a \Sigma^{(m-1)\mu}{}_{\mu\beta}\nabla_\alpha \xi^\beta \\
&\stackrel{(m)}{=} -\overset{\circ}{\nabla}_a (\operatorname{tr}_P \mathbf{Y}^{(m)}) + \operatorname{tr}_P \mathbf{Y}^{(m)} (\mathrm{r}_a - \mathrm{s}_a),
\end{aligned}
\tag{D.29}
$$

where in the last step we used (2.53) with $n^{(2)} = 0$ for $e^\alpha_a \nabla_\alpha \xi^\beta$ and recalled (D.13) and (D.3). Equation (D.22) is now obtained by simply inserting (D.27)-(D.29) into (D.24). To prove (D.23) it suffices to contract (D.22) with $n^a$ and use (2.66) and (2.57). □

We conclude the appendix with the explicit expressions for $\Sigma_{cab}$, $({}^{(3)}\Sigma)_{ca}$ and $({}^{(2,3)}\Sigma)_c$ and their contractions with $n^c$. We emphasize that this result is not contained in Proposition D.1 because, as usual, the lowest values of $m$ require a different treatment.

**Proposition D.3.** *Let $\{\mathcal{H}, \boldsymbol{\gamma}, \boldsymbol{\ell}, \ell^{(2)}\}$ be null metric hypersurface data $(\Phi, \xi)$-embedded in $(\mathcal{M}, g)$ and extend $\xi$ off $\Phi(\mathcal{H})$ by $\nabla_\xi \xi = 0$. Then,*

$$
\Sigma_{cab} = 2\overset{\circ}{\nabla}_{(a}\mathrm{Y}_{b)c} - \overset{\circ}{\nabla}_c \mathrm{Y}_{ab} + 2\mathrm{r}_c \mathrm{Y}_{ab} + \frac{1}{2}\mathrm{U}_{ab}\overset{\circ}{\nabla}_c \ell^{(2)}, \tag{D.30}
$$

$$
({}^{(3)}\Sigma)_{ca} = \mathrm{Y}^{(2)}_{ac} - \frac{1}{2}(\mathrm{r}-\mathrm{s})_a \overset{\circ}{\nabla}_c \ell^{(2)} - 2V^d{}_a \mathrm{Y}_{cd}, \tag{D.31}
$$

$$
({}^{(2,3)}\Sigma)_c = 0. \tag{D.32}
$$

*As a consequence,*

$$
\Sigma_{cab} n^c = -\pounds_n \mathrm{Y}_{ab} - 2\kappa \mathrm{Y}_{ab} + 2\overset{\circ}{\nabla}_{(a}\mathrm{r}_{b)} + \frac{1}{2} n(\ell^{(2)}) \mathrm{U}_{ab}, \tag{D.33}
$$

$$
({}^{(3)}\Sigma)_{ca} n^c = \mathrm{r}^{(2)}_a - \frac{1}{2} n(\ell^{(2)}) \Big(\mathrm{r}-\mathrm{s}\Big)_a - 2V^c{}_a \mathrm{r}_c, \tag{D.34}
$$

$$
({}^{(2,3)}\Sigma)_c n^c = 0. \tag{D.35}
$$

*Proof.* Identities (D.30)-(D.31) follow from (4.28), (4.34), (4.37) and (4.47) after setting $\mathcal{K}_{ab} = 2\mathrm{Y}_{ab}$, $\varpi_a = \frac{1}{2}\overset{\circ}{\nabla}_a \ell^{(2)}$, $\mathfrak{p} = 0$, $\varpi^{(2)}_a = 0$ (see (5.42)), $C = 1$ and $\bar{\eta} = 0$. The contraction with $n^c$ given in (D.33) is obtained from this after using $n^c \overset{\circ}{\nabla}_c \mathrm{Y}_{ab} = \pounds_n \mathrm{Y}_{ab} - 2\mathrm{Y}_{c(a}\overset{\circ}{\nabla}_{b)} n^c$ and (2.66). Finally, (D.34) and (D.35) are the contraction of (D.31)-(D.32) with $n^c$. □

E

# GENERAL EXPRESSIONS AT NULL INFINITY

In this appendix we write down explicitly the tensors (6.61)-(6.62) for every $m \geq 1$ making the leading order terms explicit. In $d = \mathfrak{n}+1$ dimensions, the definition (6.2) of the Schouten tensor in terms of the Ricci yields, after recalling (5.34),

$$L^{(m)}_{\alpha\beta} = \frac{1}{\mathfrak{n}-1}\left(R^{(m)}_{\alpha\beta} - \frac{1}{2\mathfrak{n}}\sum_{i=0}^{m-1}\binom{m-1}{i}R^{(i+1)}\pounds^{(m-1-i)}_{\xi}g_{\alpha\beta}\right). \tag{E.1}$$

From Remark 5.31 one has

$$L^{(m)}_{\alpha\beta} \overset{[m-1]}{=} \frac{1}{\mathfrak{n}-1}\left(R^{(m)}_{\alpha\beta} - \frac{1}{2\mathfrak{n}}\left(R^{(m)}g_{\alpha\beta} + (m-1)R^{(m-1)}\mathcal{K}_{\alpha\beta} + \frac{(m-1)(m-2)}{2}R^{(m-2)}\mathcal{K}^{(2)}_{\alpha\beta}\right)\right), \tag{E.2}$$

$$L^{(m)}_{\alpha\beta} \overset{[m]}{=} \frac{1}{\mathfrak{n}-1}\left(R^{(m)}_{\alpha\beta} - \frac{1}{2\mathfrak{n}}\left(R^{(m)}g_{\alpha\beta} + (m-1)R^{(m-1)}\mathcal{K}_{\alpha\beta}\right)\right), \tag{E.3}$$

and

$$L^{(m)}_{\alpha\beta} \overset{[m+1]}{=} \frac{1}{\mathfrak{n}-1}\left(R^{(m)}_{\alpha\beta} - \frac{1}{2\mathfrak{n}}R^{(m)}g_{\alpha\beta}\right), \tag{E.4}$$

$$L^{(m-1)}_{\alpha\beta} \overset{[m+1]}{=} 0. \tag{E.5}$$

In the following proposition we compute the contractions of $L^{(m)}_{\alpha\beta}$ that will be needed below. Notice that some contractions are computed up to order $m$, and some others just to order $m+1$, depending on our needs for the rest of the appendix.

**Proposition E.1.** *Let $\{\mathcal{H},\boldsymbol{\gamma},\boldsymbol{\ell},\ell^{(2)}\}$ be null metric hypersurface data $(\Phi,\xi)$-embedded in $(\mathcal{M},g)$ and extend $\xi$ off $\Phi(\mathcal{H})$ by $\nabla_\xi\xi = 0$. Let $L_{\alpha\beta}$ be the Schouten tensor, $L^{(m)}_{\alpha\beta} = \pounds^{(m-1)}_{\xi}L_{\alpha\beta}$ and $m \geq 2$. Then,*

$$\begin{aligned}
L^{(m)}_{ab} \overset{(m)}{=} \frac{1}{\mathfrak{n}-1}\Big(&-2\pounds_n \mathrm{Y}^{(m)}_{ab} - (2m\kappa + \operatorname{tr}_P \mathbf{U})\mathrm{Y}^{(m)}_{ab} - (\operatorname{tr}_P \mathbf{Y}^{(m)})\mathrm{U}_{ab} + 4P^{cd}\mathrm{U}_{c(a}\mathrm{Y}^{(m)}_{b)d}\\
&+ 4(\mathrm{s}-\mathrm{r})_{(a}\mathrm{r}^{(m)}_{b)} + 2\overset{\circ}{\nabla}_{(a}\mathrm{r}^{(m)}_{b)} + \frac{2(m-\mathfrak{n}-1)}{\mathfrak{n}}\kappa^{(m)}\mathrm{Y}_{ab}\Big)\\
+ \frac{1}{\mathfrak{n}(\mathfrak{n}-1)}\Big(&\kappa^{(m+1)} + 2\pounds_n(\operatorname{tr}_P \mathbf{Y}^{(m)}) + (2m\kappa + \operatorname{tr}_P \mathbf{U})\operatorname{tr}_P \mathbf{Y}^{(m)} + P^{ef}P^{cd}\mathrm{U}_{ec}\mathrm{Y}^{(m)}_{fd}\\
&- 2\operatorname{div}_P \mathbf{r}^{(m)} + 2P\Big((2m+1)\mathbf{r} + 3\mathbf{s}, \mathbf{r}^{(m)}\Big) + \Big(2\operatorname{tr}_P \mathbf{Y} - (m+2)n(\ell^{(2)})\Big)\kappa^{(m)}\Big)\gamma_{ab},
\end{aligned} \tag{E.6}$$

$$
\begin{aligned}
\overset{\bullet}{L}{}^{(m)}_a \overset{(m)}{=} \frac{1}{\mathfrak{n}-1}\Big( & \mathrm{r}^{(m+1)}_a + P^{bc}\overset{\circ}{\nabla}_b \mathrm{Y}^{(m)}_{ac} - 2mP^{bc}\mathrm{r}_b\mathrm{Y}^{(m)}_{ac} - \overset{\circ}{\nabla}_a(\mathrm{tr}_P\,\mathbf{Y}^{(m)}) + (\mathrm{tr}_P\,\mathbf{Y}^{(m)})(\mathrm{r}_a - \mathrm{s}_a) \\
& + (P^{bc}\mathrm{Y}_{ab} - 3V^c{}_a)\mathrm{r}^{(m)}_c + \Big(\mathrm{tr}_P\,\mathbf{Y} - \frac{m}{2}n(\ell^{(2)})\Big)\,\mathrm{r}^{(m)}_a + \frac{m-1}{2\mathfrak{n}}\kappa^{(m)}\overset{\circ}{\nabla}_a\ell^{(2)}\Big) \\
& + \frac{1}{\mathfrak{n}(\mathfrak{n}-1)}\Big(\kappa^{(m+1)} + 2\pounds_n(\mathrm{tr}_P\,\mathbf{Y}^{(m)}) + (2m\kappa + \mathrm{tr}_P\,\mathbf{U})\,\mathrm{tr}_P\,\mathbf{Y}^{(m)} + P^{ef}P^{cd}\mathrm{U}_{ec}\mathrm{Y}^{(m)}_{fd} \\
& - 2\,\mathrm{div}_P\,\mathbf{r}^{(m)} + 2P\Big((2m+1)\mathbf{r} + 3\mathbf{s}, \mathbf{r}^{(m)}\Big) + \Big(2\,\mathrm{tr}_P\,\mathbf{Y} - (m+2)n(\ell^{(2)})\Big)\kappa^{(m)}\Big)\ell_a,
\end{aligned}
\tag{E.7}
$$

$$
\overset{\bullet\bullet}{L}{}^{(m)} \overset{(m+1)}{=} \frac{1}{\mathfrak{n}-1}\left(-\,\mathrm{tr}_P\,\mathbf{Y}^{(m+1)} + \frac{\kappa^{(m+1)}}{\mathfrak{n}}\ell^{(2)}\right). \tag{E.8}
$$

*In particular,*

$$
L^{(m)}_{ab}n^b \overset{(m)}{=} \frac{1}{\mathfrak{n}-1}\left(-\pounds_n\mathrm{r}^{(m)}_a - (2(m-1)\kappa + \mathrm{tr}_P\,\mathbf{U})\mathrm{r}^{(m)}_a - \overset{\circ}{\nabla}_a\kappa^{(m)} + \frac{2(m-1)}{\mathfrak{n}}\kappa^{(m)}\mathrm{r}_a\right), \tag{E.9}
$$

$$
\begin{aligned}
\overset{\bullet}{L}{}^{(m)}_a n^a \overset{(m)}{=} \frac{1}{\mathfrak{n}(\mathfrak{n}-1)}\Bigg( & (1-\mathfrak{n})\kappa^{(m+1)} + (2-\mathfrak{n})\pounds_n(\mathrm{tr}_P\,\mathbf{Y}^{(m)}) + \Big((2m-\mathfrak{n})\kappa + \mathrm{tr}_P\,\mathbf{U}\Big)\,\mathrm{tr}_P\,\mathbf{Y}^{(m)} \\
& + (1-\mathfrak{n})P^{bc}P^{ad}\mathrm{U}_{bd}\mathrm{Y}^{(m)}_{ac} + 2P\Big(((2-\mathfrak{n})m + 1 - \mathfrak{n})\mathbf{r} + (3-2\mathfrak{n})\mathbf{s}, \mathbf{r}^{(m)}\Big) \\
& + (\mathfrak{n}-2)\,\mathrm{div}_P\,\mathbf{r}^{(m)} + \left((2-\mathfrak{n})\,\mathrm{tr}_P\,\mathbf{Y} + \frac{3\mathfrak{n} + m(\mathfrak{n}-1) - 5}{2}n(\ell^{(2)})\right)\kappa^{(m)}\Bigg),
\end{aligned}
\tag{E.10}
$$

*and*

$$
L^{(m)}_{ab} \overset{(m+1)}{=} \frac{\kappa^{(m+1)}}{\mathfrak{n}(\mathfrak{n}-1)}\gamma_{ab}, \tag{E.11}
$$

$$
\overset{\bullet}{L}{}^{(m)}_a \overset{(m+1)}{=} \frac{1}{\mathfrak{n}-1}\left(\mathrm{r}^{(m+1)}_a + \frac{\kappa^{(m+1)}}{\mathfrak{n}}\ell_a\right), \tag{E.12}
$$

$$
\overset{\bullet}{L}{}^{(m)}_a n^a \overset{(m+1)}{=} -\frac{\kappa^{(m+1)}}{\mathfrak{n}}, \tag{E.13}
$$

$$
P^{bc}L^{(m)}_{ab} \overset{(m+1)}{=} \frac{\kappa^{(m+1)}}{\mathfrak{n}(\mathfrak{n}-1)}(\delta^c_a - n^c\ell_a). \tag{E.14}
$$

*Proof.* The contraction of (E.3) with $e^\alpha_a e^\beta_b$ gives (E.6) after replacing (5.84), (6.55) and (6.56) and recalling that $\mathcal{K}_{ab} = 2\mathrm{Y}_{ab}$. Similarly, (E.7) is obtained by contracting (E.3) with $e^\alpha_a\xi^\beta$ and using (6.42), (6.55)-(6.56) and $({}^{(1)}\mathcal{K})_a = \frac{1}{2}\overset{\circ}{\nabla}_a\ell^{(2)}$ (cf. (5.42)). To prove (E.8) we simply contract (E.4) with $\xi$ twice and insert (6.41) and (6.55). Relations (E.9) and (E.10) are respectively the contractions of (E.6) and (E.7) with $n$. They are obtained after using $\gamma_{ab}n^b = \mathrm{U}_{ab}n^b = 0$, $\mathrm{s}_b n^b = 0$ and $\ell_a n^a = 1$ together with the definitions of $\kappa$, $\mathrm{r}_a$, $\kappa^{(m)}$ and $\mathrm{r}^{(m)}_a$. In addition, (E.9) uses identity (2.46) (for $n^{(2)} = 0$) applied to $\theta_a = \mathrm{r}_a$, and (E.10) employs (2.66) and (2.57). Finally, relations (E.11)-(E.14) are immediate from the previous ones by simply keeping the quantities of order $m+1$. □

The tensors $L_{ab}$, $\overset{\bullet}{L}_a$ and $\overset{\bullet\bullet}{L}$ require a separate analysis. The starting point is (E.1) with $m = 1$ (this is an exact relation), which we contract with $e^\alpha_a e^\beta_b$, $e^\alpha_a\xi^\beta$ and $\xi^\alpha\xi^\beta$. Taking into account (6.47), (6.48) and (6.49) (we do not replace $R$ from (6.54) since this will not be needed) and recalling $g_{\alpha\beta}e^\alpha_a e^\beta_b = \gamma_{ab}$, $g_{\alpha\beta}e^\alpha_a\xi^\beta = \ell_a$ and $g_{\alpha\beta}\xi^\alpha\xi^\beta = \ell^{(2)}$, the result is

$$L^{(1)}_{ab} = \frac{1}{\mathfrak{n}-1}\bigg(\mathring{R}_{(ab)} - 2\pounds_n \mathrm{Y}_{ab} - (2\kappa + \operatorname{tr}_P \mathbf{U})\mathrm{Y}_{ab} + \mathring{\nabla}_{(a}\left(\mathrm{s}_{b)} + 2\mathrm{r}_{b)}\right) - 2\mathrm{r}_a\mathrm{r}_b + 4\mathrm{r}_{(a}\mathrm{s}_{b)} - \mathrm{s}_a\mathrm{s}_b$$

$$- (\operatorname{tr}_P \mathbf{Y})\mathrm{U}_{ab} + 2P^{cd}\mathrm{U}_{d(a}\left(2\mathrm{Y}_{b)c} + \mathrm{F}_{b)c}\right) - \frac{R}{2\mathfrak{n}}\gamma_{ab}\bigg). \tag{E.15}$$

$$\dot{L}^{(1)}_a = \frac{1}{\mathfrak{n}-1}\left(\mathrm{r}^{(2)}_a - P^{bc}A_{bca} - P^{bc}(\mathrm{r}+\mathrm{s})_b(\mathrm{Y}+\mathrm{F})_{ac} + \frac{1}{2}\kappa\mathring{\nabla}_a\ell^{(2)} - \frac{1}{2}n(\ell^{(2)})(\mathrm{r}-\mathrm{s})_a - \frac{R}{2\mathfrak{n}}\ell_a\right), \tag{E.16}$$

$$\ddot{L}^{(1)} = \frac{1}{\mathfrak{n}-1}\left(-P^{ab}\mathrm{Y}^{(2)}_{ab} + P^{ab}P^{cd}(\mathrm{Y}+\mathrm{F})_{ac}(\mathrm{Y}+\mathrm{F})_{bd} + P^{ab}(\mathrm{r}-\mathrm{s})_a\mathring{\nabla}_b\ell^{(2)} - \frac{R}{2\mathfrak{n}}\ell^{(2)}\right). \tag{E.17}$$

Using these identities we can now compute $\mathcal{L}^{(m+1)}_a$ and $\dot{\mathcal{L}}^{(m+1)}$ up to order $m+1$. Taking $m$ transverse derivatives of $\mathcal{L}_\alpha := (\mathfrak{n}-1)L_{\alpha\beta}\nabla^\beta\Omega$ and using (5.34),

$$\begin{aligned}
\mathcal{L}^{(m+1)}_\alpha &= (\mathfrak{n}-1)\sum_{i=0}^{m}\binom{m}{i}L^{(i+1)}_{\alpha\beta}\pounds^{(m-i)}_\xi g^{\beta\mu}\nabla_\mu\Omega \\
&= (\mathfrak{n}-1)\sum_{i=0}^{m}\sum_{j=0}^{m-i}\binom{m}{i}\binom{m-i}{j}L^{(i+1)}_{\alpha\beta}(\pounds^{(j)}_\xi g^{\beta\mu})\nabla_\mu\pounds^{(m-i-j)}_\xi\Omega,
\end{aligned} \tag{E.18}$$

so (recall $L^{(k)}_{\alpha\beta} \overset{[k+2]}{=} 0$, by (E.5))

$$\mathcal{L}^{(m+1)}_\alpha \overset{[m+1]}{=} (\mathfrak{n}-1)\left(L^{(m+1)}_{\alpha\beta}g^{\beta\mu}\nabla_\mu\Omega + mL^{(m)}_{\alpha\beta}\left(g^{\beta\mu}\nabla_\mu\pounds_\xi\Omega + (\pounds_\xi g^{\beta\mu})\nabla_\mu\Omega\right)\right).$$

Now we note that $g^{\beta\mu}\nabla_\mu\Omega \overset{\mathscr{I}}{=} \sigma^{(1)}\nu^\beta$ (by (2.14)) and

$$\begin{aligned}
g^{\beta\mu}\nabla_\mu\pounds_\xi\Omega + (\pounds_\xi g^{\beta\mu})\nabla_\mu\Omega &\overset{\mathscr{I}}{=} g^{\beta\mu}\nabla_\mu\pounds_\xi\Omega - \mathcal{K}^{\beta\mu}\nabla_\mu\Omega \\
&\underset{(2.13)}{\overset{\mathscr{I}}{=}} \left(P^{bc}e^\beta_b e^\mu_c + \xi^\beta\nu^\mu + \xi^\mu\nu^\beta\right)\nabla_\mu\pounds_\xi\Omega - \sigma^{(1)}\nu^\mu\mathcal{K}_\mu{}^\beta \\
&\underset{(5.45)}{\overset{\mathscr{I}}{=}} \left(P^{bc}\left(\mathring{\nabla}_c\sigma^{(1)} - 2\sigma^{(1)}\mathrm{r}_c\right) + n^b\left(\sigma^{(2)} - \frac{1}{2}\sigma^{(1)}n(\ell^{(2)})\right)\right)e^\beta_b \\
&\quad + \left(\pounds_n\sigma^{(1)} + 2\sigma^{(1)}\kappa\right)\xi^\beta.
\end{aligned}$$

Hence,

$$\mathcal{L}^{(m+1)}_\alpha \underset{\mathscr{I}}{\overset{[m+1]}{=}} (\mathfrak{n}-1)\bigg(\sigma^{(1)}L^{(m+1)}_{\alpha\beta}n^b e^\beta_b + m\left(\pounds_n\sigma^{(1)} + 2\sigma^{(1)}\kappa\right)\xi^\beta L^{(m)}_{\alpha\beta}$$

$$+ m\left(P^{bc}\left(\mathring{\nabla}_c\sigma^{(1)} - 2\sigma^{(1)}\mathrm{r}_c\right) + n^b\left(\sigma^{(2)} - \frac{1}{2}\sigma^{(1)}n(\ell^{(2)})\right)\right)e^\beta_b L^{(m)}_{\alpha\beta}\bigg). \tag{E.19}$$

**Proposition E.2.** *Let* $\mathcal{L}_a^{(m+1)}$ *and* $\dot{\mathcal{L}}^{(m+1)}$ *be defined as in* (6.62). *Then,*

$$\begin{aligned}
\frac{2\mathfrak{n}}{\sigma^{(1)}}\dot{\mathcal{L}}^{(1)} &= -2(\mathfrak{n}-1)\kappa^{(2)} - 2(\mathfrak{n}-2)\pounds_n(\operatorname{tr}_P \mathbf{Y}) + 2\Big(\operatorname{tr}_P \mathbf{U} - (\mathfrak{n}-2)\kappa\Big)\operatorname{tr}_P \mathbf{Y}\\
&\quad - 2(\mathfrak{n}-1)P^{ab}P^{cd}\mathrm{U}_{ac}\mathrm{Y}_{bd} + 2(\mathfrak{n}-2)\operatorname{div}\mathbf{r} + (2\mathfrak{n}-3)\operatorname{div}\mathbf{s} - 2(2\mathfrak{n}-3)P(\mathbf{r},\mathbf{r})\\
&\quad - (4\mathfrak{n}-5)P(\mathbf{s},\mathbf{s}) - 4(2\mathfrak{n}-3)P(\mathbf{r},\mathbf{s}) + \Big(2(2\mathfrak{n}-3)\kappa + (\mathfrak{n}-1)\operatorname{tr}_P \mathbf{U}\Big)n(\ell^{(2)})\\
&\quad - \operatorname{tr}_P \mathring{R}, \qquad\qquad (\mathrm{E}.20)\\
\frac{1}{\sigma^{(1)}}\mathcal{L}_a^{(1)} &= \mathcal{R}_{ab}n^b = -\pounds_n(\mathrm{r}_b - \mathrm{s}_b) - \mathring{\nabla}_b\kappa - (\operatorname{tr}_P \mathbf{U})(\mathrm{r}_b - \mathrm{s}_b) - \mathring{\nabla}_b(\operatorname{tr}_P \mathbf{U})\\
&\quad + P^{cd}\mathring{\nabla}_c\mathrm{U}_{bd} - 2P^{cd}\mathrm{U}_{bd}s_c, \qquad\qquad (\mathrm{E}.21)
\end{aligned}$$

*and for every* $m \geq 1$,

$$\mathcal{L}_a^{(m+1)} \overset{(m+1)}{=} -\sigma^{(1)}\pounds_n\mathrm{r}_a^{(m+1)} + \Big(m\pounds_n\sigma^{(1)} - \operatorname{tr}_P \mathbf{U}\Big)\,\mathrm{r}_a^{(m+1)} - \sigma^{(1)}\mathring{\nabla}_a\kappa^{(m+1)} + \frac{m\kappa^{(m+1)}}{\mathfrak{n}}\mathring{\nabla}_a\sigma^{(1)}, \qquad (\mathrm{E}.22)$$

$$\begin{aligned}
\mathfrak{n}\dot{\mathcal{L}}^{(m+1)} \overset{(m+1)}{=}& -(\mathfrak{n}-1)\sigma^{(1)}\kappa^{(m+2)} - (\mathfrak{n}-2)\sigma^{(1)}\pounds_n(\operatorname{tr}_P \mathbf{Y}^{(m+1)}) - (\mathfrak{n}-1)\sigma^{(1)}P^{ab}P^{cd}\mathrm{U}_{ac}\mathrm{Y}_{bd}^{(m+1)}\\
&+ \Big(\Big(2m(1-\mathfrak{n}) + 2 - \mathfrak{n}\Big)\sigma^{(1)}\kappa - m\mathfrak{n}\pounds_n\sigma^{(1)} + \sigma^{(1)}\operatorname{tr}_P \mathbf{U}\Big)\operatorname{tr}_P \mathbf{Y}^{(m+1)} + \dot{\mathscr{R}}_{\mathcal{L}}^{(m)}, \qquad (\mathrm{E}.23)
\end{aligned}$$

*where* $\dot{\mathscr{R}}_{\mathcal{L}}^{(m)}$ *is an explicit tensor that depends linearly on* $\mathbf{r}^{(m+1)}$ *as well as on lower order transverse derivatives and that we do not write for simplicity (it can be easily read out by making explicit all the calculations in the proof).*

*Proof.* The definition $\mathcal{L}_\alpha := \Big(R_{\alpha\beta} - \frac{1}{2\mathfrak{n}}Rg_{\alpha\beta}\Big)\nabla^\beta\Omega$ and $\nabla_\mu\Omega \overset{\mathscr{I}}{=} \sigma^{(1)}\nu_\mu$ gives

$$\mathcal{L}_\alpha^{(1)} = \sigma^{(1)}\left(R_{\alpha\beta} - \frac{1}{2\mathfrak{n}}Rg_{\alpha\beta}\right)n^b e_b^\beta.$$

The contraction with $\xi^\alpha$ yields (E.20) after replacing (6.50) and (6.54), and that with $e_a^\alpha$ gives, after inserting (6.51), (E.21). To obtain (E.22) we contract (E.19) with $e_a^\alpha$ and use $L_{ab}^{(m)}n^b \overset{(m+1)}{=} 0$ (by (E.11)),

$$\frac{1}{\mathfrak{n}-1}\mathcal{L}_a^{(m+1)} \overset{(m+1)}{=} \sigma^{(1)}L_{ab}^{(m+1)}n^b + m\Big(\pounds_n\sigma^{(1)} + 2\sigma^{(1)}\kappa\Big)\dot{L}_a^{(m)} + mP^{bc}\Big(\mathring{\nabla}_c\sigma^{(1)} - 2\sigma^{(1)}\mathrm{r}_c\Big)L_{ab}^{(m)}.$$

This becomes (E.22) after inserting (E.9), (E.12) and (E.14). In order to prove (E.23) we contract (E.19) with $\xi^\alpha$,

$$\begin{aligned}
\frac{1}{\mathfrak{n}-1}\dot{\mathcal{L}}^{(m+1)} \overset{(m+1)}{=}& \sigma^{(1)}\dot{L}_b^{(m+1)}n^b + m\Big(\pounds_n\sigma^{(1)} + 2\sigma^{(1)}\kappa\Big)\ddot{L}^{(m)} + mP^{bc}\Big(\mathring{\nabla}_c\sigma^{(1)} - 2\sigma^{(1)}\mathrm{r}_c\Big)\dot{L}_b^{(m)}\\
&+ \left(\sigma^{(2)} - \frac{1}{2}\sigma^{(1)}n(\ell^{(2)})\right)\dot{L}_b^{(m)}n^b.
\end{aligned}$$

Inserting (E.10), (E.8), (E.12) and (E.13) and using $P^{bc}\ell_b = -\ell^{(2)}n^c$, (E.23) is established. □

Next we compute the tensors $\mathcal{Q}^{(m)}_{ab}$, $\dot{\mathcal{Q}}^{(m)}_{a}$ and $\ddot{\mathcal{Q}}^{(m)}$ to the leading order. To do that, we apply $\pounds^{(m)}_{\xi}$ to (6.57), for which we need to be able to interchange the operators $\pounds^{(m)}_{\xi}$ and $\nabla_\alpha\nabla_\beta$ acting on functions. This is achieved by applying (5.7) to $A = df$.

**Proposition E.3.** *Let $\xi \in \mathfrak{X}(\mathcal{M})$ and $m \geq 1$ an integer. Then, given any function $f$ the following identity holds*

$$\pounds^{(m)}_{\xi}\nabla_\alpha\nabla_\beta f = \nabla_\alpha\nabla_\beta(\pounds^{(m)}_{\xi}(f)) - \sum_{k=0}^{m-1}\binom{m}{k+1}\nabla_\sigma\Big(\pounds^{(k)}_{\xi}(f)\Big)\Sigma^{(m-k)\sigma}{}_{\beta\alpha}. \tag{E.24}$$

*As a consequence,*

$$\begin{aligned}\pounds^{(m)}_{\xi}\Box_g f = \sum_{i=0}^{m}\binom{m}{i}(\pounds^{(i)}_{\xi}g^{\alpha\beta})\Bigg(&\nabla_\alpha\nabla_\beta\Big(\pounds^{(m-i)}_{\xi}(f)\Big)\\ &-\sum_{k=0}^{m-i-1}\binom{m-i}{k+1}\nabla_\sigma\Big(\pounds^{(k)}_{\xi}(f)\Big)\Sigma^{(m-i-k)\sigma}{}_{\beta\alpha}\Bigg).\end{aligned} \tag{E.25}$$

*Proof.* Using (5.7) with $A = df$ as well as $\pounds_\xi d = d\pounds_\xi$, so that $A^{(k)} = d(\pounds^{(k-1)}_{\xi}(f))$, (E.24) follows. Identity (E.25) follows at once from (5.34) after inserting (E.24), i.e.

$$\pounds^{(m)}_{\xi}\Box_g f = \sum_{i=0}^{m}\binom{m}{i}(\pounds^{(i)}_{\xi}g^{\alpha\beta})\pounds^{(m-i)}_{\xi}\nabla_\alpha\nabla_\beta f.$$

□

Then,

$$\begin{aligned}\frac{1}{\mathfrak{n}-1}\mathcal{Q}^{(m+1)}_{\alpha\beta} &= \nabla_\alpha\nabla_\beta\pounds^{(m)}_{\xi}\Omega - \sum_{k=0}^{m-1}\binom{m}{k+1}\nabla_\sigma(\pounds^{(k)}_{\xi}\Omega)\Sigma^{(m-k)\sigma}{}_{\beta\alpha}\\ &+\sum_{k=0}^{m}\binom{m}{k}(\pounds^{(k)}_{\xi}\Omega)L^{(m-k+1)}_{\alpha\beta}.\end{aligned} \tag{E.26}$$

To perform the calculation it is convenient to introduce the following symmetric tensors (recall that $\Sigma^{\sigma}{}_{\beta\alpha}$ is symmetric in $\alpha,\beta$, so the same holds for $\Sigma^{(k)\sigma}{}_{\beta\alpha}$)

$$\mathrm{I}^{(m+1)}_{\alpha\beta} := \nabla_\alpha\nabla_\beta\pounds^{(m)}_{\xi}\Omega, \qquad \mathrm{II}^{(m+1)}_{\alpha\beta} := -\sum_{k=0}^{m-1}\binom{m}{k+1}\nabla_\sigma(\pounds^{(k)}_{\xi}\Omega)\Sigma^{(m-k)\sigma}{}_{\alpha\beta},$$

$$\mathrm{III}^{(m+1)}_{\alpha\beta} := \sum_{k=0}^{m}\binom{m}{k}(\pounds^{(k)}_{\xi}\Omega)L^{(m-k+1)}_{\alpha\beta}.$$

Given that $\Sigma^{(k)}$ involves at most $m+1$ derivatives of $g$ and that $\Omega = 0$, $\nabla_\mu\Omega \stackrel{\mathscr{I}}{=} \sigma^{(1)}\nu_\mu$ at $\mathscr{I}$, we can write

$$\mathrm{II}^{(m+1)}_{\alpha\beta} \stackrel{[m]}{=} -(\nabla^\sigma\Omega)\Sigma^{(m)}_{\sigma\alpha\beta} - m(\nabla^\sigma\pounds_\xi\Omega)\Sigma^{(m-1)}_{\sigma\alpha\beta} \underset{\mathscr{I}}{\stackrel{[m]}{=}} -\sigma^{(1)}\nu^\sigma\Sigma^{(m)}_{\sigma\alpha\beta} - m(\nabla^\sigma\pounds_\xi\Omega)\Sigma^{(m-1)}_{\sigma\alpha\beta}, \tag{E.27}$$

$$\mathrm{II}^{(m+1)}_{\alpha\beta} \overset{[m+1]}{\underset{\mathscr{I}}{=}} -\sigma^{(1)}\nu^{\sigma}\Sigma^{(m)}_{\sigma\alpha\beta}. \tag{E.28}$$

Similarly, $L^{(m)}$ involves at most $m+1$ derivatives of $g$, so

$$\mathrm{III}^{(m+1)}_{\alpha\beta} \overset{[m]}{=} \Omega L^{(m+1)}_{\alpha\beta} + m(\pounds_{\xi}\Omega)L^{(m)}_{\alpha\beta} + \frac{m(m-1)\pounds^{(2)}_{\xi}\Omega}{2}L^{(m-1)}_{\alpha\beta}, \tag{E.29}$$

$$\mathrm{III}^{(m+1)}_{\alpha\beta} \overset{[m+1]}{=} \Omega L^{(m+1)}_{\alpha\beta} + m(\pounds_{\xi}\Omega)L^{(m)}_{\alpha\beta}. \tag{E.30}$$

Using this and the formulas in Propositions B.1, D.1 and E.1, the computation of $\mathcal{Q}^{(m)}_{ab}$, $\dot{\mathcal{Q}}^{(m)}_{a}$ and $\ddot{\mathcal{Q}}^{(m)}$ will be straightforward. Observe again that $\ddot{\mathcal{Q}}^{(m)}$ is computed to one order less, since this is all that will be needed.

**Proposition E.4.** *Let $\mathcal{Q}^{(m)}_{ab}$, $\dot{\mathcal{Q}}^{(m)}_{a}$ and $\ddot{\mathcal{Q}}^{(m)}$ be as in (6.61) and $m\geq 2$. Then,*

$$\begin{aligned}\mathcal{Q}^{(m+1)}_{ab} \overset{(m)}{=}\ & (\mathfrak{n}-1)\sigma^{(m+1)}\mathrm{U}_{ab} + (\mathfrak{n}-1-2m)\sigma^{(1)}\pounds_{n}\mathrm{Y}^{(m)}_{ab} + 4m\sigma^{(1)}P^{cd}\mathrm{U}_{c(a}\mathrm{Y}^{(m)}_{b)d}\\ & + m\left((\mathfrak{n}-1)(\pounds_{n}\sigma^{(1)}+\sigma^{(1)}\kappa) + (\mathfrak{n}-1-2m)\sigma^{(1)}\kappa - \sigma^{(1)}\operatorname{tr}_P\mathbf{U}\right)\mathrm{Y}^{(m)}_{ab}\\ & + \frac{m}{\mathfrak{n}}\left(\sigma^{(1)}\left(\kappa^{(m+1)} + 2\pounds_{n}(\operatorname{tr}_P\mathbf{Y}^{(m)}) + (2m\kappa+\operatorname{tr}_P\mathbf{U})\operatorname{tr}_P\mathbf{Y}^{(m)} + P^{ab}P^{cd}\mathrm{U}_{ac}\mathrm{Y}^{(m)}_{bd}\right)\gamma_{ab}\right.\\ & \qquad \left. + \frac{m-1}{2}\sigma^{(2)}\kappa^{(m)}\gamma_{ab}\right) - m\sigma^{(1)}(\operatorname{tr}_P\mathbf{Y}^{(m)})\mathrm{U}_{ab} + \widetilde{\mathscr{R}}^{(m)}_{ab},\end{aligned} \tag{E.31}$$

$$\begin{aligned}\dot{\mathcal{Q}}^{(m+1)}_{a} \overset{(m)}{=}\ & (\mathfrak{n}-1)\left(\overset{\circ}{\nabla}_a\sigma^{(m+1)} - \sigma^{(m+1)}(\mathrm{r}-\mathrm{s})_a - mP^{bc}\mathrm{Y}^{(m)}_{ab}\overset{\circ}{\nabla}_c\sigma^{(1)}\right) - (\mathfrak{n}-1-m)\sigma^{(1)}\mathrm{r}^{(m+1)}_{a}\\ & + m\sigma^{(1)}\left(P^{bc}\overset{\circ}{\nabla}_b\mathrm{Y}^{(m)}_{ac} - \overset{\circ}{\nabla}_a(\operatorname{tr}_P\mathbf{Y}^{(m)}) + 2P^{bc}\mathrm{r}_b\mathrm{Y}^{(m)}_{ac} + (\operatorname{tr}_P\mathbf{Y}^{(m)})(\mathrm{r}-\mathrm{s})_a\right)\\ & + \frac{m\sigma^{(1)}}{\mathfrak{n}}\left(\kappa^{(m+1)} + 2\pounds_{n}(\operatorname{tr}_P\mathbf{Y}^{(m)}) + (2m\kappa+\operatorname{tr}_P\mathbf{U})\operatorname{tr}_P\mathbf{Y}^{(m)} + P^{bc}P^{df}\mathrm{U}_{bd}\mathrm{Y}^{(m)}_{cf}\right)\ell_a\\ & + \widetilde{\mathscr{R}}^{(m)}_{a},\end{aligned} \tag{E.32}$$

$$\ddot{\mathcal{Q}}^{(m+1)} \overset{(m+1)}{=} (\mathfrak{n}-1)\sigma^{(m+2)} - m\sigma^{(1)}\operatorname{tr}_P\mathbf{Y}^{(m+1)} + \frac{m\sigma^{(1)}\ell^{(2)}}{\mathfrak{n}}\kappa^{(m+1)}, \tag{E.33}$$

*where $\widetilde{\mathscr{R}}^{(m)}_{ab}$ and $\widetilde{\mathscr{R}}^{(m)}_{a}$ are tensors that depend on $\mathbf{r}^{(m)}$, $\sigma^{(m)}$ and lower order terms and we do not write for simplicity (they can be easily read out by performing all the calculations in the proof explicitly).*

*Proof.* To prove each identity in (E.31)-(E.33) we contract the three pieces $\mathrm{I}^{(m+1)}_{\alpha\beta}$, $\mathrm{II}^{(m+1)}_{\alpha\beta}$ and $\mathrm{III}^{(m+1)}_{\alpha\beta}$ with $e^{\alpha}_a e^{\beta}_b$, $e^{\alpha}_a\xi^{\beta}$ and $\xi^{\alpha}\xi^{\beta}$. The contraction of $\mathrm{I}^{(m+1)}_{\alpha\beta}$ with $e^{\alpha}_a e^{\beta}_b$ gives, after using identity (B.1),

$$\mathrm{I}^{(m+1)}_{ab} = \overset{\circ}{\nabla}_a\overset{\circ}{\nabla}_b\sigma^{(m)} + \pounds_{n}(\sigma^{(m)})\mathrm{Y}_{ab} + \sigma^{(m+1)}\mathrm{U}_{ab}.$$

Contracting (E.27) with $e^{\alpha}_a e^{\beta}_b$, inserting (2.13) and recalling $\Sigma^{(m-1)}_{cab} \overset{(m)}{=} 0$ (by (D.1)) gives

$$e^{\alpha}_a e^{\beta}_b \mathrm{II}^{(m+1)}_{\alpha\beta} \overset{(m)}{=} -\sigma^{(1)}\Sigma^{(m)}_{cab}n^c - m(\pounds_{n}\sigma^{(1)})({}^{(1)}\Sigma^{(m-1)})_{ab}.$$

Inserting (D.17), (D.2) and taking into account (D.5),

$$\mathrm{II}^{(m+1)}_{ab} \overset{(m)}{=} \sigma^{(1)}\pounds_{n}\mathrm{Y}^{(m)}_{ab} + \left(2m\sigma^{(1)}\kappa + m\pounds_{n}\sigma^{(1)}\right)\mathrm{Y}^{(m)}_{ab} - 2\sigma^{(1)}\overset{\circ}{\nabla}_{(a}\mathrm{r}^{(m)}_{b)} + 2\sigma^{(1)}\kappa^{(m)}\mathrm{Y}_{ab}.$$

Finally, the contraction of (E.29) with $e^\alpha_a e^\beta_b$ is (recall $\Omega \stackrel{\mathscr{I}}{=} 0$)

$$\mathrm{III}^{(m+1)}_{ab} \stackrel{(m)}{=} m\sigma^{(1)} L^{(m)}_{ab} + \frac{m(m-1)\sigma^{(2)}}{2} L^{(m-1)}_{ab},$$

where the first term is given by (E.6) and the second by (E.11). Since $\mathcal{Q}^{(m)}_{ab} = (\mathfrak{n}-1)\big(\mathrm{I}^{(m)}_{ab} + \mathrm{II}^{(m)}_{ab} + \mathrm{III}^{(m)}_{ab}\big)$ we simply need to add the three terms and simplify to get (E.31).

To prove (E.32) we repeat the same steps. Firstly, we contract $\mathrm{I}^{(m+1)}_{\alpha\beta}$ with $e^\alpha_a \xi^\beta$ and use (B.2) to get

$$e^\alpha_a \xi^\beta \mathrm{I}^{(m+1)}_{\alpha\beta} = \overset{\circ}{\nabla}_a \sigma^{(m+1)} - \sigma^{(m+1)}(\mathrm{r}-\mathrm{s})_a - V^b{}_a \overset{\circ}{\nabla}_b \sigma^{(m)}. \tag{E.34}$$

Secondly, the contraction of (E.27) with $e^\alpha_a \xi^\beta$ gives, after inserting (2.13) and recalling that $\big({}^{(1,3)}\Sigma^{(m-1)}\big)_a \stackrel{(m)}{=} 0$ (by (D.14) and (D.5)),

$$\begin{aligned} e^\alpha_a \xi^\beta \mathrm{II}^{(m+1)}_{\alpha\beta} &\stackrel{(m)}{=} -\sigma^{(1)}\big({}^{(3)}\Sigma^{(m)}\big)_{ca} n^c - m P^{cd}\big({}^{(3)}\Sigma^{(m-1)}\big)_{ca} \overset{\circ}{\nabla}_d \sigma^{(1)} - m\sigma^{(2)}\big({}^{(3)}\Sigma^{(m-1)}\big)_{ca} n^c \\ &\stackrel{(m)}{=} -\sigma^{(1)} \mathrm{r}^{(m+1)}_a + 2m\sigma^{(1)} P^{bc} \mathrm{r}_b \mathrm{Y}^{(m)}_{ac} - m P^{cd} \mathrm{Y}^{(m)}_{ca} \overset{\circ}{\nabla}_d \sigma^{(1)} \\ &\qquad + \text{terms depending on } \mathbf{r}^{(m)} \text{ and hypersurface data,} \end{aligned}$$

where in the second line we used (D.16) and (D.19). To compute $e^\alpha_a \xi^\beta \mathrm{III}^{(m+1)}_{\alpha\beta}$ we contract (E.29) with $e^\alpha_a \xi^\beta$,

$$e^\alpha_a \xi^\beta \mathrm{III}^{(m+1)}_{\alpha\beta} \stackrel{(m)}{=} m\sigma^{(1)} \overset{\bullet}{L}{}^{(m)}_a + \frac{m(m-1)\sigma^{(2)}}{2} \overset{\bullet}{L}{}^{(m-1)}_a.$$

The explicit forms of $\overset{\bullet}{L}{}^{(m)}_a$ and $\overset{\bullet}{L}{}^{(m-1)}_a$ are given in (E.7) and (E.12). Using as before $\overset{\bullet}{\mathcal{Q}}{}^{(m)}_a = (\mathfrak{n}-1) e^\alpha_a \xi^\beta \big(\mathrm{I}^{(m)}_{\alpha\beta} + \mathrm{II}^{(m)}_{\alpha\beta} + \mathrm{III}^{(m)}_{\alpha\beta}\big)$ we find (E.32) after adding the three terms and simplifying.

The proof of (E.33) is analogous. Contracting $\mathrm{I}^{(m+1)}_{\alpha\beta}$ with $\xi^\alpha \xi^\beta$ and using (B.3) gives

$$\xi^\alpha \xi^\beta \mathrm{I}^{(m+1)}_{\alpha\beta} = \sigma^{(m+2)}.$$

The corresponding term with $\mathrm{II}^{(m+1)}_{\alpha\beta}$ vanishes because the contraction of (E.28) with $\xi^\alpha \xi^\beta$ is

$$\xi^\alpha \xi^\beta \mathrm{II}^{(m+1)}_{\alpha\beta} \stackrel{(m+1)}{=} -\sigma^{(1)}\big({}^{(2,3)}\Sigma^{(m)}\big)_c n^c \stackrel{(m+1)}{=} 0,$$

where in the last equality we used (D.5) and (D.14). Finally, the term with $\mathrm{III}^{(m+1)}_{\alpha\beta}$ is obtained by contracting (E.30) with $\xi^\alpha \xi^\beta$ and inserting (E.8), which gives

$$\xi^\alpha \xi^\beta \mathrm{III}^{(m+1)}_{\alpha\beta} \stackrel{(m+1)}{=} m\sigma^{(1)} \overset{\bullet\bullet}{L}{}^{(m)} \stackrel{(m+1)}{=} \frac{m\sigma^{(1)}}{\mathfrak{n}-1}\left( -\operatorname{tr}_P \mathbf{Y}^{(m+1)} + \frac{\kappa^{(m+1)}}{\mathfrak{n}} \ell^{(2)} \right).$$

The expression $\overset{\bullet\bullet}{\mathcal{Q}}{}^{(m)} = (\mathfrak{n}-1)\xi^\alpha \xi^\beta \big(\mathrm{I}^{(m)}_{\alpha\beta} + \mathrm{II}^{(m)}_{\alpha\beta} + \mathrm{III}^{(m)}_{\alpha\beta}\big)$ yields (E.33) at once. □

As in other cases, the lowest orders $\mathcal{Q}^{(1)}$ and $\mathcal{Q}^{(2)}$ require a specific treatment (note that they are excluded from Proposition E.4 by the condition $m \geq 2$). To obtain the former we simply contract $\mathcal{Q}_{\alpha\beta} \stackrel{\mathscr{I}}{=} (\mathfrak{n}-1)\nabla_\alpha\nabla_\beta\Omega$ with $e^\alpha_a e^\beta_b$, $e^\alpha_a\xi^\beta$ and $\xi^\alpha\xi^\beta$ and use (B.1)-(B.3),

$$\mathcal{Q}^{(1)}_{ab} = \sigma^{(1)}\mathrm{U}_{ab}, \qquad \dot{\mathcal{Q}}^{(1)}_a = \mathring{\nabla}_a\sigma^{(1)} - \sigma^{(1)}(\mathrm{r}-\mathrm{s})_a, \qquad \ddot{\mathcal{Q}}^{(1)} = \sigma^{(2)}. \tag{E.35}$$

To compute the latter we evaluate the exact expression (E.26) with $m=1$ and use $\nabla_\sigma\Omega \stackrel{\mathscr{I}}{=} \sigma^{(1)}\nu_\sigma$ to get

$$\frac{1}{\mathfrak{n}-1}\mathcal{Q}^{(2)}_{\alpha\beta} \stackrel{\mathscr{I}}{=} \nabla_\alpha\nabla_\beta\pounds_\xi\Omega - \sigma^{(1)}\Sigma_{\sigma\alpha\beta}\nu^\sigma + \sigma^{(1)}L_{\alpha\beta}.$$

The contraction of this expression with $e^\alpha_a e^\beta_b$, $e^\alpha_a\xi^\beta$ and $\xi^\alpha\xi^\beta$ can be evaluated from (B.8)-(B.10), (D.33)-(D.35) and (E.15)-(E.17). The result is

$$\begin{aligned}
\mathcal{Q}^{(2)}_{ab} &= (\mathfrak{n}-3)\sigma^{(1)}\pounds_n\mathrm{Y}_{ab} + \Big((\mathfrak{n}-1)(\pounds_n\sigma^{(1)} + \sigma^{(1)}\kappa) + \sigma^{(1)}\Big((\mathfrak{n}-3)\kappa - \operatorname{tr}_P\mathbf{U}\Big)\Big)\mathrm{Y}_{ab}\\
&\quad - (\operatorname{tr}_P\mathbf{Y})\mathrm{U}_{ab} + 2\sigma^{(1)}P^{cd}\mathrm{U}_{d(a}(2\mathrm{Y}_{b)c} + \mathrm{F}_{b)c}) - 2\sigma^{(1)}(\mathfrak{n}-2)\mathring{\nabla}_{(a}\mathrm{r}_{b)} + 2\sigma^{(1)}(2\mathrm{r}_{(a}\mathrm{s}_{b)} - \mathrm{r}_a\mathrm{r}_b)\\
&\quad + \sigma^{(1)}\Big(\mathring{\nabla}_{(a}\mathrm{s}_{b)} - \mathrm{s}_a\mathrm{s}_b + \mathring{R}_{(ab)}\Big) + (\mathfrak{n}-1)\left(\mathring{\nabla}_a\mathring{\nabla}_b\sigma^{(1)} + \left(\sigma^{(2)} - \frac{1}{2}\sigma^{(1)}n(\ell^{(2)})\right)\mathrm{U}_{ab}\right)\\
&\quad - \frac{\sigma^{(1)}R}{2\mathfrak{n}}\gamma_{ab},
\end{aligned} \tag{E.36}$$

$$\begin{aligned}
\dot{\mathcal{Q}}^{(2)}_a &= -(\mathfrak{n}-2)\sigma^{(1)}\mathrm{r}^{(2)}_a + (\mathfrak{n}-1)V^c{}_a(2\sigma^{(1)}\mathrm{r}_c - \mathring{\nabla}_c\sigma^{(1)}) + (\mathfrak{n}-1)\mathring{\nabla}_a\sigma^{(2)}\\
&\quad + \left(\frac{\mathfrak{n}-2}{2}\sigma^{(1)}n(\ell^{(2)}) - (\mathfrak{n}-1)\sigma^{(2)}\right)(\mathrm{r}-\mathrm{s})_a\\
&\quad - \sigma^{(1)}\left(P^{bc}\Big(A_{bca} + (\mathrm{r}+\mathrm{s})_b(\mathrm{Y}+\mathrm{F})_{ac}\Big) + \frac{R}{2\mathfrak{n}}\ell_a - \frac{1}{2}\kappa\mathring{\nabla}_a\ell^{(2)}\right),
\end{aligned} \tag{E.37}$$

$$\begin{aligned}
\ddot{\mathcal{Q}}^{(2)} &= (\mathfrak{n}-1)\sigma^{(2)} - \sigma^{(1)}\operatorname{tr}_P\mathbf{Y}^{(2)} + \sigma^{(1)}P^{ab}P^{cd}(\mathrm{Y}+\mathrm{F})_{ac}(\mathrm{Y}+\mathrm{F})_{bd} + \sigma^{(1)}P(\mathbf{r}-\mathbf{s}, d\ell^{(2)})\\
&\quad - \frac{R}{2\mathfrak{n}}\sigma^{(1)}\ell^{(2)},
\end{aligned} \tag{E.38}$$

where $R$ is given by (6.54) and needs not be written out explicitly.

We conclude this appendix by computing the scalars $f^{(m+1)}$, obtained after applying $\pounds^{(m)}_\xi$ to $f = g^{\alpha\beta}\nabla_\alpha\Omega\nabla_\beta\Omega$ and using (5.34),

$$\pounds^{(m)}_\xi f = \sum_{k=0}^{m}\binom{m}{k}(\pounds^{(k)}_\xi g^{\alpha\beta})\sum_{j=0}^{m-k}\binom{m-k}{j}\nabla_\alpha(\pounds^{(j)}_\xi\Omega)\nabla_\beta\Big(\pounds^{(m-k-j)}_\xi\Omega\Big). \tag{E.39}$$

Again, the lowest orders $f^{(1)}$, $f^{(2)}$ and $f^{(3)}$ need a separate treatment. Obviously $f^{(1)} = 0$ because $\mathscr{I}$ is null. To compute $f^{(2)}$ we particularize (E.39) to $m=1$ so that

$$\pounds_\xi f = (\pounds_\xi g^{\alpha\beta})\nabla_\alpha\Omega\nabla_\beta\Omega + 2\nabla^\alpha\Omega\nabla_\alpha\pounds_\xi\Omega.$$

The first term is $(\pounds_\xi g^{\alpha\beta})\nabla_\alpha\Omega\nabla_\beta\Omega = 2(\sigma^{(1)})^2\kappa$ (by (E.43) and $\Omega \stackrel{\mathscr{I}}{=} 0$) and the second term is $2\sigma^{(1)}\pounds_n\sigma^{(1)}$ because $\nabla^\alpha\Omega \stackrel{\mathscr{I}}{=} \sigma^{(1)}\nu^\alpha$. Thus,

$$f^{(2)} = 2\sigma^{(1)}(\sigma^{(1)}\kappa + \pounds_n\sigma^{(1)}). \tag{E.40}$$

We now compute $f^{(3)}$. The expanded form of (E.39) for $m=2$ is

$$f^{(3)} = 2\nabla^\alpha\Omega\nabla_\alpha(\pounds^{(2)}_\xi\Omega) + 2g^{\alpha\beta}\nabla_\alpha(\pounds_\xi\Omega)\nabla_\beta(\pounds_\xi\Omega) + 4(\pounds_\xi g^{\alpha\beta})\nabla_\alpha(\pounds_\xi\Omega)\nabla_\beta\Omega \\ + (\pounds^{(2)}_\xi g^{\alpha\beta})\nabla_\alpha\Omega\nabla_\beta\Omega.$$

Now, using $\nabla^\alpha\Omega \stackrel{\mathscr{I}}{=} \sigma^{(1)}\nu^\alpha$, together with (E.43) and

$$g^{\alpha\beta}\nabla_\alpha(\pounds_\xi\Omega)\nabla_\beta(\pounds_\xi\Omega) \underset{(2.13)}{\stackrel{\mathscr{I}}{=}} P^{ab}\mathring{\nabla}_a\sigma^{(1)}\mathring{\nabla}_b\sigma^{(1)} + 2\sigma^{(2)}\pounds_n\sigma^{(1)},$$

$$(\pounds^{(2)}_\xi g^{\alpha\beta})\nabla_\alpha\Omega\nabla_\beta\Omega \stackrel{\mathscr{I}}{=} (\sigma^{(1)})^2\nu^\alpha\nu^\beta\Big(2g^{\mu\nu}\mathcal{K}_{\alpha\mu}\mathcal{K}_{\beta\nu} - \mathcal{K}^{(2)}_{\alpha\beta}\Big) \\ \underset{(2.13),(5.45)}{\stackrel{\mathscr{I}}{=}} (\sigma^{(1)})^2\Big(8P(\mathbf{r},\mathbf{r}) - 4\kappa n(\ell^{(2)}) + 2\kappa^{(2)}\Big),$$

it follows that

$$f^{(3)} = 2\sigma^{(1)}\Big(\sigma^{(1)}\kappa^{(2)} + \pounds_n\sigma^{(2)}\Big) + 2\Big(2\sigma^{(2)} - \sigma^{(1)}n(\ell^{(2)})\Big)\Big(\pounds_n\sigma^{(1)} + 2\sigma^{(1)}\kappa\Big) \\ + 8\sigma^{(1)}P\Big(\mathbf{r}, \sigma^{(1)}\mathbf{r} - d\sigma^{(1)}\Big) + 2P^{ab}\mathring{\nabla}_a\sigma^{(1)}\mathring{\nabla}_b\sigma^{(1)}. \tag{E.41}$$

We now deal with the rest of the cases in the following proposition.

**Proposition E.5.** *Let $f := g^{\alpha\beta}\nabla_\alpha\Omega\nabla_\beta\Omega$. Then, for any $m \geq 2$,*

$$f^{(m+1)} \stackrel{(m)}{=} 2\sigma^{(1)}\Big(\pounds_n\sigma^{(m)} + \sigma^{(1)}\kappa^{(m)}\Big) + 2m\Big(\pounds_n\sigma^{(1)} + 2\sigma^{(1)}\kappa\Big)\sigma^{(m)}. \tag{E.42}$$

*Proof.* The case $m=2$ is (E.41), which after ignoring terms of order one and zero becomes (E.42). For $m \geq 3$, the only terms in (E.39) that have a chance to depend on $\mathbf{Y}^{(m)}$ or $\sigma^{(m)}$ are $k=0$ (and $j=0,1,m-1,m$), $k=1$ (and $j=0,m-1$) and $k=m$, namely

$$f^{(m+1)} \stackrel{[m]}{=} \nabla^\beta\Omega\nabla_\beta(\pounds^{(m)}_\xi\Omega) + 2mg^{\alpha\beta}\nabla_\alpha(\pounds_\xi\Omega)\nabla_\beta(\pounds^{(m-1)}_\xi\Omega) \\ + 2m(\pounds_\xi g^{\alpha\beta})\nabla_\alpha(\pounds^{(m-1)}_\xi\Omega)\nabla_\beta\Omega + (\pounds^{(m)}_\xi g^{\alpha\beta})\nabla_\alpha\Omega\nabla_\beta\Omega,$$

where we used the fact that $m \geq 3$ because we have used that the terms with $k=0$ and $j=0,1,m-1,m$ are all different. When $m=2$ we would be overcounting (but it turns out that the extra factor compensates, so (E.42) is valid also for $m=2$). Now one uses $\nabla^\beta\Omega \stackrel{\mathscr{I}}{=} \sigma^{(1)}\nu^\beta$ in the first term, $\nabla^\alpha(\pounds^{(m-1)}_\xi\Omega) \stackrel{(m)}{=} \sigma^{(m)}\nu^\alpha$ in the second, and in the third and fourth

$$(\pounds_\xi g^{\alpha\beta})\nabla_\beta\Omega \stackrel{\mathscr{I}}{=} -\sigma^{(1)}\mathcal{K}^\alpha{}_\beta\nu^\beta \stackrel{(5.45)}{=} -\sigma^{(1)}\left(2P^{cd}\mathrm{r}_c + \frac{1}{2}n(\ell^{(2)})n^d\right)e^\alpha_d + 2\sigma^{(1)}\kappa\xi^\alpha, \tag{E.43}$$

$$(\pounds^{(m)}_\xi g^{\alpha\beta})\nabla_\alpha\Omega\nabla_\beta\Omega \stackrel{\mathscr{I}}{=} (\sigma^{(1)})^2(\pounds^{(m)}_\xi g^{\alpha\beta})\nu_\alpha\nu_\beta \stackrel{(m)}{=} -(\sigma^{(1)})^2\mathcal{K}^{(m)}_{ab}n^an^b = 2(\sigma^{(1)})^2\kappa^{(m)}.$$

After adding up the four terms one gets

$$f^{(m+1)} \stackrel{(m)}{=} 2\sigma^{(1)}\pounds_n\sigma^{(m)} + 2m(\pounds_n\sigma^{(1)})\sigma^{(m)} + 4m\sigma^{(1)}\kappa\sigma^{(m)} + 2(\sigma^{(1)})^2\kappa^{(m)},$$

which is (E.42) for $m \geq 3$. □